\documentclass[11pt,lot,lof]{puthesis}
\newcommand{\proquestmode}{}

\title{Plasma Physics in Strong-Field Regimes}

\submitted{September, 2018}  
\copyrightyear{2018}  
\author{Yuan Shi}
\adviser{Professor Nathaniel J. Fisch \\ and Professor Hong Qin}  
\department{Astrophysical Sciences \\Program in Plasma Physics}

\usepackage{graphicx}

\usepackage{verbatim}

\usepackage{multirow}
\usepackage{longtable}

\usepackage{booktabs}

\usepackage[round]{natbib}
\usepackage{bm}            
\DeclareMathAlphabet\mathbfcal{OMS}{cmsy}{b}{n}

\usepackage{siunitx}

\usepackage[cmex10]{amsmath}
\usepackage{amssymb}
\usepackage{mathrsfs}
\usepackage{mathtools}
\usepackage{amsthm}

\usepackage[usenames, dvipsnames]{color}
\usepackage{xcolor}
\usepackage{multirow}
\usepackage{makecell}

\usepackage{simplewick}

\usepackage{feynmp-auto}
\newcommand{\marrow}[5]{%
	\fmfcmd{style_def marrow#1
		expr p = drawarrow subpath (1/4, 3/4) of p shifted 6 #2 withpen pencircle scaled 0.4;
		label.#3(btex #4 etex, point 0.5 of p shifted 6 #2);
		enddef;}
	\fmf{marrow#1,tension=0}{#5}}

\usepackage{lettrine}
\usepackage[toc]{appendix}

\ifdefined\printmode

	\usepackage{url}

\else

	\ifdefined\proquestmode
		
		\usepackage[colorlinks=true,linkcolor=blue,citecolor=blue,
					bookmarks=true,
					bookmarksnumbered=true,
					bookmarksopen=false]{hyperref}
		\usepackage{bookmark}
		
		\usepackage[margin=1in]{geometry}
		
		\makeatletter
		\hypersetup{pdftitle=\@title,pdfauthor=\@author}
		\makeatother
	
	\else
		
		\usepackage{hyperref}
		\hypersetup{colorlinks,bookmarksnumbered}
		
		\makeatletter
		\hypersetup{pdftitle=\@title,pdfauthor=\@author}
		\makeatother
		
	\fi 
	
\fi 

\ifodd 0
\renewcommand{\maketitlepage}{}
\renewcommand*{\makecopyrightpage}{}
\renewcommand*{\makeabstract}{}

\else

\abstract{
In strong electromagnetic fields, new plasma phenomena and applications emerge, whose modeling requires analytical theories and numerical schemes that I will develop in this thesis. Based on my new results of the classical plasma model, the role of strong magnetic fields during laser-plasma interactions can now be understood.  Moreover, based my new quantum electrodynamics (QED) models for plasmas, it is now possible to understand strong-field QED effects in astrophysical environments and test them in laboratory settings.

In the classical regime, starting from megagauss magnetic fields, scattering of optical lasers becomes manifestly anisotropic. For the first time, a convenient formula for three-wave coupling coefficient in arbitrary geometry is obtained and evaluated. By solving the fluid model to the second order, I provide an alternative perspective of parametric instability and unveil how magnetic fields affect collective scattering of plasma waves. As an application, I predict that magnetic resonances can be utilized to mediate laser pulse compression. Using magnetized plasma mediation, it is not only possible to achieve higher output intensity for optical lasers with more engineering flexibility, but also possible to compress UV and soft X-ray pulses that cannot be compressed using existing techniques. Taking advantage of the emerging feasibility of strong magnetic fields, I have thus identified a pathway to next-generation powerful lasers, whose viability is supported by particle-in-cell simulations.

In even stronger magnetic fields or intense laser fields, 
relativistic quantum effects become important. At that point, plasma models based on QED are necessary. Allowing for nontrivial background fields, I develop a new formalism for QED plasmas by computing the effective action using path integrals. 
My new formalism enables simple wave dispersion relations in strongly magnetized plasmas to be obtained for the first time, based on which the modified Faraday rotation and the anharmonic cyclotron absorptions near X-ray pulsars can now be correctly interpreted. Beyond the perturbative regime, I extend real-time lattice QED to a unique tool for plasma physics, especially when collective scales overlap with relativistic-quantum scales. Applying this numerical tool to laser-plasma interactions, I demonstrate, for the first time, transition from wakefield acceleration to electron-positron pair production, when the laser fields exceed the Schwinger threshold.

}

\acknowledgements{
First and foremost, I would like to extend thanks to my advisors Nathaniel J. Fisch and Hong Qin. I am deeply indebted to Nat and Hong, not only for advising and guiding me in research and teaching, but also for encouraging me to explore other branches of physics. Their open-mindedness enabled me to find synergy between quantum field theory and plasma physics, and do research on topics that are unimaginable anywhere else in the world. I am forever thankful for the excellent role models they have provided as successful physicists, professors, and mentors.

I am deeply grateful for my collaborators Qing Jia, who carried out simulations of magnetized laser pulse compression, and Jianyuan Xiao, who coded and performed simulations of scalar-QED plasmas. It has been a privilege to work with such a talented peer group. Discussions with Qing, Jianyuan, as well as Matthew Edwards, Kenan Qu, Sebastian Meuren, and Wolf Malkin have always been inspiring and fruitful.

For this dissertation, I would like to acknowledge my readers, Ilya Dodin and Edward Startsev, for their time and constructive feedbacks. I am thankful to members of my thesis committee: Ilya Dodin, Allan Reiman and Julia Mikhailova for their time, guidance, and insightful questions.

Many people have read the work presented in this dissertation and generously provided valuable comments. I would like to extend my sincere thanks to Ilya Dodin, Daniel Ruiz, Sebastian Meuren, Alex Glasser, Ben Israeli, Xinyu Li, and those who attended my seminars and talks at the Princeton Plasma Physics Laboratory, APS DPP Meetings, Columbia University, and in Matt Kunz's journal club. As some of the chapters were originally written as articles for various journals, I would also like to thank the anonymous referees for their constructive feedback.

I would like to thank the funding agencies, without which my work would not have been possible. My research is supported by NNSA Grant No. DE-NA0002948, AFOSR Grant No. FA9550-15-1-0391, and DOE Research Grant No. DEAC02-09CH11466.

My experience working with William Tang's group for my second-year project has immensely contributed to my professional growth. I am especially appreciative of Bei Wang for teaching me the best programming practice in modern scientific computing.  

I am also grateful for the opportunity to work with Yevgeny Raitses and Ahmed Diallo for my first-year experimental project. They not only taught me what it takes to be a good experimentalist, but also made the fruitful excursion full of joy. I would like to thanks St\'ephane Mazouffre, Alex Merzhevskiy, Enrique Merino, Scott Keller, Leland Ellison, and Lahib Balika who helped me with my experiments.

Besides these formal projects, I also had the pleasure to collaborate with Alex Glasser, Jeff Lestz, Noah Mandell, Ian Ochs, and Denis St. Onge on the software \textsc{EMOOPIC}. We impressed Prof.~James Stone together, writing a fully functional PIC code in one semester.

Throughout my time at Princeton, I very much appreciate the guidance from my academic advisor, John Krommes, who spent time to meet with me listening to my concerns. John taught me to build my foundation deep before reaching high, as is perfectly exemplified through his Irreversible classes. 

I also wish to thank Barbara Sarfaty and Beth Leman for keeping the Ph.D. program organized, arranging departmental gatherings, and providing solid recommendations for local attractions. I have fond memories at the Grounds for Sculpture.

My time at Princeton would not have been so cheerful without the camaraderie with my fellow students. To Jacob Schwartz and Charles Swanson, for being fantastic roommates and officemates. To Daniel Ruiz, for inspirational discussions we had about Melrose's books. To fellow students in my cohort, Jonathan Ng, Peter Bogert, Lee Gunderson, Daniel Ruiz, Jacob Schwartz, and Charles Swanson, for the moments, hours, and even nights we studied together for Prelims and Generals. 
To Vasily Geyko and Chang Liu, whose passed their experience as graduate AIs to me, and to my students in GPPI, for challenging me to go the extra mile and helping me become a better teacher. To Lei Shi, Yao Zhou, Chang Liu, Qian Teng, Ge Dong, and Hongxuan Zhu, for all the intellectual and emotional support; I will certainly miss the delicious meals at all of our traditional Chinese holiday gatherings. 

I would like to thank Mike Campanell, who passed the GSG Assembly seat to me. I had fun working with Joshua Wallace, Rep. of the astrophysics department, setting up the first joint PPPL-Astro barbecue, after many summer barbecues organized by Jeff Parker, Eric Shi, and fellow committee members. I would also like to thank Brian Kraus, who succeeded me on the GSG seat, and did a great job in keeping plasma students connected.

I would also like to express my gratitude to Sarah Schwarz, Sandra Moskovitz, and the staff and fellows at the McGraw Center for Teaching and Learning. The McGraw programs allow me not only to improve my teaching skills, but also to present materials in a more approachable manner to a wider audience.

It was a great pleasure working with Stephanie Whetstone at Princeton Writes, who helped me prepare my invited talk at APS. Special thanks to Susanne Killian from Career Service, who helped me navigate the daunting job hunting process. Over the years, I have benefited a lot from the Princeton Writing Program. Thanks to Judith Swan, Allyson Sgro,
Sarah Budischak, Anie van Leeuwen, and Reka Daniel-Weiner for inspiring workshops and the generous spending of their time on helping me improve my writing.

I must also thank my family and friends for their support, encouragement, and company during my Ph.D. years. I am especially grateful for my parents, Lihui Feng and Xiangwu Shi, for their unconditional love and care. 
My mom is the first professor I have ever known, and my first teacher in math and science. Her intellectual curiosity, rigorous logic and work ethic have tremendously influenced my life, my view of the world, and my approach to scientific research. 
My dad is the most persevering person I have ever known. Transitioning from a career in Chinese history to one in civil law after a decade of self-studying in his spare time, my dad set an example for me to never give up in the pursuit of my dreams. 
I must also thank my parents-in-law, Lili Sun and Dongxiao Zhang, for career advice, gardening tips, and most importantly, marrying Mengru to me.

Last but not least, I want to thank my loving wife Mengru, who always jokes that I love physics more than I love her. Maybe it is true that I spend more time with physics, but she is the reason why physics can never be perfect. There is no force in physics that can explain why she followed me from Hong Kong to New England, and why I followed her from New Jersey to Connecticut. It is beyond the scope of physics why simply looking at her smiling face is sufficient to make everything else in the universe ignorable.

}

\dedication{To my parents, Lihui Feng and Xiangwu Shi\\ And to my wife, Mengru Zhang}

\fi  

\begin{document}

\makefrontmatter


\chapter{Introduction\label{ch:intro}}

\section{Motivations\label{ch:motivation}}

In strong magnetic fields or in intense laser fields, new plasma phenomena and applications emerge. Already in megagauss magnetic fields, scattering of optical lasers becomes manifestly anisotropic.     
The anisotropy is introduced by the magnetic field, which is important when the electron gyrofrequency $\Omega_e=eB/m_e$ is no longer ignorable when compared to the laser frequency and the plasma frequency. For example, a magnetic field $\sim10$ MG, corresponding to $\Omega_e\hbar\sim0.1$ eV, will noticeably alter the wave dispersion relation and the scattering cross section of optical lasers in plasmas. In low-density plasmas, the role of the strong magnetic field is largely classical. However, as plasma density increases, quantum effects may emerge when the characteristic size of electron wave function $r_0=\sqrt{2\hbar/eB}$ becomes comparable to inter-particle spacing. For example, a magnetic field $\sim10$ MG, corresponding to the magnetic de Broglie wavelength $r_0\sim 1$ nm, may already allow electrons to feel the Fermi degeneracy in solid-density plasmas. As the field strength further increases towards the Schwinger limit \mbox{$B_c\sim10^{13}$ G}, where the magnetic de Broglie wavelength shrinks to electron Compton wavelength, relativistic-quantum effects due to magnetic fields become increasingly prominent.

While strong magnetic fields on the order of Schwinger limit can only be found near compact astrophysical objects such as neutron stars, megagauss to gigagauss magnetic fields can already be produced by a number of laboratory techniques. For example, using lasers to drive plasma implosions, seed magnetic fields, either self-generated \citep{Igumenshchev14} or externally imposed \citep{Gotchev09,Knauer10}, can be amplified to tens of megagauss by magnetic flux compression. A more controllable technique produces magnetic fields of similar strengths using lasers to drive capacitor-coil targets \citep{Fujioka13,Santos15,Goyon17,Tikhonchuk2017quasistationary}. Comparable or even stronger magnetic fields can be produced by dynamo effects when solid targets are directly ablated by intense laser pulses \citep{Borghesi98,Tatarakis02,Tatarakis02Nature,Wagner04,Manuel12,Gao12,Gao15}. Using these techniques, magnetic fields may be further intensified by employing stronger drive lasers. The emerging availability of very strong magnetic fields in laboratory settings thus presents new challenges and opportunities that remain to be investigated. 

In the classical regime, the next-to-simplest phenomena in magnetized plasmas is perhaps coherent three-wave interactions. These interactions happen, for example, in magnetic confinement devices, where waves injected by antenna arrays decay to other waves \citep{Chang74,Liu86}. In the case where the wave is injected to drive current in a tokamak \citep{Fisch78,Fisch87}, there is a possibility that the lower hybrid current drive is affected by unwanted decays near the tokamak periphery \citep{Porkolab77,Cesario06}. Moreover, three-wave scattering also happens, for example, in laser implosion experiments \citep{Myatt13}, where high-intensity lasers interact with plasmas. During magnetized implosions, where the magnetic field is imposed to enhance particle confinement \citep{Gotchev09,Hohenberger12,Slutz12,Wang15,Farmer2017simulation,Barnak17}, multiple laser beams may scatter and reflect from one another via magnetic resonances. 
Understanding three-wave interactions is therefore of critical importance for avoiding deleterious effects and designing successful magnetized laser implosion experiments.

Despite of its importance, coherent three-wave scattering, well-studied in unmagnetized plasma \citep{Davidson72,Weiland77}, remains poorly understood when plasmas become magnetized. 
This situation is mostly due to the analytical difficulty when external magnetic fields are present, which substantially complicates the equations, except in the simple geometry where the participating waves are collimated either parallel or perpendicular to the magnetic field. 
Such difficulty deserves to be overcome in the midst of recent developments in strong magnetic field technologies. 
Since multiple laser beams usually propagate at angles to one another and with the magnetic field during laser-driven implosions, understanding the angular dependence of three-wave scattering in magnetized plasma becomes indispensable for making a knowledgeable choice of the experimental setups to optimize laser-plasma coupling.

In the simple collimated geometry, three kinds of theories have been developed for three-wave interactions in magnetized plasmas. 
The first kind is coupled-mode theory, which searches for normal modes of the nonlinear equations \citep{Sjolund67,Shivamoggi82}. 
The normal modes are typically linear combinations of fluctuating quantities, and the equations satisfied by normal modes are formally simple. However, these equations hide the complexity of the nonlinear problem inside their complicated coupling coefficients, from which little physical meaning has been extracted.
The second kind is nonlinear current theory, which describes three-wave parametric interaction by adding a nonlinear source term into the Maxwell's equations. 
To find the nonlinear current, the typical method is to add a constant pump wave to the equations as a background field. On top of this background, response to perturbations due to the two decay waves are expressed in terms of the coupling tensor. The coupling tensor combined with the dielectric tensor then gives an matrix equation for the decay waves, from which the nonlinear dispersion relation can be obtained. By expanding the dispersion relation near either the eigenmode or quasimode frequencies, the linear growth rate can be obtained. 
Using fluid models for nonlinear sources, parametric growth rates have been obtained for extraordinary wave pump \citep{Grebogi80,Barr84,Vyas16}, lower hybrid wave pump \citep{Sanuki77}, as well as the right-handed and left-handed circularly polarized pumps \citep{Laham98}. To capture thermal effects, a simple treatment retains only thermal corrections to the dielectric tensor \citep{Platzman68}. A more complete treatment also include thermal corrections to the coupling tensor \citep{Ram82,Aleksandrov1978principles,Boyd85}. However, beyond the simple collimated geometry, such treatment becomes so cumbersome that decades of efforts have been spent on just simplifying the expressions \citep{Stenflo70,Stenflo94,Brodin12}, with very little extractable physical results \citep{Larsson76,Stenflo04}.
Beside the coupled mode theory and the nonlinear current theory, the third kind of theory uses Lagrangian formulation. In this more systematic approach, the interaction Lagrangian is obtained either from the Low's Lagrangian \citep{Galloway71,Boyd78}, or the oscillation-center Lagrangian \citep{Dodin17} by expanding plasma response to the third order. Even with such transparent formalism, three-wave interactions in magnetized plasma, where the waves are not collimated, remains to be analyzed systematically, in generality, and in details as will be done in this thesis.

Understanding three-wave interactions in magnetized plasmas not only has implications for contemporary fusion experiments, but also enables developments of next-generation lasers.
Powerful laser pulses of high intensity, high frequency, and short duration are demanded in many applications such as inertial confinement fusion \citep{Keefe82,Lindl92,Lindl95} and single molecule imaging \citep{Neutze00,Hau-Riege07}. However, pulses generated directly from laser sources cannot meet all three requirements simultaneously. Therefore, techniques for post-processing laser pulses are necessary, which improve the pulses by amplifying their intensity \citep{Sethian95,Moses05}, converting their frequency \citep{Franken61,Bloembergen62,McPherson87,Tarasevitch07}, or shortening their duration \citep{Maine88,Milroy79,Capjack82,Guzdar96}. Industrial pulse processors use either solid or gas as gain medium, so they can only handle limited intensities before these media break down \citep{Wegner99,Canova07}. Moreover, they can only process limited frequencies up to the soft UV range before media become opaque due to single-photon ionization. While the intensity limit has been substantially increased by current research using unmagnetized plasmas as the gain medium \citep{Malkin99,Weber13,Edwards16,Edwards17}, the frequency limit remains in the UV range \citep{Clark03PhD,Malkin07,Malkin14}. Increasing the frequency limit is difficult, because higher frequencies require denser plasmas, in which deleterious effects dominate desirable processes. 
Fortunately, the requisite density can be reduced if we use magnetized plasmas instead. 
Taking advantage of the emerging feasibility of very strong magnetic fields, we can use magnetized plasmas to push the limit of laser pulse processing beyond what is possible by currently envisioned methods, as I will show in this thesis.

Outside the classical regime, the standard plasma theory, which describes plasmas as collections of point particles moving in self-consistent electromagnetic (EM) fields, become deficient. 
The conditions at which relativistic quantum effects become important may be estimated by comparing three energy scales: the energy scales of the plasmas, the energy scales of the EM fields, and the rest energy of charged particles. The energy scales of plasmas are the thermal energy $k_{\text{B}}T$, the Fermi energy $\epsilon_\text{F}$, and the plasmon energy $\hbar\omega_p$. The energy scales of wave fields are the photon energy $\hbar\omega$ and the ponderomotive energy $U_p$. The energy scales of static electric and magnetic fields are $\epsilon_\text{E}=\sqrt{eEc\hbar}$ and $\epsilon_{\text{B}}=\sqrt{eBc^2\hbar}$, respectively. Relativistic effects are important when the energy scales of either the plasmas or the EM fields are comparable to the rest energy of charged particles. Quantization effects are important when the thermal energy is low compared to other energy scales.

An example where relativistic and quantum effects are both important is the magnetosphere of an X-ray pulsar \citep{Meszaros92}. The typical magnetic fields of X-ray pulsars are on the order $B\sim10^{12}$ G. The corresponding magnetic energy $\epsilon_\text{B}\sim100$ keV is comparable to the rest energy of electrons $m_ec^2\approx511$ keV, indicating that relativistic effects are important. Moreover, the effective temperature of X-ray pulsars $k_{\text{B}}T\sim10$ keV is colder than $\epsilon_\text{B}$, indicating that quantum effects are also important. That relativistic quantum effects are both important, an inference made by comparing energy scales, is strongly supported by anharmonic cyclotron absorption features observed in spectra of X-ray pulsars \citep{Makishima90,Heindl99,Santangelo99,Heindl00,Pottschmidt05,Tsygankov06,Tsygankov07}. Since classical plasma theories cannot explain these spectral features, the presence of high-order harmonics is attributed to inelastic scatterings of photons by electrons that occupy quantized Landau levels \citep{Harding06}, and the anharmonicity is attributed to viewing geometry as well as relativistic effects \citep{Schonherr07,Nishimura13}. Despite numerous efforts, many features of cyclotron absorption lines remain to be explained \citep{Meszaros85,Freeman99,Bignami03,Schwarm17II}. The locations and shapes of these lines contain important information such as the magnetic field and plasma density of magnetospheres of X-ray pulsars. This information cannot be extracted, unless wave dispersion relations, which enter the radiation transport equations \citep{Meszaros92} that serve as the forward model in the retrieval problem \citep{Rodgers00}, are obtained for strongly magnetized plasmas. In this thesis, I will obtain, for the first time, explicit and convenient expressions of wave dispersion relations in strongly magnetized plasmas.

Another place where relativistic and quantum effects are both important is a plasma produced by ultra intense lasers interacting with a solid target. 
In such a plasma, energetic particles are produced and hard radiations are emitted. When the energy density is high enough, genuine relativistic-quantum effects such as electron-positron pairs production can happen \citep{Liang98,Gahn00,Liang15,Sarri15}. 
Even without pair production, quasistatic magnetic fields in the gigagauss range can be produced during the interactions \citep{Stamper91,Korneev15}. The corresponding magnetic energy $\epsilon_\text{B}\sim 1$ keV is comparable to the electron temperature of the plasma, indicating that quantum effects are important. Relativistic effects also turn out to be important when optical lasers are used to diagnose the plasma. This is because the frequencies of optical photons are close to wave cutoffs, if the plasma has density $n_0\sim 10^{21}\hspace{3pt}\text{cm}^{-3}$, corresponding to $\hbar\omega_p\sim 1$ eV. Due to singularities near cutoffs, small modifications of the cutoff frequencies can have large effects. Such effects have been revealed in a number of experiments \citep{Tatarakis02,Wagner04}. In these experiments, it is found that the magnetic field, determined from classical formulas, is larger when the same plasma is diagnosed by lasers with higher frequencies. This peculiar dependence of the inferred magnetic field strength on the frequencies of the diagnostic lasers indicates that systematic errors exist in classical formulas. These systematic errors can be removed only by carefully calculating how waves propagate in strongly magnetized plasmas as will be done in this thesis.

The question now is how should we model relativistic quantum plasmas? One possibility is to employ semiclassical approximations. However, semiclassical descriptions of relativistic quantum plasmas can only capture effects that are presumed to be important. For example, to model laser pair production, source terms can be inserted into classical plasma equations \citep{Berezhiani92,Kluger98,Schmidt98,Roberts02,Hebenstreit10}. 
Although effective theories of this kind can work in some situations, they have two major drawbacks. 
The first is a lack of self-consistency. For example, while it seems easy to add source terms for particles, it is far from obvious what terms should be added to the Maxwell's equations to make sure energy and momentum are both conserved during laser pair production process. 
The second major problem is the narrow range of model applicability. For example, when modeling laser pair production, it is only in a very narrow range in the parameters space that other effects such as bremsstrahlung and radiation back reaction might be ignorable. These major drawbacks make effective models, built patch by patch, somewhat unappealing.

Of course, relativistic quantum plasma theories need not be built by adding terms to equations familiar to plasma physicist \citep{Eliasson11,Bonitz2016quantum,Shi16QED}. 
It is important to remember that the now-standard plasma physics is just a semiclassical approximation of the relativistic quantum world \citep{Ruiz15}. Therefore, instead of picking up terms that were left behind by plasma physics, another approach is to start from the fundamental theory that contains everything and then add collective plasma effects. The later approach is what I will take in this thesis.
Beyond the immediate goal of establishing a formalism capable of describing relativistic
quantum plasmas, the goal of this thesis is to demonstrate that quantum field theory, in whose language the standard model of particle physics is written, and in whose language many phenomena in condensed matter physics are explained, is also a useful language for plasma physics. Since particle physics describes a few particles with high energy, condensed matter physics describes many particles with low energy, and plasma physics describes intermediate number of particles with intermediate energy, it should not be surprising that a language that is effective for both extremes is also effective in the intermediate regime. In this way, plasma physics is reconnected with other branches of physics, after a long period of isolated developments.
\section{Overview\label{ch:overview}}

In the first part of this thesis, I overcome the analytical difficulty in cold-fluid theory and obtain three-wave coupling coefficient in uniformly magnetized plasmas in the most general geometry. 
This is achieved by systematically solving the fluid-Maxwell system to second order in fluctuations, where secular terms are removed using a multiscale expansion. 
Multiscale expansion as a perturbative method is introduced in Ch.~\ref{ch:multiscale}, using ordinary differential equations as examples. When solving nonlinear differential equations, the multiscale method expands space and time scales, in addition to expanding fluctuations. These extra degrees of freedom exempt perturbative solutions from secular behaviors, whose occurrence would lead to indefinite growths that invalidate perturbative assumptions.

Applying the multiscale method to fluid-Maxwell equations, the resultant first-order equations are the usual linearized equations well known in plasma physics. In Ch.~\ref{ch:fluid-1st}, I review these familiar equations from the novel perspective of linear operators, which become necessary when solving the second-order equations. In particular, I introduce a forcing operator for magnetized plasmas, in terms of which the dispersion operator and the wave energy operator can be expressed. The forcing operator is proportional to the usual linear susceptibility, but only keeps the irreducible particle dynamics in magnetic fields. A number of highly nontrivial identities of the forcing operator will become particularly useful when solving the equations to higher orders.

Using the forcing operator, the fluid-Maxwell equations are solved to second order in the multiscale series in Ch.~\ref{ch:fluid-2nd}. Similar to the first-order electric-field equation, which can be used to determine linear wave dispersion relations, the second-order electric-field equation can be used to determine nonlinear three-wave interactions in magnetized plasmas. In terms of linear operators, the second-order electric-field equation can be written in a very compact and intuitive form: due to three-wave scattering, quasimodes are generated by beating of linear eigenmodes, and energy is redistributed among resonant waves.  
In the case where only three eigenmodes resonantly interact, the second-order electric-field equation can be reduced to the well-known three-wave amplitude equations. What is of critical importance is that during this process, a previously unknown formula for the three-wave coupling coefficient is obtained in magnetized plasmas, which can be readily evaluated in arbitrary geometry.
The general formula, derived laboriously by solving equations, becomes transparent when reformulated from the Lagrangian perspective. 

Having obtained the coupling coefficient, the behaviors of three resonantly interacting waves can be determined by solving the three-wave equations. The three waves can propagate at arbitrary angles with respect to one another and the background magnetic field, as long as the three-wave resonance conditions are satisfied.
Before the three waves overlap in spacetime, their wave envelopes advect freely at wave group velocities, which can be determined from the linear dispersion relation.
Once the wave envelopes overlap, energy exchanges between three waves start to happen at a rate determined by the coupling coefficient. 
The exact behavior depends on the details of the wave envelopes, which is determined numerically in Ch.~\ref{ch:3wave}. The three-wave equations are advection equations with nonlinear source terms, which can be readily solved using finite volume schemes. 

A special case of three-wave interaction is laser pulse compression, during which energy stored in a long pump laser is transfered to a short seed pulse and the mediating plasma wave. This phenomenon can be used to amplify and shorten the seed pulse, whereby effective compression of the pump laser is achieved. In Ch.~\ref{ch:compression}, I study laser pulse compression mediated by the upper-hybrid wave as an example, in order to demonstrate that the performance of pulse compression can be improved by applying external magnetic fields. 
For optical lasers, the improvements are largely engineering, where external magnetic fields allow better control of plasma uniformity. On the other hand, for shorter-wavelength lasers, the improvements due to alleviation of physical constraints, such as damping and instabilities, become more substantial. Due to these improvements, magnetized mediations may be used to compress intense UV and soft X-ray pulses, which cannot be compressed using other methods.  
These theoretical predictions, made by analyzing how magnetic fields change limiting effects, are confirmed by numerical simulations.

In the second part of this thesis, I contemplate scalar-QED as a toy model for relativistic quantum plasmas. 
The scalar-QED model describes charged spin-0 bosons interacting with self-consistent electromagnetic fields. Although plasmas are typically made of spin-1/2 fermions, scalar QED is the effective theory for nucleus such as ground-state deuterons, mesons such as charged pions, as well as many condensed matter systems like superconductors and superfluids \citep{Landau65}.
Since classical plasma physics takes no account of particle spin-statistics at all, I demonstrate how to build plasma models from quantum field theory using the scalar-QED model, to avoid unnecessary complications due to spin and chiral effects. 
Extension to spinor-QED plasmas, whose thermal equilibrium states have been studied extensively using thermal Green's function method \citep{Akhiezer60,Rojas79,Rojas1982absorption,Sadooghi08}, can be carried out analogously following procedures in this thesis.

Since quantum field theory is not a familiar tool for many plasma physicists, in Ch.~\ref{ch:vacuum}, I give a self-contained introduction using scalar-QED as an example. I first treat scalar fields as classical fields, using which many important results in quantum field theory can already be obtained. In particular, as a consequence of the local gauge symmetry, electromagnetic fields naturally arise when charged particles are parallel transported in spacetime. The general construction results in Yang-Mills theory \citep{Yang1954conservation}, whose simplest example is scalar QED.
On top of the classical background, quantum fluctuations can then be calculated using either the second-quantization formulation or the path-integral formulation. 
In the second-quantization approach, fields are promoted to operators. The $N$-point correlation functions can be computed in the interaction picture, from which the scattering (S) matrix elements can be extracted.
Equivalently, using the path-integral approach, fields remain functions but sample all configurations, including those that are not allowed classically. By summing up all paths weighted by their action, the $N$-point correlation functions can be calculated from the generating function.

Allowing for dynamical background fields, I extend the vacuum field theory to model relativistic quantum plasmas. In Ch.~\ref{ch:action}, I develop a general theory of wave propagation using an effective action approach. 
The wave effective action has a clear physical meaning. When waves propagate through plasmas, they interact with charged particles, whose dynamics are affected by the presence of the background fields as well as the wave fields. After all the interactions related to charged particles are summed up, what remains is the effective action of waves. This clear physical picture of the wave effective action can be translated into rigorous mathematical procedures using path integrals.
To derive the effective action, I start from the standard action of scalar QED, self-consistently factor out the background fields from the fluctuating fields, and then integrate out the charged particle fields perturbatively. 
The idea of separating classical backgrounds from quantum fluctuations is an extension to Furry's picture of strong-field QED \citep{Furry51}. In addition to external EM fields, which are treated non-perturbatively in strong-field QED  \citep{Greiner85}, I also take into account of the existence of non-perturbative background charged particle fields. The formidable task of finding S-matrix elements by calculating quantum correlation functions whose end states contain infinitely many particles is reduced by incorporating effects of background charged particle fields directly into the Lagrangian. Such an incorporation, which has been attempted phenomenologically by \cite{Shvets95}, is made rigorous and systematic in this thesis. The partition of fields into classical backgrounds and quantum fluctuations is similar to what has been done by \cite{Raicher14}. I make further progress by simplifying the Lagrangian using the self-consistency of backgrounds, and developing the classical field theory to the quantum level. 
In this way, I thoroughly clarify the role of background fields, and use bosonic plasmas as examples to demonstrate how nontrivial background fields can be treated in quantum field theory.

To test the general theory, I apply the wave effective action to study waves in unmagnetized plasmas in Ch.~\ref{ch:unmag}. 
The uniform unmagnetized bosonic plasma has been studied by a number of authors \citep{Hines78, Kowalenko85}. Results of my general theory agree with these authors' in this special case. Apart from modifying the familiar EM waves, the Langmuir wave, and the acoustic wave, relativistic quantum effects result in additional pair modes. In these longitudinal pair modes, particles and antiparticles are constantly being created and annihilated, a phenomena that only exist when both relativistic and quantum effects are taken into account.

While wave dispersion relations in unmagnetized relativistic quantum plasmas are known, this thesis (Ch.~\ref{ch:mag}) is the first to obtain useful expressions of wave dispersion relations when magnetic field are present. 
The uniform, magnetized, bosonic plasma has been studied by \cite{Witte87, Witte88, Witte90}, who can only describe wave propagation parallel to the background magnetic field. 
In addition to recovering this special results, the effective action formalism also enables descriptions of nonparallel wave propagation, which was obscure in previous studies. Many theories and models have been developed in the literature to describe waves in relativistic quantum plasmas, such as plasma response theories based on statistic Green's functions \citep{Bezzerides1972quantum,Melrose07, Melrose2012quantum}, finite temperature theories \citep{Kapusta06,Landsman87,Inagaki05}, quantum hydrodynamic models \citep{Haas11, Shukla10}, and models that are based on the historical Heisenberg-Euler effective Lagrangian \citep{Heisenberg36,Bialynicka70,Marklund06,Piazza07,Lundin09}. Nevertheless, what is being presented in this thesis is the first theory capable of demonstrating its correctness by showing that all linear modes, well-known in classical plasma physics, can be recovered when taking the classical limit in relativistic quantum results.
Moreover, concrete observable consequences are predicted for the first time. 
For example, relativistic-quantum effects can noticeably alter the dependency of Faraday rotation on the frequency of lasers. 
Near the cutoff frequency where Faraday rotation reaches maximum, the effects produce order-unity corrections already in gigagauss magnetic fields. 
In even stronger magnetic fields found near X-ray pulsars, quantum effects allows Bernstein waves to persist even when the plasma is cold, and relativistic effects shift the resonance frequencies, leading to anharmonic cyclotron absorptions features already observed in spectra of X-ray pulsars.

Finally, to study nonperturbative effects, I develop a numerical scheme for simulating relativistic-quantum plasmas in Ch.~\ref{ch:simulation}. 
Since plasmas are in the classical-statistic regime, their behaviors are adequately captured by solving the classical field equations with an ensemble of statistically equivalent initial conditions. 
This real-time lattice QED scheme provides a unique tool for simulating plasmas in strong-fields, where collective plasma scales are not well separated from relativistic-quantum scales. 
The algorithm is developed by first discretizing the scalar-QED action on a lattice,
in a way that respects both the geometric structures of exterior calculus and the U(1)-gauge symmetry. 
Taking variations of the discrete action, finite difference equations can be obtained, which can then be used to advance initial conditions in time. 
To demonstrate the capability of this numerical scheme, I apply it to two example problems. The first example is the propagation of linear waves, where the analytic wave dispersion relations is recovered using numerical spectra. The second example is an intense laser interacting with a one-dimensional plasma slab, where the natural transition from
wakefield acceleration to pair production is demonstrated for the first time, when the laser intensity exceeds the Schwinger limit.

\part{Three-wave interactions in magnetized cold-fluid plasmas}
\lettrine[lines=2, findent=3pt, lraise=0.4, loversize=1]{T}{he} first part of this thesis analyzes the effects of strong background magnetic fields on wave-wave interactions in classical plasmas. 
It is somewhat surprising that a useful general theory of coherent wave-wave interactions has never been obtained for magnetized plasmas, although magnetic fields are no stranger to plasma physics, and coherent wave-wave interactions are the next-to-leading-order effects in weakly coupled plasmas. The lack of illuminating analytical results is perhaps due to the mathematical difficulty once background magnetic fields are present. However, due to the emerging feasibility of strong magnetic fields during laser-plasma interactions, a general theory becomes necessary to describe anisotropic laser scattering beyond the special cases studied in the literature.

My work overcomes the analytical difficulty using an operator approach, which enables a systematic solution of the magnetized fluid-Maxwell's equation to the second order. The second-order solution turns out to be very simple and intuitive. In addition to recovering the well-known three-wave model, the solution provides a convenient formula for the magnetized coupling coefficient that was previously unknown. The simple formula, obtained by solving partial differential equations, becomes obvious when I reformulate it from the effective Lagrangian perspective. Using the formula, 
I have then mapped out the anisotropic three-wave scattering behavior in the most general geometry. With knowledge of the coupling coefficient, the three-wave model can then be solved to provide detailed descriptions of the coherent scattering process.

Understanding of the basic scattering physics has many implications. For example, in the application where plasmas are used as media to compress intense laser pulses, I discover that applying external magnetic fields significantly expands the range of lasers that can be compressed and increases the final pulse intensity. By choosing the strength and direction of the background magnetic field, we can now use many extra degrees of freedom to further optimize the performance of the plasma laser compressor. My theoretical identification that magnetized plasmas are superior media for laser pulse compression has since been supported by numerical simulations. This work not only opens up the new research direction of magnetized laser-plasma interactions, but also points out a pathway for generating powerful lasers beyond the attainment of existing technologies.

\chapter{Prelude: multiscale solution of ordinary differential equations\label{ch:multiscale}}

Nonlinear partial differential equations (PDEs) are the cornerstones of plasma models. These equations need to be solved, or reduced to simpler forms, before we can use them to understand the behaviors of plasmas. However, solving nonlinear PDEs is not easy. In fact, a large part of theoretical and numerical plasma physics is in essence about finding approximate solutions in various regimes. The simplest regime is perhaps the weakly-nonlinear regime, where nonlinearities are merely perturbations to the linear problem. In this regime, we can build successively better approximations using perturbation theory, starting from the general solution of the linear problem, which we know at least in principle.

In this chapter, I will introduce ideas that enable perturbative solutions in the weakly-nonlinear regime \citep[App. A]{Shi17scatter}. Since the basic ideas are not unique to PDEs, which involve many variables, I will use ordinary differential equations (ODEs), which involve only one variable, to illustrate the ideas in the simplest way possible. The equations I will focus on in this chapter are a system of hyperbolic ODEs. These equations describe oscillatory motion. Such motion can be viewed as a toy model for waves-like behaviors, which are described by the hyperbolic PDEs underlying the plasma models. The mathematical machineries and physical intuitions developed here will be used in later chapters when I study plasmas.

\section{Failure of naive perturbative solutions\label{sec:multiscale:naive}}

Let us consider the following prototypical system of ODEs, which are hyperbolic in the absence of perturbations. After some normalization, the equations can be written as
\begin{eqnarray}
	\label{eq:dx}
	\dot{x}&=&\phantom{+}y+\epsilon f(x,y),\\
	\label{eq:dy}
	\dot{y}&=&-x+\epsilon g(x,y).
\end{eqnarray}
Here $\dot{x}$ and $\dot{y}$ denote the time derivatives of $x(t)$ and $y(t)$, $f$ and $g$ are some polynomials, and $\epsilon\ll1$ is a small parameter. If we think of $x$ as the displacement and $y$ as the velocity, then the above equations describe a harmonic oscillator perturbed by external forces.

When $\epsilon=0$, the general solutions to the above system of linear ODEs describe simple harmonic oscillations:
\begin{eqnarray}
\label{eq:x00}
x_0&=&a_0e^{it}+b_0e^{-it}, \\
\label{eq:y00}
y_0&=&ia_0e^{it}-ib_0e^{-it},
\end{eqnarray}
where 
$a_0$ and $b_0$ 
are some fixed complex numbers. Now suppose $\epsilon$ is finite but small, we expect the new solutions to deviate from the above solutions only perturbatively. Naively, one may consider solving the equations using the following expansion
\begin{eqnarray}
	\label{eq:x}
	x(t)&=&x_0(t)+\epsilon x_1(t)+\epsilon^2x_2(t)+\dots,\\
	\label{eq:y}
	y(t)&=&y_0(t)+\epsilon y_1(t)+\epsilon^2y_2(t)+\dots.
\end{eqnarray}
When we write down such an expansion, we expect that $x_0\gg\epsilon x_1\gg\dots$. If this turns out to be true, then higher-order terms are just some small corrections, which can be neglected when we compute lower-order terms.

Now let us see whether this idea works. Assuming higher-order terms are subdominant, then $x_0$ and $y_0$ are still given by Eqs. (\ref{eq:x00}) and (\ref{eq:y00}). Next, since $\epsilon^1$-order terms are much larger than $\epsilon^2$-order terms, we can isolate the $\epsilon^1$-order terms and focus on solving the first order equations, which give
\begin{eqnarray}
\dot{x}_1&=&\phantom{+}y_1+f(x_0,y_0),\\
\dot{y}_1&=&-x_1+g(x_0,y_0).
\end{eqnarray}
To solve these equations, we can, for example, eliminate $y_1$ to obtain an equation for $x_1$. Taking second order derivative, we have
\begin{eqnarray}
\nonumber
\ddot{x}_1&=&\dot{y}_1+\partial_x f_0 \dot{x}_0+\partial_y f_0 \dot{y}_0\\
&=&-x_1+h_1,
\end{eqnarray}
where $h_1=g_0+\partial_x f_0 \dot{x}_0+\partial_y f_0 \dot{y}_0$ is some known source term. Using Green's functions or other techniques, we can readily solve the above inhomogeneous linear ODE. The general solution is
\begin{equation}
x_1=a_1e^{it}+b_1e^{-it}+\int_0^th_1(\tau)\sin(t-\tau)d\tau.
\end{equation}
Once we know $x_1$, we can easily compute $y_1$ using $y_1=\dot{x}_1-f_0$. Thereof, it appears that we have obtained the general solution to Eqs.~(\ref{eq:dx}) and (\ref{eq:dy}) beyond the leading order.

However, there is a potential problem. Notice that $h_1(t)$ is a functional of $x_0$ and $y_0$, and it is hence an oscillatory function in general. For a large class of problems, $h_1$ contains oscillation at frequency $\omega=\pm1$. If this is the case, we can write $h_1(\tau)\propto2\cos(\tau+\theta)+\dots$, where $\theta$ is some phase. Then the integral
\begin{eqnarray}
\nonumber
\int_0^th_1(\tau)\sin(t-\tau)d\tau&\propto&\int_0^t d\tau[\sin(t+\theta)+\sin(t-2\tau-\theta)+\dots]\\
&=&t \sin(t+\theta)+\dots.
\end{eqnarray}
The term $t \sin(t+\theta)$ is problematic, because it can grow indefinitely. This type of behavior is known as secular growth \citep{Debnath11}, which renders our naive perturbative scheme invalid beyond $t\sim1/\epsilon$. At later time $t\gg1/\epsilon$, the first order term $\epsilon x_1\gg x_0$. This situation invalidates our assumption that $x_0\gg\epsilon x_1\gg\dots$, making it illegitimate to ignore $\epsilon^1$-order terms when solving the $\epsilon^0$-order equations.

\section{Multiscale expansion: general structure of solutions\label{sec:multiscale:general}}

The key to obtain a valid perturbative solution is removing the secular term. This can be accomplished by a number of methods. For example, the Poincar\'e-–Lindstedt method allows the oscillation frequency $\omega=\omega_0+\epsilon\omega_1+\dots$ to be perturbed. Perhaps a more systematic way to introduce such an effect it to recognize that there can be multiple time scales in the problem \citep{Davidson72}. In addition to the expansion in amplitudes [Eqs.~(\ref{eq:x}) and (\ref{eq:y})], let us also introduce an expansion in time 
\begin{eqnarray}
	\label{eq:t}
	t&=&t_0+\frac{1}{\varepsilon} t_1+\frac{1}{\varepsilon^2}t_2+\dots,\\
	\label{eq:dt}
	\partial_t&=&\partial_{0}+\varepsilon\partial_{1}+\varepsilon^2\partial_{2}+\dots.
\end{eqnarray}
In such an expansion, different time scales are regarded as independent, and one unit of the slow time scale $t_n$ worths $1/\varepsilon^n$ units of the fastest time scale $t_0$. 
In general, the expansion parameter $\varepsilon$ for the time scales needs not be the same as the expansion parameter $\epsilon$ for the amplitudes. In what follows, I will only consider the special case $\varepsilon=\epsilon$. This criteria defines the weakly-nonlinear regime. Substituting the amplitude and time expansions into the equations, and collect terms according to their orders in $\epsilon$, we can obtain a series of equations. In what follows, without loss of generality, let us focus on real-valued solutions.

\subsection{Zeroth order\label{sec:multiscale:general:0th}}

Similar to the naive solution, the $\epsilon^0$-order equations are simply the equations for a simple harmonic oscillator in the phase space:
\begin{eqnarray}
	\partial_0x_0-y_0&=&0,\\
	\partial_0y_0+x_0&=&0.
\end{eqnarray}
For real valued $x$ and $y$, the general solution is 
\begin{eqnarray}
	\label{eq:x0}
	x_0&=&a_0e^{it_0}+\text{c.c.},\\
	\label{eq:y0}
	y_0&=&b_0e^{it_0}+\text{c.c.},
\end{eqnarray}
where $b_0=ia_0$ and c.c. stands for complex conjugate. If we truncate the solution at this order, then $x$ and $y$ oscillate harmonically with constant amplitude. However, in multiscale expansion [Eqs.~(\ref{eq:t}) and (\ref{eq:dt})], the complex amplitude $a_0=a_0(t_1,t_2,\dots)$ is in general a function of slow variables. When we move to higher orders, we will obtain equations that describe how perturbations $\epsilon f(x,y)$ and $\epsilon g(x,y)$ cause the amplitude $a_0$ to vary on slow time scales. 
\subsection{First order\label{sec:multiscale:general:1st}}

The first order equations start to couple perturbations on different time scales. Collecting $\epsilon^1$-order terms, we have
\begin{eqnarray}
	\label{eq:e1x}
	\partial_1x_0+\partial_0x_1-y_1-f_0&=&0,\\
	\label{eq:e1y}
	\partial_1y_0+\partial_0y_1+x_1-g_0&=&0,
\end{eqnarray}
where $f_0=f(x_0,y_0)$ and $g_0=g(x_0,y_0)$, in which $x_0$ and $y_0$ are given by Eqs.~(\ref{eq:x0}) and (\ref{eq:y0}). The above two equations contain three unknowns $x_1,y_1$, and $\partial_1a_0$. Therefore, we can use the extra degree of freedom to remove secular terms. 

To remove the secular term, let us first separate variables $x_1$ and $y_1$. Taking $\partial_0$ derivatives on both sides of Eqs.~(\ref{eq:e1x}) and (\ref{eq:e1y}), and using the zeroth- and first-order equations, we obtain two decoupled equations
\begin{eqnarray}
	\label{eq:2x1}
	\partial_0^2x_1+x_1+2\partial_1y_0&=&u_1,\\
	\label{eq:2y1}
	\partial_0^2y_1+y_1-2\partial_1x_0&=&v_1.
\end{eqnarray}
The source terms $u_1$ and $v_1$ are known. They are explicit functions of the fastest time scale $t_0$, while implicitly depend on other time scales through the coefficient $a_0$:
\begin{eqnarray}
	\label{eq:u1}
	u_1[a_0]&=&\partial_0f_0+g_0,\\
	\label{eq:v1}
	v_1[a_0]&=&\partial_0g_0-f_0.
\end{eqnarray}
Substituting $x_0$ and $y_0$ into polynomials $f$ and $g$, we can write $f_0=\sum_nf_{0n}e^{int_0}+$c.c., and $g_0=\sum_ng_{0n}e^{int_0}+$c.c., where $f_{0n}$ and $g_{0n}$ are some functionals of $a_0$. Then the source terms can be expanded similarly using Fourier series as $u_1=\sum_n u_{1n}e^{int_0}+$c.c. and $v_1=\sum_n v_{1n}e^{int_0}+$c.c., where 
\begin{eqnarray}
u_{1n}&=&g_{0n}+inf_{0n}, \\
v_{1n}&=&-f_{0n}+ing_{0n},
\end{eqnarray}
are some functionals of $a_0$. These Fourier coefficients of the source terms are what we need when solving the first-order equations.

Now we are ready to solve the $\epsilon^1$-order equations (\ref{eq:2x1}) and (\ref{eq:2y1}). Matching coefficients of Fourier exponents, we can split the equations into two sets. The first set of equations govern how the amplitude $a_0$ evolves on the slow time scale $t_1$, which can be written as $\partial_1x_0=-\frac{1}{2}(v_{11}e^{it_0}+\text{c.c.})$, or $\partial_1y_0=\frac{1}{2}(u_{11}e^{it_0}+\text{c.c.})$. These two equations are essentially the same, as can be seen from the relations between $x_0$ and $y_0$, as well as the definitions of $u_{11}$ and $v_{11}$. In terms of $a_0$, the amplitude equation can be written as
\begin{equation}
	\label{eq:d1a0}
	\partial_1a_0=\frac{1}{2}(f_{01}-ig_{01}),
\end{equation}
where the right-hand-side (RHS) is some known functional of $a_0$. Integrating this first order ODE, $a_0$ becomes a known function of $t_1$. In other words, due to perturbations, the otherwise constant amplitude $a_0$ now varies on the slow time scale $t_1$. Alternatively, Eq.~(\ref{eq:d1a0}) can be regarded as the non-secular condition. By satisfying this condition, we have thus removed the secular terms from the first-order equations.

Having taken out terms oscillating at the fundamental frequency $\omega=\pm1$, we remove resonant forcing terms from the first order equations (\ref{eq:2x1}) and (\ref{eq:2y1}). What remain are equations governing $x_1$ and $y_1$, which can be written as
\begin{eqnarray}
	\label{eq:x1s}
	\partial_0^2x_1+x_1&=&\sum_{n\ne 1}u_{1n}e^{int}+\text{c.c.},\\
	\label{eq:y1s}
	\partial_0^2y_1+y_1&=&\sum_{n\ne 1}v_{1n}e^{int}+\text{c.c.}.
\end{eqnarray}
Now that the harmonic oscillators on the left-hand-side (LHS) are no longer driven resonantly, secular terms do not arise, and the secular-free solutions can be readily obtained:
\begin{eqnarray}
	\label{eq:x1}
	x_1=a_1e^{it_0}+\sum_{n\ne 1}\frac{u_{1n}}{1-n^2}e^{int_0}+\text{c.c.},\\
	\label{eq:y1}
	y_1=b_1e^{it_0}+\sum_{n\ne 1}\frac{v_{1n}}{1-n^2}e^{int_0}+\text{c.c.}.
\end{eqnarray}
Notice that although harmonics of the fundamental frequency are not eigenmode of the linear system, they can now be generated through the nonlinear source terms. In the above general solution, the amplitudes $a_1$ and $b_1$ are clearly related. From the $\epsilon^1$-order equations, 
\begin{equation}
	b_1=ia_1-\frac{1}{2}(f_{01}+ig_{01}),
\end{equation}
which is completely determined once $a_0$ and $a_1$ are fixed. Notice that in expansion Eq.~(\ref{eq:x}), we can always redefine $a_0+\epsilon a_1\rightarrow a_0'$. Hence it is sufficient to set the amplitude $a_1=0$. In this way, we obtain an $x$-major solution, in the sense that the amplitude of $e^{it_0}$ in $x$ is precisely given by $a_0$, whereas amplitude of $e^{it_0}$ in $y$ is given by the summation $b_0+\epsilon b_1+\dots$. In other words, we can always carry out resummation and renormalization, such that $a=a_0$ is the exact amplitude, while $b$ is given by a perturbative series. Alternatively, by setting $b_1=0$, we can obtain a $y$-major solution, which I shall not pursue here. 
\subsection{Second order\label{sec:multiscale:general:2nd}}

To show the general structure of the multiscale expansion, it is instructive to carry out the solution to the next order. The $\epsilon^2$-order equations are
\begin{eqnarray}
	\label{eq:e2x}
	\partial_2x_0+\partial_1x_1+\partial_0x_2-y_2-f_1&=&0,\\
	\label{eq:e2y}
	\partial_2y_0+\partial_1y_1+\partial_0y_2+x_2-g_1&=&0,
\end{eqnarray}
where $f_1=x_1\partial_xf_0+y_1\partial_yf_0$ and $g_1=x_1\partial_xg_0+y_1\partial_yg_0$. From the previous order, $a_0$ is now a known function of $t_1$. After setting $a_1=0$ in the $x$-major solution, $x_1$ and $y_1$, and thereof $f_1$ and $g_1$ are known functionals of $a_0$. In other words, there are only three unknowns $x_2$, $y_2$ and $\partial_2a_0$ in the above equations. 

Similar to what we have done before, we can use the extra degree of freedom to remove secular terms. Separating variables $x_2$ and $y_2$, we can rewrite the equations as
\begin{eqnarray}
	\label{eq:2x2}
	\partial_0^2x_2+x_2+2\partial_2y_0&=&u_2,\\
	\label{eq:2y2}
	\partial_0^2y_2+y_2-2\partial_2x_0&=&v_2.
\end{eqnarray}
These equations are structurally the same as Eqs.~(\ref{eq:2x1}) and (\ref{eq:2y1}), except now the subscript ``2" replaces the subscript ``1".  The source terms, albeit different than $u_1$ and $v_1$, are again functionals of $a_0$ only:
\begin{eqnarray}
	\label{eq:su2}
	u_2[a_0]&=&\partial_0f_1+g_1+\partial_1^2x_0-2\partial_1y_1-\partial_1 f_0,\\
	\label{eq:sv2}
	v_2[a_0]&=&\partial_0g_1-f_1+\partial_1^2y_0+2\partial_1x_1-\partial_1 g_0.
\end{eqnarray}
These source terms may look complicated. Nevertheless, keeping in mind that $f$ and $g$ are polynomials, we can always write $f_1=\sum_nf_{1n}e^{int_0}+$c.c., and $g_1=\sum_ng_{1n}e^{int_0}+$c.c.. Consequently, the source terms can always be written in the form $u_2=\sum_n u_{2n}e^{int_0}$+c.c. and $v_2=\sum_n v_{2n}e^{int_0}$+c.c.. After some algebra, it is not difficult to find $v_{21}=iu_{21}=i\partial_1^2a_0+ig_{11}-\partial_1g_{01}-f_{11}$. Moreover, for $n\ge2$, we have 
\begin{eqnarray}
\label{eq:u2n}
u_{2n}&=&i n f_{1n}-\partial_1f_{0n}+g_{1n}-\frac{2\partial_1v_{1n}}{1-n^2}, \\
\label{eq:v2n}
v_{2n}&=&i n g_{1n}-\partial_1g_{0n}-f_{1n}+\frac{2\partial_1u_{1n}}{1-n^2}.
\end{eqnarray}
These Fourier coefficients are what we need when solving the second-order equations. The above general formula can usually be simplified substantially, when the multiscale method is applied to specific problems of interest.

To solve the $\epsilon^2$-order equations (\ref{eq:2x2}) and (\ref{eq:2y2}), we can use similar procedure to split the equations into two sets. The first set of equations can be written as a single equation governing how the amplitude $a_0$ evolve on the slower time scale:
\begin{equation}
	\label{eq:d2a0}
	\partial_2a_0=\frac{1}{2}(f_{11}-ig_{11})-\frac{i}{4}\partial_1(f_{01}+ig_{01}).
\end{equation}
Regarding $t_0$ and $t_1$ as parameters, this equation is a first order ODE for $a_0(t_2)$, which can usually be integrated. The second sets of equations are similar to Eqs.~(\ref{eq:x1s}) and (\ref{eq:y1s}), with $u_{1n}$ and $v_{1n}$ replaced by $u_{2n}$ and $v_{2n}$, respectively. The solutions to these secular-free equations are similar to  Eqs.~(\ref{eq:x1}) and (\ref{eq:y1}) with the order index ``1" replaced by the order index ``2". Similar to the first-order solutions, the second-order amplitudes $a_2$ and $b_2$ are related by the $\epsilon^2$-order equations according to 
\begin{equation}
	\label{eq:b2}
	b_2=ia_2-\frac{1}{2}(f_{11}+ig_{11})-\frac{i}{4}\partial_1(f_{01}+ig_{01}).
\end{equation}
To obtain the $x$-major solution, we again set $a_2$ to zero. By the obvious analogy between the \mbox{$\epsilon^1$- and $\epsilon^2$-order} equations, the above multiscale procedures can be readily extended to higher orders in the perturbation series.

\section{Example problems: perturbative match exact\label{sec:multiscale:examples}}

In this section, I will demonstrate the multiscale method using two example problems, for which exact solutions are known. By comparing the exact solutions with the perturbative solutions, we can get a better sense of how the multiscale method works in practice. Before getting into details, it is helpful to summarize the method. First, we expand both amplitudes and time scales, and convert a system of ODEs (\ref{eq:dx})-(\ref{eq:dy}) to a hierarchy of equations. At each order, one set of equations are the secular-free conditions [e.g. Eq.~(\ref{eq:d1a0})], which govern how amplitudes of faster oscillations vary on slower time scales. The other set of equations describe non-resonant driven oscillations [e.g. Eq.~(\ref{eq:x1})], through which higher harmonics can be generated and phase locked with the fundamental mode of oscillation.

\subsection{A linear problem\label{sec:multiscale:examples:linear}}

Consider the linear problem $f(x,y)=-x$ and $g(x,y)=0$. The equations can be written in the matrix form:
\begin{equation}
\left( \begin{array}{c}
\dot{x} \\ \dot{y}
\end{array} \right)
=\left( \begin{array}{cc}
-\epsilon & 1 \\
-1 & 0
\end{array} \right)
\left( \begin{array}{c}
x \\ y
\end{array} \right).
\end{equation}
One eigenvalue of the matrix is $\lambda=-\epsilon/2+ i\sqrt{1-(\epsilon/2)^2}$, and the other eigenvalue is its complex conjugate. Then the $x$-major general solution is
\begin{eqnarray}
\label{eq:eg1_exact_x}
\nonumber
x\!&=&\!\alpha e^{\lambda t}+\text{c.c.} \\
&\simeq&\alpha \exp(-\frac{\epsilon t}{2}-\frac{i\epsilon^2t}{8}+\dots)e^{it}+\text{c.c.}, \\
\label{eq:eg1_exact_y}
\nonumber
y\!&=&\!(\lambda+\epsilon)\alpha e^{\lambda t}+\text{c.c.} \\
&\simeq&(i+\frac{\epsilon}{2}-\frac{i\epsilon^2}{8}+\dots)\alpha \exp(-\frac{\epsilon t}{2}-\frac{i\epsilon^2t}{8}+\dots)e^{it}+\text{c.c.},
\end{eqnarray}
where 
$\alpha$ is some complex number determined by initial conditions. Now let us compare the expansion of the exact solution with the perturbative solution.

To $\epsilon^0$ order, following Eqs.~(\ref{eq:x0}) and (\ref{eq:y0}), the lowest-order perturbative solution is
\begin{eqnarray}
x_0&=&a_0e^{it_0}+\text{c.c.},\\
y_0&=&ia_0e^{it_0}+\text{c.c.}.
\end{eqnarray}
Substituting these into polynomials $f$ and $g$, we have $f_0=-a_0e^{it_0}+\text{c.c.}$ and $g_0=0$. In other words, $f_{01}=-a_0$, and all other Fourier coefficients are zero. Using Eqs.~(\ref{eq:u1}) and (\ref{eq:v1}), we can immediately find $u_{11}=-ia_0$, $v_{11}=a_0$, and $u_{1n}=v_{1n}=0$ for all $n\ge2$.

Next, to $\epsilon^1$ order, the amplitude equation (\ref{eq:d1a0}) becomes $\partial_1 a_0=-a_0/2$, which can be easily integrated to give
\begin{equation}
\label{eq:eg1_a01}
a_0=\alpha \exp(-\frac{t_1}{2}),
\end{equation}
where $\alpha\in\mathbb{C}$ is determined by initial conditions. Since $u_{1n}=v_{1n}=0$ are trivial for all $n\ge2$, the secular-free first-order equations (\ref{eq:x1s}) and (\ref{eq:y1s}) are also trivial. For $x$-major solution, $a_1=0$ and $b_1=a_0/2$, therefore
\begin{eqnarray}
x_1&=&0,\\
y_1&=&\frac{a_0}{2}e^{it_0}+\text{c.c.},
\end{eqnarray}
where $a_0$ is given by Eq.~(\ref{eq:eg1_a01}) at this order. Substituting these into polynomials $f$ and $g$, we immediately find $f_1=g_1=0$. Using the first-order amplitude equation, together with Eqs.~(\ref{eq:su2}) and (\ref{eq:sv2}), the source terms can be easily found, whose Fourier coefficients $v_{21}=iu_{21}=ia_0/4$, and $u_{2n}=v_{2n}=0$ for all $n\ge2$. 

Finally, to $\epsilon^2$ order, the amplitude equation (\ref{eq:d2a0}) becomes $\partial_2 a_0=-ia_0/8$. Again, this equation can be easily integrated
\begin{equation}
\label{eq:eg1_a02}
a_0=a_0(t_1)\exp(-\frac{it_2}{8})=\alpha \exp(-\frac{t_1}{2}-\frac{it_2}{8}),
\end{equation}
where we have used the solution Eq.~(\ref{eq:eg1_a01}) at the faster time scale as the initial condition for the slower time scale. The secular-free second-order equations are again trivial, because $u_{2n}=v_{2n}=0$ for all $n\ge2$. Setting $a_2=0$ for $x$-major solution, we can use Eq.~(\ref{eq:b2}) to find $b_2=-ia_0/8$. Therefore, we have
\begin{eqnarray}
x_2&=&0,\\
\label{eq:eg1_y2}
y_2&=&-\frac{ia_0}{8}e^{it_0}+\text{c.c.},
\end{eqnarray}
where $a_0$ is now given by Eq.~(\ref{eq:eg1_a02}). The multiscale procedure can be analogously carried out to higher orders, which I will not pursue here. Notice that in this linear problem, frequency of the oscillation is always independent of its amplitude. Due to perturbations, the amplitude receives corrections [e.g. Eq.~(\ref{eq:eg1_y2})], and varies on slow time scales [e.g. Eq.~(\ref{eq:eg1_a02})]. Such behaviors are not specific to this problem, and they are universal features to all linear problems.

Having solved the equations order by order, we can now sum up terms in the perturbation series to find expressions for the final solution:
\begin{eqnarray}
\nonumber
x&=&x_0+\epsilon x_1+\epsilon^2 x_2+\dots \\
\nonumber
&=&a_0 e^{it_0}+0+0+\dots+\text{c.c.}\\
&\simeq&\alpha\exp(-\frac{t_1}{2}-\frac{it_2}{8})e^{it_0}+\dots+\text{c.c.},\\
\nonumber
y&=&y_0+\epsilon y_1+\epsilon^2 y_2+\dots \\
\nonumber
&=&ia_0 e^{it_0}+\epsilon\frac{a_0}{2}e^{it_0}-\epsilon^2\frac{ia_0}{8}e^{it_0}+\dots+\text{c.c.}\\
&\simeq&(i+\frac{\epsilon}{2}-\frac{i\epsilon^2}{8}+\dots)\alpha\exp(-\frac{t_1}{2}-\frac{it_2}{8})e^{it_0}+\dots+\text{c.c.}.
\end{eqnarray}
The above perturbation series are identical to the Taylor series of the exact solution [Eqs.~(\ref{eq:eg1_exact_x}) and (\ref{eq:eg1_exact_y})], once we restore the time scales $t_0=t$, $t_1=\epsilon t$, and $t_2=\epsilon^2 t$. The multiscale method thus correctly recovers the exact solution.
\subsection{A nonlinear problem\label{sec:multiscale:examples:nonlinear}}

Now let us consider a nonlinear example, where $f(x,y)=0$ and $g(x,y)=-x+2x^3$. The exact solutions in this case are the Jacobi elliptic functions \citep[Ch.~22]{NIST:DLMF}. To convert the equation to the standard form, we can eliminate $y$ and then $x$ satisfies
\begin{equation}
\label{eq:nonlinear}
\ddot{x}+(1+\epsilon)x-2\epsilon x^3=0.
\end{equation}
In this standard form, the modulus of the elliptic function $k^2=\epsilon=1-k'^2$ can be easily identified. The two linearly independent solutions are the Jacobi elliptic functions $\text{sn}(t,k)$ and $\text{cd}(t,k)$. The general solution can be expressed in terms of these two basis functions. Here, for simplicity, let us consider a special solution with $x(t=0)=0$ and $\dot{x}(t=0)=1$. Given these initial conditions, the solution is
\begin{eqnarray}
\label{eq:eg2_exact_x}
x(t)&=&\text{sn}(t,k).
\end{eqnarray}
Since $y(t)=\dot{x}(t)=\text{cn}(t,k)\,\text{dn}(t,k)$ do not enter the perturbations $f$ and $g$, I will focus on $x(t)$ in this example. 

To compare with perturbative solutions, let us expand the exact solution Eq.~(\ref{eq:eg2_exact_x}) using Fourier series
\begin{eqnarray}
x=\frac{2\pi}{kK}\sum_{n=0}^{\infty}\frac{q^{n+1/2}\sin[(2n+1)\tau]}{1-q^{2n+1}}.
\end{eqnarray}
Here, $K=K(k)$ is the complete elliptic integral of the first kind. Since $\epsilon\ll1$, the elliptic integral can be expanded as $K\simeq\frac{\pi}{2}(1+\frac{\epsilon}{4}+\frac{9\epsilon^2}{64}+\dots)$.
Denoting $K'=K(k')$, then the nome $q=\exp(-\pi K'/K)$ can be expanded as $q\simeq\frac{1}{16}(\epsilon+\frac{\epsilon^2}{2}+\dots)$.
With these Taylor series, the exact solution can be written as
\begin{eqnarray}
\label{eq:eg2_x_series}
\nonumber
x&=&\Big(1+\frac{\epsilon}{16}+\frac{7\epsilon^2}{256}+\dots\Big)\sin\tau\\
&+&\Big(\frac{\epsilon}{16}+\frac{\epsilon^2}{32}+\dots\Big)\sin 3\tau\\
\nonumber
&+&\Big(\frac{\epsilon^2}{256}+\dots\Big)\sin 5\tau+\dots.
\end{eqnarray}
Finally, the normalized time $\tau=\frac{\pi t}{2K}$, which can be expanded as
\begin{equation}
\label{eq:eg2_t_series}
\tau=t(1-\frac{\epsilon}{4}-\frac{5\epsilon^2}{64}+\dots).
\end{equation} 
In this form, we can readily compare the exact solution with the perturbative solution. In what follows, I will demonstrate how to find the pertubative solution by applying the multiscale procedure described in the previous section.

To $\epsilon^0$ order, following Eq.~(\ref{eq:x0}), the lowest order solution is again $x_0=a_0e^{it_0}+\text{c.c.}$.
If we truncate at this order, then using the initial conditions, it is easy to see $a_0=-i/2$ is a constant. However, when we move on to higher orders, the amplitude $a_0$ will vary on slow time scales as we shall see. To prepare for the next-order solution, substituting the above zeroth-order solution into polynomials $f$ and $g$, we have $f_0=0$ and $g_0=(6|a_0|^2-1)a_0e^{it_0}+2a_0^3 e^{3it_0}+$c.c.. In other words, the only nonzero Fourier coefficients are $g_{01}=(6|a_0|^2-1)a_0$ and $g_{03}=2a_0^3$. Using Eqs.~(\ref{eq:u1}) and (\ref{eq:v1}), we can immediately find $u_{11}=g_{01}$, $u_{13}=g_{03}$, $v_{11}=ig_{01}$, $v_{13}=3ig_{03}$, and all other Fourier coefficients are zero.

Next, to $\epsilon^1$ order, the amplitude equation (\ref{eq:d1a0}) becomes $\partial_1 a_0=\frac{i}{2}(1-6|a_0|^2)a_0$. It is easy to see $\partial_1|a_0|^2=0$. Thus, evolution of $a_0$ on the $t_1$ time scale is a pure phase rotation. After integrating the first-order amplitude equation, 
\begin{equation}
\label{eq:eg2_a01}
a_0=\alpha \exp\Big[\frac{it_1}{2}\Big(1-6|\alpha|^2\Big)\Big],
\end{equation}
where 
$\alpha\in\mathbb{C}$ will be determined by initial conditions later. Since only $u_{13}\ne0$ in the secular-free first-order equations (\ref{eq:x1s}), the $x$-major solution is
\begin{eqnarray}
x_1&=&-\frac{a_0^3}{4}e^{3it_0}+\text{c.c.}.
\end{eqnarray}
Substituting it into polynomials $f$ and $g$, we have $f_1=0$ and $g_1=g_{11}e^{it_0}+g_{13}e^{3it_0}+g_{15}e^{5it_0}+$c.c., where $g_{11}=-3a_0|a_0|^2/2$, $g_{13}=(1-12|a_0|^2)a_0^3/4$, and $g_{15}=-3a_0^5/2$. Here, it is more convenient to find the Fourier coefficients $u_{2n}$ and $v_{2n}$ of the source terms using Eqs.~(\ref{eq:u2n}) and (\ref{eq:v2n}). Substituting in the first-order amplitude equation, the four relevant Fourier coefficients are $u_{23}=a_0^3(21|a_0|^2-4)/2$, $u_{25}=-3a_0^5/2$, $v_{23}=3ia_0^3(9|a_0|^2/2-1)$, and $v_{25}=-15ia_0^5/2$. Apart from $v_{21}=iu_{21}$, which are not needed at the next order, all other Fourier coefficients are zero.

Finally, to $\epsilon^2$ order, the amplitude equation (\ref{eq:d2a0}) becomes $\partial_2 a_0=-\frac{ia_0}{8}(1-12|a_0|^2+30|a_0|^4)$. We see $\partial_1|a_0|^2=0$, so the evolution of $a_0$ on the $t_2$ time scale is again a pure phase rotation. Integrating the second-order amplitude equation, we can easily obtain
\begin{equation}
\label{eq:eg2_a02}
a_0=a_0(t_1)\exp\Big[-\frac{it_2}{8}\Big(1-12|\alpha|^2+30|\alpha|^4\Big)\Big],
\end{equation}
where $a_0(t_1)$ is given by Eq.~(\ref{eq:eg2_a01}). Again, the initial condition for the slower time scale is given by the solution at the faster time scale. As for the secular-free second-order equation, which contains two nontrivial terms, the $x$-major solution is
\begin{eqnarray}
x_2&=&\frac{a_0^3}{16}\big(4-21|a_0|^2\big)e^{3it_0}+\frac{a_0^5}{16}e^{5it_0}+\text{c.c.},
\end{eqnarray}
where $a_0$ is now given by Eq.~(\ref{eq:eg2_a02}). We see nonlinearities allow modes to beat, and thereof produce successively higher-order harmonics in the perturbation series. The above multiscale procedure can be repeated to find higher order corrections, which I will not elaborate here.

Having solved the equations order by order, we can now sum up terms in the perturbation series to find an expression for the final solution:
\begin{eqnarray}
\label{eq:nonlinear_x}
\nonumber
x&=&x_0+\epsilon x_1+\epsilon^2 x_2+\dots \\
\nonumber
&=&a_0 e^{it_0} +\epsilon\big(-\frac{a_0^3}{4}e^{3it_0}\big) +\epsilon^2\big[\frac{a_0^3}{16}\big(4-21|a_0|^2\big)e^{3it_0}+\frac{a_0^5}{16}e^{5it_0}\big]+\dots+\text{c.c.}\\
&=&\alpha e^{i\tau}+\big[-\frac{\epsilon}{4} +\frac{\epsilon^2}{16}\big(4-21|\alpha|^2\big)\big]\alpha^3e^{3i\tau}+\frac{\epsilon^2}{16}\alpha^5e^{5i\tau} +\dots+\text{c.c.},
\end{eqnarray}
where the normalized time
\begin{equation}
\label{eq:nonlinear_t}
\tau=t_0+\frac{t_1}{2}\Big(1-6|\alpha|^2\Big)-\frac{t_2}{8}\Big(1-12|\alpha|^2+30|\alpha|^4\Big)+\dots,
\end{equation}
is obtained using the second-order solution [Eq.~(\ref{eq:eg2_a02})] for the slowly varying amplitude $a_0$ . The above expansions give the general solution to the nonlinear problem Eq.~(\ref{eq:nonlinear}). The general solution has two features that worth mentioning. First, from Eq.~(\ref{eq:nonlinear_x}), we see nonlinearities generate higher harmonics, which have definite phase relations with the fundamental mode of oscillation. Second, from Eq.~(\ref{eq:nonlinear_t}), we see nonlinearities introduce frequency shifts, which depend in the amplitude of oscillations. Although the exact phase relations and frequency shifts depend on specific forms of nonlinearities, these two features are universal for nonlinear problems.

In the end, let use determine the complex amplitude $\alpha$ for the initial value problem $x(t=0)=0$ and $\dot{x}(t=0)=1$. After identifying $t_0=t$, $t_1=\epsilon t$, and $t_2=\epsilon^2 t$, it is obvious that $\alpha=-iA$ is purely imaginary. Then, the real amplitude $A$ satisfies the algebraic equation $1=2A +\frac{\epsilon}{2}(2-9A^2)A -\frac{\epsilon^2}{4}(1-9A^2+14A^4) +O(\epsilon^3)$. Equations of this type can be solved using Kruskal-Newton's method \citep{White10}. After identifying the dominant balance, roots of the algebraic equation can be iteratively approximated to higher order in the perturbation series. To $\epsilon^2$ order, one solution is 
\begin{equation}
A=\frac{1}{2}\big(1+\frac{\epsilon}{16}+\frac{7\epsilon^2}{256}+\dots\big).
\end{equation}
Substituting $\alpha=-iA$ into the general solution Eqs.~(\ref{eq:nonlinear_x}) and (\ref{eq:nonlinear_t}), it is straightforward to check that the perturbative solution matches expansions of the exact solution [Eqs.~(\ref{eq:eg2_x_series}) and (\ref{eq:eg2_t_series})]. Again, we see the multiscale method correctly produces the exact solution as demanded.

\chapter{Linear waves in magnetized cold-fluid plasmas\label{ch:fluid-1st}}

Having developed the mathematical machinery and the physical intuition, we can now apply the multiscale method to describe three-wave scattering in magnetized plasmas. As mentioned in Ch.~\ref{ch:intro}, three-wave scattering in magnetized plasmas is largely an uncharted territory. Therefore, many groundbreaking results can already be obtained by analyzing the cold-fluid plasma model. 

The fluid model is a reduced model of plasmas, which can be obtained by taking moments of the more fundamental kinetic equation. When the plasma thermal speeds are much slower than characteristic speeds of interest, kinetic effects are of little importance. Moreover, when the plasma Debye length is much smaller than spatial scales of interest, each species in the plasma behaves collectively like charged fluid. In reality, these fluids are usually viscous due to collisions. However, dissipative effects may be ignored if the collisional mean free path is much larger and the collision frequency is much smaller than the spatial-temporal scales in the problem. All these conditions can be satisfied when plasma temperature is low and the density is moderate. In this restricted corner of the parameter space, plasmas are well described by the collisionless cold-fluid model. 

Before discussing scattering of waves, it is helpful to review what waves are there in the system. The linear eigenmodes in collisionless, magnetized, cold-fluid plasmas are well known \citep{Stix92}. In this chapter, I will review properties of linear waves from the perspective of linear operators \citep{Shi17scatter}. This new perspective will equip us with tools that will become crucial later.  

\section{The cold-fluid model of plasmas\label{sec:fluid-1st:model}}

\subsection{Fluid-Maxwell's equations\label{sec:fluid-1st:model:equation}}

The cold-fluid equations are the first two moments of the kinetic equation. Taking the first moment gives the continuity equation:
\begin{eqnarray}
\label{eq:continuity}
\partial_t n_{s}&=&-\nabla\cdot(n_{s}\mathbf{v}_{s}).
\end{eqnarray}
The continuity equation describes the conservation of particles of species \textit{s}, whose density is $n_s$ and average velocity is $\mathbf{v}_{s}$. Taking the second moment gives the momentum equation:
\begin{eqnarray}
\label{eq:momentum}
\partial_t\mathbf{v}_{s}&=&-\mathbf{v}_{s}\cdot\nabla\mathbf{v}_{s} +\frac{e_s}{m_s}(\mathbf{E}+\mathbf{v}_{s}\times\mathbf{B}).
\end{eqnarray}
The momentum equation governs how the velocity field $\mathbf{v}_{s}$ change due to both the advection and the Lorentz force, where $e_s$ and $m_s$ are the charge and mass of individual particles of species \textit{s}. Here, we are concerned with non-relativistic fluids. 
It is instructive to count the degrees of freedom. For each species, the cold fluid is completely characterize by the scalar density fields $n_s$ and the vector velocity fields $\mathbf{v}_{s}$. Therefore, there are in total four scalar components, which are completely constrained by the above four fluid equations.

Since plasmas are constituted of charged particles, they not only respond to electromagnetic fields, but also influence how electric and magnetic fields evolve. The electric field $\mathbf{E}$ evolves according to the Maxwell-Amp\`ere's law:
\begin{eqnarray}
\label{eq:Ampere}
\partial_t\mathbf{E}&=& c^2\nabla\times\mathbf{B}-\frac{1}{\epsilon_0}\sum_{s}e_sn_{s}\mathbf{v}_{s},
\end{eqnarray}
where the current density is contributed by all charged species in the system. The initial condition of the time evolution is given by the Gauss' law $\epsilon_0\nabla\cdot\mathbf{E}=\sum_{s}e_sn_{s}$. Once this condition is satisfied initially, the continuity equation [Eq.~(\ref{eq:continuity})] and the Maxwell-Amp\`ere's law [Eq.~(\ref{eq:Ampere})] guarantee that the Gauss' law will be satisfied for all time. On the other hand, the magnetic field $\mathbf{B}$ evolves according to the Faraday's law:
\begin{eqnarray}
\label{eq:Faraday}
\partial_t\mathbf{B}&=&-\nabla\times\mathbf{E},
\end{eqnarray}
which has no explicit dependence on the plasma behavior. The initial condition for this time evolution is the last Maxwell's equation, namely, $\nabla\cdot\mathbf{B}=0$. This Gauss's law for magnetism will be satisfied for all time, once it is satisfied initially.
\subsection{Multiscale expansion\label{sec:fluid-1st:model:expansion}}

The fluid-Maxwell equations [Eqs.~(\ref{eq:continuity})-(\ref{eq:Faraday})] are a system of nonlinear hyperbolic partial differential equations. Such a system of equations are in general difficult to solve. Nevertheless, when fluctuations near an equilibrium are small, nonlinearities may be regarded as perturbations, and the equations may be solved perturbatively. To see when nonlinearities may be regarded as perturbations, we can normalize the equations such that all quantities become dimensionless numbers. For example, we may normalize time to the plasma frequency $\omega_p$ and distance to the skin depth $c/\omega_p$. We may further normalize mass to electron mass $m_e$, charge to elementary charge $e$, density to unperturbed density $n_{s0}$, and velocity to the speed of light $c$. Finally, we can normalize electric field to the unmagnetized wave-breaking field $m_ec\omega_p/e$ and normalize magnetic field to $m_e\omega_p/e$.  With the above normalizations, the fluid-Maxwell equation can be written in a dimensionless form. In this form, nonlinearities are products of small numbers and are therefore even smaller, provided that the perturbations are small in the above units.

In the absence of nonlinearities, the general solution to the fluid-Maxwell system is a spectrum of linear waves with constant amplitudes. Now imagine we have the magic to ramp up nonlinearities adiabatically, then waves will start to scatter one another. Due to weak scattering, amplitudes of waves will evolve slowly in space and time. This physical picture may be translated into a formal mathematical procedure. Formally, to solve the fluid-Maxwell equations peturbatively, it is helpful to keep track of terms by inserting an auxilliary small parameter $\lambda\ll1$ in the perturbation series, and let the adiabatic parameter $\lambda\rightarrow 1$ in the end, mimicking the adiabatic ramping up of nonlinearities. The electric field, magnetic field, density, and velocity can be expanded using asymptotic series:
\begin{eqnarray}
\label{eq:expandE}
\mathbf{E}&=&\mathbf{E}_0+\lambda\mathbf{E}_1+\lambda^2\mathbf{E}_2+\dots,\\
\mathbf{B}&=&\mathbf{B}_0+\lambda\mathbf{B}_1+\lambda^2\mathbf{B}_2+\dots,\\
n_s&=&n_{s0}+\lambda n_{s1}+\lambda^2n_{s2}+\dots,\\
\label{eq:expandV}
\mathbf{v}_s&=&\mathbf{v}_{s0}+\lambda\mathbf{v}_{s1}+\lambda^2\mathbf{v}_{s2}+\dots.
\end{eqnarray}
In quasineutral plasma, which satisfies $\sum_se_sn_{s0}=0$, a self-consistent equilibrium is given by $\mathbf{E}_0=\mathbf{0}$ and $\mathbf{v}_{s0}=\mathbf{0}$, whereas the background magnetic field $\mathbf{B}_0$ and densities $n_{s0}$ can take nonzero constant values. From this equilibrium, a family of equilibria can be obtained by boosting to different inertial frames. Apart from this somewhat trivial family, there also exist many nontrivial self-consistent background states, in which the fields are not constants. In this thesis, I will avoid such unnecessary complications, and develop an essential picture for three-wave scattering using uniformly magnetized plasmas. 
Notice that no assumption is made regarding the nature of the higher-order terms. For example, the average $\langle \mathbf{v}_{s2}\rangle$ is not assumed to be zero. In other words, $\mathbf{v}_{s0}$ is not assumed to be the mean field. The only requirement is that $\mathbf{E}_0$, $\mathbf{B}_0$, $n_{s0}$, and $\mathbf{v}_{s0}$ form a self-consistent background in the absence of other fluctuations.

To remove the secular terms from the perturbation series, let us also expand both the space and the time, following what we have done in Ch.~\ref{ch:multiscale}. In the weakly-nonlinear regime, the multiscale expansions are 
\begin{eqnarray}
x^i&=&x^i_{(0)}+\frac{1}{\lambda} x^i_{(1)}+\frac{1}{\lambda^2}x^i_{(2)}+\dots,\\
t&=&t_{(0)}+\frac{1}{\lambda} t_{(1)}+\frac{1}{\lambda^2}t_{(2)}+\dots,
\end{eqnarray}
where $x^i$ is the $i$-th components of vector $\mathbf{x}$. In the above expansion, $x^i_{(0)}$ is the shortest spatial scale. In comparison, one unit of $x^i_{(1)}$ is $1/\lambda$ times longer that one unit of $x^i_{(0)}$, and so on. Similarly, $t_{(0)}$ is the fastest time scale, and one unit of $t_{(n)}$ is $1/\lambda^n$ times longer that one unit of $t_{(0)}$. In the above multiscale expansion, different spatial and temporal scales are regarded as independent, namely, 
\begin{eqnarray}
\label{eq:multispace}
\partial_i^{(a)}x^j_{(b)}=\delta_i^j\delta^{(a)}_{(b)},\\
\label{eq:multitime}
\partial_t^{(a)}t_{(b)}=\delta^{(a)}_{(b)}.
\end{eqnarray}
Using the chain rule, the total spatial and temporal derivatives are
\begin{eqnarray}
\label{eq:expand_dx}
\partial_i&=&\partial_i^{(0)}+\lambda\partial_i^{(1)}+\lambda^2\partial_i^{(2)}+\dots,\\
\label{eq:expand_dt}
\partial_t&=&\partial_{t(0)}+\lambda\partial_{t(1)}+\lambda^2\partial_{t(2)}+\dots.
\end{eqnarray}
Using these multiscale expansions, together with expansions in field amplitudes (\ref{eq:expandE})-(\ref{eq:expandV}), secular terms can be removed and the perturbative solution will be well-behaved. 
\subsection{First order equations\label{sec:fluid-1st:model:1st}}

To obtain first order equations, we expand fields, space, and time in fluid-Maxwell equations, and collect all the $O(\lambda)$ terms:
\begin{eqnarray}
\label{eq:B1}
\partial_{t(0)}\mathbf{B}_1&=&-\nabla_{(0)}\times\mathbf{E}_1,\\
\label{eq:V1}
\partial_{t(0)}\mathbf{v}_{s1}&=&\frac{e_s}{m_s}(\mathbf{E}_1+\mathbf{v}_{s1}\times\mathbf{B}_0),\\
\label{eq:n1}
\partial_{t(0)} n_{s1}&=&-n_{s0}\nabla_{(0)}\cdot\mathbf{v}_{s1},\\
\label{eq:E1}
\Box^{(0)}_{ij}E_1^j&=&-\frac{1}{\epsilon_0}\sum_{s}e_sn_{s0}\partial_{t(0)}v^i_{s1}.
\end{eqnarray} 
Here, I have written the equations in the order in which I am going to use them. The electric field equation (\ref{eq:E1}) is obtained by substituting the Faraday's law (\ref{eq:Faraday}) into the Maxwell-Amp$\grave{\text{e}}$re's equation (\ref{eq:Ampere}), and then making the multiscale expansion. This procedure introduces the zeroth order differential operator
\begin{equation}
\label{eq:Dij}
\Box^{(0)}_{ij}:=(\partial_{t(0)}^2-c^2\nabla_{(0)}^2)\delta_{ij}+c^2\partial_i^{(0)}\partial_j^{(0)}.
\end{equation}
This operator is the d'Alembert wave operator projected in the transverse direction. This is a manifestation that the vacuum EM waves are transverse waves. 

Since the first order equations are linear, the general solution is a superposition of plane waves. In the weakly coupled regime, let us consider a collection of waves, whose spectra are well separated in the Fourier space. Then, the electric field can be expressed as
\begin{equation}
\label{eq:E1k}
\mathbf{E}_1=\frac{1}{2}\sum_{\mathbf{k}\in\mathbb{K}_1}\mathbf{\mathbfcal{E}}_{\mathbf{k}}^{(1)} e^{i\theta_{\mathbf{k}}},
\end{equation}
where $\mathbf{\mathbfcal{E}}_{\mathbf{k}}^{(1)}(t_{(1)},\mathbf{x}_{(1)};t_{(2)},\mathbf{x}_{(2)};\dots)$ is the slowly-varying complex wave amplitude, and $\theta_{\mathbf{k}}=\mathbf{k}\cdot\mathbf{x}_{(0)}-\omega_{\mathbf{k}}t_{(0)}$ is the fast-varying wave phase. The summation of wave vector $\mathbf{k}$ is over a discrete spectrum $\mathbb{K}_1$. 
In order for $\mathbf{E}_1\in\mathbb{R}^3$ to be a real-valued vector, two conditions must be satisfied. First, whenever the spectrum $\mathbb{K}_1$ contains $\mathbf{k}$, it must also contain $-\mathbf{k}$, with $\omega_{-\mathbf{k}}=-\omega_{\mathbf{k}}$ flips its sign such that the direction of wave propagation is the same. Second, the amplitude $\mathbf{\mathbfcal{E}}_{\mathbf{k}}^{(1)}$ must satisfy the reality condition $\mathbf{\mathbfcal{E}}_{-\mathbf{k}}^{(1)}=\mathbf{\mathbfcal{E}}_{\mathbf{k}}^{(1)*}$, where $*$ denotes the complex conjugate. 
Therefore, it is natural to introduce the following notations:
\begin{eqnarray}
\label{eq:notationz}
\mathbf{z}_{-\mathbf{k}}&=&\mathbf{z}_{\mathbf{k}}^{*},\\
\label{eq:notationa}
\alpha_{-\mathbf{k}}&=&-\alpha_{\mathbf{k}},
\end{eqnarray}
for any complex vector $\mathbf{z}\in\mathbb{C}^3$ and real scalar $\alpha\in\mathbb{R}$ that are labeled with subscript $\mathbf{k}$. For example, the complex vector $\mathbf{\mathbfcal{E}}_{-\mathbf{k}}=\mathbf{\mathbfcal{E}}_{\mathbf{k}}^{*}$, and the real scalar $\theta_{-\mathbf{k}}=-\theta_{\mathbf{k}}$. Using the above notations, the reality condition is conveniently built into the symbols. In the spectral expansion [Eq.~(\ref{eq:E1k})], it is tempting to write the summation over discrete wave vector $\mathbf{k}$ as an integral over some continuous spectrum. However, such a treatment will be very cumbersome due to double counting, because wave amplitude $\mathbf{\mathbfcal{E}}_{\mathbf{k}}$, which can vary on slow spatial and temporal scales, already has an spectral width. 

The first order magnetic field $\mathbf{B}_1$, velocity field $\mathbf{v}_{s1}$, and density field $n_{s1}$ can be expressed in terms of the first-order electric field $\mathbf{E}_1$, by solving Eqs.~(\ref{eq:B1})-(\ref{eq:n1}). These linear PDEs are particularly easy to solve in the Fourier space. Substituting spectral expansion Eq.~(\ref{eq:E1k}) into the first-order fluid-Maxwell equations, we immediately find
\begin{eqnarray}
\label{eq:Bwave}
\mathbf{B}_1&=&\frac{1}{2}\sum_{\mathbf{k}\in\mathbb{K}_1}\frac{\mathbf{k}\times\mathbf{\mathbfcal{E}}^{(1)}_\mathbf{k}}{\omega_{\mathbf{k}}}e^{i\theta_{\mathbf{k}}},\\
\label{eq:vwave}
\mathbf{v}_{s1}&=&\frac{ie_s}{2m_s}\sum_{\mathbf{k}\in\mathbb{K}_1}\frac{\mathbb{F}_{s,\mathbf{k}}\mathbf{\mathbfcal{E}}^{(1)}_\mathbf{k}}{\omega_{\mathbf{k}}}e^{i\theta_{\mathbf{k}}},\\
\label{eq:nwave}
n_{s1}&=&\frac{ie_sn_{s0}}{2m_s}\sum_{\mathbf{k}\in\mathbb{K}_1}\frac{\mathbf{k}\cdot\mathbb{F}_{s,\mathbf{k}}\mathbf{\mathbfcal{E}}^{(1)}_\mathbf{k}}{\omega_{\mathbf{k}}^2}e^{i\theta_{\mathbf{k}}}.
\end{eqnarray}

\subsection{The forcing operator\label{sec:fluid-1st:model:operatorF}}

To solve the cold momentum equation of the form $\hat{\mathbf{v}}=\mathbfcal{E}+i\beta\hat{\mathbf{v}}\times\mathbf{b}$, I have introduced the forcing operator, such that $\hat{\mathbf{v}}=\mathbb{F}\mathbfcal{E}$. In constant background magnetic field, the forcing operator \mbox{$\mathbb{F}_{s,\mathbf{k}}:\mathbb{C}^{3}\rightarrow\mathbb{C}^{3}$} is a linear map.
Here, the subscripts $s$ and $\mathbf{k}$ are merely labels that will only become useful later when I discuss multiple species and waves. 
As shown in Appendix.~\ref{ch:append:F}, the linear map acts on any complex vector $\mathbf{z}\in\mathbb{C}^{3}$ by the following coordinate-independent rule\footnote[1]{In Cartesian coordinate, the forcing operator can be expressed using three of the Gell-Mann matrices as shown by \cite{Daniel17}.}
\begin{equation}
	\label{eq:F}
	\mathbb{F}_{s,\mathbf{k}}\mathbf{z}:=\gamma_{s,\mathbf{k}}^2[\mathbf{z}+i\beta_{s,\mathbf{k}}\mathbf{z}\times\mathbf{b}-\beta_{s,\mathbf{k}}^2(\mathbf{z}\cdot\mathbf{b})\mathbf{b}].
\end{equation}
Here, $\mathbf{b}$ is the unit vector in the $\mathbf{B}_0$ direction
, $\gamma_{s,\mathbf{k}}^2:=1/(1-\beta_{s,\mathbf{k}}^2)$ is the magnetization factor, $\beta_{s,\mathbf{k}}:=\Omega_{s}/\omega_{\mathbf{k}}$ is the magnetization ratio, and $\Omega_{s}=e_sB_0/m_s$ is the gyro frequency of species $s$. It is clear from Eq.~(\ref{eq:vwave}) that the forcing operator $\mathbb{F}_{s,\mathbf{k}}$ is related to the linear electric susceptibility $\chi_{s,\mathbf{k}}$ by
\begin{equation}
	\label{eq:chi}
	\chi_{s,\mathbf{k}}=-\frac{\omega_{ps}^2}{\omega^2_{\mathbf{k}}}\mathbb{F}_{s,\mathbf{k}},
\end{equation}
where $\omega^2_{ps}=e_s^2n_{s0}/\epsilon_0m_s$ is the plasma frequency of species $s$. 
Notice that in the limit $B_0\rightarrow 0$, the forcing operator $\mathbb{F}_{s,\mathbf{k}}\rightarrow\mathbb{I}$ becomes the identity operator, and $\chi_{s}$ becomes the cold unmagnetized susceptibility.

While the susceptibility $\chi_{s,\mathbf{k}}$ is typically used in linear theories, the forcing operator $\mathbb{F}_{s,\mathbf{k}}$ will be extremely useful when we solve the second order equations. Therefore, let us observe a number of important properties of this operator. For brevity, I will suppress the subscripts $s$ and $\mathbf{k}$, with the implied understanding that all quantities have the same subscript. First, by construction, the forcing operator satisfies the following vector identity: 
\begin{equation}
\label{eq:Fvector}
\mathbb{F}\mathbf{z}=\mathbf{z}+i\beta(\mathbb{F}\mathbf{z})\times\mathbf{b}.
\end{equation}
This identity guarantees that the velocity field $\mathbf{v}_{s1}$, given by Eq.~(\ref{eq:vwave}), satisfies the first order momentum equation (\ref{eq:V1}). Second, $\mathbb{F}$ is a self-adjoint operator with respect to the inner product $\langle \mathbf{w},\mathbf{z}\rangle:=\mathbf{w}^\dagger\mathbf{z}$,
\begin{equation}
\label{eq:Fadj}
\mathbf{w}^\dagger\mathbb{F}\mathbf{z}=(\mathbb{F}\mathbf{w})^\dagger\mathbf{z},
\end{equation} 
for all complex vectors $\mathbf{z}, \mathbf{w}\in\mathbb{C}^{3}$. Using this property, we can move $\mathbb{F}$ from acting on one vector to acting on the other vector in an inner product pair. In other words, after choosing a basis, the matrix representation of the linear operator $\mathcal{F}^\dagger=\mathcal{F}$ is Hermitian. Third, using its definition, $\mathbb{F}$ satisfies an obvious identity
\begin{equation}\label{eq:-F}
\mathbb{F}(-\omega)=\mathbb{F}^*(\omega),
\end{equation}
where $*$ denotes complex conjugation. This property can also be written as $\mathbb{F}_{-\mathbf{k}}=\mathbb{F}^*_{\mathbf{k}}$, which is consistent with the notation Eq.~(\ref{eq:notationz}). The forcing operator depends on $\mathbf{k}$ implicitly through $\omega_\mathbf{k}$, where the explicit $\omega$-dependence comes from $\beta(\omega)$.

In addition to the aforementioned properties, the forcing operator also satisfies two nontrivial identities, which are proven in Appendix~\ref{ch:append:F}. First, the square of the forcing operator, namely the product of two forcing operators of the same frequency, satisfies  
\begin{equation}
\label{eq:F2}
\mathbb{F}^2=\mathbb{F}-\omega\frac{\partial\mathbb{F}}{\partial\omega}.
\end{equation}
This identity allows one to express derivatives of the forcing operator in terms of its polynomials, which is very convenient for numerical evaluations. When two frequencies $\omega_1$ and $\omega_2$ are involved, we have the quadratic identity
\begin{equation}\label{eq:F12}
(\beta_1-\beta_2)\mathbb{F}_1\mathbb{F}_2=\beta_1\mathbb{F}_1-\beta_2\mathbb{F}_2.
\end{equation}
This identity allows one to express products of forcing operators in terms of their linear combinations. As a corollary, we see the product $\mathbb{F}_1\mathbb{F}_2=\mathbb{F}_2\mathbb{F}_1$ always commute. Moreover, combining with property Eq.~(\ref{eq:-F}), the above identity can generate a number of other similar identities, which I will not list here. These properties of the forcing operator will enable important simplifications when we solve the second-order equations.

\subsection{The dispersion operator\label{sec:fluid-1st:model:operatorD}}

Finally, there is one more first-order equation that we haven't used, which is the electric-field equation (\ref{eq:E1}). Substituting in the spectral expansion for $\mathbf{E}_1$ [Eq.~(\ref{eq:E1k}] and $\mathbf{v}_{s1}$ [Eq.~(\ref{eq:vwave})], we obtain the first order electric field equation in the momentum space:
\begin{equation}
	\label{eq:E1Fourier}
	\omega_{\mathbf{k}}^2\mathbf{\mathbfcal{E}}^{(1)}_\mathbf{k}+c^2\mathbf{k}\times(\mathbf{k}\times\mathbf{\mathbfcal{E}}^{(1)}_\mathbf{k})=\sum_s\omega_{ps}^2\mathbb{F}_{s,\mathbf{k}}\mathbf{\mathbfcal{E}}^{(1)}_\mathbf{k},
\end{equation}
which must be satisfied for individual wave vector $\mathbf{k}$ in the spectrum. This equation constrains the relations between the wave amplitude $\mathbf{\mathbfcal{E}}^{(1)}_{\mathbf{k}}$, the wave frequency $\omega_{\mathbf{k}}$, and the wave vector $\mathbf{k}$. In operator form, this equation can be written as $\mathbb{D}_{\mathbf{k}}\mathbf{\mathbfcal{E}}^{(1)}_\mathbf{k}=0$, where the dispersion operator is the following linear map:
\begin{equation}
	\label{eq:Dk}
	\mathbb{D}_{\mathbf{k}}=(\omega_{\mathbf{k}}^2-c^2\mathbf{k}^2)\mathbb{I} +c^2\mathbf{k}\mathbf{k}-\sum_s\omega_{ps}^2\mathbb{F}_{s,\mathbf{k}},
\end{equation}
where $\mathbb{I}$ is the identity operator, and $\mathbf{k}\mathbf{k}$ is proportional to the projection operator. The operator equation has nontrivial solutions if and only if the dispersion operator $\mathbb{D}_{\mathbf{k}}$ has a nontrivial kernel. The degeneracy condition requires that the wave vector $\mathbf{k}$ and wave frequency $\omega_{\mathbf{k}}$ satisfy the linear dispersion relation $\det \mathbb{D}(\mathbf{k},\omega_{\mathbf{k}})=0$. When the dispersion relation is indeed satisfied, the nontrivial kernel is the vector space spanned by wave polarizations. 

So far, everything has been derived in a coordinate-free manner using the language of linear operators. Now, in order to solve the electric field equation (\ref{eq:E1Fourier}), it is useful to choose a coordinate system, in which the dispersion operator can then be represented by some matrix. A convenient choice is the usual Cartesian basis $(\mathbf{x}, \mathbf{y}, \mathbf{z})$, where the $z$-axis is chosen to align with the uniform background magnetic field $\mathbf{B}_0$. In this coordinate, the forcing operator $\mathbb{F}_{s,\mathbf{k}}$ has matrix representation
\begin{eqnarray}
\mathcal{F}_{s,\mathbf{k}} =
\left( \begin{array}{ccc}
\gamma^2_{s,\mathbf{k}} & i\beta_{s,\mathbf{k}}\gamma^2_{s,\mathbf{k}} & 0 \\
-i\beta_{s,\mathbf{k}}\gamma^2_{s,\mathbf{k}} & \gamma^2_{s,\mathbf{k}} & 0 \\
0 & 0 & 1
\end{array} \right).
\end{eqnarray}
Having fixed the $z$-axis, we can rotate the coordinate system, such that the wave vector is in the $xz$-plane. Then $\mathbf{k}=(k_\perp,0,k_\parallel)=k(\sin\theta,0,\cos\theta)$, where $\theta$ is the angle between $\mathbf{k}$ and $\mathbf{b}$. In this coordinate system, the dispersion operator $\mathbb{D}_{\mathbf{k}}$ is represented by the matrix
\begin{equation}
\mathcal{D}_{\mathbf{k}} =
\left( \begin{array}{ccc}
\omega_{\mathbf{k}}^2-c^2k_\parallel^2-\sum_s\omega_{ps}^2\gamma^2_{s,\mathbf{k}} & i\sum_s\omega_{ps}\beta_{s,\mathbf{k}}\gamma^2_{s,\mathbf{k}} & c^2k_\perp k_\parallel \\
-i\sum\omega_{ps}^2\beta_{s,\mathbf{k}}\gamma^2_{s,\mathbf{k}} & \omega_{\mathbf{k}}^2-c^2k^2-\sum_s\omega_{ps}^2\gamma^2_{s,\mathbf{k}} & 0 \\
c^2k_\perp k_\parallel & 0 & \omega_{\mathbf{k}}^2-c^2k^2-\sum_s\omega_{ps}^2
\end{array} \right).
\end{equation}
Using Stix's notation \citep{Stix92}, we can obtain a more compact expression
\begin{eqnarray}
\label{eq:Dkw}
\frac{\mathcal{D}}{\omega^2} =
\left( \begin{array}{ccc}
S-n^2\cos^2\theta & -iD & n^2\sin\theta\cos\theta \\
iD & S-n^2 & 0 \\
n^2\sin\theta\cos\theta & 0 & P-n^2\sin^2\theta
\end{array} \right),
\end{eqnarray}
where $n=ck/\omega$ is the refractive index. The components of the dielectric tensor are
\begin{eqnarray}
\label{eq:StixS}
S&=&1-\sum_s\frac{\omega_{ps}^2}{\omega^2-\Omega_s^2},\\
\label{eq:StixD}
D&=&\sum_s\frac{\Omega_s}{\omega}\frac{\omega_{ps}^2}{\omega^2-\Omega_s^2},\\
\label{eq:StixP}
P&=&1-\sum_s\frac{\omega_{ps}^2}{\omega^2},
\end{eqnarray}
where I have omitted the subscript $\mathbf{k}$. The expressions for $S$ and $D$ can be simplified in quasi-neutral plasmas. In particular, in quasi-neutral electron-ion plasmas, $n_e=Z_in_i$, so $\Omega_i\omega_{pe}^2+\Omega_e\omega_{pi}^2=0$ and $\Omega_i^2\omega_{pe}^2+\Omega_e^2\omega_{pi}^2+\omega_p^2\Omega_e\Omega_i=0$, where $\omega_p^2=\sum_s\omega_{ps}^2$ is the total plasma frequency squared. 

In addition to the above Cartesian coordinate, there is another special orthonormal coordinate system. This is because in magnetized plasmas, there are two special vectors. One is the background magnetic field $\mathbf{B}_0$, and the other is the wave vector $\mathbf{k}$. While the $(\mathbf{x}, \mathbf{y}, \mathbf{z})$ coordinate is chosen to align with $\mathbf{B}_0$, 
a more convenient basis 
is the wave basis $(\hat{\mathbf{k}}, \mathbf{y}, \hat{\mathbf{k}}\times\mathbf{y})$, which is chosen to align with $\mathbf{k}$. In general, these two special coordinates systems are related by the linear transformation $(\hat{\mathbf{k}}, \mathbf{y}, \hat{\mathbf{k}}\times\mathbf{y})=(\mathbf{x}, \mathbf{y}, \mathbf{z})R$, where the rotation matrix is given by
\begin{eqnarray}
R=
\left( \begin{array}{ccc}
\sin\theta & 0 & -\cos\theta \\
0 & 1 & 0 \\
\cos\theta & 0 &\sin\theta
\end{array} \right).
\end{eqnarray}
Under coordinate transformation, the matrix representation of a linear operator is transformed by $\mathcal{D}\rightarrow \mathcal{D}'=R^{-1}\mathcal{D}R$. Therefore, in the wave basis, the dispersion operator $\mathbb{D}_{\mathbf{k}}$ is represented by the matrix
\begin{eqnarray}
\label{eq:Dkwk}
\frac{\mathcal{D}'}{\omega^2} =
\left( \begin{array}{ccc}
S\sin^2\theta+P\cos^2\theta & -iD\sin\theta & (P-S)\sin\theta\cos\theta \\
iD\sin\theta  & S-n^2 & -iD\cos\theta \\
(P-S)\sin\theta\cos\theta & iD\cos\theta & S\cos^2\theta+P\sin^2\theta-n^2
\end{array} \right).
\end{eqnarray}
In this representation, it is easy to see that waves in magnetized plasmas are neither longitudinal nor transverse in general, because $\mathcal{D}'_{12}=-\mathcal{D}'_{21}\ne0$ and $\mathcal{D}'_{13}=\mathcal{D}'_{31}\ne0$ except at special angles. A special case is the unmagnetized plasma, in which $S=P$ and $D=0$, so the matrix $\mathcal{D}'_{\mathbf{k}}$ becomes diagonal. 

\section{Wave dispersion relations and polarizations\label{sec:fluid-1st:eigen}}

In the remaining part of this chapter, I will abbreviate the first-order Fourier amplitude $\mathbfcal{E}_\mathbf{k}^{(1)}$ as $\mathbfcal{E}$. This simplified notation cause no confusion here, now that we are only discussing one linear wave with a given wave vector $\mathbf{k}$. In the next chapter, I will restore the full notation, when we discuss nonlinear scattering of many waves. Here, to solve the linear equation $\mathcal{D}'\mathbfcal{E}=0$ in the wave frame, it is convenient to express the electric field as $\mathbfcal{E}=\mathcal{E}_{k}\hat{\mathbf{k}} +\mathcal{E}_{y}\mathbf{y} +\mathcal{E}_{\times}\hat{\mathbf{k}}\times\mathbf{y}$, where the subscripts now denote the components of the electric field.

\subsection{Parallel propagation\label{sec:fluid-1st:dispersion:para}}
 
When wave propagates parallel to the magnetic field, namely, when $\mathbf{k}\parallel\mathbf{B}_0$ and $\theta=0^\circ$, the dispersion matrix becomes
\begin{eqnarray}
\label{eq:Dkwk_para}
\frac{\mathcal{D}'}{\omega^2} =
\left( \begin{array}{ccc}
P & 0 & 0 \\
0 & S-n^2 & -iD \\
0 & iD & S-n^2
\end{array} \right).
\end{eqnarray}
We see at this special angle, the longitudinal mode decouples from the transverse modes, and satisfies the dispersion relation
\begin{equation}
P=0.
\end{equation}
In cold plasma, this is simply the Langmuir wave $\omega^2=\omega_p^2$ propagating along the background magnetic field, with the wave electric field 
\begin{equation}
\mathbfcal{E}=\mathcal{E}_k\hat{\mathbf{k}},
\end{equation}
polarized along the wave vector. By Faraday's law, the wave magnetic field  $\mathbf{B}\propto\mathbf{k}\times\mathbfcal{E}=\mathbf{0}$. Therefore, the longitudinal mode is purely electrostatic and is unaffected by the background magnetic field.

On the other hand, the transverse modes are electromagnetic, and satisfy the dispersion relation $(S-n^2)^2=D^2$. Using Stix's notation, one EM wave satisfies
\begin{equation}
n^2=R:=S+D.
\end{equation}
Solving the matrix equation $\mathcal{D}'\mathbfcal{E}=0$, this transverse mode is polarized in such a way that
\begin{equation}
\mathcal{E}_{\times}=i\mathcal{E}_{y}.
\end{equation}
This mode is right-handed circularly polarized, and is therefore denoted, in Stix's notation, as the R wave. Naturally, the other EM wave is left-handed circularly polarized. The L wave satisfies the dispersion relation
\begin{equation}
n^2=L:=S-D,
\end{equation}
and the polarization is such that 
\begin{equation}
\mathcal{E}_{\times}=-i\mathcal{E}_{y}.
\end{equation}
Notice that $R(-\omega)=L(\omega)$, so the R wave and the L wave are images of one another under time reversal. From Eqs.~(\ref{eq:StixS}) and (\ref{eq:StixD}), we see $R$ and $L$ have poles at cyclotron frequencies. Therefore, these waves are split into $N_s+1$ branches in general, where $N_s$ is the number of cyclotron resonances in the system. Two of these branches asymptote to the two vacuum EM waves, and the other branches asymptote to cyclotron resonances.
\subsection{Perpendicular propagation\label{sec:fluid-1st:dispersion:perp}}
  
When wave propagates perpendicular to the magnetic field, namely, when $\mathbf{k}\perp\mathbf{B}_0$ and $\theta=90^\circ$, the dispersion matrix becomes
\begin{eqnarray}
\label{eq:Dkwk_perp}
\frac{\mathcal{D}'}{\omega^2} =
\left( \begin{array}{ccc}
S & -iD & 0 \\
iD& S-n^2 & 0 \\
0 & 0 & P-n^2
\end{array} \right).
\end{eqnarray}
We see at this special angle, a transverse EM mode decouples from two other modes, and satisfies the dispersion relation
\begin{equation}
n^2=P.
\end{equation}
This dispersion relation can be written as $\omega^2=\omega_p^2+c^2k^2$, which is the same dispersion relation of EM waves in unmagnetized plasmas. Therefore, this wave is called the ordinary (O) wave, whose electric field is
\begin{equation}
\mathbfcal{E}=\mathcal{E}_\times\mathbf{b},
\end{equation}
is polarized along the $\hat{\mathbf{k}}\times\mathbf{y}=\mathbf{b}$ direction. In this linear mode, charged particles move along the background magnetic field. Therefore, they do not feel the $\mathbf{v}\times\mathbf{B}_0$ force, and the magnetic field has no effect on the linear wave dispersion relation.

The other two modes have wave electric field in the $ky$-plane. Since $\mathbfcal{E}\perp\mathbf{B}_0$, charged particles 
feel the $\mathbf{v}\times\mathbf{B}_0$ force, and the longitudinal and transverse motion are hybridized by the dispersion relation
\begin{equation}
\label{eq:n2perp}
n^2=\frac{RL}{S},
\end{equation} 
which is symmetric under time reversal $\omega\rightarrow-\omega$. Solving the matrix equation $\mathcal{D}'\mathbfcal{E}=0$, the wave polarization is such that
\begin{equation}
\mathbfcal{E}\propto D\;\hat{\mathbf{k}}-iS\;\mathbf{y}.
\end{equation}
Since the RHS of Eq.~(\ref{eq:n2perp}) has $N_s$ poles, where $N_s$ is again the number of charged species, there are $N_s+1$ branches. Among these, one branch satisfies $\omega\rightarrow ck$ as $ck\rightarrow\infty$. In this limit, $S\rightarrow1$ and $D\rightarrow0$, and the wave asymptotes to purely transverse EM wave. However, at finite frequency, this EM wave deviates from the unmagnetized EM wave, and is therefore called the extraordinary (X) wave. The other $N_s$ branches of Eq.~(\ref{eq:n2perp}) asymptotes to hybrid resonances, which satisfies $\omega\rightarrow \omega_r$ as $ck\rightarrow\infty$, where $S(\omega_r)=0$. In this limit, the waves asymptote to purely longitudinal waves. For example, in two-species electron-ion plasmas, there are two hybrid waves. The higher-frequency wave is usually referred to as the upper-hybrid wave, whose frequency is mostly determined by electron frequencies [Eq.~(\ref{eq:w_UH})]. On the other hand, the lower-frequency wave is contributed by both electrons and ions [Eq.~(\ref{eq:w_LH})], and is usually called the lower-hybrid wave.

\subsection{Oblique propagation\label{sec:fluid-1st:dispersion:oblique}}

\begin{figure}[t]
	\centering
	\includegraphics[angle=0,width=0.7\textwidth]{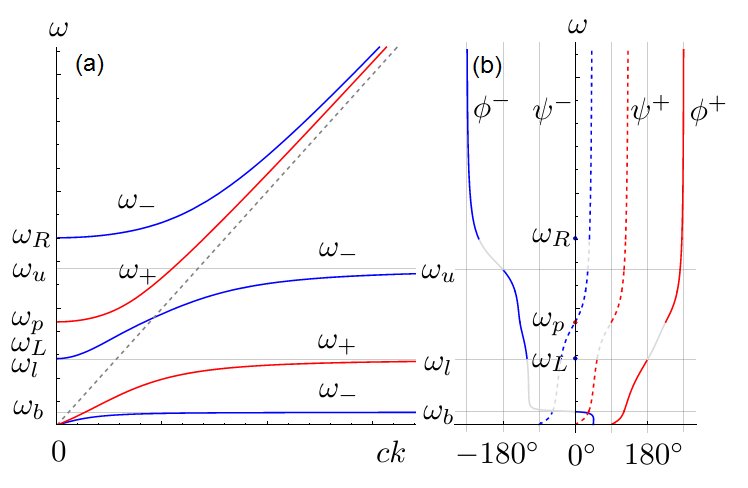}
	\caption[Oblique propagation in magnetized cold-fluid plasma]{Linear wave dispersion relations (a) and polarization angles (b) at $\theta=45^\circ$ in a cold electron-ion plasma with $m_i/m_e=10$ and $|\Omega_e|/\omega_{pe}=1.2$. Both the $n^2_+$ (red) and the $n^2_-$ (blue) solutions contain an electromagnetic-like branch and electrostatic-like branches. The electromagnetic-like branches asymptote to vacuum light wave $\omega\rightarrow ck$ when $ck\rightarrow\infty$, where the waves become transverse ($\phi\rightarrow90^\circ$ mod $180^\circ$). The electrostatic-like branches asymptote to resonances $\omega\rightarrow \omega_r$ as $ck\rightarrow\infty$, where the waves become longitudinal ($\phi\rightarrow 0^\circ$ mod $180^\circ$). The waves are in general elliptically polarized ($\psi\ne0^\circ$ mod $90^\circ$), except at special angles. }
	\label{fig:Linear}
\end{figure}

When waves propagate at general angles, all three modes are mixed, and the dispersion relation can be found by setting $\det\mathcal{D}'=0$. Using Stix's notation, the dispersion relation can be written as
\begin{equation}
\label{eq:disp}
An^4-Bn^2+C=0,
\end{equation}
where the dispersion coefficients 
\begin{eqnarray}
A&=&S\sin^2\theta+P\cos^2\theta,\\
B&=&RL\sin^2\theta+PS(1+\cos^2\theta),\\
C&=&PRL.
\end{eqnarray}
In cold plasma, these coefficients are independent of the wave vector. Moreover, they are invariant under time reversal. Therefore, the dispersion coefficients are functions of $\omega^2$ only, and we can use the quadratic equation (\ref{eq:disp}) to express $n^2$ as functions of $\omega^2$:
\begin{eqnarray}
n_{\pm}^2=\frac{B\pm F}{2A},
\end{eqnarray}
where $F^2=B^2-4AC=(RL-PS)^2\sin^4\theta+4P^2D^2\cos^2\theta$. Since $F^2\ge0$, the two solutions $n_{\pm}^2$ are always real. However, $n_{\pm}^2$ are not always positive. In fact, as we have seen in the special cases, each of these two solutions contains poles that cut the dispersion relation into many separate branches. For example, in a two-species quasi-neutral plasma (Fig.~\ref{fig:Linear}a), the dispersion relation contains two electromagnetic-like branches, for which $\omega\rightarrow ck$ as $ck\rightarrow\infty$, as well as four electrostatic-like branches, for which $\omega\rightarrow \omega_r$ as $ck\rightarrow\infty$, where $\omega_r$ is some resonance frequencies\footnote[1]{The lowest-frequency branch with $\omega=0$ is trivial in cold magnetized plasmas. }.

The resonance frequencies are special to cold magnetized plasmas. These are the poles of the refractive index, at which $n^2(\omega_r^2)=\infty$. 
From solutions to the quadratic equation, we see resonance frequencies satisfy
\begin{equation}
\label{eq:coefA}
A(\omega_r^2)=0.
\end{equation}
In a plasma with $N_s$ species, $S(\omega^2)$ contains $N_s$ poles at cyclotron frequencies $\omega^2=\Omega_s^2$, and $P(\omega^2)$ contains a single pole at $\omega^2=0$. Hence, Eq.~(\ref{eq:coefA}) has $N_s+1$ positive solutions, giving $N_s+1$ resonances that are angle dependent $\omega_r^2=\omega_r^2(\theta)$. 
When $\theta\rightarrow0^\circ$, $\omega_r$ approaches the cyclotron frequencies and the plasma frequency. On the other hand, when $\theta\rightarrow90^\circ$, $\omega_r$ approaches the hybrid frequencies, as well as zero. At general angles, the resonance frequencies take values in between these special values. For example, in two-species plasmas (Fig.~\ref{fig:Linear}), there are three nontrivial resonances: the upper resonance $\omega_u$, the lower resonance $\omega_l$, and the bottom resonance $\omega_b$. Their general formulas, as well as asymptotic expressions are given in Appendix~\ref{ch:append:resonance}. The asymptotic expressions will be useful later when we analytically evaluate the wave energy coefficient and the three-wave scattering coefficient. 

The other set of special frequencies are the cutoff frequencies $\omega_c$, which are very robust even when thermal effects are taken into account. The cutoffs are zeros of the refractive index, at which $n^2(\omega_c^2)=0$. From Eq.~(\ref{eq:disp}), we see cutoff frequencies satisfy
\begin{equation}
\label{eq:coefC}
C(\omega_c^2)=0.
\end{equation}
In a plasma with $N_s$ species, $RL$ contains $N_s$ poles at cyclotron frequencies $\omega^2=\Omega_s^2$ and an additional pole at $\omega^2=0$ if the plasma is not quasi-neutral. Since $\omega^2=\omega_p^2$ is always an zero of $P$, Eq.~(\ref{eq:coefC}) has $N_s+1$ solutions in quasi-neutral plasmas, and $N_s+2$ solutions in non-neutral plasmas. Notice that in addition to emanating from the cutoffs, dispersion branches $\omega(ck)$ can also emanate from zero frequency. The low-frequency limit of these gapless branches are the fast magnetohydrodynamic (MHD) wave and the Alfv\'en wave. If the temperature is nonzero, then there is an additional branch, corresponding to the slow MHD wave, which is now trivial ($\omega=0$) in the cold case. For example, in quasi-neutral two-species plasma (Fig.~\ref{fig:Linear}a), three waves emanate from the three cutoffs at $\omega_R$, $\omega_p$, and $\omega_L$, while two waves emanate from $\omega=0$. On these MHD branches, since $\omega\rightarrow0$ as $ck\rightarrow0$, the refractive index $n^2=c^2k^2/\omega^2$ is in general nonzero. 

Finally, let us determine the polarization of waves. In spherical coordinate, the wave electric field $\mathcal{E}_k=\mathcal{E}\cos\phi$, $\mathcal{E}_y=-i\mathcal{E}\sin\phi\cos\psi$, and $\mathcal{E}_\times=\mathcal{E}\sin\phi\sin\psi$. Using row operations to solve the matrix equation $\mathcal{D}'\mathbfcal{E}=0$, it is easy to find that the polarization angles satisfy
\begin{eqnarray}
\label{eq:polarization_psi}
\tan\psi&=&\frac{Sn^2-RL}{n^2D\cos\theta}, \\
\label{eq:polarization_phi}
\tan\phi&=&\frac{P\cos\theta}{(n^2-P)\sin\theta\sin\psi}.
\end{eqnarray}
Notice that $\mathcal{E}_\times/\mathcal{E}_y=i\tan\psi$ is purely imaginary except in special cases. Therefore, waves are elliptically polarized in general. Also notice that the polarization ray $\hat{\mathbfcal{E}}$ is invariant under transformations $(\phi,\psi)\rightarrow(\phi\pm180^\circ,\psi)$ and  $(\phi,\psi)\rightarrow(-\phi,\psi\pm180^\circ)$. Therefore, the polarization angles should be interpreted up to these identity transformations. Finally, notice that $\psi_\pm$ for the $n^2_\pm$ solutions satisfies the identity $\tan\psi_+\tan\psi_-=-1$. Hence, polarizations of the two eigenmodes with the same frequency are always orthogonal in the transverse plane. Using two-species plasma as an example (Fig.~\ref{fig:Linear}b), we see when frequencies approach resonances, the waves becomes longitudinal ($\phi=0^\circ \mod 180^\circ$). On the other hand, the wave becomes transverse ($\phi=90^\circ \mod 180^\circ$) when frequencies approach infinity. At other frequencies, the waves are usually neither transverse nor longitudinal.

\section{Energy of linear waves\label{sec:fluid-1st:energy}}

\subsection{The energy operator\label{sec:fluid-1st:energy:general}}

To introduce one more operator that will be useful for solving the second order equations, let us calculate the wave energy density. The average energy carried by linear waves can be found by summing up average energy carried by fields and particles. For a single linear wave, after averaging on $t_{(0)}$ and $\mathbf{x}_{(0)}$ scales, the wave energy density in the absence of background plasma flow is
\begin{equation}
U=U_E+U_B+U_V.
\end{equation}
The average energy density contained in the oscillating electric field is
\begin{eqnarray}
\label{eq:U_E}
U_E=\frac{\epsilon_0}{2}\langle\mathbf{E}_1^2\rangle_{(0)}=\frac{\epsilon_0}{4}\mathbfcal{E}^*\cdot\mathbfcal{E},
\end{eqnarray}
where $\mathbfcal{E}$ is the slowly varying Fourier amplitude of the plane wave. Using Faraday's law to express the wave magnetic field in terms of the wave electric field [Eq.~(\ref{eq:Bwave})], the averaged energy density contained in the magnetic field of the plane wave can be written as
\begin{eqnarray}
\label{eq:U_B}
\nonumber
U_B&=&\frac{1}{2\mu_0}\langle\mathbf{B}_1^2\rangle_{(0)}=\frac{\epsilon_0c^2}{4}\frac{\mathbf{k}\times\mathbfcal{E}^*}{\omega_{\mathbf{k}}}\cdot\frac{\mathbf{k}\times\mathbfcal{E}}{\omega_{\mathbf{k}}} \\
\nonumber
&=&\frac{\epsilon_0c^2}{4\omega_{\mathbf{k}}^2} \mathbfcal{E}^*\cdot(\mathbf{k}^2\mathbb{I} -\mathbf{k}\mathbf{k})\mathbfcal{E}\\
&=&\frac{\epsilon_0}{4}\mathbfcal{E}^*\cdot(\mathbb{I} -\sum_s\frac{\omega_{ps}^2}{\omega_{\mathbf{k}}^2}\mathbb{F}_{s,\mathbf{k}})\mathbfcal{E}.
\end{eqnarray}
To obtain the last equality, I have used the fact that $\mathbb{D}_{\mathbf{k}}\mathbfcal{E}=\mathbf{0}$ is a linear eigenmode, where the dispersion tensor is given by Eq.~(\ref{eq:Dk}). In the vacuum, where the density of charged species is zero, the plasma frequency $\omega_{ps}^2=0$. In this case, the energy of an vacuum EM wave is equally partitioned between electric and magnetic fields. In contrast, when plasma is present, $U_E$ and $U_B$ are different in general. Notice that the cold forcing operator $\mathbb{F}_{s,\mathbf{k}}$ is Hermitian [Eq.~(\ref{eq:Fadj})]. Hence, the second term in Eq.~(\ref{eq:U_B}) is always real-valued. However, it can be either positive or negative, because $\mathbb{F}_{s,\mathbf{k}}$ is not positive definite. In fact, $\mathbb{F}_{s,\mathbf{k}}$ has a pole at the cyclotron resonance, and its three eigenvalues are $1, 1/(1-\beta_{s,\mathbf{k}})$, and $1/(1+\beta_{s,\mathbf{k}})$. Therefore, although the magnetic energy density $U_B\ge0$ is always bounded from below, it can be larger or smaller than the electric energy density $U_E$, depending on the wave frequency. 
Finally, using the forcing operator to express particle's oscillation velocity in terms of the wave electric field [Eq.~(\ref{eq:vwave})], the average wave kinetic energy density carried by particles' oscillatory motion can be written as
\begin{eqnarray}
\label{eq:U_v}
\nonumber
U_V&=&\frac{1}{2}\sum_sn_{s0}m_s\langle\mathbf{v}_{s1}^2\rangle_{(0)}=\frac{\epsilon_0}{4} \sum_s\frac{\omega_{ps}^2}{\omega_{\mathbf{k}}^2} (\mathbb{F}_{s,\mathbf{k}}\mathbfcal{E})^\dagger (\mathbb{F}_{s,\mathbf{k}}\mathbfcal{E}) \\
&=&\frac{\epsilon_0}{4} \sum_s\frac{\omega_{ps}^2}{\omega_{\mathbf{k}}^2} \mathbfcal{E}^*\cdot \mathbb{F}_{s,\mathbf{k}}^2\mathbfcal{E}, 
\end{eqnarray}
where I have used the self-adjoint property [Eq.~(\ref{eq:Fadj})] of the forcing operator. Not surprisingly, this kinetic energy exists only when charged particles are present $(\omega_{ps}^2\ne0)$. Similar to the magnetic energy density, the kinetic energy density $U_V\ge0$ can be large or small, depending on the wave frequency. Substantially more complicated than the unmagnetized case, now the total wave energy can be distributed in a large variety of ways among the electric, magnetic, and kinetic energy.


Now, summing up the electric, magnetic, and kinetic contributions, the total wave energy density can be expressed in terms of the wave electric field amplitude as
\begin{eqnarray}
\label{eq:U}
\nonumber
U&=&\frac{\epsilon_0}{4}\mathbf{\mathbfcal{E}}^{*}\cdot\Big[2\mathbb{I}+\sum_s\frac{\omega_{ps}^2}{\omega_{\mathbf{k}}^2} \Big(\mathbb{F}_{s,\mathbf{k}}^2- \mathbb{F}_{s,\mathbf{k}}\Big)\Big]\mathbfcal{E} \\
&=& \frac{\epsilon_0}{4}\mathbf{\mathbfcal{E}}^{*}\cdot \Big(2\mathbb{I} -\sum_s\frac{\omega_{ps}^2}{\omega_\mathbf{k}} \frac{\partial\mathbb{F}_{s,\mathbf{k}}}{\partial\omega_\mathbf{k}}\Big) \mathbfcal{E} \\
\nonumber
&=&\frac{\epsilon_0}{4}\mathbf{\mathbfcal{E}}^{*}\cdot \frac{1}{\omega_\mathbf{k}}\frac{\partial(\omega_\mathbf{k}^2\epsilon_\mathbf{k})}{\partial\omega_\mathbf{k}} \mathbfcal{E}.
\end{eqnarray}
To obtain the second equality, I have used the derivative property of the forcing operator [Eq.~(\ref{eq:F2})]. On the last line, $\epsilon_\mathbf{k}=\mathbb{I}+\sum_s\chi_{s,\mathbf{k}}$ is the dielectric tensor, where the linear susceptibility $\chi_{s,\mathbf{k}}$ is related to the forcing operator by Eq.~(\ref{eq:chi}). In the above expression of wave energy density, the key term is the wave energy operator, which is defined by
\begin{eqnarray}
	\label{eq:Hk}
	\mathbb{H}_{\mathbf{k}}&:=&2\mathbb{I}-\sum_s\frac{\omega_{ps}^2}{\omega_\mathbf{k}} \frac{\partial\mathbb{F}_{s,\mathbf{k}}}{\partial\omega_\mathbf{k}}.
\end{eqnarray}
The energy operator $\mathbb{H}_{\mathbf{k}}$ not only introduces a compact formula for wave energy, but also introduces the following alternative form of the first-order electric field equation. Taking $\partial/\partial k_l$ derivative on both side of Eq.~(\ref{eq:E1Fourier}), we obtain
\begin{equation}
	\label{eq:dE1}
	\frac{\partial\omega_{\mathbf{k}}}{\partial k_l}\omega_\mathbf{k}\mathbb{H}_{\mathbf{k}}^{ij}\mathcal{E}^{j}
	=c^2(2k_l\delta_{ij}-k_i\delta_{jl}-k_j\delta_{il})\mathcal{E}^{j}.
\end{equation}
This alternative form of the first-order electric-field equation is expressed in terms of wave energy instead of the wave dispersion. Although one may not think of using this equation to solve for $\mathbfcal{E}$, it will become useful when we solve the second-order equations. 

\subsection{The wave energy coefficient\label{sec:fluid-1st:energy:example}}

The energy of a linear wave depends on three factors: the wave amplitude, the wave polarization, and the wave dispersion relation. For a linear wave, since one can arbitrarily scale the wave amplitude, the quadratic dependence of the wave energy on the wave amplitude is somewhat trivial. Using the wave energy operator, we can write
\begin{equation}
	U=\frac{\epsilon_0}{4}\mathbf{\mathbfcal{E}}^{*}_\mathbf{k}\cdot \mathbb{H}_{\mathbf{k}}\mathbfcal{E}_{\mathbf{k}} =\frac{1}{2}\epsilon_0u_{\mathbf{k}}|\mathcal{E}_\mathbf{k}|^2,
\end{equation} 
we see the wave energy in plasmas is proportional to the wave energy in the vacuum. What is nontrivial is the proportionality coefficient, namely, the wave energy coefficient
\begin{eqnarray}
\label{eq:uk}
\nonumber
u_{\mathbf{k}}&:=&\frac{1}{2}\mathbf{e}_{\mathbf{k}}^\dagger\mathbb{H}_{\mathbf{k}}\mathbf{e}_{\mathbf{k}}\\
&=&1+\sum_s\frac{\omega_{ps}^2}{2\omega_\mathbf{k}^2} \Big(\mathbf{f}_{s,\mathbf{k}}^\dagger \mathbf{f}_{s,\mathbf{k}}- \mathbf{e}_{\mathbf{k}}^\dagger\mathbf{f}_{s,\mathbf{k}}\Big).
\end{eqnarray}
Here $\mathbf{f}_{s,\mathbf{k}}=\mathbb{F}_{s,\mathbf{k}}\mathbf{e}_{\mathbf{k}}$, and $\mathbf{e}_{\mathbf{k}}$ is the unit polarization vector of the wave, whose general formula is given by Eqs.~(\ref{eq:polarization_psi}) and (\ref{eq:polarization_phi}) in spherical coordinate. 
When evaluating the wave energy coefficient, instead of using its definition directly, it is usually more convenient to use the second line as a formula. 
In fact, using the second line, we can attribute the total wave energy coefficient $u_{\mathbf{k}}=u_{\mathbf{k},E}+u_{\mathbf{k},B}+u_{\mathbf{k},V}$ to the electric field contribution $u_{\mathbf{k},E}=1/2$, the magnetic field contribution $u_{\mathbf{k},B}=1/2-\sum_s\omega_{ps}^2\mathbf{e}_{\mathbf{k}}^\dagger\mathbf{f}_{s,\mathbf{k}}/2\omega_{\mathbf{k}}^2$, and the kinetic contribution $u_{\mathbf{k},V}=\sum_s\omega_{ps}^2\mathbf{f}_{s,\mathbf{k}}^\dagger\mathbf{f}_{s,\mathbf{k}}/2\omega_{\mathbf{k}}^2$, which can be evaluated separately.

Let us observe a number of special cases for the wave energy coefficient. First, it is obvious that in vacuum, $u_{\mathbf{k}}=1$. Second, in cold unmagnetized plasma, the forcing operator $\mathbb{F}_{s,\mathbf{k}}=\mathbb{I}$. Therefore, $\mathbf{f}_{s,\mathbf{k}}=\mathbf{e}_{\mathbf{k}}$, and the last two terms in Eq.~(\ref{eq:uk}) cancel, which gives $u_{\mathbf{k}}=1$ as in the vacuum case. We see for unmagnetized waves, the magnetic energy is reduced by exactly the same amount as gained by charged particles. 
More generally, for EM waves in magnetized plasmas, when $\omega\rightarrow\infty$, the forcing operator $\mathbb{F}_{s,\mathbf{k}}\rightarrow\mathbb{I}$. Hence, the wave energy coefficient for high-frequency EM waves in magnetized plasma $u_{\mathbf{k}}\simeq1$. This is expected, because particles do not have time to respond in high-frequency EM waves, regardless of whether there is a background magnetic field or not. Since there is little energy contained in particles' motion, the wave energy coefficient is simply its vacuum value. 
However, for low frequency EM waves in magnetized plasmas, particles gyro motion superimposes with their wave motion. Therefore, particles usually gain a different amount of energy than what is lost by the wave magnetic field.

In addition to quasi-transverse waves, the other set of special waves are quasi-longitudinal waves. From Sec.~\ref{sec:fluid-1st:eigen}, we know these waves are electrostatic resonances, whose frequency $\omega\rightarrow\omega_r$, and polarization $\mathbf{e}\rightarrow\hat{\mathbf{k}}$. In this case, it is easy to compute the wave energy coefficient in the $(\mathbf{x}, \mathbf{y}, \mathbf{z})$ coordinate. From the definition of the forcing operator, $\mathbb{F}\hat{\mathbf{k}}=\gamma^2(\hat{\mathbf{k}}-i\beta\sin\theta\mathbf{y}-\beta^2\cos\theta\mathbf{b})$,
where I have abbreviated subscripts for simplicity. Then, using $\gamma^2(1-\beta^2)=1$, the inner products
\begin{eqnarray}
\hat{\mathbf{k}}\cdot\mathbb{F}\hat{\mathbf{k}}&=&\gamma^2(1-\beta^2\cos^2\theta),\\
(\mathbb{F}\hat{\mathbf{k}})^\dagger(\mathbb{F}\hat{\mathbf{k}})&=&\gamma^4(1+\beta^2)\sin^2\theta+\cos^2\theta,
\end{eqnarray}
which gives $\mathbf{f}^\dagger \mathbf{f}- \mathbf{e}^\dagger\mathbf{f}=2\gamma^4\beta^2\sin^2\theta$. The total wave energy coefficient for longitudinal waves is thereof
\begin{equation}
\label{eq:u_coef_L}
u=1+\sum_s\frac{\omega_{ps}^2}{\omega^2} \gamma_s^4\beta_s^2\sin^2\theta.
\end{equation}
Here, I have restored the species subscript, while keeping the wave subscript $\mathbf{k}$ suppressed for simplicity. Notice that although $\gamma^2$ may be negative, $\gamma^4$ is always positive. Therefore, the wave energy coefficient $u\ge1$ is always larger than or equal to its vacuum value.

As examples, consider wave propagation perpendicular $(\theta=90^\circ)$ to the background magnetic field in a two-species electron-ion plasma. In this case, there are two longitudinal waves, namely, the upper-hybrid (UH) and lower-hybrid (LH) waves. For the UH wave, since its frequency $\omega_{UH}^2\simeq\omega_p^2+\Omega_e^2$ [Eq.~(\ref{eq:w_UH})] is much larger than ion frequencies, ion contribution is subdominant. In fact, $\beta_e^2=\Omega_e^2/\omega_{UH}^2$ so $\gamma_e^2\simeq \omega_{UH}^2/\omega_{LH}^2$, while $\beta_i^2\simeq 0$ so $\gamma_i^2\simeq 1$. Then, for the UH wave, the energy coefficients are
\begin{equation}
\label{eq:u_coef_UH}
u_E=\frac{1}{2}, \; u_B\simeq 0, \; u_V\simeq\frac{1}{2}+\frac{\Omega_e^2}{\omega_p^2}\;\Rightarrow\;u\simeq\frac{\omega_{UH}^2}{\omega_p^2}>1.
\end{equation}
We see plasma diamagnetism completely consumes the wave magnetic energy, and stores it in terms of kinetic energy, which is contributed by both wave motion and gyro motion. Notice that in strong magnetic field, the kinetic energy can be much larger than the electric field energy. 
On the other hand, for the LH wave, since its frequency $\omega_{LH}^2\simeq\omega_p^2|\Omega_e|\Omega_i/\omega_{UH}^2$ [Eq.~(\ref{eq:w_LH})], we have $\beta_e^2=\Omega_e^2/\omega_{LH}^2\gg1$ so $\gamma_e^2\simeq -\omega_{LH}^2/\Omega_{e}^2\ll1$, and $\beta_i^2\simeq \Omega_i^2/\omega_{LH}^2\ll1$ so $\gamma_i^2\simeq 1$. Therefore, for the LH wave,
\begin{equation}
\label{eq:u_coef_LH}
u_E=\frac{1}{2}, \; u_B\simeq 0, \; u_V\simeq\frac{1}{2}+\frac{\omega_p^2}{\Omega_e^2}\;\Rightarrow\;u\simeq\frac{\omega_{UH}^2}{\Omega_e^2}>1.
\end{equation}
Again, in longitudinal waves, plasma diamagnetism completely consumes the wave magnetic energy, and stores it in terms of kinetic energy. Contrary to the UH case, here for the LH wave, the ion kinetic energy coefficient $u_{Vi}\simeq\omega_{UH}^2/2\Omega_e^2$ is slightly larger than the electron kinetic energy coefficient $u_{Ve}\simeq\omega_{p}^2/2\Omega_e^2$. Moreover, the kinetic energy is now much larger than the electric field energy if the magnetic field is weak.

Finally, consider the examples when electrostatic waves propagate parallel ($\theta\sim0^\circ$) to the background magnetic field in a cold-fluid electron-ion plasma. First, for the Langmuir wave, $\beta_e^2=\Omega_e^2/\omega_{p}^2$ so $\gamma_e^2= \omega_{p}^2/(\omega_p^2-\Omega_{e}^2)$, while $\beta_i^2\simeq 0$ so $\gamma_i^2\simeq 1$. Since all magnetization ratios and magnetization factors are finite, we can simply set $\sin\theta=0$. Then $\mathbf{e}^\dagger\mathbf{f}=1$ and $\mathbf{f}^\dagger \mathbf{f}=1$ are essentially unmagnetized. Therefore, for Langmuir wave,
\begin{equation}
u_E=\frac{1}{2}, \; u_B=0, \; u_V=\frac{1}{2}\;\Rightarrow\;u=1.
\end{equation}
However, for the cyclotron waves, we have to be more careful. Although it may seem, from Sec.~\ref{sec:fluid-1st:dispersion:para}, that cyclotron waves are transverse, they are in fact longitudinal. This is because for cyclotron wave, $\gamma_s^2\rightarrow\infty$ when $\omega^2\rightarrow\Omega_s^2$ along the dispersion surface $\omega=\omega(\mathbf{k})$. This singularity has to be carefully treated by properly taking the limit $\theta\rightarrow0$. When $\omega_p^2\ne\Omega_e^2$, we can use the asymptotic expression Eq.~(\ref{eq:wl_para}). Then, $\beta_e^2\simeq1+\omega_{p}^2\sin^2\theta/(\omega_{p}^2-\Omega_e^2)$ so $\gamma_e^2\simeq(\Omega_e^2-\omega_{p}^2)/(\omega_{p}^2\sin^2\theta)$ blows up, while $\beta_i^2\simeq 0$ so $\gamma_i^2\simeq 1$ is trivial. Therefore, $\mathbf{e}_i^\dagger\mathbf{f}_i=\mathbf{f}_i^\dagger \mathbf{f}_i=1$ are trivial, $\mathbf{e}_e^\dagger\mathbf{f}_e\simeq\Omega_e^2/\omega_p^2$ is finite, whereas $\mathbf{f}_e^\dagger \mathbf{f}_e\simeq2(\Omega_e^2-\omega_{p}^2)^2/(\omega_{p}^4\sin^2\theta)$ approaches infinity. We see, for electron-cyclotron wave, almost all the wave energy is contained in the resonant electron-cyclotron motion:
\begin{equation}
u_E=\frac{1}{2}, \; u_B\simeq0, \; u_V=\frac{(\Omega_e^2-\omega_{p}^2)^2}{\omega_{p}^2\Omega_e^2\sin^2\theta}\;\Rightarrow\;u\simeq u_V\sim\infty.
\end{equation}
Similarly, we can use the asymptotic expression Eq.~(\ref{eq:wb_para}) to compute the ion-cyclotron wave energy. In this case, we have trivial electron terms $\beta_e^2=\Omega_e^2/\Omega_{i}^2\gg1$ so $\gamma_e^2\simeq-\Omega_i^2/\Omega_{e}^2\ll1$, and nontrivial ion terms $\beta_i^2\simeq1+\Omega_i\tan^2\theta/|\Omega_{e}|$ so $\gamma_i^2\simeq-|\Omega_{e}|/(\Omega_i\tan^2\theta)$. Therefore, $\mathbf{e}_e^\dagger\mathbf{f}_e\simeq\mathbf{f}_e^\dagger \mathbf{f}_e\simeq1$ are trivial, $\mathbf{e}_i^\dagger\mathbf{f}_i\simeq\Omega_e/\Omega_i$ is large but finite, whereas $\mathbf{f}_i^\dagger \mathbf{f}_i\simeq2\Omega_e^2/(\Omega_{i}^2\sin^2\theta)$ approaches infinity. Therefore, for ion-cyclotron wave,
\begin{equation}
u_E=\frac{1}{2}, \; u_B\simeq 0, \; u_V=\frac{\omega_{pi}^2\Omega_e^2}{\Omega_i^4\sin^2\theta}\;\Rightarrow\;u\simeq u_V\sim\infty.
\end{equation}
Again, we see almost all the wave energy is contained in the resonant ion-cyclotron motion. The cancellation between large terms in $u_B$ is somewhat subtle, but since the wave is quasi-longitudinal, the oscillating magnetic field contains little energy compared to the kinetic energy. The above wave energy coefficients will be useful later, when we discuss special cases of three-wave scattering in the next chapter.

\chapter{Magnetized three-wave interactions\label{ch:fluid-2nd}}

Having developed a comprehensive picture of linear waves in cold-magnetized plasmas, we are now ready to study their nonlinear interactions. In the weakly-coupled regime, the linear eigenmode structures are largely retained. In other words, weak nonlinearities slowly change the amplitudes of waves, without significantly altering the wave dispersion relation. In this regime, where nonlinearities can be treated as perturbations, the lowest-order interactions are three-wave interactions. The notion of three-wave interaction does not mean that there can only be three waves in the system. Instead, it means waves are coupled in triplets. Through a network of wave triplets, a large number of waves can become interconnected.

The traditional physical picture of coherent three-wave interaction is that two waves nonlinearly interact and generate the third waves as a beat wave \citep{Drake74}. This picture allows one to develop the concept of nonlinear dispersion relation, namely, the dispersion relation of a seed wave on the background of a pump wave. 
Calculating growth rates of parametric instabilities is a powerful method for understanding three-wave interactions, at least in unmagnetized plasmas. However, in magnetized plasmas, as mentioned in the introduction (Ch.~\ref{ch:intro}), the algebra becomes so complicated that only the simplest situation, namely, when all three waves are collimated and propagate either parallel or perpendicular to the background magnetic field, is understood after decades of efforts. With emerging feasibility of very strong magnetic fields during laser-plasma interactions, it is imperative that we understand three-wave interaction in magnetized plasmas with more generality. 

A breakthrough is made in \cite{Shi17scatter}, where we develop a much more tractable mathematical description of three-wave interactions based on a different physical picture. Instead of breaking the apparent three-wave exchange symmetry, the pump wave, the seed wave, and the beat wave can be treated on equal footing \citep{Davidson72}. This perspective removes the subjective choices, because no wave is more special than the other two waves in a three-wave triplet. In fact, from the Lagrangian perspective, three-wave interactions arise from a cubic term in the effective Lagrangian. Since a wave is nothing but an extremum of the quadratic part of the effective Lagrangian, the Lagrangian itself cares little about which wave plays what role during three-wave interactions. In this chapter, I will first carry out multiscale solution to the second order (Sec.~\ref{sec:fluid-2nd:general}) and obtain three-wave amplitude equations (Sec.~\ref{sec:fluid-2nd:example}). After elaborating on the coupling coefficient (Sec.~\ref{sec:fluid-2nd:coefficient}), I will then recast the interactions from a variational principle perspective (Sec.~\ref{sec:fluid-2nd:Lagrangian}), which enables a more convenient and intuitive understanding. This chapter is devoted to deriving the three-wave amplitude equations and the essential coupling coefficient contained therein. The actual solution to the equations will be discussed in the next chapter. 

\section{Multiscale solution next to leading order\label{sec:fluid-2nd:general}}

The fluid-Maxwell's equations are a PDE analogy of the hyperbolic ODEs discussed in Ch.~\ref{ch:multiscale}. To the lowest order, the linearized equations describe uncoupled linear waves, which we have discussed in Ch.~\ref{ch:fluid-1st}. To the next order in the perturbation series, we need to remove secular terms in order to obtain a well-behaved solution. In this section, I will remove secular terms by carrying out the multiscale procedure to the second order. Relying on the intuition built in Sec.~\ref{sec:multiscale:general:1st}, we expect to find an amplitude equation as the secular-free condition, as well as a driven oscillator equation that describes how second harmonics, which are not normal modes of the linear system, are generated due to nonlinearities.

\subsection{Second-order equations\label{sec:fluid-2nd:2nd:equation}}

In multiscale expansion, recall that the amplitudes are expanded by Eqs.~(\ref{eq:expandE})-(\ref{eq:expandV}) and the spatial and temporal scales are expanded by Eqs.~(\ref{eq:expand_dx}) and (\ref{eq:expand_dt}). Substituting these expansions into the cold-fluid model [Eqs.~(\ref{eq:continuity})-(\ref{eq:Faraday})], we can get a hierarchy of equations ordered by the auxiliary adiabatic parameter $\lambda\ll1$. To obtain the second-order equations, we can collect all the $O(\lambda^2)$ terms in the asymptotic expansions. The resulting second-order equations are 
\begin{eqnarray}
\label{eq:B2}
\partial_{t(0)}\mathbf{B}_2&=&-\partial_{t(1)}\mathbf{B}_1-\nabla_{(1)}\times\mathbf{E}_1-\nabla_{(0)}\times\mathbf{E}_2,\\
\label{eq:v2}
\partial_{t(0)}\mathbf{v}_{s2}&=&-\partial_{t(1)}\mathbf{v}_{s1}-\mathbf{v}_{s1}\cdot\nabla_{(0)}\mathbf{v}_{s1}+\frac{e_s}{m_s}\Big(\mathbf{v}_{s1}\times\mathbf{B}_1+\mathbf{E}_2+\mathbf{v}_{s2}\times\mathbf{B}_0\Big),\\
\label{eq:n2}
\partial_{t(0)} n_{s2}&=&-\partial_{t(1)} n_{s1}-\nabla_{(0)}\cdot(n_{s1}\mathbf{v}_{s1})-n_{s0}\big(\nabla_{(1)}\cdot\mathbf{v}_{s1}+\nabla_{(0)}\cdot\mathbf{v}_{s2}\big),\\
\label{eq:E2}
\Box^{(0)}_{ij}E_2^j&=&-\Box^{(1)}_{ij}E_1^j-\frac{1}{\epsilon_0}\sum_{s}e_s\Big[n_{s0}\partial_{t(1)}v^i_{s1}+\partial_{t(0)}(n_{s1}v^i_{s1})+n_{s0}\partial_{t(0)}v^i_{s2}\Big].
\end{eqnarray}
Again, the electric field equation (\ref{eq:E2}) is obtained by substituting Faraday's law into the Maxwell-Amp$\grave{\text{e}}$re's equation. In doing so, I introduce the first-order differential operator
\begin{eqnarray}
	\Box^{(1)}_{ij}:&=&2\big(\partial_{t(0)}\partial_{t(1)}-c^2\partial_l^{(0)}\partial_l^{(1)}\big)\delta_{ij}+c^2\big(\partial_i^{(0)}\partial_j^{(1)}+\partial_i^{(1)}\partial_j^{(0)}\big).
\end{eqnarray}
This operator mixes fast and slow scales, and will govern how wave amplitudes vary on the slow scales due to interactions that happen on the fast scales.

Although the second-order equations look awfully more complicated than the first-order equations, they are nevertheless quite easy to solve. This is because although the second-order equations are nonlinear in $\mathbf{B}_1$, $\mathbf{v}_{s1}$, and $n_{s1}$, they are linear in $\mathbf{E}_2$, $\mathbf{B}_2$, $\mathbf{v}_{s2}$, and $n_{s2}$. Therefore, we may solve for these second-order perturbations from the essentially linear equations, regarding nonlinearities in first-order perturbations as source terms. The general solution to such a system of inhomogeneous linear equations is again a superposition of plane waves. Let us write the second order electric field
\begin{equation}
	\label{eq:E2k}
	\mathbf{E}_2=\frac{1}{2}\sum_{\mathbf{k}\in\mathbb{K}_2}\mathbf{\mathbfcal{E}}_{\mathbf{k}}^{(2)} e^{i\theta_{\mathbf{k}}}.
\end{equation}
Similar to the first-order expansion [Eq.~(\ref{eq:E1k})], here $\theta_{\mathbf{k}}$ is the fast wave phase, and the complex Fourier amplitude $\mathbf{\mathbfcal{E}}_{\mathbf{k}}^{(2)}(t_{(1)},\mathbf{x}_{(1)};t_{(2)},\mathbf{x}_{(2)};\dots)$ can be slowly-varying. Notice that the summation is now carried over a discrete spectrum $\mathbb{K}_2$ of second-order fluctuations, which is different from the first-order spectrum $\mathbb{K}_1$. However, $\mathbb{K}_2$ is not arbitrary once $\mathbb{K}_1$ is given. In fact, as we will see later, $\mathbb{K}_2$ is completely determined by $\mathbb{K}_1$ when we solve the second-order electric-field equation. For now, it is sufficient to think of $\mathbb{K}_2$ as a set of wave vectors, which contains $-\mathbf{k}$ whenever $\mathbf{k}\in\mathbb{K}_2$, such that the reality condition for $\mathbf{E}_2$ is satisfied. 

To obtain a $\mathbf{E}$-major solution, let us express $\mathbf{B}_2$ in terms of $\mathbf{E}_2$. Plugging in expressions for the first order fluctuations Eqs.~(\ref{eq:E1k}) and (\ref{eq:Bwave}) into the second order Faraday's law [Eq.~(\ref{eq:B2})], the second-order magnetic field can be expressed as
\begin{eqnarray}
	\label{eq:Bwave2}
	\nonumber
	\mathbf{B}_2&=&\frac{1}{2}\sum_{\mathbf{k}\in\mathbb{K}_2}\frac{\mathbf{k}\times\mathbf{\mathbfcal{E}}^{(2)}_\mathbf{k}}{\omega_{\mathbf{k}}}e^{i\theta_{\mathbf{k}}}\\
	&+&\frac{1}{2}\sum_{\mathbf{k}\in\mathbb{K}_1}\Big(\frac{\nabla_{(1)} \times\mathbf{\mathbfcal{E}}_{\mathbf{k}}^{(1)}}{i\omega_\mathbf{k}} +\frac{\mathbf{k}\times\partial_{t(1)}\mathbf{\mathbfcal{E}}_{\mathbf{k}}^{(1)}}{i\omega_\mathbf{k}^2}\Big)e^{i\theta_{\mathbf{k}}}.
\end{eqnarray}
The first line has the same structure as $\mathbf{B}_1$, except now the summation is over the second order spectrum $\mathbb{K}_2$. The second line involves slow derivatives of the first order amplitude $\mathbf{\mathbfcal{E}}^{(1)}_\mathbf{k}$. These derivatives, still unknown at this step, can be determined later, once we solve the amplitude equation from the second-order electric-field equation. 

Similarly, the second-order velocity $\mathbf{v}_{s2}$ can be solved from the second-order momentum equation [Eq.~(\ref{eq:v2})]. One way of solving this equation is by first taking the Fourier transform on $t_{(0)}$ and $\mathbf{x}_{(0)}$ scales. Then, in the Fourier space, the resultant algebraic equation can be readily solved using the vector identity of the forcing operator [Eq.~(\ref{eq:Fvector})]. After taking the inverse Fourier transform, the second-order velocity can be expressed as
\begin{eqnarray}
	\label{eq:vwave2}
	\nonumber
	\mathbf{v}_{s2}&=&\frac{ie_s}{2m_s}\sum_{\mathbf{k}\in\mathbb{K}_2} \frac{\mathbb{F}_{s,\mathbf{k}}\mathbfcal{E}_{\mathbf{k}}^{(2)}}{\omega_\mathbf{k}}e^{i\theta_{\mathbf{k}}} \\
	\nonumber
	&+&\frac{e_s}{2m_s}\sum_{\mathbf{k}\in\mathbb{K}_1}\frac{\mathbb{F}^2_{s,\mathbf{k}}\partial_{t(1)}\mathbfcal{E}_{\mathbf{k}}^{(1)}}{\omega_\mathbf{k}^2}e^{i\theta_{\mathbf{k}}}\\
	&-&\frac{e_s^2}{4m_s^2}\!\sum_{\mathbf{q},\mathbf{q}'\in\mathbb{K}_1} \frac{\mathbb{F}_{s,\mathbf{q}+\mathbf{q}'}(\mathbf{L}^{s}_{\mathbf{q},\mathbf{q}'}\!+\!\mathbf{T}^{s}_{\mathbf{q},\mathbf{q}'})}{\omega_\mathbf{q}+\omega_\mathbf{q}'}e^{i\theta_{\mathbf{q}}+i\theta_{\mathbf{q}'}}.
\end{eqnarray}
The first two lines of the above expression are analogous to those in Eq.~(\ref{eq:Bwave2}) for $\mathbf{B}_2$. The third line comes from beating of nonlinearities. In particular, the $\mathbf{v}_{s1}\times\mathbf{B}_1$ nonlinearity introduce a longitudinal beating 
\begin{equation}
	\label{eq:L}
	\mathbf{L}^{s}_{\mathbf{q},\mathbf{q}'}=\frac{(\mathbb{F}_{s,\mathbf{q}}\mathbfcal{E}_{\mathbf{q}}^{(1)})\times(\mathbf{q}'\times\mathbfcal{E}_{\mathbf{q}'}^{(1)})}{\omega_\mathbf{q}\omega_{\mathbf{q}'}}.
\end{equation}
In addition, the Euler derivative $\mathbf{v}_{s1}\cdot\nabla_{(0)}\mathbf{v}_{s1}$, which is responsible for generating turbulence in neutral fluids, gives rise to a turbulent beating
\begin{equation}
	\label{eq:T}
	\mathbf{T}^{s}_{\mathbf{q},\mathbf{q}'}=\frac{(\mathbb{F}_{s,\mathbf{q}}\mathbfcal{E}_{\mathbf{q}}^{(1)})(\mathbf{q}\cdot\mathbb{F}_{s,\mathbf{q}'}\mathbfcal{E}_{\mathbf{q}'}^{(1)})}{\omega_\mathbf{q}\omega_{\mathbf{q}'}}.
\end{equation}
The third line in Eq.~(\ref{eq:vwave2}) may be simplified using the quadratic property (\ref{eq:F12}) of the forcing operator. This simplification will be done later when we discuss interaction of three waves in the next section.
Notice that the average $\langle \mathbf{v}_{s2} \rangle_0$ is in general nonzero due to the zero-frequency beating terms on the third line of Eq.~(\ref{eq:vwave2}). These terms are well behaved because when $\mathbf{q}'\rightarrow\mathbf{q}$ both the denominator and the numerator go to zero (Appendix~\ref{ch:append:scattering}), and their ratio is finite. For example, consider an unmagnetized plasma, then $\mathbf{L}_{\mathbf{p},\mathbf{q}}+\mathbf{T}_{\mathbf{p},\mathbf{q}}+\mathbf{L}_{\mathbf{q},\mathbf{p}}+\mathbf{T}_{\mathbf{q},\mathbf{p}}=(\mathbfcal{E}_{\mathbf{p}} \cdot \mathbfcal{E}_{\mathbf{q}})(\mathbf{p}+\mathbf{q})/\omega_\mathbf{p}\omega_\mathbf{q}$, where I have abbreviated the species and the order indexes for brevity. Then, for a single linear wave with wave vector $\mathbf{k}$, the average velocity $\langle \mathbf{v}_{2} \rangle_0=(e^2/4m^2)(\mathbfcal{E}_\mathbf{k}\cdot\mathbfcal{E}_\mathbf{k}^*/\omega_\mathbf{k}^2)(\mathbf{k}/\omega_\mathbf{k})$, which is in agreement with \cite{Liu2015nonlinear}.

Next, we can express the second-order density $n_{s2}$ in terms of $\mathbf{E}_2$ using the second-order continuity equation [Eq.~(\ref{eq:n2})]. Although the expression for $n_{s2}$ is not essential for studying three-wave scattering, I present it here to introduce the quadratic response. The expression for $n_{s2}$ will also become useful when one studies four-wave or even higher-order interactions. Using similar method for solving $\mathbf{v}_{s2}$, the second-order density is
\begin{eqnarray}
	\label{eq:nwave2}
	\nonumber
	n_{s2}&=&\frac{e_sn_{s0}}{2m_s}\Bigg[\sum_{\mathbf{k}\in\mathbb{K}_2}\frac{i\mathbf{k}\cdot\mathbb{F}_{s,\mathbf{k}}\mathbfcal{E}^{(2)}_\mathbf{k}}{\omega_{\mathbf{k}}^2}e^{i\theta_{\mathbf{k}}}\\
	\nonumber
	&+&\!\sum_{\mathbf{k}\in\mathbb{K}_1}\!\bigg(\!\frac{\mathbf{k}\!\cdot\!(\mathbb{F}_{s,\mathbf{k}}\!+\!\mathbb{F}^2_{s,\mathbf{k}})\partial_{t(1)}\mathbfcal{E}_{\mathbf{k}}^{(1)}}{\omega_\mathbf{k}^3}\!+\!\frac{\nabla_{(1)}\!\cdot\!\mathbb{F}_{s,\mathbf{k}}\mathbfcal{E}_{\mathbf{k}}^{(1)}}{\omega_\mathbf{k}^2}\!\bigg)\!e^{i\theta_{\mathbf{k}}}\Bigg]\\
	&-&\frac{e_s^2n_{s0}}{4m_s^2}\!\sum_{\mathbf{q},\mathbf{q}'\in\mathbb{K}_1} \frac{(\mathbf{q}+\mathbf{q}')\cdot\mathbf{R}^{s}_{\mathbf{q},\mathbf{q}'}}{(\omega_\mathbf{q}+\omega_\mathbf{q}')^2}e^{i\theta_{\mathbf{q}}+i\theta_{\mathbf{q}'}}.
\end{eqnarray}
The above three lines are in analogy to those of $\mathbf{v}_{s2}$ in Eq.~(\ref{eq:vwave2}). In the third line, the quadratic response is
\begin{equation}
	\label{eq:R}
	\mathbf{R}^{s}_{\mathbf{q},\mathbf{q}'}=\mathbb{F}_{s,\mathbf{q}+\mathbf{q}'} (\mathbf{L}^{s}_{\mathbf{q},\mathbf{q}'}+\mathbf{T}^{s}_{\mathbf{q},\mathbf{q}'}) +(1+\frac{\omega_\mathbf{q}}{\omega_\mathbf{q}'})\mathbf{C}^{s}_{\mathbf{q},\mathbf{q}'},
\end{equation}
where the longitudinal beating $\mathbf{L}^{s}_{\mathbf{q},\mathbf{q}'}$ and the turbulent beating $\mathbf{T}^{s}_{\mathbf{q},\mathbf{q}'}$ are given by Eqs.~(\ref{eq:L}) and (\ref{eq:T}). The third term, proportional to $\mathbf{C}^{s}_{\mathbf{q},\mathbf{q}'}$, comes from the divergence of the nonlinear current $\nabla_{(0)}\cdot(n_{s1}\mathbf{v}_{s1})$, which introduces the current beating 
\begin{equation}
	\label{eq:C}
	\mathbf{C}^{s}_{\mathbf{q},\mathbf{q}'}=\frac{(\mathbb{F}_{s,\mathbf{q}}\mathbfcal{E}_{\mathbf{q}}^{(1)})(\mathbf{q}'\cdot\mathbb{F}_{s,\mathbf{q}'}\mathbfcal{E}_{\mathbf{q}'}^{(1)})}{\omega_\mathbf{q}\omega_{\mathbf{q}'}}.
\end{equation}
Notice here the inner product $\mathbf{q}'\cdot\mathbb{F}_{s,\mathbf{q}'}\mathbfcal{E}_{\mathbf{q}'}^{(1)}$ is now with $\mathbf{q}'$, in contrast to turbulent beating $\mathbf{T}^{s}_{\mathbf{q},\mathbf{q}'}$, in which the inner product $\mathbf{q}\cdot\mathbb{F}_{s,\mathbf{q}'}\mathbfcal{E}_{\mathbf{q}'}^{(1)}$ is with $\mathbf{q}$ instead. 
This makes the physics of these two types of beating fundamentally different.

Finally, we can use the second-order electric field equation [Eq.~(\ref{eq:E2})] to obtain an equation that only involves electric perturbations. Substituting in first-order spectral expansions [Eqs.~(\ref{eq:E1k}), (\ref{eq:vwave}), and (\ref{eq:nwave})] and second-order spectral expansions [Eqs.~(\ref{eq:E2k}) and (\ref{eq:vwave2})], and then simplifying using Eqs.~(\ref{eq:F2}) and (\ref{eq:dE1}), the second-order electric-field equation can then be put into a rather simple and intuitive form
\begin{eqnarray}
	\label{eq:E2s}
	\sum_{\mathbf{k}\in\mathbb{K}_2}\mathbb{D}_{\mathbf{k}}\mathbfcal{E}^{(2)}_\mathbf{k}e^{i\theta_{\mathbf{k}}} +i\sum_{\mathbf{k}\in\mathbb{K}_1}\omega_{\mathbf{k}}\mathbb{H}_{\mathbf{k}}d_{t(1)}^{\mathbf{k}}\mathbfcal{E}^{(1)}_\mathbf{k} e^{i\theta_{\mathbf{k}}} =\frac{i}{2}\sum_{s,\mathbf{q},\mathbf{q}'\in\mathbb{K}_1}\mathbf{S}^{s}_{\mathbf{q},\mathbf{q}'}e^{i\theta_{\mathbf{q}}+i\theta_{\mathbf{q}'}}.
\end{eqnarray}
The LHS describes two types of behaviors in a collection of waves, and these behaviors are consequences of three-wave scattering on the RHS, which is given by
\begin{equation}
	\label{eq:S}
	\mathbf{S}^{s}_{\mathbf{q},\mathbf{q}'}=\frac{e_s\omega_{ps}^2}{2m_s}\Big(\mathbf{R}^{s}_{\mathbf{q},\mathbf{q}'}+\mathbf{R}^{s}_{\mathbf{q}',\mathbf{q}}\Big),
\end{equation}
where the quadratic response $\mathbf{R}^{s}_{\mathbf{q},\mathbf{q}'}$ is given formally by Eq.~(\ref{eq:R}) and explicitly by Eq.~(\ref{eq:R_explicit}).  
In the first term on the LHS of Eq.~(\ref{eq:E2s}), the dispersion tensor $\mathbb{D}_{\mathbf{k}}=\mathbb{D}^*_{-\mathbf{k}}$ is defined by Eq.~(\ref{eq:Dk}). Without three-wave scattering, this term would have required that the second-order fluctuations also satisfy the linear dispersion relation $\mathbb{D}_{\mathbf{k}}\mathbfcal{E}^{(2)}_\mathbf{k}=0$. However, now due to the presence of $\mathbf{S}^{s}_{\mathbf{q},\mathbf{q}'}$, the second-order fluctuations are no longer eigenmodes of the linear system. 
In the second term on the LHS of Eq.~(\ref{eq:E2s}), the wave energy operator $\mathbb{H}_{\mathbf{k}}=\mathbb{H}^*_{-\mathbf{k}}$ is defined by Eq.~(\ref{eq:Hk}), and $d_{t(1)}^{\mathbf{k}}=d_{t(1)}^{-\mathbf{k}}$ is the advective derivative on the slow $t_{(1)}$ and $\mathbf{x}_{(1)}$ scales defined as
\begin{equation}
	\label{eq:dt1}
	d_{t(1)}^{\mathbf{k}}:=\partial_{t(1)}+\frac{\partial\omega_{\mathbf{k}}}{\partial\mathbf{k}}\cdot\nabla_{(1)},
\end{equation}
which advects the wave envelope at the wave group velocity $\mathbf{v}_g=\partial\omega_{\mathbf{k}}/\partial\mathbf{k}$. In the absence of three-wave scattering, linear waves pass through each other without any interaction, and their amplitudes remain unchanged. However, due to three-wave scattering $\mathbf{S}^{s}_{\mathbf{q},\mathbf{q}'}$, waves may now start to exchange energy, and their amplitudes evolve slowly in a way that is consistent with their energy exchange. 
\subsection{The three-wave scattering tensor\label{sec:fluid-2nd:2nd:tensor}}

The three-wave scattering strength [Eq.~(\ref{eq:S})] is a bilinear functional $\mathbf{S}^{s}_{\mathbf{q},\mathbf{q}'}=\mathbf{S}^{s}[\mathbfcal{E}_{\mathbf{q}}, \mathbfcal{E}_{\mathbf{q}'}]$. Mathematically, it is a rank $(2,1)$-tensor, which linearly maps two vectors to another vector. 
The scattering strength $\mathbf{S}^{s}_{\mathbf{q},\mathbf{q}'}$ is proportional to the density $n_{s0}$. This is intuitive because three-wave scattering cannot happen in the vacuum. Hence, all three-wave scatterings come from charged particle response, which is additive when the scattering is coherent and thereof proportional to the density. 
Also notice that $\mathbf{S}^{s}_{\mathbf{q},\mathbf{q}'}$ is proportional to the charge-to-mass ratio times the plasma frequency squared. This is intuitive because $e_s/m_s$ is the coefficient by which charged particles respond to the electric field, on top of a linear response whose rate is determined by the plasma frequency. 
The total coherent three-wave scattering in the plasma is a linear superposition of scattering from all charged-species. This is also intuitive, because electric fields due to scattering from different charged species interfere to give the total scattered field. 

Let us observe a number of additional properties of the scattering strength $\mathbf{S}^{s}_{\mathbf{q},\mathbf{q}'}$. First, by construction, the scattering strength is symmetric with respect to $\mathbf{q},\mathbf{q}'$, namely,
\begin{equation}
	\mathbf{S}^{s}_{\mathbf{q},\mathbf{q}'}=\mathbf{S}^{s}_{\mathbf{q}',\mathbf{q}}.
\end{equation}
In addition, using notation (\ref{eq:notationz}) and (\ref{eq:notationa}), the reality condition for $\mathbf{S}_{\mathbf{q},\mathbf{q}'}$ is
\begin{equation}
	\mathbf{S}_{\mathbf{q},\mathbf{q}'}^{s*}=-\mathbf{S}^{s}_{-\mathbf{q},-\mathbf{q}'}.
\end{equation}
Moreover, it turns out that the scattering strength $\mathbf{S}^{s}_{\mathbf{q},\mathbf{q}'}$ satisfies the important identity
\begin{equation}
	\label{eq:SDC}
	\mathbf{S}^{s}_{\mathbf{q},-\mathbf{q}}=\mathbf{0}.
\end{equation}
This identity, proven in Appendix~\ref{ch:append:scattering}, guarantees that if $\mathbb{K}_1$ does not contain $\omega_{\mathbf{k}}=0$ mode, then no zero-frequency beat mode will arise for the electric perturbations. 
If a zero-frequency beat mode did arise, then any change in the wave amplitude would be faster then this zero-frequency mode. This situation would then violate the multiscale assumptions in the weakly-coupled regime. Fortunately, such violation does not happen. Due to the identity (\ref{eq:SDC}), the weakly-coupled regime is a self-consistent regime, and the multiscale assumptions can always be satisfied when wave amplitudes are sufficiently small.

Finally, let us find an explicit formula for the scattering strength $\mathbf{S}^{s}_{\mathbf{q},\mathbf{q}'}$. For simplicity, I will suppress the species index $s$, with the implied understanding that all terms are associated with the same species. Next, I will abbreviate the wave vector index $\mathbf{k}_2$ and $\mathbf{k}_3$ as ``2" and ``3". For example, the frequencies $\omega_{\mathbf{k}_2}$ and $\omega_{\mathbf{k}_3}$ will be abbreviated as $\omega_2$ and $\omega_3$, respectively. Moreover, since the summation of the two waves appears in the quadratic response tensor [Eq.~(\ref{eq:R})], I will denote $\omega_1:=\omega_2+\omega_3$ and $\mathbf{k}_1:=\mathbf{k}_2+\mathbf{k}_3$. Then, using the quadratic identity of the forcing operator [Eq.~(\ref{eq:F12})], we can obtain a simple expression for $\mathbf{S}^{s}_{\mathbf{k}_2,\mathbf{k}_3}$. Using the abbreviated notations, 
\begin{eqnarray}
	\label{eq:S23}
	\nonumber
	\mathbf{S}_{2,3}&=&\frac{e\omega_p^2\omega_1}{2m\omega_2\omega_3}\Big[\frac{(\mathbfcal{E}_3\cdot\mathbb{F}_2\mathbfcal{E}_2)(\mathbb{F}_1\mathbf{k}_3) +(\mathbfcal{E}_2\cdot\mathbb{F}_3\mathbfcal{E}_3)(\mathbb{F}_1\mathbf{k}_2)}{\omega_1}\\
	&&\hspace{38pt}+\frac{(\mathbb{F}_3\mathbfcal{E}_3)(\mathbf{k}_1\cdot\mathbb{F}_2\mathbfcal{E}_2) -(\mathbb{F}_1\mathbfcal{E}_3)(\mathbf{k}_3\cdot\mathbb{F}_2\mathbfcal{E}_2)}{\omega_2}\\
	\nonumber
	&&\hspace{38pt}+\frac{(\mathbb{F}_2\mathbfcal{E}_2)(\mathbf{k}_1\cdot\mathbb{F}_3\mathbfcal{E}_3) -(\mathbb{F}_1\mathbfcal{E}_2)(\mathbf{k}_2\cdot\mathbb{F}_3\mathbfcal{E}_3)}{\omega_3}\Big].
\end{eqnarray}
One may be puzzled by the above formula\footnote[1]{In fact, the secular-free identity [Eq.~(\ref{eq:SDC})] can be immediately proven using this formula, together with the self-adjoint property of the forcing operator [(\ref{eq:Fadj})].}.
After all, why $\mathbf{S}_{2,3}$ is given by those six particular combinations of vectors $\mathbb{F}_{\mathbf{q}}\mathbfcal{E}_{\mathbf{q}'}$ and $\mathbb{F}_{\mathbf{q}}\mathbf{\mathbf{q}'}$, weighted by inner products $\mathbfcal{E}_\mathbf{q}\!\cdot\!\mathbb{F}_{\mathbf{q}'}\mathbfcal{E}_{\mathbf{q}'}$ and $\mathbf{q}\!\cdot\!\mathbb{F}_{\mathbf{q}'}\mathbfcal{E}_{\mathbf{q}'}$? Why the terms have the signs they have, and why are they divided by one frequency $\omega_i$ but not another? At first glance, there seems to be no obvious pattern, other than the built-in symmetry $2\leftrightarrow3$. However, when I discuss the Lagrangian of three-wave interaction later in Sec.~\ref{sec:fluid-2nd:Lagrangian}, the formula for $\mathbf{S}_{2,3}$ will become obvious.

\subsection{The on-shell and off-shell equations\label{sec:fluid-2nd:2nd:shell}}

The second-order electric field equation (\ref{eq:E2s}) is a PDE analogy of Eq.~(\ref{eq:2x1}) for ODEs. Similar to what we have done before, this equation can be split into two parts. The first part is the secular-free condition, which describes how oscillation amplitude evolves on the slow scale. The second part is secular-free and describes non-resonant driven oscillations. 
For $\mathbf{E}$-major solution, the first-order spectrum contains all the on-shell waves, which satisfy the dispersion relation $\det\mathbb{D}(\mathbf{k},\omega_\mathbf{k})=0$ for all $\mathbf{k}\in\mathbb{K}_1$. While the second-order spectrum $\mathbb{K}_2$ contains all the off-shell waves. These off-shell quasi-modes do not satisfy the linear dispersion relation, and their amplitude is driven by the beating of two on-shell waves.

The only difference is that now there are many linear waves in the system, so we need to use Eq.~(\ref{eq:E2s}) to constrain the spectrum $\mathbb{K}_2$. Since the Fourier exponents $e^{i\theta_{\mathbf{k}}}$ are orthogonal on the fast scale, in order to satisfy the second-order electric field equation, Fourier coefficients must match on both sides. 
To match the spectrum and obtain a $\mathbf{E}$-major solution, the second-order spectrum is
\begin{equation}
	\label{eq:K2}
	\mathbb{K}_2=(\mathbb{K}_1^0\bigoplus\mathbb{K}_1^0)\setminus\mathbb{K}_1^0,
\end{equation}
where the set $\mathbb{K}_1^0:=\mathbb{K}_1\bigcup\{\mathbf{0}\}$, and the notation $B\setminus A$ denotes the relative complement of set $A$ with respect to set $B$. The direct sum of two sets $G_1,G_2\subseteq G$, where $G$ is an additive group, is defined by $G_1\bigoplus G_2:=\{g_1+g_2|g_1\in G_1, g_2\in G_2\}$. Here, we can exclude the zero vector $\mathbf{0}$ from the second-order spectrum $\mathbb{K}_2$, because of the secular-free property Eq.~(\ref{eq:SDC}). To obtain the $\mathbf{E}$-major solution, I have also excluded vectors that are already contained in the first order spectrum $\mathbb{K}_1$, such that $\mathbb{K}_2$ only contains off-shell waves.

The secular-free conditions of the second-order electric-field equation (\ref{eq:E2s}) are the amplitude equations, which are analogous to Eq.~(\ref{eq:d1a0}) and can take one of the following two forms. In one scenario, the first-order spectrum $\mathbb{K}_1$ contains waves that are resonant with the $\mathbf{k}$-wave. In other words, for given $\mathbf{k}\in\mathbb{K}_1$, there exist $\mathbf{q},\mathbf{q}'\in\mathbb{K}_1$ such that $\theta_{\mathbf{k}}=\theta_{\mathbf{q}}+\theta_{\mathbf{q}'}$. In this case, by matching Fourier coefficients, the amplitude equation for the resonant wave is
\begin{eqnarray}
\label{eq:onE1Hk_23}
\omega_{\mathbf{k}}\mathbb{H}_{\mathbf{k}}d_{t(1)}^\mathbf{k}\mathbfcal{E}_{\mathbf{k}}^{(1)} =\sum_{s}\mathbf{S}^{s}_{\mathbf{q},\mathbf{q}'},
\end{eqnarray} 
where he factor $1/2$ has been canceled by the symmetry property $2\mathbf{S}^{s}_{\mathbf{q},\mathbf{q}'}=\mathbf{S}^{s}_{\mathbf{q},\mathbf{q}'}+\mathbf{S}^{s}_{\mathbf{q}',\mathbf{q}}$.
In the other scenario, the first-order spectrum $\mathbb{K}_1$ contains no wave resonant with the $\mathbf{k}$-wave. In this case, the amplitude equation for the non-resonant wave is
\begin{eqnarray}
\label{eq:onE1Hk_1}
\omega_{\mathbf{k}}\mathbb{H}_{\mathbf{k}}d_{t(1)}^\mathbf{k}\mathbfcal{E}_{\mathbf{k}}^{(1)} =\mathbf{0}.
\end{eqnarray} 
Since the energy operator $\mathbb{H}_{\mathbf{k}}$ is a positive definite Hermitian matrix, it is non-degenerate. Therefore, the solution to the above matrix equation is trivial $d_{t(1)}^\mathbf{k}\mathbfcal{E}_{\mathbf{k}}^{(1)}=0$. In other words, the Fourier amplitude satisfies the advection equation on the slow scale. Consequently, the wave envelop remains constant in reference frames that travel at the wave group velocity. 

The secular-free part of the second-order electric field equation (\ref{eq:E2s}) determines the non-resonantly driven amplitude $\mathbfcal{E}_{\mathbf{k}}^{(2)}$, which is analogous to Eq.~(\ref{eq:x1s}). Since the matrix $\mathbb{D}_{\mathbf{k}}$ is invertible for all $\mathbf{k}\in\mathbb{K}_2$, we can immediately solve the matrix equation, and obtain
\begin{equation}
	\label{eq:E2Dk}
	\mathbfcal{E}_{\mathbf{k}}^{(2)}=i\mathbb{D}^{-1}_{\mathbf{k}}\sum_{s}\mathbf{S}^{s}_{\mathbf{q},\mathbf{q}'},
\end{equation}
where $\mathbf{q},\mathbf{q}'\in\mathbb{K}_1$ are such that $\mathbf{k}=\mathbf{q}+\mathbf{q}'\in\mathbb{K}_2$. 
Again, the factor $1/2$ is canceled by the symmetry property of $\mathbf{S}^{s}_{\mathbf{q},\mathbf{q}'}$. 
The above expression is a PDE analogy of Eq.~(\ref{eq:x1}), where we used the multiscale method to find the $x$-major solution for the ODEs. Substituting the amplitude equations and the off-shell solutions to Eqs.~(\ref{eq:Bwave2}), (\ref{eq:vwave2}), and (\ref{eq:nwave2}), we have thus obtained a formal $\mathbf{E}$-major solution to the fluid-Maxwell system beyond the leading order.

To illustrate the abstract notations introduced above, let us consider the simplest example where the spectrum $\mathbb{K}_1$ contains only one on-shell wave, namely, $\mathbb{K}_1=\{\mathbf{k},-\mathbf{k}\}$. In this case, the second order spectrum $\mathbb{K}_2=\{2\mathbf{k},-2\mathbf{k}\}$ simply contains the second harmonic. Matching the Fourier exponents, the on-shell equation is trivial:
\begin{equation}
	\omega_{\mathbf{k}}\mathbb{H}_{\mathbf{k}}d_{t(1)}^{\mathbf{k}}\mathbfcal{E}^{(1)}_\mathbf{k}=\mathbf{0}.
\end{equation}
Next, matching coefficients of the other Fourier exponent, we obtain the off-shell equation for the second harmonic:
\begin{equation}
	\mathbb{D}_{2\mathbf{k}}\mathbfcal{E}^{(2)}_{2\mathbf{k}}=i\sum_{s}\mathbf{S}^{s}_{\mathbf{k},\mathbf{k}}.
\end{equation}
After inverting the matrix $\mathbb{D}_{2\mathbf{k}}$, this equation gives the amplitude of the second harmonic in terms of the amplitude of the linear wave. Moreover, since the complex amplitude $\mathbfcal{E}^{(2)}_{2\mathbf{k}}$ also encodes the phase information, the above equation also tells how the second harmonic is phase-locked with the fundamental.

\section{Interactions between three on-shell waves\label{sec:fluid-2nd:example}}

In this section, I will discuss the simplest nontrivial example, where the first-order spectrum $\mathbb{K}_1$ contains exactly three resonant on-shell waves. Without loss of generality, the resonance condition $\theta_{\mathbf{k}_1}=\theta_{\mathbf{k}_2}+\theta_{\mathbf{k}_3}$ can be written in components as
\begin{eqnarray}
\label{eq:resonantK}
\mathbf{k}_{1}&=&\mathbf{k}_{2}+\mathbf{k}_{3},\\
\label{eq:resonantW}
\omega_{\mathbf{k}_1}&=&\omega_{\mathbf{k}_2}+\omega_{\mathbf{k}_3},
\end{eqnarray}
where all $\omega$'s are all positive. In this case, the spectrum $\mathbb{K}_1=\{\mathbf{k}_{1}, \mathbf{k}_{2}, \mathbf{k}_{3}, (\mathbf{k}\rightarrow-\mathbf{k})\}$. Using Eq.~(\ref{eq:K2}), we find the second order spectrum $\mathbb{K}_2=\{2\mathbf{k}_{1}, 2\mathbf{k}_{2}, 2\mathbf{k}_{3}, \mathbf{k}_{1}+\mathbf{k}_{2}, \mathbf{k}_{2}-\mathbf{k}_{3}, \mathbf{k}_{3}+\mathbf{k}_{1}, (\mathbf{k}\rightarrow-\mathbf{k})\}$. Notice that resonant waves, such as $\mathbf{k}_{1}=\mathbf{k}_{2}+\mathbf{k}_{3}$, are not contained in the second order spectrum $\mathbb{K}_2$. In this way, all perturbative corrections to the first-order amplitude $\mathbfcal{E}^{(1)}_{\mathbf{k}}$ are accounted for by its slow derivatives.

\subsection{Electric field equations\label{sec:fluid-2nd:example:waves}}

Using the second-order electric field equation (\ref{eq:E2s}), we can extract component equations by matching coefficients of Fourier exponents. The off-shell components are passive, in the sense that they are completely determined by on-shell waves, and do not affect how on-shell waves behave. There are twelve off-shell equations, appearing in six conjugate pairs. Among these, three pairs govern the production of second harmonics $2\mathbf{k}_1$, $2\mathbf{k}_2$, and $2\mathbf{k}_3$: 
\begin{eqnarray}
\mathbb{D}_{2\mathbf{k}_1}\mathbfcal{E}^{(2)}_{2\mathbf{k}_1}&=&i\sum_{s}\mathbf{S}^{s}_{\mathbf{k}_1,\mathbf{k}_1},\\
\mathbb{D}_{2\mathbf{k}_2}\mathbfcal{E}^{(2)}_{2\mathbf{k}_2}&=&i\sum_{s}\mathbf{S}^{s}_{\mathbf{k}_2,\mathbf{k}_2},\\
\mathbb{D}_{2\mathbf{k}_3}\mathbfcal{E}^{(2)}_{2\mathbf{k}_3}&=&i\sum_{s}\mathbf{S}^{s}_{\mathbf{k}_3,\mathbf{k}_3}.
\end{eqnarray}
The other three pairs of equations govern quasi-modes produced by off-shell beatings: 
\begin{eqnarray}
\mathbb{D}_{\mathbf{k}_{1}+\mathbf{k}_{2}}\mathbfcal{E}^{(2)}_{\mathbf{k}_{1}+\mathbf{k}_{2}} &=&i\sum_{s}\mathbf{S}^{s}_{\mathbf{k}_1,\mathbf{k}_2}, \\
\mathbb{D}_{\mathbf{k}_{2}-\mathbf{k}_{3}}\mathbfcal{E}^{(2)}_{\mathbf{k}_{2}-\mathbf{k}_{3}} &=&i\sum_{s}\mathbf{S}^{s}_{\mathbf{k}_2,-\mathbf{k}_3},\\
\mathbb{D}_{\mathbf{k}_{3}+\mathbf{k}_{1}}\mathbfcal{E}^{(2)}_{\mathbf{k}_{3}+\mathbf{k}_{1}} &=&i\sum_{s}\mathbf{S}^{s}_{\mathbf{k}_3,\mathbf{k}_1}.
\end{eqnarray}
Since the dispersion tensor $\mathbb{D}_\mathbf{k}$ for off-shell quasi-modes are non-degenerate, the second order amplitudes $\mathbfcal{E}^{(2)}_{\mathbf{k}}$ can be found by simply inverting the above matrix equations, which gives the second-order Fourier amplitudes in terms of the first-order Fourier amplitudes.

On the other hand, the on-shell equations are active, in the sense that they affect the behavior of one another. There are six on-shell equations, three of which are complex conjugation of the following three on-shell equations: 
\begin{eqnarray}
\label{eq:onE1}
\omega_{\mathbf{k}_1}\mathbb{H}_{\mathbf{k}_1}d_{t(1)}^{\mathbf{k}_1}\mathbfcal{E}^{(1)}_{\mathbf{k}_1} &=&\sum_{s}\mathbf{S}^{s}_{\mathbf{k}_2,\mathbf{k}_3},\\
\label{eq:onE2}
\omega_{\mathbf{k}_2}\mathbb{H}_{\mathbf{k}_2}d_{t(1)}^{\mathbf{k}_2}\mathbfcal{E}^{(1)}_{\mathbf{k}_2} &=&\sum_{s}\mathbf{S}^{s}_{\mathbf{k}_1,-\mathbf{k}_3},\\
\label{eq:onE3}
\omega_{\mathbf{k}_3}\mathbb{H}_{\mathbf{k}_3}d_{t(1)}^{\mathbf{k}_3}\mathbfcal{E}^{(1)}_{\mathbf{k}_3} &=&\sum_{s}\mathbf{S}^{s}_{\mathbf{k}_1,-\mathbf{k}_2}.
\end{eqnarray}
In the above equations, the LHS are basically advections of wave envelopes at group velocities, while the RHS, as we shall see next, govern redistribution of wave actions due to three-wave scattering. In what follows, I will focus on these more interesting on-shell equations. Since only first-order amplitudes $\mathbfcal{E}_{\mathbf{k}_j}^{(1)}$ are involved, I will suppress the order superscript ``$(1)$", abbreviate the wave index ``$\mathbf{k}_j$"  as ``$j$", and denote $-j$ as $\bar{j}$ for simplicity. Whenever the species index $s$ is suppressed, all terms are associated with the same species.

\subsection{Action conservation among three resonant waves\label{sec:fluid-2nd:example:action}}

Using the simplified notations and the formula for $\mathbf{S}_{2,3}$ [Eq.~(\ref{eq:S23})], it is a straightforward calculation to show that whenever three waves are in resonance [Eqs.~(\ref{eq:resonantK}) and (\ref{eq:resonantW})], their scattering strengths satisfy identities
\begin{eqnarray}
\label{eq:actionS12}
\frac{\mathbfcal{E}_{1}\cdot\mathbf{S}^{*}_{2,3}}{\omega_{1}^2}+\frac{\mathbfcal{E}^{*}_{2}\cdot\mathbf{S}_{1,\bar{3}}}{\omega_{2}^2}&=&0,\\
\label{eq:actionS23}
\frac{\mathbfcal{E}^{*}_{2}\cdot\mathbf{S}_{1,\bar{3}}}{\omega_{2}^2} -\frac{\mathbfcal{E}^{*}_{3}\cdot\mathbf{S}_{1,\bar{2}}}{\omega_{3}^2}&=&0.
\end{eqnarray}
The expression for $\mathbf{S}_{1,\bar{3}}$ can be obtained easily from $\mathbf{S}_{2,3}$ using the replacement rule $1\rightarrow 2$, \mbox{$2\rightarrow 1$}, $3\rightarrow-3$, where the minus sign is interpreted using notations (\ref{eq:notationz}) and (\ref{eq:notationa}). Similarly, to obtain the expression for $\mathbf{S}_{1,\bar{2}}$ from $\mathbf{S}_{2,3}$, we can replace $1\rightarrow 3, 2\rightarrow 1, 3\rightarrow-2$ in Eq.~(\ref{eq:S23}). Having obtained expressions for $\mathbf{S}_{2,3}$, $\mathbf{S}_{1,\bar{3}}$, and $\mathbf{S}_{1,\bar{2}}$, we can then use the self-adjoint property of the forcing operator [Eq.~(\ref{eq:Fadj})] to verify the above identities.

Now that the three scattering strengths are related, the slow change of the three wave amplitudes are also related. On both sides of the on-shell equations (\ref{eq:onE1})-(\ref{eq:onE3}), taking inner products with the Fourier amplitudes $\mathbfcal{E}_\mathbf{k}^*$, it is easy to see that Eqs.~(\ref{eq:actionS12}) and (\ref{eq:actionS23}) give rise to the following action conservation laws: 
\begin{eqnarray}
	\label{eq:action12}
	d_{t}\frac{U_{1}}{\omega_{1}}+d_{t}\frac{U_{2}}{\omega_{2}}&=&0,\\
	\label{eq:action23}
	d_{t}\frac{U_{3}}{\omega_{3}}-d_{t}\frac{U_{2}}{\omega_{2}}&=&0.
\end{eqnarray}
Here, $U_{j}$ is the energy density of linear wave ``$j$", which can be computed using Eq.~(\ref{eq:U}). In a quantum-mechanical language, the wave energy density $U=n\hbar\omega$, where $\hbar\omega$ is the energy of each wave quantum and $n$ is the density of wave quanta. Then, $U/\omega$ is proportional to $n$, and is usually called wave action density in classical physics. The first conservation law (\ref{eq:action12}) implies that the total number of wave quanta in the incident wave $\omega_1$ and the scattered wave $\omega_2$ is a constant. This is intuitive because, in the absence of damping, whenever a quantum of the $\omega_1$ mode is annihilated, it is consumed to create a quantum of the $\omega_2$ mode. Analogously, the second conservation law (\ref{eq:action23}) implies that whenever a quantum of the $\omega_2$ mode is created, a quantum of the $\omega_3$ mode must also be created by the three-wave process ``1"$\rightleftharpoons$``2"+``3". As a corollary of wave action conservation, the total wave energy is also conserved during resonant three-wave interactions:
\begin{equation}
	\label{eq:action123}
	d_{t}U_{1}+d_{t}U_{2}+d_{t}U_{3}=0.
\end{equation}
This local energy conservation law can be obtained by linearly combining Eqs.~(\ref{eq:action12}) and (\ref{eq:action23}), and then use the resonance condition for frequencies [Eq.~(\ref{eq:resonantW})]. The conservation of wave energy is also intuitive, because in the absence of damping and other waves, three-wave scattering can only redistribute energy among the three waves.
\subsection{Three-wave equations\label{sec:fluid-2nd:example:equation}}

When we are not concerned with the vector dependence of the complex wave amplitude $\mathbfcal{E}$, the on-shell equations (\ref{eq:onE1})-(\ref{eq:onE3}) can be written as three scalar equations, called the three-wave equations. To remove the vector dependence, let us decompose $\mathbfcal{E}=\mathbf{e}\varepsilon$, where $\mathbf{e}$ is the complex unit vector satisfying $\mathbf{e}^\dagger\mathbf{e}=1$. This decomposition is not unique due to the U(1) symmetry $\mathbf{e}\rightarrow e^{i\alpha}\mathbf{e}$ and $\varepsilon\rightarrow e^{-i\alpha}\varepsilon$. By requiring that the scalar amplitude $\varepsilon\in\mathbb{R}$ is real-valued, the symmetry group of the decomposition is reduced to the $\mathbb{Z}_2$ symmetry $\varepsilon\rightarrow-\varepsilon$. With such a decomposition, the convective derivative of the vector wave amplitude 
\begin{equation}
	d_{t}\mathbfcal{E}=\mathbf{e}d_{t}\varepsilon+\varepsilon d_{t}\mathbf{e}.
\end{equation}
In other words, the change in the vector amplitude can be decomposed into the change in the scalar amplitude and the change due to the rotation of the complex unit vector.

The three-wave equations can be put into a particularly simple form by normalizing the scalar amplitudes. It is natural that 
the wave energy coefficient $u_\mathbf{k}$, defined by Eq.~(\ref{eq:uk}), will come into the normalization, because the on-shell equations involve the wave energy operator $\mathbb{H}_\mathbf{k}$. Taking inner product with $\mathbf{e}_\mathbf{k}^*$ on both sides of the on-shell equations (\ref{eq:onE1Hk_23}) and averaging the result with its Hermitian conjugate, we have
\begin{equation}
u_\mathbf{k}d_{t}\varepsilon_\mathbf{k}+\frac{1}{2}\varepsilon_\mathbf{k} d_{t}u_\mathbf{k}=\frac{1}{4\omega_\mathbf{k}}\sum_s\big(\mathbf{e}_\mathbf{k}^\dagger\mathbf{S}_{\mathbf{q},\mathbf{q}'}^s+\text{c.c.}\big),
\end{equation} 
where the factor $1/4=1/2\times1/2$ comes from the definition of the wave energy coefficient, as well as averaging with the Hermitian conjugate. On the LHS of the above expression, $\varepsilon u^{1/2}$ emerges as a particularly convenient combination. Let us nondimensionalize the electric field by electron mass and charge, and define the normalized wave scalar amplitude as
\begin{equation}
	\label{eq:ak}
	a_\mathbf{k}:=\frac{e\varepsilon_\mathbf{k}u_\mathbf{k}^{1/2}}{m_ec\omega_\mathbf{k}},
\end{equation}
which is usually a very small dimensionless number $a_\mathbf{k}\ll1$ in the weakly coupled regime.
Since the scalar electric field amplitude $\varepsilon_\mathbf{k}$ is real-valued, the normalized wave amplitude $a_\mathbf{k}$ is also real-valued. Then, the on-shell equation can be written as
\begin{eqnarray}
\label{eq:aLHS}
	d_{t}a_\mathbf{k}&=&\frac{e}{4m_ec u_\mathbf{k}^{1/2}} \sum_s\Big(\frac{\mathbf{e}_\mathbf{k}^\dagger\mathbf{S}^s_{\mathbf{q},\mathbf{q}'}}{\omega_\mathbf{k}^2} +\text{c.c.}\Big).
\end{eqnarray}
From this equation, we see only the real part of $\mathbf{e}^\dagger\mathbf{S}$ affects how the scalar amplitude change, while the imaginary part affects how the phase of $\mathbf{e}$ rotates on the complex unit sphere. Notice that the relative phases of the three waves are important. By changing the relative phases, $\mathbf{e}^\dagger\mathbf{S}$ can be tunned from purely real to purely imaginary. When $\mathbf{e}^\dagger\mathbf{S}$ is purely real, the scalar amplitude changes at the fastest rate, and this happen when the three waves are synchronized to drive charged particles constructively, which allows maximum energy exchange between the three waves. On the contrary, when waves are destructively synchronized, $\mathbf{e}^\dagger\mathbf{S}$ is purely imaginary. In this case, nonlinear plasma response cancels so that there is no energy exchange between the three waves.

Having obtained the generic form of the normalized scalar amplitude equation (\ref{eq:aLHS}), we can convert all three on-shell equations (\ref{eq:onE1})-(\ref{eq:onE3}) into this form. Using action conservation laws [Eqs.~(\ref{eq:actionS12}) and (\ref{eq:actionS23})], the RHS of the on-shell equations are originated from a single scattering term:
\begin{equation}
	\label{eq:aRHS}
	\frac{e_s\omega_{ps}^2}{2m_sc}\frac{\varepsilon_1\varepsilon_2\varepsilon_3}{\omega_1\omega_2\omega_3}\Theta^s:=-\frac{\mathbfcal{E}_{1}\cdot\mathbf{S}^{*}_{2,3}}{\omega_{1}^2} =\frac{\mathbfcal{E}^{*}_{2}\cdot\mathbf{S}_{1,\bar{3}}}{\omega_{2}^2} =\frac{\mathbfcal{E}^{*}_{3}\cdot\mathbf{S}_{1,\bar{2}}}{\omega_{3}^2}.
\end{equation}
The normalized scattering strength $\Theta^s=\Theta^s_r+i\Theta^s_i$ contains both real and imaginary parts. 
Since the phase of $\Theta^s$ can be shifted, for example, by redefining $\mathbfcal{E}_{1}\rightarrow\mathbfcal{E}_{1}\exp(i\alpha)$, only the absolute value of $\Theta^s$ is physically significant.
Using the above results, the change in scalar amplitude is described by the following three-wave equations:
\begin{eqnarray}
	\label{eq:3waves1}
	d_{t}a_1&=&-\frac{\Gamma}{\omega_1}a_2a_3,\\
	\label{eq:3waves2}
	d_{t}a_2&=&\phantom{+}\frac{\Gamma}{\omega_2}a_3a_1,\\
	\label{eq:3waves3}
	d_{t}a_3&=&\phantom{+}\frac{\Gamma}{\omega_3}a_1a_2.
\end{eqnarray}
In these equations, $d_t$ is again the advective derivative at the respective wave group velocity. The essential parameter in the three-wave equations is the coupling coefficient, which is given by
\begin{equation}
\label{eq:coupling}
\Gamma=\sum_s\frac{Z_s\omega_{ps}^2\Theta^s}{4M_s(u_1u_2u_3)^{1/2}},
\end{equation}
where $Z_s:=e_s/e$ and $M_s:=m_s/m_e$ are the normalized charge and mass of species $s$. Notice that when density $n_{s0}\rightarrow 0$, coupling due to species $s$ vanishes as expected, because there is no three-wave scattering in vacuum and the coherent scattering is additive. The sign of $\Gamma$ is insignificant, because of the residual $\mathbb{Z}_2$ symmetry $a_j\rightarrow-a_j$. However, the relative signs of the above three equations are important. The equation for the highest frequency $\omega_1$ wave always has the opposite sign as the equations for lower frequencies $\omega_2$ and $\omega_3$ waves.

The three-wave equations manifestly state that wave action and energy are conserved during the interactions. The wave action $U/\omega=\epsilon_0u\varepsilon^2/2\omega$ is proportional to the number of wave quanta. With a convenient normalization, the wave action density is proportional to
\begin{equation}
\mathcal{I}_\mathbf{k}:=\frac{\epsilon_0E_s}{2ec}\omega_{\mathbf{k}}a_{\mathbf{k}}^2,
\end{equation} 
where $E_s=m_e^2c^3/2\hbar$ is the Schwinger critical field. To see the above definition has correct units, notice that the combination $\epsilon_0E_s\omega/ec$ has the units of the number density, as can be seen from the Gauss' law $\nabla\cdot \mathbf{E}=en/\epsilon_0$. 
Since the Schwinger field $E_s$ is an extremely large electric field, the normalized scalar amplitude $a_\mathbf{k}\sim 1$ corresponds to an extremely large wave amplitude. In the weakly coupled non-relativistic regime, the normalized scalar amplitude $a_\mathbf{k}\ll 1$.  
Now, using the definition of wave action, the three-wave equations (\ref{eq:3waves1})-(\ref{eq:3waves3}) immediately give the action conservation laws:
\begin{eqnarray}
\label{eq:I12}
d_t\mathcal{I}_1+d_t\mathcal{I}_2&=&0,\\
\label{eq:I23}
d_t\mathcal{I}_2-d_t\mathcal{I}_3&=&0,
\end{eqnarray} 
where $d_t$ is again the advective derivative at the respective wave group velocity. Using the action conservation laws and the frequency resonance condition, the conservation law of energy density $U=\mathcal{I}\hbar\omega$ can be written as
\begin{equation}
d_t\omega_1\mathcal{I}_1+d_t\omega_2\mathcal{I}_2+d_t\omega_3\mathcal{I}_3=0.
\end{equation}
The action conservation laws and the energy conservation law will be very useful when we solve the three-wave equations in the next chapter. Here, it is worth mentioning that action and energy are conserved because we have ignored dissipation and other interactions. When other processes are present, the energy in the three-wave system can leak out, either by conversion to other waves or by dissipation through damping.

\section{The coupling coefficient and parametric growth rates\label{sec:fluid-2nd:coefficient}}

While the three-wave equations always take the same form [Eqs.~(\ref{eq:3waves1})-(\ref{eq:3waves3})], what distinguishes the interactions between one three-wave triplet from the interactions of another three-wave triplet is the coupling coefficient [Eq.~(\ref{eq:coupling})]. In other words, the coupling coefficient encodes all the physical details that lead to an interaction strength, while the three-wave equations generically describe the outcome once the interaction strength is given. It is possible that two completely different three-wave triplets interact with the same strength, and it is also possible that two very similar three-wave triplets interact with quite different strength. In any case, the three-wave equations only care about the coupling coefficient, the wave frequencies, and the wave group velocities. The solutions to three-wave equations will be discussed in the next chapter.

In this section, I will focus on the coupling coefficient and the parametric growth rates as immediate experimental observables. I will first discuss the general results in Sec.~\ref{sec:fluid-2nd:coefficient:formula}, and then give two sets of examples, where the participating waves are either quasi-transverse (\textit{T}) or quasi-longitudinal (\textit{L}). In these special situations, the wave dispersion relation and polarization are greatly simplified, and evaluating the formula asymptotically becomes relatively easy. Although there are four different three-wave triplets $\{T,T,T\}$, $\{T,T,L\}$, $\{T,L,L\}$, and $\{L,L,L\}$ in general, only two of these triplets can couple resonantly. From Sec.~\ref{sec:fluid-1st:dispersion:oblique}, we know that the \textit{T} waves are electromagnetic waves with $\omega\gg\omega_p,|\Omega_e|$, while the \textit{L} waves are electrostatic waves with $\omega\rightarrow\omega_r$, for some resonance $\omega_r\sim\omega_p,|\Omega_e|,\Omega_i$. Since the frequency of a \textit{T} wave is much higher than the frequency of an \textit{L} wave, only the following two types of interactions can match resonance conditions\footnote[1]{In warm plasmas, scattering that involves the three MHD waves and the Bernstein waves are also interesting cases. These warm cases do not necessarily fit into either the $TTL$ or $LLL$ scenarios being considered here.}:
\begin{eqnarray}
&&T\rightleftharpoons T+L,\\
&&L\rightleftharpoons L+L.
\end{eqnarray}
A typical scenario for the \textit{TTL} interaction is the scattering of lasers. For example, an incident lasers is scattered inelastically by some plasma waves and thereafter propagates in some other direction with a shifted frequency. This \textit{TTL} scenario will be discussed in Sec.~\ref{sec:fluid-2nd:coefficient:TTL}. Similarly, a typically scenario for the \textit{LLL} interaction is the decay of a plasma wave launched by some antenna array. This scenario will be discussed in Sec.~\ref{sec:fluid-2nd:coefficient:LLL} in details.

\subsection{The general formula\label{sec:fluid-2nd:coefficient:formula}}

In the general formula for the coupling coefficient $\Gamma$ [Eq.~(\ref{eq:coupling})], there are two sets of wave-dependent terms. The denominator measures how energetically expensive it is to excite the linear waves, and the numerator measures how large the scattering strength is. These two factors compete to determine the coupling coefficient. 

The set of terms in the denominator are the wave energy coefficients $u_j$, whose general formula is given by Eq.~(\ref{eq:uk}). The coupling coefficient $\Gamma\propto1/\sqrt{u_j}$, because the three-wave interaction is an energy-exchange process.
When a quantum of the wave energy $U\propto u\varepsilon^2$ is exchanged, the wave amplitudes change by $\varepsilon\propto1/\sqrt{u}$. A very small wave energy coefficient $u$ means that most wave energy is contained in the electric field. Therefore, the energy exchange efficiently alters the wave electric field, and the interaction is energetically cheap. On the other hand, a very large wave energy coefficient $u$ means that most energy is either kinetic or magnetic. Consequently, the energy exchange has small effect on the wave electric field, and the interaction is energetically expensive.

The set of terms in the numerator of the coupling coefficient $\Gamma$ are the normalized scattering strengths $\Theta^s$. The contributions from different charged species are additive, because the total coherently scattered electric field is the linear superposition of all the scattering from individual charged species. When the complex-valued scattering strengths $\Theta^s$ from different species are aligned in the complex plane, they constructively add and lead to a large total scattering. In the opposite scenario, $\Theta^s$ from different species can also destructively interfere and lead to a small total scattering. To find an explicit formula for $\Theta^s$, which is introduced in Eq.~(\ref{eq:aRHS}), we can use Eq.~(\ref{eq:S23}), the formula for $\mathbf{S}_{2,3}$. Using the self-adjoint property of the forcing operator, the normalized scattering strength can be written as the summation of the strengths of six scattering channels:
\begin{eqnarray}
\label{eq:Theta3}
\nonumber
\Theta^s&=&\Theta_{1,\bar{2}\bar{3}}^s+\Theta_{\bar{2},\bar{3}1}^s+\Theta_{\bar{3},1\bar{2}}^s\\
&+&\Theta_{1,\bar{3}\bar{2}}^s+\Theta_{\bar{2},1\bar{3}}^s+\Theta_{\bar{3},\bar{2}1}^s,
\end{eqnarray}
where the normalized scattering strength due to each channel is given by the simple formula
\begin{equation}
\label{eq:Thetaijl}
\Theta_{i,jl}^s=\frac{1}{\omega_j}(c\mathbf{k}_i\cdot\mathbf{f}_{s,j})(\mathbf{e}_i\cdot\mathbf{f}_{s,l}).
\end{equation}
As defined before, here $\mathbf{e}_j$ is the complex unit polarization vector and $\mathbf{f}_{s,j}:=\mathbb{F}_{s,\mathbf{k}_j}\mathbf{e}_j$. The abbreviated notation $\bar{j}$ is interpreted by $\mathbf{k}_{\bar{j}}=-\mathbf{k}_j$, as well as the notations Eqs.~(\ref{eq:notationz}) and (\ref{eq:notationa}). In Sec.~\ref{sec:fluid-2nd:Lagrangian}, I will show that the normalized scattering strength $\Theta^s$ is related to the reduced scattering matrix element of the quantized theory. Then, the above six scattering channels simply correspond to the $3!=6$ permutations when contracting a single interaction vertex in the quantized theory.

It is instructive to count how many degrees of freedom the three-wave coupling coefficient $\Gamma$ contains. For each wave, its 4-momentum is constrained by one dispersion relation. Once the 4-momentum is fixed, the wave polarization is determined by the dispersion tensor up to the wave amplitude, which $\Gamma$ does not dependent on. Therefore, for each wave, there are three degrees of freedom. Now that the resonant conditions give another four constrains, there are in total $3\times3-4=5$ independent variables. Therefore, in the absence of additional symmetry, the three-wave coupling coefficient $\Gamma$ is a function of five independent variables in a given plasma.

Without solving the three-wave equations, a number of experimental observables can already be extracted from the coupling coefficient. For example, $\Gamma$ can be related to the growth rate of parametric instabilities. Consider the parametric decay instability where a pump wave with frequency $\omega_1$ decays into two waves with frequencies $\omega_2$ and $\omega_3$. Suppose the pump has constant amplitude $a_1$, and the decay waves have no spatial variation. Then solving the linearized three-wave equations, we find $a_2$ and $a_3$ grow exponentially with rate
\begin{equation}
\label{eq:GrowthRate}
\gamma_0=\frac{|\Gamma a_1|}{\sqrt{\omega_2\omega_3}}.
\end{equation}
The experimentally observed growth rate will be somewhat different from $\gamma_0$ due to wave damping. Wave damping, both collisional and collisionless, can be taken into account by inserting a phenomenological damping term $\nu a$ into the LHS of the three-wave equations. Solving the linearized equations, the growth rate, modified by wave damping, is
\begin{equation}
\label{eq:GrowthRateDamped}
\gamma=\sqrt{\gamma_0^2+\Big(\frac{\nu_2-\nu_3}{2}\Big)^2}-\frac{\nu_2+\nu_3}{2},
\end{equation} 
where $\nu_2$ and $\nu_3$ are the phenomenological damping rates of the two decay products. 
In addition to wave damping, the experimentally observed growth rate can also be modified by frequency mismatch $\delta\omega=\omega_1-\omega_2-\omega_3$. When the frequency mismatch is much smaller than the spectral width of waves, the three waves can still couple almost resonantly. To find the growth rate in the presence of small $\delta\omega$, we can promote the amplitude $a$ to be complex and change variable $\alpha_j:=a_je^{-it\delta\omega/2}$ for $j=2$ and $3$. This change of variable is equivalent to modifying the damping rates to $\nu'_2:=\nu_2+i\delta\omega/2$ and $\nu_3^{'*}:=\nu_3-i\delta\omega/2$. Therefore, the growth rate of parametric decay instability, modified by both weak damping and small frequency mismatch is
\begin{equation}
\gamma'=\sqrt{\gamma_0^2+\Big(\frac{\nu_2-\nu_3+i\delta\omega}{2}\Big)^2}-\frac{\nu_2+\nu_3}{2}.
\end{equation} 
The frequency mismatch $\delta\omega$ not only introduces amplitude modification, but also results in phase modification. In the following discussions, we shall only be concerned with the growth rate $\gamma_0$ as the observable, ignoring wave damping and frequency mismatch.
To get a sense of how large this growth rate is, we can compare it with Raman backscattering $\gamma_0=\gamma_R\mathcal{M}$, where the normalized growth rate is
\begin{equation}
\label{eq:GrwothM}
\mathcal{M}=2\frac{|\Gamma|}{\omega_p^2}\Big(\frac{\omega_p^3}{\omega_1\omega_2\omega_3}\Big)^{1/2},
\end{equation}
and $\gamma_R=\sqrt{\omega_1\omega_p}|a_1|/2$ the backward Raman growth rate in an unmagnetized plasma of the same density. In experiments, most signal will come from the largest growth rate, for which wave phases are synchronized. In other words, for given pump wave and decay products, the largest growth rate is attained when the wave phases are such that $|\sum_s\omega_{ps}^2Z_s\Theta_r^s/M_s|=|\sum_s\omega_{ps}^2Z_s\Theta^s/M_s|$. When this condition is met, the scattering strengths align in the complex plane and constructively add, so that the synchronized decay quickly dominate unsynchronized decays to give the dominate signal in experiments.

\begin{figure}[b]
	\centering
	\includegraphics[angle=0,width=0.4\textwidth]{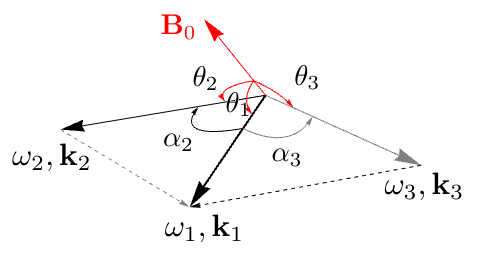}
	\caption[General geometry of three-wave interaction]{The most general geometry of three-wave scattering in a uniform plasma with a constant magnetic field. The three wave vectors $\mathbf{k}_1=\mathbf{k}_2+\mathbf{k}_3$ are in the same plane, and are at angles $\theta$'s with respect to the magnetic field.}
	\label{fig:ThreeWave}
\end{figure}

To evaluate the normalized growth rate $\mathcal{M}$, we can mimic what happens in experiments, where the general three-wave scattering geometry is depicted in Fig.~\ref{fig:ThreeWave}. In experiments, we can control the incident pump wave, whose frequency is $\omega_1$ and direction of propagation is $\hat{\mathbf{k}}_1$. Given these control variables and cold plasma parameters, the pump wave can be a superposition of the two eigenmodes with the same frequency. One eigenmode have larger wavelength and smaller wave vector $k_1^-$, and the other eigenmode has smaller wavelength and larger wave vector $k_1^+$, whenever $\omega_1$ is not in a spectral gap. 
Suppose we set up an experiment to select one of the eigenmodes, then the wave vector $\mathbf{k}_1$ and the wave polarization $\mathbf{e}_1$ of the pump wave are fixed. We can then make observations in the $\hat{\mathbf{k}}_2$ direction. In particular,
the frequency $\omega_2$ of the scattered wave can be measured using some probe or spectrometer, and the polarization $\mathbf{e}_2$ can be selected using some polarizer or filters placed along the $\hat{\mathbf{k}}_2$ direction. 
Whenever $\omega_3=\omega_1-\omega_2$ and $\mathbf{k}_3=\mathbf{k}_1-\mathbf{k}_2$ correspond to an eigenmode of the plasma, resonant three-wave scattering will happen, and the spectrometer will display a peak centered at $\omega_2$, whose height is related to the growth rate. 
This experimental procedure is exactly followed in my computer program that is used to numerically evaluate the growth rate. As a final check, among the 5 degrees of freedom contained in the formula for $\Gamma$, I have now fixed 3 degrees of freedom by choosing the pump wave, and fixed 2 other degrees of freedom by choosing the direction of observation.


\begin{figure}[!b]
	\centering
	\includegraphics[angle=0,width=0.78\textwidth]{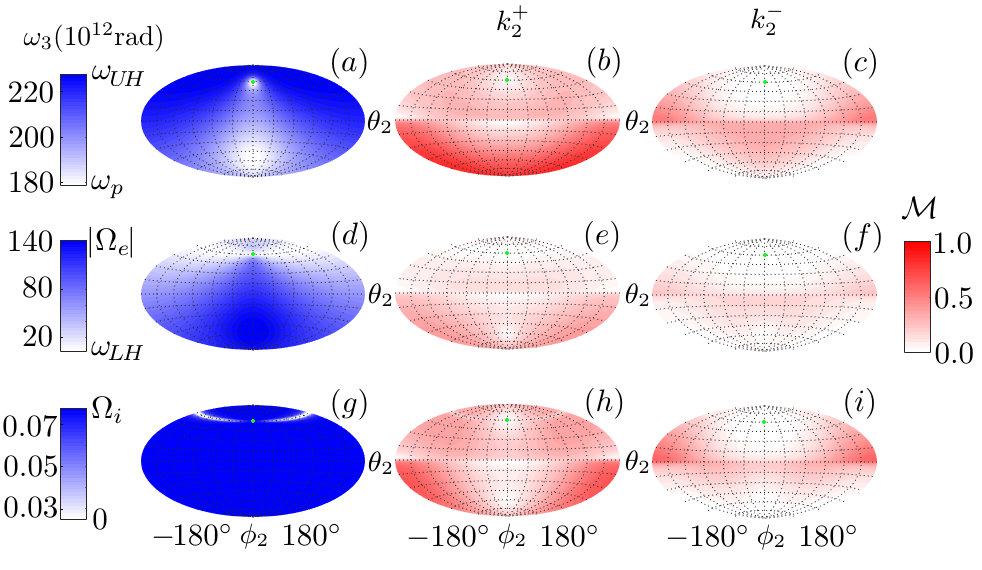}
	\caption[Scattering of oblique pump laser in magnetized hydrogen plasma]{Coherent scattering of an incident laser in $k_1^+$ eigenmode, propagating ($\theta_1=30^\circ$,\mbox{$\phi_1=0^\circ$}, green dots) in a magnetized cold hydrogen plasma with $\omega_p\approx\omega_1/10$ and $|\Omega_e|\approx0.8\omega_p$. 
		For scattering off the \textit{u}-branch waves (upper panel), the frequency downshift (a) is between $\omega_p$ and $\omega_{UH}$. The normalized growth rate $\mathcal{M}^+$ of $k_2^+$ eigenmode (b) is suppressed in polarization-forbidden regions near the equatorial plane ($\theta_2\approx90^\circ$), while the normalized growth rate $\mathcal{M}^-$ of $k_2^-$ eigenmode (c) is polarization-forbidden in forward ($\theta_2=\theta_1, \phi_2=0^\circ$) and backward ($\theta_2=180^\circ-\theta_1, \phi_2=\pm180^\circ$) directions. 
		For scattering off the \textit{l}-branch waves (middle panel), the frequency downshift (d) is between $\omega_{LH}$ and $|\Omega_e|$. In addition to polarization forbidden regions, the growth rate $\mathcal{M}^+$ (e) and $\mathcal{M}^-$ (f) are suppressed in interference-forbidden regions where electron and ion scattering cancel (near $\theta_2\approx\theta_1$), as well as in energy-forbidden regions where $\omega_3\approx|\Omega_e|$. 
		Finally, the incident laser can scatter off the \textit{b}-branch waves (bottom panel). The frequency downshift (g) is between zero and $\Omega_i$, and the growth rate $\mathcal{M}^+$ (h) and $\mathcal{M}^-$ (i) are suppressed in polarization-forbidden regions, as well as in energy-forbidden regions.}
	\label{fig:Stokes}
\end{figure}

As an example, I evaluate the normalized growth rate of a 1.06-$\mu$m Nd:glass laser in a magnetized hydrogen plasma, when the incident laser propagates at polar angle $\theta_1=30^\circ$ in $k_1^+$ eigenmode (Fig.~\ref{fig:Stokes}). I take the density of the fully ionized plasma to be $n_0=10^{19}\, \text{cm}^{-3}$, which is typical for gas jet plasmas. In addition, I take the magnetic field \mbox{$B_0=8.12$~MG}, achievable using current technology. In this plasma, the laser frequency is $\omega_1\approx10\omega_p$, $|\Omega_e|\approx0.8\omega_p$, and the magnetic field plays an important role in coherent Stokes scattering. In two-species cold plasmas, there exist three branches of magnetized plasma waves, each resulting in a different angular dependence of the normalized growth rate. 
First, in this over-dense plasma ($\omega_p>|\Omega_e|$), the upper (\textit{u}) branch is the Langmuir wave when $\mathbf{k}_3\parallel\mathbf{B}_0$. The \textit{u} branch becomes the upper-hybrid (UH) wave when $\mathbf{k}_3\perp\mathbf{B}_0$, whose frequency $\omega_{UH}\simeq\sqrt{\omega_p^2+\Omega_e^2}$ in the large-$k_3$ limit. 
For scattering off the \textit{u} branch, the frequency downshift (Fig.~\ref{fig:Stokes}a) is between $\omega_p$ and $\omega_{UH}$. For $k_2^+$ eigenmode, backscattering is favored while scattering perpendicular to $\hat{\mathbf{k}}_1$, where $\mathbf{e}_1^\dagger\mathbf{e}_2\approx0$, is forbidden (Fig.~\ref{fig:Stokes}b). On the contrary, the polarization of $k_2^-$ eigenmode is such that forward and backward scattering are forbidden, while perpendicular scattering is allowed (Fig.~\ref{fig:Stokes}c). 
Second, the laser can scatter from the lower (\textit{l}) branch plasma wave. The \textit{l} branch is the electron-cyclotron wave when $\mathbf{k}_3\parallel\mathbf{B}_0$, and it becomes the lower-hybrid (LH) wave when $\mathbf{k}_3\perp\mathbf{B}_0$, whose frequency $\omega_{LH}\simeq\sqrt{|\Omega_e|\Omega_i}\omega_p/\omega_{UH}$ in the large-$k_3$ limit.
For scattering from the \textit{l} branch, the frequency downshift (Fig.~\ref{fig:Stokes}d) is between $\omega_{LH}$ and $|\Omega_e|$. In addition to polarization-forbidden regions, both $k_2^+$ (Fig.~\ref{fig:Stokes}e) and $k_2^-$ (Fig.~\ref{fig:Stokes}f) scattering encounter special angles where electron and ion scattering destructively interfere and therefore cancel one another (near $\theta_2\approx\theta_1$).
Finally, the bottom (\textit{b}) branch is the Alfv\'en wave in the small-$k_3$ limit, and become the ion-cyclotron wave in the large-$k_3$ limit. For scattering off the bottom (\textit{b}) branch, the frequency downshift (Fig.~\ref{fig:Stokes}g) is between zero and $\Omega_i$. Both $k_2^+$ scattering (Fig.~\ref{fig:Stokes}h) and $k_2^-$ scattering (Fig.~\ref{fig:Stokes}i) encounter energy forbidden regions near $\theta_2\approx\theta_1$, where plasma waves are energetically too expensive to excite. 
Away from these polarization, interference, and energy forbidden regions where $u_i\gg|\Theta^s|$, coherent Stokes scattering from magnetized plasma waves have growth rates comparable to that of Raman backscattering.

In the most general cases, the growth rate needs to be evaluated numerically. In what follows, I will discuss the special cases of \textit{TTL} and \textit{LLL} scattering, where simple analytical expression can be obtained. For convenience, let me summarize the key formula here. The normalized growth rate $\mathcal{M}$ is given by Eq.~(\ref{eq:GrwothM}), which is proportional to the coupling coefficient $\Gamma$ given by Eq.~(\ref{eq:coupling}). The most important term in the numerator of $\Gamma$ is the scattering strength $\Theta^s$, which is given by Eqs.~(\ref{eq:Theta3}) and (\ref{eq:Thetaijl}). The most important term in the denominator of $\Gamma$ is the wave energy coefficient $u$, which is given by Eq.~(\ref{eq:uk}). When computing both $\Theta^s$ and $u$, we need the forcing operator $\mathbb{F}$, which is given by Eq.~(\ref{eq:F}). Finally, we need to ensure that the resonance conditions Eqs.~(\ref{eq:resonantK}) and (\ref{eq:resonantW}) are satisfied by three otherwise arbitrary on-shell waves. 
\subsection{Special case: scattering of transverse waves\label{sec:fluid-2nd:coefficient:TTL}}

Consider the decay of a pump laser ($\omega_1$) into a scattered laser ($\omega_2$) and a plasma wave ($\omega_3$). Since the frequency $\omega_{1,2}\gg\Omega_s$, the magnetization ratio $\beta_{1,2}\simeq 0$ and the magnetization factor $\gamma_{1,2}\simeq 1$ for any species. Consequently, the forcing operator $\mathbb{F}_{1,2}\simeq\mathbb{I}$ is approximately the identity operator, and the lasers are therefore transverse electromagnetic waves. As for the plasma wave, using the quasi-longitudinal approximation $\mathbf{e}_3\simeq\hat{\mathbf{k}}_3$, the inner products is purely real:
\begin{equation}
	\label{eq:TTL_inner}
	\hat{\mathbf{k}}_{3}\cdot\hat{\mathbf{f}}^*_{s,3}\simeq\hat{\mathbf{k}}_{3}\cdot\mathbb{F}_{s,3}\hat{\mathbf{k}}_3=\gamma_{s,3}^2(1-\beta_{s,3}^2\cos^2\theta_3),
\end{equation}
where $\theta_3$ is the angle between $\mathbf{k}_3$ and $\mathbf{b}$ as shown in Fig.~\ref{fig:ThreeWave}, and $\hat{\mathbf{k}}_3$ is the unit vector along $\mathbf{k}_3$ direction. With these basic setup, we can readily evaluate the growth rate.

Let us first calculate the wave energy coefficients Eq.~(\ref{eq:uk}), which enters the denominator of the coupling coefficient. Since $\mathbb{F}_{1,2}\simeq\mathbb{I}$, the wave energy coefficients for the lasers are simply
\begin{equation}
	\label{eq:TTL_U12}
	u_1\simeq u_2\simeq 1.
\end{equation}
As for the quasi-longitudinal plasma wave, using Eq.~(\ref{eq:u_coef_L}), the wave energy coefficient 
\begin{equation}
	\label{eq:TTL_U3}
	u_3\simeq 1+\sum_s\frac{\omega_{ps}^2}{\omega_3^2}\gamma_{s,3}^4\beta_{s,3}^2\sin^2\theta_3.
\end{equation}
Notice that, $u_3$ is always positive, although $\gamma_{s,3}^2$ can be either positive or negative, depending on whether $\beta_{s,3}$ is either smaller or larger than one.

To find the normalized scattering strength Eq.~(\ref{eq:Theta3}), which enters the numerator of $\Gamma$, we again use the fact $\omega_{1,2}\gg\omega_3$. Since the wave vectors are comparable in magnitudes, the dominant terms of the coupling strength are
the two terms proportional to $1/\omega_3$, 
if the inner product $\mathbf{e}_1\cdot\mathbf{f}_{2}^*\simeq\mathbf{f}_1\cdot\mathbf{e}_{2}^*\simeq\mathbf{e}_1\cdot\mathbf{e}_{2}^*$ is of oder unity. Using the resonance condition $\mathbf{k}_1-\mathbf{k}_2=\mathbf{k}_3$, the dominant term of the\textit{TTL} scattering strength is
\begin{equation}
	\label{eq:TTL_THETAr}
	\Theta^s\simeq-\frac{ck_3}{\omega_3} (\hat{\mathbf{k}}_{3}\cdot\mathbb{F}_{s,3}\hat{\mathbf{k}}_3) (\mathbf{e}_1\cdot\mathbf{e}_{2}^*),
\end{equation}
where the inner product $\hat{\mathbf{k}}_{3}\cdot\mathbb{F}_{s,3}\hat{\mathbf{k}}_3$ is given explicitly by Eq.~(\ref{eq:TTL_inner}). Now that we have simplified both the denominator and the numerator of Eq.~(\ref{eq:coupling}), a simple formula for the three-wave coupling coefficient $\Gamma$ can be obtained.

Having obtained an explicit formula for the coupling coefficient, we can use it to obtain expressions for $\mathcal{M}_T$, the normalized growth rate of the \textit{TTL} scattering. Clearly, $\mathcal{M}_T$ is proportional to the coupling coefficient $\Gamma=\omega_p^2\mu/4$ up to some kinematic factor
\begin{equation}
	\label{eq:MT}
	\mathcal{M}_T=\frac{1}{2}\Big(\frac{\omega_p^3}{\omega_1\omega_2\omega_3}\Big)^{1/2}\mu_{T},
\end{equation}
where the normalized coupling coefficient $\mu_T$ is given by
\begin{equation}
	\label{eq:uT}
	\mu_T\simeq \sum_s\frac{Z_s}{M_s}\frac{\omega_{ps}^2}{\omega_p^2}\frac{ck_3}{\omega_3}\frac{\hat{\mathbf{k}}_{3}\cdot\mathbb{F}_{s,3}\hat{\mathbf{k}}_3}{u_3^{1/2}}.
\end{equation}
In the unmagnetized limit $B_0\rightarrow 0$, 
we have $\beta_3\rightarrow 0$ and $\gamma_3\rightarrow 1$. Since ion mass is much larger than electron mass, we have $\mu_T\rightarrow -ck_3/\omega_3$. Moreover, since the lasers can only couple through the Langmuir wave in cold unmagnetized plasma, we have $\omega_3\rightarrow\omega_p$. Then, the normalized growth rate $\mathcal{M}_T\rightarrow ck_3/2\sqrt{\omega_1\omega_2}$. Finally, in backward scattering geometry $ck_3=ck_1+ck_2\simeq\omega_1+\omega_2\simeq2\omega_0$, where we have denoted $\omega_0:=\omega_1\simeq\omega_2$. We see $\mathcal{M}_T\rightarrow 1$ in the unmagnetized limit as expected. 

The normalized growth rate becomes particularly simple when waves propagate at special angles. For example, consider the situation where the three waves propagate along the magnetic field $\mathbf{B}_0$, and the plasma wave $\omega_3=\omega_p$ is the Langmuir wave. Since $\gamma_{s,3}^2$ remains finite as $\theta_3\rightarrow 0$, the normalized growth rate for collimated parallel wave propagation is
\begin{equation}
	\label{eq:MTP}
	\mathcal{M}_{T\parallel}^{P}\simeq-\frac{1}{2}\frac{ck_3}{\sqrt{\omega_1\omega_2}},
\end{equation}
where we have used $M_i\gg1$ to drop the summation over species. The above is exactly the same as the unmagnetized result \citep{Davidson72,Laham98}, which is expected because the plasma wave is not affected by the parallel magnetic field. 

To give another simple example, consider the situation where the three waves are collimated and propagate perpendicular to the magnetic field $\mathbf{B}_0$. In cold electron-ion plasma, there are two \textit{L} waves in the perpendicular direction: the upper-hybrid (UH) wave and the lower-hybrid (LH) wave. Let us first consider scattering mediated by the UH wave $\omega_3\simeq\omega_{UH}\simeq\sqrt{\omega_p^2+\Omega_e^2}$. In this situation, the magnetization factor $\gamma_{3,e}^2\simeq\omega_{UH}^2/\omega_p^2$ and $\gamma_{3,i}^2\simeq 1$. Since $M_i\gg1$, the dominant contribution for both the wave energy coefficient and the scattering strength comes from electrons. The wave energy coefficient $u_3\simeq\omega_{UH}^2/\omega_p^2$, and the normalized coupling coefficient $\mu_T\simeq-ck_3/\omega_p$. Therefore, the normalized growth rate for collimated perpendicular wave propagation mediated by the UH wave is
\begin{equation}
	\label{eq:MTUH}
	\mathcal{M}_{T\perp}^{UH}\simeq-\frac{1}{2}\frac{ck_3}{\sqrt{\omega_1\omega_2}}\bigg(\frac{\omega_p}{\omega_{UH}}\bigg)^{1/2}.
\end{equation}
Similarly, let us consider scattering mediated by the LH wave $\omega_3\simeq\omega_{LH}\simeq\sqrt{|\Omega_e|\Omega_i}\omega_p/\omega_{UH}$. Since the LH frequency satisfies $\Omega_i\ll\omega_{LH}\ll|\Omega_e|$, the magnetization ratios $\beta_{3,e}\gg1$ and \mbox{$\beta_{3,i}\ll1$}. Consequently, the magnetization factor $\gamma_{3,e}\simeq-1/\beta_{3,e}^2$ and $\gamma_{3,i}\simeq 1$. When $\omega_p\sim|\Omega_e|$ are comparable, electron contributions again dominate. The wave energy coefficient $u_3\simeq\omega_{UH}^2/\Omega_e^2$, and the normalized coupling coefficient $\mu_T\simeq ck_3\omega_{LH}/\omega_{UH}|\Omega_e|$. Hence, the normalized growth rate for LH wave mediation in the collimated perpendicular geometry is
\begin{equation}
	\label{eq:MTLH}
	\mathcal{M}_{T\perp}^{LH}\simeq\frac{1}{2}\frac{ck_3}{\sqrt{\omega_1\omega_2}} \frac{\omega_p^{3/2}\omega_{LH}^{1/2}}{\omega_{UH}|\Omega_e|}.
\end{equation}
The above examples recover the results of \cite{Grebogi80}, who analyze the same problem in the restricted geometry where the waves are collimated and propagate perpendicular to the magnetic field.

Having reproduced well-known results, let us evaluate the normalized growth rate in more general geometry, where the waves are not collimated and propagate at angles with respect to the magnetic field. The normalized growth rate can be evaluated using the following procedure, mimicking what happens in an actual experiment where the plasma density and magnetic field strength are known. First, we shine a laser with frequency $\omega_1$ into the plasma at some angle $\theta_1$ with respect to the magnetic field. Then the wave vector $k_1$ is known from the dispersion relation. Second, we observe the scattered laser using some detector placed at angle $\theta_2$ with respect to the magnetic field, and point the detector at angle $\alpha_2$ with respect to the incoming laser. Suppose the detector can measure the frequency $\omega_2$ of the scattered laser, then from this frequency information, we immediately know $k_2$ from the dispersion relation, as well as $\omega_3=\omega_1-\omega_2$ from the resonance condition. Next, we can calculate $k_3=\sqrt{k_1^2+k_2^2-2k_1k_2\cos\alpha_2}$, and determine $\theta_3$ by inverting $\omega_3=\omega_r(\theta_3)$, where $\omega_r$ is the angle-dependent resonance frequency. Using this procedure, the normalized growth rate can be readily evaluated. Conversely, when plasma density and magnetic field are unknown, we may use information measured from laser scattering experiments to fit plasma parameters.

\subsubsection{Parallel pump}\label{sec:TTLPara}

\begin{figure}[t]
	\centering
	\includegraphics[angle=0,width=0.55\textwidth]{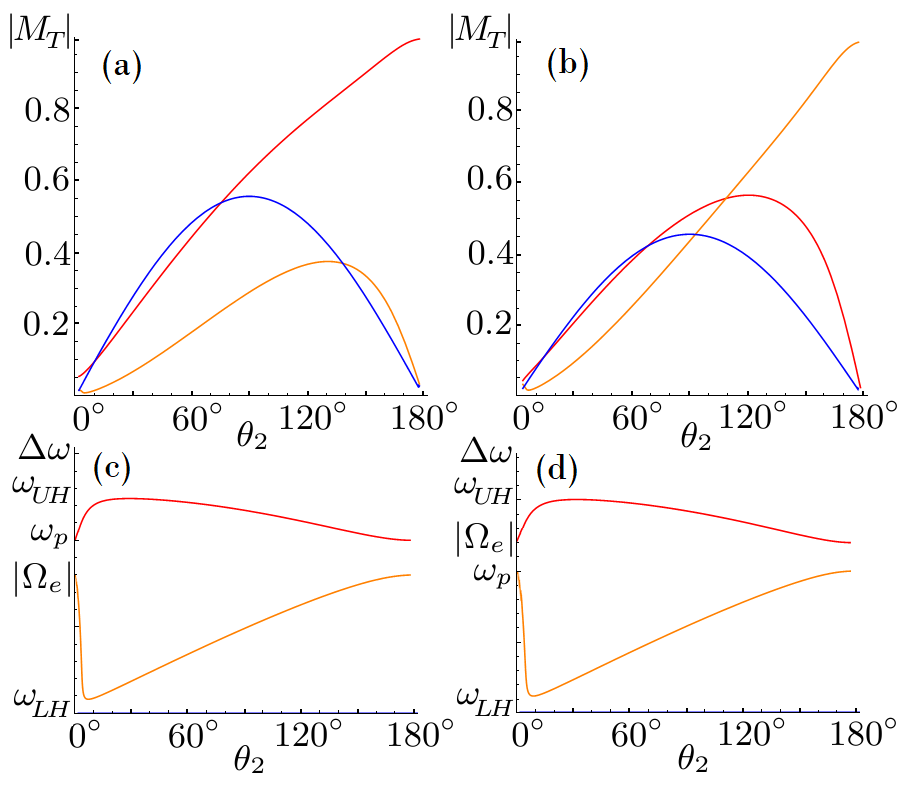}
	\caption[Scattering of parallel pump laser in magnetized hydrogen plasma]{Scattering of a parallel pump laser in uniform hydrogen plasmas. 
	The pump laser has frequency $\omega_1/\omega_p\!=\!10$, and the scattered laser propagates at angle $\theta_2$ with respect to $\mathbf{k}_1\!\parallel\!\mathbf{B}_0$. The laser can scatter from the upper resonance (red), the lower resonance (orange), and the bottom resonance (blue). When the plasma is over-dense, e.g.~$|\Omega_e|/\omega_p=0.8$ (a, c), the upper resonance is Langmuir-like, while the lower and bottom resonances are cyclotron-like; when the plasma is under-dense, e.g.~$|\Omega_e|/\omega_p=1.2$ (b, d), the lower resonance is Langmuir-like, while the upper and bottom resonances are cyclotron-like. For Langmuir-like resonance, the frequency shift (c, d) $\Delta\omega\rightarrow\omega_p$, and the normalized growth rate (a, b) is monotonically increasing; while for cyclotron-like resonances, $\Delta\omega\rightarrow|\Omega_e|,\Omega_i$, and the normalized growth rate $|\mathcal{M}_{T}|$ peaks at intermediate $\theta_2$, while becoming zero for exact backscattering. See text for how $|\mathcal{M}_{T}|$ scales with plasma parameters.
	}
	\label{fig:TTL_Para}
\end{figure}

To demonstrate how to evaluate the normalized growth rate $\mathcal{M}_{T}$, consider the example where the incident laser propagates along the magnetic field, while the scattered laser propagates at some angle $\theta_2$. In this case $\alpha_2=\theta_2$, and by cylindrical symmetry, $\mathcal{M}_{T}$ depends on only one free parameter $\theta_2$. In Fig.~\ref{fig:TTL_Para}, I plot the normalized growth rate and Stokes frequency shift in a hydrogen plasma with $\omega_1/\omega_p=10$. When there are only two charged species, as in the case of the hydrogen plasma, there are three electrostatic resonances the lasers can scatter from (Fig.~\ref{fig:Resonance}). The first resonance is the upper resonance, whose frequency asymptotes to the upper-hybrid frequency $\omega_{UH}$ when $\theta_3\rightarrow\pi/2$. When scattered from the upper resonance (red curves), the scattered laser is frequency down-shifted ($\Delta\omega=\omega_1-\omega_2$) by the largest amount. The second resonance is the lower resonance, whose frequency asymptotes to the lower-hybrid frequency $\omega_{LH}$ when $\theta_3\rightarrow\pi/2$. When scattered from the lower resonance (orange curves), the scattered laser is frequency-shifted by either $|\Omega_e|$ in over-dense plasma ($|\Omega_e|<\omega_p$), or by $\omega_p$ in under-dense plasma ($|\Omega_e|>\omega_p$), when $\theta_3\rightarrow0$. The third resonance is the bottom resonance, whose frequency asymptotes to $0$ when $\theta_3\rightarrow\pi/2$. When scattered from the bottom resonance (blue curves), the scattered laser is frequency-shifted by at most $\Omega_i$ when $\theta_3\rightarrow0$. Since $\Omega_i$ is much smaller than other frequency scales, the frequency shift $\Delta\omega$ for scattering off the bottom resonance is not discernible in Fig.~\ref{fig:TTL_Para}c and Fig.~\ref{fig:TTL_Para}d. As for the normalized growth rate (upper panels), we see $\mathcal{M}_{T}\rightarrow1$  when the laser is backscattered from the Langmuir resonance with $\Delta\omega\rightarrow\omega_p$, while $\mathcal{M}_{T}\rightarrow 0$ when the laser is scattered from the cyclotron resonances with $\Delta\omega\rightarrow|\Omega_e|,\Omega_i$. For Langmuir-like resonance, $\mathcal{M}_{T}$ increases monotonically with $\theta_2$. In contrast, for cyclotron-like resonances, $\mathcal{M}_{T}$ peaks at intermediate $\theta_2$, and becomes zero for exact backscattering.

To better understand the angular dependence of the normalized growth rate $\mathcal{M}_{T}$, let us find its asymptotic expressions. In the limit $\omega_{1,2}\gg\omega_3$, the wave vector $k_2/k_1\simeq 1$ and $k_3/k_1\simeq 2\sin(\theta_2/2)$. At finite angle $\theta_2>0$, we can approximate $\theta_3\simeq(\pi-\theta_2)/2$
. For even larger $\theta_2$, we can also approximate the resonance frequency $\omega_3$ using Eqs.~(\ref{eq:wu_para})-(\ref{eq:wb_para}), because $\theta_3\sim 0$ is now small. These asymptotic geometric relations will be useful next when we evaluate the coupling coefficient.

First, consider scattering off the Langmuir-like resonance $\omega_3\sim\omega_p$. Since $\gamma_{3,s}$ is finite, the lowest-order angular dependence comes from $k_3$. Take the limit $\theta_3\rightarrow 0$, we get Eq.~(\ref{eq:MTP}). Now retain the angular dependence of $k_3$, we can grossly approximate
\begin{equation}
	|\mathcal{M}_{T}^p|\simeq\sin\frac{\theta_2}{2}.
\end{equation}
This approximation is of course very crude, but it captures the monotonic increasing feature for scattering off the Langmuir-like resonance. In fact, the above result becomes a very good approximation when the magnetic field $B_0\rightarrow0$. In this unmagnetized limit, we recover the angular dependence of Raman scattering.

Second, consider scattering off the electron-cyclotron-like resonance $\omega_3\sim|\Omega_e|$. Notice that in this case, the magnetization factor $\gamma_{3,e}^2\gg 1$ for small $\theta_3$. Nevertheless, since both the numerator and the denominator contains this factor, $\mathcal{M}_{T}$ remains finite. For electrons, the magnetization ratio $\beta_{3,e}\simeq1$. Using Eq.~(\ref{eq:wl_para}), which is valid when $\omega_p\ne|\Omega_e|$, the magnetization factor $\gamma_{3,e}^2\simeq(\Omega_e^2-\omega_p^2)/(\omega_p^2\sin^2\theta_3)$. In comparison, $\beta_{3,i}\ll 1$ and $\gamma_{3,i}^2\simeq1$. Hence the dominant contribution comes from electrons. Substituting these into formula Eq.~(\ref{eq:MT}), we see to leading order the normalized growth rate is
\begin{equation}
	\label{eq:MTe}
	|\mathcal{M}_{T}^e|\simeq\frac{1}{2}\bigg(\frac{\omega_p}{\omega_3}\bigg)^{1/2}\sin\theta_2,
\end{equation}
where $\omega_3$ as function of $\theta_2$ is given by Eq.~(\ref{eq:wl_para}), with $\theta_3\simeq(\pi-\theta_2)/2$. From Eq.~(\ref{eq:MTe}), we see $|\mathcal{M}_{T}^e|$ reaches maximum when the laser is scattered almost perpendicularly to the magnetic field. The maximum value scales roughly as $|\mathcal{M}_{T}^e|\sim\sqrt{\omega_p/|\Omega_e|}/2$, which can be very large in weakly magnetized plasmas, as long as the cold-fluid approximation remains valid. Away from $\theta_2\sim\pi/2$, the normalized growth rate $|\mathcal{M}_{T}^e|$ falls off to zero. This falloff is expected, because exciting cyclotron resonance is energetically forbidden. 

In the end, consider scattering off ion-cyclotron-like resonance $\omega_3\sim\Omega_i$. In this case, the ion contribution to the wave energy coefficient is no longer negligible, because $\beta_{3,i}\simeq1$ and $\gamma_{3,i}^2\simeq\Omega_e/\Omega_i\tan^2(\theta_2/2)\gg1$, as can be seen from Eq.~(\ref{eq:wb_para}). The scattering strength is still dominated by electrons, for which $\beta_{3,e}\gg1$, and $\gamma_{3,e}^2\simeq-1/\beta_{3,e}^2\ll1$. Substituting these into Eq.~(\ref{eq:MT}), the normalized growth rate is
\begin{equation}
	\label{eq:MTi}
	|\mathcal{M}_{T}^i|\simeq\frac{1}{2}\bigg(\frac{\phantom{.}\omega_p\phantom{.}\Omega_i}{|\Omega_e|\omega_3}\bigg)^{1/2}\sin\theta_2.
\end{equation}
We see the above result is rather similar to Eq.~(\ref{eq:MTe}), except that $\omega_3\sim\Omega_i$ has very weak angular dependence. Therefore, $|\mathcal{M}_{T}^i|$ is very well approximated by Eq.~(\ref{eq:MTi}). The normalized growth rate peaks almost at $\theta_2=\pi/2$, reaching a maximum $|\mathcal{M}_{T}^i|\sim\sqrt{\omega_p/|\Omega_e|}/2$, which can be very large in weakly magnetized plasmas. Similar to the electron cyclotron case, $|\mathcal{M}_{T}^i|$ falls off to zero for parallel scattering due to energy suppression.

\subsubsection{Perpendicular pump}\label{sec:TTLPerp}
\begin{figure}[!t]
	\centering
	\includegraphics[angle=0,width=0.6\textwidth]{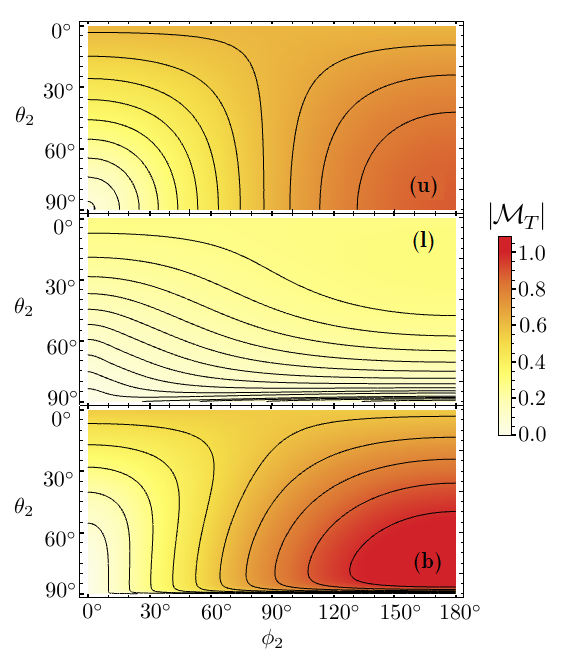}
	\caption[Scattering of perpendicular pump laser in magnetized hydrogen plasma 2D]{Normalized growth rate $|\mathcal{M}_{T}|$ for scattering of a perpendicular pump laser ($\mathbf{k}_1\!\perp\!\mathbf{B}_0$) in a uniform hydrogen plasma with $\omega_1/\omega_p=10$ and $|\Omega_e|/\omega_p=0.8$. In spherical coordinate, the scattered laser propagates at polar angle $\theta_2$ with respect to $\mathbf{B}_0$, and azimuthal angle $\phi_2$ measured from $\mathbf{k}_1$. The laser can scatter from the upper resonance (u), in which case backscattering is the strongest scattering mode. Alternatively, the laser can scatter off the lower resonance (l). In this case, one maximum of $|\mathcal{M}_{T}|$ is attained for backscattering, and another maximum is attained when the scattered laser propagate almost perpendicular to the incident laser along the magnetic field. Finally, the laser can scatter off the bottom resonance (b). In this case, exact backscattering is suppressed while nearly backward scattering is strong.}
	\label{fig:TTL_Perp}
\end{figure}

Consider the other special case where the pump laser propagates perpendicular to the magnetic field. In this geometry, it is natural to plot the normalized growth rate $|\mathcal{M}_{T}|$ in spherical coordinate (Fig.~\ref{fig:TTL_Perp}), where the polar angle $\theta_2$ is measured from the magnetic field $\mathbf{B}_0$, and the azimuthal angle $\phi_2$ is measured from the wave vector $\mathbf{k}_1$. By symmetry of this setup, it is obvious that $\mathcal{M}_{T}(\phi_2,\theta_2) =\mathcal{M}_{T}(\phi_2,\pi-\theta_2) =\mathcal{M}_{T}(-\phi_2, \theta_2)$. Therefore, it is sufficient to consider the range $\theta_2\in[0,\pi/2]$ and $\phi_2\in[0,\pi]$. By matching the $\mathbf{k}$ resonance, we can read $\theta_3$ from the spherical coordinates $(\phi_2,\theta_2)$, and thereafter read the frequency shift $\omega_3$ from Fig.~\ref{fig:Resonance}. As for the growth rate, in electron-ion plasma, 
when scattered from the upper resonance (Fig.~\ref{fig:TTL_Perp}u), backscattering has the largest growth rate. While for scattering off the lower resonance (Fig.~\ref{fig:TTL_Perp}l), $|\mathcal{M}_{T}|$ reaches maximum for both backscattering and nearly parallel scattering, where the scattered laser propagates almost parallel to the magnetic field. In comparison, for scattering off the bottom resonance  (Fig.~\ref{fig:TTL_Perp}b), the normalized growth rate peaks for nearly backward scattering, and falls to zero for exact backscattering.

To better understand the angular dependence of the normalized growth rate, let us consider its asymptotic expressions for two special cases. The first special case is when all waves lie in the plane perpendicular to the magnetic field, namely, when $\theta_2=90^\circ$. In this case, the angle $\theta_3$ is fixed to $90^\circ$, and the frequency of the plasma resonances are also fixed to $\omega_{UH}$, $\omega_{LH}$, or zero. Therefore, the angular dependence only comes from $k_3$. In the limit $\omega_{1,2}\gg\omega_3$, we have $k_3\simeq2k_1\sin(\phi_2/2)$. Using Eqs.~(\ref{eq:MTUH}) and (\ref{eq:MTLH}), it is easy to see, for scattering off UH and LH waves in the perpendicular plane, the growth rates are
\begin{eqnarray}
	|\mathcal{M}_{T\perp}^{UH}|&\simeq&\bigg(\frac{\omega_p}{\omega_{UH}}\bigg)^{1/2}\sin\frac{\phi_2}{2},\\
	|\mathcal{M}_{T\perp}^{LH}|&\simeq&\frac{\omega_p^{3/2}\omega_{LH}^{1/2}}{\omega_{UH}|\Omega_e|}\sin\frac{\phi_2}{2}.
\end{eqnarray}
Now let us calculate $\mathcal{M}_{T\perp}^{b}$ for scattering off the bottom resonance. Using asymptotic expression Eq.~(\ref{eq:wb_perp}) for $\omega_3$, we see although the magnetization ratio $\beta_{3,s}\rightarrow\infty$, the product $\beta_{3,s}\cos\theta_3$ remains finite as $\theta_3\rightarrow\pi/2$. Since the magnetization factor $\gamma_{3,s}\simeq-1/\beta_{3,s}^2\ll1$, it is easy to see $\mathcal{M}_{T\perp}^{b}\propto\sqrt{\omega_3}$, which goes to zero when $\theta_3\rightarrow\pi/2$. Hence, scattering off the bottom resonance in the perpendicular plane is completely suppressed:
\begin{equation}
	|\mathcal{M}_{T\perp}^{b}|=0.
\end{equation}
Consequently, exact backscattering from the bottom resonance is also suppressed.

\begin{figure}[!b]
	\centering
	\includegraphics[angle=0,width=0.6\textwidth]{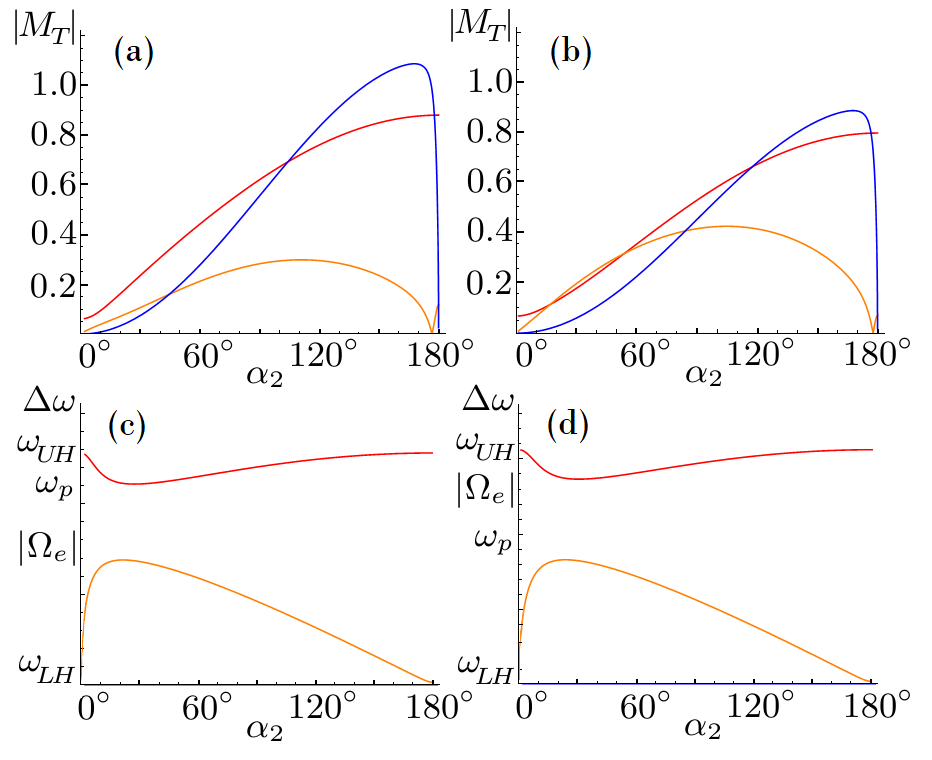}
	\caption[Scattering of perpendicular pump laser in magnetized hydrogen plasma 1D]{Scattering of a perpendicular pump laser with $\omega_1/\omega_p=10$ in a uniform hydrogen plasma. Figure (a) can be obtained from Fig.~\ref{fig:TTL_Perp} by taking a one dimensional cut along the unit sphere using the plane spanned by $\mathbf{k}_1\perp\mathbf{B}_0$. The scattered laser, propagating at angle $\alpha_2$ with respect to $\mathbf{k}_1$, can scatter from the upper resonance (red), the lower resonance (orange), and the bottom resonance (blue). Both the normalized growth rate $|\mathcal{M}_{T}|$ (a, b) and the Stokes frequency shifts $\Delta\omega$ (c, d) behave qualitatively the same in over-dense plasma, e.g.~$|\Omega_e|/\omega_p=0.8$ (a, c), and in under-dense plasma, e.g.~$|\Omega_e|/\omega_p=1.2$  (b, d). As $\alpha_2$ increases from $0^\circ$ to $180^\circ$, $|\mathcal{M}_{T}|$ increases monotonically for scattering from the upper resonance. For scattering off the lower resonance, $|\mathcal{M}_{T}|$ hit zero near $\alpha_2\sim176^\circ$, where electron and ion contributions exactly cancel, and then increase to finite value at exact backscattering. In contrast, when the laser is scattered from the bottom resonance, $|\mathcal{M}_{T}|$ strongly peaks near $\alpha_2\sim170^\circ$, and becomes zero for exact backward scattering. See text for how $|\mathcal{M}_{T}|$ scales with plasma parameters.}
	\label{fig:TTL_Perp1D}
\end{figure}

To see how $\mathcal{M}_{T}^{b}$ climbs up from zero, consider the second special case where $\mathbf{k}_2$ is in the plane spanned by $\mathbf{k}_1$ and $\mathbf{b}$ (Fig.~\ref{fig:TTL_Perp1D}). In this case, it is more natural to consider $\mathcal{M}_{T}$ as function of $\alpha_2$, the angle between $\mathbf{k}_1$ and $\mathbf{k}_2$. Let us find the asymptotic expression of $\mathcal{M}_{T}^{b}$ when $\alpha_2\sim\pi$. In this limit, we have $\theta_3\sim\pi/2$, and the resonance frequency $\omega_3$ can be approximated by Eq.~(\ref{eq:wb_perp}). Then, the magnetization ratios $\beta_{3,e}^2\simeq\Omega_e^2/\Omega_i^2+|\Omega_e|/(\Omega_i\cos^2\theta_3)$ and $\beta_{3,i}^2\simeq1+\Omega_i/(|\Omega_e|\cos^2\theta_3)$. Consequently, the magnetization factors can be well approximated by $\gamma_{3,e}^2\simeq-1/\beta_{3,e}^2$ and $\gamma_{3,i}^2\simeq-|\Omega_e|\cos^2\theta_3/\Omega_i$. Moreover, since $\omega_{1,2}\gg\omega_3$, the angle $\theta_3\simeq\alpha_2/2$ and the wave vector $k_3\simeq2k_1\sin(\alpha_2/2)$. Substituting these into formula Eq.~(\ref{eq:MT}), we see when $\alpha_2\sim\pi$, the normalized growth rate is
\begin{equation}
	|\mathcal{M}_{T}^{b}|^2\simeq\frac{[\zeta(1+\zeta\cos^2\frac{\alpha_2}{2})]^{3/2}\sin^2\frac{\alpha_2}{2}\cos\frac{\alpha_2}{2}}{r^3+r[1+\zeta(1+\zeta\cos^2\frac{\alpha_2}{2})^2]\sin^2\frac{\alpha_2}{2}},
\end{equation}
where $r:=|\Omega_e|/\omega_p$ and $\zeta:=M_i/Z_i\gg1$. To see the lowest-order angular dependence, we can use a cruder but simpler approximation $|\mathcal{M}_{T}^{b}|^2\simeq\zeta^{1/2}\cos(\alpha_2/2)/r$. We see $|\mathcal{M}_{T}^{b}|$ increases sharply from zero away from exact backscattering. 
In the other limit $\alpha_2\sim0$, we can use Eq.~(\ref{eq:MTi}), and the normalized growth rate is
\begin{equation}
	|\mathcal{M}_{T}^{b}|\simeq\frac{\sin^2\frac{\alpha_2}{2}}{r^{1/2}}\Big(1-\frac{1}{\zeta}\tan^2\frac{\alpha_2}{2}\Big)^{-3/4}.
\end{equation} 
We see scattering from the bottom resonance can be strong when the plasma is weakly magnetized, as long as the scattering angle is away from exact forward or backward scattering.

In summary, the \textit{TTL} scattering in magnetized plasma is mostly due to density beating [Eq.~(\ref{eq:TTL_THETAr})], and the modification due to the magnetic field can be represented by the normalized growth rate $\mathcal{M}_{T}$. In magnetized plasmas, cyclotron-like resonances, in addition to the Langmuir-like resonance, contribute to the scattering of the \textit{T} waves. When scattered from the Langmuir-like resonance, both the wave energy coefficient and the scattering strength are finite. Therefore in this case, the angular dependence of $\mathcal{M}_{T}$ comes mostly from $k_3$, which reaches maximum for backscattering. In contrast, for scattering from cyclotron-like resonances, both the scattering strength and the wave energy coefficient can blow up. Their ratio, $\mathcal{M}_{T}$, goes to zero when the scattering angles are such that the \textit{L} wave frequency approaches either zero or the cyclotron frequencies. In addition, $\mathcal{M}_{T}$ can also become zeros at special angles where scattering from electrons and ions exactly cancel. Away from these special angles, scattering from cyclotron-like resonances, which increases with decreasing magnetic field, typically have growth rates that are comparable to scattering from Langmuir-like resonances. When the plasma parameters are known, we can determine the angular dependence of $\mathcal{M}_{T}$ using Eq.~(\ref{eq:MT}). This knowledge can be used to choose injection angles of two lasers such that their scattering is either enhanced or suppressed. Conversely, by measuring angular dependence of $\mathcal{M}_{T}$ in laser scattering experiments, one may be able to fit plasma parameters to match Eq.~(\ref{eq:MT}). This provides a diagnostic method from which the magnetic field, as well as the plasma density and composition can be measured.

\subsection{Special case: scattering of longitudinal waves\label{sec:fluid-2nd:coefficient:LLL}}

Now let us consider the other scenario where the three-wave scattering happens between three resonant quasi-longitudinal waves. This happens, for example, when we launch an electrostatic wave into the plasma by some antenna arrays. When the wave power is strong enough to overcome damping, namely, when the damped growth rate [Eq.~(\ref{eq:GrowthRateDamped})] is positive, the pump wave may subsequently decay to two other waves that satisfies the resonance conditions. The decay waves are not necessarily electrostatic, but for the purpose of illustrating the general results in Sec.~\ref{sec:fluid-2nd:coefficient:formula}, I will only give examples where the two decay waves are also electrostatic. 

The coupling strength between three \textit{L} waves can be simplified using the approximation that the waves are quasi-longitudinal. Substituting $\mathbf{e}_i\simeq\hat{\mathbf{k}}_i$ into Eq.~(\ref{eq:Thetaijl}) and using the resonance condition (\ref{eq:resonantW}), the normalized scattering strength for \textit{LLL} scattering is
\begin{eqnarray}
	\label{eq:THETAL3}
	\nonumber
	\Theta^s\simeq&-&\frac{ck_1\omega_1}{\omega_2\omega_3}(\hat{\mathbf{k}}_1 \cdot\mathbb{F}_{s,2}^*\hat{\mathbf{k}}_2)(\hat{\mathbf{k}}_1\cdot\mathbb{F}_{s,3}^*\hat{\mathbf{k}}_3)\\
	&+&\frac{ck_2\omega_2}{\omega_3\omega_1}(\hat{\mathbf{k}}_2 \cdot\mathbb{F}_{s,1}\hat{\mathbf{k}}_1)(\hat{\mathbf{k}}_2\cdot\mathbb{F}_{s,3}^*\hat{\mathbf{k}}_3)\\
	\nonumber
	&+&\frac{ck_3\omega_3}{\omega_1\omega_2}(\hat{\mathbf{k}}_3 \cdot\mathbb{F}_{s,1}\hat{\mathbf{k}}_1)(\hat{\mathbf{k}}_3\cdot\mathbb{F}_{s,2}^*\hat{\mathbf{k}}_2),
\end{eqnarray}
where $k_i:=|\mathbf{k}_i|$ is the magnitude of the wave vector, and $\hat{\mathbf{k}}_i$ is the unit vector along $\mathbf{k}_i$ direction. It is easy to recognize that $\hat{\mathbf{k}}_i\cdot(\mathbb{F}_{s,j}/\omega_j)\hat{\mathbf{k}}_j$ is the projection of quiver velocity $\hat{\mathbf{v}}_j$ in $\hat{\mathbf{k}}_i$ direction. 
The first term in $\Theta^s$ is proportional to the rate of creating wave 1 by annihilating waves 2 and 3, the second term is proportional to the rate of annihilating waves 3 and $\bar{1}$ to create wave $\bar{2}$, and the last term can be interpreted similarly. The interference between these processes determines the overall scattering strength.

Having obtained expressions for the normalized scattering strength [Eq.~(\ref{eq:THETAL3})] and wave energy [Eq.~(\ref{eq:TTL_U3})], we can immediately evaluate the coupling coefficient [Eq.~(\ref{eq:coupling})], and find expressions for the parametric growth rate. In this case, since the pump wave is not an EM wave, it no longer makes sense to compare the growth rate with the backward Raman growth rate. Instead, we can choose a different normalization, and write the linear growth rate $\gamma_0$ [Eq.~(\ref{eq:GrowthRate})] of the parametric decay instability as
\begin{equation}
	\label{eq:GrowthRateL3}
	\gamma_0=\gamma_L|\mathcal{M}_L|,
\end{equation}
where $\gamma_L$ is purely determined by the pump wave as
\begin{equation}
	\gamma_L=\frac{1}{2}ck_1|a_1|.
\end{equation}
The normalized growth rate for \textit{LLL} scattering is then
\begin{equation}
	\label{eq:ML}
	\mathcal{M}_L=\frac{\omega_p}{2ck_1}\bigg(\frac{\omega_p^2}{\omega_2\omega_3}\bigg)^{1/2}\mu_L,
\end{equation}
which is proportional to the coupling coefficient $\Gamma=\omega_p^2\mu/4$ up to a kinematic factor. In the \textit{LLL} approximation, the normalized coupling coefficient is
\begin{equation}
	\label{eq:uL}
	\mu_L\simeq\sum_s\frac{Z_s}{M_s}\frac{\omega_{ps}^2}{\omega_p^2}\frac{\Theta^s_r}{(u_1u_2u_3)^{1/2}},
\end{equation}
where $\Theta^s_r$ is the real part of Eq.~(\ref{eq:THETAL3}). Again, notice that when the density of species $s$ goes to zero, its contribution to $\mu_L$ also goes to zero as expected.

To evaluate the normalized growth rate $\mathcal{M}_L$, we can use the following procedure to mimic what happens in an actual experiment. Suppose we know the species density and magnetic field, then we know what resonances are there in the plasma. We can then launch a pump wave at resonance frequency $\omega_1$ using some antenna array. The antenna array not only injects a wave at the given frequency, but also selects the wave vector $k_1$ and the wave direction $\theta_1$. 
To observe the decay waves, we can place a probe at some angle $\theta_2$ with respect to the magnetic field, and some azimuthal angle $\phi_2$ in a spherical coordinate. The probe can measure fluctuations of the plasma potential and therefore inform us about the wave frequency $\omega_2$. Then, we immediately know $\omega_3=\omega_1-\omega_2$ from the three-wave resonance condition. Moreover, since the third wave is a magnetic resonance, the frequency $\omega_3$ constrains the angle $\theta_3$ at which the third wave can propagate. However, a simple probe cannot measure the wave vector, so we will have to solve $k_2$ and $k_3$ from the resonance condition (\ref{eq:resonantK}), which can be written in components as
\begin{eqnarray}
	\label{eq:k32}
	&&k_3^2=k_1^2+k_2^2-2k_1k_2\cos\alpha_2,\\
	\label{eq:kpara}
	&&k_3\cos\theta_3=k_1\cos\theta_1-k_2\cos\theta_2.
\end{eqnarray}
Here $\alpha_2=\alpha_2(\theta_1,\theta_2,\phi_2)$ is the angle between $\mathbf{k}_1$ and $\mathbf{k}_2$. The above system of quadratic equations have two solutions in general. This degeneracy comes from the symmetry $2\leftrightarrow3$, because we cannot distinguish whether the probe is measuring wave 2 or wave 3, both of which are electrostatic resonances. If the solutions $k_2$ and $k_3$ are both real and positive, the three-wave resonance conditions can be satisfied. Then, three-wave decay will happen once the pump amplitude $a_1$ exceeds the damping threshold, for which the damped growth rate [Eq.~(\ref{eq:GrowthRateDamped})] becomes positive. In other words, we control $\omega_1$ and $\mathbf{k}_1$ by the antenna array, measure $\omega_2$ using probes, and infer $\omega_3$, $\mathbf{k}_2$, and $\mathbf{k}_3$ by solving resonance conditions. With these information, the analytical formula for the normalized growth rate $\mathcal{M}_L$ can be readily evaluated.

\subsubsection{Parallel pump}\label{sec:LLLPara}
To demonstrate how to evaluate the normalized growth rate $\mathcal{M}_L$, consider the example where the pump wave is launched along the magnetic field ($\theta_1=0$). In an electron-ion plasma, this geometry allows the antenna to launch three electrostatic waves: the Langmuir wave, the electron-cyclotron wave, or the ion-cyclotron wave. In the regime where $\omega_p\sim|\Omega_e|\sim|\omega_p-\Omega_e|\gg\Omega_i$, four decay modes are allowed by the resonance conditions: $u\rightarrow l+l$, $l\rightarrow l+l$, $l\rightarrow l+b$, and $b\rightarrow b+b$, where I have labeled waves by the resonance branch they belong to, and $u$, $l$, and $b$ denote the upper, lower and bottom resonances, respectively. 

\begin{figure}[t]
	\centering
	\includegraphics[angle=0,width=0.65\textwidth]{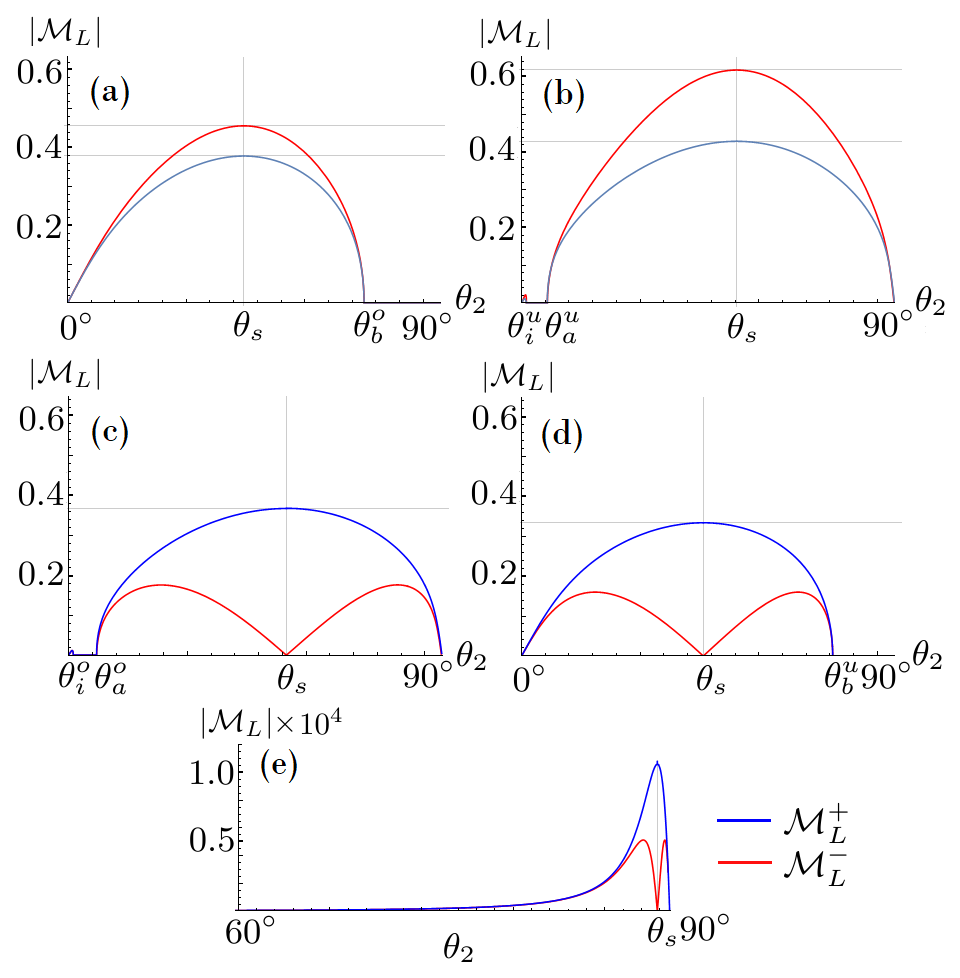}
	\caption[Scattering of parallel electrostatic pump in magnetized hydrogen plasma]{Scattering of a parallel electrostatic pump wave in uniform hydrogen plasmas, when observed at angle $\theta_2$ with respect to $\mathbf{k}_1\!\parallel\!\mathbf{B}_0$.  At each $\theta_2$, due to the degeneracy $2\leftrightarrow3$, the wave vector has two possible values $k_2^{\pm}$, corresponding to $\mathcal{M}_{L}^+$ (blue, $\theta_3>90^\circ$) and $\mathcal{M}_{L}^-$ (red, $\theta_3<90^\circ$), which satisfies $\mathcal{M}_{L}^+(\theta_2)=\mathcal{M}_L^-(\pi-\theta_2)$. The pump wave can be the Langmuir wave (a, b); the electron-cyclotron wave (c, d); and the ion-cyclotron wave (e). The normalized growth rate attains local extrema for symmetric scattering, where the two decay waves have the same frequency $\omega_r(\theta_s)=\omega_1/2$. In over-dense plasma, e.g.~$|\Omega_e|/\omega_p=0.8$ (a, c), $u\rightarrow l,l$ happens for $\theta_2<\theta_b^o$, where $\omega_l(\theta_b^o)=\omega_p-|\Omega_e|$; $l\rightarrow l,l$ happens for $\theta_2>\theta_a^o$, where $\omega_l(\theta_a^o)=|\Omega_e|-\omega_{LH}$; and $l\rightarrow l_2,b_3$ happens for $\theta_2<\theta_i^o$, where $\omega_l(\theta_i^o)=|\Omega_e|-\Omega_i$. In under-dense plasma, e.g.~$|\Omega_e|/\omega_p=1.2$ (b, d), $u\rightarrow l,l$ happens for $\theta_2<\theta_b^u$, where $\omega_l(\theta_b^u)=|\Omega_e|-\omega_p$; $l\rightarrow l,l$ happens for $\theta_2>\theta_a^u$, where $\omega_l(\theta_a^u)=\omega_p-\omega_{LH}$; and $l\rightarrow l_2,b_3$ happens for $\theta_2<\theta_i^u$, where $\omega_l(\theta_i^u)=\omega_p-\Omega_i$. Regardless of plasma density (e), $b\rightarrow b,b$ can always happen, for which the growth rate peaks near $\theta_s\sim88^\circ$, where the decay is symmetrical. The gray lines indicate the symmetric angles and the asymptotic maxima obtained in the text.}
	\label{fig:LLL_Para}
\end{figure}

First, let us consider the case where the pump wave is the Langmuir wave (Fig.~\ref{fig:LLL_Para}a, \ref{fig:LLL_Para}b). In this case, the magnetization factor $\gamma_1$ is finite, the wave energy coefficient $u_1=1$, and \mbox{$\mathbb{F}_{s,1}\hat{\mathbf{k}}_1=\hat{\mathbf{k}}_1$}. The normalized scattering strength (\ref{eq:THETAL3}) contains the following four simple inner products:  $(\hat{\mathbf{k}}_1\cdot\mathbb{F}_{s,2}^*\hat{\mathbf{k}}_2) =(\hat{\mathbf{k}}_2\cdot\mathbb{F}_{s,1}\hat{\mathbf{k}}_1)=\cos\theta_2$; $(\hat{\mathbf{k}}_1\cdot\mathbb{F}_{s,3}^*\hat{\mathbf{k}}_3)= (\hat{\mathbf{k}}_3\cdot\mathbb{F}_{s,1}\hat{\mathbf{k}}_1)=\cos\theta_3$, as well as two other inner products $(\hat{\mathbf{k}}_2\cdot\mathbb{F}_{s,3}^*\hat{\mathbf{k}}_3)=\cos\theta_2\cos\theta_3-\gamma_{s,3}^2\sin\theta_2\sin\theta_3$; and $(\hat{\mathbf{k}}_3\cdot\mathbb{F}_{s,2}^*\hat{\mathbf{k}}_2)=\cos\theta_3\cos\theta_2-\gamma_{s,2}^2\sin\theta_3\sin\theta_2$. Substituting these inner products into Eq.~(\ref{eq:THETAL3}), and using the resonance condition (\ref{eq:kpara}), the normalized scattering strength can be immediately found. 
In the above expressions, $\theta_2$ is the independent variable, and $\omega_2$ is measured. Then, we can determine $\theta_3$ from $\omega_3(\theta_3)=\omega_1-\omega_2$ using Eq.~(\ref{eq:resonance}), and solve for $k_2$ and $k_3$ from Eqs.~(\ref{eq:k32}) and (\ref{eq:kpara}). Finally, with the above information, the normalized matrix element $\mathcal{M}_L$ can be readily evaluated.

When pumped at the Langmuir frequency ($\omega_1=\omega_p$), the resonance conditions constrain the plasma parameters and angles at which the three-wave decay can happen. In over-dense plasma (e.g. Fig.~\ref{fig:LLL_Para}a), the Langmuir wave is in the upper resonance, so the resonance condition can be satisfied only if $\omega_p<2|\Omega_e|$. Having satisfied this condition, the $u\rightarrow l+l$ decay can happen if $\theta_2<\theta_b^o$, where $\theta_b^o$ is the angle such that $\omega_l(\theta_b^o)=\omega_p-|\Omega_e|$. In comparison, in under-dense plasma (e.g. Fig.~\ref{fig:LLL_Para}b), the Langmuir wave is in the lower resonance, and therefore can always decay. One decay mode is $l\rightarrow l+l$, which can happen for $\theta_2>\theta_a^u$, where $\omega_l(\theta_a^u)=\omega_p-\omega_{LH}$. Another decay mode is $l\rightarrow l+b$. When $\omega_2=\omega_l$, this decay mode happens for $0<\theta_2<\theta_i^u$, where $\omega_l(\theta_i^u)=\omega_p-\Omega_i$; whereas when $\omega_2=\omega_b$, this decay mode can happen at any $\theta_2$. Finally, using the symmetry $2\leftrightarrow3$, the constrains on $\theta_3$ can be readily deduced.

For Langmuir-wave pump, the normalized growth rate reaches maximum for symmetric decay, where $\omega_2=\omega_3=\omega_p/2$. Let us find the asymptotic expression of $\mathcal{M}_L$ in the symmetric case, so as to get a sense of how the normalized growth rate scales with plasma parameters. The symmetric angle $\theta_s$ can be solved from Eq.~(\ref{eq:resonance}). Using $\omega_p\sim|\Omega_e|\gg \Omega_i$, I find $\sin^2\theta_s\simeq3[1-\omega_p^2/(4\Omega_e^2)]/4$. Then, the wave energy coefficient $u_{2}=u_3\simeq1+3\omega_p^2/(4\Omega_e^2-\omega_p^2)$, where the sub-dominant ion contribution in Eq.~(\ref{eq:TTL_U3}) has been dropped. To solve for the degenerate wave vectors in the symmetric case, it is more convenient to consider the two limits:  $\theta_2=\theta_s-\phi, \theta_3=\theta_s+\phi$, and $\theta_2=\theta_s-\phi, \theta_3=\pi-\theta_s-\phi$, and then let $\phi\rightarrow0$. Solving Eqs.~(\ref{eq:k32}) and (\ref{eq:kpara}) for the wave vectors, the two solutions are $k_2^-/k_1\simeq1/(2\cos\theta_s)$ and $k_2^+/k_1\simeq\sin\theta_s/(2\sin\phi)$. For the $k_2^-$ solution, all terms are finite, and the normalized scattering strength is dominated by electron contribution $\Theta_e^-\simeq-3ck_1[1+\omega_p^2/(2\Omega_e^2)]/(4\omega_p)$. Consequently, the normalized growth rate
for symmetric $k^-$ scattering is
\begin{equation}
	\label{eq:MLp-}
	\mathcal{M}_L^-\Big(\omega_p\rightarrow\frac{\omega_p}{2},\frac{\omega_p}{2}\Big)\simeq\frac{3}{4}\Big(1-\frac{\omega_p^2}{4\Omega_e^2}\Big).
\end{equation}
Notice that this decay mode can happen only if $|\Omega_e|\ge\omega_p/2$. To see what happens to the $k_2^+$ solution, we need to keep the dominant terms, and expand $\omega_2\simeq\omega_p/2-\omega_s'\phi$ and $\omega_3\simeq\omega_p/2+\omega_s'\phi$, where the angular derivative 
of lower resonance $\omega_l(\theta)$ can be evaluated at the symmetric angle using Eq.~(\ref{eq:resonance}) to be $\omega_s'/\omega_p\simeq-2\Omega_e^2\sin(2\theta_s)/(2\Omega_e^2+\omega_p^2)$. Since ion terms does not contain singularity, the normalized scattering strength is again dominated by electrons $\Theta_e^+\simeq3ck_1[1+5\omega_p^2/(4\Omega_e^2)]/(8\omega_p)$. Consequently, the normalized growth rate
for symmetric $k^+$ scattering is
\begin{equation}
	\label{eq:MLp+}
	\mathcal{M}_L^+\Big(\omega_p\rightarrow\frac{\omega_p}{2},\frac{\omega_p}{2}\Big)\simeq -\frac{\mathcal{M}_L^-}{2}\Big(1+\frac{3\omega_p^2/2}{\omega_p^2+2\Omega_e^2}\Big),
\end{equation}
where $\mathcal{M}_L^-$ is given by Eq.~(\ref{eq:MLp-}). Since $\omega_p\le2|\Omega_e|$, it is easy to see that $|\mathcal{M}_L^+|$ is always smaller than $|\mathcal{M}_L^-|$. Moreover, wave damping tends to be smaller for the $k_2^-$ solution. Therefore, the dominant decay mode in experiments will be the $k^-$ mode, where the two decay waves propagate symmetrically at angle $\theta_s$ with respect to the parallel pump wave.

Second, let us consider the case where the pump wave is the electron-cyclotron wave (Fig.~\ref{fig:LLL_Para}c, \ref{fig:LLL_Para}d). In this case, $\beta_{e,1}\sim1$ and the magnetization factor $\gamma_{e,1}^2\simeq(\Omega_e^2/\omega_p^2-1)/\sin^2\theta_1$ approaches infinity, so the dominant contribution comes from electrons. Keeping track of dominant terms as \mbox{$\theta_1\rightarrow0$} and using small angle expansion Eq.~(\ref{eq:wu_para}), the inner products 
$(\hat{\mathbf{k}}_2\cdot\mathbb{F}_{e,1}\hat{\mathbf{k}}_1 )\simeq\mp\gamma_{e,1}^2\sin\theta_1\sin\theta_2$, and  $(\hat{\mathbf{k}}_3\cdot\mathbb{F}_{e,1}\hat{\mathbf{k}}_1) \simeq\pm\gamma_{e,1}^2\sin\theta_1\sin\theta_3$.
The other four inner products that enters Eq.~(\ref{eq:THETAL3}) are the same as before. Keeping terms $\propto1/\sin\theta_1$, the leading term of the normalized scattering strength 
can be readily found.
Although the normalized scattering strength is divergent as $\theta_1\rightarrow0$, the normalized growth rate remains finite. This is because the divergence in $\Theta_e$ cancels the divergence in the wave energy coefficient $u_1\simeq(\omega_p^2-\Omega_e^2)^2/(\omega_p^2\Omega_e^2\sin^2\theta_1)$, which enters the denominator of $\mathcal{M}_L$. Following procedure in the first example, the normalized growth rate can be readily obtained.

When the amplitude of an intense electron-cyclotron pump\footnote[1]{The longitudinal wave considered here is the asymptotic wave that satisfies $\omega_1\rightarrow|\Omega_e|$, when $\theta_1\rightarrow 0^\circ$ and $k_1\rightarrow\infty$.}
exceeds the damping threshold, a number of decay modes are possible. In over-dense plasma (e.g. Fig.~\ref{fig:LLL_Para}c), the electron-cyclotron wave is in the lower resonance, and three-wave decay is always possible. One decay mode is $l\rightarrow l+l$, which can happen for $\theta_2>\theta_a^o$, where $\omega_l(\theta_a^o)=|\Omega_e|-\omega_{LH}$. Another decay mode is $l\rightarrow l+b$, which can happen for any $\theta_2$ if $\omega_2=\omega_b$, and can happen for $0<\theta_2<\theta_i^o$ if $\omega_2=\omega_l$, where $\omega_l(\theta_i^o)=|\Omega_e|-\Omega_i$. In comparison, in under-dense plasma (e.g. Fig.~\ref{fig:LLL_Para}d), the electron-cyclotron wave is in the upper resonance. The resonance condition can be satisfied if $|\Omega_e|<2\omega_p$, and $u\rightarrow l+l$ decay can happen if $\theta_2<\theta_b^u$, where $\omega_l(\theta_b^u)=|\Omega_e|-\omega_p$. We see the angular constrains for electron-cyclotron pump decay is in reciprocal to that of the Langmuir pump.

For electron-cyclotron pump, the normalized growth rate crosses zero and therefore vanishes for symmetric $k^-$ decay, while reaching a maximum for symmetric $k^+$ decay. Let us find the asymptotic expression for $\mathcal{M}_L^+$ to get a sense of how the normalized growth rate scales with plasma parameters. Again, we can find the symmetric angle $\theta_s$ from Eq.~(\ref{eq:resonance}), which gives $\sin^2\theta_s\simeq3[1-\Omega_e^2/(4\omega_p^2)]/4$. Then, the wave energy coefficients $u_{2}=u_3\simeq2(1+2\omega_p^2/\Omega_e^2)/3$. To find the leading behavior of the scattering strength, consider the limit $\theta_2=\theta_s-\phi, \theta_3=\pi-\theta_s-\phi$, and let $\phi\rightarrow0$. In this limit, the wave vector $k_2^+/k_1\simeq\sin\theta_s/(2\sin\phi)\rightarrow\infty$, and the frequencies can be expanded by $\omega_2\simeq\omega_p/2-\omega_s'\phi$ and $\omega_3\simeq\omega_p/2+\omega_s'\phi$, where the angular derivative $\omega_s'$ can again be solved from Eq.~(\ref{eq:resonance}) to be $\omega_s'/\Omega_e\simeq2\omega_p^2\sin(2\theta_s)/(\Omega_e^2+2\omega_p^2)$. Keeping the dominant terms as $\phi\rightarrow0$, the normalized scattering strength $|\Theta_e^+|\simeq ck_1\sin(2\theta_s)(1-r^2)(1-r^2/4)/(\sin\theta_1\Omega_e)$, where $r:=|\Omega_e|/\omega_p$. Since the ion contributions are subdominant, the normalized growth rate for symmetric $k+$ scattering is
\begin{equation}
	\label{eq:MLe+}
	\Big|\mathcal{M}_L^+\Big(\!\Omega_e\!\rightarrow\!\frac{\Omega_e}{2},\!\frac{\Omega_e}{2}\!\Big)\Big|\!\simeq\! \frac{r}{4}\frac{\sqrt{(3\!-\!3r^2/4)^{3}(1\!+\!3r^2/4)}}{2+r^2}.
\end{equation}
We see $\mathcal{M}_L^+$ is nonzero for $0<r<2$, and reaches a maximum of $\sim0.38$ when $r\sim0.92$. The normalized growth rate can be related to the decay rate in experiments, once wave damping is taken into account. 

Finally, let us consider the case where the frequency of the electrostatic pump wave asymptotes to the ion-cyclotron frequency (Fig.~\ref{fig:LLL_Para}e). Since $\Omega_i$ is much smaller than any other characteristic wave frequencies, the only possible decay mode is $b\rightarrow b+b$. Such decay can happen for any angle $\theta_2$, because the resonance conditions can always be satisfied. Similar to what happens in the previous example, the normalized growth rate $\mathcal{M}_L$ changes sign and therefore vanishes for symmetric $k^-$ decay, while reaching a maximum for symmetric $k^+$ decay. Now let us give an estimate of the maximum value of $\mathcal{M}_L^+$. Since the magnetization factor $\gamma_{1,i}^2\simeq\zeta/\tan^2\theta_1\rightarrow\infty$, where \mbox{$\zeta:=M_i/Z_i\gg1$}, the ion terms dominate. The divergent inner products are $(\hat{\mathbf{k}}_2\cdot\mathbb{F}_{i,1}\hat{\mathbf{k}}_1)\simeq\mp\gamma_{i,1}^2\sin\theta_1\sin\theta_2$ and $(\hat{\mathbf{k}}_3\cdot\mathbb{F}_{i,1}\hat{\mathbf{k}}_1)\simeq\pm\gamma_{i,1}^2\sin\theta_1\sin\theta_3$. The other four inner products are finite and similar to what we had before. Using these inner products and keeping the leading terms, the normalized scattering $|\Theta_i^+|\simeq ck_1\Omega_e^2\cos\theta_s/(2\Omega_i^3\sin\theta_1)$, where I have expanded near the symmetric angle as before, with $\omega_s'\simeq9\Omega_e\sin(2\theta_s)/16$. The symmetric angle, which is very close to $\pi/2$, can be estimated from Eq.~(\ref{eq:wb_perp}) to be $\cos^2\theta_s\simeq\Omega_i/(3|\Omega_e|)$. The wave energy coefficients $u_1\simeq\omega_p^2|\Omega_e|/(\Omega_i^3\sin^2\theta_1)$, and $u_2=u_3\simeq16\omega_p^2/(9\Omega_i|\Omega_e|)$. Substituting these results into Eq.~(\ref{eq:uL}), the normalized symmetric $k^+$ decay rate is
\begin{equation}
	\label{eq:MLi+}
	\Big|\mathcal{M}_L^+\Big(\Omega_i\!\rightarrow\!\frac{\Omega_i}{2},\!\frac{\Omega_i}{2}\Big)\Big|\simeq \frac{3\sqrt{3}}{32}\frac{\Omega_i}{\omega_p}.
\end{equation}
We see in a typical plasma where $\omega_p\gg\Omega_i$, the decay mode $b\rightarrow b+b$ is orders of magnitude weaker than the other decay modes. Nevertheless, when compared with the pump frequency $\omega_1=\Omega_i$, the growth rate of the three-wave decay instability is not necessarily small.

\subsubsection{Perpendicular pump}\label{sec:LLLPerp}
In this subsection, I use another set of examples to illustrate how to evaluate the normalized growth rate $\mathcal{M}_L$, by considering the cases where the pump wave propagates perpendicular to the magnetic field. In this geometry, the pump frequency can either be the upper-hybrid frequency $\omega_{UH}$, or the lower hybrid frequency $\omega_{LH}$, in an electron-ion plasma. For three-wave decay to happen, the frequency resonance condition [Eq.~(\ref{eq:resonantW})] must be satisfied. Since the lower hybrid frequency $\omega_{LH}\gg\Omega_i$, it is not possible to match the frequency resonance condition with a \textit{LH} pump wave in a uniform plasma. By similar consideration, for a \textit{UH} pump wave, the decay mode $u\rightarrow u+u$ is also forbidden. However, other decays modes of the \textit{UH} pump are possible. Using expression $\omega_{UH}^2\simeq\omega_p^2+\Omega_e^2$, we see that $u\rightarrow u+b$ is always possible; $u\rightarrow u+l$ is possible if $2/\sqrt{\zeta}\lesssim r\lesssim \sqrt{\zeta}/2$, where $\zeta=M_i/Z_i\gg1$ is the normalized charge-to-mass ratio for ions; and $u\rightarrow l+l$ is possible only if $1/\sqrt{3}\leq r\leq \sqrt{3}$. Here, $r=|\Omega_e|/\omega_p$ is the ratio of electron-cyclotron frequency to the plasma frequency. In this section, I will consider $r$ in the range where all three decay modes are possible. 

In addition to the frequency condition, the wave vector resonance conditions [Eq.~(\ref{eq:resonantK})] must also be satisfied for three-wave decay to happen. To see when this condition can be satisfied in this perpendicular geometry, it is convenient to discuss in the spherical coordinate where the polar angle $\theta$ is measured from the magnetic field $\mathbf{b}$, and the azimuthal angle $\phi$ is measured from $\mathbf{k}_1$. In this spherical coordinate, the wave vectors $\mathbf{k}_2$ and $\mathbf{k}_3$ are constrained on the two cones spanning angles $\theta_2, \pi-\theta_2$ and $\theta_3, \pi-\theta_3$. 
Then, $\mathbf{k}_2$ and $\mathbf{k}_3$ can reside along the lines generated by cutting the two cones with a plane passing through $\mathbf{k}_1$. When $|\cos\theta_2|>|\cos\theta_3|$, the plane starts to intercept both cones when 
$|\cos\phi_3|\ge|\cos\phi_c|$, where the critical angle $\sin\phi_c=\tan\theta_2/\tan\theta_3$. When the strict inequality holds, for each $\mathbf{k}_3$, there are two solutions to $\mathbf{k}_2$ such that the resonance conditions is satisfied. By the exchange symmetry $2\leftrightarrow3$, we immediately know what happens when $|\cos\theta_2|<|\cos\theta_3|$. 
The resonance condition [Eq.~(\ref{eq:resonantK})] constrains where in the $\theta_2$-$\phi_2$ plane can the normalized growth rate $\mathcal{M}_L$ take nonzero values.

\begin{figure}[!b]
	\centering
	\includegraphics[angle=0,width=0.70\textwidth]{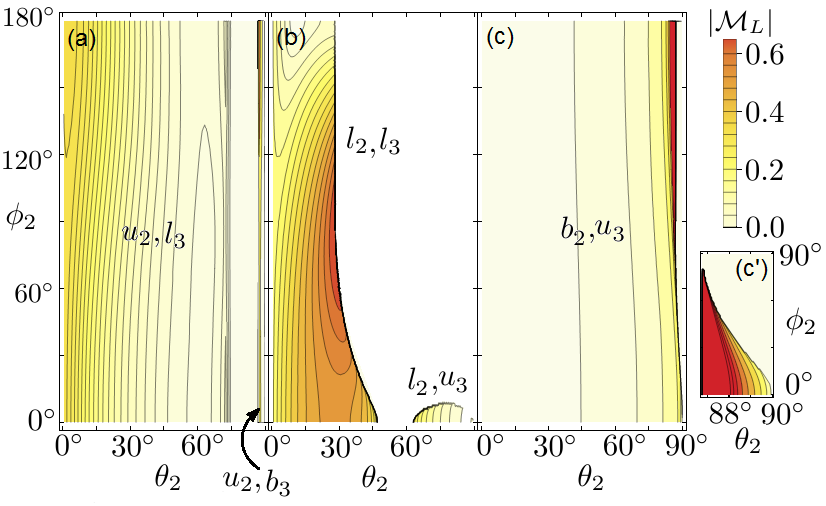}
	\caption[Scattering of perpendicular electrostatic pump in magnetized hydrogen plasma]{Normalized growth rate $|\mathcal{M}_L|$ when pumped by an upper-hybrid wave ($\mathbf{k}_1\!\perp\!\mathbf{B}_0$) in a uniform hydrogen plasma with $|\Omega_e|/\omega_p=1.2$. The growth rates are observed at polar angle $\theta_2$ with respect to $\mathbf{B}_0$ and azimuthal angle $\phi_2$ with respect to $\mathbf{k}_1$. When $\omega_2$ is on the upper resonance (a), the $u\rightarrow u_2,l_3$ decay can happen for $\theta_2<\theta_u^a$, where $\omega_u(\theta_u^a)=\omega_{UH}-\omega_{LH}$. In this case, an important decay channel has $\omega_2\sim|\Omega_e|$ propagating almost parallel to $\mathbf{B}_0$ in the backward direction, and $\omega_3\gg\omega_{LH}$ propagating almost perpendicular to $\mathbf{B}_0$ in the forward direction. This region corresponds to the $l_2,u_3$ region in (b), in which $\omega_2$ is on the lower resonance instead. The other decay mode is $u\rightarrow u_2,b_3$, which can happen in the narrow strip $\theta_2>\theta_u^b$ in (a), where $\omega_u(\theta_u^b)=\omega_{UH}-\Omega_i$. Equivalently, exchanging the labels to $b_2,u_3$, this decay mode can happen in the colored region in (c), in which $\omega_2$ is on the bottom resonance instead. For this decay mode, the dominant decay channel has $\omega_2\sim\omega_{UH}$ propagating almost perpendicular to $\mathbf{B}_0$ in the forward direction, and $\omega_3\sim\Omega_i$ propagating either in the forward or backward direction. The last decay mode is $u\rightarrow l_2,l_3$, which corresponds to the large colored region in (b). For this decay mode, the dominant decay channel is the symmetric decay, where $\omega_2\sim\omega_3\sim\omega_{UH}/2$ and both waves propagate at angles with $\mathbf{B}_0$ in the forward direction.}
	\label{fig:LLL_Perp}
\end{figure}

Having matched the resonance conditions, the normalize growth rate in the polar coordinate can be readily evaluated (Fig.~\ref{fig:LLL_Perp}). To understand the angular dependence of $\mathcal{M}_L$, it is useful to notice that due to the exchange symmetry $\mathcal{M}_L(2,3)=\mathcal{M}_L(3,2)$, the normalized growth rate $\mathcal{M}_L(\theta_2,\phi_2)$ in one region can be mapped to $\mathcal{M}_L(\theta_2',\phi_2')$ in anther region. To be more specific, when $\omega_2$ is on the upper resonance (Fig.~\ref{fig:LLL_Perp}a), the normalized growth rate $\mathcal{M}_L$ is nonzero in two regions. The first region is $\theta_2<\theta_u^a$, where $\omega_u(\theta_u^a)=\omega_{UH}-\omega_{LH}$. In this region, the decay mode $u_1\rightarrow u_2+l_3$ is allowed, where $\omega_3$ is on the lower resonance. By the exchange symmetry, this region can be mapped to the island on the bottom right corner of Fig.~\ref{fig:LLL_Perp}b, in which $\omega_2$ is on the lower resonance instead. The other region in Fig.~\ref{fig:LLL_Perp}a where $\mathcal{M}_L$ is nonzero is the narrow strip $\theta_2>\theta_u^b$, where $\omega_u(\theta_u^b)=\omega_{UH}-\Omega_i$. In this region, the decay mode $u_1\rightarrow u_2+b_3$ is allowed, where $\omega_3$ is on the bottom resonance. Exchanging $2\leftrightarrow3$, this region corresponds to the case where $\omega_2$ is on the bottom resonance instead (Fig.~\ref{fig:LLL_Perp}c). The remaining decay mode is $u_1\rightarrow l_2+l_3$, where both decay waves are on the lower resonance. This decay mode is allowed within the large region on the left of Fig.~\ref{fig:LLL_Perp}b. This region has a straight boundary at $\theta_2=\theta_l^m$, where $\omega_l(\theta_l^m)=\omega_{UH}/2$. To the left of this boundary, we have $\theta_2<\theta_3$, so there is only one solution for $k_2$. To the right of this boundary, we have $\theta_2>\theta_3$, so both  $k_2^-$ and $k_2^+$ solutions exist as long as $\sin\phi_2<\tan\theta_3/\tan\theta_2$. Whenever both solutions exist, Fig.~\ref{fig:LLL_Perp} shows the $k^-$ branch, which has weaker damping. In those degenerate cases, the $k^+$ branch is usually comparable with the $k^-$ branch. An exception is inserted in Fig.~\ref{fig:LLL_Perp}c', where the $k^+$ branch is dominant for $u_1\rightarrow b_2+u_3$ decay, corresponding to the forward scattering of the \textit{UH} pump with little frequency shift.

For the $u\rightarrow u_2+l_3$ decay (Fig.~\ref{fig:LLL_Perp}a), one important decay channel has $\omega_2\sim|\Omega_e|$ propagating almost parallel to $\mathbf{b}$ in the backward direction ($\phi_2=180^\circ$), and 
the other decay wave propagating almost perpendicular to $\mathbf{b}$ in the forward direction ($\phi_3=0^\circ$). To see how does $\mathcal{M}_L$ scales with plasma parameters, let us find its asymptotic expression when $\theta_2\rightarrow0$. In this limit $\omega_2\rightarrow|\Omega_e|$, so the magnetization factor $\gamma_{2,e}^2$ is divergent. Then, the dominant terms of the coupling strength [Eq.~(\ref{eq:THETAL3})] comes from the $\mathbb{F}_{e,2}$ terms. The divergent inner products are $(\hat{\mathbf{k}_1}\cdot\mathbb{F}_{e,2}^*\hat{\mathbf{k}_2})\simeq-\gamma_{e,2}^2\sin\theta_2$ and $(\hat{\mathbf{k}_3}\cdot\mathbb{F}_{e,2}^*\hat{\mathbf{k}_2})\simeq-\gamma_{e,2}^2\sin\theta_2\sin\theta_3$, and we also need the finite inner products $(\hat{\mathbf{k}_1}\cdot\mathbb{F}_{e,3}^*\hat{\mathbf{k}_3})\simeq\gamma_{e,3}^2\sin\theta_3$ and $(\hat{\mathbf{k}_3}\cdot\mathbb{F}_{e,1}\hat{\mathbf{k}_1})\simeq\gamma_{e,1}^2\sin\theta_3$. Then, the leading term of the normalized scattering strength is $\Theta_e\simeq ck_1\gamma_{e,1}^2\gamma_{e,2}^2\gamma_{e,3}^2 (\omega_1^2-\omega_3^2)\sin\theta_2\sin\theta_3/(\omega_1\omega_2\omega_3)$, where I have used the resonance condition $k_3\sin\theta_3=k_1$. The angle $\theta_3$ can be estimated from Eq.~(\ref{eq:resonance}) using $\omega_3\gg\Omega_i$, which gives $\sin^2\theta_3\simeq(\omega_3^2-\omega_p^2)(\omega_3^2-\Omega_e^2)/(\omega_p^2\Omega_e^2)$. Then, the wave energy coefficient $u_3\simeq(2\omega_3^2-\omega_{UH}^2)(\omega_3^2-\Omega_e^2)$. As for the other two wave energy coefficients, using previous results, we know $u_1=\omega_{UH}^2/\omega_p^2$ and $u_2\simeq(\Omega_e^2-\omega_p^2)^2/(\Omega_e^2\omega_p^2\sin^2\theta_2)$. Substituting these into Eqs.~(\ref{eq:ML}) and (\ref{eq:uL}), we can find the normalized growth rate for this decay mode to be
\begin{equation}
	\label{eq:MLul}
	\Big|\mathcal{M}_L\big(\omega_{UH}\!\rightarrow\!|\Omega_e|,\omega_3\big)\Big|\!\simeq\! \frac{\omega_3(\omega_3+\omega_{UH})}{\omega_p\sqrt{2(\omega_{UH}^2-2\omega_3^2)}},
\end{equation}
where $\omega_3=\omega_{UH}-|\Omega_e|$ is the resonance frequency. From previous discussion, we know this decay mode can happen as long as $1/\sqrt{3}\leq r\lesssim\sqrt{\zeta}/2$. Within this parameter range, it is easy to see that Eq.~(\ref{eq:MLul}) decreases monotonically with increasing magnetic field. The maximum value $\mathcal{M}_L=\sqrt{3}/2$ is attained at $r=1/\sqrt{3}$, where $\omega_3=|\Omega_e|=\omega_{UH}/2$ such that the decay is symmetric.

For the $u\rightarrow l_2+l_3$ decay (Fig.~\ref{fig:LLL_Perp}b), the dominant decay channel is the symmetric decay, where $\omega_2=\omega_3=\omega_{1}/2$. In the symmetric decay geometry, we have $\theta_3=\pi-\theta_2$ and $\phi_3=-\phi_2$. Then, the wave vector resonance condition becomes $k_2=k_3=k_1/(2\sin\theta_2\cos\phi_2)$. The symmetric decay angle $\theta_2=\theta_s$ can be estimated from Eq.~(\ref{eq:resonance}) using $\omega_2=\omega_{UH}/2\gg\Omega_i$, which gives $\cos^2\theta_s\simeq3\omega_{UH}^4/(16\omega_p^2\Omega_e^2)$. Since the frequencies are far away from cyclotron frequencies, all the magnetization factors are finite. Then, the inner products $(\hat{\mathbf{k}_1}\cdot\mathbb{F}_{s,2}^*\hat{\mathbf{k}_2}) \simeq\gamma_{s,2}^2(\cos\phi_2+i\beta_{s,2}\sin\phi_2)\sin\theta_2$, $(\hat{\mathbf{k}_2}\cdot\mathbb{F}_{s,1}\hat{\mathbf{k}_1})\simeq\gamma_{s,1}^2(\cos\phi_2+i\beta_{s,1}\sin\phi_2)\sin\theta_2$,  $(\hat{\mathbf{k}_3}\cdot\mathbb{F}_{s,2}^*\hat{\mathbf{k}_2})\simeq-1+\gamma_{s,2}^2\sin^2\theta_2(2\cos^2\phi_2+i\beta_{s,2}\sin2\phi_2-\beta_{s,2}^2)$, and by exchanging $2\leftrightarrow 3$, we can easily find the other three inner products. Substituting these inner products into Eq.~(\ref{eq:THETAL3}), the normalized scattering strength becomes particularly simple when $\phi_2\rightarrow\pi/2$. In this limit $k_2,k_3\rightarrow\infty$, but the products $k_2\cos\phi_2=-k_3\cos\phi_3$ remains finite. Keeping nonzero terms as $\phi_2\rightarrow\pi/2$, the scattering strength simplifies to $\Theta_e^+\simeq- 2ck_1\omega_{UH}^3/[\omega_p^2(3\Omega_e^2-\omega_p^2)]$. The electron terms also dominate the wave energy coefficients $u_2=u_3\simeq2\omega_{UH}^2/(3\Omega_e^2-\omega_p^2)$. Gathering the above results, the normalized growth rate for symmetric $k^+$ scattering is
\begin{equation}
	\label{eq:MLll}
	\Big|\mathcal{M}_L^+\big(\omega_{UH}\!\rightarrow\!\frac{\omega_{UH}}{2},\frac{\omega_{UH}}{2}\big)\Big|\!\simeq\! \frac{\omega_p}{\omega_{UH}}.
\end{equation}
The above special value of $\mathcal{M}_L$ is approximately the maximum in Fig.~\ref{fig:LLL_Perp}b, where $\theta_2=\theta_s$ and $\phi_2=90^\circ$. Notice that this special case is singular in wave vector $k_2,k_3\rightarrow\infty$, and hence will be suppressed by wave damping. Therefore, the dominant decay channels observed in experiment will happen at smaller angle $\phi_2<90^\circ$ in the symmetric decay geometry.

Finally, for the $u\rightarrow b_2+u_3$ decay (Fig.~\ref{fig:LLL_Perp}c), the dominant decay channel has $\omega_2\sim\omega_{UH}$ propagating almost perpendicular to $\mathbf{b}$ in the forward direction, and $\omega_3\sim\Omega_i$ propagating either in the forward or backward direction. As an example, let us consider symmetric forward scattering where $\phi_2=\phi_3=0$ and $\theta_2=\pi-\theta_3=\theta_s$. In this geometry, $k_2^-=k_3^-=k_1/(2\sin\theta_s)$. Since $\theta_s\sim\pi/2$, we can estimate the symmetric angle using asymptotic expressions Eqs.~(\ref{eq:wu_perp}) and (\ref{eq:wb_perp}). Substituting these expressions into the frequency resonance condition [Eq.~(\ref{eq:resonantW})], I obtain $\cos^2\theta_s\simeq2\Omega_i\omega_{UH}^3/(\Omega_e^2\omega_p^2)\sim0$, where I have used that $\omega_p^2|\Omega_e|/(2\omega_{UH}^3)\lesssim0.2$ is always a small number. Then the wave energy $u_2\simeq u_1=\omega_{UH}^2/\omega_p^2$, and $u_3\simeq\omega_p^2[1+2\omega_{UH}^3/(\omega_p^2|\Omega_e|)]^2/(\Omega_i|\Omega_e|)$. Now that the magnetization factors are all finite, the inner products are simply $(\hat{\mathbf{k}_1}\cdot\mathbb{F}_{s,2}^*\hat{\mathbf{k}_2}) \simeq\gamma_{s,2}^2\sin\theta_2$, $(\hat{\mathbf{k}_2}\cdot\mathbb{F}_{s,1}\hat{\mathbf{k}_1})\simeq\gamma_{s,1}^2\sin\theta_2$,  $(\hat{\mathbf{k}_3}\cdot\mathbb{F}_{s,2}^*\hat{\mathbf{k}_2})\simeq\cos\theta_3\cos\theta_2+\gamma_{2,s}^2\sin\theta_3\sin\theta_2$, and the three other inner products can be obtained by exchanging $2\leftrightarrow 3$. Again, the scattering is mostly due to electrons, for which $\gamma_{e,1}^2\simeq\gamma_{e,2}^2\simeq\omega_{UH}^2/\omega_p^2$ and $\gamma_{e,3}^2\simeq-\omega_3^2/\Omega_e^2\ll\cos\theta^2_s$. Therefore, the dominant term comes from the second line of Eq.~(\ref{eq:THETAL3}), which gives the scattering strength $\Theta_e^-\simeq- ck_1\Omega_i\omega_{UH}^5/(\omega_3\Omega_e^2\omega_p^4)$. Substituting these results into Eqs.~(\ref{eq:ML}) and (\ref{eq:uL}), we immediately see that the normalized growth rate for forward scattering is
\begin{equation}
	\label{eq:MLub}
	\Big|\mathcal{M}_L^-\big(\omega_{UH}\!\rightarrow\!\omega_{UH},\Omega_i\big)\Big|\!\simeq\! \frac{\omega_p}{4\sqrt{\omega_{UH}|\Omega_e|}}\bigg(\frac{\omega_3}{\Omega_i}\bigg)^{1/2},
\end{equation}
where $\omega_3=\omega_b(\theta_s)\sim\Omega_i$ can be obtained from Eq.~(\ref{eq:wb_perp}). Using the above result, we can also find the symmetric nearly backward scattering $\mathcal{M}_L^+$ by replacing the coefficient $1/4$ with $k_2^+/(2k_1)$. The symmetric nearly backward scattering channel has divergent $k_2^+$, and therefore can have very large growth rate in the absence of damping.

\section{Effective Lagrangian of three-wave interactions\label{sec:fluid-2nd:Lagrangian}}

In the previous section, I elaborated on the general formula for the three-wave coupling coefficient, which is obtained by solving the fluid-Maxwell equations to second order using multiscale expansion. Evaluating the general formula may be somewhat tedious but otherwise straightforward, especially when it is done using a computer program. As we have seen, the general formula is very powerful. It not only recovers special results known in the literature, but also provides previously unknown insights to three-wave scattering in magnetized plasmas in arbitrary geometry. In this section, I will further deepen the insight from a variational principle perspective. After obtaining the interaction Lagrangian by reverse engineering, the physical mechanism of the three-wave interactions will become very transparent. I will first show that the three on-shell equations can be derived from a classical three-wave Lagrangian. More importantly, I will then show that all terms in the classical interaction Lagrangian arise from a single term in the quantized Lagrangian. 

\subsection{Classical Lagrangian for the three-wave equations\label{sec:fluid-2nd:Lagrangian:classical}}

To write down the Lagrangian, it is more convenient to use the gauge field $A^{\mu}$ instead of the electric or magnetic fields. Since we will later quantize the Lagrangian, it is convenient to use the temporal gauge $A^{0}=0$. In temporals gauge, the electric field is related to the vector potential by 
\begin{eqnarray}
	i\mathbf{A}_{\mathbf{k}}=\frac{\mathbfcal{E}_{\mathbf{k}}}{\omega_{\mathbf{k}}}.
\end{eqnarray} 
It is easy to check that the subscript notation is consistent with Eqs.~(\ref{eq:notationz}) and (\ref{eq:notationa}): $\mathbf{A}_{-\mathbf{k}}=-i\mathbfcal{E}_{-\mathbf{k}}/\omega_{-\mathbf{k}} =i\mathbfcal{E}_{\mathbf{k}}^*/\omega_{\mathbf{k}}=\mathbf{A}_{\mathbf{k}}^*$. In the natural units $\hbar=c=\epsilon_0=1$, which I will use in this section, the vector potential has the dimension of energy $M$. Similarly, we can dimensionalize the wave energy operator by
\begin{equation}
	i\Lambda_{\mathbf{k}}:=\omega_{\mathbf{k}}\mathbb{H}_{\mathbf{k}}.
\end{equation}
The energy operator has the dimension of energy $M$ as it should, and it respects the notation $\Lambda_{-\mathbf{k}}=\Lambda_{\mathbf{k}}^*$. Finally, we will also need the displacement operator, which is defined as
\begin{eqnarray}
i\Pi_{s,\mathbf{k}}:=\frac{\mathbb{F}_{s,\mathbf{k}}}{\omega_{\mathbf{k}}},
\end{eqnarray}
where $\mathbb{F}_{s,\mathbf{k}}$ is the forcing operator defined by Eq.~(\ref{eq:F}). The displacement operator $\Pi_{s,\mathbf{k}}$ has the dimension of distance $M^{-1}$, and satisfies the reality condition $\Pi_{s,-\mathbf{k}}=\Pi_{s,\mathbf{k}}^*$. Moreover, using Eq.~(\ref{eq:vwave}), when this operator acts on $\mathbf{A}_{\mathbf{k}}$, the combination $e\Pi\mathbf{A}/m=i\mathbf{v}/\omega=\mathbf{d}$ is the displacement of a charged species in response to the wave perturbation. Notice that since $\mathbb{F}_{s,\mathbf{k}}$ is self-adjoint, the displacement operator $\Pi_{s,\mathbf{k}}$ is anti-self-adjoint.

In terms of the above operators, we can rewrite the on-shell equations in a form that resembles the Euler-Lagrangian equation. Using the formula for the scattering strength $\mathbf{S}_{2,3}^s$ [Eqs.~(\ref{eq:S23})] and the definition of the normalized scattering strength $\Theta_s$ [Eq.~(\ref{eq:aRHS})], the on-shell equation (\ref{eq:onE1}) can be written in the abbreviated notation as
\begin{eqnarray}
\label{eq:A1}
\mathbf{A}_1^\dagger i\Lambda_1d_{t(1)}\mathbf{A}_1 \hspace{-7pt}&=&\hspace{-7pt}\sum_s\frac{\mathbfcal{E}_1^\dagger\mathbf{S}_{2,3}^s}{\omega_1^2} =-iA_1^*A_2A_3\sum_s \frac{e_s\omega_{ps}^2}{2m_sc}\Theta_s^* \\
\nonumber
\hspace{-7pt}&=&\hspace{-7pt}i\sum_s\!\frac{e_s\omega_{ps}^2}{2m_s}\!\Big[ \omega_2(\mathbf{k}_3\!\cdot\!\Pi_{s,\bar{1}}\mathbf{A}_{\bar{1}})(\mathbf{A}_3\!\cdot\!\Pi_{s,2}\mathbf{A}_2)+ \omega_3(\mathbf{k}_2\!\cdot\!\Pi_{s,\bar{1}}\mathbf{A}_{\bar{1}})(\mathbf{A}_2\!\cdot\!\Pi_{s,3}\mathbf{A}_3) \\
\nonumber
&&\hspace{37pt}
+\omega_3(\mathbf{k}_{\bar{1}}\!\cdot\!\Pi_{s,2}\mathbf{A}_2)(\mathbf{A}_{\bar{1}}\!\cdot\!\Pi_{s,3}\mathbf{A}_3)+ \omega_{\bar{1}}(\mathbf{k}_3\!\cdot\!\Pi_{s,2}\mathbf{A}_2)(\mathbf{A}_3\!\cdot\!\Pi_{s,\bar{1}}\mathbf{A}_{\bar{1}}) \\
\nonumber
&&\hspace{37pt}
+\omega_2(\mathbf{k}_{\bar{1}}\!\cdot\!\Pi_{s,3}\mathbf{A}_3)(\mathbf{A}_{\bar{1}}\!\cdot\!\Pi_{s,2}\mathbf{A}_2)+ \omega_{\bar{1}}(\mathbf{k}_2\!\cdot\!\Pi_{s,3}\mathbf{A}_3)(\mathbf{A}_2\!\cdot\!\Pi_{s,\bar{1}}\mathbf{A}_{\bar{1}})
\Big],
\end{eqnarray}
where $A_j=|\mathbf{A}_j|$ is the norm of the wave amplitude. The abbreviated notation $\bar{j}$ is interpreted by $\mathbf{k}_{\bar{j}}=-\mathbf{k}_j$, as well as the notations Eqs.~(\ref{eq:notationz}) and (\ref{eq:notationa}). To obtain the last equality, I have used the anti-self-adjoint property of the displacement operator $\Pi_{j}$, such that $\Pi_{j}$ always act on $\mathbf{A}_j$ that shares the same wave index $j$. Unlike the six terms in $\mathbf{S}_{2,3}^s$, the six terms in the above expression have identical structures. The permutation patterns of the normalized scattering strength $\Theta_s$ [Eqs.~(\ref{eq:Theta3}) and (\ref{eq:Thetaijl})] thereof emerge.  

Having defined the necessary operators and rewritten the on-shell equation in a Euler-Lagrangian form, we can now write down the classical three-wave action. The classical action $S_c$ contains the free waves as well as their interactions:
\begin{equation}
	\label{eq:Sc}
	S_{c}=\int d^4x_{(1)}(\mathcal{L}_{c0}+\mathcal{L}_{cI}),
\end{equation}
where the integrations over space and time are on the slow scales $x_{(1)}$ and $t_{(1)}$. The first term $\mathcal{L}_{c0}$ is the Lagrangian of freely advecting wave envelopes:
\begin{equation}
	\mathcal{L}_{c0}=\sum_{j=1}^3\mathbf{A}_j^\dagger i\Lambda_jd_{t(1)}\mathbf{A}_j,
\end{equation}
where the complex amplitude $\mathbf{A}_j(x_{(1)},t_{(1)})$ is a function of the slow spatial and temporal scales, and the advective derivative $d_{t(1)}$ is defined by Eq.~(\ref{eq:dt1}). This Lagrangian is the spin-1 analogy to the Lagrangian of massless spin-$1/2$ Dirac fermions $\mathcal{L}=\bar{\psi}i\gamma^{\mu}\partial_\mu\psi$, where the vector potential $\mathbf{A}$ is analogous to the Dirac spinor $\psi$, and the energy matrix $\Lambda$ is analogous to the Dirac matrices $\gamma^\mu$. It is easy to show that $\mathcal{L}_{c0}$ gives rise to a real-valued action $S_{c0}$ after integration by part. The second term in the classical action [Eq.~(\ref{eq:Sc})] is the three-wave interaction Lagrangian 
\begin{equation}
	\label{eq:LcI}
	\mathcal{L}_{cI}=-i(\Xi-\Xi^*),
\end{equation}
which is obviously real-valued. Using Eq.~(\ref{eq:A1}), the cubic interaction is given by
\begin{equation}
	\label{eq:Xi}
	\Xi[\mathbf{A}_1,\mathbf{A}_2^*,\mathbf{A}_3^*]=A_1 A_2^* A_3^*\sum_s\frac{e_s\omega_{ps}^2}{2m_sc}\Theta_s.
\end{equation}
Clearly, the three-wave Lagrangian has mass dimension $M^4$, and hence the action $S_{c}$ is dimensionless in the natural unit as expected. 
Now that we have written down the Lagrangian, we can find the classical equations of motion by taking variations with respect to $\mathbf{A}_1$, $\mathbf{A}_2$, and $\mathbf{A}_3$, or equivalently, their independent complex conjugates. Using the anti-self-adjoint property of the displacement operator, it is straightforward to verify that the three on-shell equations (\ref{eq:onE1})-(\ref{eq:onE3}) are the resultant Euler-Lagrangian equations.

The classical three-wave Lagrangian $\mathcal{L}_c=\mathcal{L}_{c0}+\mathcal{L}_{cI}$ has U(1) symmetries, which lead to the action conservation laws. For example, the Lagrangian is invariant under the following global U(1) transformation:
\begin{eqnarray}
\mathbf{A}_1\rightarrow e^{i\alpha}\mathbf{A}_1,\hspace{5pt}
\mathbf{A}_2\rightarrow e^{i\alpha}\mathbf{A}_2,\hspace{5pt}
\mathbf{A}_3\rightarrow \mathbf{A}_3,
\end{eqnarray}
where $\alpha$ is an arbitrary real constant. Under the above transformation, the infinitesimal variation of the Lagrangian is zero $\delta\mathcal{L}_{c}=0$, while the infinitesimal variation $\delta\mathbf{A}_1=i\alpha\mathbf{A}_1$, $\delta\mathbf{A}_2=i\alpha\mathbf{A}_2$, and $\delta\mathbf{A}_3=0$, giving rise to a Noether's current. In fact, we have an even stronger symmetry $\delta\Xi=0$ for any $\alpha$. Therefore this U(1) symmetry leads to the identity
\begin{equation}
	\mathbf{A}_1\cdot\frac{\delta\Xi}{\delta\mathbf{A}_1}-\mathbf{A}_2^*\cdot\frac{\delta\Xi}{\delta\mathbf{A}_2^*}=0,
\end{equation}
which is exactly the action conservation law Eq.~(\ref{eq:actionS12}). Using similar arguments, other action conservation laws can also be derived from U(1) symmetries.

\subsection{Quantized Lagrangian and the scattering matrix\label{sec:fluid-2nd:Lagrangian:quantum}}

The large number of terms contained in the classical Lagrangian can be reduced when we quantized the Lagrangian, in which the gauge field becomes real valued. Before introducing the quantized Lagrangian, it is helpful to review the second quantization notations, which will be discussed in more details in Sec.~\ref{sec:vacuum:quantization}. For simplicity, I will omit the subscripts for the slow spatial and temporal variables $x_{(1)}$ and $t_{(1)}$, with the implied understanding that all spatial and temporal dependences are on the full scales. Let us promote the gauge field $\mathbf{A}$ to a quantized operator
\begin{equation}
	\label{eq:Ahat}
	\hat{\mathbf{A}}:=\int\frac{d^3\mathbf{k}}{(2\pi)^3}\frac{1}{\sqrt{2\omega_{\mathbf{k}}}}\Big(\mathbf{e}_{\mathbf{k}}\hat{a}_{\mathbf{k}}e^{-ikx}+\mathbf{e}_{\mathbf{k}}^*\hat{a}^\dagger_{\mathbf{k}}e^{ikx}\Big),
\end{equation}
where $kx:=\omega_{\mathbf{k}}t-\mathbf{k}\cdot\mathbf{x}$ is the Minkowski inner product, $\mathbf{e}_\mathbf{k}$ is the unit polarization vector, and the summation over branches of the dispersion relation is implied. The annihilation operator $\hat{a}_{\mathbf{k}}$ and the creation operator $\hat{a}^\dagger_{\mathbf{k}}$ satisfies the canonical commutation relations for bosons, where the nontrivial commutator is
\begin{equation}
	[\hat{a}_{\mathbf{p}},\hat{a}^\dagger_{\mathbf{k}}]=(2\pi)^3\delta^{(3)}(\mathbf{p}-\mathbf{k}).
\end{equation} 
Using the standard normalization, the single boson state is
\begin{equation}
	|\mathbf{k}\rangle:=\sqrt{2\omega_{\mathbf{k}}}\hat{a}^\dagger_{\mathbf{k}}|0\rangle,
\end{equation}
where $|0\rangle$ is the vacuum state. Then we have the following Wick contraction
\begin{equation}
	\contraction{}{\hat{\mathbf{A}}}{|}{\mathbf{k}}
	\hat{\mathbf{A}}|\mathbf{k}\rangle=\mathbf{e}_\mathbf{k}e^{-ikx}|0\rangle.
\end{equation}
Let us also promote the displacement operator for species $s$ to act on the operator $\hat{\mathbf{A}}$ by
\begin{equation}
	\hat{\Pi}_{s}\hat{\mathbf{A}}:=\int\frac{d^3\mathbf{k}}{(2\pi)^3}\frac{1}{\sqrt{2\omega_{\mathbf{k}}}}\Big(-i\frac{\mathbb{F}_{s,\mathbf{k}}\mathbf{e}_{\mathbf{k}}}{\omega_{\mathbf{k}}} \hat{a}_{\mathbf{k}}e^{-ikx} +i\frac{\mathbb{F}_{s,\mathbf{k}}^*\mathbf{e}_{\mathbf{k}}^*}{\omega_{\mathbf{k}}} \hat{a}^\dagger_{\mathbf{k}}e^{ikx}\Big),
\end{equation}
where the sign of the second term comes from notation Eq.~(\ref{eq:notationa}). Taking time derivative of the displacement operator, we can also obtain the velocity operator $\partial_t(\hat{\Pi}_{s}\hat{\mathbf{A}})$, which is proportional to the current operator. 

Now we are ready to write down the quantized Lagrangian, which contains a kinetic term and a single cubic interaction term:
\begin{equation}
	\label{eq:Lagrangian}
	\mathcal{L}=\hat{\mathbf{A}}^{\dagger}i\Lambda d_t \hat{\mathbf{A}} -\sum_s\frac{e_s\omega_{ps}^2}{2m_s}(\hat{\Pi}_{s}\hat{\mathbf{A}})_i(\partial_i\hat{\mathbf{A}}_j)\partial_t(\hat{\Pi}_{s}\hat{\mathbf{A}})_j.
\end{equation}
Here, the $i$ and $j$ indices in the second term are the spatial indices, and summation over repeated indices is assumed. The first term $\mathcal{L}_0$ closely resembles the kinetic term of quantum electrodynamics (QED), with the Dirac spinor replaced by the gauge field, and the Dirac gamma matrices replaced by the $\Lambda$ energy matrix. The second term $\mathcal{L}_I$ is the three-wave interaction Lagrangian, which is nonvanishing only if the background density of some species $s$ is nonzero. 
Notice that the three-wave interaction is nonrenormalizable, which is not unexpected in an effective field theory.

To make sense of the quantized Lagrangian, notice that in terms of physical quantities, the displacement $\mathbf{d}=e\hat{\Pi}\hat{\mathbf{A}}/m$, and the current density $\mathbf{J}=e^2n_0\partial_t(\hat{\Pi}\hat{\mathbf{A}})/m$. Therefore, the three-wave interaction Lagrangian is of the form $\mathcal{L}_I\propto \mathbf{d}^i(\partial_i\mathbf{A}_j)J^j$, where the displacement and current density are determined by linear response. Although one may not have guessed this form of the interaction Lagrangian, it makes very intuitive sense. In the absence of the third wave, the electromagnetic field interacts with the particle fields through $\mathbf{A}_jJ^j$ in the temporal gauge; now when the third wave is present, it modulates the medium through which the electromagnetic field advects, giving rise to the $\mathbf{d}^i(\partial_i\mathbf{A}_j)J^j$ interaction. In this interaction term, there is no reason why a particular wave should only be responsible for $\mathbf{d}$, $\mathbf{A}$, or $\mathbf{J}$. Therefore, the three waves can switch their roles, and the total interaction is given by the linear superpositions of all possible permutations.

To see how the quantized Lagrangian, with the linear superposition principle built in, gives rise to the classical Lagrangian, let us compute the scattering (\textit{S}) matrix element of three-wave decay $\mathbf{k}_1\rightarrow\mathbf{k}_2+\mathbf{k}_3$. The \textit{S}-matrix element is
\begin{equation}
	\langle\mathbf{k}_2,\mathbf{k}_3|i\mathcal{L}_I|\mathbf{k}_1\rangle=i\mathcal{M} e^{i(k_2+k_3-k_1)x},
\end{equation}
in which the three-wave resonance conditions are enforced. The reduced matrix element $i\mathcal{M}$ can be represented using Feynman diagrams:
\begin{fmffile}{w3}
	\begin{eqnarray}
		\label{eq:w3}
		i\mathcal{M}=
		\begin{gathered}
			\begin{fmfgraph*}(50,55)  
				\fmfkeep{w3}
				\fmfleft{i1}
				\fmfright{o2,o3}
				\fmf{photon}{i1,v1}
				\fmf{photon}{v2,o2}
				\fmf{photon}{v3,o3}
				\fmf{fermion}{v1,v3}
				\fmf{plain}{v1,v2}
				\fmfdot{v1,v2,v3}
				\fmfv{label=$1$,label.angle=-120,label.dist=6}{v1}
				\fmfv{label=$2$,label.angle=-120,label.dist=6}{v2}
				\fmfv{label=$3$,label.angle=120,label.dist=6}{v3}
			\end{fmfgraph*}
		\end{gathered}+\text{5 permutations}.
	\end{eqnarray}
\end{fmffile}Since there are three external boson lines, each connecting to one of the three vertices, there are in total $3!=6$ Feynman diagrams. In the above Feynman diagram, interaction vertex to which "1" is connected to is the usual QED vertex, whereas vertices "2" and "3" appear only when there are background particle fields \citep{Shi16QED}. The arrow between vertices "1" and "3" indicates the direction of momentum flow during the interaction, and also labels which vertex the $\partial_t$ derivative is acting on. The above Feynman diagram corresponds to the particular Wick contraction
\begin{eqnarray}
	\label{eq:Feynman}
	\begin{gathered}
		\nonumber
		\fmfreuse{w3}
	\end{gathered}\hspace{-10pt}&=&-
	\contraction[1.5ex]{i\frac{e_s\omega_{ps}^2}{2m_s} \langle}{\mathbf{k}_2}{,\mathbf{k}_3|(\hat{\Pi}}{_{s}\hat{\mathbf{A}})}
	\bcontraction{i\frac{e_s\omega_{ps}^2}{2m_s} \langle\mathbf{k}_2,}{\mathbf{k}_3}{|(\hat{\Pi}_{s}\hat{\mathbf{A}})_j(\partial_j\hat{\mathbf{A}}_l)\partial_t(\hat{\Pi}}{_{s}\hat{\mathbf{A}}}
	\contraction{i\frac{e_s\omega_{ps}^2}{2m_s} \langle\mathbf{k}_2,\mathbf{k}_3|(\hat{\Pi}_{s}\hat{\mathbf{A}})_j(\partial}{_j\hat{\mathbf{A}}_l}{)\partial_t(\hat{\Pi}_{s}\hat{\mathbf{A}})_l|}{\mathbf{k}}
	i\frac{e_s\omega_{ps}^2}{2m_s} \langle\mathbf{k}_2,\mathbf{k}_3|(\hat{\Pi}_{s}\hat{\mathbf{A}})_j(\partial_j\hat{\mathbf{A}}_l)\partial_t(\hat{\Pi}_{s}\hat{\mathbf{A}})_l|\mathbf{k}_1\rangle\\
	&=&-i\frac{e_s\omega_{ps}^2}{2m_s}\big(i\frac{\mathbb{F}_{s,2}^*\mathbf{e}_{2}^*}{\omega_{2}}\!\big)_j ( i\mathbf{k}_1^j\mathbf{e}_{1}^l)(i\omega_3) \big(i\frac{\mathbb{F}_{s,3}^*\mathbf{e}_{3}^*}{\omega_{3}}\!\big)_l\\
	\nonumber
	&=&i\frac{e_s\omega_{ps}^2}{2m_sc}\Theta_{1,\bar{2}\bar{3}}^s.
\end{eqnarray} 
Summing with the other five Feynman diagrams, the reduced \textit{S}-matrix element in the quantum theory is related to the normalized scattering strength in the classical theory by the simple relation\footnote[1]{Notice that here $\mathcal{M}$ denotes the reduced \textit{S}-matrix element, which is different from the normalized growth rate we discussed in the previous section.}
\begin{equation}
	\label{eq:MTheta}
	\mathcal{M}=\sum_s\frac{e_s\omega_{ps}^2}{2m_sc}\Theta^s.
\end{equation} 
From the Lagrangian perspective, the classical three-wave coupling is related to the quantized interaction through the \textit{S} matrix: 
\begin{equation}
	i\Xi=A_1A_2^*A_3^*\langle\mathbf{k}_2,\mathbf{k}_3|i\mathcal{L}_I|\mathbf{k}_1\rangle e^{i(k_1-k_2-k_3)x}.
\end{equation}
Using the above relation, 
we can immediately recover the classical three-wave coupling by computing the \textit{S}-matrix element in the quantized theory. Alternatively, one may simply regard Lagrangian (\ref{eq:Lagrangian}) as a classical Lagrangian, and substitute Eq.~(\ref{eq:Ahat})
as the spectral expansion of the gauge field. Then, after integrating over spacetime, $\int d^4x\exp[i(k_1-k_2-k_3)x]=(2\pi)^4\delta^{(4)}(k_1-k_2-k_3)$ will select out the six resonate terms from the interaction Lagrangian, which give rise to the classical cubic interaction terms.

Now that we understand how the classical theory and the quantized theory are connected, we may postulate that the three-wave coupling always arises from the $\mathbf{d}^i(\partial_i\mathbf{A}_j)J^j$ term in the effective Lagrangian, regardless of the plasma model that is used to calculate the linear response. 
In the cold-fluid model, the linear response is expressed in terms of the cold forcing operator $\mathbb{F}$. By modifying this operator to include thermal or even quantum effects, and plugging it into the formalism I have developed, the three-wave scattering strength may be evaluated immediately. Having obtained the normalized scattering strength, as well as the wave energy coefficients in that particular plasma model, one can then compute the three-wave coupling coefficient using Eq.~(\ref{eq:coupling}). I have thus conjectured a prescription for computing three-wave coupling, without the need for going through the perturbative solution of the equations. The coupling coefficient then enters the three-wave equation, which governs the evolution of the envelopes of the three waves, which I will discuss next.

\chapter{Behaviors of three-wave equations\label{ch:3wave}}

In the previous section, I derived the three-wave equations [Eqs.~(\ref{eq:3waves1})-(\ref{eq:3waves3})] as a special case of the second-order electric-field equation. The three-wave equations contain a number of parameters, such as the wave frequencies and the wave group velocities\footnote[1]{Unlike in unmagnetized plasmas, the group velocity of a higher-frequency EM wave can now be smaller than that of a lower-frequency EM wave. Consequently, the behaviors of the three waves can be very different, even when coupling coefficients are the same.}, 
which can be readily obtained from the dispersion relation of linear waves. In addition, they contain an essential parameter, the coupling coefficient, for which I have obtained a convenient formula [Eq.~(\ref{eq:coupling})]. Once these parameters are evaluated for three resonantly interacting waves, the behavior of the three waves in a given plasma can then be determined by solving the three-wave equations. 

The three-wave equations are hyperbolic PDEs with an action-conserving nonlinear coupling. This system of equations has been studied extensively in the literature. For the homogeneous problem, where the spatial derivatives are zero, the equations become a system of nonlinear ODEs, and the general solution is given by the Jacobi elliptic functions \citep{Jurkus60,Armstrong62}. Similarly, in one dimension, the steady state problem, where the time derivatives are zero, can also be solved in terms of the Jacobi elliptic functions \citep{Harvey75}. As a trivial extension, traveling wave solutions in one spatial dimension can also be found using the coordinate transform $\xi=x-vt$ \citep{Armstrong70,Nozaki73,Ohsawa74}. In addition to these periodic solutions, the nonlinear three-wave equations also has compact solutions, such as the $N$-soliton solutions \citep{Zakharov75,Turner88}. The most general solution to the initial value problem in three spatial dimensions may be constructed using the inverse scattering method \citep{Ablowitz74,Reiman78,Kaup79,Kaup81}. Exact solutions can also be constructed using other formalisms \citep{Gilson98,Calogero05,Degasperis06}. Beyond obtaining analytical solutions, the deeper geometrical structure and group structures of three-wave equations have been analyzed \citep{Tondo85,Martina89,Alber98}, and algorithms have been developed to obtain numerical solutions \citep{Degasperis11,Skjaeraasen11}.

In this chapter, to give a sense of how three-resonantly interacting waves behave, I will be content with a brief review of simple analytical results, and numerical algorithms for solving the three-wave equations. First, I will review analytical solutions to the homogeneous problem, and develop an action-conserving algorithm. Next, I will discuss the one-dimensional problem, which admits a simple soliton solution. For more general cases, I will solve the three-wave equations numerically, by combining the upwind finite volume scheme with the strong-stability preserving Runge-Kutta time advance method. 

\section{Temporal solution of the homogeneous problem\label{sec:3wave:3wave:temporal}}

When the wave envelopes have no spatial dependence, the three-wave equations become three nonlinearly coupled ODEs
 \begin{eqnarray}
 \label{eq:da1}
 \dot{a}_1&=&-\frac{\Gamma}{\omega_1}a_2a_3,\\
 \dot{a}_2&=&\phantom{+}\frac{\Gamma}{\omega_2}a_3a_1,\\
 \label{eq:da3}
 \dot{a}_3&=&\phantom{+}\frac{\Gamma}{\omega_3}a_1a_2,
 \end{eqnarray}
 where the frequency $\omega_j$ is positive, $\dot{a}_j$ denotes the slow time derivative, and the wave envelope $a_j$ is real-valued for all $j=1,2,3$. In this section, I will discuss solutions to this system of equations.

\subsection{Parametric instability at the linear stage\label{sec:3wave:3wave:temporal:linear}}

Parametric instability is an instability that is characteristic of the linearized three-wave equations. For parametric decay, the pump amplitude $a_1$ is approximated as a constant and $\dot{a}_1$ is approximately zero. Then, taking second derivative
\begin{eqnarray}
\ddot{a}_2\simeq\frac{\Gamma}{\omega_2}a_1\dot{a}_3=\frac{\Gamma^2a_1^2}{\omega_2\omega_3}a_2.
\end{eqnarray}
The general solution to this linear ODE is
\begin{equation}
a_2(t)=\frac{1}{2}\Big(a_{20}+\frac{\dot{a}_{20}}{\gamma_0}\Big)e^{\gamma_0 t} +\frac{1}{2}\Big(a_{20}-\frac{\dot{a}_{20}}{\gamma_0}\Big)e^{-\gamma_0 t},
\end{equation} 
where $a_{20}=a_2(t=0)$ and $\dot{a}_{20}=\dot{a}_2(t=0)$ are the initial conditions. The exponential growth rate, which was given in Eq.~(\ref{eq:GrowthRate}), is again 
\begin{equation}
\gamma_0=\frac{|\Gamma a_1|}{\sqrt{\omega_2\omega_3}}.
\end{equation} 
Following a similar procedure, we can obtain the solution for $a_3$, which is formally similar to the above solution. When $t\rightarrow\infty$, the exponential growth dominates. However, shortly after the amplitudes of the decay products grow, the approximation of a constant $a_1$ ceases to be valid. At that point, the three-wave equation enters the nonlinear stage, which I will discuss next.
\subsection{General solution at the nonlinear stage\label{sec:3wave:3wave:temporal:nonlinear}}

At the nonlinear stage, the general solution can be obtained using the action conservation laws. Up to some constant, one has $\mathcal{I}_j=\omega_ja_j^2$. From Eqs.~(\ref{eq:I12}) and (\ref{eq:I23}), we have the following constants of motion
\begin{eqnarray}
	S_3=\mathcal{I}_1+\mathcal{I}_2>0,\\
	S_2=\mathcal{I}_1+\mathcal{I}_3>0.
\end{eqnarray}
Geometrically, these equations describe two cylinders in the $(a_1,a_2,a_3)$ space, and the system dynamics is thereof constrained along the intersections of these two cylinders. In fact, due to the temporal resonance condition $\omega_1=\omega_2+\omega_3$, the intersection curves also lie on the energy ellipsoid
\begin{equation}
U=\omega_1\mathcal{I}_1+\omega_2\mathcal{I}_2+\omega_3\mathcal{I}_3>0.
\end{equation} 
This energy constraint is similar to that of freely-rotating rigid bodies, which are described by the Euler's equations that resemble the temporal three-wave equations.

\begin{figure}[t]
	\centering
	\includegraphics[angle=0,width=0.8\textwidth]{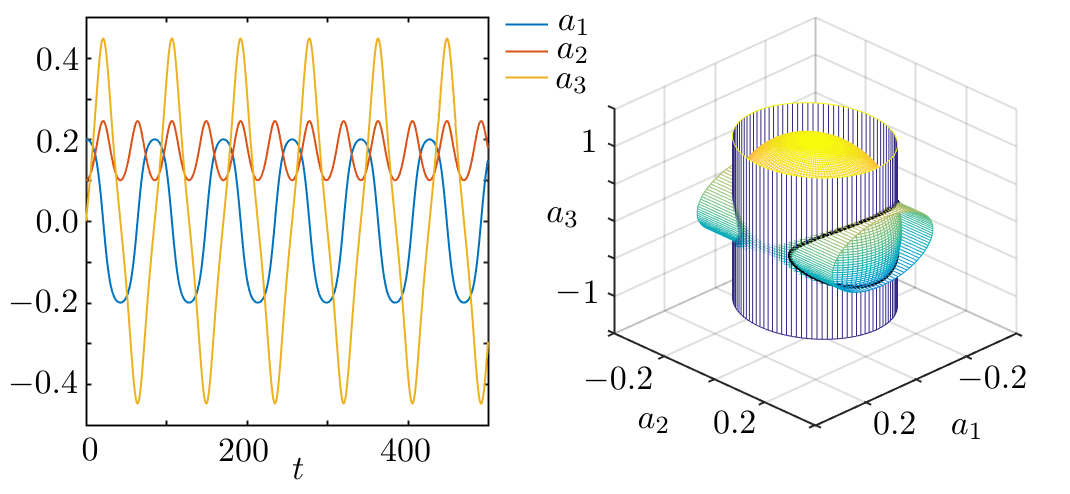}
	\caption[Temporal solution of three-wave equations]{An exact solution of the temporal three-wave equations [Eqs.~(\ref{eq:da1})-(\ref{eq:da3})] when the coupling coefficient $\Gamma=2$. The wave frequencies are $\omega_1=10$, $\omega_2=8$, and $\omega_3=2$. In time domain (left), the solutions are oscillatory. The periods of the nonlinear oscillations are determined by initial conditions. In the configuration space (right), the trajectory (black) lies on the intersection of the energy ellipsoid $U=\sum_j\omega_j^2a_j^2$ with the two action cylinders $S_3=\omega_1a_1^2+\omega_2a_2^2$ and $S_2=\omega_1a_1^2+\omega_3a_3^2$.}
	\label{fig:homo}
\end{figure}

Using the above constants of motion, we can decouple the three equations. For example, the amplitude $a_1$ satisfies
\begin{eqnarray}
\label{eq:a1}
\nonumber
(\dot{a}_1)^2&=&\frac{\Gamma^2}{\omega_1^2}a_2^2a_3^2\\
\nonumber
&=&\frac{\Gamma^2}{\omega_1^2}\bigg(\frac{S_3-\omega_1a_1^2}{\omega_2}\bigg)\bigg(\frac{S_2-\omega_1a_1^2}{\omega_3}\bigg)\\
&=&\frac{R}{\omega_1}(S_2-\omega_1a_1^2)(S_3-\omega_1a_1^2)
\end{eqnarray}
where $R=\Gamma^2/(\omega_1\omega_2\omega_3)$ has the units of frequency. Since $S_2,S_3>0$ are constants, the above equation can be solved in terms of the Jacobi elliptic function \citep{Jurkus60}. To put Eq.~(\ref{eq:a1}) into the standard form, we can rescale both the time $t=\tau\eta$ and the amplitude $\alpha_1(\tau)=\lambda a_1(t)$, so that
\begin{equation}
	(\dot{\alpha}_1)^2=(1-k^2\alpha_1^2)(1-\alpha_1^2).
\end{equation}
In this form, we can immediately recognize the modulus of the elliptic function $k^2=S_2/S_3$. The required normalization for time is $\eta^2=1/(R S_3)$, and the required scaling for the amplitude is $\lambda^2=\omega_1/S_2$. The general solution to the above equation is the Jacobi elliptic function $\alpha_1(\tau)=\text{sn}(\tau+\tau_0,k)$, where $\tau_0$ is a phase shift determined by initial conditions. When $S_2<S_3$, the modulus $k<1$, and the solution to $a_1$ is
\begin{equation}
	a_1(t)=\frac{a_{10}\text{cn}(\tau)\text{dn}(\tau)+\eta\dot{a}_{10} \text{sn}(\tau)}{1-k^2\lambda^2a_{10}^2\text{sn}^2(\tau)},
\end{equation}
where $a_{10}$ and $\dot{a}_{10}$ are the initial value and derivative of $a_1$ at $t=0$. To find the solution when $S_2>S_3$, we can using the formula for transforming the modulus $k\rightarrow1/k$ for the Jacobi elliptic functions. Similarly, the amplitude $a_2$ and $a_3$ can be expressed in terms of the Jacobi elliptic functions or obtained directly from the action conservation laws
\begin{eqnarray}
	a_2(t)=\frac{a_{20}\text{dn}(\tau)+\eta\dot{a}_{20}\text{sn}(\tau)\text{cn}(\tau)}{1-k^2\lambda^2a_{10}^2\text{sn}^2(\tau)},\\
	a_3(t)=\frac{a_{30}\text{cn}(\tau)+\eta\dot{a}_{30}\text{sn}(\tau)\text{dn}(\tau)}{1-k^2\lambda^2a_{10}^2\text{sn}^2(\tau)},
\end{eqnarray}
where all the elliptic functions have the same modulus $k^2$. The rescaling parameters $\eta$ and $\lambda$ are the same as before. An example of the exact solution is plotted in Fig.~\ref{fig:homo}, where the coupling coefficient $\Gamma=2$ in some units. The wave frequencies $\omega_1=10$, $\omega_2=8$, and $\omega_3=2$. The initial condition $a_{10}=0.2$, $a_{20}=0.1$, and $a_{30}=0.0$. Such an exact solution will be useful for validating numerical solutions of the three-wave equations.

\subsection{Numerical scheme with exact action conservation\label{sec:3wave:3wave:temporal:numerical}}

As a warm up excise for numerical solutions of the spacetime problem, let us solve the three-wave equations (\ref{eq:da1})-(\ref{eq:da3}) numerically. Notice that simple algorithms operating on amplitudes $a_j$ cannot exactly satisfy the action conservation laws. To see why, consider a two-step method $a_j^{n+1}=a_j^n+\Delta_j^n$, where $a_j^n$ is the numerical approximation at time step $n$, and $\Delta_j^n$ is the difference. Then, it can be shown by straightforward calculations that the change in action $I_1^n+I_2^n$ cannot be canceled by choosing $\Delta_j^n$ is a way that is consistent with the differential equation. 
In fact, unless we double the configuration space into the phase space \citep{Hairer06}, there is no volume-conserving sympletic algorithm for the three-wave equations in the $(a_1,a_2,a_3)$ configuration space, such that the actions would be conserved exactly. To see why, notice that a symplectic structure, defined on the cotangent bundle, does not exist when the manifold is odd dimensional. In what follows, I will develop an \textit{ad hoc} algorithm, such that the exact action conservation laws can be satisfied. For this purpose, it is more helpful to write the equation in another form
\begin{eqnarray}
	d_t a_1^2&=&-\frac{2\Gamma}{\omega_1}a_1a_2a_3,\\
	d_t a_2^2&=&\phantom{+}\frac{2\Gamma}{\omega_2}a_1a_2a_3,\\
	d_t a_2^2&=&\phantom{+}\frac{2\Gamma}{\omega_3}a_1a_2a_3.\\
\end{eqnarray}
Using this form of the three-wave equations, even the simple Euler's method for time advance will conserves action exactly
\begin{equation}
	(a_j^{n+1})^2=(a_j^{n})^2+s_j\frac{2\Gamma h}{\omega_j}a_1^na_2^na_3^n,
\end{equation}
where $s_1=-1$, $s_{2,3}=+1$ are the signs, and $h$ is a fixed time step size. Since the changes in the actions exactly cancel, the above first-order numerical scheme exactly satisfies the action conservation laws. 

The only problem is that the above algorithm advances $(a_j^{n})^2$, while throwing away the sign information of $a_j$, which is needed on the RHS. However, this problem can be solved using the following numerical scheme.
First, notice that whenever $a_j=0$, the derivative $\dot{a}_j$ must be nonzero in order for the solution to be nontrivial. In other words, whenever $(a_j^{n})^2$ touches zero, the amplitude $a_j$ must cross zero, and thereby flip sign. To see what happens to the other two amplitudes when one amplitude crosses zero, consider the example $a_1(t_0)=0$. Using the ODEs, the first-order derivatives $\dot{a}_2=\dot{a}_3=0$, while the second-order derivatives $\ddot{a}_2$ and $\ddot{a}_3$ are nonzero at $t_0$ if the solution is nontrivial. Therefore, zero crossing points of $a_1$ correspond to extrema of $a_2$ and $a_3$. 
Now that we understand what happen at zero-crossing points, we can use the following reflection scheme to step across the zero-crossing point: $a_1\rightarrow-a_1$, $a_{2,3}\rightarrow a_{2,3}$. In this way, the action is exactly conserved during zero crossing. Using Taylor expansion near $t_0$, suppose reflection maps time $t_0-\Delta t/2\rightarrow t_0+\Delta t/2$, then amplitudes at previous time step is related to the values at the zero-crossing point by
\begin{eqnarray}
	a_1^{n-1}&=&\frac{\Gamma\Delta t}{2\omega_1}a_{20}a_{30},\\
	a_2^{n-1}&=&a_{20}-\frac{\Gamma^2\Delta t^2}{8\omega_1\omega_2}a_{20}a_{30}^2,\\
	a_3^{n-1}&=&a_{30}-\frac{\Gamma^2\Delta t^2}{8\omega_1\omega_3}a_{30}a_{20}^2,
\end{eqnarray}
where $a_{20}, a_{30}$ are the extrema of $a_2$ and $a_3$, attained at $t_0$. From the above three equations, we can solve for the three unknowns, namely, the values of the extrema $a_{20}$ and $a_{30}$:
\begin{eqnarray}
a_{20}&=&\frac{1}{2}\Big[a_{2}^{n-1}+s_2\sqrt{(a_{2}^{n-1})^2+\frac{2\omega_1}{\omega_2}(a_{1}^{n-1})^2}\Big], \\
a_{30}&=&\frac{1}{2}\Big[a_{3}^{n-1}+s_3\sqrt{(a_{3}^{n-1})^2+\frac{2\omega_1}{\omega_3}(a_{1}^{n-1})^2}\Big], 
\end{eqnarray}
as well as the adaptive time step $\Delta t$ for zero crossing:
\begin{equation}
\Delta t =\frac{2\omega_1}{\Gamma}\frac{a_1^{n-1}}{a_{20}a_{30}},
\end{equation}
where $s_{2,3}$ are the signs of $a_{2,3}$ at $t_0$.
The above zero-crossing procedure is invoked when $a_1$ approaches zero, and $|a_1|<\epsilon A_1$. Here, $\epsilon\ll1$ is some threshold and $A_1$ is the maximum value of $a_1$ on the energy ellipsoid. If the zero-crossing procedure yields $\Delta t<h$ smaller than the fixed time step $h$, then a stepping is made. Otherwise, an Euler time stepping is made with halved time step $h\rightarrow h/2$ before repeating the zero-crossing procedure. The algorithm works in similar ways when either $a_2$ or $a_3$ crosses zero instead. This zero-crossing algorithm introduces phase errors due to the error in the adaptive time step $\Delta t$. This is in the same spirit as symplectic algorithms, which trade amplitude error for phase error.

\begin{figure}[t]
	\centering
	\includegraphics[angle=0,width=0.8\textwidth]{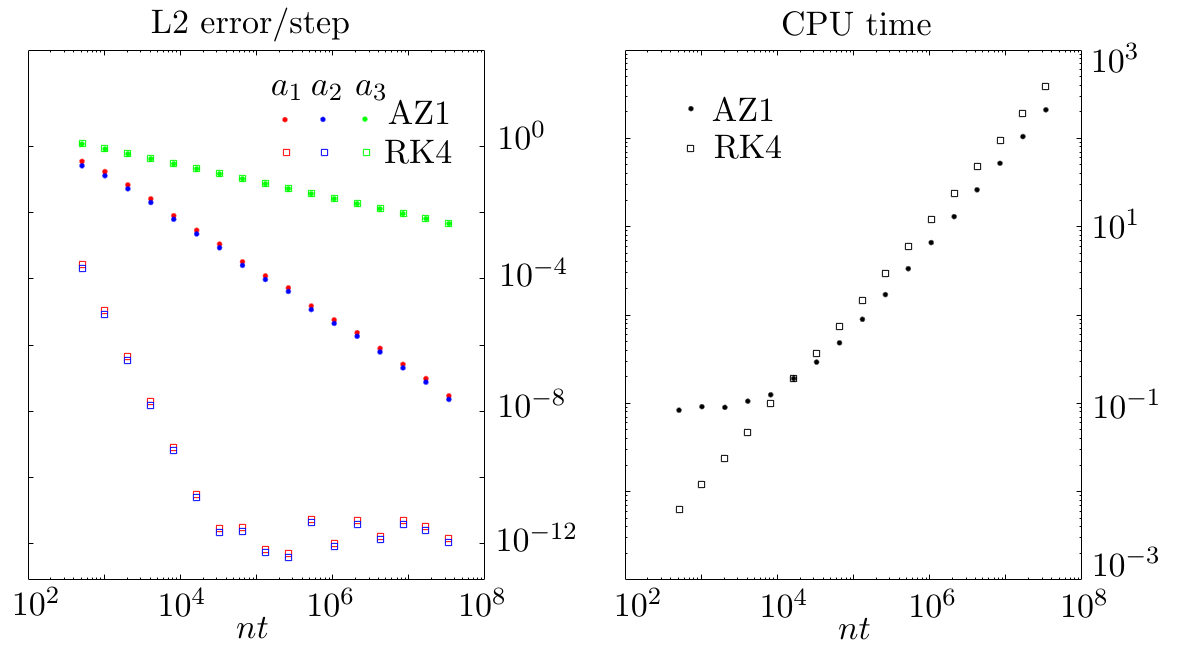}
	\caption[Error convergence and computational cost in temporal problem]{Error convergence (left) and computation cost (right) of the AZ1 and RK4 algorithms when applied to the problem $\Gamma=2$, with initial conditions $a_{10}=20, a_{20}=0$, and $a_{30}=-1$, and frequencies $\omega_1=10$, $\omega_2=8$, and $\omega_3=2$. The final time $t_f=10$ is fixed when increasing $nt$, the total number of time steps.}
	\label{fig:performance}
\end{figure}

To test the first-order Action-conserving Zero-crossing (AZ1) algorithm, consider the example $\Gamma=2$, with the initial conditions\footnote[1]{When deriving the three-wave equations, the normalized amplitudes are assumed to be small. However, once the differential equations are obtained, there is no restriction on what values the amplitudes can take. Mathematically, one can always rescale the frequencies, such that the amplitudes becomes small. } 
$a_{10}=20, a_{20}=0$, and  $a_{30}=-1$, and frequencies $\omega_1=10$, $\omega_2=8$, and $\omega_3=2$. The performance of the AZ1 method is compared with the standard fourth-order Runge-Kutta (RK4) method in Fig.~\ref{fig:performance}. 
First, the errors between numerical solutions and the exact solution converge as $\Delta t\rightarrow0$ for both algorithms. The convergence rates for both $a_1$ and $a_2$ are consistent with the order of the algorithm. However, surprisingly enough, the errors for $a_3$ decrease at a much slower rate. Perhaps even more surprisingly, the convergence rates of $a_3$ are the same for both algorithms. Such an interesting behavior might be attributed to the following factors. The amplitude $a_3$ is special because $\omega_3$ is much smaller than $\omega_1$ and $\omega_2$. Therefore, $a_3$ changes at the fastest rate and gains the largest error. Due to the nonlinearity, the linear error estimation might be inaccurate for the three-wave problem.
Second, in terms of the computational cost, the time cost of RK4 scales linearly with the number of time steps. In comparison, the time cost for the AZ1 algorithm is nonlinear. This is because the computational time spent for zero-crossing is much longer than the time spent for an ordinary time step. Since the number of zero-crossing points are fixed for a fixed problem, time spent fore zero-crossing is roughly constant. Therefore, for a small number of time steps, the time cost is dominated by the iterative zero-crossing detections. Whereas for a larger number of time steps, the zero-crossing time becomes subdominant. Hence, the time cost for AZ1 initially remains roughly constant, and then increases linear with the number of time steps. 

\begin{figure}[t]
	\centering
	\includegraphics[angle=0,width=0.8\textwidth]{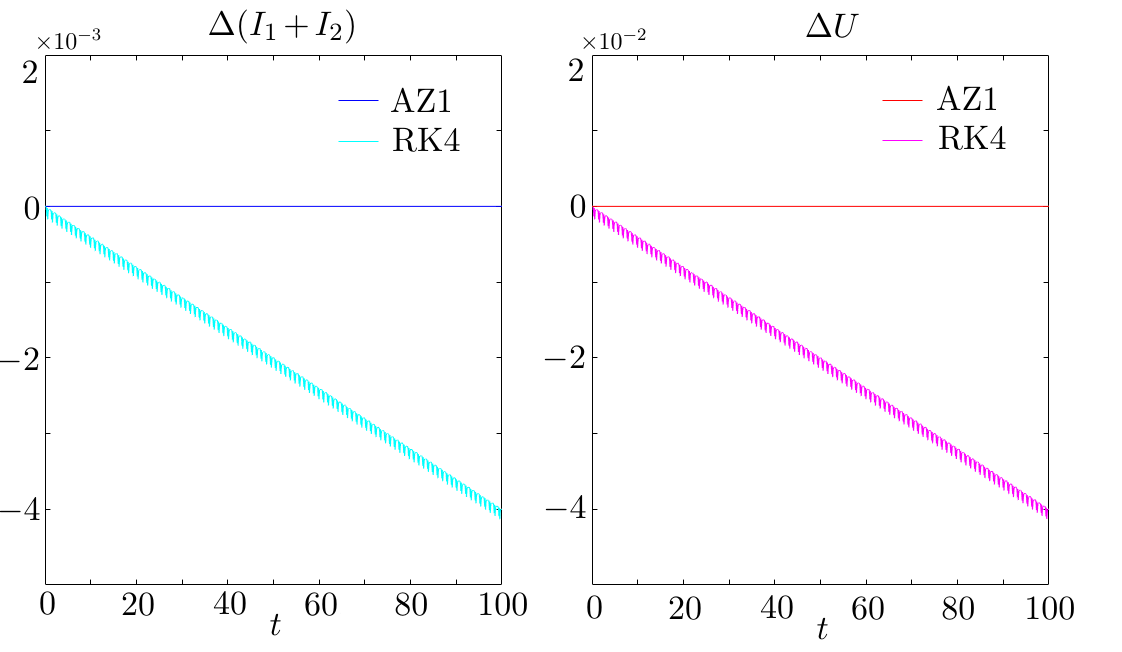}
	\caption[Action error and energy error in temporal problem]{Action error (left) and energy error (right) as function of time for fixed time step size. The AZ1 algorithm has no amplitude error. In comparison, the RK4 algorithm dissipates action and energy secularly.}
	\label{fig:Action_tf1e2_dt5e3}
\end{figure}

By design, the AZ1 algorithm has no amplitude error (Fig.~\ref{fig:Action_tf1e2_dt5e3}), while the RK4 algorithm tends to dissipate the total action and energy. However, the AZ1 algorithm has phase error. Since AZ1 is only first order while RK4 is fourth order, the L2 error norm of AZ1 converges slower than that of the RK4 algorithm. Hence, if one is very concerned with amplitude error, then AZ1 has definite advantage. On the other hand, if one is more concerned with the L2 error, then AZ1 is not necessarily a better choice.

\section{Solutions of the spacetime problem\label{sec:3wave:3wave:spacetime}}

Having considered the homogeneous problem in zero spatial dimensional, now let us move on to solve the one-dimensional problem
\begin{eqnarray}
\label{eq:a11D}
(\partial_t+v_1\partial_x)a_1&=&-\frac{\Gamma}{\omega_1}a_2a_3,\\
(\partial_t+v_2\partial_x)a_2&=&\phantom{+}\frac{\Gamma}{\omega_2}a_3a_1,\\
\label{eq:a31D}
(\partial_t+v_3\partial_x)a_3&=&\phantom{+}\frac{\Gamma}{\omega_3}a_1a_2,
\end{eqnarray}
where $a_j$ is the real-valued wave envelope for the wave with positive frequency $\omega_j$, whose group velocity $v_j$ can either be positive or negative. Before I discuss these equations, it is worth pointing out that the vector group velocity $\mathbf{v}_g=\partial\omega/\partial\mathbf{k}$ is not necessarily in the same direction as the wave vector $\mathbf{k}$. Therefore, the resonance condition $\mathbf{k}_1=\mathbf{k}_2+\mathbf{k}_3$ does not imply that the group velocities have any definite relations. Conversely, here in this one-dimensional problem, although the waves are chosen to have aligned group velocities, it does not mean that their wave vectors are necessarily aligned. 

\subsection{A brief review of the soliton solutions\label{sec:3wave:3wave:spacetime:analytical}}

An exact solution to the above system of hyperbolic PDEs is the soliton solution \citep{Nozaki73,Ohsawa74}. The soliton solution is of the form $a_j(x,t)=\alpha_j(\xi)$, where $\xi=x-\lambda t$ is the soliton coordinate, and $\lambda$ is the soliton speed. In such a soliton solution, envelopes of the three waves are mutually locked by nonlinearities, and move together at the same speed. 

Let us substitute the soliton anzatz into Eqs.~(\ref{eq:a11D})-(\ref{eq:a31D}), then the equation can be converted to the zero-dimensional case. The self-consistency condition of the soliton speed is given by
\begin{equation}
	(\lambda-v_1)(\lambda-v_2)=\frac{\Gamma^2\kappa^2}{\omega_1\omega_2},
\end{equation}
where $\kappa$ is an arbitrary parameter determining the soliton amplitude. From the above equation, it is easy to see either $\lambda>\max(v_1,v_2)$, or $\lambda<\min(v_1,v_2)$. Moreover, suppose we have $\lambda>\max(v_1,v_2,v_3)$ or $\lambda<\min(v_1,v_2,v_3)$, then soliton solutions exist
\begin{eqnarray}
	\alpha_1(\xi)&=&-\alpha_\infty\tanh(R\xi),\\
	\alpha_2(\xi)&=&\alpha_{20}\text{sech}(R\xi),\\
	\alpha_3(\xi)&=&\kappa R\text{sech}(R\xi),
\end{eqnarray}
where $\alpha_\infty$ is the asymptotic value of $\alpha_1$, the amplitude $\alpha_{20}/\alpha_\infty=\Gamma\kappa/[\omega_2(v_2-\lambda)]$, and the scaling factor $R^2=\Gamma^2\alpha_\infty^2/[\omega_2\omega_3(v_2-\lambda)(v_3-\lambda)]>0$. It is a straightforward calculation to verify that the above expressions give an exact solution to the one-dimensional problem. 

\begin{figure}[b]
	\centering
	\includegraphics[angle=0,width=0.75\textwidth]{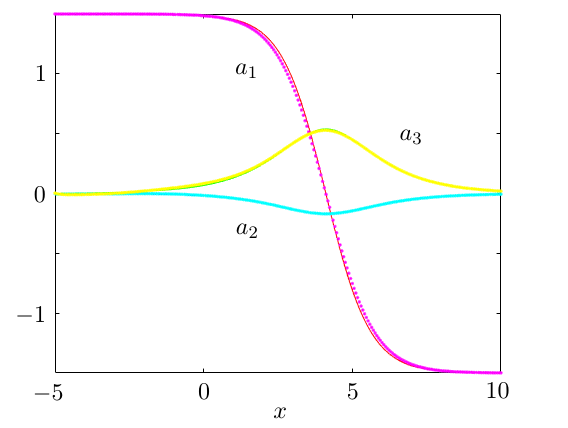}
	\caption[Exact and numerical soliton solution]{At final time $t_f=12$, the exact soliton solutions (lines) are matched by the numerical solutions (dots) obtained using the UW1-SSPRK3 algorithm. The initial conditions for the soliton problem are given by the analytical formula, with $\Gamma=2$, $v_{1}=0.8, v_{2}=-0.9$, and $v_{3}=0.1$. The numerical algorithm is then used to advance the initial conditions in time by $nt=256$ time steps. The number of spatial grid points $nx=256$, and the CFL number is $\Delta t/\Delta x=0.8$.}
	\label{fig:soliton_final}
\end{figure}

It is remarkable that although the advection velocities are in general different and can even have opposite signs, the three solitons always travel at the same speed $\lambda$, which is either faster or slower than all the advection velocities. This is a vivid illustration of how nonlinearities can dramatically change the behavior of otherwise linear waves. On one hand, the linear advection terms tend to maintain the shapes of wave envelopes and move them at the wave group velocities. On the other hand, the nonlinear three-wave interactions tend to alter the shape of the envelopes, growing the envelope of one wave where the other two waves coincide, while diminishing the envelope of ones wave where the other two waves mismatch. 
The final behaviors of the three waves are determined by these two competing effects, and the soliton solution is the special case where the three wave envelopes are locked by special matching conditions. This exact soliton solution will be used to validate numerical solutions.

\subsection{Numerical solutions of three-wave equations\label{sec:3wave:3wave:spacetime:numerical}}

The preferred method for numerically solving hyperbolic PDEs is the finite volume method, which solves the equations in their integral form. In the integral form, flux conservation becomes explicit even in the presence of discontinuities. For example, for action $\mathcal{I}_1=\omega_1a_1^2$, we can average its equation within a cell, which is centered at $x_k$ and has width $\Delta x$. Integrating on both sides,
\begin{eqnarray}
\nonumber
0&=&\int_{x_k-\frac{\Delta x}{2}}^{x_k+\frac{\Delta x}{2}}dx\big(\partial_t\mathcal{I}_1+v_1\partial_x\mathcal{I}_1+2\Gamma a_1a_2a_3\big)\\
&=&\Delta x\partial_t\langle \mathcal{I}_1\rangle_k+v_1\big(\mathcal{I}_{1,k+\frac{1}{2}}-\mathcal{I}_{1,k-\frac{1}{2}}\big)+2\Gamma\Delta x \langle a_1a_2a_3\rangle_k,
\end{eqnarray}
where $\langle w \rangle_k:=\frac{1}{\Delta x}\int_{x_k-\Delta x/2}^{x_k+\Delta x/2}w(x)dx$ denotes the cell average, and $w_{k+1/2}(t)=w(t,x_k+\Delta x/2)$ denotes the value of $w$ on the cell boundary. We can further average in the time cell $[t_n,t_{n+1}]$, then the equation becomes
\begin{eqnarray}
\nonumber
0&=&\int_{t_n}^{t_{n+1}}dt\big[\Delta x\partial_t\langle \mathcal{I}_1\rangle_k+v_1\big(\mathcal{I}_{1,k+\frac{1}{2}}-\mathcal{I}_{1,k-\frac{1}{2}}\big)+2\Gamma\Delta x \langle a_1a_2a_3\rangle_k \big]\\
&=&\Delta x\Big(\langle \mathcal{I}_1\rangle_k^{n+1}-\langle \mathcal{I}_1\rangle_k^n\Big) +v_1\Delta t\Big(\bar{\mathcal{I}}_{1,k+\frac{1}{2}}^{n+\frac{1}{2}}-\bar{\mathcal{I}}_{1,k-\frac{1}{2}}^{n+\frac{1}{2}}\Big) +2\Gamma\Delta x\Delta t \overline{\langle a_1a_2a_3\rangle}_k^{n+\frac{1}{2}},
\end{eqnarray}
where $\bar{w}^{n+1/2}:=\frac{1}{\Delta t}\int_{t_n}^{t_{n+1}}w(t)dt$ denotes the time average. The above integral form of the equation is exact, and we can formally express the cell average at the next time step as
\begin{eqnarray}
	\label{eq:I1_integral}
	\langle \mathcal{I}_1^{n+1}\rangle_k=\langle \mathcal{I}_1^{n}\rangle_k- \frac{v_1\Delta t}{\Delta x}(\bar{\mathcal{I}}_{1,k+1/2}^{n+1/2}-\bar{\mathcal{I}}_{1,k-1/2}^{n+1/2}) -2\Gamma\Delta t\langle\overline{a_1a_2a_3}\rangle_k^{n+1/2},
\end{eqnarray}
However, time-averaged values of $\mathcal{I}_1$ on the cell boundaries are required on the RHS. These values are unfortunately not known, unless we had already solved the equation. The finite volume scheme seeks to approximate these unknowns, thereby construct an approximate solution to the integral equation.

To approximate the time averaged values at cell boundaries, we can use the standard upwind schemes \citep{Durran10}. The upwind schemes use information from upstream to approximate values downstream using interpolations. Such schemes not only respect the causality, but also enhance numerical stability. When the advection velocity $v_1>0$, the first-order upwind approximation
\begin{eqnarray}
	\bar{\mathcal{I}}_{1,k+1/2}^{n+1/2}&\simeq&\langle \mathcal{I}_1\rangle_k^{n},\\
	\bar{\mathcal{I}}_{1,k-1/2}^{n+1/2}&\simeq&\langle \mathcal{I}_1\rangle_{k-1}^{n}.
\end{eqnarray}
For the $v_1<0$ case, analogous upwind formula can be written down using information that is upstream of the advection. In addition to the flux through cell boundaries, the nonlinear problem also need the cell average of products $\langle\overline{a_1a_2a_3}\rangle_k^{n+1/2}$. To lowest order, the average of products may be approximated by the product of averages:
\begin{eqnarray}
	\langle\overline{a_1a_2a_3}\rangle_k^{n+1/2}\simeq\langle a_1\rangle_k^n \langle a_2\rangle_k^n \langle a_3\rangle_k^n.
\end{eqnarray}
Then the integral equation (\ref{eq:I1_integral}) can be used to advance the cell averages in time. Similar schemes can be used to advance $\mathcal{I}_2$ and $\mathcal{I}_3$ in time, and it is a straightforward calculation to check that the actions $\mathcal{I}_1+\mathcal{I}_2$ and $\mathcal{I}_2-\mathcal{I}_3$ are constants of advection. 

The above numerical scheme, solving for $\mathcal{I}_j$ instead of $a_j$, is similar the AZ1 scheme I developed for the zero-dimensional problem (Sec.~\ref{sec:3wave:3wave:temporal:numerical}). Again, we need some zero-crossing procedure, because the nonlinear terms is proportional to $a_j$ instead of $\mathcal{I}_j$. The sign of $a_j$ is significant, which is unfortunately disregarded when we time advance $\mathcal{I}_j$. In principle, we could laboriously construct a similar zero-crossing procedure as before. However, such a procedure will have phase errors, even though the amplitude error is exactly zero. From the L2 error norm perspective, the first-order action-conserving time advance may not be a more favorable choice over standard higher-order methods. Therefore, in what follows, I will reconfigure the finite volume scheme to solve for amplitudes $a_j$ directly.

\begin{figure}[b]
	\centering
	\includegraphics[angle=0,width=0.75\textwidth]{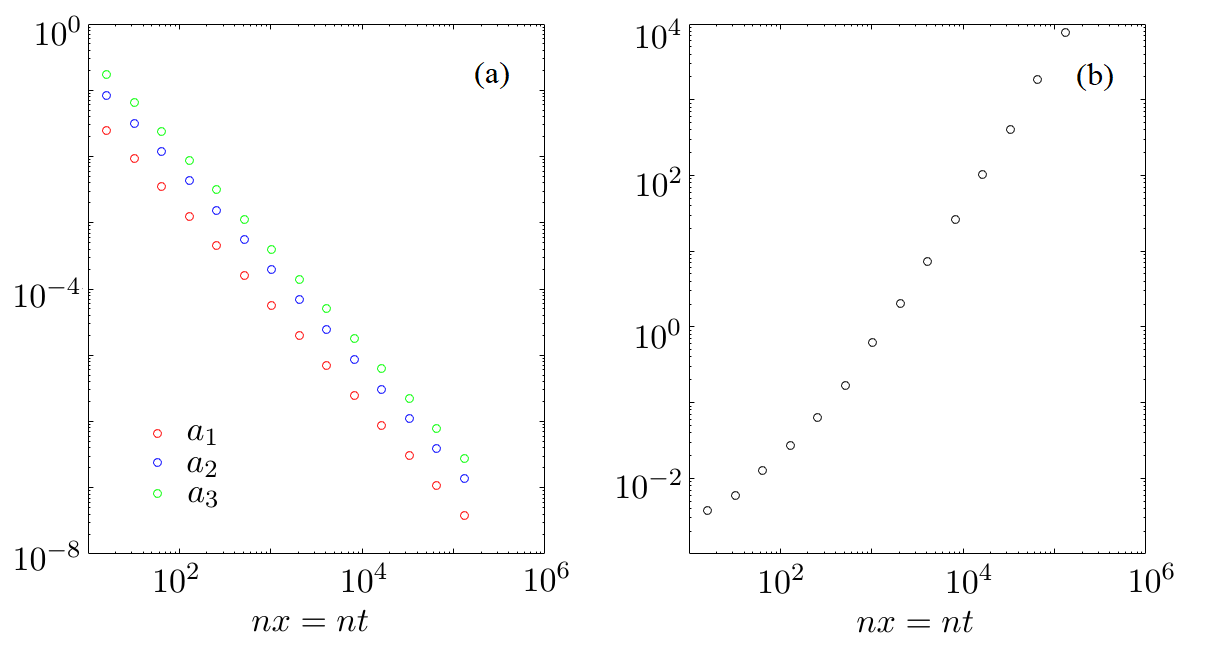}
	\caption[Error convergence and computational cost in soliton problem]{As the resolution increases, the L2 error norm decreases (a) and computation cost increases (b), when the UW1-SSPRK3 algorithms is applied to the soliton problem with $\Gamma=2$, $v_{1}=0.8, v_{2}=-0.9$, and $v_{3}=0.1$. The final time $t_f=12$ and domain size $L=15$ are fixed. The number of time steps $nt$ and number of spatial grid points $nx$ are increased together such that CFL$=0.8$ is fixed.}
	\label{fig:Performance_UW1_SSPRK3}
\end{figure}

Using the first-order upwind scheme in finite volume method, the spatially averaged amplitude $\langle a_1\rangle_k$ approximately satisfies
\begin{equation}
\dot{\langle a_1\rangle}_k\simeq-v_1^+\frac{\langle a_1\rangle_k-\langle a_1\rangle_{k-1}}{\Delta x}-v_1^-\frac{\langle a_1\rangle_{k+1}-\langle a_1\rangle_{k}}{\Delta x}-\frac{\Gamma}{\omega_1}\langle a_2\rangle_k\langle a_3\rangle_k,
\end{equation}  
where $\dot{w}$ denotes the time derivative. The upwind velocities $v_1^+=\max(v_1,0)$ and $v_1^-=\min(v_1,0)$. The differential equations for $\langle a_2\rangle_k$ and $\langle a_3\rangle_k$ are analogous to the above equation. Then, we obtain a system of ODEs of the form $\dot{\mathbf{A}}=F(\mathbf{A})$. Here $F$ is a generic function, and $\mathbf{A}$ is the matrix $\mathbf{A}=(\mathbf{a}_1,\mathbf{a}_2,\mathbf{a}_3)$, where $\mathbf{a}_j$ is the one dimensional array of cell averages $\langle a_j\rangle_k$. To advance this system of ODEs in time, we can use the strong-stability-preserving (SSP) third-order Runge-Kutta (RK3) method \citep{Shu88}. The SSPRK3 method uses a convex combination of Euler steps to reduce oscillatory behavior in the linear advection problems. The SSPRK3 scheme for the generic ODE is
\begin{eqnarray}
	\mathbf{A}^{(1)}&=&\mathbf{A}^n+\Delta tF(\mathbf{A}^n),\\
	\mathbf{A}^{(2)}&=&\frac{3}{4}\mathbf{A}^n+\frac{1}{4}[\mathbf{A}^{(1)}+\Delta tF(\mathbf{A}^{(1)})],\\
	\mathbf{A}^{n+1}&=&\frac{1}{3}\mathbf{A}^n+\frac{2}{3}[\mathbf{A}^{(2)}+\Delta tF(\mathbf{A}^{(2)})].
\end{eqnarray}
The upwind scheme combined the above time advance method gives an algorithm (UW1-SSPRK3) for the three-wave equations in one spatial dimension. The UW1-SSPRK3 algorithm can be readily improved in two directions using methods well-established in the literature. First, the upwind scheme can be extended to higher spatial dimensions, and thereby allows solution to the higher-dimensional problems. Second, both the upwind scheme and the ODE time advance can be replaced by schemes with higher-order accuracy. Here, I will be content with the UW1-SSPRK3 algorithm.

First, let us validate the algorithm and its implementation by applying it to the soliton problem, for which we know the exact solution (Sec.~\ref{sec:3wave:3wave:spacetime:analytical}). For example, let us take $\Gamma=2$, $v_{1}=0.8$, \mbox{$v_{2}=-0.9$}, and $v_{3}=0.1$, and compare the numerical solutions with the exact solutions (Fig.\ref{fig:soliton_final}). With the Courant-Friedrichs-Lewy (CFL) number $\Delta t/\Delta x=0.8$, and the number of time steps $nt$ equals to the number of spatial grid points $nx=256$, the numerical solutions (dots) match the exact solutions (lines) with only a small amount of diffusive errors. Moreover, the L2 error norms decrease with increasing $nt=nx$ (Fig.\ref{fig:Performance_UW1_SSPRK3}a), indicating that the numerical solutions converge to the exact solutions when the resolution is sufficiently high. In this convergence study, the domain size $L=15$, the final time $t_f=12$, and CFL=0.8 are fixed. As the resolution increases, the computational cost scales roughly as $n^2$ (Fig.\ref{fig:Performance_UW1_SSPRK3}b), where $n=nx=nt$ is the number of grid points for a fixed CFL number.  

\begin{figure}[!t]
	\centering
	\includegraphics[angle=0,width=0.73\textwidth]{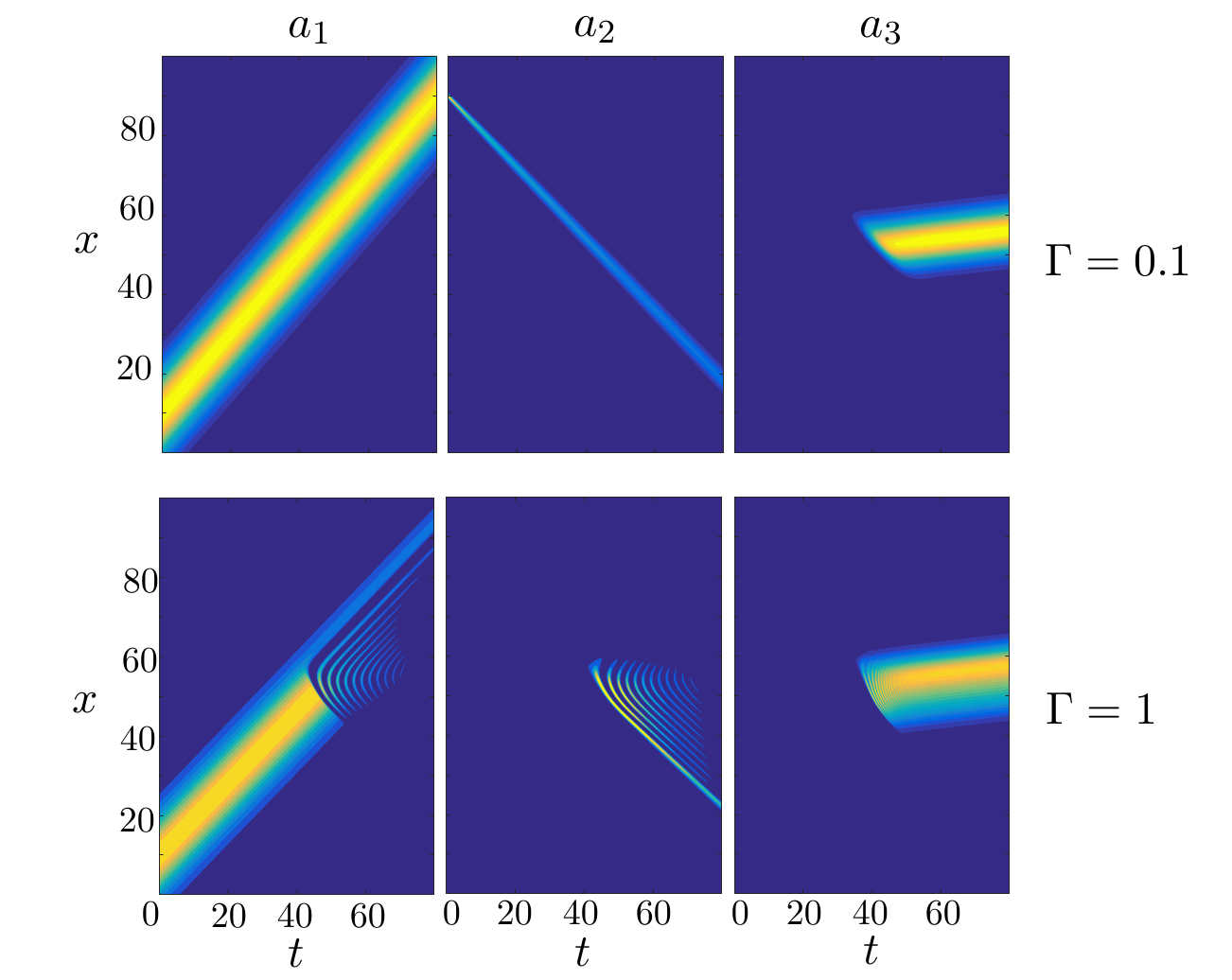}
	\caption[Numerical solution of back scattering problem]{Backscattering of a pump laser ($a_1$, left) by a counter propagating seed pulse ($a_2$, middle), during which a plasma wave ($a_3$, right) is excited. When the coupling coefficient is small (upper panel), the two counter-propagating lasers pass through each other without much interaction. When the coupling coefficient is large (lower panel), the two counter-propagating lasers interact strongly. The energy in the pump laser is transfered to the seed pulse and the plasma wave, whose amplitudes are greatly amplified. The color scale in each sub-figure is normalized.}
	\label{fig:BS}
\end{figure}

Having verified the code implementation, we can now apply it to cases where simple analytical solutions do not exist. As the first example, consider the situation where a large amplitude Gaussian pump laser is scattered by a counter-propagating Gaussian seed laser (Fig.~\ref{fig:BS}). In this example, the longer pump laser propagate at $v_1=1.0$ with initial maximum amplitude $a_1=10$, and the shorter seed pulse propagate at $v_2=-0.8$ with initial maximum amplitude $a_2=1$. When the coupling coefficient is small (upper panel, $\Gamma=0.1$), the two lasers pass through each other without much interaction, and leave behind an excited plasma wave whose group velocity is chosen to be $v_3=0.1$. On the other hand, when the coupling coefficient is large (lower panel, $\Gamma=1$), the two lasers interact strongly via the mediating plasma wave. A large fraction of the pump lasers is consumed to amplify the initially weak seed pulse. After the two lasers leave the interaction region, the seed pulse is amplified by orders of magnitude with a somewhat shortened durations. Moreover, the Gaussian pulse structure is altered, and the main pulse is now followed by a train of short pulses generated during the nonlinear interactions.

\begin{figure}[b]
	\centering
	\includegraphics[angle=0,width=0.73\textwidth]{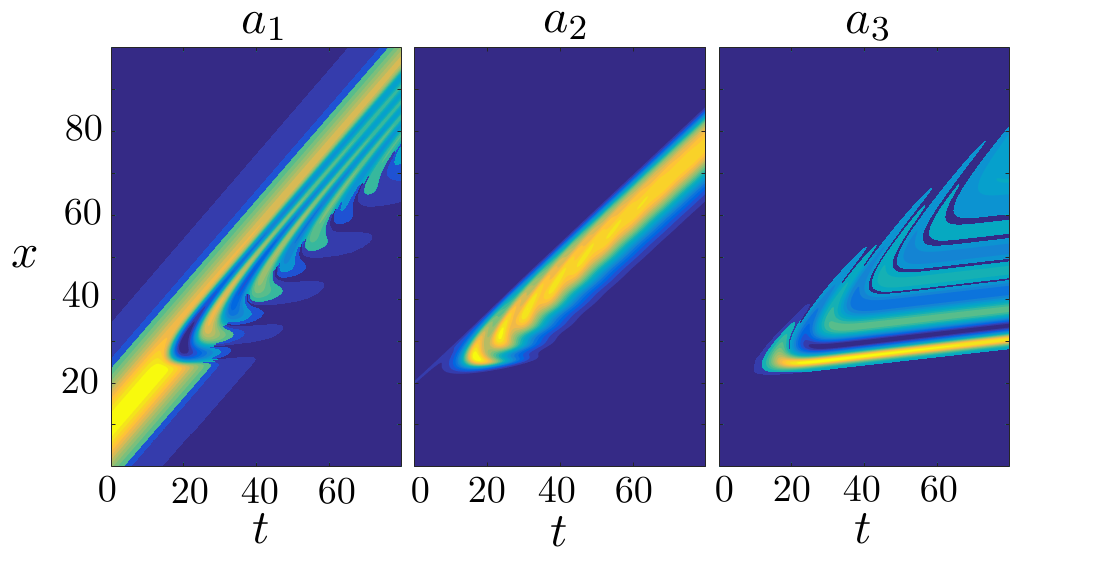}
	\caption[Numerical solution of forward scattering problem]{Forward scattering of a pump laser ($a_1$, left), catching up with a slower seed pulse ($a_2$, middle), during which a plasma wave ($a_3$, right) is excited. Although the coupling coefficient $\Gamma=0.2$ is small, the pump laser spends a long time passing the seed pulse. Consequently, after the pump finally passes the seed, a large fraction of energy is exchanged, leaving behind a wake of plasma waves. The color scale in each sub-figure is normalized.}
	\label{fig:FS}
\end{figure} 

As another example, consider the situation where the large amplitude Gaussian pump laser \mbox{($v_1=1.0$)} catches up with the seed laser ($v_2=0.8$) from behind, during which forward scattering happens (Fig.~\ref{fig:FS}). Since the two lasers propagate in the same direction with similar group velocities, they can spend a long time interacting with one another, compensating for a relatively small coupling coefficient $\Gamma=0.2$. Once the lasers start to overlap spatially, they interact and excite a plasma wave. However, the plasma wave has much smaller group velocity ($v_3=0.1$). Before the plasma wave has much time to grow, the two lasers already run ahead and leave the plasma wave behind. Since the front of the $a_1$ pulse never spends any time with the plasma wave, it advects with very little change. On the other hand, the back of the $a_1$ pulse travels through the excited plasma wave together with $a_2$, and thereby gets strongly modified by the three-wave interactions. In comparison, the seed pulse $a_2$, traveling at a slower group velocity than the pump, always spends some time with the two other waves. Therefore, ever since the seed get caught up by the pump, its envelope is altered from tail to head, until the entire pump passes through the seed. Finally, the slow plasma wave $a_3$ is generated and passed by both the pump and the seed lasers. 
The amplitude of the plasma wave oscillates, similar to what we have seen in the zero-dimensional problem.

\chapter{Application: laser pulse compression in magnetized plasmas\label{ch:compression}}

From numerical examples in the previous chapter, we see that plasma waves can mediate energy transfer between two lasers. In particular, energy stored in a long pump laser can be transfered via a plasma wave to a seed pulse, whose intensity is amplified and duration is shortened. 
The way pulse compression works has similarities to parametric amplification, where a nonlinear medium is used to couple three waves, whose interactions are governed by the three-wave equations. However, in parametric amplification, the third wave is usually an idler electromagnetic wave that leaves the system, whereas in pulse compression, the third wave is usually an electrostatic plasma wave that stays in the medium. The plasma medium absorbs the energy and momentum difference between the two lasers, and becomes excited to higher energy level after the interaction. This is analogous to laser amplification, where an atom is pumped to higher energy level, and then emit a photon when stimulated by another. However, pulse compression is not quite the same as laser amplification. Unlike in laser amplification, where the atom can be excited by the pump laser alone, in pulse compression, the plasma wave cannot be excited unless a seed, which can either be deliberately injected or spontaneously excited, is also present. Instead of amplifying the seed by stimulated emission, pulse compression amplify the seed by stimulated scattering of the pump laser.

In this chapter, I will analyze laser pulse compression as an application of three-wave interactions. Although pulse compression in unmagnetized plasmas has long been contemplated as a promising technique for producing powerful laser pulses beyond the attainment of other methods, my work is the first to identify that magnetized plasmas are more suitable media for pulse compression \citep{Shi17laser}. In this chapter, I will first review existing techniques for producing high-intensity short-wavelength lasers, and then argue why magnetized pulse compression expands the range of lasers that can be produced. By considering limiting effects, improvements enabled by magnetizing the plasma media will be estimated. Simple analytical estimations have since been supported by numerical simulations in collaboration with \cite{Jia17}, which is just a beginning of the new research direction of magnetized laser-plasma interactions.

\section{Why pulse compression and how does it work?\label{sec:compression:motivation}}

\subsection{Why is there a need for pulse compression?\label{sec:compression:motivation:why}}

Laser pulses of high intensity, high frequency, and short duration are demanded in many applications. One important application is in inertial confinement fusion \citep{Keefe82}, where powerful laser pulses are used to ignite fusion fuels. The ignition, achieved when the fuel is compressed and heated to fusion conditions, requires lasers of high intensity and high frequency. High intensity is needed in order to achieve strong ablation, which compresses a fusion fuel pellet by rocket-like blow off of its surface materials; high frequency is needed in order for lasers to be able to penetrate the plasma formed during the ablation, which will strongly reflect the lasers unless the laser frequency is above the plasma frequency. Since lasers of both high intensity and high frequency are not available \citep{Lindl92}, one has to trade frequency for intensity in the direct-drive mode, or vice versa in the indirect-drive mode. In the direct-drive mode \citep{Craxton15}, lasers are shone directly onto the fuel pellets to achieve compression. While the demand for high intensity can be readily met by current technologies \citep{Regan16,Goncharov17}, the requirement for high frequency is not satisfied. In an alternative approach, the indirect-drive mode \citep{Lindl95}, lasers are shone on a metal cylinder to radiate thermal X rays, which are then used to compress fuel pellets. While this mode easily meets the demand for high frequency \citep{Kauffman94,Glenzer11}, it struggles to meet the demand for high intensity, because the thermal conversion process is very inefficient. Up to now, neither the direct-drive mode nor the indirect-drive mode has succeeded in igniting fusion fuels in experiments \citep{Bose16,Betti15}. Therefore, either the lack of frequency in the direct-drive mode, or the lack of intensity in the indirect-drive mode must be addressed, before we can harness the energy produced by inertial confinement fusion.

Another area that demands next-generation lasers is material science, where laser pulses can be used to probe the dynamical structure of materials. This application demands intense laser pulses of high frequency and short duration, because the wavelength of the pulses set the length scale of the smallest structure that can be resolved, and the duration of the pulses set the time scale of the fastest dynamical process that can be captured. For example, to probe structures of biological molecules \citep{Kendrew58}, whose sizes are of the order of nanometer, the laser frequency needs to be in the X-ray range. When X-ray photons pass through the molecules, they are scattered by various structural components, which leave their fingerprints in the diffraction pattern of the X rays. Unfortunately, the diffraction signal from a single photon scattered off a single molecule is statistically insignificant, so either a large number of molecules or a large number of photons are required. 
To ensure that small signals interfere constructively, one can arrange a large number of molecules in identical state in the form of a crystal. This is a conventional and widely-used approach, called X-ray crystallography, which has been very successful and led to the discovery of DNA double helix \citep{Franklin53}, as well as the structure of a large number of proteins \citep{Abrahams94,Baranovskiy08}.
However, this conventional approach is intrinsically static, and cannot be used to probe functioning biological molecules, whose structures are dynamical. To capture the evolving structure of biological molecules, alternative approaches are needed. One approach takes snapshots of a single molecule using a large number of phase coherent photons in the form of X-ray laser pulses \citep{Neutze00,Chapman11}. To avoid motion blurring of snapshots, the duration of the X-ray pulses needs to be shorter than tens of femtoseconds, on which time scale molecules disintegrate due to radiation damage. While recent development in free electron lasers can already provide X-ray sources with enough intensity \citep{Ishikawa12}, the duration of X-ray pulses needs to be shortened by at least ten times \citep{Hau-Riege07} before we can utilize these pulses to probe the evolving structures of biological molecules.

The aforementioned applications demand high-intensity, high-frequency, and short-duration pulses that cannot be produced by laser sources directly. Therefore, after being produced, laser pulses need to be sent through post-processors that can amplify their intensity, convert their frequency, or reduce their duration. One type of laser processor is pulse amplifier, which intensifies a short seed pulse using the energy provided by either optical \citep{Moses05} or electrical \citep{Sethian95} pumps. Conventional amplifiers, using pumped solid or gas as gain media, can process limited intensity up to \mbox{$\sim10^{12}\;\text{W/cm}^2$}, above which the media become saturated or even damaged \citep{Stuart95,Canova07}. To amplify short pulses to even higher intensity, other techniques have to be used. The current state-of-the-art technique is Chirped Pulse Amplification (CPA), which can tolerate intensity up to \mbox{$\sim10^{14}\; \text{W/cm}^2$} \citep{Maine88}. This technique handles high intensity by first stretching the pulse to longer duration and lower intensity, such that it can be safely processed by conventional amplifiers. The amplified long pulse is then compressed, usually by a pair of gratings, to recover its original short duration. Processing even higher intensities using the CPA technique becomes impractical, because it requires scaling up the size of gratings, as well as the size of conventional amplifiers, whose linear dimensions are already on the order of meters \citep{Miller04}. Moreover, this technique is incapable of processing high-frequency lasers beyond the visible range. At higher frequencies, no solid-state amplifier is available, because host materials become opaque and there is no known impurity ion that can provide the necessary resonances \citep{Koechner13}. Gas media cannot provide high-frequency resonances either \citep{Davis14}, because a single photon whose frequency is above the soft UV range already has sufficient energy to ionize the gas. In addition to the lack of suitable amplifiers, there is also a lack of gratings needed at the compression stage of CPA, because gratings, made of solid-state materials, are too fragile to survive radiation damages induced by high-intensity and high-frequency lasers \citep{Canova07}. For these reasons, current amplifier technologies can only process laser pulses with unfocused intensities on the order of \mbox{$10^{14}\; \text{W/cm}^2$} and frequencies on the order of a few eV. 

To produce higher-frequency pulses, the commonly used techniques work by converting lasers of high intensity but low frequency to lasers of lower intensity but higher frequency. Laser frequency conversion can be achieved by harmonic generation inside nonlinear crystals \citep{Franken61,Ghimire11}, near solid surfaces \citep{Bloembergen62,Dromey06}, and in gaseous media \citep{McPherson87,Seres05,Gohle05}. In these techniques, intense lasers with fundamental frequency $\omega$ are used to pump media into the nonlinear regime to generate second harmonics with doubled frequencies $2\omega$, third harmonics with tripled frequency $3\omega$, and so on. Harmonic generation inside solid media can only produce high-frequency pulses up to the UV range, beyond which most crystals become opaque, with intensities limited by radiation damage to \mbox{$\sim10^{10} \;\text{W/cm}^2$} \citep{Wegner99}. The frequency and intensity limits are extended by harmonic generation near solid surfaces, where a reflective setup instead of a transmissive setup is used. In the reflective geometry, high-frequency harmonics propagate freely back into the vacuum, and the damaged surfaces keep on radiating harmonics even after being broken down to plasmas \citep{Carman81,Bezzerides82}. The surface plasma mirrors can radiate high-order harmonics up to the soft X-ray range when driven into the relativistic regime by $\text{EW/cm}^2$ pump lasers \citep{Tarasevitch07}. However, instead of selectively generating a single harmonic, a full spectrum of harmonics are generated with very low efficiency \citep{Zepf98,Ishizawa99}. Moreover, the conversion efficiency rolls off very rapidly with the harmonic order \citep{Linde96}, and becomes extremely low for high-oder harmonics, limiting the intensity of high-frequency pulses to \mbox{$\sim10^{10}\;\text{W/cm}^2$}. The situation is similar in gas media, where the harmonic spectrum rolls off exponentially for low-order harmonics in the driven-dipole regime \citep{Huillier92,Krause92} and reaches a plateau for high-order harmonics in the tunneling-recollision regime \citep{Corkum93}, before terminating at some cutoff frequency \citep{Krause92PRL}. When driven by intense lasers with $\text{PW/cm}^2$ intensity, the efficiency in the plateau region is on the order of $10^{-5}$ \citep{Kim05}, making the intensity of high-frequency pulses again on the order of \mbox{$10^{10}\;\text{W/cm}^2$}. Although a number of techniques is being developed to improve the efficiency \citep{Tamaki99,Bartels00,Shiner09}, frequency-conversion technologies using unselective harmonic generation will still be limited to \mbox{$\sim10^{12}\; \text{W/cm}^2$} in intensity and \mbox{$\sim 100$ eV} in frequency in the near future.    
\subsection{Unmagnetized Raman and Brillouin compressions\label{sec:compression:motivation:unmag}}

To go beyond limitations of the aforementioned techniques, current research contemplates using unmagnetized plasmas, a media that is already ionized and broken down, to amplify and compress short laser pulses \citep{Malkin99,Weber13}. In unmagnetized plasmas, pulse compression is mediated by either the Langmuir wave or the acoustic wave \citep{Forslund75}, the two eigenmodes of unmagnetized plasmas. When pulse compression is mediated by the Langmuir wave, the energy is transfered from the pump laser to the seed via stimulated Raman backscattering \citep{Milroy79,Capjack82}, during which the nonlinear response of electrons dominate. Alternatively, when pulse compression is mediated by the acoustic wave, the energy is transfered through stimulated Brillouin backscattering \citep{Guzdar96,Andreev06}, during which both electrons and ions contribute to the nonlinear response. 

Using Raman or Brillouin backscattering, the maximum pulse intensity can be reached when the most intense pump is used to amplify the seed for the longest time. The maximum intensity of the pump is limited by wavebreaking \citep{Malkin14E,Lancia10}, which happens when the amplitude of the plasma wave exceeds the threshold. Beyond the wavebreaking threshold, energy stored in the coherent wave motion is dissipated in random turbulent motion, and the collapsed plasma wave can no longer mediate energy transfer effectively. The other way of reaching high intensity, using the longest amplification time, is limited by plasma instabilities. When intense lasers propagate through plasmas, they suffer from numerous instabilities they excite, such as the modulational instability, filamentation instability and forward scattering instability \citep{Malkin99,Weber13}. Although these instabilities grow much slower than the seed pulse, they nevertheless compete with the pulse compression process and eventually limit the maximum time that the pulse can be amplified by the pump. These two limits, the maximum pump intensity and the longest amplification time, determines the maximum unfocused pulse intensity theoretically achievable in unmagnetized plasmas to $\sim10^{18}\; \text{W/cm}^2$, which is four orders of magnitude higher than the current industrial limit.   

Although pulse compression using unmagnetized plasmas can in principle produce very intense pulses, this technique is limited by the availability of suitable plasma targets of necessary density, temperature, and uniformity. High plasma density is required to avoid wavebreaking. Typically, to couple the pump and seed lasers effectively, the ratio of plasma wave frequency to the laser frequency needs to be on the order of $0.1$, which set the required plasma density to be $\sim10^{19}\; \text{cm}^{-3}$ and the required plasma temperature to be $\sim 10$ eV, for compressing lasers of $\sim 1$ eV frequency and $\sim10^{14}\; \text{W/cm}^2$ intensity. To produce such plasmas, the laboratory standard is to use a high-pressure hydrocarbon gas jet, which is ionized and heated by a separate laser \citep{Ping04,Cheng05,Ping09}. However, plasmas produced in such a way are usually very turbulent and nonuniform. The lack of uniformity jeopardizes resonant coupling between the pump and the seed laser, resulting in very pool efficiency of only a few percent, much lower than $>30\%$ efficiency predicted by theory and simulations \citep{Ping02,Yampolsky08,Lancia16}. Moreover, in order to compress lasers of higher frequency or intensity, plasmas of higher density and temperature are required, for which no technology is currently available. To produce plasma targets with density higher than $\sim10^{20}\; \text{cm}^{-3}$, a method using dense aerosol jet has been envisioned \citep{Hay13,Ruiz14}. However, reaching high temperature and sufficient uniformity with these targets is considerably more challenging and is yet to be demonstrated experimentally.

Beside engineering challenges, lasers that can be compressed using unmagnetized plasmas is theoretically limited to $\sim100$ eV due to wave damping \citep{Malkin07}. The lasers are damped due to inverse bremsstrahlung, during which the laser energy is transfered to particles whose motion is randomized by collisions. During this process, usable energy in the form of laser fields is converted to thermal energy that is ineffectual. 
This process competes with the desired energy transfer from the pump laser to the seed pulse and can even drain all the available energy when damping becomes strong. To reduce collisional damping in high-density plasmas, which are required to compress high-frequency lasers, one can heat up the plasma wherein the collision frequency is reduced. However, this manipulation unwittingly increases collisionless damping \citep{Landau46}, an additional mechanism through which the plasma waves are damped. Plasma waves are damped collisionlessly by loosing energy to particles that are trapped in the waves. Since more particles become trapped as plasma temperature increases, collisionless damping is increased as one tries to decrease collisional damping by heating up the plasma. These two damping mechanism squeeze out the operation window in the plasma density-temperature space when the laser frequency approaches $\sim100$ eV. Above this frequency, either collisional or collisionless damping becomes strong, so pulse compression in unmagnetized plasmas is not possible, even if technologies for making suitable plasma targets were available.

\subsection{Pulse compression in magnetized plasmas\label{sec:compression:motivation:mag}}

Now that strong magnetic fields start to become feasible (Ch.~\ref{ch:intro}), and we start to understand three-wave interactions in magnetized plasmas (Ch.~\ref{ch:fluid-2nd}), it is natural to ask whether magnetic fields can help extend the frequency and intensity range of laser pulse compression. It turns out, as we shall see, that the answer is yes. 

By applying an external magnetic field, many more waves, such as the MHD waves, the hybrid waves and the Bernstein waves, become available for mediating laser pulse compression. In resonances provide by these waves, contribution from plasma density and temperature are partially replaced by the contribution from magnetic fields. Such a replacement reduces the dependencies on internal plasma parameters, and allows the use of external fields to control the performance of pulse compression. Consequently, the engineering flexibility is increased in the optical and UV range using magagauss magnetic fields, and the operation window is expanded into the X-ray regime whenever fields of several gigagauss become available. Although gigagauss fields are challenging, magnetized plasmas at least provide a theoretical opportunity to compress lasers that otherwise could not be compressed at all.  
  
\begin{figure}[t]
	\centering
	\includegraphics[angle=0,width=0.4\textwidth]{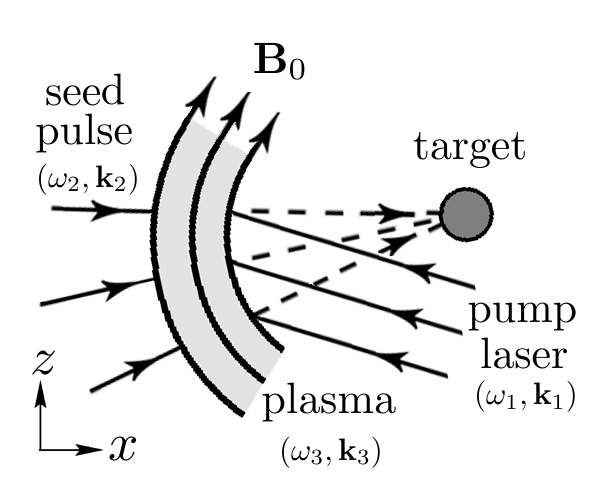}
	\caption[Schematics of magnetized pulse compression]{Amplifying and focusing a seed pulse by stimulated backscattering of a pump laser in magnetized plasma.}
	\label{fig:schematics}
\end{figure}

In what follows, I will examine pulse compression mediated by the upper-hybrid (UH) wave as an example. The UH wave is just one of the many waves that can be utilized for pulse compression in magnetized plasmas. While analyzing all viable waves is beyond the scope of this thesis, the UH-wave example suffices to demonstrate that external magnetic fields are beneficial for pulse compression. 
The UH wave is the cold limit of the lowest-order electron Bernstein wave propagating perpendicular to the background magnetic field (Sec.~\ref{sec:fluid-1st:dispersion:perp}). 
The setup that the magnetic field is nonparallel to the direction of laser propagation is different from what has been considered by \cite{Vij16,Shoucri16,Luan16}, and lends itself naturally to the main application where the amplified pulse is focused onto a distant target (Fig.~\ref{fig:schematics}). 
During three-wave interactions, the mostly transverse lasers can be polarized either in the X mode or the O mode. When both the pump and the seed are in the same mode, they interact strongly through the UH wave with a coupling coefficient that is indifferent to the laser polarization (Sec.~\ref{sec:fluid-2nd:coefficient:TTL}). The UH wave is an almost longitudinal wave with frequency $\omega_{UH}\simeq\sqrt{\omega_p^2+\Omega_e^2}$, where the external magnetic field partially replaces the role of plasma density in the three-wave resonance conditions. In other words, suppose we are given a pump laser and a seed laser, and the task is to find a plasma target to mediate their resonant interactions. Then, the plasma density required to match the resonance conditions can be reduced if we apply a magnetic field transverse to the direction of laser propagation.

The reduction of the required plasma density has immediate engineering benefits. First, challenging technology for producing high-density plasmas can now be substituted by available technologies for generating strong magnetic fields (Ch.~\ref{ch:intro}). The plasma density required to compress 1-$\mu$m pulses using unmagnetized plasmas is $\sim10^{19}\; \text{cm}^{-3}$, which is already at the verge of what is feasible with gas jet plasmas. To compress shorter wavelength lasers using unmagnetized plasmas, denser plasma targets, such as foams and aerosol jets, remain to be developed. Allowing dense plasmas to be replaced by magnetic fields thus relaxes the engineering challenges. Second, uniformity of the plasma target becomes more controllable when magnetic fields supply the resonance frequency. While it is difficult to control the internal plasma density, adding an external magnetic field introduces an extra control variable, which may be adjusted to maintain the three-wave resonant conditions, and tune the performance of laser pulse compressors.

Moreover, UH mediation gives relative advantage to the pulse compression process over competing effects. When the UH wave ($a_3$) mediates resonant energy transfer between a given pump laser ($a_1$) and a given seed pulse ($a_2$) , the lower plasma density results in a slower linear growth rate. Using Eq.~(\ref{eq:MTUH}), the growth rate for backward scattering $\gamma_0\simeq\sqrt{\omega_3\omega_0}|a_1|/2\gamma_3$, 
where $\omega_0\simeq\omega_{1,2}\gg\omega_{UH}$ is the laser frequency, and $\gamma_3=\omega_{UH}/\omega_p>1$ is the electron magnetization factor. 
Since UH-wave mediation has smaller growth rate, it takes longer time, and equivalently longer pump laser and plasma lengths, to achieve the same compression of the seed pulse. However, what is of critical importance is that although the amplification rate is reduced, the damping rates and growth rates of competing instabilities are reduced even more, as we shall analyze in details in the next section.

Without the competing effects, laser pulse compression mediated by the UH wave works in the same way as Raman compression \citep{Malkin99}, except for a different growth rate.  
At the linear stage of the amplification, the pump amplitude $a_1=a_{10}$ is approximately constant. The solution to the linearized three-wave equations describes an exponential growth and broadening of the seed pulses, with linear growth rate $\gamma_0=K |a_{10}|$. 
When pump depletion becomes significant, the interaction enters the nonlinear compression stage. At this stage, a self-similar attractor solution exists. Using the Lamb transformation, the wave envelopes $a_1\simeq a_{10}\cos(u/2)$, $a_2\simeq-\partial_\xi u/K\sqrt{2}$, and $a_3\simeq\sqrt{2}a_{10}\sin(u/2)$, where $u(\xi\tau)$ satisfies the sine-Gordon equation $\partial_\xi\partial_\tau u=\sin u$, with $\tau:=\gamma_0^2t$ and $\xi:=x/c+t$. For $u(0)\ll1$, the self-similar solution grows from its initial value and then oscillates about $\pi$, forming the so-called $\pi$-pulse solution. 
Transforming into the self-similar coordinate $z:=2\sqrt{\xi\tau}$, the equation becomes $\partial_z^2 u+\partial_z u/z=\sin u$. The solution to this equation describes a pulse train formed at the nonlinear stage of the compression. The leading spike amplitude $a_2\propto t$ grows linearly with time, while its duration $\Delta t_2\propto 1/t$ shrinks linearly with time, where the proportionality constants depend on initial conditions. Suppose the initial conditions of the three-wave equations are such that the solution enters the nonlinear stage after $\Lambda_0$ linear exponentiations, then the pulse intensity and duration
\begin{eqnarray}
\label{eq:I2}
I_2&\simeq& 2I_{10}\Big(\frac{2\gamma_0 t}{\Lambda_0}\Big)^2, \\
\label{eq:dt2}
\Delta t_2&\simeq&\frac{\Lambda_0}{\gamma_0^2 t}.
\end{eqnarray}
In the absence of competing instabilities, this nonlinear compression process will continue until the seed pulse transits the entire length of the pump laser, and emerges as a train of amplified pulses with shortened durations. 

From the experimental standpoint, pulse compression works by counter-propagating the pump laser and the seed pulse in a suitable plasma target. Denote $t_M$ the maximum amplification time, then we can prepare a plasma target whose length is $L\simeq t_Mc$. At time $t=-t_M$, the pump laser enters the plasma from its right boundary, and by time $t=0$ the front of the pump laser reached the plasma left boundary. At that moment, we let the seed pulse enter the plasma from left, whereby the pulse compression process starts. The seed pulse counter-propagates with the pump laser through the plasma target, and is amplified and then compressed by three-wave interactions. The pulse compression process continues until $t=t_M$, at which point the compressed seed pulse has traversed the entire length of the plasma target and exits the plasma from its right boundary, where the tail of the pump laser has just entered.

\section{Limitations due to competing effects\label{sec:compression:limits}}

The above physical picture of pulse compression is based on the assumption that no other process occurs other than the three-wave interactions we demand. However, in reality, a number of effects, which we have ignored when deriving the three-wave equations, may turn out to be competitive with three-wave interactions. In this section, I will analyze the most competitive effects identified by \cite{Malkin99} for unmagnetized compressions, with emphasis on the new ingredient of a transverse background magnetic field.

\subsection{Instabilities: plasma wave and EM wave \label{sec:compression:limits:instabilities}}

The first category of competing effects are instabilities that we have ignored when using perturbation theory to solve for three-wave interactions. Within this category, two effects are identified as the most competitive for the unmagnetized scenario. The first effect is wavebreaking of the plasma wave, which limits the maximum pump intensity. The second effect is the modulational instability of the EM waves, which limits the maximum amplification time. Although these effects are not necessarily the most competitive in the magnetized scenario, as we will see later from PIC simulations, here I will focus on analyzing how magnetic fields change these two effects.

\subsubsection{Breaking of the plasma wave}

When deriving the three-wave equations, we have assumed that the waves are linear in the absence of three-wave interactions. This assumption fails when the amplitude of the wave becomes large. In particular, the amplitude of the plasma wave becomes nonperturbative if the wave electric field becomes comparable to the wavebreaking field $E_b=m_ec\omega_p/e$ (Sec.~\ref{sec:fluid-1st:model:expansion}). At the beginning of the pulse compression process, the plasma wave grows from zero amplitude, where the linear wave approximation is well justified. However, as the seed pulse getting amplified and compressed, it will grow a larger-amplitude plasma wave within a shorter period of time.
However, the exact growth rate of the plasma wave is determined by the pump amplitide. With a small pump amplitude, even a highly compressed seed takes a long time to generate a large-amplitude plasma wave. On the other hand, with a large pump amplitude, even a weak seed can quickly produce a plasma wave whose amplitude exceeds the wakebreaking limit. Therefore, the requirement that plasma wave remains unbroken primarily constrains the maximum pump intensity that can be employed for pulse compression.

In magnetized plasmas, the wavebreaking intensity is modified by the Lorentz force, and the UH wave breaks when the electron quiver velocity in the $\mathbf{k}_3$-direction $v_{q}$ exceeds the wave phase velocity $v_p$. This condition can be derived rigorously by solving the simplified nonlinear fluid equations \citep{Karmakar16}, which shows that the gradient of the density perturbation becomes infinite at the wavebreaking limit. This is intuitive because when particles quiver at a faster longitudinal velocity than the wave phase velocity, particles will outrun the wave, causing the density to pile up. Using the first-order linear response [Eq.~(\ref{eq:vwave})], the longitudinal quiver velocity of electrons in the UH wave is
\begin{equation}
v_q\simeq\frac{eE_3\omega_{UH}}{m_e\omega_p^2}=\gamma_3 a_3 c,
\end{equation}
where $\gamma_3=\omega_{UH}/\omega_p>1$ is the electron magnetization factor.
The plasma-wave amplitude $a_3$ that appears in the three-wave equation is properly normalized by Eq.~(\ref{eq:ak}), where I have used the UH wave energy coefficient [Eq.~(\ref{eq:u_coef_UH})]. While the quiver velocity is modified, the phase velocity of the plasma wave is the same as in the unmagnetized case. In the backscattering geometry, the phase velocity
\begin{equation}
v_p=\frac{\omega_3}{k_3}\simeq\frac{c\omega_3}{2\omega_0},
\end{equation}
where $\omega_0\simeq\omega_{1,2}$ is again the laser frequency. Then, the condition that wavebreaking does not happen can be expressed in terms of the plasma wave amplitude as
\begin{equation}
a_3\lesssim \frac{\omega_3}{2\omega_0\gamma_3}.
\end{equation} 
We see for fixed pump and seed lasers, a smaller amplitude plasma wave is allowed before the wavebreaking threshold is reached, when we replace plasma density with a transverse background magnetic field. 
 
If wavebreaking were truly a limiting effect, then it constrains the maximum allowable pump laser intensity. The condition $v_{q}\lesssim v_p$, which guarantees that the UH wave remains unbroken, can be rewritten in terms of a constraint on the pump intensity $I_1=8I_c|a_1|^2$, where  $I_c=n_cm_ec^3/16$, and $n_c=\epsilon_0m_e\omega_0^2/e^2$ is the critical density. Using the Manley-Rowe relation for three-wave interactions, suppose all the wave action $\mathcal{I}_1\propto \omega_1a_1^2$ of the pump is converted to the wave action $\mathcal{I}_3\propto \omega_3a_3^2$ of the plasma wave, then the maximum amplitude of the UH wave $|a_3|\le\sqrt{\omega_0/\omega_3}|a_1|$. Therefore, a sufficient condition that the UH wave remains unbroken is that the pump intensity
\begin{equation}\label{eq:WaveBreaking}
	I_1\lesssim I_c\Big(\frac{\omega_3}{\omega_0}\Big)^3\gamma_3^{-2}.
\end{equation}
Here, a factor of two is canceled since the pump is of twice the length of the plasma. When more plasma density is replaced by magnetic field in $\omega_3$, less number of particles remain to carry the energy of the UH wave, giving rise to the $\gamma_3^{-2}$ reduction.

Although wavebreaking is a condition that the linear wave approximation breaks down, it does not necessarily mean that pulse compression cannot happen in the wavebreaking regime. In fact, even in the unmagnetized scenario, it is viable to use a pump laser whose intensity exceeds the wavebreaking threshold \citep{Yampolsky08}. 
This is because in the counter-propagating geometry, the pump and the seed continue to encounter in a fresh region of the plasma. The large amplitude plasma wave excited previously is left behind, and a new mediating plasma wave always needs to be grown from zero amplitude in the interaction region. 
Within the interaction region, the plasma wave, which is yet to be broken, can efficiently mediate energy transfer, even when the pump intensity exceeds the wavebreaking threshold. The overall amplification efficiency remains high, until the pump intensity far exceeds the threshold \citep{Toroker14,Edwards15}. At that point, the plasma wave quickly reaches the wavebreaking amplitude within the interaction region. 
This situation is detrimental in unmagnetized plasmas, because once the plasma wave reaches wavebreaking, the infinite density gradient collapses and the plasma wave looses its coherence. Since a well-defined mediating wave no longer exists, energy transfer from the pump to the seed is thereof impeded. 
However, in magnetized plasmas, the plasma wave remains coherent even when its amplitude exceeds the wavebreaking threshold \citep{Jia17}. This is because the transverse background magnetic field provides an additional restoring force. As the density piles up in the plasma wave, the wave electric field crosses the background magnetic field to generate a large $\mathbf{E}\times \mathbf{B}$ shear on the wavefront. The strong shear reduces the density gradient and stabilizes the large-amplitude wave, so that a coherent mediating wave persists well beyond the wavebreaking limit, as we shall see later from the PIC simulations.

\subsubsection{Relativistic modulational instability}

Similar to the plasma wave, the EM waves also suffer from instabilities when their amplitudes become large. One of the fastest growing instability is the modulational instability, whereby small irregularities on the wave envelope exacerbate. The modulational instability can happen for many reasons, and the most important reason for intense lasers in tenuous plasmas is the relativistic nonlinearity. Relativistic effects increase the effective mass of particles, when they oscillate transversely in a large amplitude EM wave. Consequently, the plasma frequency becomes smaller where the EM wave is stronger. Since the group velocity of the EM wave is negatively correlated with the plasma frequency, the EM pulse propagates faster where the pulse is stronger. This nonlinear effect causes modulations on the EM wave envelope to grow in the longitudinal direction, leading to the relativistic modulational instability. Moreover, it causes self-focusing of the EM wave in the transverse direction, where initially small inhomogeneities can grow into pronounced filaments. 

The growth rate of modulational instability can be estimated using the following heuristic arguments. First, for a high-frequency EM wave, the group velocity is close to the vacuum speed of light. To the next order, using dispersion relations when the wave propagates perpendicular to the background magnetic field (Sec.~\ref{sec:fluid-1st:dispersion:perp}), the group velocity of the O wave $v_g/c=1-\omega_p^2/2\omega_0^2$, 
and the group velocity of the X wave $v_g/c=1-\omega_p^2/2\omega_0^2+3\omega_p^2\Omega_e^2/2\omega_0^4-\omega_p^2/8\omega_0^4+O(\omega_*/\omega_0)^6$, 
where $\omega_*=\omega_p$ or $\Omega_e$ is the frequency scale of the plasma. In the limit $\omega_0\gg\omega_*$, the X wave $v_g$ is well approximated by the much simpler expression for the O wave $v_g$, which is independent of the background magnetic field. 
Second, due to relativistic effects, both the plasma frequency $\omega_p^2\simeq e^2n_{e0}/\epsilon_0m_e\gamma_q$ and the gyro frequency $\Omega_e=eB_0/m_e\gamma_q$ are decreased by the relativistic factor $\gamma_q=1/\sqrt{1-\beta_q^2}$, where $\beta_q^2=\langle v_q^2/c^2\rangle$ is due to the quiver motion. In weakly relativistic EM waves, $\gamma_q\simeq1+\beta_q^2/2$. Using the linear velocity response [Eq.~(\ref{eq:vwave})], the averaged quiver velocity
\begin{equation}
\beta_q^2\simeq\frac{a^2}{2}+O\Big(\frac{\omega_*}{\omega_0}\Big)^2,
\end{equation} 
where $a$ is the normalized amplitude of the EM wave. Since the characteristic plasma frequencies now depend on the wave amplitude, so does the group velocity. The $v_g$ of an infinitesimal EM wave differs from the $v_g$ of a finite amplitude EM wave by
\begin{equation}
\frac{\Delta v_g}{c}\simeq\frac{a^2}{8}\frac{\omega_p^2}{\omega_0^2}+O\Big(\frac{\omega_*}{\omega_0}\Big)^4.
\end{equation}
Suppose the envelope of the EM wave has some modulations, then bumps on the envelope that have larger $a$ will propagate at faster velocities than the dips on the envelope that have smaller $a$. Consequently, the peaks will outruns the troughs, causing the wave envelope to break up. The growth rate of the modulational instability may be estimated by
\begin{equation}
\Gamma_M\simeq\frac{\Delta v_g}{\lambda}\simeq\frac{\omega_3^2}{8\omega_0}a^2\gamma_3^{-2},
\end{equation}  
where $\lambda=c/\omega_0$ is the vacuum wavelength of the EM waves. The above estimation is of course very crude, but it suffice to give an idea for the modulational instability as a limiting effect. Notice that while the amplification rate $\gamma_0$ is reduced by $\gamma_3>1$ when plasma density is replaced by the background magnetic field, the growth rate of the modulational instability is reduced by an additional factor of $\gamma_3$. Therefore, the magnetic field gives pulse compression a relative advantage.

The maximum time that pulse compression can occur uninterrupted is limited to a few inverse growth rates of the modulational instability. It is a straightforward calculation to check that the amplification rate $\gamma_0\gg\Gamma_M$ even at the wavebreaking intensity. Therefore, the modulational instability does not prevent the amplification from reaching the nonlinear stage. Since $\gamma_0\propto|a_1|$ while $\Gamma_M\propto|a_1|^2$, we can always reduce the pump intensity such that the demanded three-wave interactions prevail over the unwanted modulational instability. On the other hand, for the seed pulse, if we want to achieve high final output intensity, then we cannot get away with its modulational instability. Adding the lowest-order relativistic nonlinearity $a_2^3$ into the three-wave equations \citep{Malkin14saturation}, the leading spike intensity saturates after
\begin{equation}
\label{eq:tM}
t_M\simeq(12\delta\Lambda_0^2)^{1/3}\frac{\gamma_3^{4/3}}{\omega_3a_{10}^{4/3}},
\end{equation}
where $\delta=\int \Gamma_Mdt \sim 1$ is the accumulated phase shift, $\Lambda_0$ is the number of linear exponentiations before the nonlinear stage is reached, and $a_{10}$ is the initial pump amplitude. The largest pulse compression is attained at the maximum compression time $t_M$, which gives the highest leading spike intensity $I_2\approx 16I_c (3\delta/\Lambda_0)^{2/3} (2a_{10})^{4/3}\gamma_3^{2/3}\omega_0/\omega_3$ and the shortest spike duration
$\Delta t_2\approx 2(2\Lambda_0/3\delta)^{1/3}a_{10}^{-2/3}\gamma_3^{2/3}/\omega_0$. Ramping up the pump intensity while keeping plasma parameters fixed, the maximum output intensity is reached using the most intense pump allowed by wavebreaking, which gives $I_2\le16I_c(3\delta/2\Lambda_0)^{2/3}\gamma_3^{-2/3}\omega_3/\omega_0$. Alternatively, optimizing plasma parameters while keeping lasers fixed, the maximum output intensity is reached using the smallest possible $\omega_3$ allowed by wavebreaking, which gives $I_2\le 8 I_c(3\delta a_{10}/2\Lambda_0)^{2/3}$, independent of $\gamma_3$.  
Notice that this output intensity could have been achieved using unmagnetized plasmas, if wavebreaking and longitudinal modulational instability were the only limiting effects.

\subsection{Damping: collisional and collisionless\label{sec:compression:limits:damping}}

The second category of competing effects are wave damping. Using the ideal cold-fluid model, we ignored collisions between particles that constitute the fluid and thereof ignored collisional damping of waves in plasmas. Moreover, the cold-fluid model does not capture kinetic effects. Since plasma particles can have a distribution of velocities instead of a single fluid velocity,  the same wave is Doppler-shifted by different amounts when seen by particles with different velocities. As a consequence, the phase mixing of the particles' response leads to collisionless Landau damping of plasma waves. In this section, I will discuss these two damping mechanisms in magnetized plasmas.

\subsubsection{Collisional damping}
Collisional damping occurs for both the EM waves and the UH wave. When collisions happen, the coherent quiver motion of charged particles is randomized, and the wave energy that would otherwise just oscillate between fields and particles are now irreversibly lost. Consequently, the plasma is heated up and the waves are damped. Within the cold-fluid framework, this collisional damping phenomenon may be modeled by adding Drude-type collision terms into the momentum equations. However, for an order-of-magnitude estimate, here it is sufficient to use a heuristic argument. 

The dominant contribution comes from collisions between different species, during which current is dissipated. On the other hand, collisions between the same species does not change current, so their contributions to wave damping can be ignored to the lowest order. In an electron-ion plasma, the intra-species collision frequency is
\begin{equation}
\nu_{ei}\simeq n_{0}\frac{Z^2e^4\Lambda}{4\pi\epsilon_0^2m_e^2v^3},
\end{equation}
where $n_0$ is the plasma density, $Z$ is the ion charge, $\Lambda$ is the Coulomb logarithm, and $v$ is the characteristic velocity of \textit{e--i} collisions. In a classical plasma where Fermi statistics is insignificant, the characteristic velocity contains contributions from both the thermal motion $v_T$ and the quiver motion $v_q$. To obtain an upper bound of the collision frequency, we can ignore wave motion, and take $v\sim v_T$ up to some constants. The Coulomb logarithm $\Lambda$ may be adjusted to account for effects of Fermi degeneracy and magnetization, but to lowest order, the collision frequency is proportional to density.

To determine the damping rate of the EM waves due to \textit{e--i} collisions, consider the following simple estimation. Suppose the quiver motion of an electron is fully randomized by a single \textit{e--i} collision, then the amount of wave energy carried by the electron is completely lost. However, the wave energy carried by electric and magnetic fields still remains. Therefore, the wave only loses the kinetic fraction $u_V$ of its total energy $u$. During the inverse Bremsstrahlung, the energy damping rates of the pump laser ($a_1$) and the seed laser ($a_2$) can be estimated by
\begin{equation}
\nu_{1,2}\simeq\nu_{ei}\frac{u_V}{u}\simeq\nu_{ei}\frac{\omega_p^2}{2\omega_{1,2}^2},
\end{equation}  
where I have used Eq.~(\ref{eq:uk}) for the wave energy coefficient, with the assumption that the laser frequencies are much higher than the characteristic frequencies of the plasma. Notice that the inverse-Bremsstrahlung damping rate is proportional to $n_0^2$ and thereof decreases rapidly when the plasma density decreases.

The collisional damping rate of the quasi-longitudinal plasma wave can be estimated similarly. Again, suppose the wave kinetic energy is completely randomized by a single \textit{e--i} collision, then using Eq.~(\ref{eq:u_coef_UH}), the collisional damping rate of the UH wave ($a_3$) is
\begin{equation}
\nu_{3c}\simeq\nu_{ei}\frac{u_V}{u}\simeq\nu_{ei}\Big(1-\frac{\omega_p^2}{2\omega_{UH}^2}\Big).
\end{equation}
When plasma density is replaced by the magnetic field in UH-mediated pulse compression, not only is the collision rate $\nu_{ei}\propto n_0$ reduced, but the fraction $\omega_p/\omega_{UH}$ is also reduced, resulting in a rapid decrease of the collisional damping rate of the mediating plasma wave. 

In order for pulse compression to work, collisional damping must be subdominant when compared to three-wave interactions. In particular, the plasma wave must persist until the lasers traverse the interaction region. Denote $\Delta t_2$ the duration of the seed pulse, which is roughly the interaction time between the pump and the seed, then damping of the plasma wave must be weak enough such that
\begin{equation}
\label{eq:collision3}
\nu_{3c}\Delta t_2\lesssim 1.
\end{equation}
Moreover, the lasers must be able to penetrate the plasma with little energy loss. Since the EM waves need to propagate through the plasma by the maximum amplification time $t_M$, the lasers need to survive collisional damping on this time scale
\begin{equation}\label{eq:collision12}
\nu_{1,2}t_M\lesssim 1.
\end{equation}
When plasma density is replaced by the background magnetic field, collisional damping is reduced, resulting in higher pulse compression efficiency. Using Eq.~(\ref{eq:tM}) for the maximum amplification time $t_M$, the collisional damping of the UH wave and the EM waves are alleviated by $\gamma_3^{-4/3}$ and $\gamma_3^{-8/3}$, respectively. When less amplification time is used, the pulse duration becomes longer, so the constraints become more strict for the UH wave while less strict for the lasers.

\subsubsection{Collisionless damping}

Collisionless damping in magnetized plasmas is a controversial subject. Numerous regimes exist where the wave frequency, the gyro frequency, and the trapping frequency are ordered differently. It is not the goal of this thesis to clarify controversies in the literature. For the purpose of pulse compression, it is suffice to notice that collisionless damping is due to phase mixing. Therefore, to lowest order, the damping rage $\nu_{3l}\propto n_0$ decreases when the density of the plasma is replaced by the background magnetic field.

To see why collisionless damping becomes very complicated when a transverse magnetic field is present, let me first briefly review collisionless damping in unmagnetized plasmas. By solving the linearized Vlasov-Poisson's equation, \cite{Landau46} computed damping rate as the imaginary part of the wave frequency, which arises when averaging the distribution function over the Doppler pole using the Landau contour. Beyond the linear theory, collisionless damping is a robust phenomenon due to phase mixing, whereby regularity is transfered from the spatial electric field to the velocity space trapped particles \citep{Mouhot11}. Without using these formal mathematical treatments, the damping rate can be simply estimated using the following heuristic argument. 

Consider a test particle moving in a prescribed electrostatic wave with $E(x,t)=E_0\cos(kx-\omega t)$. The Lagrangian of the test particle is $L=m\dot{x}^2/2-e\phi$, where $\phi$ is the electrostatic potential. Using normalized variables, 
the Hamiltonian can be written as $\varepsilon=\dot{\xi}^2/2-r^2\sin \xi$, where $\xi=kx-\omega t$ is the wave coordinate, and $r=v_t/v_p$ is the ratio of the trapping velocity $v_t=\sqrt{eE_0/mk}$ over the phase velocity $v_p=\omega/k$. When the test particle has energy $\varepsilon>r^2$ in the co-moving frame, it has enough kinetic energy to overcome the potential barrier. On the other hand, when $\varepsilon<r^2$, the test particle does not have enough kinetic energy and becomes trapped in a potential well. Changing coordinate back to the lab frame, there exists some $\Delta v\sim v_t$, such that particles with velocity $|v-v_p|<\Delta v$ are trapped particles, while particles with $|v-v_p|>\Delta v$ are passing particles. 

In a plasma, instead of a single particle, there exists a velocity distribution of particles, whose averaged response determines the wave behavior. Suppose the electrostatic wave can be somehow setup, then for passing particles, their averaged velocity $\langle v^2\rangle=v^2$ is unchanged by the wave. Therefore, passing particles do not exchange energy with the wave on average. In contrast, for trapped particles, their averaged velocity $\langle v^2\rangle=v_p^2$. Therefore, on average, a trapped particle with initial velocity $v>v_p$ looses energy to the wave, while a trapped particle with $v<v_p$ gains energy from the wave. By energy conservation, the wave damps if there are more energy-gaining particles than energy-losing particles, whereas the wave grows in the opposite case. Denote $f(v)$ the one-dimensional distribution function with $\int dvf(v)=n_0$. When $\Delta v\ll v_p$, the energy exchange $\Delta U\sim mv_pf'(v_p)(\Delta v)^3$ occurs on the trapped particle bouncing time scale $\Delta t\sim m\Delta v/eE_0$. Therefore, the wave damping rate is approximately $\nu\sim\omega_p v_p^2 f'(v_p)$. This estimation gives the Landau damping rate up to some order-unity numerical constants.

Now let us add a transverse background magnetic field to the test particle picture. Suppose the magnetic field is in the $z$-direction, and the UH wave propagates in the $x$-direction, then in the Landau gauge, the Lagrangian of the test particle is $L\propto(v_x^2+v_y^2)/2+\Omega x v_y+v_t^2\sin(kx-\omega t)$, where $\Omega$ is the gyro frequency and $v_t$ is again the trapping velocity. Unlike the unmagnetized case, the effective potential now depends on the particle's velocity. By the translational symmetry in the $y$-direction, one Euler-Lagrange equation yields a constant of motion $u=v_y+\Omega x$. The other equation can be written as $\ddot{\chi}=-R^2\chi+r^2\cos(\chi-\tau)$, 
where the normalized coordinates $\chi=kx+\omega u/\Omega v_p$ and $\tau=\omega t+\omega u/\Omega v_p$. Since there are three characteristic frequencies in the problem, the behavior of the equation is governed by two dimensionless ratios, namely, the ratio of gyro frequency over the wave frequency $R=\Omega/\omega$, and the ratio of the trapping frequency over the wave frequency $r=v_t/v_p$. In addition to having an extra parameter, the magnetized case is substantially more complicated for two reasons. First, due to the $\mathbf{E}\times \mathbf{B}$ drift, there is no simple coordinate transform, such as $\xi=\chi-\tau$, by which the force can be made time-independence. Second, the relative phase between the gyro motion and the wave motion is important, and the initial conditions can dramatically affect the particle's behavior, even when the initial velocities are the same. Consequently, there are many different types of chaotic orbits, and the particle can hop between these orbits with a slight change of initial conditions \citep{Dodin2011surfatron}. Thus, we are unfortunately stuck at just the first step of a heuristic estimation. 
  
Many attempts have been made in the literature to circumvent the aforementioned difficulties. A naive linear kinetic calculation using unperturbed gyro orbits in the velocity space integral yields exactly zero damping rate when wave propagate perpendicular to the background magnetic field \citep{Stix92}. While this result might be a reasonable approximation when $R\gg r$, it is invalid in the opposite case \citep{Sukhorukov97}, because in the limit $B_0\rightarrow 0$ the result fails to recover the nonzero Landau damping rate. Therefore, it is necessary to take both wave motion and gyro motion into account. 

In the regime gyro motion dominates wave motion, collisionless damping can happen due to stochastic heating \citep{Karney78,Karney79}. In this regime, a gyrating particle, whose perpendicular velocity $v_\perp>v_p$, receives small random kicks from the wave at two points along its gyro orbit, where the particle is in Landau resonance with the wave. When the kicks are small, particle's diffusion in the energy space is well correlated with its diffusion in the configuration space. On the other hand, larger kicks destroy the phase of gyration, making the particle's motion stochastic. 
The stochastic motion occurs within a window in the phase space, wherein a plateau of the distribution function is formed.
During the formation of the plateau, if the distribution function is such that more particles gain energy than loss energy, then the wave damps in the same way as in the unmagnetized case, except now the time scale for the plateau formation is the stochastic diffusion time. Assuming that good estimations for the stochastic window and the diffusion coefficient can be obtained, the wave damping rate can then be estimated.

In the opposite regime, where the wave motion dominates the gyromotion, and collisionless damping can happen due to surfatron acceleration \citep{Sagdeev73,Dawson83}. In this regime, magnetic field is considered as an initially-small perturbation to the unmagnetized picture. In the unmagnetized picture, resonant particles are trapped in the wave, and bounce back and forth along the $E_x$ direction in the wave trough. In the co-moving frame, which moves across the transverse background magnetic field $B_z$ at wave phase velocity $v_p$, the trapped particles see an electric field $E_y=v_pB_z$, which accelerates trapped particles along the wave front. Due to this secular surfatron acceleration in the $y$-direction, trapped particles quickly gain energy. After a few bounces, the $v_y\times B_z$ force becomes sufficient for particles to overcome the $E_x$ barrier and thereafter become untrapped. In other words, the surfatron acceleration mechanism converts trapped particles with averaged velocity $v_p$ to passing particles with averaged velocity $\sim E_x/B_z$. When the magnetic field is weak, this conversion results in a large energy gain for the particles, and consequently a strong damping of the wave.

Estimating the damping rate is very difficult even in the above two regimes where the physical picture is relatively clear. Here, as a very rough estimation, notice that the UH wave frequency is typically comparable to the gyro frequency. 
Hence an electron having perpendicular velocity close to $v_p$ sees an almost-constant wave electric field. 
In such an electric field, the electron may gain or loss energy to the wave, depending on the relative phase of wave motion and gyro motion. 
The phase mixing process causes the UH wave to damp on a Maxwellian background with rate $\nu_L\sim\sqrt{\pi}(v_p/v_T)^3\exp(-v_p^2/v_T^2)\omega_p^2/\omega_3$, where $v_T$ is the thermal velocity. 
Since linear wave requires $v_p>v_T$, the sufficient condition that collisionless damping is weak may be approximated as
\begin{equation}
\frac{\nu_{3L}}{\omega_3}\approx\sqrt{\pi}(3/2)^{3/2}e^{-v_p^2/v_T^2}\gamma_3^{-2}\ll 1.
\end{equation}
As $\omega_3\rightarrow|\Omega|$, the electron density vanishes, so there are fewer electrons to participate in the phase mixing, and collisionless damping thereof becomes smaller.

\section{Operation windows\label{sec:compression:window}}

The limiting effects define an operation window, within which efficient pulse compression is theoretically possible. In unmagnetized plasma, the four limiting effects discussed in the previous section give the tightest constraints. These constraints rule out regions in the parameter space, where the pulse compression process is interrupted. If these regions do not cover the entire parameter space, then we are left with a viable operation window, within which we can avoid all competing effects and use three-wave interactions to compress laser pulses. Now with an external magnetic field, the limiting effects are changed with respect to the pulse compression process. In other words, the magnetic field gives an extra degree of freedom, so that for a given seed pulse, the parameters we can tweak are now the pump laser frequency $\omega_0$ and intensity $I_1$, and plasma density $n_0$, temperature $T$, magnetic field $B_0$, and the plasma length $L$. By adjusting the extra control variable $\gamma_3$, the unmagnetized operation window can thus be expanded.

\subsection{Laser parameter space: wavelength-intensity window\label{sec:compression:window:laser}}

Consider the two-dimensional projection of the multi-dimensional operation window to the pump laser parameter space. For simplicity, I will plot sharp boundaries for the operation window by replacing constraints of the type $x\ll y$ by more definite constraints of the type $x/y<0.1$. It is worth noting that the operation window does not in fact have sharp boundaries in the sense that pulse compression is possible on one side and impossible on the other side. Instead, the pulse compression efficiency makes a smooth transition from low to high when crossing the boundary from the outside to the inside of the operation window. The sharp boundaries merely give a sense where the transition happens in the parameter space. To plot the boundaries, we need to solve a number of inequalities imposed by the four limiting effects, which can be done numerically. Let us introduce an boolean function $C(\omega_0,I_1)$, whereas $C=1$ means all constraints are satisfied, while $C=0$  means at least one constraint is not satisfied. We can scan the $\omega_0$-$I_1$ space with the boolean function, and whenever $C=1$, we can then use the solution to the three-wave equations as an objective function to maximize the output pulse intensity $I_2$, by choosing among viable plasma parameters.

\begin{figure}[t]
	\centering
	\includegraphics[angle=0,width=0.70\textwidth]{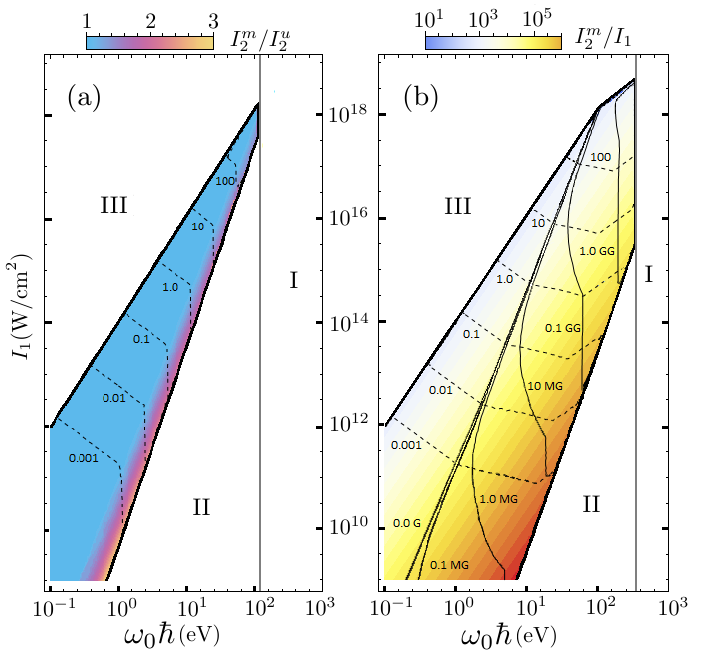}
	\caption[Operation windows in pump laser parameter space]{Operation windows in pump laser parameter space (colored regions). Regions I-III are excluded by limiting effects and fluid model constraints. (a) The operation window when $B_0=0$. The color scale compares the maximum output achievable in the magnetized case $I_2^m$ versus that in the unmagnetized case $I_2^u$. The dashed contours (in units of $10^{20}\,\text{cm}^{-3}$) are plasma density necessary for achieving $I_2^u$. (b) The expanded operation window when $B_0\ge 0$. The color scale is the gain $I_2^m/I_1$. By applying optimal magnetic fields (solid contours), plasma densities (dashed contours) necessary for achieving $I_2^m$ are now reduced.}
	\label{fig:regime}
\end{figure}

For example, consider pulse compression in hydrogen plasmas (Fig.~\ref{fig:regime}), where species matters because the ion charge affects the collision frequency.
First, notice that the unmagnetized operation window (Fig.~\ref{fig:regime}a, colored region) can be maximally expanded to the magnetized window (Fig.~\ref{fig:regime}b, colored region), when the optimal external magnetic fields (Fig.~\ref{fig:regime}b, black contours) are applied. In these figures, region I is excluded because collisionless damping becomes strong while keeping the plasma condition $n_e\lambda_D^3\gg1$; region II is excluded, because both damping mechanisms are strong; region III is excluded because the wavebreaking limit is exceeded while keeping $\omega_3\ll\omega_0$. A pump laser can be compressed using plasmas when the laser intensity is not too large and the laser frequency is not too high. Although the range of lasers that can be compressed by plasmas is still bounded, the range is already orders of magnitude larger than before.

Second, the maximum achievable output intensity can be increased by applying the optimal magnetic field. In Fig.~\ref{fig:regime}a, the colors denote the ratio of the maximum output intensity $I_2^u$ in the unmagnetized case to $I_2^m$ in the magnetized case. As can be seen from the figure, if a laser whose parameters are such that it can already be compressed using unmagnetized plasmas, then applying a magnetic field can only improve the final intensity by a factor of a few. This improvement is enabled by the alleviation of the modulational instability, so that pulse compression can proceed for a longer time. While improvements are moderate within the unmagnetized operation window, the most dramatic improvements are in the region where applying an external magnetic field enables compression of lasers that could not be compressed before. In Fig.~\ref{fig:regime}b, the colors denote the amplification gain, namely, the ratio of the output pulse intensity $I_2^m$ over the pump laser intensity $I_1$. As can be seen from the figure, amplifications by orders of magnitude are possible even in the region where unmagnetized compression does not work due to strong damping. 

Finally, the necessary plasma density is reduced when external magnetic fields are applied. In Fig.~\ref{fig:regime}, the dashed contours are the requisite plasma density (in units of $10^{20}\,\text{cm}^{-3}$) for optimizing the output intensity. In unmagnetized plasmas, the density needs to be high enough to provide sufficient coupling and avoid wavebreaking. On the other hand, the density cannot be too large, otherwise less energy is distributed to the seed pulse during three-wave interactions and damping also becomes strong. Now when we magnetize the plasma, the same resonance condition can be satisfied with less plasma density. Since the resonance condition is the same, the energy distribution during three-wave interactions also remains the same. Although the coupling is reduced, the limiting effects are reduced more. In particular, the reduction of the modulation instability allows longer amplification time, and the reduction of wave damping reduces energy loss. The density reduction not only allows an increase of the output intensity, but also relaxes engineering constraints. As mentioned earlier, achieving high-density uniform plasmas is challenging with current technologies. Therefore, replacing density with available magnetic fields now enables compression of lasers that were technologically challenging to compress before. 
This technological advantage makes it beneficial to use magnetized plasma within the unmagnetized window, even when it does not improve the maximum output intensity. 

\subsection{Plasma parameter space: density-temperature window\label{sec:compression:window:plasma}}

The multi-dimensional operation window can also be projected to the plasma parameter space. This projection informs us what plasma targets we need to set up in order to achieve efficient pulse compression. As before, the operation window only gives a rough indication where the transition from efficient to inefficient pulse compression happens. In what follows, I will give two examples. The first example is the compression of a soft X-ray laser, which cannot be compressed using unmagnetized plasmas due to strong damping. The second example is the compression of a UV laser, for which magnetic field helps to improve performance and relax engineering requirements.

\begin{figure}[!b]
	\centering
	\includegraphics[angle=0,width=0.70\textwidth]{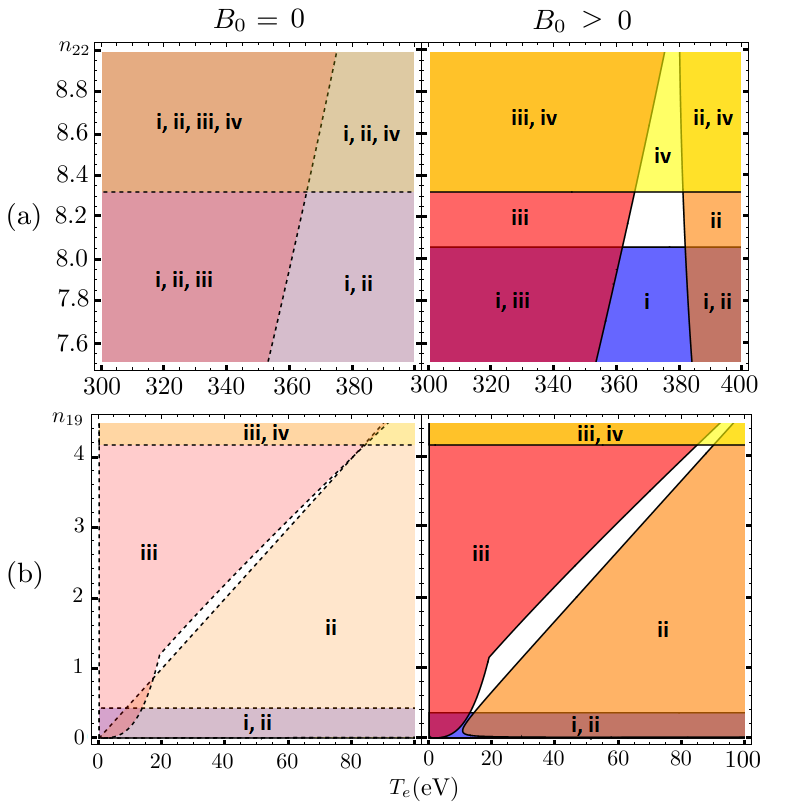}
	\caption[Operation windows in plasma parameter space]{Operation windows in plasma parameter space (white regions). The colored regions, possibly overlapping, are excluded by wavebreaking (blue-i), collisionless (orange-ii) or collisional damping (red-iii), and $\omega_3/\omega_0>0.1$ (yellow-iv). The exclusions in unmagnetized plasmas (left) are larger than those in magnetized plasmas (right). (a) Soft x ray laser with $I_1=10^{18}\hspace{3pt} \text{W/cm}^2$ and $\omega_0\hbar=250\,\text{eV}$. $B_0=1.5$ GG. (b) KrF laser with $I_1=10^{13}\,\text{W/cm}^2$ and $\omega_0\hbar=5\,\text{eV}$. $B_0=5$ MG. $n_{19}$ and $n_{22}$ are $n_e$ in the units of $10^{19}\,\text{cm}^{-3}$ and $10^{22}\,\text{cm}^{-3}$.}
	\label{fig:examples}
\end{figure}

First, to illustrate the expanded regime made possible through  magnetized plasma, consider the very ambitious, and speculative, compression of soft X-ray pulses. 
For example, X-ray pulses produced at the Linac Coherent Light Source have 2-6 mJ in energy, 5-500 fs in duration, and focal spot $\sim 10\,\mu\text{m}^2$  \citep{Bostedt13}, corresponding to intensity $\sim 10^{18}\, \text{W/cm}^2$. 
Since the photon energy in these pulses is in the range 250 eV--10 keV, efficient pulse compression using unmagnetized plasmas is not possible (Fig.~\ref{fig:regime}). 
However, the inefficient compression using unmagnetized plasmas \citep{Sadler15} can be made efficient by applying a magnetic field on the order of gigagauss (Fig.~\ref{fig:examples}a) using hydrogen plasmas. 
Such a field is of course huge, but in principle achievable over the small volumes; for compressing a 500 fs pulse, a plasma length of only 0.3 mm is needed. 
The strong magnetic field reduces necessary plasma density and therefor reduces wave damping, making it theoretically possible to compress picosecond X-ray pulses to femtosecond (Table~\ref{table:parameters}). 
In this example, the magnetic field opens up the otherwise closed operation window.

\begin{table}[t]
	\centering
	\begin{tabular}{c|cc|cc|ccc}
		\toprule
		\multirow{2}{*}{Pump}& \multicolumn{2}{c|}{Plasma} & \multicolumn{2}{c|}{Pulse} &\multicolumn{3}{c}{Compression}\\
		\cline{2-3} \cline{4-5} \cline{6-8} 
		&$B_0$ &$\min n_e$ &$\max I_2/I_1$ &$\Delta t_2$ &$\gamma_3$ &$t_M$ &$\omega_3/\omega_0$\\
		\midrule
		\makecell{$250\;\text{eV}, I_{18}$} & 1.5 GG & 8.1 $n_{22}$ & $2.3\!\times\! 10^3$ & 0.5 fs & 1.9 & 0.9 ps & 8.1\% \\
		\hline
		\multirow{2}{*}{$5\;\text{eV}, I_{13}$} & 0 G & 8.9 $n_{18}$ & $1.9\!\times\! 10^4$ & 54 fs & 1.0 & 0.8 ns & 2.2\% \\
		& 5 MG & 3.6 $n_{18}$ & $2.7\!\times\! 10^4$ & 65 fs & 1.3 & 1.3 ns & 1.8\% \\
		\bottomrule		
	\end{tabular}
	\caption[Magnetized UV and X-ray laser pulse compression]{Key parameters for examples given in Fig.~\ref{fig:examples}, assuming the initial pulse duration is not much longer than $\Delta t_2$, and the initial pulse intensity is such that $\Lambda_0\approx 6$. For soft X-ray pulses, applying a magnetic field opens up the otherwise closed operation window. For UV pulses, applying a magnetic field reduces the necessary plasma density and enables more intense and longer outputs. }
	\label{table:parameters}
\end{table}

To illustrate the use of  magnetized plasma in a more practicable example, consider the compression of UV lasers. 
For example, KrF pulses produced at the Nike laser facility have kilojoules energy with nanoseconds duration \citep{Obenschain96}. 
These pulses can be focused on a spot of size $\sim\!0.01\, \text{cm}^2$, reaching peak intensity $\sim\! 10^{14}\, \text{W/cm}^2$. 
The average intensity, however, falls in the range $10^{12}$--$10^{13}\, \text{W/cm}^2$. 
Since the photon energy of the KrF laser is $\sim5$ eV, the unmagnetized operation window is about to close when the laser intensity is at the lower end (Fig.~\ref{fig:regime}). However, the narrow unmagnetized window can be expanded by applying a megagauss magnetic field (Fig.\ref{fig:examples}b), when hydrocarbon plasmas ($\text{C}_3\text{H}_8, Z_{\text{eff}}\!\approx\! 2.36$) are used. 
In the expanded operation window, the minimum plasma density is reduced, which enables the output pulse to have larger intensity and longer duration (Table~\ref{table:parameters}).
In this example, less density is required and more intense output can be produced using magnetized plasma.

The required plasma parameters in the above two examples are very challenging, but in principle achievable, using current technologies for strong magnetic fields. As mentioned in Ch.~\ref{ch:intro},
one technique generates magnetic field by driving capacitor coil targets with intense lasers. In a number of experiments \citep{Fujioka13,Santos15}, generation of megagauss magnetic field, which is uniform on millimeter scale and quasi-static on nanosecond scale, has been demonstrated. 
Another technique generates magnetic field by ablating solid targets with intense laser pulses \citep{Wagner04,Tatarakis02,Borghesi98}. 
This technique can produce plasmas with $\sim 10^{21}\, \text{cm}^{-3}$ density and magnetic fields on the order of gigagauss, when picoseconds pulses with \mbox{$\sim 1\,\mu$m} wavelength and $\sim 10^{20}\, \text{W/cm}^2$ intensity are used in experiments. 
The density and magnetic field produced near the solid surfaces are uniform on micrometer scale and quasi-static on picosecond scale. The usefulness of strong magnetic fields provides an additional justification to further development of these magnetic field technologies.

\section{Particle-in-cell validations\label{ch:PIC}}

From theoretical analysis in the previous sections, it is clear that a transverse external magnetic field can improve the performance of laser pulse compression. However, the order-of-magnitude estimations cannot quantify the exact improvements. In addition, the one-by-one analysis of limiting effects may not capture the complete picture when all processes are at play. Moreover, we have hitherto only considered limiting effects that are known to be stringent for unmagnetized compressions, and it is not guaranteed that no other effects can impose stricter conditions. Therefore, it is helpful to carry out comprehensive numerical simulations for the magnetized compression process.

The two well-established methods for simulating kinetic plasmas where high-frequency processes are modeled directly are the electromagnetic Vlasov simulations and the electromagnetic particle-in-cell (PIC) simulations. 
The Vlasov simulations directly solve the coupled Vlasov-Maxwell's equations as partial differential equations. The distribution function $f(t,\mathbf{x},\mathbf{v})$ lives on a phase space grid, while $\mathbf{E}(t,\mathbf{x})$ and $\mathbf{B}(t,\mathbf{x})$ live on a configuration space grid. Plasmas influence $\mathbf{E}(t,\mathbf{x})$ and $\mathbf{B}(t,\mathbf{x})$ through charge and current densities, which are computed as velocity space integrals; while $\mathbf{E}(t,\mathbf{x})$ and $\mathbf{B}(t,\mathbf{x})$ directly act on plasmas through the velocity space advection term. Special attentions need to be paid when solving the hyperbolic PDEs to prevent numerical oscillations and ensure positivity of the distribution function.
In comparison, the PIC simulations exploit the Klimontovich formalism, and represent the distribution function by a collection of super-particles. The electromagnetic fields again live on a spatial grid, and are solved from the Maxwell's equations as PDEs. On the other hand, the $j$-th super-particle's phase space coordinates $(\mathbf{x}_j(t),\mathbf{v}_j(t))$ are functions of time only, and they are solved from the Newton's equation with the Lorentz force as ODEs. Since particles and fields live on different domains, some deposition scheme is necessary to translate particles' position and velocity to charge and current densities on the grid, and some interpolation schemes is necessary to use field values on the grid to compute forces on particles that live in the continuum. 
Neither the Vlasov nor the PIC scheme is perfect. They merely provide approximate solutions to the plasma kinetic model, in which other effects such as collisions may be later added. Although there is no guarantee for long-time accuracy, solutions by both schemes are expected to converge to the correct results for sufficiently high resolutions.

Many implementations of the PIC scheme are available. In fact, I led a collaborative software-engineering project \textsc{EMOOPIC}, which implemented the PIC scheme using C++ with 3D MPI and OpenMP parallelizing capability. The \textit{Electro-Magnetic Object-Oriented Particle-In-Cell} code uses the standard relativistic Boris algorithm \citep{Boris70} to solve the Newton's equation, the standard Yee's algorithm \citep{Yee66} to solve the Maxwell's equations, and a first-order deposition and interpolation scheme to interconnect particles and fields. The \textsc{EMOOPIC} code can handle a suite of initial and boundary conditions, and is equipped with particle-sorting options to accelerate computations. The code has been successfully tested in a few example problems, and the latest release can be downloaded from GitHub \citep{EMOOPIC}, which is accompanied by a detailed documentation. However, a more established code that has been thoroughly benchmarked is perhaps more suitable for the purpose of simulating magnetized pulse compression, a setup that had never been investigated before. In collaboration with \cite{Jia17}, a set of PIC simulations are carried out using the \textsc{EPOCH} code \citep{Arber2015} in one dimension to model pulse compression with a transverse magnetic field. In what follows, I will highlight findings of these simulations.

\subsection{Fluid effects: growth and saturation\label{sec:PIC:fluid}}

\begin{figure}[!b]
	\centering
	\includegraphics[angle=0,width=0.7\textwidth]{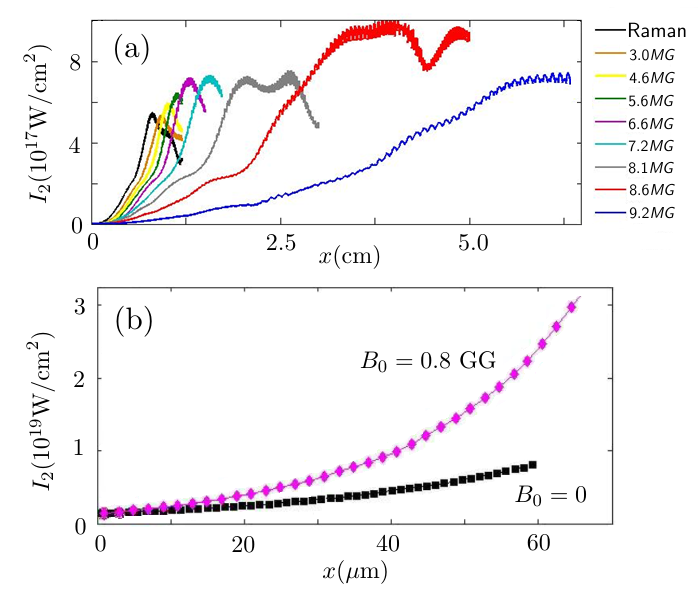}
	\caption[Compression of 1-$\mu$m and $10$-nm pump lasers]{Applying a transverse magnetic field improves the performance of plasma-based laser pulse compression, as shown here in terms of the peak intensity as a function of the distance of propagation, obtained using 1D PIC simulations. 
		For a 1-$\mu$m optical pulse (a), using a longer plasma and an optimal magnetic field (red line), the final pulse intensity is twice of what is achievable using unmagnetized Raman (black line). The initial pump intensity $I_{10}=3.5\times10^{14} \;\text{W/cm}^2$, and the 1.1-$\mu$m seed has initial intensity $I_{20}=1.8\times10^{13} \;\text{W/cm}^2$ and initial duration \mbox{$\Delta t_{20}=33$ fs}. 
		For a $10$-nm X-ray pulse (b), replacing plasma density with a transverse magnetic field on gigagauss scale alleviates strong damping. Consequently, magnetized pulse compression becomes possible (purple), while unmagnetized amplification can barely work (black). The pump intensity \mbox{$I_{10}=1.4\times10^{18} \;\text{W/cm}^2$}, and the $11$-nm seed pulse has $I_{20}=1.4\times10^{18} \;\text{W/cm}^2$ and $\Delta t_{20}=1.5$ fs. }
	\label{fig:PIC}
\end{figure}

The prediction that applying a moderate magnetic field improves the performance of laser pulse compression has been verified using PIC simulations. First, consider the use of a 1.0-$\mu$m pump laser, with constant initial intensity $I_{10}=3.5\times10^{14} \;\text{W/cm}^2$, to compress a counter-propagating 1.1-$\mu$m seed pulse, whose initial intensity $I_{20}=1.8\times10^{13} \;\text{W/cm}^2$ and initial duration \mbox{$\Delta t_{20}=33$ fs}. Given the pump and the seed lasers, we apply a magnetic field transverse to the direction of laser propagation, and reduce the plasma density accordingly to maintain the resonance conditions (Fig.~\ref{fig:PIC}a). When there is no magnetic field (black line), pulse compression is mediated by Raman backscattering. After the initial exponential growth, the seed pulse enters the nonlinear compression stage, until its intensity saturates at $I_{2}\approx5.5\times10^{17} \;\text{W/cm}^2$ due to the modulational instability. As we increase the magnetic field (color lines), the growth becomes slower, but the saturation is delayed. The net consequence is that the attainable final pulse intensity increases with the magnetic field, until an optimal field $B\approx 8.6$ MG is reached (red line), where the final pulse intensity is about twice of what is achievable using Raman compression. When a stronger magnetic field is applied (blue line), the seed pulse loses a substantial amount of energy to the wakefield, which inhibits further increase of the pulse intensity.

In addition to improving the performance in the optical regime, applying a magnetic field enables compression of short-wavelength pulses that cannot be compressed using unmagnetized plasmas. For example, a $10$-nm soft X-ray laser is at the verge of what can be amplified using Raman compression. At even shorter wavelength, collisional damping becomes too strong. The total damping could have been alleviated by increasing the plasma temperature, if it were not due to collisionless damping, which increases with the plasma temperature. Therefore, the operation window in the plasma parameter space is almost closed (Fig.~\ref{fig:regime}). In one-dimensional PIC simulations with the collision module of the EPOCH code turned on \citep{Jia16a}, the $11$-nm seed pulse, whose initial intensity $I_{20}=1.4\times10^{18} \;\text{W/cm}^2$ and initial duration $\Delta t_{20}=1.5$ fs, barely grows when pumped at $I_{10}=1.4\times10^{18} \;\text{W/cm}^2$ (Fig.~\ref{fig:PIC}b, black). 
However, keeping the same plasma temperature $T_e=200$ eV and $T_i=1$ eV, while applying a $0.8$-GG magnetic field and reduce the plasma density in such a way that the frequency of the plasma wave remains fixed, 
the effective growth rate becomes much larger (purple). This is because although the undamped growth rate $\gamma_0\propto n_e^{1/2}$ is reduced in lower density plasmas, the collisionless damping $\nu_{3}\propto n_e$ and the collisional damping $\nu_{1,2}\propto n_e^2$ are reduced more substantially. Therefore, faster effective growth is possible when we magnetize the plasma medium, using which compression of soft X-ray pulses beyond the reach of previous methods becomes possible.

\subsection{Kinetic effects: wakefields and wavebreaking\label{sec:PIC:kinetic}}

\begin{figure}[b]
	\centering
	\includegraphics[angle=0,width=0.6\textwidth]{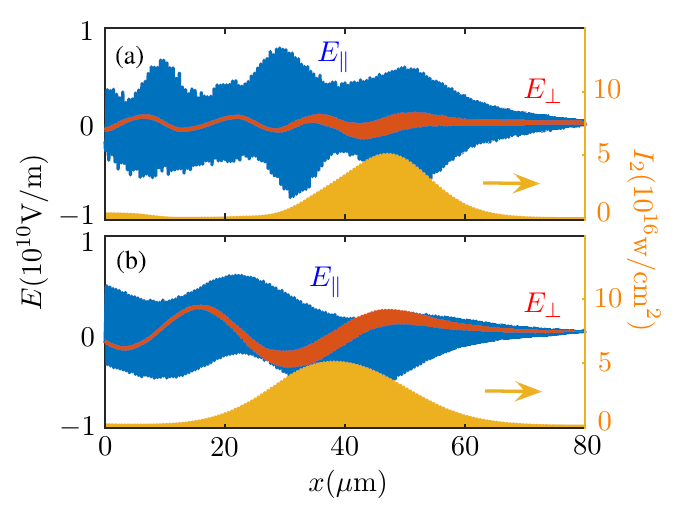}
	\caption[Wakefield generation in magnetized pulse compression]{Left-propagating electric fields (left axis) when the right-propagating 1.1-$\mu$m pulse (right axis) reaches relativistic intensity $a_2\sim 1$, whereby wakefield starts to be excited. The silhouettes outline the wave envelopes, inside which the colored regions are covered by fast oscillations on the wavelength scale. When the magnetization factor is $\gamma_3=1.42$ (a), the wakefield excited by the intense pulse (orange) is largely electrostatic, where the longitudinal field $E_\parallel$ (blue) dominates the transverse field $E_\perp$ (red).  When magnetization increases to $\gamma_3=2.42$ (b), the $E_\perp$ component starts to become comparable to $E_\parallel$. The magnetized wakefield contains more energy, causing the amplified pulse to loss energy more rapidly.}
	\label{fig:wakefield}
\end{figure}

From simulations of 1-$\mu$m pulse compression, we see that while moderate magnetic fields delay modulational saturation and improve final pulse intensity, applying too large a magnetic field results in a decrease of the pulse intensity due to wakefield generation. 
When an intense laser pulses propagates in a plasma, its ponderomotive force expels plasma electrons and excites plasma wakefield. In a weakly magnetized plasma (Fig.~\ref{fig:wakefield}a), the wakefield is similar to the unmagnetized wakefield, in which the longitudinal component $E_\parallel$ dominates. However, in a strongly magnetized plasma (Fig.~\ref{fig:wakefield}b), the transverse component $E_\perp$ grows to comparable strengths. Therefore, the field energy density contained in the wakefield increases when the magnetization factor $\gamma_3$ becomes larger.
In addition, the scale of the wakefield $\sim c/\omega_p$ is comparable to the plasma skin depth. When more plasma density is replaced by the magnetic field, the skin depth increases. Hence, not only is a larger energy density contained in the wavefield, but the wakefield also extends for a larger spatial region when $\gamma_3$ increases.
Moreover, wakefield energize plasma electrons and transfer its energy to the kinetic energy of fast particles. In the unmagnetized case, electrons are accelerated only in the longitudinal direction. Whereas in the magnetized case, the wakefield also accelerates electrons in the transverse direction.
Consequently, when the magnetization factor increases, both the field energy and the kinetic energy contained in the wakefield increases, to which the amplified laser pulse losses more energy. 

\begin{figure}[b]
	\centering
	\includegraphics[angle=0,width=0.5\textwidth]{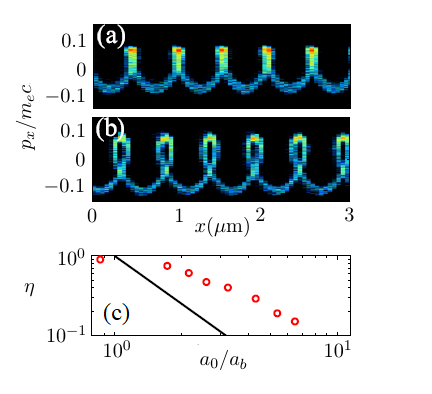}
	\caption[Wavebreaking and phase-mixing effects on compression efficiency]{Due to resilience of magnetized plasma waves to wavebreaking, pulse compression efficiency is higher than expected. In a plasma with $\gamma_3=2.24$, a set of 1-$\mu$m pump lasers with $a_1$ above the wavebreaking intensity $a_b$ are used to compress the 1.1-$\mu$m seed to $a_2\sim0.5$. When $a_1\approx2a_b$, spikes are formed in the electron phase space (a), which indicate infinite density gradient and thereof wavebreaking. However, the plasma wave remains intact until $a_1\approx6a_b$, where phase space islands form (b), and the plasma wave becomes strongly damped. The pulse compression efficiency (c, red circles) is well above the wavebreaking expectation (c, black line). The efficiency remains high after the pump amplitude exceeds the wavebreaking threshold, and $\eta$ diminishes only after an even higher phase mixing threshold is surpassed.} 
	\label{fig:wavebreaking}
\end{figure}

The energy loss due to electromagnetic wakefield generation may be partially compensated by the resilience of magnetized plasmas to wavebreaking, whereby a stronger pump laser can be employed to supply energy to the growing seed pulse. As mentioned in Sec.~\ref{sec:compression:limits:instabilities}, although wavebreaking is an important limiting effect in unmagnetized compression, the magnetized plasma wave can still mediate pulse compression efficiently beyond the wavebreaking threshold. 
This is because the transverse magnetic field $B_0$ provides an additional restoring force, and the large $E_\parallel\times B_0$ shear stabilizes the plasma wave beyond the wavebreaking threshold. For example, for compression of 1-$\mu$m lasers in a plasma with $\gamma_3=2.24$, the wavebreaking pump amplitude $a_b\approx5\times 10^{-3}$. When pumped by $a_1\approx2a_b$, the plasma wave breaks, which is manifested in the electron phase space density plot (Fig.~\ref{fig:wavebreaking}a).
The plasma density piles up at the spikes, reaching theoretically infinite density gradient. However, instead of collapsing, the plasma continues to mediate pulse compression, and the efficiency $\eta\approx70\%$ (Fig.~\ref{fig:wavebreaking}c, red circles) remains well above the wavebreaking expectation (Fig.~\ref{fig:wavebreaking}c, black line).
Although the pump intensity is not limited by wavebreaking, it is bounded by a higher phase-mixing threshold, beyond which the plasma wave becomes strongly damped. For example, when the compression is pumped by $a_1\approx6a_b$, the large amplitude plasma wave can no longer be sustained by the background magnetic field, and phase space islands form (Fig.~\ref{fig:wavebreaking}b). Although $\eta\approx10\%$ is still well above the wavebreaking expectation, the collapsing plasma wave can no longer efficiently mediate energy transfer (Fig.~\ref{fig:wavebreaking}c). We see phase mixing, which happens at amplitudes higher than wavebreaking, is the stringent effect that limits the maximum allowable pump intensity in magnetized laser pulse compression.

Simulations of 1-$\mu$m pulse compression in magnetized plasmas confirm the theoretical expectation that replacing plasma density with background magnetic fields on magagauss scale relatively suppresses instability and damping, and allows the use of slightly longer plasma targets to improves the performance of laser pulse compression. Moreover, the PIC simulations identify wakefield generation as an important mechanism that disfavor the use of magnetic fields that are too strong. This limitation can however be circumvented due to the resilience of magnetized plasmas to wavebreaking. Until a higher phase-mixing threshold is reached, the pump can be strengthened to efficiently compress the pulse to intensities that are not achievable using unmagentized plasmas.

\part{QED plasma theory and simulations}
\lettrine[lines=2, findent=3pt, lraise=0.4, loversize=1]{W}{hen} strong electromagnetic fields or energetic particles are present, we enter the realm of relativistic quantum physics. For example, in the vacuum, electron-positron pairs can be
created by strong electric fields beyond the Schwinger limit \mbox{$E_s\sim10^{18}$ V/m}, which corresponds to a magnetic field \mbox{$B_s\sim10^9$ T}. 
These strong fields also noticeably alter plasma behaviors, leading to anharmonic cyclotron absorption features observed in spectra of X-ray pulsars.
Moreover, when high-energy-density plasmas are present, relativistic quantum effects become important even when fields are orders-of-magnitude smaller than the Schwinger fields. For example, when ultra-intense lasers interact with solid targets, electron-positron pairs can be produced, whose experimentally observed energy spectra remain to be explained by better models.

In this part of my thesis, I develop a relativistic quantum model for plasma physics, starting directly from quantum field theory. In particular, I extend quantum electrodynamic (QED), which is the \textit{ab initio} standard model for electromagnetic interactions, to a model also for plasmas. In the standard QED, only a few charged particles and a few photons are involved, and what is of concern is usually relativistic quantum interactions between these particles. On the other hand, in the standard plasma physics, many particles and waves are present, and the focus is usually on the collective behavior of the medium. Now, in order to model systems like neutron star magnetospheres and high-energy-density laser plasmas, I develop a model for QED plasmas, 
which is applicable when both relativistic-quantum and collective effects are important.

As a toy model, I contemplate scalar QED, which describes spin-0 charged bosons, such as deuteron ions, charged pions, and Cooper pairs. Although plasmas are typically made of spin-1/2 charge fermions, classical plasma physics takes no account of particle spin-statistics at all, which is valid when one is not concerned with spin effects and when the plasma is not Fermi degenerate. Keeping in mind that the developments for spinor QED can proceeds analogously, the scalar-QED plasma model enables a clean demonstration of collective effects, without the complication due to spin and chiral effects associated with Dirac fermions. As a proof-of-principle demonstration that QED can be extended into a many-body theory, just as quantum mechanics did for condensed matter physics, I will focus on scalar-QED plasmas. 
\chapter{Prerequisite: quantum electrodynamics in vacuum\label{ch:vacuum}}

%

In this chapter, I will give a condensed introduction to quantum field theory, which is the foundation of modern descriptions of elementary particles and their interactions, as elaborated by many text books, for example, \cite{Peskin95,Schwartz2014quantum}. 
Quantum field theory used to be in the tool box of many first-generation plasma physicists, who later decided that relativistic quantum effects are neither important for understanding astrophysical plasmas like the sun, nor achieving thermonuclear fusion using magnetic confinement. After decades of isolated developments, many plasma physicists nowadays are probably not familiar with quantum field theory.
However, the scope of plasma physics has been expanded beyond the realm of classical physics. In recent decades, X-ray and gamma-ray telescopes have unveiled rich spectral features of neutron stars, providing valuable data for strong-field astrophysics. Moreover, developments of intense lasers have paved ways not only for achieving fusion through inertial confinement, but also for studying high-energy-density matter previously inaccessible by experiments. With these technological developments, it becomes necessary that plasma physicists pick up tools that have been sharpened in particle physics and condensed matter physics and use them to farm the fertile land that so far has not been fully explored.

Quantum field theory is perhaps the most successful theory in physics, based on which three out of four fundamental interactions can be described to an astonishing precision. As an example of quantum field theory, quantum electrodynamics arises when one combines quantum mechanics and special relativity. These two building blocks of QED are quite simple. One of the building blocks is the U(1) symmetry of quantum mechanics, which requires that physics remains the same when an arbitrary complex phase is added to the wave function. The other building block is the Lorentz symmetry of special relativity, which requires that physical laws remain the same in any inertial reference frame. However, putting these two building blocks together results in nontrivial consequences. 
One consequence is that if charged particles exist, then there must also exist electromagnetic field; otherwise the theory cannot satisfy the local U(1) symmetry and the Lorentz symmetry simultaneously. Perhaps an even more profound consequence is that if one type of charged particles exist, then there must also exist another type of charged particles that have exactly the same mass but the opposite charge. These amazing facts of QED, together with its quantitative predictions, have been proven experimentally with extremely high accuracy for a wide range of conditions. It is the indisputable facts that QED, with its very simple building blocks, can already explain every aspects of electromagnetic interactions, including strange things such as the anomalous magnetic moment of electrons and the Lamb shift of hydrogen energy levels, 
that made people believe QED to be a cornerstone of our understanding of the universe. This same cornerstone will be useful in order to develop a theory of relativistic quantum plasmas. 

\section{Classical field theory\label{sec:vacuum:field}}

The name ``quantum field theory" usually presumes that fields are promoted to operators and thereof become quantized. However, this notion of second quantization is narrower than necessary. In fact, all observables in quantum field theory can be computed without quantizing the fields. Quantization is therefore a method instead of the essence of field theory. In this section, I will first work with field theory from the perspective of equations. For convenience, I will use the natural units $\hbar=c=\epsilon_0=1$, whereby all dimensional quantities have units of energy scales. 

\subsection{Classical field equation\label{sec:vacuum:field:free}}

First, let us consider a free complex scalar field $\phi$, which satisfies the Klein--Gordon (KG) equation. The KG equation is the simplest relativistic extension of the Schrödinger's equation, and it describes particles that satisfy the relativistic energy-momentum relation $E^2=p^2+m^2$. Taking the ansatz $\phi\propto\exp(-iEt+ipx)$ that particles behave like waves, then we can replace energy $E\rightarrow i\partial_t$ and momentum $p_j\rightarrow -i\partial_j$. The energy-momentum relation can then be casted into a PDE, namely, the KG equation
\begin{equation}
\label{eq:KG}
(\partial_t^2-\nabla^2+m^2)\phi=0.
\end{equation} 
Not surprisingly, since the KG equation is built with the wave ansatz, the general solution is a linear superposition of plane waves
\begin{equation}
\label{eq:phi0}
\phi(x)=\int\frac{d^3\mathbf{p}}{(2\pi)^3}\frac{1}{\sqrt{2E_\mathbf{p}}} \Big(a_{\mathbf{p}}e^{-ipx}+b^\dagger_{\mathbf{p}}e^{ipx}\Big),
\end{equation}
where $E_\mathbf{p}=\sqrt{\mathbf{p}^2+m^2}$ is the positive energy associated with momentum $\mathbf{p}$, and $px:=p_\mu x^\mu=E_\mathbf{p}t-\mathbf{p}\cdot\mathbf{x}$ is the Minkowski inner product with the metric $g^{\mu\nu}=\text{diag}(1,-1,-1,-1)$. In the above spectral expansion, $a_{\mathbf{p}}$ and $b^\dagger_{\mathbf{p}}$ are some complex functions of $\mathbf{p}$, where $b^\dagger_{\mathbf{p}}$ denotes the Hermitian conjugate of $b_{\mathbf{p}}$. The Fourier integral is normalized such a way for later convenience. 

To develop the free-field theory to an interacting-field theory, it is helpful to take an variational principle perspective. From this perspective, the KG equation is the least-action trajectory of the action $S_0=\int d^4x\, \mathcal{L}_0$, with the Lagrnagian density of the free field
\begin{equation}
\label{eq:L0_phi}
\mathcal{L}_0=\partial_\mu\phi^*\partial^\mu\phi-m^2\phi^*\phi,
\end{equation}
where $\phi^*$ is the complex conjugate of $\phi$. 
The Lagrangian $\mathcal{L}_0$ is invariant under the global U(1) transformation $\phi\rightarrow\phi'=\exp(i\alpha)\phi$, where $\alpha$ is an arbitrary real constant. This U(1) symmetry requires that $\phi$ and $\phi^*$ appear in pairs. Moreover, the action $S_0$ is invariant under the Lorentz transformation $x^\mu\rightarrow x^{'\mu}=\Lambda^{\mu}_{\nu}x^\nu$, under which the complex scalar field is transformed by under the pullback $\phi(x)\rightarrow\phi'(x)=\phi(\Lambda^{-1}x)$. This Lorentz symmetry requires that all spacetime indices are properly contracted. 
In addition to the kinetic term and the mass term in $\mathcal{L}_0$, interaction terms are also allowed by the U(1) symmetry and the Lorentz symmetry. For example, consider the Lagrangian with a $\phi^4$-interaction term
\begin{eqnarray}
\label{eq:L_phi4}
\nonumber
\mathcal{L}&=&\mathcal{L}_0+\mathcal{L}_I \\
&=&\partial_\mu\phi^*\partial^\mu\phi-m^2\phi^*\phi-\frac{\lambda}{4}(\phi^*\phi)^2.
\end{eqnarray}  
This interaction Lagrangian is natural in the sense that the coupling coefficient $\lambda$ is a dimensionless number. To see why $\lambda$ is dimensionless, notice that in the natural units, space and time have mass dimension $M^{-1}$. In order for the action is a dimensionless number, the Lagrangian density needs to have mass dimension $M^4$, which can be satisfied if the complex scalar field have mass dimension $M$. Therefore, the $\phi^4$ interaction Lagrangian has the correct dimension of $M^4$, and the coupling coefficient is thereof dimensionless.  

To obtain the classical equation of motion for the $\phi$ field, we demand the classical field to be the trajectory that extremize the action $S$. Integrating by part, we can rewrite the action $S=\int d^4x [\phi^*(-\partial_\mu\partial^\mu-m^2)\phi-\frac{\lambda}{4}(\phi^*\phi)^2]$. In this form, it is easy to take variation with respect to $\phi^*$. The action is extremized when $\phi$ satisfies $\delta S/\delta\phi^*(x)=0$, namely,
\begin{equation}
\label{eq:KG4}
(\partial^2+m^2)\phi=-\frac{\lambda}{2}\phi^*\phi^2,
\end{equation}
where $\partial^2=\partial_\mu\partial^\mu$. This equation recovers the KG equation [Eq.~(\ref{eq:KG})] when $\lambda\rightarrow0$. When $\lambda\ll1$, Eq.~(\ref{eq:KG4}) may be solved perturbatively using the Born approximation with the asymptotic expansion
\begin{equation}
\phi=\phi_0+\lambda\phi_1+\lambda^2\phi_2+\dots.
\end{equation}
To $\lambda^0$ order, $\phi_0$ simply satisfies the KG equation, whose general solution is given by Eq.~(\ref{eq:phi0}). To $\lambda^1$ order, $\phi^1$ satisfies the driven KG equation
\begin{equation}
(\partial^2+m^2)\phi_1=-\frac{\lambda}{2}\phi_0^*\phi_0^2,
\end{equation}
where the forcing term on the RHS is known from the previous order.
Suppose we know the inverse of the differential operator $(\partial^2+m^2)G(x,x')=-i\delta^{(4)}(x-x')$, then the solution to the above equation can be easily constructed as follows:
\begin{equation}
\label{eq:phi1}
\phi_1(x)=-\frac{i\lambda}{2}\int d^4x' \phi_0^*(x')\phi_0^2(x') G(x,x').
\end{equation}
This construction will encounter secular terms, as we have seen in Ch.~\ref{ch:multiscale}, where the field can grow indefinitely. Instead of using a multiscale expansion, in quantum field theory, removing secular terms is usually done using a procedure called renormalization, which we shall discuss later. For now, it is sufficient to recognize that we can systematically obtain the perturbative solution order by oder, if we know the inverse of the differential operator.
\subsection{The Green's function\label{sec:vacuum:field:green}}

The inverse of the differential operator is called the Green's function, which satisfies
\begin{equation}
\label{eq:dG}
(\partial^2+m^2)G(x,x')=-i\delta^{(4)}(x-x'),
\end{equation}
where the factor $-i$ in front of the delta function is inserted for later convenience. Notice that the Green's function is not unique unless we specifies the boundary conditions. This is because if $G$ is a Green's function, then $G+\phi_0$ is also a Green's function, where $\phi_0$ satisfies the KG equation. Using the identify $\int d^4p\exp[ip(x-x')]=(2\pi)^4\delta^{(4)}(x-x')$, the Green's function can be represented by the integral
\begin{equation}
	G(x,x')=\int\frac{d^4p}{(2\pi)^4}\frac{ie^{-ip(x-x')}}{p^2-m^2},
\end{equation}
where $p^2=p_\mu p^\mu$ is again the Minkowski  inner product. The integral encounters poles at $p^2=m^2$, and the different integration contours for getting around the poles thereof give different Green's functions that satisfy different boundary conditions. 

A particularly important Green's function is the one that respects causality. From Eq.~(\ref{eq:phi1}), we see the value of $\phi_1(x)$ at one spacetime coordinate $x$ is determined by values of $\phi_0(x')$ at another spacetime coordinate $x'$, weighted by the Green's function $G(x,x')$. 
From special relativity, we expect that the classical information at $x'$ can influence the field at $x$ if their separation $x-x'$ is timelike. On the other hand, we expect that the classical information cannot propagate faster than the speed of light, so that $\phi_1(x)$ should have no dependence on $\phi_0(x')$, if the separation $x-x'$ is spacelike. 
In classical field theory, the present value is determined by the past values through the advanced Green's function $G_A$, and the present value affects future values through the retarded Green's function $G_R$. However, in quantum field theory, quantum fluctuations enable particles to take arbitrary trajectories that are not allowed classically. For example, in a particular realization, a quantum particle can travel faster than the speed of light or even travel backward in time. 
As we shall see later, the Green's function that respects causality in the quantum sense is the Feynman Green's function
\begin{equation}
\label{eq:GF}
G_F(x,x')=\int\frac{d^4p}{(2\pi)^4}\frac{ie^{-ip(x-x')}}{p^2-m^2+i\epsilon},
\end{equation}
where $\epsilon>0$ is an arbitrarily small parameter, prescribing how to move the integration contour away from the poles.  The above integral representation can be evaluated using Bessel functions (Appendix~\ref{ch:append:GreenF}), and an explicit formula for the Feynman Green's function can be written in terms of the correlation function as
\begin{equation}
\label{eq:GF_D}
G_F(x,x')=\theta(t-t')D(x-x')+\theta(t'-t)D(x'-x),
\end{equation}
where $\theta$ is the Heaviside step function, and the correlation function
\begin{equation}
\label{eq:Corr_D}
D(x)=\int \frac{d^3\mathbf{p}}{(2\pi)^3}\frac{e^{-ipx}}{2E_{\mathbf{p}}}.
\end{equation}
When the separation is spacelike, the correlation function $\propto \rho^{-3/2}e^{-\rho}$ is suppressed exponentially for large spatial separations
\begin{equation}
D(x-x')=\frac{m^2}{4\pi^2}\frac{K_1(\rho)}{\rho},
\end{equation}
where $K_\nu(z)$ is the modified Bessel functions of the second kind, and the normalized proper length $\rho=m\sqrt{(\mathbf{x}-\mathbf{x}')^2-(t-t')^2}$. On the other hand, when the separation is timelike, the correlation function $\propto \tau^{-3/2}e^{-i\tau}$ behaves like a decaying wave for large temporal separations
\begin{equation}
D(x-x')=\frac{im^2}{8\pi}\frac{H_1^{(2)}(\tau)}{\tau},
\end{equation}
where $H_\nu^{(2)}(z)$ is the Hankel function of the second kind, and the normalized proper time $\tau=m\sqrt{(t-t')^2-(\mathbf{x}-\mathbf{x}')^2}[\theta(t-t')-\theta(t'-t)]$ keeps the original sign of the time separation. The correlation functions inside and outside the light cone are connected by analytic continuation (Fig.~\ref{fig:Analytic_Continuation}), and it is thus continuous on the light cone $(t-t')^2=(\mathbf{x}-\mathbf{x}')^2$. However, due to the Heaviside step function, the derivative of the Green's function $G_F$ is not continuous, giving rise to the delta function in Eq.~(\ref{eq:dG}). We see there is an important distinction between the classical and the quantum world. In the classical world, information cannot propagate faster than the speed of light. On the other hand, propagation faster than the speed of light is allowed in the quantum world, but the probability is exponentially suppressed.
\subsection{Interacting fields\label{sec:vacuum:field:interaction}}

Now that we have a formula for the Green's function, we can in principle evaluate integrals, such as Eq.~(\ref{eq:phi1}), to obtain perturbative solutions to the field equation. In scattering theory, one is usually concerned with the asymptotic behaviors of the wave functions. For example, consider the case where $\phi_0$ contains two incoming waves with momentum $p_1$ and $p_2$, and an out going wave with momentum $p_3$, where the wave function asymptotes to
\begin{equation}
\phi_0(x)\sim\frac{1}{r}(e^{ip_1x}+e^{ip_2x}+e^{-ip_3x}),
\end{equation} 
for some large proper distance $r\gg1/m$ measured from the lab origin. Then, in Eq.~(\ref{eq:phi1}), the products $\phi_0^2\phi_0^*\sim\frac{2}{r^3}\exp[i(p_1+p_2-p_3)x]+\dots$, where the factor of ``2" comes from $\phi_0^2$. After carrying out the convolution integral with the Green's function, the first-order field contains many terms, including
\begin{equation}
\phi_1(x)\sim\frac{-i\lambda}{r}e^{-ip_4x}+\dots,
\end{equation} 
where the 4-momentum $p_4=p_1+p_2-p_3$. We see the scattering amplitude
\begin{equation}
i\mathcal{M}(p_1p_2\rightarrow p_3p_4)=-i\lambda+\dots.
\end{equation}
Knowing the probability amplitude, we can then compute the scattering cross section, which is proportional to $|\mathcal{M}|^2$ up to some phase space factors. Similarly, it is possible to compute scattering cross sections of other processes by first solving the field equation perturbatively and then extracting the asymptotic behavior of the wave function.    
  
Apart from the $\phi^4$ interaction, local U(1) symmetry of the complex scalar field requires that the complete theory also contains additional fields and interactions. To see why, let us consider how derivatives transform under the local U(1) symmetry
\begin{equation}
\label{eq:U1_phi}
\phi(x)\rightarrow\phi'(x)=e^{ie\alpha(x)}\phi(x),
\end{equation} 
where $\alpha(x)$ is an arbitrary real scalar field, and $e$ is the charge of the field $\phi(x)$ under the group action. In particular, when the charge $e=0$, the field $\phi$ does not transform. 
Under the above transformation, the mass term $m^2\phi^*\phi$ and the interaction term $\frac{\lambda}{4}(\phi^*\phi)^2$ are both invariant. If $\alpha$ was just a constant, then the kinetic term $\partial_\mu\phi^*\partial^\mu\phi$ would also be invariant. However, now that $\alpha(x)$ is a function, the derivative is transformed by
\begin{equation}
\partial_{\mu}\phi\rightarrow e^{ie\alpha}(\partial_{\mu}\phi+ie\partial_{\mu}\alpha\,\phi),
\end{equation}
which does not look particularly nice. In fact, the partial derivative is defined as the limit \mbox{$\partial_{\mu}\phi=\lim_{\Delta_\mu\rightarrow 0}\frac{1}{\Delta_\mu}[\phi(x+\Delta_\mu)-\phi(x)]$}. Since the two points $\phi(x+\Delta_\mu)$ and $\phi(x)$ are transformed differently under the local U(1) symmetry, the partial derivative does not have a well-defined transformation.

In order to define a properly transformed derivative, we need some comparator $U(x,y)$, such that under symmetry transformation 
\begin{eqnarray}
\label{eq:G_phi}
\phi(x)&\rightarrow& g(x)\phi(x), \\
\label{eq:G_U}
U(x,y)&\rightarrow& g(x)U(x,y)g^{-1}(y),
\end{eqnarray}
for all element $g$ in the Lie group $G$. Under local transformation, $g(x)$ is some smooth section of the fiber bundle $P\times G$, where $P$ is the spacetime manifold. In this way, the combination $U(x,y)\phi(y)$ transforms as $\phi(x)$, and we can then compare values at two different points $x,y\in P$ in a well defined manner. In particular, we can define a well-transformed derivative
\begin{equation}
\label{eq:Dphi}
D_\mu(x)\phi(x)=\lim_{\Delta_\mu\rightarrow 0}\frac{1}{\Delta_\mu}\Big[U(x,x+\Delta_\mu)\phi(x+\Delta_\mu)-\phi(x)\Big].
\end{equation}
This derivative is called the gauge covariant derivative, which transforms as the field $\phi(x)$ by $D\phi\rightarrow D'\phi'=gD\phi$. From how the comparator transform, it is clear that the covariant derivative is transformed by
\begin{equation}
\label{eq:D_transform}
D(x)\rightarrow D'(x)=g(x)D(x)g^{-1}(x).
\end{equation}
When defining the covariant derivative, we want the comparator $U(x,x)=\mathbb{I}\in G$ to be the group identity. Then, there exist some Lie-algebra-valued 1-form $A$, such that the comparator can be locally expressed as the trace of the path-ordered ($\mathcal{P}$) exponential map
\begin{equation}
\label{eq:comparator}
U_\gamma(x,y)=\text{tr}\,\mathcal{P}\exp\Big[ie\int_y^x A_\mu(z)dz^\mu\Big],
\end{equation}
which in usually called the Wilson's line, where $\gamma$ is a smooth curve $\gamma: [0,1]\rightarrow P$ on the spacetime manifold $P$, such that $\gamma(0)=x$ and $\gamma(1)=y$.
The Lie algebra valued 1-form $A$ is usually called the gauge field, and it is a smooth section of $T^*P\times \mathfrak{g}$, where $T^*P$ is the cotangent bundle of the spacetime manifold $P$, and $\mathfrak{g}$ is the Lie algebra associated with the Lie group $G$. 
Substituting the comparator $U_\gamma$ into Eq.~(\ref{eq:Dphi}), the covariant derivative can be written in terms of the exterior derivative $d$ and the gauge 1-form $A$ in a coordinate-independent way
\begin{equation}
\label{eq:connection}
D=d-ieA,
\end{equation}
which becomes $D_\mu=\partial_{\mu}-ieA_\mu$ in a local coordinate patch of the spacetime manifold. 
From the transformation rule of the covariant derivative [Eq.~(\ref{eq:D_transform})], it is easy to see that the transformation rule of $A$ under the action of $g\in G$ is 
\begin{equation}
\label{eq:A_transform}
ieA\rightarrow ieA'=g(ieA)g^{-1}+(dg)g^{-1}.
\end{equation} 
Using the Newton-Leibniz formula for the line integral $\exp(\int_{a}^bdx\; x^{-1})=\exp[\ln(ba^{-1})]=ba^{-1}$, and the cyclic identity $\text{tr}(abc)=\text{tr}(bca)$ of the trace operator, we see the Wilson's line [Eq.~(\ref{eq:comparator})] transforms as the comparator [Eq.~(\ref{eq:G_U})]. 
The covariant derivative is a linear operator $D(\phi_1+\phi_2)=D\phi_1+D\phi_2$, and satisfies the  Leibniz's rule $D(f\phi_1)=fD\phi_1+df\otimes \phi_1$ for any smooth sections $\phi_1$ and $\phi_2$ and smooth function $f$. 
Therefore, in the language of differential geometry, the covariant derivative defines a connection on the fiber bundle whose connection 1-form is the gauge field and the parallel  transport $D(x)U(x,y)=0$ is induced by the comparator. 
From the physics standpoint, it is sufficient to recognize that under local U(1)-gauge transformation
\begin{equation}\label{Gauge}
\phi\rightarrow e^{ie\alpha}\phi,\quad A_{\mu}\rightarrow A_{\mu}+\partial_{\mu}\alpha,
\end{equation}
the gauge covariant derivative $D_\mu\phi\rightarrow e^{ie\alpha}D_\mu\phi$ 
transforms as the complex scalar field. Therefore, a kinetic term of the form $(D_\mu\phi)^*(D^\mu\phi)$ is what we need to ensure that the Lagrangian remains invariant under the local U(1) symmetry.

From the above discussion, we see that in order for charged fields to be dynamical, it is necessary that the gauge field is also dynamical. To see what term in the Lagrangian governs the dynamics of the gauge field, notice that the Wilson's line $U_\gamma$ [Eq.~(\ref{eq:comparator})] defines a parallel transport on the fiber bundle that is path dependent. In the mathematical language, the connection $D$ [Eq.~(\ref{eq:connection})] has curvature. 
To see the effect of curvature, one way is to compare using two paths, or equivalently around a loop
\begin{equation}
U_{\partial S}=\text{tr}\mathcal{P}\exp\big(ie\oint_{\partial S} A\big)=\text{tr}\mathcal{P}\exp\big(ie\int_{S} dA\big),
\end{equation}
which is usually called the Wilson's loop. Using properties of the exterior derivative and trace, it is easy to see that the Wilson's loop is invariant under the gauge transformation Eq.~(\ref{eq:A_transform}). Alternatively, instead of transporting around a finite size loop, we can compare infinitesimal transport along two directions. For example, we can ask what is the difference if we first transport in $X$ direction and then in $Y$ direction, versus if we first transport in $Y$ direction and then in $X$ direction, where $X$ and $Y$ are two vector fields. The infinitesimal difference is given by the curvature function 
\begin{equation}
\label{eq:F_curvatureXY}
-ieF(X,Y)=D_XD_Y-D_YD_X-D_{[X,Y]},
\end{equation}
where the factor $-ie$ is inserted by convention, $D$ is the connection [Eq.~(\ref{eq:connection})], and $[X,Y]$ is the Lie bracket, which measures the intrinsic non-communicativeness of the two vector fields $X$ and $Y$. The difference $-ieF(X,Y)$ then measures the infinitesimal curvature of the parallel transport itself. Since $F(X,Y)=-F(Y,X)$ is a skew-linear map, it is a differential 2-form, known as the curvature form of the connection $D$. To find a formula for the curvature form, we can contract $F$ with two vector fields and evaluate Eq.~(\ref{eq:F_curvatureXY}). Equivalently, we can consider how the curvature 2-form, which is a skew-symmetric generalization of the second-order derivative, acts on smooth sections of the fiber bundle. Taking the covariant derivative twice, the curvature 2-form acts on a smooth section $\phi$ by
\begin{eqnarray}
\nonumber
-ieF\phi&=&D\wedge D\phi\\
\nonumber
&=&d^2\phi-ied(A\phi)-ieA\wedge d\phi-e^2A\wedge A\phi\\
&=&-ie\big(dA-ieA\wedge A\big)\phi,
\end{eqnarray}
where $A\wedge B=[A,B]$ denotes the exterior product on $T^*P$ and the Lie bracket on $\mathfrak{g}$. To obtain the last line, I have used properties of the exterior derivative $d^2=0$ and $d(a\wedge b)=(da)\wedge b+(-1)^a a\wedge(db)$, where $(-1)^a$ is the degree of the differential form $a$. In particular, for 1-form $A$, the sign is $(-1)^1=-1$. Since the above expression holds for any smooth section, we have thus obtain a formula for the curvature 2-form
\begin{equation}
\label{eq:eq:F_curvature}
F=dA-ieA\wedge A.
\end{equation}
To see how the curvature 2-form transform under the group action, we can use the transformation rule for $D$ [Eq.~(\ref{eq:D_transform})]. Then $-ieF\rightarrow -ieF'=D'\wedge D'=(gDg^{-1})\wedge(gDg^{-1})=gD\wedge Dg^{-1}$, so that the curvature form is transformed under the group action by
\begin{equation}
\label{eq:F_transform}
F\rightarrow gFg^{-1}
\end{equation}
Equivalently, this transformation rule can also be obtained using the transformation rule for $A$ [Eq.~(\ref{eq:A_transform})].
In the special case $G=U(1)$ is an abelian group, the Lie algebra is trivial $[A,A]=0$, so we simply have $F=dA$. In local coordinate, $A=A_\mu dx^\mu$, and $F=dA=\partial_\nu A_\mu dx^\nu\wedge dx^\mu=\frac{1}{2}(\partial_{\mu}A_\nu-\partial_{\nu}A_\mu)dx^\mu\wedge dx^\nu$, namely, components of the curvature tensor for U(1)-gauge theory are
\begin{equation}
F_{\mu\nu}=\partial_{\mu}A_\nu-\partial_{\nu}A_\mu,
\end{equation}
which are nothing but the gauge-invariant electromagnetic fields that transform trivially under the group action by $F_{\mu\nu}\rightarrow F_{\mu\nu}$. Hence, we see electromagnetic fields are in fact components of the curvature tensor of the U(1) connection. Moreover, by straightforward calculations, one can show that the relativistic Newton's equation for point charged particle under the Lorentz force is in fact the geodesic equation $\ddot{x}^{\mu}=\Gamma^{\mu}_{\alpha\beta}\dot{x}^\alpha\dot{x}^\beta$, where the Christoffel symbols $\Gamma^{\mu}_{\alpha\beta}$ is now due to the curvature of electromagnetic fields. In Einstein's language, instead of treating particle's motion using Newton's second law, one can also think of the particle as a freely moving particle traveling along geodesics on a curved manifold.

Now we have all the terms needed in the action $S=\int d^4x\,\mathcal{L}$. To summarize, first, we start with a massive charged scalar field, whose mass term $\phi^*\phi$ is in fact the norm of the scalar field. Second, we demand that the complex scalar field to have kinetic energy. Then, the local U(1)-gauge symmetry requires that kinetic term of $\phi$ to be $(D_\mu\phi)^*g^{\mu\nu}(D_\nu\phi)$, which is in fact the Minkowski norm of the covariant derivative $g(D\phi,D\phi)$. Third, upon introducing the covariant derivative, we inevitably need a gauge field $A_\mu$. Analogously, we can write the kinetic term of the gauge field as the norm of the curvature 2-form $g(F,F)=\text{tr}(F\wedge\star F)$, where $\star F$ denotes the hodge dual of $F$. It is obvious from the transformation rule Eq.~(\ref{eq:F_transform}) that $g(F,F)$ is gauge invariant. Finally, up to some normalization convention, the Lagrangian density can be written as
\begin{eqnarray}
\label{eq:SQED_Lagrangian}
\mathcal{L}=(D_{\mu}\phi)^{*}(D^{\mu}\phi)-m^2\phi^{*}\phi - U(\phi^*\phi)-\frac{1}{4}F_{\mu\nu}F^{\mu\nu},
\end{eqnarray}
where $U(\phi^*\phi)$ is some potential of the complex scalar field.
The physics model defined by the above Lagrangian with $U=0$ is usually called scalar QED. 
Let us check the mass dimension of terms in the scalar QED Lagrangian.
In the natural units, the action is a dimensionless number. Since space and time have mass dimension $M^{-1}$, the Lagrangian must have mass dimension $M^4$. As mentioned earlier, the $\phi$ field has mass dimension $M$, so the kinetic term $|D\phi|^2$ has the correct dimension $M^4$. 
In the covariant derivative, the gauge field $A$ has mass dimension $M$, so its kinetic term $F\wedge\star F$ also has the correct dimension $M^4$. Another term that has the same dimension is $F\wedge F$. However, this term is skew symmetric and violate the discrete time-reversal and parity symmetry of the Lorentz group. 
The quadratic term $|\phi|^2$ has mass dimension $M^2$, and after multiplying by $m^2$, the mass term $m^2|\phi|^2$ thereof has the correct dimension. Notice that a simple mass term of the gauge field of the form $m^2A_\mu A^\mu$ is not allowed by the gauge symetry. 
%
%
The scalar-QED Lagrangian [Eq.~(\ref{eq:SQED_Lagrangian})] is the most general renormalizable Lagrangian that respects both the local U(1)-gauge symmetry and the Lorentz symmetry, where the Lorentz group is represented by the scalar field.

The classical equations of motion of scalar QED are the Euler--Lagrangian equations, which extremize the action $\delta S=0$. For the scalar field, regarding $\phi$ and $\phi^*$ as independent fields, then the scalar field satisfies the Euler--Lagrangian equation
\begin{eqnarray}
\label{eq:EOM_phi}
\nonumber
0&=&\partial_{\mu}\frac{\partial\mathcal{L}}{\partial(\partial_\mu\phi^*)}-\frac{\partial\mathcal{L}}{\partial \phi^*}\\
\nonumber
&=&\partial_{\mu}(D^\mu\phi)-\big[ ieA_\mu(D^\mu\phi)-m^2\phi-\frac{\lambda}{2}\phi^*\phi^2\big]\\
&=&D_\mu D^\mu\phi+m^2\phi+\frac{\lambda}{2}\phi^*\phi^2,
\end{eqnarray}
which transforms by an overall factor of $\exp(ie\alpha)$, and is thereof invariant under local U(1)-gauge symmetry, as well as the Lorentz symmetry. 
The equation for $\phi^*$ is the complex conjugate of the above equation. Similarly, taking variation with respect to $A_\nu$, The Euler--Lagrangian equation for the gauge field
\begin{eqnarray}
\label{eq:EOM_A}
\nonumber
0&=&\partial_{\mu}\frac{\partial\mathcal{L}}{\partial(\partial_\mu A_\nu)}-\frac{\partial\mathcal{L}}{\partial A_\nu}\\
\nonumber
&=&\partial_{\mu}\big(-\partial^\mu A^\nu+\partial^\nu A^\mu\big)-\big[ie\phi^*(D^\mu\phi)-ie\phi(D^\mu\phi)^*\big]\\
&=&-\partial_\mu F^{\mu\nu}+J^\nu.
\end{eqnarray}
These are the two nontrivial Maxwell's equations, namely, the Gauss' law and the Maxwell--Amp\`ere's law,  with the 4-current density
\begin{equation}
\label{eq:Jmu}
J^{\mu}=\frac{e}{i}[\phi^*(D^\mu\phi)-\phi(D^\mu\phi)^*\big].
\end{equation}
The other two other Maxwell's equations, namely, $\nabla\cdot\mathbf{B}=0$ and the Faraday's law, are simply the geometric identity $d^2F=0$. It is obvious that both the current density $J^\mu$ and the Maxwell's equations are invariant under local U(1)-gauge transformations.

By Noether's theorem, symmetries of the action correspond to conserved quantities. Now for the scalar-QED model, under the global U(1) symmetry where $\alpha$ is a constant, the infinitesimal transformation is $\delta\phi=ie\alpha\phi$. Using the Euler--Lagrangian equation, the infinitesimal change of the Lagrangian under the global U(1) symmetry is
\begin{eqnarray}
\nonumber
0=\delta\mathcal{L}&=&\frac{\partial\mathcal{L}}{\partial \phi}\delta\phi+\frac{\partial\mathcal{L}}{\partial(\partial_\mu\phi)}\delta\partial_\mu\phi+\text{c.c.}\\
\nonumber
&=&\partial_{\mu}\Big(\frac{\partial\mathcal{L}}{\partial(\partial_\mu\phi)}\delta\partial_\mu\phi +\text{c.c.}\Big)\\
&=&\partial_{\mu}\big[(D^\mu\phi)^*ie\alpha\phi+ \text{c.c.}\big].
\end{eqnarray}
We see the conserved Noether's current is proportional to the 4-current density [Eq.~(\ref{eq:Jmu})]. Using Eq.~(\ref{eq:EOM_phi}) of the $\phi$ field, it is a straightforward calculation to check that $\partial_{\mu}J^\mu=0$ is guaranteed. Equivalently, using Eq.~(\ref{eq:EOM_A}), $\partial_{\mu}J^\mu=\partial_\mu\partial_\nu F^{\mu\nu}=0$, because $F^{\mu\nu}$ is an antisymmetric tensor while the derivatives $\partial_\mu\partial_\nu$ are symmetric. In addition to the charge current, another Noether's current comes from the translational symmetry. Under translation $x\rightarrow x-a$, functions transform by pullback $f(x)\rightarrow f(x+a)$, and the infinitesimal transformation is $\delta f=a^\mu\partial_\mu f$. In particular, using the Euler--Lagrangian equations, the Lagrangian function transforms under the translational symmetry by
\begin{eqnarray}
\nonumber
0&=&\delta\mathcal{L}-a^\mu\partial_\mu\mathcal{L}\\
&=&a^\nu\partial_{\mu}\Big[\frac{\partial\mathcal{L}}{\partial(\partial_\mu\phi)}\partial_\nu\phi+ \frac{\partial\mathcal{L}}{\partial(\partial_\mu\phi^*)}\partial_\nu\phi^* + \frac{\partial\mathcal{L}}{\partial(\partial_\mu A_\lambda)}\partial_\nu A_\lambda-\delta^{\mu}_{\nu}\mathcal{L}\Big].
\end{eqnarray}
Then, up to some constant, the term in the bracket $\hat{\mathcal{T}}^{\mu}_{\phantom{\mu}\nu}$ is the conserved Noether's current. 
Let us add some terms to it and construct a symmetric tensor that is well-transformed under the Lorentz symmetry and the U(1)-gauge symmetry. The symmetrized stress-energy tensor
\begin{eqnarray}
\label{eq:stress-energy}
\mathcal{T}^{\mu\nu}&=&(D^{\mu}\phi)^*(D^{\nu}\phi)+(D^{\mu}\phi)(D^{\nu}\phi)^*+F^{\mu\lambda}g_{\lambda\sigma}F^{\sigma\nu}-g^{\mu\nu}\mathcal{L}.
\end{eqnarray}
The time-time component $\mathcal{T}^{00}=|D_{0}\phi|^2+|D_i\phi|^2+m^2|\phi|^2+\frac{\lambda}{4}|\phi|^4+\frac{1}{2}(\mathbf{E}^2+\mathbf{B}^2)$ is the Hamiltonian density, and the space-time component $\mathcal{T}^{0i}=(D^0\phi)^*(D^i\phi)+(D^i\phi)^*(D^0\phi)+(\mathbf{E}\times\mathbf{B})^i$ is the momentum density. 
Therefore, the symmetric tensor Eq.~(\ref{eq:stress-energy}) is in fact the stress-energy tensor, whose space-space component $\mathcal{T}^{ij}$ are the stress tensor. To see energy and momentum are conserved, notice that the stress-energy tensor is related to the Noether's current by
\begin{eqnarray}
\nonumber
\mathcal{T}^{\mu}_{\phantom{\mu}\nu}&=&\hat{\mathcal{T}}^{\mu}_{\phantom{\mu}\nu}+(D^\mu\phi)^*ieA_\nu\phi-(D^\mu\phi)ieA_\nu\phi^*-F^{\mu\lambda}(\partial_\lambda A_\nu)\\
&=&\hat{\mathcal{T}}^{\mu}_{\phantom{\mu}\nu}+\partial_\lambda (A_\nu F^{\lambda\mu})+A_\nu (J^\mu-\partial_\lambda F^{\lambda\mu}).
\end{eqnarray}
Using the Maxwell's equation [Eq.~(\ref{eq:EOM_A})], the last two terms cancel, and the difference between $\mathcal{T}^{\mu}_{\phantom{\mu}\nu}$ and $\hat{\mathcal{T}}^{\mu}_{\phantom{\mu}\nu}$ is a total derivative. Moreover, since $\partial_{\mu}\hat{\mathcal{T}}^{\mu}_{\phantom{\mu}\nu}=0$ is a conserved current, we have $\partial_{\mu}\mathcal{T}^{\mu}_{\phantom{\mu}\nu}=\partial_{\mu}\partial_\lambda (A_\nu F^{\lambda\mu})=0$, because $F^{\lambda\mu}$ is an antisymmetric tensor, while the derivatives $\partial_{\mu}\partial_\lambda$ are symmetric. Therefore, energy and momentum of scalar QED are both conserved locally at the classical level, which can also be shown by straightforward calculations using the Euler--Lagrangian equations.

\section{Second quantization\label{sec:vacuum:quantization}}

In this section, I will approach the field theory using the standard method of second quantization. This method is developed from our familiarities with quantum harmonic oscillators, which can be described in terms of the creation and annihilation operators instead of the wave functions. Since quantum field theory is usually concerned with states that are asymptotically free, the exact details of wave functions are usually not important. Therefore, second quantization, which promotes classical fields to operators, hides unnecessary details and enables a conceptually clean treatment of the field theory. 

\subsection{Canonical quantization\label{sec:vacuum:quantization:canonical}}


To see how to quantized the fields, let us mimic what is done for quantum harmonic oscillators. The simple oscillator field is a function of time only. Hence, let represent the complex scalar field using Fourier expansion
\begin{equation}
\label{eq:phi_tx}
\phi(t,\mathbf{x})=\int \frac{d^3\mathbf{p}}{(2\pi)^3} e^{i\mathbf{p}\cdot\mathbf{x}}\phi_\mathbf{p}(t).
\end{equation}
Substituting the Fourier expansion into the KG equation [Eq.~(\ref{eq:KG})], then each time-dependent Fourier amplitude must satisfy
\begin{equation}
(d_t^2+\mathbf{p}^2+m^2)\phi_\mathbf{p}(t)=0.
\end{equation} 
This is the equation for a simple harmonic oscillator, whose displacement is $\phi_\mathbf{p}(t)$ and frequency is $E_\mathbf{p}=\sqrt{\mathbf{p}^2+m^2}$. The Lagrangian of the oscillator is $L_\mathbf{p}=\dot{\phi}^\dagger_\mathbf{p} \dot{\phi}_\mathbf{p} -E_\mathbf{p}^2\phi_\mathbf{p}^\dagger\phi_\mathbf{p}$. Instead of the Lagrangian, the usual quantum mechanics works with the Hamiltonian
\begin{equation}
\label{eq:H_pt}
H_\mathbf{p}(t)=\dot{\phi}_\mathbf{p}^\dagger(t) \dot{\phi}_\mathbf{p}(t) +  E_\mathbf{p}^2 \phi_\mathbf{p}^\dagger(t) \phi_\mathbf{p}(t),
\end{equation}
where $\dot{\phi}_\mathbf{p}:=d_t \phi_\mathbf{p}$ denotes the time derivative. In the Hamiltonian formalism, the canonical momentum of $\phi_\mathbf{p}$ is $\pi_\mathbf{p} =\partial L_\mathbf{p}/\partial \dot{\phi}_\mathbf{p}=\dot{\phi}_\mathbf{p}^\dagger$. Now that $\phi_\mathbf{p}$ and $\pi_\mathbf{p}$ are a pair of canonical coordinates in the phase space, analogous to the commutation relation $[x,p_x]=i$, we can impose canonical quantization rules
\begin{eqnarray}
\label{eq:phi_pi}
&&[\phi_\mathbf{p}(t), \pi_\mathbf{q}(t)]=i(2\pi)^3\delta^{(3)}(\mathbf{p}-\mathbf{q}),\\
&&[\phi_\mathbf{p}(t), \pi_\mathbf{q}^\dagger(t)]=0,\\
&&[\phi_\mathbf{p}(t), \phi_\mathbf{q}(t)]=[\phi_\mathbf{p}(t), \phi_\mathbf{q}^\dagger(t)]=0, \\
\label{eq:phi_phi}
&&[\pi_\mathbf{p}(t), \pi_\mathbf{q}(t)]=[\pi_\mathbf{p}(t), \pi_\mathbf{q}^\dagger(t)]=0.
\end{eqnarray}
where $(2\pi)^3$ is a phase space factor accompanying the Fourier transform, and the delta function is reminiscent of the fact that waves with different momentum are independent. To obtain other commutation relations, notice that for operators, the Hermitian conjugate $(AB)^\dagger=B^\dagger A^\dagger$. Therefore, the Hermitian conjugation of the commutator $[A,B]^\dagger=-[A^\dagger,B^\dagger]$ is the negative of the commutator of Hermitian conjugates.

Having quantized the Fourier amplitudes as simple harmonic oscillators, the canonical quantization rules for fields in the configuration space can thereof be determined. Since the canonical momentum of the $\phi$ field is $\pi=\partial\mathcal{L}_0/\partial\dot{\phi}=\dot{\phi}^\dagger$, the canonical momentum has Fourier expansion
\begin{equation}
\label{eq:pi_tx}
\pi(t,\mathbf{x})=\int \frac{d^3\mathbf{p}}{(2\pi)^3} e^{-i\mathbf{p}\cdot\mathbf{x}}\pi_\mathbf{p}(t).
\end{equation}
Notice that the negative sign in the Fourier exponent is necessary in order for $\pi=\dot{\phi}^\dagger$ to be consistent with $\pi_\mathbf{p} =\dot{\phi}_\mathbf{p}^\dagger$. Then, the equal-time commutation relation
\begin{eqnarray}
\label{eq:canonical_phi_pi}
\nonumber
[\phi(t,\mathbf{x}),\pi(t,\mathbf{y})]&=&\int \frac{d^3\mathbf{p}}{(2\pi)^3} \frac{d^3\mathbf{q}}{(2\pi)^3} e^{i\mathbf{p}\cdot\mathbf{x}-i\mathbf{q}\cdot\mathbf{y}} [\phi_\mathbf{p}(t), \pi_\mathbf{q}(t)] \\
\nonumber
&=&\int \frac{d^3\mathbf{p}}{(2\pi)^3} \frac{d^3\mathbf{q}}{(2\pi)^3} e^{i\mathbf{p}\cdot\mathbf{x}-i\mathbf{q}\cdot\mathbf{y}} i(2\pi)^3\delta^{(3)}(\mathbf{p}-\mathbf{q}) \\
&=&i\delta^{(3)}(\mathbf{x}-\mathbf{y}).
\end{eqnarray}
Following similar calculations, it is easy to verify that the other commutation relations are trivial. For example, the equal-time commutations
\begin{equation}
[\phi(t,\mathbf{x}),\phi(t,\mathbf{y})]=0.
\end{equation}
We see the quantization rules in the configuration space is also canonical, where the displacement $\phi(t,\mathbf{x})$ and the momentum $\pi(t,\mathbf{x})$ are now the canonical coordinate pair in the phase space. 

Having promoted fields to operators, the dynamics in the phase space is determined by the Hamiltonian. For free fields, the Hamiltonian can be found by summing up contributions from all oscillators in the momentum space. Substituting the inverse Fourier transforms of Eqs.~(\ref{eq:phi_tx}) and (\ref{eq:pi_tx}) into the momentum space Hamiltonian [Eq.~(\ref{eq:H_pt})], we have
\begin{eqnarray}
	\label{eq:H0}
	\nonumber
	H_0(t)&=&\int \frac{d^3\mathbf{p}}{(2\pi)^3} H_\mathbf{p}(t) \\
	\nonumber
	&=& \int \frac{d^3\mathbf{p}}{(2\pi)^3} d^3\mathbf{x} d^3\mathbf{y} \Big[\pi(t,\mathbf{y})\pi^\dagger(t,\mathbf{x}) + \phi^\dagger(t,\mathbf{y})\phi(t,\mathbf{x}) \Big(\nabla_\mathbf{x}\cdot\nabla_\mathbf{y}+m^2 \Big)\Big] e^{i\mathbf{p}\cdot(\mathbf{y}-\mathbf{x})} \\
	&=& \int d^3\mathbf{x} \Big[ \pi(t,\mathbf{x})\pi^\dagger(t,\mathbf{x})+ \nabla\phi^\dagger(t,\mathbf{x})\cdot\nabla\phi(t,\mathbf{x}) +m^2\phi^\dagger(t,\mathbf{x})\phi(t,\mathbf{x}) \Big].
\end{eqnarray}
Alternatively, the Hamiltonian can be computed from the Lagrangian using the Legendre transformation. From the Lagrangian density $\mathcal{L}_0$, the Hamiltonian density of the free field is
\begin{eqnarray}
\nonumber
\mathcal{H}_0&=&\frac{\partial\mathcal{L}_0}{\partial\dot{\phi}}\dot{\phi} +\dot{\phi}^\dagger\frac{\partial\mathcal{L}_0}{\partial\dot{\phi}^\dagger}-\mathcal{L}_0 \\
&=&\dot{\phi}^\dagger\dot{\phi}+(\nabla\phi^\dagger)\cdot(\nabla\phi)+m^2\phi^\dagger\phi.
\end{eqnarray} 
After identifying the canonical momentum of the $\phi$ field as $\pi=\partial\mathcal{L}_0/\partial\dot{\phi}=\dot{\phi}^\dagger$, and the canonical momentum of the $\phi^\dagger$ field as $\pi^\dagger=\partial\mathcal{L}_0/\partial\dot{\phi}^\dagger=\dot{\phi}$, the total Hamiltonian $H_0(t)=\int d^3\mathbf{x} \mathcal{H}_0(t,\mathbf{x})$ is then in agreement with the Fourier space computation.

As a functional of the quantized fields, the Hamiltonian is now also a quantized operator. In the Heisenberg picture, the equation of motion for the operator $\mathcal{O}$ is given by the Heisenberg equation $i\partial_t\mathcal{O}=[\mathcal{O}, H]$, where $H$ is the Hamiltonian operator. For free fields, the Hamiltonian is simply $H_0$. Using the identity $[A,BC]=[A,B]C+B[A,C]$, the Heisenberg equation for the $\phi$ field 
\begin{eqnarray}
\label{eq:dt_phi}
\nonumber
i\partial_t\phi(t,\mathbf{x})&=&[\phi(t,\mathbf{x}), H_0(t)]\\
\nonumber
&=&\int d^3\mathbf{y} \pi^\dagger(t,\mathbf{y}) [\phi(t,\mathbf{x}), \pi(t,\mathbf{y})]\\
&=&i \pi^\dagger(t,\mathbf{x}),
\end{eqnarray}  
which recovers the relation $\pi^\dagger=\dot{\phi}$. Similarly, after integration by part, the Heisenberg equation for the $\pi^\dagger$ field can be easily found
\begin{eqnarray}
\label{eq:dt_pi}
\nonumber
i\partial_t\pi^\dagger(t,\mathbf{x})&=&[\pi^\dagger(t,\mathbf{x}), H_0(t)]\\
\nonumber
&=&\int d^3\mathbf{y} (-\nabla^2\phi(t,\mathbf{y})+m^2\pi(t,\mathbf{y})) [\pi^\dagger(t,\mathbf{x}), \phi^\dagger(t,\mathbf{y})]\\
&=&i (\nabla^2-m^2)\phi(t,\mathbf{x}),
\end{eqnarray}  
which recovers the KG equation after identifying $\pi^\dagger=\dot{\phi}$. In other words, $\phi$ as an operator also satisfies the KG equation.

Since the free $\phi$ field satisfies the KG equation, the general solution to this operator-valued partial differential equation is 
\begin{equation}
\label{eq:phi0_opt}
\phi(x)=\int\frac{d^3\mathbf{p}}{(2\pi)^3}\frac{1}{\sqrt{2E_\mathbf{p}}} \Big(a_{\mathbf{p}}e^{-ipx}+b^\dagger_{\mathbf{p}}e^{ipx}\Big).
\end{equation}
The above expression is identical to Eq.~(\ref{eq:phi0}), except now $a_{\mathbf{p}}$ and $b^\dagger_{\mathbf{p}}$ are some spacetime-independent Heisenberg picture operators. To see what commutation relations these operators must satisfy, we can used the canonical quantization rules in the momentum space. First, taking Fourier transform, we have
\begin{eqnarray}
\phi_\mathbf{p}(t)&=&\int d^3\mathbf{p} e^{-i\mathbf{p}\cdot\mathbf{x}} \phi(t,\mathbf{x})=\frac{1}{\sqrt{2E_\mathbf{p}}} \Big(a_{\mathbf{p}}e^{-iE_\mathbf{p}t}+b^\dagger_{-\mathbf{p}}e^{iE_\mathbf{p}t}\Big), \\
\pi^\dagger_\mathbf{p}(t)&=&\int d^3\mathbf{p} e^{-i\mathbf{p}\cdot\mathbf{x}} \dot{\phi}(t,\mathbf{x})=-i\sqrt{\frac{E_\mathbf{p}}{2}} \Big(a_{\mathbf{p}}e^{-iE_\mathbf{p}t}-b^\dagger_{-\mathbf{p}}e^{iE_\mathbf{p}t}\Big). 
\end{eqnarray}
Second, mimicking what is done for simple quantum harmonic oscillators, we can solve for operators $a_{\mathbf{p}}$ and $b^\dagger_{-\mathbf{p}}$ in terms of the displacement operator $\phi_\mathbf{p}(t)$ and the momentum operator $\pi^\dagger_\mathbf{p}(t)$ as
\begin{eqnarray}
a_{\mathbf{p}}&=&\Big[\sqrt{\frac{E_\mathbf{p}}{2}}\phi_\mathbf{p}(t) +\frac{i}{\sqrt{2E_\mathbf{p}}}\pi^\dagger_\mathbf{p}(t)\Big] e^{iE_\mathbf{p}t}, \\
b^\dagger_{-\mathbf{p}}&=&\Big[\sqrt{\frac{E_\mathbf{p}}{2}}\phi_\mathbf{p}(t) -\frac{i}{\sqrt{2E_\mathbf{p}}}\pi^\dagger_\mathbf{p}(t)\Big] e^{-iE_\mathbf{p}t}.
\end{eqnarray}
Finally, using commutation rules for operators $\phi_\mathbf{p}(t)$ and $\pi^\dagger_\mathbf{p}(t)$ [Eqs.~(\ref{eq:phi_pi})-(\ref{eq:phi_phi})], it is a straightforward calculation to find the commutation rules for operators $a_{\mathbf{p}}$ and $b^\dagger_{-\mathbf{p}}$
\begin{eqnarray}
\label{eq:ab}
&&[a_{\mathbf{p}}, a^\dagger_{\mathbf{q}}]=[b_{\mathbf{p}}, b^\dagger_{\mathbf{q}}]=(2\pi)^2 \delta^{(3)}(\mathbf{p}-\mathbf{q}), \\
&&[a_{\mathbf{p}}, b^\dagger_{\mathbf{q}}]=[a_{\mathbf{p}}, b_{\mathbf{q}}]=0, \\
&&[a_{\mathbf{p}}, a_{\mathbf{q}}]=[b^\dagger_{\mathbf{p}}, b^\dagger_{\mathbf{q}}]=0,
\end{eqnarray}
and other commutation relations are Hermitian conjugations of the above rules. The operators $a^\dagger_{\mathbf{p}}$ and $b^\dagger_{\mathbf{p}}$ are called the creation operators, and the operators $a_{\mathbf{p}}$ and $_{\mathbf{p}}$ are called the annihilation operators.

To see the physical meaning of the creation and annihilation operators, we can compute a number of observables (Appendix~\ref{ch:append:HPQ}). First, the Hamiltonian, namely the total energy operator [Eq.~(\ref{eq:energy_opt})], can be expressed as 
\begin{equation}
\label{eq:H0_energy}
H_0=\int\frac{d^3\mathbf{p}}{(2\pi)^3} E_\mathbf{p} (a^\dagger_{\mathbf{p}}a_{\mathbf{p}}+b^\dagger_{\mathbf{p}}b_{\mathbf{p}}).
\end{equation}
The total energy equals to the sum of energy carried by all the particles \mbox{$H_0=\sum_\mathbf{p} E_\mathbf{p} (N^a_\mathbf{p}+N^b_\mathbf{p})$}, where $N^a_\mathbf{p}$ is the number of particles of type $a$, $N^b_\mathbf{p}$ is the number of particles of type $b$, and both types of particles have energy $E_\mathbf{p}$ when their momentum is $\mathbf{p}$. 
Therefore, we see $a^\dagger_{\mathbf{p}}a_{\mathbf{p}}=N^a_\mathbf{p}$ and $b^\dagger_{\mathbf{p}}b_{\mathbf{p}}=N^b_\mathbf{p}$ are the number operators.
To confirm the interpretation that $\mathbf{p}$ is related to the momentum of particles, we can calculate the momentum operator [Eq.~(\ref{eq:momentum_opt})], which can be expressed as
\begin{equation}
\mathbf{P}=\int\frac{d^3\mathbf{p}}{(2\pi)^3} \mathbf{p} (a^\dagger_{\mathbf{p}}a_{\mathbf{p}}+b^\dagger_{\mathbf{p}}b_{\mathbf{p}}).
\end{equation}
As expected, the total momentum $\mathbf{P}$ is the sum of momentum carried by all particles $\mathbf{P}=\sum_\mathbf{p} \mathbf{p} (N^a_\mathbf{p}+N^b_\mathbf{p})$. 
From the above discussion, it is apparent that type $a$ and type $b$ particles have the same mass. Nevertheless, they are different types of particles, because they carry the opposite charge. To see this, we can compute the charge operator [Eq.~(\ref{eq:charge_opt})], which can be expressed as
\begin{equation}
Q=e\int\frac{d^3\mathbf{p}}{(2\pi)^3} (-a^\dagger_{\mathbf{p}}a_{\mathbf{p}}+b^\dagger_{\mathbf{p}}b_{\mathbf{p}}).
\end{equation}
We see type $a$ particles have negative charge $-e$ as ``electrons", while type $b$ particles have positive charge $+e$ as ``positrons", except that these particle, satisfying commutation relations, are now charged bosons. Since the complex scalar field $\phi$ has no intrinsic degree of freedom, the intrinsic angular momentum is trivially zero. Therefore, we see the complex scalar field describes spin-0 charged bosons, whose only quantum numbers are the mass $m$ and the charge $\pm e$. It is conventional to refer to $a$ as particles, then $b$ are antiparticles, which have the same mass but the opposite charge.
Up to this point, quantization does not play an essential role and the preceding results can also be derived from the classical field theory, which already provides useful machineries for plasma physics \citep{Dodin2014Geometric}.

Now that we understand the physical meaning of the creation and annihilation operators, we can construct the quantum Hilbert space. Similar to what is done for quantum harmonic oscillators, using the bra-–ket notation, we can define the vacuum state $|0\rangle$ as the null space of the annihilation operators
\begin{equation}
\label{eq:ab_vacuum}
a_{\mathbf{p}}|0\rangle=b_{\mathbf{p}}|0\rangle=0.
\end{equation}
Then, the one-particle and one-antiparticle eigenstates with momentum $\mathbf{p}$ can be created out of the vacuum by their creation operators
\begin{eqnarray}
|\mathbf{p}\rangle&=&\sqrt{2E_\mathbf{p}}a^\dagger_{\mathbf{p}}|0\rangle,\\
\label{eq:ab_one}
|\bar{\mathbf{p}}\rangle&=&\sqrt{2E_\mathbf{p}}b^\dagger_{\mathbf{p}}|0\rangle.
\end{eqnarray}
The normalization of the state is chosen such that the expectation value $\langle\mathbf{q}|\mathbf{p}\rangle=2E_\mathbf{p}(2\pi)^3\delta^{(3)}(\mathbf{p}-\mathbf{q})$ is Lorentz invariant. This combination is invariant because under Lorentz transformations, the energy is boosted by the $\gamma$ factor, while the momentum space volume is contracted by the same $\gamma$ factor.
The quantum Hilbert space is a complete inner product space. The completeness of one-particle states can be expressed in terms of the identity operator
\begin{equation}
\mathbb{I}=\int\frac{d^3\mathbf{p}}{(2\pi)^3}\frac{|\mathbf{p}\rangle\langle\mathbf{p}|}{2E_\mathbf{p}},
\end{equation}
and the identity operator can similarly be written in terms of one-antiparticle states. 
Having constructed the single-boson states, we can construct the Fock space of many identical particles by repeatedly acting the creation operators on the vacuum, similar to how it is done for quantum harmonic oscillators.

Finally, we can now also give an interpretation of the $\phi$ field as a quantized operator. First, similar to how it is done in quantum mechanics, the momentum eigenstates have wave-function representations. By projecting a one-particle state $|\mathbf{p}\rangle$ to the configuration space, one find that its Lorentz-invariant wave function is $\langle\mathbf{p}|\mathbf{x}\rangle=e^{ipx}$. This is a plane wave propagating with wave vector $\mathbf{p}$, whose frequency $E_\mathbf{p}$ is positive in our Fourier convention. Similarly, projecting the one-antiparticle state to configuration space, its wave function $\langle\bar{\mathbf{p}}|\mathbf{x}\rangle=e^{-ipx}$, which has the same direction of propagation as the one-particle state but a negative energy.  
Now using its spectral expansion, when $\phi^\dagger$ acts on the vacuum state
\begin{equation}
\label{eq:phi_dagger_x}
\phi^\dagger(x)|0\rangle=\int\frac{d^3\mathbf{p}}{(2\pi)^3} \frac{1}{2E_\mathbf{p}} e^{ipx} |\mathbf{p}\rangle =\int\frac{d^3\mathbf{p}}{(2\pi)^3}\frac{|\mathbf{p}\rangle\langle\mathbf{p}|\mathbf{x}\rangle}{2E_\mathbf{p}} =|\mathbf{x}\rangle.
\end{equation}
we see the field $\phi^\dagger(x)$ creates a particle at position $\mathbf{x}$ out of the vacuum. Similarly, we can show that $\phi(x)$ creates an antiparticle at position $\mathbf{x}$ out of the vacuum $\phi(x)|0\rangle=|\bar{\mathbf{x}}\rangle$. On the other hand, the meaning of $\phi(x)|\mathbf{y}\rangle$ is not as simple, because $\phi$ annihilates the particle while creates an antiparticle, so the outcome is a linear superposition of the vacuum state and a pair state. In the next subsection, I will introduce notations that can simplify the above physical picture.
\subsection{Feynman propagator and Wick contraction\label{sec:vacuum:quantization:propagator}}

The canonical quantization procedure imposes rules on equal-time commutation relations. As a next step, it is a natural question to ask what happens when the time is different. Using the spectral expansion for $\phi$ [Eq.~(\ref{eq:phi0_opt})], and the commutation relations of the creation and annihilation operators, we have
\begin{eqnarray}
\nonumber
[\phi(x),\phi^\dagger(y)]&=&\int\frac{d^3\mathbf{p}}{(2\pi)^3} \frac{d^3\mathbf{q}}{(2\pi)^3} \frac{1}{2\sqrt{E_\mathbf{p}E_\mathbf{q}}}\Big[a_{\mathbf{p}}e^{-ipx}+b^\dagger_{\mathbf{p}}e^{ipx}, a^\dagger_{\mathbf{q}}e^{iqy}+b_{\mathbf{q}}e^{-iqy}\Big]\\
\nonumber
&=&\int\frac{d^3\mathbf{p}}{(2\pi)^3}  \frac{1}{2E_\mathbf{p}}\Big(e^{ip(y-x)}-e^{ip(x-y)}\Big)\\
&=&D(x-y)-D(y-x),
\end{eqnarray}
where $D(x)$ is the correlation function [Eq.~(\ref{eq:Corr_D})]. Similarly, we can express other commutation relations in terms of the correlation function. 
When placed inside the vacuum bracket, it is easy to see that the correlation function can in fact be written as
\begin{equation}
D(x-y)=\langle 0|\phi(x)\phi^\dagger(y)|0\rangle=\langle 0|\phi^\dagger(x)\phi(y)|0\rangle.
\end{equation}
We see the two-point correlation function is related to the probability amplitude of some two-point processes.
To be more precise, we can write $\langle 0|\phi(x)\phi^\dagger(y)|0\rangle=\langle 0|\phi(x)|\mathbf{y}\rangle$. As mentioned earlier, $\phi(x)|\mathbf{y}\rangle$ is a mixture of the vacuum state and a pair state, so $D(x-y)$ is the projection of this mixed state on the vacuum state. 

To further clarify, one can ask the following physical question about vacuum fluctuations: what is the probability amplitude that a boson emerges at one spacetime coordinate out of the vacuum and then disappears back into the vacuum at some other spacetime coordinate? This question can be experimentally answered, if we put up a detector that sees a boson pops out of the vacuum at time $t=0$ and then vanishes at some later time $t=T$. To compute the probability amplitude of this process, we need to introduce the time-ordering operator $\mathcal{T}$, which instructs us to place operators at later time after those at earlier times. The probability amplitude is then the two-point function
\begin{eqnarray}
\label{eq:two-point}
\nonumber
\langle 0|\mathcal{T}\phi(x)\phi^\dagger(y)|0\rangle&:=& \theta(x^0-y^0)\langle 0|\phi(x)\phi^\dagger(y)|0\rangle + \theta(y^0-x^0)\langle 0|\phi^\dagger(y)\phi(x)|0\rangle\\
\nonumber
&=&\theta(x^0-y^0)D(x-y)+\theta(y^0-x^0)D(y-x)\\
&=&G_F(x,y),
\end{eqnarray} 
where $G_F$ is the Feynman Green's function [Eq.~(\ref{eq:GF_D})]. We see that $G_F$ is the probability amplitude of propagating a vacuum pair fluctuation from one point to another. The time-ordered two-point function [Eq.~(\ref{eq:two-point})] is therefore also called the Feynman propagator. As discussed in Sec.~\ref{sec:vacuum:field:green}, propagations between timelike separations have wavelike amplitude, while propagations between spacelike separations are exponentially suppressed due to causality. 

In fact, the time-ordering operator does not need to be placed inside a vacuum bracket, and it can instead be expressed in terms of Wick contractions and the normal-ordering operator. The normal-ordering operator $\mathcal{N}$ rearranges products of creation and annihilation operators, such that all creation operators are placed on the left of the annihilation operators. For example, $\mathcal{N}(a_\mathbf{p}a^\dagger_\mathbf{q}b^\dagger_\mathbf{k}) =a^\dagger_\mathbf{q}b^\dagger_\mathbf{k}a_\mathbf{p}$. Since the creation operators commute among themselves, the normal-ordering operator is uniquely defined. Using the normal-ordering operator, we can write
\begin{eqnarray}
\nonumber
\phi(x)\phi^\dagger(y)&=&\mathcal{N}\{\phi(x)\phi^\dagger(y)\} +\int\frac{d^3\mathbf{p}}{(2\pi)^3} \frac{d^3\mathbf{q}}{(2\pi)^3} \frac{1}{2\sqrt{E_\mathbf{p}E_\mathbf{q}}}\big[a_{\mathbf{p}}, a^\dagger_{\mathbf{q}}\big] e^{iqy-ipx}\\
&=&\mathcal{N}\{\phi(x)\phi^\dagger(y)\}+D(x-y).
\end{eqnarray}
Then the time-ordered products of two fields
\begin{eqnarray}
\label{eq:T_phi_phi}
\nonumber
\mathcal{T}\phi(x)\phi^\dagger(y)&=&\theta(x^0-y^0)\phi(x)\phi^\dagger(y) + \theta(y^0-x^0) \phi^\dagger(y)\phi(x)\\
&=&\mathcal{N}\{\phi(x)\phi^\dagger(y)\}+G_F(x,y),
\end{eqnarray}
where I have used the property of the Heaviside step function  $\theta(t)+\theta(-t)=1$. Once the above is placed inside the vacuum bracket, $\langle 0|\mathcal{N}\{\phi(x)\phi^\dagger(y)\}|0\rangle=0$, 
and we immediately recover Eq.~(\ref{eq:two-point}).
The Feynman Green's function can also be written in a notation where the fields are explicit. This notation is called the Wick contraction, which is defined on any two operators $\hat{A}$ and $\hat{B}$ such that
\begin{equation}
\contraction{}{\hat{A}}{}{\hat{B}}
\hat{A}\hat{B}=\mathcal{T}(\hat{A}\hat{B})-\mathcal{N}(\hat{A}\hat{B}),
\end{equation}
where the time-ordering operator only applies to the case where both operators $\hat{A}$ and $\hat{B}$ have time dependence. Using the above definition, Eq.~(\ref{eq:T_phi_phi}) immediately gives
\begin{equation}
\contraction{}{\phi}{(x)}{\phi}
\phi(x)\phi^\dagger(y)=
\contraction{}{\phi}{(y)}{\phi}
\phi^\dagger(y)\phi(x)=
G_F(x,y).
\end{equation} 
In other words, we have the operator identity
\begin{equation}
\mathcal{T}\phi(x)\phi^\dagger(y)=\mathcal{N}\{\phi(x)\phi^\dagger(y)+ 
\contraction{}{\phi}{(x)}{\phi}
\phi(x)\phi^\dagger(y)
\}.
\end{equation}
More generally, for products of more than two fields, for example when $n$ fields are involved, it is possible to prove by mathematical induction the following Wick's theorem
\begin{equation}
\label{eq:Wick}
\mathcal{T}\{\phi(x_1)\phi^\dagger(x_2)\dots\phi(x_n)\}=\mathcal{N}\{\phi(x_1)\phi^\dagger(x_2)\dots\phi(x_n)+\text{all possible contractions}\},
\end{equation}
where the phrase ``all possible contractions" means the sum of all possible ways of contracting one pair, two pairs, and up to $\lfloor n/2\rfloor$ pairs. The Wick's theorem enables easy calculation of time-ordered $n$-point Green's function such as $\langle 0|\mathcal{T}\phi(x_1)\phi^\dagger(x_2)\dots\phi(x_n)|0\rangle$, which will appear when interactions between otherwise free fields are present.

Using the Wick contraction, we can also compute other quantities. For example, it is a straightforward calculation to show that 
\begin{equation}
\label{eq:two-point-trivial}
\contraction{}{\phi}{(x)}{\phi}
\phi(x)\phi(y)=
\contraction{}{\phi}{(x)}{\phi}
\phi^\dagger(x)\phi^\dagger(y)=0.
\end{equation}
Moreover, we can contract fields with states. For example, we can annihilate one-particle state $|\mathbf{p}\rangle$ with field $\phi$ using the Wick contraction
\begin{eqnarray}
\nonumber
\contraction{}{\phi}{(x)|}{\mathbf{p}}
\phi(x)|\mathbf{p}\rangle&=&
\contraction{\sqrt{2E_\mathbf{p}}}{\phi}{(x)}{a}
\sqrt{2E_\mathbf{p}}\phi(x)a^\dagger_\mathbf{p}|0\rangle \\
\nonumber
&=&\sqrt{2E_\mathbf{p}}\Big(\mathcal{T}\big\{\phi(x)a^\dagger_\mathbf{p}\big\}-\mathcal{N}\big\{\phi(x)a^\dagger_\mathbf{p}\big\}\Big) |0\rangle\\
\nonumber
&=&\sqrt{2E_\mathbf{p}}\int\frac{d^3\mathbf{q}}{(2\pi)^3}  \frac{1}{\sqrt{2E_\mathbf{q}}} [a_\mathbf{q},a^\dagger_\mathbf{p}] e^{-iqx} |0\rangle \\
&=&e^{-ipx} |0\rangle,
\end{eqnarray}
where $\mathcal{T}\big\{\phi(x)a^\dagger_\mathbf{p}\big\}=\phi(x)a^\dagger_\mathbf{p}$ because $a^\dagger_\mathbf{p}$ has no time dependence.
Similarly, we can annihilate one-antiparticle state $|\bar{\mathbf{p}}\rangle$ with field $\phi^\dagger$ 
\begin{equation}
\contraction{}{\phi}{^\dagger(x)|}{\bar{\mathbf{p}}}
\phi^\dagger(x)|\bar{\mathbf{p}}\rangle=e^{-ipx} |0\rangle.
\end{equation}
On the other hand, it is a straightforward calculation to show that
\begin{eqnarray}
\contraction{}{\phi}{(x)|}{\bar{\mathbf{p}}}
\phi(x)|\bar{\mathbf{p}}\rangle
=
\contraction{}{\phi}{^\dagger(x)|}{\mathbf{p}}
\phi^\dagger(x)|\mathbf{p}\rangle=0.
\end{eqnarray}
Combining the above results with Eq.~(\ref{eq:phi_dagger_x}), we see the $\phi^\dagger$ field creates particles and annihilate antiparticles, while the $\phi$ field creates antiparticles and annihilate particles. 
\subsection{The interaction picture\label{sec:vacuum:quantization:smatrix}}

So far, I have only discussed free fields. 
In this case, the fields can be canonically quantized in the same way as simple harmonic oscillators, and the quantum states are well defined.
However, when the Hamiltonian $H=H_0+V$ becomes more complicated, solutions to the Heisenberg equations no longer describe simple harmonic oscillations. In this case, even if we can solve the nonlinear equations, it is not obvious how we should quantize the fields. 
Moreover, it is not obvious how we should define quantum states, because the quantum Hilbert space is intrinsically a vector space whereas the solution space of nonlinear equations is usually not a vector space.
The difficulties of both quantizing the fields and defining quantum states render most nonlinear problems intractable with second quantization.

The special case we know how to deal with is when $V$ can be considered small in some sense, for which perturbation methods can be applied. In this case, the quantum theory is almost free, except for some weak interactions that happen only within some localized spacetime regions. 
With this physical picture in mind, we can quantized the fields as if they were free, and treat nonlinearities as couplings between these otherwise noninteracting fields. 
Moreover, we can define the notion of asymptotic states, in the sense that they asymptote to eigenstates of simple harmonic oscillators at infinity. These states starts to deform when they come closer to each other and interact. The exact details of how they deform do not matter in a collider experiment, because a detector can measure well-defined states only  when they have flown apart and become asymptotically free again. Therefore, all we need to compute is the transition amplitude from some incoming asymptotic states to some outgoing asymptotic states. These amplitudes can be thought of as elements of some scattering matrix, called the S matrix.

An S matrix element has a simple representation in the Schr\"odinger picture, where the initial state $|i\rangle$ is evolved by the time-evolution operator $U$, so that the transition amplitude to the final state $|f\rangle$ is simply the projection $\langle f|U|i\rangle$. 
To be more precise, in the Schr\"odinger picture, the time-evolution operator $U_S(t,t')$ is an unitary operator that evolves the state from an earlier time $t'$ to a later time $t$ by the full Hamiltonian operator. In other words, the Schr\"odinger equation $i\partial_t|\psi\rangle_S=H|\psi\rangle_S$ is solved by $|\psi(t)\rangle_S=U_S(t,t')|\psi(t')\rangle_S$. Then $U_S(t,t')$, with initial condition $U_S(t,t)=1$, must satisfy the equation
\begin{equation}
i\partial_tU_S(t,t')=H(t)U_S(t,t').
\end{equation}
It is easy to check that this operator equation can be solved by the expansion
\begin{eqnarray}
U_S(t,t')=1+(-i)\int_{t'}^{t}dt_1H(t_1)+(-i)^2\int_{t'}^{t}dt_1 \int_{t'}^{t_1}dt_2 H(t_1)H(t_2)+\dots .
\end{eqnarray}
Using the Fubini's theorem, integrations inside the simplexes can be converted to integration over cubes divided by the symmetry factor
\begin{equation}
\int_{t'}^{t}dt_1 \dots \int_{t'}^{t_{n-1}}dt_{n} H(t_1)\dots H(t_n)=\frac{1}{n!} \int_{t'}^{t}dt_1 \dots dt_n\mathcal{T}\{H(t_1) \dots H(t_n)\},
\end{equation}
where $\mathcal{T}$ is the time-ordering operator and $n!$ is the symmetry factor. Then, the expansion of $U_S(t,t')$ can be written in terms of the time-ordered exponential
\begin{equation}
\label{eq:US}
U_S(t,t')=\mathcal{T}\Big\{ \exp\Big[-i\int_{t'}^{t}H(\tau) d\tau\Big] \Big\}.
\end{equation}
The time-evolution operator satisfies the group properties
\begin{eqnarray}
\label{eq:US_product}
U_S(t_1,t_2)U_S(t_2,t_3)=U_S(t_1,t_3), \\
\label{eq:US_inverse}
U_S^\dagger(t,t')=U_S^{-1}(t,t')=U_S(t',t).
\end{eqnarray}
Unfortunately, since we do not know how the full Hamiltonian should act on states that are only defined asymptotically, the above symbolic solution is of little practical use. What we do know, on the other hand, is how the free Hamiltonian $H_0$ acts on asymptotic states. Therefore, we need to convert the above treatment from the Schr\"odinger picture to a different picture called the interaction picture.

In the interaction picture 
there is an interaction region where fields are coupled, and outside this region fields become free, just as what happens in a collider experiment. The incoming asymptotic state $|i\rangle$ is brought into the interaction region by the free Hamiltonian, and similarly the outgoing asymptotic state $|f\rangle$ is brought away from the interaction region by the free Hamiltonian. Since we know how the free Hamiltonian acts on asymptotic states, the remaining task is to find out what happens within the interaction region, which we shall determine using perturbation theory.

From the mathematical perspective, in the interaction picture, dynamics associated with the free Hamiltonian $H_0$ is factored out, so that we can focus on the interaction Hamiltonian $V$. To avoid confusion, let us denote the time-independent Schr\"odinger operators as $\phi_S$ and $\pi_S$. Then, the Schr\"odinger picture Hamiltonian $H(\phi_S,\pi_S,t)= H_0(\phi_S,\pi_S)+V(\phi_S,\pi_S,t)$, and the Schr\"odinger picture states $|\psi(t)\rangle_S$ evolves according to the Schr\"odinger equation $i\partial_t|\psi(t)\rangle_S=H(\phi_S,\pi_S,t)|\psi(t)\rangle_S$. Now in the interaction picture, states and operators are related to the Schr\"odinger picture states and operators by
\begin{eqnarray}
\label{eq:int_psi}
|\psi(t)\rangle_I&=&e^{iH_0(\phi_S,\pi_S)t}|\psi(t)\rangle_S,\\
\label{eq:int_O}
\mathcal{O}_I(t)&=&e^{iH_0(\phi_S,\pi_S)t}\mathcal{O}_S(t) e^{-iH_0(\phi_S,\pi_S)t}.
\end{eqnarray}
This is different from the Heisenberg picture, because now only the free part $H_0$ is involved, instead of the full Hamiltonian. Since $\exp(iH_0t)$ can be expanded as a polynomial of $H_0$, it commutes with $H_0$. Then, the interaction picture wave function satisfies
\begin{eqnarray}
\label{eq:int_Schrodinger}
\nonumber
i\partial_t|\psi(t)\rangle_I&=&-H_0|\psi(t)\rangle_I+e^{iH_0t} H(t) |\psi(t)\rangle_S\\
&=&V_I(t)|\psi(t)\rangle_I,
\end{eqnarray}
where $V_I(t)=V(\phi_I,\pi_I,t)$ is the interaction picture operator. The above equation is mathematically identical to the Schr\"odinger equation, except that the trivial dynamics associated with $H_0$ has now been removed. Similarly, the interaction picture operator satisfies
\begin{eqnarray}
\label{eq:int_Heisenberg}
\nonumber
i\partial_t\mathcal{O}_I(t)&=&-H_0e^{iH_0t}\mathcal{O}_S(t)e^{-iH_0t}+ie^{iH_0t}\partial_t\mathcal{O}_S(t)e^{-iH_0t} +e^{iH_0t}\mathcal{O}_S(t)H_0e^{-iH_0t}\\
&=&[\mathcal{O}_I(t),H_0]+i(\partial_t\mathcal{O})_I.
\end{eqnarray}
In other words, the interaction picture operator is trivially evolved by the free Hamiltonian $H_0(\phi_I,\pi_I)=\exp[iH_0(\phi_S,\pi_S)t] H_0(\phi_S,\pi_S)\exp[-iH_0(\phi_S,\pi_S)t]$. 
We see the interaction picture is a mixture of the Schr\"odinger picture and the Heisenberg picture, where states are evolved by $V_I$ according to the Schr\"odinger equation, while operators are evolved by $H_0$ according to the Heisenberg equation.

Analogous to how we convert $\mathcal{O}_S$ to $\mathcal{O}_I$, we can define the interaction picture time-evolution operator 
\begin{equation}
\label{eq:UI_US}
U_I(t,t')=e^{iH_0t} U_S(t,t') e^{-iH_0t'}.
\end{equation}
Notice that now the left and right actions by $H_0$ are at different moments of time. The time-evolution operator apparently satisfies the initial condition $U_I(t,t)=1$. Moreover, it satisfies the interaction picture Schr\"odinger equation
\begin{eqnarray}
\nonumber
i\partial_tU_I(t,t')&=&-H_0 e^{iH_0t} U_S(t,t') e^{-iH_0t'} +e^{iH_0t} H(t) U_S(t,t') e^{-iH_0t'}\\
\nonumber
&=&e^{iH_0t} V(t) U_S(t,t') e^{-iH_0t'}\\
&=&V_I(t) U_I(t,t').
\end{eqnarray} 
Similar as before, this equation can be solved by the time-ordered exponential
\begin{equation}
\label{eq:UI}
U_I(t,t')=\mathcal{T}\Big\{ \exp\Big[-i\int_{t'}^{t}V_I(\tau) d\tau\Big] \Big\}.
\end{equation}
The time-evolution operator satisfies group properties similar to Eqs.~(\ref{eq:US_product}) and (\ref{eq:US_inverse}). However, what is of critical importance is that now it only involves interaction-picture operators, which we know how to quantize and how they should act on asymptotic states.

Now both the time-evolution operator and the quantum states are defined in the same interaction picture, we are ready to compute S-matrix elements. It is important to recognize that the spectral expansion for the $\phi$ field  [Eq.~(\ref{eq:phi0_opt})] is a solution to the Heisenberg equations (\ref{eq:dt_phi}) and (\ref{eq:dt_pi}). Since these equations only involve the free Hamiltonian, they are special cases of the interaction picture operator equation (\ref{eq:int_Heisenberg}), with $\mathcal{O}=\phi$ and $\pi$, respectively. Therefore, the free field given by Eq.~(\ref{eq:phi0_opt}) is in fact the interaction-picture operator $\phi_I$, and the quantum states defined by Eqs.~(\ref{eq:ab_vacuum})-(\ref{eq:ab_one}) are in fact interaction-picture asymptotic states.  As an example, let us compute the transition amplitude from the incoming state $|i\rangle=|\mathbf{p}_1\mathbf{p}_2\rangle$ to the outgoing state $|f\rangle=|\mathbf{p}_3\mathbf{p}_4\rangle$ due to the $\phi^4$ interaction [Eq.~(\ref{eq:L_phi4})]. Since the interaction Lagrangian $L_I=-\int d^3\mathbf{x}\frac{\lambda}{4}(\phi^\dagger\phi)^2$, the interaction Hamiltonian $V_I=\int d^3\mathbf{x}\frac{\lambda}{4}(\phi_I^\dagger\phi_I)^2$. Expanding the time-ordered exponential perturbatively, and using Wick's theorem [Eq.~(\ref{eq:Wick})], the S-matrix element can be computed by
\begin{fmffile}{smatrix}
	\begin{eqnarray}
	\label{eq:smatrix}
	\nonumber
	\langle\mathbf{p}_3\mathbf{p}_4|S|\mathbf{p}_1\mathbf{p}_2\rangle&=&\langle\mathbf{p}_3\mathbf{p}_4| \mathcal{T} \exp [-i\int d^4x \frac{\lambda}{4}(\phi_I^\dagger\phi_I)^2 ]  |\mathbf{p}_1\mathbf{p}_2\rangle\\
	&=&\langle\mathbf{p}_3\mathbf{p}_4| 1-\frac{i\lambda}{4} \mathcal{T} \int d^4x (\phi_I^\dagger\phi_I)^2 +\dots |\mathbf{p}_1\mathbf{p}_2\rangle \\
	\nonumber
	&=&\Big(1+
	\begin{gathered}
		\begin{fmfgraph*}(30,35) 
			\fmfkeep{eight}
			\fmfleft{i}
			\fmfright{o}
			\fmf{phantom}{i,v}
			\fmf{phantom}{v,o}
			\fmf{plain}{v,v}
			\fmf{plain,left=90}{v,v}
			\fmfdot{v}
			\fmfv{label=$z$,label.angle=180,label.dist=6}{v}
		\end{fmfgraph*}
	\end{gathered}
	\hspace{-2pt}+\dots\Big)\Big(\hspace{10pt}
	\begin{gathered}
		\begin{fmfgraph*}(20,25)  
		\fmfkeep{1324}
		\fmftop{i1,d1,o1}
		\fmfbottom{i2,d2,o2}

		\fmf{plain}{i1,o1}
		\fmf{plain}{i2,o2}
		\fmfv{label=1,label.angle=180,label.dist=2}{i1}
		\fmfv{label=2,label.angle=180,label.dist=2}{i2}
		\fmfv{label=3,label.angle=0,label.dist=2}{o1}
		\fmfv{label=4,label.angle=0,label.dist=2}{o2}
		\end{fmfgraph*}
	\end{gathered}
	\hspace{10pt}+\hspace{10pt}
	\begin{gathered}
		\begin{fmfgraph*}(20,25)  
		\fmfkeep{1423}
		\fmftop{i1,d1,o1}
		\fmfbottom{i2,d2,o2}

		\fmf{plain}{i1,o2}
		\fmf{plain}{i2,o1}
		\fmfv{label=1,label.angle=180,label.dist=2}{i1}
		\fmfv{label=2,label.angle=180,label.dist=2}{i2}
		\fmfv{label=3,label.angle=0,label.dist=2}{o1}
		\fmfv{label=4,label.angle=0,label.dist=2}{o2}
		\end{fmfgraph*}
	\end{gathered}	
	\hspace{10pt}+\hspace{10pt}
	\begin{gathered}
		\begin{fmfgraph*}(20,25)  
		\fmfkeep{phi4}
		\fmfleft{i1,i2} 
		\fmfright{o1,o2}
		\fmf{plain}{i1,v,o1}
		\fmf{plain}{i2,v,o2}
		\fmfdot{v}
		\fmfv{label=2,label.angle=180,label.dist=6}{i1}
		\fmfv{label=1,label.angle=180,label.dist=6}{i2}
		\fmfv{label=4,label.angle=0,label.dist=6}{o1}
		\fmfv{label=3,label.angle=0,label.dist=6}{o2}
		\fmfv{label=$x$,label.angle=180,label.dist=8}{v}
		\end{fmfgraph*}
	\end{gathered}
	\hspace{10pt}+\dots\Big).
	\end{eqnarray}
\end{fmffile}Now let me elaborate the meaning of the last line, which is written in terms of Feynman diagrams that represent different ways of contracting operators using the Wick's theorem. Terms in the first bracket are contractions where the external states are not involved. For example, the figure-eight diagram
\begin{equation}
\label{eq:eight}
\begin{gathered}
\fmfreuse{eight}
\end{gathered}
=
\contraction{-2\frac{i\lambda}{4} \int d^4z }{\phi^\dagger}{}{\phi}
\contraction{-2\frac{i\lambda}{4} \int d^4z \phi^\dagger \phi }{\phi^\dagger}{}{\phi}
-2\frac{i\lambda}{4} \int d^4z \phi^\dagger \phi \phi^\dagger \phi
=-\frac{i\lambda}{2} \int d^4z G_F(z,z) G_F(z,z) ,
\end{equation}
where the vertex labeled by $z$ indicates where the interaction takes place, the factor of ``2" comes from two equivalent ways of contracting the two pairs of $\phi^\dagger\phi$ fields, and each internal line is associated with the Feynman propagator.
The second set of terms are contraction that involve external states. For example, the parallel-line diagram
\begin{equation}
\begin{gathered}
\fmfreuse{1324}
\end{gathered}
\hspace{10pt}=
\contraction{\langle}{\mathbf{p}}{{}_3\mathbf{p}_4|}{\mathbf{p}}
\bcontraction{\langle\mathbf{p}_3}{\mathbf{p}}{{}_4|\mathbf{p}_1}{\mathbf{p}}
\langle\mathbf{p}_3\mathbf{p}_4|\mathbf{p}_1\mathbf{p}_2\rangle
=2E_1 2E_2(2\pi)^6\delta^{(3)}(\mathbf{p}_1-\mathbf{p}_3) \delta^{(3)}(\mathbf{p}_2-\mathbf{p}_4),
\end{equation}
corresponds to the situation where no interaction happens between the identical particles, such that the $|\mathbf{p}_1\rangle$ state can be identified with the $|\mathbf{p}_3\rangle$ state, and the $|\mathbf{p}_2\rangle$ state can be identified with the $|\mathbf{p}_4\rangle$ state, where the identifications are enforced by the delta functions. 
Apart from the two terms where no interaction happens, other terms are contractions that involve both internal vertexes and external states. For example, the cross diagram
\begin{eqnarray}
\nonumber
\begin{gathered}
\fmfreuse{phi4}
\end{gathered}
\hspace{10pt}
&=&
\bcontraction[2ex]{-2\cdot 2 \frac{i\lambda}{4} \int d^4x \langle}{\mathbf{p}}{{}_3\mathbf{p}_4|\phi^\dagger \phi }{\phi}
\bcontraction{-2\cdot 2 \frac{i\lambda}{4} \int d^4x \langle\mathbf{p}_3}{\mathbf{p}}{{}_4|}{\phi}
\contraction[2ex]{-2\cdot 2 \frac{i\lambda}{4} \int d^4x \langle\mathbf{p}_3\mathbf{p}_4|\phi^\dagger}{\phi}{\phi^\dagger \phi|\mathbf{p}_1}{\mathbf{p}}
\contraction{-2\cdot 2 \frac{i\lambda}{4} \int d^4x \langle\mathbf{p}_3\mathbf{p}_4|\phi^\dagger\phi \phi^\dagger}{\phi}{|}{\mathbf{p}}
-2\cdot 2 \frac{i\lambda}{4} \int d^4x \langle\mathbf{p}_3\mathbf{p}_4|\phi^\dagger \phi \phi^\dagger \phi|\mathbf{p}_1\mathbf{p}_2\rangle\\
\nonumber
&=&-i\lambda\int d^4x e^{-i(p_1+p_2-p_3-p_4)x}\\
&=&-i\lambda(2\pi)^4\delta^{(4)}(p_1+p_2-p_3-p_4),
\end{eqnarray} 
where the delta function enforces that the 4-momentum of the incoming states are the same as the 4-momentum of the outgoing states, and thereof ensures energy and momentum conservation during the interaction process. 

From the above example, we see the S matrix can be symbolically decomposed into two parts ``$S=\mathbf{1}+iT$", and each part can be represented by Feynman diagrams, which can be evaluated with a set of rules called the Feynman rules. The part ``$\mathbf{1}$" does not contribute to the coupling between incoming states and outgoing states, while the part ``$iT$" is responsible for transitions between states. This later part is thereof called the transition matrix, or simply the T matrix. The T matrix always contains the factor $ (2\pi)^4\delta^{(4)}(\sum_ip_i-\sum_fp_f)$, because physical transitions between the initial and final states always conserve energy and momentum. Therefore, it is convenient to define the reduced matrix element $\mathcal{M}$, such that $iT=i\mathcal{M} (2\pi)^4\delta^{(4)}(\sum_ip_i-\sum_fp_f)$. For example, with an overload of the Feynman diagrams, the reduced matrix element 
\begin{fmffile}{phi4v}
	\begin{eqnarray}
	\label{eq:phi4v}
	i\mathcal{M}&=&
	\begin{gathered}
		\begin{fmfgraph*}(20,25)  
		\fmfkeep{phi4v}
		\fmfleft{i1,i2} 
		\fmfright{o1,o2}
		\fmf{plain}{i1,v,o1}
		\fmf{plain}{i2,v,o2}
		\fmfdot{v}
		\end{fmfgraph*}
	\end{gathered}
	+\dots
    =-i\lambda+\dots .
	\end{eqnarray}
\end{fmffile}We see the $\phi^4$ diagram can be attributed with the value of $-i\lambda$. This attribution is the Feynman rule for the interaction vertex. Similarly, the Feynman rule attributes the propagator to internal lines 
\begin{fmffile}{phi4G}
	\begin{eqnarray}
	\label{eq:phi4G}
	\begin{gathered}
		\begin{fmfgraph*}(20,25)  
		\fmfkeep{propagator_x}
		\fmfleft{i} 
		\fmfright{o}
		\fmf{plain}{i,v,o}
		\fmfdot{i,o}
		\fmfv{label=$x$,label.angle=180,label.dist=6}{i}
		\fmfv{label=$y$,label.angle=0,label.dist=6}{o}
		\end{fmfgraph*}
	\end{gathered}
	\hspace{15pt}
	=G_F(x,y),
	\hspace{25pt}
	\begin{gathered}
		\begin{fmfgraph*}(20,25)  
		\fmfkeep{propagator_p}
		\fmfleft{i} 
		\fmfright{o}
		\fmf{plain}{i,v,o}
		\fmfdot{i,o}
		\fmfv{label=$p$,label.angle=90,label.dist=6}{v}
		\end{fmfgraph*}
	\end{gathered}	
	\hspace{10pt}
	=\frac{i}{p^2-m^2+i\epsilon},
	\end{eqnarray}
\end{fmffile}in the configuration space and the momentum space, respectively. Finally, the Feynman rule attributes wave functions to external lines, where the configuration space and the momentum space wave functions are
\begin{fmffile}{phi4f}
	\begin{eqnarray}
	\label{eq:phi4f_in}
	&&
	\begin{gathered}
		\begin{fmfgraph*}(20,25)  
		\fmfkeep{phi_ext_x_in}
		\fmfleft{i} 
		\fmfright{o}
		\fmf{plain,label=$\rightarrow$}{i,o}
		\fmfdot{o}
		\fmfv{label=$x$,label.angle=0,label.dist=6}{o}
		\fmfv{label=$p$,label.angle=180,label.dist=6}{i}
		\end{fmfgraph*}
	\end{gathered}
	\hspace{15pt}
	=e^{-ipx},
	\hspace{25pt}
	\begin{gathered}
		\begin{fmfgraph*}(20,25)  
		\fmfkeep{phi_ext_p_in}
		\fmfleft{i} 
		\fmfright{o}
		\fmf{plain}{i,o}
		\fmfdot{o}
		\end{fmfgraph*}
	\end{gathered}	
	\hspace{10pt}
	=1, \\
	\label{eq:phi4f_out}
	&&
	\begin{gathered}
		\begin{fmfgraph*}(20,25)  
		\fmfkeep{phi_ext_x_out}
		\fmfleft{i} 
		\fmfright{o}
		\fmf{plain,label=$\rightarrow$}{i,o}
		\fmfdot{i}
		\fmfv{label=$x$,label.angle=180,label.dist=6}{i}
		\fmfv{label=$p$,label.angle=0,label.dist=6}{o}		
		\end{fmfgraph*}
	\end{gathered}
	\hspace{15pt}
	=e^{ipx},
	\hspace{33pt}
	\begin{gathered}
		\begin{fmfgraph*}(20,25)  
		\fmfkeep{phi_ext_p_out}
		\fmfleft{i} 
		\fmfright{o}
		\fmf{plain}{i,o}
		\fmfdot{i}
		\end{fmfgraph*}
	\end{gathered}	
	\hspace{10pt}
	=1,
	\end{eqnarray}
\end{fmffile}where Eq.~(\ref{eq:phi4f_in}) is for incoming states, and Eq.~(\ref{eq:phi4f_out}) is for outgoing state.
With the above rules, Feynman diagrams can be evaluated either in the configuration space by integrating over all interaction points, or evaluated in the Fourier space by integrating all internal momentum and imposing momentum conservation at each vertex. If multiple ways contracting a term exist, then the results need to be devided by the symmetry factor.  Computing the T matrix is usually sufficient for determining the scattering cross sections, which are proportional to the reduced matrix element $|\mathcal{M}|^2$ up to some phase space volume. The cross sections can be directly related to experimental observables in collider experiments, which I will not elaborate here.

Although vaguely speaking, experimental observables are mostly contained in the T matrix, the symbolic expression ``$S=\mathbf{1}+iT$" needs some clarifications. As we have seen, the term ``$\mathbf{1}$" is in fact nontrivial, because the physical vacuum is different from the interaction picture vacuum, and fluctuations such as Eq.~(\ref{eq:eight}) can occurs. Now let us look into vacuum fluctuations in some more details.

To compute properties of the fluctuating vacuum, we can calculate time-ordered two-point function similar to Eq.~(\ref{eq:two-point}), where now the vacuum state becomes the Schr\"odinger picture ground state, and the fields become the Heisenberg-picture operators in the full quantum theory. Since we only know how to compute in the interaction picture, we need to convert both the state and the operators to the interaction picture. 
First, to see how the physical ground state $|\Omega\rangle$ is related to the interaction-picture vacuum $|0\rangle$, we can use the completeness relation $1=\sum_n|\lambda_n\rangle\langle\lambda_n|$, where $|\lambda_n\rangle$ denotes the Schr\"odinger-picture eigenstate of the full Hamiltonian $H|\lambda_n\rangle=E_n|\lambda_n\rangle$. Then, the time evolution [Eq.~(\ref{eq:US})] of the interaction-picture vacuum $|0\rangle$ can be expanded using the eigenstates
\begin{eqnarray}
\nonumber
U_S(t,T)|0\rangle
&=&U_S(t,T)\sum_n|\lambda_n\rangle\langle\lambda_n|0\rangle\\
&=&e^{-i(t-T)E_0}|\Omega\rangle\langle\Omega|0\rangle+\sum_{n\ne 0}e^{-i(t-T)E_n}|\lambda_n\rangle\langle\lambda_n|0\rangle,
\end{eqnarray}
where $t$ is an arbitrary reference time.
Since excited states have energy $E_n>E_0$, we can project out the ground state by taking the limit $T\rightarrow -\infty(1-i\epsilon)$, assuming the density of states does not grow too fast when $n\rightarrow\infty$. Then, the interaction-picture ground state $|\Omega\rangle_I$, which is different from the interaction-picture vacuum $|0\rangle$ due to quantum fluctuations, can be represented by
\begin{eqnarray}
\label{eq:vacuum_S}
\nonumber
|\Omega\rangle_I&=&e^{iH_0t}|\Omega\rangle\\
\nonumber
&=&\lim_{T\rightarrow -\infty(1-i\epsilon)} \Big(e^{-i(t-T)E_0}\langle\Omega|0\rangle\Big)^{-1} e^{iH_0t}U_S(t,T)|0\rangle\\
&=&\lim_{T\rightarrow -\infty(1-i\epsilon)} \Big(e^{-i(t-T)E_0}\langle\Omega|0\rangle\Big)^{-1} U_I(t,T)|0\rangle,
\end{eqnarray}
where I have used Eq.~(\ref{eq:H0_energy}) for the free Hamiltonian and the definition Eq.~(\ref{eq:ab_vacuum}) for the interaction picture vacuum, which gives $H_0|0\rangle=0$ and thereof $\exp(-iH_0T)|0\rangle=|0\rangle$. 
Second, the Heisenberg-picture operator is related to the interaction-picture operator by
\begin{eqnarray}
\label{eq:phi_Heisenberg}
\nonumber
\phi(x)&=&U_S^\dagger(x^0,t)\phi_S(x)U_S(x^0,t)\\
\nonumber
&=&U_S^\dagger(x^0,t)e^{-iH_0t}\phi_I(x) e^{iH_0t}U_S(x^0,t)\\
&=&e^{-iH_0t}U_I^\dagger(x^0,t)\phi_S(x) U_I(x^0,t)e^{iH_0t},
\end{eqnarray}
where I have used Eq.~(\ref{eq:UI_US}) to convert the Schr\"odinger picture time evolution $U_S$ to $U_I$, the interaction picture time evolution. 
Having expressed both the state [Eq.~(\ref{eq:vacuum_S})] and the operator [Eq.~(\ref{eq:phi_Heisenberg})] in the interaction picture, we can now compute the vacuum expectation value of the two-point function. Without loss of generality, suppose $x^0>y^0$, then
\begin{eqnarray}
\nonumber
\langle\Omega|\phi(x)\phi^\dagger(y)|\Omega\rangle
\hspace{-8pt}&=&\hspace{-8pt}\lim_{T\rightarrow \infty}\hspace{-5pt} \frac{\langle 0|U_I(T,t) U_I^\dagger(x^0,t)\phi_I(x)U_I(x^0,t) U_I^\dagger(y^0,t)\phi_I(y)U_I(y^0,t) U_I(t,-T)|0\rangle}{e^{-2iTE_0}|\langle\Omega|0\rangle|^2}\\
&=&\hspace{-8pt}\lim_{T\rightarrow \infty}\hspace{-5pt} \frac{\langle 0|U_I(T,x^0)\phi_I(x)U_I(x^0,y^0) \phi_I(y)U_I(y^0,-T)|0\rangle}{e^{-2iTE_0}|\langle\Omega|0\rangle|^2},
\end{eqnarray}
where I have used group properties of the time-evolution operator [Eqs.~(\ref{eq:US_product}) and (\ref{eq:US_inverse})], and canceled the interconnecting terms $\exp(iH_0t)\exp(-iH_0t)=1$. The normalization factor in the denominator can be canceled by the normalization of the ground state
\begin{equation}
1=\langle\Omega|\Omega\rangle=\frac{\langle 0|U_I(T,-T)|0\rangle}{e^{-2iTE_0}|\langle\Omega|0\rangle|^2}. 
\end{equation} 
Finally, notice that the operator products are time ordered. Using Eq.~(\ref{eq:UI}) for $U_I$, and taking the limit $T\rightarrow +\infty(1-i\epsilon)$, we can obtain a simple formula for the two-point function
\begin{equation}
\langle \Omega|\mathcal{T}\phi(x)\phi^\dagger(y)|\Omega\rangle=\frac{\langle 0|\mathcal{T}\phi_I(x) \phi_I(y) \exp[-i\int dt V_I(t)]|0\rangle}{\langle 0|\mathcal{T}\exp[-i\int dt V_I(t)]|0\rangle},
\end{equation}
which can be readily expanded using Feynman diagrams and evaluated using Feynman rules. Notice that the term in the denominator exactly gives the terms in the first bracket of Eq.~(\ref{eq:smatrix}). Therefore, we see the meaning of ``$\mathbf{1}$" in the symbolic expression ``$S=\mathbf{1}+iT$" is in fact the vacuum persistence amplitude, which is the amplitude that the interaction-picture vacuum does not end up in excited states amids all the quantum fluctuations.

Similarly, the precise meaning of ``$iT$" is the transition amplitude times persistence amplitudes of quantum states, which is the amplitude that an interaction-picture state does not end up in some other states due to quantum fluctuations. 
In other words, an interaction-picture state can fluctuate. Nevertheless, as long as the state remains the same when it enters the interaction region, these fluctuations do not matter. Therefore, physical transition amplitudes, which are related to experimental measurables, should remove these fluctuations by dividing them from the total amplitude. 
From a diagram perspective, an matrix element, such as Eq.~(\ref{eq:phi4v}), can have bubbles on external legs that represent fluctuations of the incoming and outgoing states. These bubble diagrams are irrelevant to the physical transition amplitude. To compute physical transition amplitudes, we will only need the so-called amputated diagrams, which are obtained by cutting away bubbles so that external legs become simple lines. 
The precise mathematical statement of the above result is the LSZ reduction formula, which is first proven by Lehman, Symanzik, and Zimmermann \citep{Lehmann1955}. 

In this section, the second-quantization formalism is developed using the scalar field theory as an example. For Dirac fermions and gauge bosons, similar developments can be made except for somewhat different spins-statistics and symmetry groups. The basic idea is nevertheless the same: we first quantized free fields, then compute correlation functions in the interaction picture, and finally obtain the S matrix using the LSZ reduction formula. Here, I will not elaborate on the technical details, which can be found in many textbooks. 
To conclude this section, let me list the Feynman rules for scalar QED, using which I will develop a theory for relativistic quantum plasmas in the following chapters. The scalar-QED Lagrangian is given by Eq.~(\ref{eq:SQED_Lagrangian}), which can be written explicitly as
\begin{eqnarray}
\label{eq:SQED}
\nonumber
\mathcal{L}&=&\partial_{\mu}\phi^\dagger\partial^\mu\phi-m^2\phi^\dagger\phi-\frac{1}{4}F_{\mu\nu}F^{\mu\nu}\\
&+&ieA^\mu(\phi^\dagger\partial_{\mu}\phi-\partial_{\mu}\phi^\dagger\, \phi)+e^2A_\mu A^\mu\phi^\dagger\phi 
\end{eqnarray} 
where terms on the first line are the Lagrangian of free complex scalar field and the free U(1) gauge field, while terms on the second line are the interaction Lagrangian. For convenience, let me denote $\mathcal{L}_{e\phi^2A}=ieA^\mu(\phi^\dagger\partial_{\mu}\phi-\partial_{\mu}\phi^\dagger\, \phi)$ and $\mathcal{L}_{e^2\phi^2A^2}=e^2A_\mu A^\mu\phi^\dagger\phi$, 
where the quantum fields should be interpreted as the interaction picture operators. First, similar to scalar field wave functions Eqs.~(\ref{eq:phi4f_in}) and (\ref{eq:phi4f_out}), the gauge field wave functions can be attributed to external lines
\begin{fmffile}{Af}
	\begin{eqnarray}
	\label{eq:Af_in}
	&&
	\begin{gathered}
		\begin{fmfgraph*}(20,25)  
		\fmfkeep{A_ext_x_in}
		\fmfleft{i} 
		\fmfright{o}
		\fmf{photon,label=$\rightarrow$}{i,o}
		\fmfdot{o}
		\fmfv{label=$x$,label.angle=0,label.dist=6}{o}
		\fmflabel{$k,\mu$}{i}
		\end{fmfgraph*}
	\end{gathered}
	\hspace{25pt}
	=\epsilon_\mu(k)e^{-ikx},
	\hspace{25pt}
	\begin{gathered}
		\begin{fmfgraph*}(20,25)  
		\fmfkeep{A_ext_k_in}
		\fmfleft{i} 
		\fmfright{o}
		\fmf{photon,label=$k$}{i,o}
		\fmfdot{o}
		\fmflabel{$\mu$}{i}
		\end{fmfgraph*}
	\end{gathered}	
	\hspace{15pt}
	=\epsilon_\mu(k), \\
	\label{eq:Af_out}
	&&
	\begin{gathered}
		\begin{fmfgraph*}(20,25)  
		\fmfkeep{A_ext_x_out}
		\fmfleft{i} 
		\fmfright{o}
		\fmf{photon,label=$\rightarrow$}{i,o}
		\fmfdot{i}
		\fmfv{label=$x$,label.angle=180,label.dist=6}{i}	
		\fmflabel{$k,\mu$}{o}
		\end{fmfgraph*}
	\end{gathered}
	\hspace{25pt}
	=\epsilon^\dagger_\mu(k)e^{ikx},
	\hspace{33pt}
	\begin{gathered}
		\begin{fmfgraph*}(20,25)  
		\fmfkeep{A_ext_p_out}
		\fmfleft{i} 
		\fmfright{o}
		\fmf{photon,label=$k$}{i,o}
		\fmfdot{i}
		\fmflabel{$\mu$}{o}
		\end{fmfgraph*}
	\end{gathered}	
	\hspace{15pt}
	=\epsilon^\dagger_\mu(k),
	\end{eqnarray}
\end{fmffile}where $\epsilon_\mu(k)$ is the unit polarization vector of the gauge boson, which can have up to three differently polarized eigenstates in general. Second, similar to the scalar field propagator [Eq.~(\ref{eq:phi4G})], the gauge boson propagator is attributed to internal lines
\begin{fmffile}{AG}
	\begin{eqnarray}
		\label{eq:AG}
		\begin{gathered}
			\begin{fmfgraph*}(20,25)  
				\fmfkeep{A_propagator_x}
				\fmfleft{i} 
				\fmfright{o}
				\fmf{photon}{i,o}
				\fmfdot{i,o}
				\fmflabel{$x,\mu$}{i}
				\fmflabel{$y,\nu$}{o}
			\end{fmfgraph*}
		\end{gathered}
		\hspace{25pt}
		=\Lambda_F^{\mu\nu}(x,y),
		\hspace{25pt}
		\begin{gathered}
			\begin{fmfgraph*}(20,25)  
				\fmfkeep{A_propagator_k}
				\fmfleft{i} 
				\fmfright{o}
				\fmf{photon,label=$k$}{i,o}
				\fmfdot{i,o}
				\fmflabel{$\mu$}{i}
				\fmflabel{$\nu$}{o}
			\end{fmfgraph*}
		\end{gathered}	
		\hspace{10pt}
		=\frac{-i}{k^2+i\epsilon}\Big[g^{\mu\nu}-(1-\xi)\frac{k^\mu k^\nu}{k^2}\Big],
	\end{eqnarray}
\end{fmffile}where $\xi$ is the Faddeev--Popov gauge parameter. The Feynman Green's function for the gauge boson satisfies
\begin{equation}
\label{eq:dG_A}
\mathbb{D}_{\mu\nu}(x)\Lambda_F^{\nu\rho}(x,y)=-i\delta^\rho_\mu\delta^{(4)}(x-y),
\end{equation}
where $\mathbb{D}_{\mu\nu}$ is the dispersion operator. In the vacuum theory, $\mathbb{D}_{\mu\nu}=(\partial_{\mu}\partial_\nu-\partial^2g_{\mu\nu})$, and its spatial components is precisely Eq.~(\ref{eq:Dij}). With the minus sign $A^i=-A_i$, the sign convention for the gauge boson Green's function is the opposite of what is used for the scalar field Green's function [Eq.~(\ref{eq:phi4G})], because the gauge boson is usually quantized in the temporal gauge $A^0=0$. Quantization in this gauge can readily ensure positivity of the Hamiltonian, because $g_{ij}=-\delta_{ij}$ have the same sign and terms with the opposite sign $g_{0\nu}=\delta_{0\nu}$ do not contribute. Quantizations in other gauges are also viable, as long as a gauge condition is fixed. Without fixing a gauge, the operator $\mathbb{D}_{\mu\nu}$ has a nontrivial kernel and is thereof not invertible. Fixing the gauge using the Faddeev--Popov procedure introduces the gauge parameter $\xi$ in Eq.~(\ref{eq:AG}), which disappears when computing observables that are gauge-invariant.
Finally, the scalar-QED Lagrangian [Eq.~(\ref{eq:SQED})] has two interaction terms. 
To find out the Feynman rules, let us compute two matrix elements. One matrix element is 
\begin{eqnarray}
\nonumber
\langle\mathbf{p}_2|i\mathcal{T}\!\int\! d^4x\mathcal{L}_{e\phi^2A}|\mathbf{p}_1\mathbf{k}\rangle
\!&=&\!
\contraction[1.5ex]{i\int d^4x \langle}{\mathbf{p}}{{}_2ieA^\mu}{\phi^\dagger}
\contraction{i\int d^4x \langle\mathbf{p}_2|ieA^\mu\phi^\dagger\partial_{\mu}}{\phi}{|}{\mathbf{p}}
\bcontraction{i\int d^4x \langle\mathbf{p}_2|ie}{A}{{}^\mu\phi^\dagger\partial_{\mu}\phi|\mathbf{p}_1}{\mathbf{k}}
i\int d^4x \langle\mathbf{p}_2|ieA^\mu\phi^\dagger\partial_{\mu}\phi|\mathbf{p}_1\mathbf{k}\rangle
-
\contraction[1.5ex]{\langle}{\mathbf{p}}{{}_2ieA^\mu\partial_{\mu}}{\phi^\dagger}
\contraction{\langle\mathbf{p}_2|ieA^\mu\partial_{\mu}\phi^\dagger\, }{\phi}{|}{\mathbf{p}}
\bcontraction{\langle\mathbf{p}_2|ie}{A}{{}^\mu\partial_{\mu}\phi^\dagger\, \phi|\mathbf{p}_1}{\mathbf{k}}
\langle\mathbf{p}_2|ieA^\mu\partial_{\mu}\phi^\dagger\, \phi|\mathbf{p}_1\mathbf{k}\rangle\\
\nonumber
\!&=&\!-e\int\! d^4x \!\Big[e^{ip_2x}\epsilon^\mu e^{-ikx}(-ip_1^\mu)e^{-ip_1x}\!-\!(ip_2^\mu )e^{ip_2x}\epsilon^\mu e^{-ikx}e^{-ip_1x} \Big]\\
\!&=&\!\int d^4x \epsilon^\mu e^{i(p_2-p_1-k)x}ie(p_1+p_2)^\mu.
\end{eqnarray}
Taking away integration of wave functions, we see the interaction vertex can be attributed with the following momentum-space Feynman rule
\begin{fmffile}{Aphi2}
	\begin{eqnarray}
	\label{eq:Aphi2}
	\begin{gathered}
		\begin{fmfgraph*}(30,30) 
		\fmfkeep{Aphi2}
		\fmftop{i1,o} 
		\fmfbottom{i2}
		\fmf{plain}{i1,v}
		\fmf{plain}{v,o}
		\fmf{photon,label=$k$}{i2,v}
		\fmfdot{v}
		\fmflabel{$p_1$}{i1}
		\fmflabel{$\mu$}{i2}
		\fmflabel{$p_2$}{o}
		\marrow{a}{up}{top}{}{i1,v}
		\marrow{b}{up}{top}{}{v,o}
		\end{fmfgraph*}
	\end{gathered}
	\hspace{25pt}
	=ie(p_1+p_2)^\mu.
	\end{eqnarray}
\end{fmffile}The matrix element due to the $\mathcal{L}_{e^2\phi^2A^2}$ interaction can be computed similarly, which gives the Feynman rule for the other scalar-QED vertex
\begin{fmffile}{A2phi2}
	\begin{eqnarray}
	\label{eq:A2phi2}
	\begin{gathered}
	\begin{fmfgraph*}(30,35)  
	\fmfkeep{A2phi2}
	\fmfleft{i1,i2} 
	\fmfright{o1,o2}
	\fmf{plain}{i2,v,o2}
	\fmf{photon}{i1,v,o1}
	\fmfdot{v}
	\fmflabel{$\mu$}{i1}
	\fmflabel{$\nu$}{o1}
	\end{fmfgraph*}
	\end{gathered}
	\hspace{25pt}
	=2ie^2g^{\mu\nu},
	\end{eqnarray}
\end{fmffile}where the factor of two comes from two ways of contracting the gauge field. With the above Feynman rules, we can readily evaluate all matrix elements of the scalar-QED model.

\section{Path integral formulation\label{sec:vacuum:pathintegral}}

In the previous section, quantum field theory is treated using the standard second quantization formulation, 
which accounts for quantum uncertainty using the commutation relations.
However, introducing noncommutativeness is not the only way of incorporating the quantum uncertainty. 
A more intuitive way 
is perhaps using the path integral formulation. In this alternative formulation, quantum uncertainty is manifested through random trajectories of particles.
Unlike in classical physics where a particle travels along the definite trajectory that extremizes the action, quantum fluctuations allow a particle to wiggle around the extrema and travel along random trajectories that are not allowed in classical physics. 
If no trajectory is forbidden in an experimental setup, then after multiple realizations of the experiment, the averaged trajectory will be close to the classical trajectory. However, interesting things happen when some trajectories are blocked, for example, in a double-slit experiment. In this later situation, averaging the quantum trajectories no longer recovers the classical trajectory, and genuine quantum phenomena, such as double-slit interference of a single electron, can then happen.

By carrying out the path integral, namely, allowing particles to take all possible trajectories and then averaging them with the weighting factor $e^{iS}$, all experimentally observed behaviors of quantum particles can be explained \citep{Feynman1985qed}. Not only does the path integral formulation explains quantum behaviors, but it also explains why our daily experience is mostly classical. This is because without special setups to block the classical path, the classical trajectory, which extremizes the action $S$, is the saddle point of the path integral and thereof has the dominate contribution. The insight that classical physics and quantum physics can be unified through path integral can be extended from the time domain to the entire spacetime. The result of such an extension is the path integral formulation of quantum field theory, which I will discuss in this section.

\subsection{Path integrals in quantum mechanics\label{sec:vacuum:pathintegral:formulation}}

\begin{figure}[t]
	\centering
	\includegraphics[angle=0,width=0.35\textwidth]{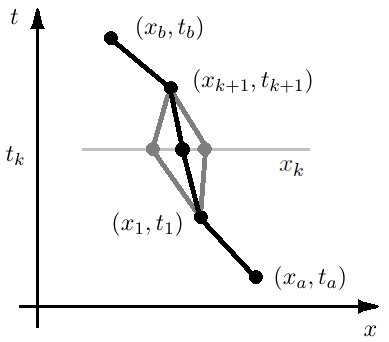}
	\caption[Path integral of all possible trajectories]{A classical particle travels along the minimal-action trajectory (black), while a quantum particle can wiggle around (gray) and sample all possible states. At given time $t_k$, the particle's location $x_k$ can be anywhere, with a probability amplitude $\exp(iS)$ determined by the action $S=\int dt L(x,\dot{x})$ of the trajectory $x(t)$. The weighted average of all trajectories gives the expected trajectory, where the classical contribution usually dominates. }
	\label{fig:Path_int}
\end{figure}

To see that path-integral formulation is equivalent to second quantization, let us first consider the motion of a single particle. Suppose we are interested in the following question: what is the probability that a particle, initially found at location $x_a$ at time $t_a$, appears at the final location $x_b$ at time $t_b$? The answer given by the Schr\"odinger picture is that the probability amplitude is $\langle x_b|\mathcal{T}\exp(-i\int_{t_a}^{t_b} dt H)|x_a\rangle$, where $\mathcal{T}$ is the time-ordering operator and $H$ is the full Hamiltonian. Now instead of inquiring only the initial and final states, suppose we are also interested in finding out what happens in between, namely, we want to know what trajectory the particle takes to move from $x_a$ to $x_b$. Then, we can make a series of observations at time $t_0<t_1<\dots<t_{N-1}<t_N$, when we find the particle at locations $x_0, x_1, \dots, x_{N-1}$ and $x_{N}$, where the initial coordinate $(t_0, x_0)=(t_a, x_a)$ and the final coordinate $(t_{N}, x_{N})=(t_b, x_b)$. Taking the limit $\Delta t_k=t_k-t_{k-1}\rightarrow 0$ while keeping the sum $\sum_{k=1}^{N}\Delta t_k=t_b-t_a$ fixed, the transition amplitude can be partitioned as
\begin{eqnarray}
\label{eq:path_partition}
\nonumber
\langle x_b|\mathcal{T}\exp(-i\int_{t_a}^{t_b} dt H)|x_a\rangle
&=&\lim_{\Delta t\rightarrow 0}\langle x_b|\mathcal{T}\exp(-i\sum_{k=1}^{N} \Delta t_k H)|x_a\rangle \\
&=&\lim_{\Delta t\rightarrow 0} \int \prod_{k=1}^{N} dx_k \langle x_{k}|\mathcal{T}\exp(-i\Delta t_k H)|x_{k-1}\rangle,
\end{eqnarray}
where I have inserted the completeness of quantum states $\mathbb{I}=\int dx_k |x_k\rangle\langle x_k|$ at each time when an observation is made (Fig.~\ref{fig:Path_int}). The above amplitude is the transition amplitude along the path $(t_0,x_0), (t_1,x_1), \dots,$ and $(t_N,x_{N})$, where the end points are fixed, but the interior points are allowed to vary. In the limit $N\rightarrow\infty$, the discrete path becomes a smooth trajectory.

To evaluate the transition amplitude along an infinitesimal path, let us explicitly write the Hamiltonian operator $H=H(\hat{x},\hat{p})$, where $\hat{p}$ and $\hat{x}$ are the momentum and position operators in the second quantization formulation. The position eigenstate $|x\rangle$ is an eigenstate of the position operators $\hat{x}|x\rangle=x|x\rangle$. Then, using the orthonormal condition of the eigenvectors, the projection $\langle x_k|f(\hat{x})|x_{k-1}\rangle$ can be expressed in the momentum space as
\begin{eqnarray}
\nonumber
\langle x_k|f(\hat{x})|x_{k-1}\rangle
&=&f\big(\frac{x_k+x_{k-1}}{2}\big)\delta(x_k-x_{k-1})\\
\nonumber
&=&\int\frac{dp_k}{2\pi}e^{ip_k(x_k-x_{k-1})}f\big(\frac{x_k+x_{k-1}}{2}\big).
\end{eqnarray}
Similarly, denote the momentum space eigenstate as $|p\rangle$, which satisfies $\hat{p}|p\rangle=p|p\rangle$. Since the momentum operator has the configuration-space representation $\hat{p}=-i\partial_x$, the momentum eigenstate wave function $\langle x|p\rangle$ satisfies the equation $-i\partial_x \langle x|p\rangle =p\langle x|p\rangle$. It is thereof easy to see that the projection $\langle x|p\rangle=e^{ipx}$. Inserting the completeness condition in momentum space $\mathbb{I}=\int \frac{dp_k}{2\pi} |p_k\rangle\langle p_k|$, the projection $\langle x_k|g(\hat{p})|x_{k-1}\rangle$ can be expressed as
\begin{eqnarray}
\nonumber
\langle x_k|g(\hat{p})|x_{k-1}\rangle&=&\langle x_k|\int \frac{dp_k}{2\pi} |p_k\rangle\langle p_k|g(\hat{p})|x_{k-1}\rangle\\
\nonumber
&=&\langle x_k|\int \frac{dp_k}{2\pi} |p_k\rangle\langle p_k|g(p_k)|x_{k-1}\rangle\\
\nonumber
&=&\int\frac{dp_k}{2\pi}e^{ip_k(x_k-x_{k-1})}g(p_k).
\end{eqnarray}
If the Hamiltonian is separable, namely, $H(x,p)=f(p)+g(x)$, as is the case for simple harmonic oscillators, then we can apply the above tricks to the transition amplitude along an infinitesimal path
\begin{eqnarray}
\label{eq:path_infinitesimal}
\nonumber
\langle x_{k}|\mathcal{T}\exp[-i\Delta t_k H(\hat{x},\hat{p})]|x_{k-1}\rangle&=&\langle x_{k}|1-i\Delta t_k H(\hat{x},\hat{p})+\dots|x_{k-1}\rangle\\
\nonumber
&=&\int\frac{dp_k}{2\pi}\exp\big[ip_k(x_k-x_{k-1})-i\Delta t_kH\big(\frac{x_k+x_{k-1}}{2},p_k\big)\big]\\
&\rightarrow&\int\frac{dp_k}{2\pi}\exp\big\{i\int_{t_{k-1}}^{t_k}dt\big[p_k\dot{x}_k-H\big(x_k,p_k\big)\big]\big\},
\end{eqnarray}
where the last line is obtained by taking the limit $\Delta t_k\rightarrow 0$. It is easy to recognize that the inverse Legendre transformation $p\dot{x}-H=L$ is the Lagrangian, and what sits in the exponential is simply the infinitesimal action. Therefore, the Schr\"odinger picture time evolution can be related to summation of $\exp(iS)$ over all possible paths.  
Here, it is worth mentioning what happens when $H(x,p)$ is not separable. In this case, the ordering of operators $\hat{x}$ and $\hat{p}$ matters in the quantized Hamiltonian $H(\hat{x},\hat{p})$. A particular ordering is call the Weyl ordering, in which $\hat{x}$ and $\hat{p}$ appears symmetrically so that the product reads the same from left and right. 
If the Hamiltonian operator $H(\hat{x},\hat{p})$ is Weyl-ordered, it will simply be replaced by the Hamiltonian function $H(x,p)$ in the path-integral formula.
When the Hamiltonian operator is not Weyl-ordered, it can be converted to an auxiliary Hamiltonian $\tilde{H}(\hat{x},\hat{p})$ that is Weyl-ordered. This is because an arbitrary product can always be converted to a linear combination of Weyl-ordered products by commuting $\hat{x}$ and $\hat{p}$. For example, $\hat{p}\hat{x}^2=(\hat{p}\hat{x}^2+\hat{x}^2\hat{p})/2+[\hat{p},\hat{x}]\hat{x}$, where $[\hat{p},\hat{x}]$ is simply a complex number so both terms on the RHS are Weyl-ordered.
After Weyl ordering, the auxiliary Hamiltonian will in general contain additional terms, and what appears in the path-integral formula will be $\tilde{H}(x,p)$ instead of the original Hamiltonian function.

After adding up the contributions from infinitesimal paths, the transition amplitude of a quantum particle from the initial coordinate $(t_a, x_a)$ to the final coordinate $(t_b, x_b)$ can be computed. Using the path partition Eq.~(\ref{eq:path_partition}) and the infinitesimal contribution Eq.~(\ref{eq:path_infinitesimal}), the initial to final transition amplitude can be written as
\begin{eqnarray}
\label{eq:path_1D}
\langle x_b|\mathcal{T}\exp\big[-i\int_{t_a}^{t_b} dt H(\hat{x},\hat{p})\big]|x_a\rangle
%
%
&=&\int_{x_a}^{x_b} \mathscr{D} x(t) \exp\big[i\int_{t_{a}}^{t_b}dt L(x,\dot{x})\big],
\end{eqnarray}
where 
$\mathscr{D} x(t)=\lim_{N\rightarrow\infty} \prod_{k=1}^{N} \frac{dx_k dp_k}{2\pi}$
denote the path integral of all possible paths $x(t)$ that satisfy the boundary conditions $x(t_a)=x_a$ and $x(t_b)=x_b$. 
The above expression connects the Hamiltonian second quantization formulation and Lagrangian path integral formulation of quantum mechanics. 

To see how path integral works in practice, consider the simple example where the particle is free. In this case, the Hamiltonian $H=p^2/2m$ only contains a kinetic term. First, using quantum mechanics, we have
\begin{eqnarray}
\nonumber
\langle x_b|\mathcal{T}\exp[-i(t_b-t_a) H]|x_a\rangle&=&\langle x_b|\int \frac{dp}{2\pi} |p\rangle\langle p|\mathcal{T}\exp\big[-i(t_b-t_a) \frac{\hat{p}^2}{2m}\big]|x_a\rangle \\
\nonumber
&=&\int \frac{dp}{2\pi} e^{ipx_b}\exp\Big[-i(t_b-t_a)\frac{p^2}{2m}\Big] e^{-ipx_a}\\
&=&\sqrt{\frac{m}{2\pi i (t_b-t_a)}}\exp\Big[i\frac{m(x_a-x_b)^2}{2(t_b-t_a)}\Big],
\end{eqnarray}
where the time is slightly rotated $t\rightarrow(1-i\epsilon)t$ away from the real axis to make the Gaussian integral converge. The above result gives the wave function $\Phi(x_b,t_b)$, which satisfies the Schr\"odinger equation $i\partial_{t_b}\Psi=-\frac{1}{2m}\partial_{x_b}\Psi$ and the initial condition $\Phi(x_b,t_a)=\delta(x_b-x_a)$. The wave packet spreads from an initial delta function to an imaginary Gaussian distribution, whose standard deviation increases with time. The same result can be obtained using the path integral. From its definition
\begin{eqnarray}
\label{eq:eg_path}
\nonumber
\int_{x_a}^{x_b} \mathscr{D} x(t) \exp\big[i\int_{t_{a}}^{t_b}dt L(x,\dot{x})\big]&=&\lim_{\Delta t\rightarrow 0} \int \prod_{k=1}^{N} dx_k\frac{dp_k}{2\pi} \exp\Big[i\Delta t\Big(p_k\dot{x}_k-\frac{p_k^2}{2m}\Big)\Big]\\
\nonumber
&=&\lim_{\Delta t\rightarrow 0} \int \prod_{k=1}^{N} dx_k \sqrt{\frac{m}{2\pi i \Delta t}} \exp\Big[i\frac{m(x_{k+1}-x_k)^2}{2\Delta t}\Big] \\
&=&\sqrt{\frac{m}{2\pi i (t_b-t_a)}}\exp\Big[i\frac{m(x_a-x_b)^2}{2(t_b-t_a)}\Big].
\end{eqnarray}
The integral series can be computed by choosing $N=2^n$ for some integer $n$ and then successively using the following pairwise reduction
\begin{eqnarray}
\nonumber
&&\int dx_k \exp\Big\{\frac{im}{2\Delta t}\Big[(x_{k+1}-x_k)^2+(x_{k}-x_{k-1})^2\Big]\Big\}\\
\nonumber
&=&\int dx_k \exp\Big\{\frac{im}{2\Delta t}\Big[2\Big(x_k-\frac{x_{k+1}+x_{k-1}}{2}\Big)^2+\frac{(x_{k+1}-x_{k-1})^2}{2}\Big]\Big\}\\
\nonumber
&=&\sqrt{\frac{i\pi\Delta t}{m}} \exp\Big[i\frac{m(x_{k+1}-x_{k-1})^2}{4\Delta t}\Big].
\end{eqnarray}
To make the Gaussian integral converge, time is again rotated in the complex plane $\Delta t\rightarrow(1-i\epsilon)\Delta t$. After one round of pairwise reduction, the number of intermediate points $N=2^n\rightarrow 2^{n-1}$ is reduced by half, while the time step $\Delta t\rightarrow 2\Delta t$ is doubled. Therefore, after $n$ rounds of pairwise reductions, Eq.~(\ref{eq:eg_path}) can then be obtained.

The usefulness of the path-integral formulation comes from the mathematical developments, whereby functional integrations can be computed directly without starting from their definition. The functional integration can be computed either in the configuration space or in the Fourier space \citep{Grosche1998handbook}. For example, in the configuration space, since the paths satisfy the boundary conditions $x(t_a)=x_a$ and $x(t_b)=x_b$, it is convenient to subtract the classical contribution $x_c(t)$ and then integrating over quantum fluctuations. In other words, the trajectory can be decomposed as $x(t)=x_c(t)+\xi(t)$, where the quantum fluctuation satisfies $\xi(t_a)=\xi(t_b)=0$. Moreover, we can also shift the time $t\rightarrow t-t_a$, and denote $T=t_b-t_a$. Then, in our simple example $L=m\dot{x}^2/2$, the classical trajectory is simply free streaming $x_c(t)=x_a+vt$, 
where the constant velocity $v=(x_b-x_a)/T$, and the action $S=\int dt L$ can be written as
\begin{eqnarray}
\nonumber
S&=&\int_0^T dt\,\frac{1}{2}m(v^2+2v\dot{\xi}+\dot{\xi}^2)\\
\nonumber
&=&\frac{1}{2}mv^2T+mv[\xi(T)-\xi(0)]+\int_0^T dt\,\frac{1}{2}m\dot{\xi}^2 \\
&=&S_c-\frac{m}{2}\int_0^T dt\, \xi d^2_t\xi.
\end{eqnarray}
On the last line, $S_c$ is the classical action, the second terms vanishes because of the boundary conditions $\xi(0)=\xi(T)=0$, and the third term has been rewritten using integration by part, whose boundary terms also vanish. Then, changing the integration path, the functional integral can be factored as
\begin{eqnarray}
\label{eq:path_free}
\int_{x_a}^{x_b} \mathscr{D}x e^{iS}&=&e^{iS_c}\int_{x_a}^{x_b} \mathscr{D}\xi \exp\Big[-\frac{m}{2i}\int_0^T \xi (-d^2_t)\xi \Big],
\end{eqnarray}
where $-d^2_t\sim\omega^2>0$ is a positive-definite elliptic operator. 
The Gaussian functional integral is an infinite-dimensional generalization of the $N$-dimensional Gaussian integral. For a positive-definite $N\times N$ symmetric matrix $A$, the Gaussian integral
\begin{eqnarray}
\label{eq:Gaussian0}
\int \prod_{i=1}^{N} dx_i e^{-\frac{1}{2}x_pA_{pq}x_q}=\prod_{i=1}^{N}\sqrt{\frac{\pi}{\lambda_i}}=\Big(\det \frac{A}{2\pi}\Big)^{-\frac{1}{2}},
\end{eqnarray} 
where $\lambda_i>0$ are the eigenvalues of the matrix. More generally, when integrated with a polynomial of even order $2n$, the Gaussian integral
\begin{eqnarray}
\label{eq:Gaussian1}
\int\! \prod_{i=1}^{N}\! dx_i\! \prod_{j=1}^{2n}x^{k_j}\! e^{-\frac{1}{2}x_pA_{pq}x_q}\!=\!\Big(\!\det \frac{A}{2\pi}\!\Big)^{-\frac{1}{2}} \hspace{-5pt}\frac{1}{2^{n} n!}\hspace{-5pt}\sum_{\sigma\in S_{2n}}\hspace{-5pt} (\!A^{-1}\!)^{k_{\sigma(1)}k_{\sigma(2)}}\!\dots\! (\!A^{-1}\!)^{k_{\sigma(2n-1)}k_{\sigma(2n)}},
\end{eqnarray} 
where $S_{2n}$ is the permutation group of $2n$ elements. As a generalization of the multivariate Gaussian integral, the Gaussian functional integral
\begin{equation}
\label{eq:Gaussian_path}
\int_{x_a}^{x_b} \mathscr{D}\xi \exp\Big[-\alpha\int_0^T \xi (-d^2_t)\xi \Big]=\gamma(\alpha)\big[\det(-d_t^2)\big]^{-1/2},
\end{equation}
where $\gamma(\alpha)$ is some normalization factor. To compute the functional determinant, notice that for a matrix $A$, we have the identity $\det A=\exp[\text{tr}(\ln A)]=\exp\sum_n\ln\lambda_n$, where $\lambda_n$ are the eigenvalues. To compute the trace of $\ln A$, we can use the zeta function
\begin{equation}
\zeta^{r}_{A}(s):=\text{tr} A^{-rs}=\sum_n\frac{1}{\lambda_n^{rs}},
\end{equation}
where $r$ is an arbitrary parameter, on which the final result does not depend. Take derivative $d_s$ of the zeta function and then take the limit $s\rightarrow 0$, 
$\zeta^{r'}_{A}(s)=-\sum_n r\ln\lambda_n/\lambda_n^{rs}\rightarrow-r\sum_n\ln\lambda_n$ is well-behaved. Therefore, the determinant of the matrix $A$ can be written in terms of the companion zeta function as
\begin{equation}
\label{eq:detA}
\det A=\exp\big[-\frac{1}{r}\zeta^{r'}_{A}(0)\big],
\end{equation}
which is a relation also holds for the functional determinant. Now let us return to the example $A=-d_t^2$, and solve the eigenvalue problem $A\xi_n=\lambda_n\xi_n$. The eigenvector $\xi_n(t)$ of the operator are functions that satisfies the Dirichlet boundary conditions $\xi(0)=\xi(T)=0$. It is easy to see that the eigenvectors are $\xi_n(t)\propto\sin(n\pi t/T)$, and the eigenvalues are $\lambda_n=(n\pi/T)^2$. For simplicity, denote $\lambda_n=cn^2$ where $c$ is a constant, then the companion zeta function
\begin{equation}
\zeta^{r}_{A}(s)=c^{-rs}\sum_{n=1}^{\infty}\frac{1}{n^{2rs}}=c^{-rs}\zeta(2rs),
\end{equation}
where $\zeta$ is the Riemann zeta function. The derivative $\zeta^{r'}_{A}(s)=rc^{-rs}[-(\ln c) \zeta(2rs)+2\zeta'(2rs)]$, whose value at $s=0$ is $\zeta^{r'}_{A}(0)=r[\ln(c)/2-\ln(2\pi)]$, where I have used the special values of the Riemann zeta function $\zeta(0)=-\frac{1}{2}$ and $\zeta'(0)=-\frac{1}{2}\ln2\pi$. Using Eq.~(\ref{eq:detA}), the determinant $\det A=2\pi c^{-1/2}$. Notice that when we rescale the matrix $A\rightarrow \alpha A$, the eigenvalue is scaled by the same factor $\lambda\rightarrow \alpha\lambda$, so the determinant $\det(\alpha A)=\alpha^{-1/2}\det A$. This scaling law of functional determinant is very different from that of $N\times N$ matrices, which is scaled by $\alpha^N$. In any case, such a scaling only affects the overall normalization. Now that $c=(\pi/T)^2$, the functional determinant
\begin{equation}
\det(-d_t^2)=2T.
\end{equation}
Substituting this result into Eq.~(\ref{eq:Gaussian_path}), then the path integral Eq.~(\ref{eq:path_free}) becomes
\begin{equation}
\int \mathscr{D}x e^{iS}=\gamma\Big(\frac{m}{2i}\Big)\frac{1}{\sqrt{2(t_b-t_a)}} \exp\Big[i\frac{m(x_a-x_b)^2}{2(t_b-t_a)}\Big],
\end{equation}
which equals to our earlier calculations [Eq.~(\ref{eq:eg_path})] up to some normalization factor, which contains no information regarding the dynamical process. Normalization factors in functional integrals can be difficult to compute, unless the measure of the functional space is treated carefully. Fortunately, normalization can usually be determine by some other ways, for example, by normalizing the total probability to one. Moreover, when we compute physical observables using path integrals, the normalization factors can usually be canceled as we shall see next.  
\subsection{Path integral in quantum field theory\label{sec:vacuum:pathintegral:Gaussian}}

Now we can extend the path integral formulation from quantum mechanics in the time domain to quantum field theory defined on the entire spacetime. In the time domain, we have seen that the transition amplitude of $x(t)$ from $(t_a,x_a)$ to $(t_b,x_b)$ can be computed both using second quantization in the Hamiltonian formalism, and using path integral in the Lagrangian formalism [Eq.~(\ref{eq:path_1D})]. Following similar steps, the transition amplitude of a real scalar field $\phi(x)$ from the field configuration $\phi_a(x)$ at time $t_a$ to another field configuration $\phi_b(x)$ at some later time $t_b$, can be computed using both formulations
\begin{equation}
\label{eq:path_real}
\langle \phi_b|\mathcal{T}\exp\big[-i\int_{t_a}^{t_b} dt H(\phi,\pi)\big]|\phi_a\rangle
=\int_{\phi_a}^{\phi_b} \mathscr{D} \phi \exp\big[i\int_{t_{a}}^{t_b}d^4x \mathcal{L}(\phi,\partial_\mu\phi)\big],
\end{equation}
where $\pi$ is the canonical momentum of $\phi$, and $\mathcal{L}$ is the Lagrangian density. If the field is complex, then we can treat $\phi$ and $\phi^*$ as independent fields, and the integration would be carried over $\mathscr{D} \phi\mathscr{D} \phi^*$. Here, for simplicity, I will first illustrate with a real scalar field.

In the usual quantum field theory, one is more concerned with scattering amplitudes than the actual field configuration $\phi(x)$. For this purpose, as discussed in Sec.~\ref{sec:vacuum:quantization:smatrix}, the S-matrix elements can be extracted from correlation functions using the LSZ reduction formula. Therefore, the task now is to compute $N$-point functions using path integrals. For example, let us consider 
the two-point path integral
\begin{equation}
\nonumber
I=\int_{\phi_a}^{\phi_b} \mathscr{D} \phi\; \phi(x_1)\phi(x_2)e^{iS_a^b},
\end{equation}   
where $\phi(x_1)$ and $\phi(x_2)$ are the field values at the two spacetime coordinates $x_1$ and $x_2$, and $S_a^b=\int_{t_{a}}^{t_b}d^4x \mathcal{L}$. Without loss of generality, let us suppose that $t_a<x^0_1<x^0_2<t_b$. Since the path integral sums over all possible field configurations, we can first restrict the field at two configurations $\phi_1$ and $\phi_2$, and then integrate over $\phi_1$ and $\phi_2$. In other words, in addition to the initial $\phi_a$ and final $\phi_b$ configurations, we make two additional observations during the field evolution, where we find the field is $\phi_1$ at $x_1^0$ and $\phi_2$ at $x_2^0$. Then, the path integral can be written as products of three sections
\begin{eqnarray}
\nonumber
I&=&\int \mathscr{D}\phi_1 \mathscr{D}\phi_2\; \phi(x_1)\phi(x_2)\int_{\phi_2}^{\phi_b} \mathscr{D}\phi \;e^{iS_2^b} \int_{\phi_1}^{\phi_2} \mathscr{D}\phi\; e^{iS_1^2} \int_{\phi_a}^{\phi_1} \mathscr{D}\phi\; e^{iS_a^1}\\
\nonumber
&=&\int \mathscr{D}\phi_1 \mathscr{D}\phi_2 \phi(x_1)\phi(x_2) \langle\phi_b|U_S(t_b,x_2^0)|\phi_2\rangle \langle\phi_2|U_S(x_2^0,x_1^0)|\phi_1\rangle \langle\phi_1|U_S(x_1^0,t_a)|\phi_a\rangle,
\end{eqnarray}
where I have used Eq.~(\ref{eq:path_real}) to convert path integrals to transition amplitudes, and $U_S$ is the Schr\"odinger picture time-evolution operator [Eq.~(\ref{eq:US})]. The field configuration $|\phi\rangle$ is the eigenstate of the Schr\"odinger picture operator $\phi_S(x)|\phi\rangle=\phi(x)|\phi\rangle$. Absorbing the eigenvalue $\phi(x_i)$ into the transition amplitude, and using the completeness condition of eigenstates $\mathbb{I}=\int \mathscr{D}\phi |\phi\rangle\langle\phi|$, we can write
\begin{eqnarray}
\nonumber
I&=&\vspace{-5pt}\int \mathscr{D}\phi_1 \mathscr{D}\phi_2 \langle\phi_b|U_S(t_b,x_2^0)\phi_S(x_2)|\phi_2\rangle \langle\phi_2|U_S(x_2^0,x_1^0)\phi_S(x_1)|\phi_1\rangle \langle\phi_1|U_S(x_1^0,t_a)|\phi_a\rangle\\
\nonumber
&=&\langle\phi_b|U_S(t_b,x_2^0)\phi_S(x_2)U_S(x_2^0,x_1^0)\phi_S(x_1)U_S(x_1^0,t_a)|\phi_a\rangle\\
\nonumber
&=&\langle\phi_b|U_S(t_b,t)\phi_H(x_2)\phi_H(x_1)U_S(t,t_a)|\phi_a\rangle.
\end{eqnarray}
To avoid confusion, I have inserted the subscripts to emphasize that 
$\phi_H$ is the Heisenberg-picture operator [Eq.~(\ref{eq:phi_Heisenberg})], which is related to the Schr\"odinger-picture operator by $\phi_H(x)=U_S(t,x^0)\phi_S(x)U_S(x^0,t)$.
To obtain the last line, I have used goup properties of the time-evolution operator Eqs.~(\ref{eq:US_product})-(\ref{eq:US_inverse}), and inserted an arbitrary reference time $t$. Denote the Heisenberg-picture state $|\phi(t)\rangle_H=U_S(t,x^0)|\phi(x)\rangle$, which reverses the time evolution of the Schr\"odinger picture state such that $|\phi(t)\rangle_H$ is fixed at the reference time $t$ and hence does not evolve. Then, the two-point path integral is
\begin{equation}
\label{eq:path_2pt}
\int_{\phi_a}^{\phi_b} \mathscr{D} \phi\; \phi(x_1)\phi(x_2)\exp\Big(i\int_{t_{a}}^{t_b}d^4x \mathcal{L}\Big) =\phantom{I}_H\langle\phi_b|\mathcal{T}\phi_H(x_1)\phi_H(x_2)|\phi_a\rangle_H,
\end{equation}
where the assumption that $x_2^0>x_1^0$ is now manifested by the time-ordering operator $\mathcal{T}$. We see the two-point path integral equals to the transition amplitude from the initial to final states when we make two intermediate observations. The above formula connects Lagrangian path-integral formulation, where fields are functions, to Hamiltonian second-quantization formulation, where fields are operators.

To compute S-matrix elements using the LSZ reduction formula, we need the notion of asymptotic vacuum. Similar to how we projected out the vacuum in Eq.~(\ref{eq:vacuum_S}), we can push the boundary condition $t_a$ to the remote past $t_a\ll t$ of the reference time. Inserting a complete set of eigenstates of the full Hamiltonian $\mathbb{I}=\sum_n|n\rangle\langle n|$, and sending $t_a\rightarrow-\infty(1-i\epsilon)$, the boundary state
\begin{eqnarray}
\nonumber
|\phi_a\rangle_H=U_S(t,t_a)|\phi_a\rangle&=&U_S(t,t_a)\sum_n|n\rangle\langle n|\phi_a\rangle\\
\nonumber
&=&\sum_ne^{-iE_n(t-t_a)}|n\rangle\langle n|\phi_a\rangle \\
&\rightarrow&e^{iE_0t_a}|\Omega\rangle\langle \Omega|\phi_a\rangle.
\end{eqnarray}
Similarly, sending the other boundary to the remote future, $t_b\rightarrow+\infty(1-i\epsilon)$, the field configuration is also dominated by the physical vacuum
\begin{equation}
\phantom{I}_H\langle\phi_b|\rightarrow e^{-iE_0t_b}\langle \phi_b|\Omega\rangle\langle \Omega|.
\end{equation}
With the vacuum as the asymptotic state in both the remote past and the remote future, the two-point function
\begin{eqnarray}
\nonumber
\phantom{I}_H\langle\phi_b|\mathcal{T}\phi_H(x_2)\phi_H(x_1)|\phi_a\rangle_H&\rightarrow& \langle\phi_b|\Omega\rangle \langle\Omega|\phi_a\rangle e^{-iE_0(t_b-t_a)} \langle\Omega|\mathcal{T}\phi_H(x_2)\phi_H(x_1)|\Omega\rangle \\
\nonumber
&=&\langle\phi_b|U_S(t_b,t)U_S(t,t_a)|\phi_a\rangle \langle\Omega|\mathcal{T}\phi_H(x_2)\phi_H(x_1)|\Omega\rangle \\
\nonumber
&=&\int_{\phi_a}^{\phi_b} \mathscr{D} \phi\; \exp\Big(i\int_{t_{a}}^{t_b}d^4x \mathcal{L}\Big) \langle\Omega|\mathcal{T}\phi_H(x_2)\phi_H(x_1)|\Omega\rangle, 
\end{eqnarray}
where I have used the normalization of the physical vacuum $\langle\Omega|\Omega\rangle=1$ to rewrite the normalization factor in terms of the path integral. Taking the limit $t_b\rightarrow+\infty$ and $t_a\rightarrow-\infty$, substituting the above expression into Eq.~(\ref{eq:path_2pt}), we thus obtain a formula for the vacuum two-point function for the real scalar field in terms of path integrals
\begin{equation}
\label{eq:path_2pt_vacuum}
\langle\Omega|\mathcal{T}\phi_H(x_1)\phi_H(x_2)|\Omega\rangle = \frac{\int \mathscr{D} \phi\; \phi(x_1)\phi(x_2)\exp\Big(i\int d^4x \mathcal{L}\Big)}{\int \mathscr{D} \phi\; \exp\Big(i\int d^4x \mathcal{L}\Big)}.
\end{equation}
Now the boundary conditions are pushed to infinity and the Lagrangian density is integrated over the entire spacetime. As promised earlier, the normalization factors of the functional integrals cancel, and physical results are independent of the functional measure. The above formula, derived using the real scalar field, also holds for fields of other types.

\subsection{Feynman rules from path integrals\label{sec:vacuum:pathintegral:Feynman}}

To illustrate the path-integral formulation, let me first use the example of the free complex scalar field, whose Lagrangian is given by Eq.~(\ref{eq:L0_phi}). Suppose the field vanishes at infinity, then using integration by part, the action 
\begin{equation}
iS_0=i\int d^4x\mathcal{L}_0=-\int d^4x \phi^*i(\partial^2+m^2)\phi=-\frac{1}{2}\int d^4x \Phi^T \left( \begin{array}{cc}
0 & D \\
D & 0 \end{array} \right)\Phi,
\end{equation} 
which can be regarded as the action of a real scalar field $\Phi^T=(\phi,\phi^*)$ with the differential operator appearing as the off-diagonal components of an extended symmetric matrix. The differential operator $D=i(\partial^2+m^2-i\epsilon)$, where a small positive number $\epsilon>0$ is inserted to ensure that the operator is positive definite. To compute the Gaussian functional integral, we can mimic the following integrals in the complex plane
\begin{eqnarray}
&&\int dz d\bar{z}\; e^{-\bar{z} a z}=\frac{\pi}{a},\\
&&\int dz d\bar{z}\; z\bar{z} e^{-\bar{z} a z}=\frac{\pi}{a}a^{-1},\\
&&\int dz d\bar{z}\; z^2 e^{-\bar{z} a z}=\int dz d\bar{z}\; \bar{z}^2 e^{-\bar{z} a z}=0,
\end{eqnarray}
which are natural extensions of Gaussian integrals in the real vector space [Eq.~(\ref{eq:Gaussian0}) and (\ref{eq:Gaussian1})], if we regard $\mathbf{x}=(z,\bar{z})$ as a real vector and double the size the matrix. Denoting the inverse of the differential operator $D$ as $G(x,x')$, which satisfies $DG(x,x')=\delta^{(4)}(x-x')$, then up to some normalization of the functional determinant, the functional Gaussian integrals
\begin{eqnarray}
\label{eq:path_Gaussian0}
\int \mathscr{D} \phi\mathscr{D} \phi^*\; \exp\Big(-\int d^4 z \phi^* D \phi\Big)\hspace{-5pt}&=&\hspace{-5pt}(\det D)^{-1}, \\
\label{eq:path_Gaussian1}
\int \mathscr{D} \phi\mathscr{D} \phi^*\; \phi(x)\phi^*(y) \exp\Big(-\int d^4 z \phi^* D \phi\Big)\hspace{-5pt}&=&\hspace{-5pt}(\det D)^{-1}\frac{1}{2}\Big[G(x,y)+G(y,x)\Big]. \hspace{10pt}
\end{eqnarray}
To see what the inverse $G$ is, we need to solve the differential equation. Since the operator $D$ is translational invariant, the inverse $G(x,y)=G(x-y)$. Then, taking Fourier transform $\int d^4x\exp[ip(x-y)]$, the Fourier space inverse
\begin{equation}
\hat{G}(p)=\frac{i}{p^2-m^2+i\epsilon},
\end{equation}
which is exactly the momentum space Feynman propagator for the scalar field [Eq.~(\ref{eq:phi4G})]. Indeed, after taking the inverse Fourier transform, the function $G(x,y)=G(y,x)$ is exactly the Feynman Green's function [Eq.~(\ref{eq:GF})]. In other words, for the free complex scalar field, the two-point function
\begin{equation}
\langle 0|\mathcal{T}\phi_H(x)\phi^\dagger_H(y)|0\rangle = \frac{\int \mathscr{D} \phi\mathscr{D} \phi^*\;\phi(x)\phi^*(y) e^{iS_0}}{\int \mathscr{D} \phi\mathscr{D} \phi^*\; e^{iS_0}}=G_F(x,y),
\end{equation}
which agrees with our previous result [Eq.~(\ref{eq:two-point})], which was obtained using second quantization. 
Previously, the prescription of how to integrate around the poles in $G_F$ was imposed by causality. Here, the same prescription is given by imposing that the differential operator $D$ is positive definite so that the Gaussian integrals converge. Moreover, from the path integral formulation, we see it is natural to use the convention where the Green's function is normalized such that $(\partial^2+m^2)G(x,x')=-i\delta^{(4)}(x-x')$, which is different from the typical mathematical definition by a factor of $-i$. Analogous to multivariate Gaussian integral, the two other two-point functions are zero
\begin{equation}
\int \mathscr{D} \phi\mathscr{D} \phi^*\;\phi(x)\phi(y) e^{iS_0}=\int \mathscr{D} \phi\mathscr{D} \phi^*\;\phi^*(x)\phi^*(y) e^{iS_0}=0,
\end{equation} 
which agree with our earlier results Eq.~(\ref{eq:two-point-trivial}). Therefore, two-point Gaussian path integrals of free fields are equivalent to Wick's contractions in the second quantization formulation.

Next, let us compute the propagator of the free U(1) gauge field. Due to the gauge symmetry, the U(1) gauge field contains a redundant degree of freedom. This redundancy requires that we fix a gauge in the second quantization formulation, which usually uses the temporal gauge in order to ensure the positivity of the Hamiltonian. Similarly, the gauge redundancy needs to be reduced in the path integral formulation, which usually uses the Lorentz gauge, in order to ensure that the functional integrals converge. To see why this is necessary, notice that the free action
\begin{eqnarray}
\nonumber
iS_0=-\frac{i}{4}\int d^4x F_{\mu\nu} F^{\mu\nu}&=&-\frac{i}{2}\int d^4x \big[(\partial_{\mu}A_\nu)(\partial^{\mu}A^\nu)-(\partial_{\mu}A_\nu)(\partial^{\nu}A^\mu) \big]\\
&=&-\frac{1}{2}\int d^4x A^\mu i\big(\partial_{\mu}\partial_{\nu}-\partial^2g_{\mu\nu}\big)A^\nu,
\end{eqnarray}
is a quadratic form with a degenerate matrix $iD_{\mu\nu}=i(\partial_{\mu}\partial_{\nu}-\partial^2g_{\mu\nu})$. The matrix has a nontrivial kernel, which is the set of functions $\alpha$ such that $iD_{\mu\nu}\partial^\nu\alpha=0$. This situation is equivalent to having a zero eigenvalue in the matrix of multivariate Gaussian integrals, in which case the integrals diverge. To solve this problem, we need to remove the kernel from the integration domain. For functional integrals, this can be achieved using the Faddeev--Popov procedure \citep{Faddeev1967}.  Instead of integrating over the entire functional space, we can integrate in the quotient space $A^{\mu}\sim A^\mu+\partial^\mu\alpha$, with equivalent classes represented by field configurations that satisfy the generalized Lorenz-gauge condition $\partial_{\mu} A^{\mu}=\omega$. To enforce the gauge condition, we can insert the identity
\begin{equation}
\label{eq:1_gauge}
1=\int \mathcal{D}\alpha \delta(\partial_{\mu} A^{\mu}-\omega) \det\Big(\frac{\delta \partial_{\mu} A^{\mu}}{\delta\alpha}\Big).
\end{equation}
The above identity is the infinite dimensional generalization of the delta function identity in multivariate calculus
\begin{equation}
\nonumber
1=\int\prod_{i=1}^{n}dx^i \delta^{(n)}[\mathbf{f}({\mathbf{x}})-\mathbf{x}_0] \det\Big(\frac{\partial f_i}{\partial x_j}\Big),
\end{equation} 
where the determinant of the Jacobian compensates for the change of measure when inverting $\mathbf{x}=\mathbf{f}^{-1}(\mathbf{x}_0)$. The Jacobian of the functional change of variable is $\delta \partial_{\mu} \mathcal{A}^{\mu}/\delta\alpha=\partial^2$, which is simply a numerical factor. The functional delta function can be absorbed by inserting another Gaussian integral identity
\begin{equation}
\label{eq:1_Gaussian}
1=\gamma(\xi)\int \mathcal{D}\omega \exp\Big(-i\int d^4x\frac{\omega^2}{2\xi}\Big),
\end{equation}
where $\xi$ is an arbitrary parameter and $\gamma(\xi)$ is the normalization factor of the functional Gaussian integral. Inserting both identities Eqs.~(\ref{eq:1_gauge}) and (\ref{eq:1_Gaussian}), the functional integral can be written as the infinity $\int \mathcal{D}\alpha$ times the integration in the quotient space
\begin{eqnarray}
\nonumber
\int\! \mathcal{D} A \mathcal{O}(A)e^{iS_0}\!&=&\!\gamma(\xi)\!\int\!  \mathcal{D}\alpha \mathcal{D} A \mathcal{D}\omega \exp\Big(\!-i\!\int d^4x\frac{\omega^2}{2\xi}\Big) \delta(\partial_{\mu} A^{\mu}\!-\!\omega) \det\Big(\frac{\delta \partial_{\mu} A^{\mu}}{\delta\alpha}\Big)  \mathcal{O}(A)e^{iS_0}\\
\nonumber
\!&=&\!\gamma(\xi)\det(\partial^2)\Big(\int \mathcal{D}\alpha\Big) \int \mathcal{D} A \mathcal{O}(A)\exp\Big[i\int d^4x\Big(\mathcal{L}_0-\frac{1}{2\xi}(\partial_{\mu} A^{\mu})^2\Big)\Big],
\end{eqnarray}
where $\mathcal{O}(A)$ is any gauge invariant functional of $A$. 
Since the three terms in front are just normalization factors that will be canceled in the $N$-point function, what the above Faddeev--Popov procedure does to U(1) gauge field\footnote[1]{For other gauge groups, such as the $SU(N)$ group, the Faddeev--Popov procedure can introduce a functional determinant that depends on dynamical fields. In this case, other terms, such as the Faddeev--Popov ghost fields, are also added to the Lagrangian.}
is transforming the Lagrangian $\mathcal{L}_0\rightarrow\tilde{\mathcal{L}}_0(\xi) =\mathcal{L}_0-\frac{1}{2\xi}(\partial_{\mu} A^{\mu})^2$. Then the operator in the Gaussian integral becomes $iD_{\mu\nu}\rightarrow i\tilde{D}_{\mu\nu}(\xi)=i[(1-\frac{1}{\xi})\partial_{\mu}\partial_{\nu} -(\partial^2-i\epsilon)g_{\mu\nu}]$, which becomes nondegenerate. 
Now that the functional integrals converge, we can use the property, analogous Eq.~(\ref{eq:Gaussian1}), of the Gaussian integral to compute the two-point function of the free gauge field
\begin{equation}
\label{eq:2pt_A}
\langle 0|\mathcal{T}A^\mu_H(x) A^\nu_H(y)|0\rangle = \frac{\int \mathscr{D} A \;A^\mu(x)A^\nu(y) e^{i\tilde{S}_0}}{\int \mathscr{D} A e^{i\tilde{S}_0}}=\Lambda_F^{\mu\nu}(x,y).
\end{equation}
To find the Green's function, which satisfies $i\tilde{D}_{\mu\nu}(\xi)\Lambda_F^{\nu\sigma}(x,y)=\delta^{\rho}_{\mu}\delta^{(4)}(x-y)$, it is easy to solve the equation in the momentum space
\begin{equation}
\label{eq:Green_A_xi}
i\tilde{D}_{\mu\nu}\hat{\Lambda}_F^{\nu\sigma}(k)=-i\big[(1-\frac{1}{\xi})k_{\mu}k_{\nu} -(k^2+i\epsilon)g_{\mu\nu}\big]\hat{\Lambda}_F^{\nu\sigma}(k)=\delta^{\sigma}_{\mu}.
\end{equation}
Since there are only two Lorentz invariant symmetric tensors $g_{\mu\nu}$ and $k_\mu k_\nu$ in this problem, the inverse must be a linear combination of these two tensors. It is straightforward to compute the coefficients, and the momentum space Feynman Green's function
\begin{equation}
\hat{\Lambda}_F^{\nu\sigma}=\frac{-i}{k^2+i\epsilon}\Big[g^{\mu\nu}-(1-\xi)\frac{k^\mu k^\nu}{k^2}\Big],
\end{equation}
which is exactly the Feynman propagator of the gauge field Eq.~(\ref{eq:AG}) with the correct pole prescription. The special value $\xi=0$ is the Landau gauge, and $\xi=1$ is the Feynman gauge, which usually makes calculations simple in practice. However, it is advisable to keep the gauge parameter $\xi$ in the calculation, whose cancellation can be used as a criteria to check whether correct results are obtained when computing gauge-invariant quantities.

Finally, apart from propagators of free fields, let me use the $\phi^4$ theory [Eq.~(\ref{eq:L_phi4})] to illustrate how to obtain Feynman rules for interaction vertexes using path integrals. When the coupling coefficient $\lambda$ is small, we can compute the functional integral perturbatively
\begin{eqnarray}
\nonumber
\int \mathscr{D} \phi\mathscr{D} \phi^*\;e^{iS}&=&\int \mathscr{D} \phi\mathscr{D} \phi^*\;e^{iS_0} \Big[1-\frac{i\lambda}{4}\int d^4z (\phi\phi)^*+\dots\Big]\\
\nonumber
&=&\big[\det i(\partial^2+m^2)\big]^{-1}\Big(1+ 
\begin{gathered}
\fmfreuse{eight}
\end{gathered}
\hspace{-5pt}+\dots
\Big).
\end{eqnarray}
The figure-eight diagram corresponds to the Gaussian integral $\int dz d\bar{z}\; (z\bar{z})^2 e^{-\bar{z} a z} =(\pi/a)2a^{-2}$. Now that $a$ is the operator $D=i(\partial^2+m^2-i\epsilon)$, whose inverse is the Feynman Green's function $G_F$, the figure-eight diagram is given by functional Gaussian integral as
\begin{equation}
\begin{gathered}
\fmfreuse{eight}
\end{gathered}
\hspace{-5pt}=-\frac{i\lambda}{4}\int d^4 z 2 G_F(z,z)G_F(z,z),
\end{equation}
which is identical to Eq.~(\ref{eq:eight}), which was obtained using second quantization. Similarly, we can compute the two-point path integral
\begin{fmffile}{phi2pt}
	\begin{eqnarray}
	\label{eq:phi2pt}
	\nonumber
	\int \mathscr{D} \phi\mathscr{D} \phi^*\; \phi(x)\phi^*(y)e^{iS}&=&\int \mathscr{D} \phi\mathscr{D} \phi^*\;e^{iS_0} \phi(x)\phi^*(y)\Big[1-\frac{i\lambda}{4}\int d^4z (\phi\phi)^*+\dots\Big]\\
	\nonumber
	&=&(\det D)^{-1}\Big[
	\hspace{12pt}
	\begin{gathered}
	\fmfreuse{propagator_x}
	\end{gathered}
	\hspace{10pt}+\Big(
	\hspace{10pt}
	\begin{gathered}
	\fmfreuse{propagator_x}
	\end{gathered}
	\hspace{10pt}\times\hspace{-3pt}
	\begin{gathered}
	\fmfreuse{eight}
	\end{gathered}
	\hspace{-8pt}+\hspace{5pt}
	\begin{gathered}
	\begin{fmfgraph*}(30,40) 
	\fmfkeep{phi_1loop}
	\fmfleft{i}
	\fmfright{o}
	\fmf{plain}{i,v}
	\fmf{plain}{v,o}
	\fmf{plain}{v,v}
	\fmfdot{i,o,v}
	\fmfv{label=$x$,label.angle=90,label.dist=6}{i}
	\fmfv{label=$y$,label.angle=90,label.dist=6}{o}
	\fmfv{label=$z$,label.angle=-90,label.dist=6}{v}
	\end{fmfgraph*}
	\end{gathered}
	\hspace{5pt}\Big)+\dots\Big],
	\end{eqnarray}
\end{fmffile}where the figure-eight diagram represents the same Gaussian integral as before, and the line diagram represent the free propagator as in Eq.~(\ref{eq:phi4G}). The 1-loop diagram represent the following terms in the Gaussian integral
\begin{equation}
	\label{eq:phi_1loop}
	\begin{gathered}
	\fmfreuse{phi_1loop}
	\end{gathered}
	\hspace{5pt}=
	-\frac{i\lambda}{4}\int d^4z\; 4G_F(x,z)G_F(z,z)G_F(z,y),
\end{equation}
where the factor ``4" comes from the four permutations that give the same term in the Gaussian integral similar to Eq.~(\ref{eq:Gaussian1}). From the above example, we see functional Gaussian integrals can be represented by Feynman diagrams, in which an internal line is associated with the propagator $G_F$, and an interaction vertex $z$ is associated with the integral $-i\lambda\int d^4z$. These Feynman rules are identical to what we have obtained in Sec.~\ref{sec:vacuum:quantization:smatrix} using second quantization. After properly accounting for the symmetry factor, the path integrals can then be evaluated using the Feynman rules.

\section{Beyond lowest order: renormalization\label{sec:vacuum:renormalization}}

Using either the second-quantization or the path-integral formulations, what we need to do in the end is computing Feynman diagrams. Once the Feynman rules are figured out, computing diagrams is a rather mechanical task. However, when computing higher-order diagrams, in which loops appear, simply following the Feynman rules yields infinities. This is a typical situation facing perturbation theories, where secular terms arise beyond the leading order, rendering naive higher-order perturbative solutions invalid. When solving the classical field equations as PDEs, the secular terms can be removed by multiscale expansions discussed in Ch.~\ref{ch:multiscale}, which rescales space and time to absorb the infinity. Similarly, infinities in perturbative solutions of quantum field theory can be absorbed by a technique called renormalization, which I will briefly discuss in this section.

\subsection{Basic idea: subtract infinity by rescaling\label{sec:vacuum:renormalization:regularization}}

Although many different approaches to renormalization have been developed in the second half of the 20th century, they share the same idea that peturbative infinities may be removed by rescaling, namely, renormalizing. To see how this idea comes about, let us compute the 1-loop propagator Eq.~(\ref{eq:phi_1loop}). Using the Fourier representation of the Green's function, the 1-loop contribution to the propagator is
\begin{eqnarray}
\nonumber
\begin{gathered}
\fmfreuse{phi_1loop}
\end{gathered}
\hspace{5pt}
&=&-i\lambda\int d^4z \frac{d^4p}{(2\pi)^4} \frac{d^4k}{(2\pi)^4}\frac{d^4q}{(2\pi)^4} \hat{G}_F(p)e^{-ip(x-z)} \hat{G}_F(k) \hat{G}_F(q)e^{-ip(z-y)}\\
&=&-i\lambda\int \frac{d^4p}{(2\pi)^4} \frac{d^4k}{(2\pi)^4} \hat{G}_F(k) \hat{G}^2_F(p) e^{-ip(x-y)}, 
\end{eqnarray}
where the $z$ and $q$ integrals have been carried out. Notice that the $k$ integral is divergent, because it scales as $d^4k\hat{G}_F(k)\sim k^4 k^{-2}$, which goes to infinity when we integrate over the entire momentum space. For now, let us bear with this infinity problem and carry on. In the Fourier space, the 1-loop diagram can be written as the product
\begin{fmffile}{phi2ptp}
	\begin{eqnarray}
	\label{eq:phi2pt_p}
	\begin{gathered}
		\begin{fmfgraph*}(30,40) 
		\fmfkeep{phi_1loop_pxy}
		\fmfleft{i}
		\fmfright{o}
		\fmf{plain,label=p}{i,v}
		\fmf{plain,label=p}{v,o}
		\fmf{plain,label=k}{v,v}
		\fmfdot{i,o,v}
		\end{fmfgraph*}
	\end{gathered}
	\hspace{5pt}&=&-i\lambda \hat{G}^2_F(p) \int \frac{d^4k}{(2\pi)^4}\hat{G}_F(k)=\hat{G}^2_F(p)\times	
	\begin{gathered}
		\begin{fmfgraph*}(30,35) 
		\fmfkeep{phi_1loop_p}
		\fmfleft{i}
		\fmfright{o}
		\fmf{plain,label=p}{i,v}
		\fmf{plain,label=p}{v,o}
		\fmf{plain}{v,v}
		\fmfdot{v}
		\end{fmfgraph*}
	\end{gathered}
	\hspace{3pt},
	\end{eqnarray}
\end{fmffile}
The above diagram is the first term in the following series of diagrams:
\begin{fmffile}{1PIphi}
	\begin{eqnarray}
	\label{eq:1PIphi}
	\begin{gathered}
		\begin{fmfgraph*}(45,30)
		\fmfkeep{1PIphi}
		\fmfleft{i}
		\fmfright{o}
		\fmf{plain,label=p}{i,v}
		\fmf{plain,label=p}{v,o}
		\fmfv{d.sh=circle,d.f=empty,d.si=.5w}{v}
		\fmfv{label=1PI,label.angle=180,label.dist=0}{v}
		\end{fmfgraph*}
	\end{gathered}
	=
	\begin{gathered}
	\fmfreuse{phi_1loop_p}
	\end{gathered}
	\hspace{2pt}+\hspace{2pt}
	\begin{gathered}
		\begin{fmfgraph*}(40,35)
		\fmfleft{i}
		\fmfright{o}
		\fmf{plain,label=p}{i,v1}
		\fmf{plain}{v1,v2}
		\fmf{plain,label=p}{v2,o}
		\fmf{plain,left,tension=0.1}{v1,v2,v1}
		\fmfdot{v1,v2}
		\end{fmfgraph*}
	\end{gathered}
	\hspace{2pt}+\dots
	=-i\mu^2(p^2).
	\end{eqnarray}
\end{fmffile}On the LHS, the summed diagram is called the 1-particle-irreducible (1PI) diagram. This name comes from the fact that the diagrams involve one external particle and each diagram in the series is irreducible. A diagram is said to be irreducible, if it cannot be made disconnected by cutting a single internal propagator. In other words, if a connected diagram can be cut into two separate subdiagrams by removing a single internal line, then the diagram is said to be reducible.
On the RHS of Eq.~(\ref{eq:1PIphi}), $\mu^2$ denotes the 1PI amplitude that only depends on the Lorentz scalar $p^2$. Then, the Fourier space two-point function $\hat{G}^{(2)}(p)$ can be expanded by the geometric series
\begin{fmffile}{Rphi}
	\begin{eqnarray}
	\label{eq:Rphi}
	\nonumber
	\hat{G}^{(2)}(p)=
	\begin{gathered}
	\begin{fmfgraph*}(45,35)
	\fmfleft{i}
	\fmfright{o}
	\fmfdot{i,o}
	\fmf{plain,label=p}{i,v}
	\fmf{plain,label=p}{v,o}
	\fmfv{d.sh=circle,d.f=empty,d.si=.5w,b=(0.5,,0.5,,0.5)}{v}
	\end{fmfgraph*}
	\end{gathered}
	&=&
	\begin{gathered}
	\fmfreuse{propagator_p}
	\end{gathered}
	\hspace{2pt}+\hspace{2pt}
	\begin{gathered}
		\begin{fmfgraph*}(45,35)
		\fmfleft{i}
		\fmfright{o}
		\fmfdot{i,o}
		\fmf{plain,label=p}{i,v}
		\fmf{plain,label=p}{v,o}
		\fmfv{d.sh=circle,d.f=empty,d.si=.5w}{v}
		\fmfv{label=1PI,label.angle=180,label.dist=0}{v}
	\end{fmfgraph*}
	\end{gathered}
	\hspace{2pt}+\hspace{2pt}
	\begin{gathered}
		\begin{fmfgraph*}(70,35)
		\fmfleft{i}
		\fmfright{o}
		\fmf{plain,label=p}{i,v1}
		\fmf{plain}{v1,v2}
		\fmf{plain,label=p}{v2,o}
		\fmfdot{i,o}
		\fmfv{d.sh=circle,d.f=empty,d.si=.28w}{v1}
		\fmfv{label=1PI,label.angle=180,label.dist=0}{v1}
		\fmfv{d.sh=circle,d.f=empty,d.si=.28w}{v2}
		\fmfv{label=1PI,label.angle=180,label.dist=0}{v2}
		\end{fmfgraph*}
	\end{gathered}
	\hspace{2pt}+\dots\\
	\nonumber
	&=&\hat{G}_F(p)+(-i\mu)\hat{G}^2_F(p)+(-i\mu)^2\hat{G}^3_F(p)+\dots\\
	&=&\frac{\hat{G}_F(p)}{1+i\mu^2\hat{G}_F(p)}=\frac{i}{p^2-m^2-\mu^2+i\epsilon}.
	\end{eqnarray}
\end{fmffile}Notice that the two-point function is the Green's function of the effective action $\Gamma^{(2)}G^{(2)}=\mathbb{I}$, where $\Gamma^{(2)}=i(\partial^2+m^2)+\dots$ is the tree-level action plus higher-order interactions. We see the loop diagrams effectively shift the mass of the particle $m^2\rightarrow m^2+\mu^2$, where the mass shift $\mu^2(p^2)$ is energy-dependent. In other words, due to the $\phi^4$ interaction, the observed effective mass of the particle is different from its bare mass by an amount that depends on the energy scale. This is the same physical effect that the dispersion relation of a particle is altered by its interactions with a medium, 
except now the interactions are self-interactions due to the $\phi^4$ coupling.

Having understood that loop effects are equivalent to shifting parameters in the Lagrangian, let us quantify the amount of infinity, such that they can be shifted away later. One way to quantify the infinity is using dimensional regularization. In this approach, instead of integrating in the four dimensional spacetime, we first carry out the Wick rotation $k^0=ik^0_E$ such that the integration is transformed to the Euclidean space. Next, we integrate in a hypothetical $d$-dimensional space where $d=4-\epsilon$, such that the integral $d^dk\hat{G}_F(k)\sim k^{d-2}$ becomes convergent. After obtaining $d$-dimensional results, we can then take the limit $\epsilon\rightarrow 0$. Although the limits will go to infinity, we can now quantify the divergence using the Laurent series $\sim\sum_{n=-k}^\infty c_{n}\epsilon^{n}$. For example, in the 1-loop diagram Eq.~(\ref{eq:phi2pt_p}), the divergent momentum integral in d-dimension
\begin{eqnarray}
\nonumber
\int \frac{d^4k}{(2\pi)^4}\frac{i}{k^2-m^2}&\rightarrow&\int \frac{d^dk_E}{(2\pi)^d}\frac{1}{k_E^2+m^2} \\
\nonumber
&=&\int_0^{+\infty} \frac{S_{d-1}k^{d-1}dk}{(2\pi)^d}\frac{1}{k^2+m^2}\\
\nonumber
&=&\frac{S_{d-1}}{2(2\pi)^d (m^2)^{1-d/2}} \int_0^{+\infty} \frac{t^{d/2-1}}{t+1} dt,
\end{eqnarray}
where $S_{d-1}=2\pi^{d/2}/\Gamma(d/2)$ is the surface area of the $(d-1)$-dimensional unit sphere. The remaining integral is the beta function in its standard form
\begin{equation}
\label{eq:beta}
B(x,y)=\int _0^{+\infty} \frac{t^{x-1}}{(t+1)^{x+y}} dt=\frac{\Gamma(x)\Gamma(y)}{\Gamma(x+y)},
\end{equation}
where $\Gamma(z)$ is the gamma function $\Gamma(z+1)=z\Gamma(z)$. Then, the 1-loop diagram
\begin{eqnarray}
\label{eq:phi_1loop_dim}
\begin{gathered}
\fmfreuse{phi_1loop_p}
\end{gathered}
\hspace{2pt}=-i\lambda \frac{\Gamma(1-d/2)}{(4\pi)^{d/2} (m^2)^{1-d/2}},
\end{eqnarray}
where $1-d/2=-1+\epsilon/2$. Taking the limit $\epsilon\rightarrow 0$, the gamma function $\Gamma(\epsilon)=\epsilon^{-1}-\gamma+O(\epsilon)$, where $\gamma$ is the Euler's constant. In the same limit, the mass term can be expanded by $z^\epsilon=1+\epsilon\ln z+O(\epsilon^2)$. Using these expansions, Eq.~(\ref{eq:phi_1loop_dim}) can be expressed as Laurent series, in which the divergent term is $i\lambda m^2/8\pi^2\epsilon$. We have thereof quantified the loop infinity using dimensional regularization.

The infinities, quantifiable using dimensional regularization, can be shifted away by redefining parameters in the Lagrangian. For example, in the above example, infinity in $\mu^2(p^2)$ can be canceled if we subtract the same infinity from the mass term $m^2$ such that the combination \mbox{$m^2+\mu^2(p^2)$} corresponds to the finite physical mass. In retrospect, there is no particular reason why the Lagrangian has to be normalized in the standard form Eq.~(\ref{eq:L_phi4}). The normalizations are thus degrees of freedom we can exploit. Now let us use the multiplicative renormalization scheme with dimensional regularization, where scaling factors are kept explicit in the Lagrangian 
\begin{eqnarray}
\label{eq:L_phi4_renormalize}
\nonumber
\mathcal{L}&=&Z_\phi \partial_\mu\phi^*\partial^\mu\phi-Z_m m^2\phi^*\phi-\frac{\lambda}{4}Z_\lambda M^\epsilon (\phi^*\phi)^2\\
&=&\partial_\mu\phi_B^*\partial^\mu\phi_B-m_B^2\phi^*\phi-\frac{\lambda_B}{4} (\phi_B^*\phi_B)^2.
\end{eqnarray} 
The second line is identical to the standard Lagrangian Eq.~(\ref{eq:L_phi4}) except for the subscript $B$, which indicates that the terms are in fact their bare values. These bare values are related to the renormalized values by
\begin{eqnarray}
\label{eq:Zphi}
\phi_B&=&Z_\phi^{1/2}\phi,\\
\label{eq:Zm}
m_B^2&=&\frac{Z_m}{Z_\phi}m^2,\\
\label{eq:Zlambda}
\lambda_B&=&\frac{Z_\lambda}{Z_\phi^2}M^{\epsilon}\lambda.
\end{eqnarray}
The term $M^\epsilon$ is inserted to make the scaling factor $Z_\lambda$ dimensionless. To see why the term is needed, notice that in dimensional regularization, in order for the action $S=\int d^dx\mathcal{L}$ to be dimensionless, the field has mass dimension $[\phi]=M^{d/2-1}$. Therefore, the interaction term, which has mass dimension $[\phi^4]=M^{d-\epsilon}$, needs to be compensated by $M^\epsilon$.
The above scaling factors can be expanded as power series of the normalized coupling constant 
\begin{eqnarray}
\label{eq:Z_lambda}
Z_\lambda&=&1+\delta_\lambda\hspace{3pt}=1+a_1\lambda+a_2\lambda^2+\dots,\\
\label{eq:Z_m}
Z_m&=&1+\delta_m=1+b_1\lambda+b_2\lambda^2+\dots,\\
\label{eq:Z_phi}
Z_\phi&=&1+\delta_z\hspace{3pt}=1+z_1\lambda+z_2\lambda^2+\dots.
\end{eqnarray} 
When the normalized coupling $\lambda\rightarrow 0$, the field theory becomes free, so there is no interaction and thereof no need for perturbative corrections. In this case, we can either second-quantize or path-integrate the theory in ways discussed in previous sections. 
On the other hand, for small but finite coupling $\lambda$, divergent perturbative corrections arise, which can be removed by allowing the expansion coefficients $c_i=\sum_{j=-k}^\infty c_{ij}\epsilon^j$ to contain infinities. In the expansion $Z=1+\delta$, the ``1" part can be treated in the usual way, while the ``$\delta$" part can be regarded as extra interactions. It is easy to see two additional interactions arise, and the following momentum space Feynman rules can be attributed to the counter terms
\begin{fmffile}{counter}
	\fmfcmd{
		path quadrant, q[], otimes;
		quadrant = (0, 0) -- (0.5, 0) & quartercircle & (0, 0.5) -- (0, 0);
		for i=1 upto 4: q[i] = quadrant rotated (45 + 90*i); endfor
		otimes = q[1] & q[2] & q[3] & q[4] -- cycle;
	}
	\fmfwizard
	\begin{eqnarray}
	\label{eq:counter2}
	\begin{gathered}
		\begin{fmfgraph*}(20,25) 
		\fmfkeep{counter2}
		\fmfleft{i}
		\fmfright{o}
		\fmf{plain,label=p}{i,v}
		\fmf{plain,label=p}{v,o}		
		\fmfv{d.sh=otimes,d.f=empty,d.si=.3w}{v}
		\end{fmfgraph*}
	\end{gathered}
	\hspace{5pt}&=&i(p^2\delta_\phi-m^2\delta_m), \\
	\label{eq:counter4}
	\begin{gathered}
		\begin{fmfgraph*}(20,25) 
		\fmfkeep{counter4}
		\fmfleft{i1,i2}
		\fmfright{o1,o2}
		\fmf{plain}{i1,v,o1}
		\fmf{plain}{i2,v,o2}
		\fmfv{d.sh=otimes,d.f=empty,d.si=.3w}{v}
		\end{fmfgraph*}
	\end{gathered}
	\hspace{5pt}&=&-i\lambda\delta_\lambda.
	\end{eqnarray}
\end{fmffile}
By renormalizing the Lagrangian [Eq.~(\ref{eq:L_phi4_renormalize})], we introduce the above counter terms into the Lagrangian, which can be regarded as additional interactions. These additional interactions can then be used to cancel infinities in loop diagrams.

Before dealing with specific infinities, let us classify what types of infinities can possibly arise. As we have seen earlier, infinities come from loop integrals. Suppose there are $L$ loops in a diagram, then the $L$ integrals, each contributes $\sim d^dk$ in $d$ dimension, have total momentum dimension $k^{dL}$. This momentum in the numerator is canceled by the momentum in the denominator, which is provided by the propagator $\sim k^{-2}$. Hence, if an $L$-loop diagram contains $P$ propagators, the integral is superficially divergent with degree 
\begin{equation}
D=dL-2P.
\end{equation}
Using simple graph theory, the number of loops $L$ and the number of internal lines $P$ can be related to the number of external lines $N$ and the number of vertexes $V$ of the diagram. First, a loop exist if two vertexes are connected by more than one internal lines. Therefore, discarding all external lines, when we remove one internal line together with one of its vertex, the number of loops is reduced by one. By induction, the graph satisfies
\begin{equation}
L=P-V+1.
\end{equation}
Second, a line in the digram is connected to some vertex. Each external line connects to only one vertex, whereas each internal line connects to two vertexes. Since each vertex emanates $n$ lines in the $\phi^n$ theory, the number of lines and vertexes are related by
\begin{equation}
nV=N+2P.
\end{equation}
Expressing $L$ and $P$ in terms of $V$ and $N$, the superficial degree of divergence of the diagram can be written as
\begin{equation}
D=d-\Big[d-n\Big(\frac{d}{2}-1\Big)\Big]V-\Big(\frac{d}{2}-1\Big)N.
\end{equation}
Notice that in $\phi^n$ theory in $d$-dimension, the mass dimension of the coupling coefficient is \mbox{$[\lambda]=M^{d-n(d/2-1)}$}, whose exponent is precisely the coefficient of $V$ in the superficial degree of divergence. 
When $\lambda$ has positive mass dimension, namely when $d-n(d/2-1)>0$, the coefficient of $V$ is negative. In this case, when a diagram contains more vertexes, the diagram becomes less divergent. In other words, when computing higher order diagrams in the perturbation series, the momentum integral becomes more convergent. The perturbation series is well-behaved in this case, and the theory is thereof said to be super-renormalizable. 
On the other hand, when $\lambda$ has negative mass dimension, namely when $d-n(d/2-1)<0$, the coefficient of $V$ is positive. In this case, when a diagram contains more vertexes, the diagram becomes more divergent. In other words, when computing higher order diagrams in the perturbation series, the momentum integral becomes more divergent. The perturbation series is thus ill-behaved, and the theory is said to be non-renormalizable. 
Finally, when $\lambda$ is dimensionless, namely when $d-n(d/2-1)=0$, the $D$ is independent of $V$. In this case, the degree of divergence only depends on the number of external lines $N$, and the divergence is the same to all orders of the perturbation series. In this case, the theory is said to be renormalizable.

According to the above classification, the $\phi^4$ theory in four dimensional spacetime is renormalizable, and contains three types of infinities that can be sifted away by imposing three renormalization conditions. Now that $d=n=4$, the superficial degree of divergence $D=4-N$. Since the theory is invariant under the symmetry $\phi\rightarrow-\phi$, only diagrams with even number of external lines can have nonzero amplitudes. In other words, the possible values of $N$ such that a digram is divergent is $N=0, 2$, and $4$. When $N=0$, there is no incoming or outgoing states, so the diagram is not related to the \textit{S} matrix. The $N=0$ diagrams, such as Eq.~(\ref{eq:eight}), contribute to zero-point function $G^{(0)}=\langle\Omega|\Omega\rangle$, which gives vacuum fluctuations.
When $N=2$, the diagrams are of the form Eq.~(\ref{eq:Rphi}), which give perturbative corrections to the propagator. In other words, $N=2$ diagrams contribute to the two-point function $G^{(2)}(x,y)=\langle\Omega|\mathcal{T}\phi(x)\phi^*(y)\Omega\rangle$, which is the Green's function of the effective action $\Gamma^{(2)}$ of the interaction theory.
In the effective action $\Gamma^{(2)}$, parameters of the Lagrangian are shifted by loop diagrams. 
To subtract infinities by rescaling the parameters, we need to impose renormalization conditions. For example, we can impose the physical renormalization condition to $N=2$ diagrams such that the renormalized mass $m$ in the Lagrangian is the physical rest mass of the particle
\begin{eqnarray}
\label{eq:condition_N2}
\begin{gathered}
\fmfreuse{1PIphi}
\end{gathered}
=0, 
\hspace{10pt} 
\frac{d}{dp^2} \bigg(
\begin{gathered}
\fmfreuse{1PIphi}
\end{gathered}
\bigg)
=0, \hspace{15pt}\text{at}\hspace{5pt} p^2=m^2.
\end{eqnarray}
The first condition requires that the location of the propagator pole is simply $p^2=m^2$ at the physical mass of the particle. The second condition requires that the residue of the pole is simply $i$, so that the normalization of the propagator also remains physical. 
Finally, the remaining divergent diagrams are the $N=4$ diagrams. These diagrams contribute to the 4-point function $G^{(4)}(x_1,x_2,y_1,y_2)=\langle\Omega|\mathcal{T}\phi(x_1)\phi(x_2)\phi^*(y_1)\phi^*(y_2)|\Omega\rangle$. By the LSZ reduction formula, the momentum space 4-point function $\hat{G}^{(4)}(p_1,p_2,p_3,p_4)=\hat{\Gamma}^{(4)}(p_1,p_2,p_3,p_4)\prod_{i=1}^4\hat{G}^{(2)}(p_i)$, where $\hat{\Gamma}^{(4)}$ is the 1PI amplitude. 
To subtract infinities from $\hat{\Gamma}^{(4)}$, we can impose the physical renormalization condition that the coupling coefficient $\lambda$ in the renormalized Lagrangian equals to the physical coupling at rest
\begin{fmffile}{phi4}
	\begin{eqnarray}
	\label{eq:condition_N4}
	\begin{gathered}
		\begin{fmfgraph*}(40,45) 
		\fmfkeep{phiN4}
		\fmfleft{i1,i2}
		\fmfright{o1,o2}
		\fmf{plain}{i1,v}
		\fmf{plain}{i2,v}
		\fmf{plain}{v,o1}
		\fmf{plain}{v,o2}
		\fmfv{d.sh=circle,d.f=empty,d.si=.65w}{v}
		\fmfv{label=1PI,label.angle=180,label.dist=0}{v}
		\end{fmfgraph*}
	\end{gathered}
	\hspace{5pt}&=&-i\lambda, \hspace{15pt} \text{at}\hspace{5pt} s=4m^2, t=u=0.
	\end{eqnarray}
\end{fmffile}Here $s=(p_1+p_2)^2=(p_3+p_4)^2$, $t=(p_1-p_3)^2-(p_2-p_4)^2$, and $u=(p_1-p_4)^2=(p_2-p_3)^2$ are the Mandelstam variables, where $p_1$ and $p_2$ are incoming 4-momentum whereas $p_3$ and $p_4$ are the outgoing 4-momentum. Three renormalization conditions, such as those given above, are necessary and sufficient to determine the three scaling factors $Z_\lambda$, $Z_m$ and $Z_\phi$.   

\subsection{Renormalization of $\phi^4$ theory at 1-loop level\label{sec:vacuum:renormalization:1-loop}}

Now let us consider renormalization of the propagator, namely, the $N=2$ diagrams. Using the 1-loop integral Eq.~(\ref{eq:phi_1loop_dim}) and the counter term Eq.~(\ref{eq:counter2}), the $\lambda^1$-order 1PI amplitude contains two diagrams
\begin{eqnarray}
\nonumber
-i\mu^2&=&
\hspace{38pt}
\begin{gathered}
\fmfreuse{phi_1loop_p}
\end{gathered}
\hspace{33pt}+\hspace{25pt}
\begin{gathered}
\fmfreuse{counter2}
\end{gathered} 
\hspace{32pt}+\dots
\\
&=& -i\lambda \frac{\Gamma(1-d/2)}{(4\pi)^{d/2} (m^2)^{1-d/2}}+ i\lambda(p^2z_1-m^2b_1)+\dots
\end{eqnarray}
where $z_1$ and $b_1$ are the first-order coefficients in expansions Eqs.~(\ref{eq:Z_m}) and (\ref{eq:Z_phi}). Using the two renormalization conditions Eq.~(\ref{eq:condition_N2}), the two expansion coefficients
\begin{eqnarray}
z_1&=&0,\\
b_1&=&-\frac{\Gamma(1-d/2)}{(4\pi)^{d/2} (m^2)^{2-d/2}}\simeq \frac{1}{8\pi^2}\epsilon^{-1}+O(1).
\end{eqnarray}
Therefore, the 1-loop correction to the propagator vanishes, and the 1PI amplitude
\begin{equation}
-i\mu^2=O(\lambda^2).
\end{equation}
This is a special feature of $\phi^4$ theory, and the 1PI amplitude will receive corrections starting from two-loop diagrams at second order in the perturbation series. 

Next, let us consider normalization of the interaction vertex, namely, the $N=4$ diagrams. At 1-loop level, the $\phi^4$ interaction contains the following diagrams
\begin{fmffile}{interaction}
	\begin{eqnarray}
	\label{eq:interaction}
	\nonumber
	-i\Lambda&=&
	\begin{gathered}
	\fmfreuse{phiN4}
	\end{gathered}
	=
	\begin{gathered}
	\fmfreuse{phi4v}
	\end{gathered}
	+
	\begin{gathered}
	\fmfreuse{counter4}
	\end{gathered}
	+
	\begin{gathered}
		\begin{fmfgraph*}(30,35) 
		\fmfkeep{phi4_2s}
		\fmftop{i1,o1}
		\fmfbottom{i2,o2}
		\fmf{plain}{i1,v1,i2}
		\fmf{plain}{o1,v2,o2}
		\fmf{plain,left,tension=0.4}{v1,v2,v1}
		\fmfdot{v1,v2}
		\end{fmfgraph*}
	\end{gathered}
	+
	\begin{gathered}
		\begin{fmfgraph*}(30,35) 
		\fmfkeep{phi4_2t}
		\fmfleft{i1,i2}
		\fmfright{o1,o2}
		\fmf{plain}{i1,v1,o1}
		\fmf{plain}{i2,v2,o2}
		\fmf{plain,left,tension=0.4}{v1,v2,v1}
		\fmfdot{v1,v2}
		\end{fmfgraph*}
	\end{gathered}
	+	
	\begin{gathered}
		\begin{fmfgraph*}(30,35) 
		\fmfkeep{phi4_2u}
		\fmfleft{i1,i2}
		\fmfright{o1,o2}
		\fmf{plain}{i1,v1}
		\fmf{plain,tension=0.2}{v1,o2}
		\fmf{plain}{i2,v2}
		\fmf{plain,tension=0.2}{v2,o1}
		\fmf{plain,left,tension=0.1}{v1,v2,v1}
		\fmfdot{v1,v2}
		\end{fmfgraph*}
	\end{gathered}
	+\dots \\
	&=& -i\lambda(1+a_1\lambda)+(-i\lambda)^2\big[iV(s)+ iV(t)+ iV(u)\big] +O(\lambda^3),
	\end{eqnarray}
\end{fmffile}where $V(p^2)$ denotes the 1-loop integral with loop momentum $p$. The first loop diagram is the $s$-channel diagram with loop momentum $s=(p_1+p_2)^2$, where two incoming particles merge to produce a pair of virtual particles, which then annihilate to produce the outgoing particles. The second loop diagram is the $t$-channel diagram with loop momentum $t=(p_1-p_3)^2$,  where the two incoming particles scatter by exchanging a pair of virtual particles. The third diagram is the $u$-channel diagram with loop momentum $u=(p_1-p_4)^2$, which is similar to the $t$-channel diagram except that the two outgoing particles are exchanged. 
Using the Feynman trick
\begin{equation}
\label{eq:FeynmanTrick}
\frac{1}{AB}=\int_0^1\frac{dx}{[xA+(1-x)B]^2},
\end{equation}
where the dummy variable $x$ is called the Feynman parameter, the 1-loop integral can be computed as follows
\begin{eqnarray}
\nonumber
iV(p^2)&=&\int \frac{d^4k}{(2\pi)^4}\frac{i}{k^2-m^2}\frac{i}{(k+p)^2-m^2}\\
\nonumber
&=&-\int_0^1 dx\int \frac{d^4k}{(2\pi)^4} \frac{1}{\{(1-x)(k^2-m^2)+x[(k+p)^2-m^2]\}^2}\\
\nonumber
&=&-\int_0^1 dx\int \frac{d^4l}{(2\pi)^4} \frac{1}{[l^2+x(1-x)p^2-m^2]^2}.
\end{eqnarray}
On the last line, the integral variable is changed to $l=k+xp$. For large momentum, the integral scales as $d^4l/l^{4}\sim 1$ and is thereof divergent. The divergence can be quantified using dimensional regularization, where we first rotate $l^0=il_E^0$ to the imaginary axis and then integrate in a hypothetical $d=4-\epsilon$ dimensional Euclidean space. Denoting $\Delta^2=m^2-x(1-x)p^2$, then the 1-loop integral
\begin{eqnarray}
\label{eq:V_1loop}
\nonumber
V(p^2)&=&-\int_0^1 dx\int \frac{d^dl_E}{(2\pi)^d} \frac{1}{(l_E^2+\Delta^2)^2}\\
\nonumber
&=&-\int_0^1 dx\int_0^{+\infty} \frac{S_{d-1} l^{d-1} dl}{(2\pi)^d} \frac{1}{(l^2+\Delta^2)^2}\\
\nonumber
&=&-\frac{1}{(4\pi)^{d/2}\Gamma(d/2)}\int_0^1 dx \frac{1}{\Delta^{4-d}}\int_0^{+\infty} dt \frac{t^{d/2-1}}{(t+1)^2}\\
&=&-\frac{\Gamma(\epsilon/2)}{(4\pi)^{2-\epsilon/2}}\int_0^1 dx \frac{1}{[m^2-x(1-x)p^2]^{\epsilon/2}},
\end{eqnarray} 
where $S_{d-1}$ is again the surface area of the $(d-1)$-dimensional unit sphere, and I have used the beta function Eq.~(\ref{eq:beta}) to carry out the $t=l^2/\Delta^2$ integral.
Substituting the above result Eq.~(\ref{eq:V_1loop}) into the 1PI interaction diagram Eq.~(\ref{eq:interaction}), we have thus obtained a formula for the second order scattering amplitude. The 1-loop infinity contained in $V(p^2)$ can be subtracted by the infinity in the 1-loop counter term $a_1$. Using the physical renormalization condition Eq.~(\ref{eq:condition_N4}), the coefficient
\begin{equation}
a_1=-V(4m^2)-2V(0)\simeq \frac{3}{8\pi^2} \epsilon^{-1}+O(1).
\end{equation}
Substituting the above result into Eq.~(\ref{eq:interaction}), the renormalized 4-point 1PI amplitude
\begin{eqnarray}
\label{eq:Lambda}
-i\Lambda\!=\!-i\lambda\!-\!\frac{i\lambda^2}{16\pi^2}\!\int_0^1\!dx \ln\!\Big|\! \big(\frac{1\!-\!x(1\!-\!x)s/m^2}{1\!-4x	(1\!-\!x)}\big)\!\big(\!1\!-\frac{\!x(1\!-\!x)t}{m^2}\big) \!\big(\!1\!-\frac{\!x(1\!-\!x)u}{m^2}\big)\!\Big|\!+O(\lambda^3), \hspace{5pt}
\end{eqnarray}
in the limit $\epsilon\rightarrow 0$. We have thus obtained a well-behaved scattering amplitude to second order in $\lambda$. In the center of mass frame, the scattering is symmetric $p_1=(E,\mathbf{p})$, $p_2=(E,-\mathbf{p})$, $p_3=(E,\mathbf{p}')$, and $p_4=(E,-\mathbf{p}')$. Then the Mandelstam variables $s=4E^2$, $t=-2(E^2-m^2)(1-\cos\theta)$, and $u=-2(E^2-m^2)(1+\cos\theta)$, where $\theta=\langle \mathbf{p}, \mathbf{p}'\rangle$ is the angle between the incoming and outgoing momentum. We see the energy and angle dependences in the second-order amplitude do not cancel. In other words, while $-i\Lambda$ is independent of the incoming and outgoing momentum at tree level, the amplitude starts to depend on the energy scale at 1-loop level, because the interactions now involve creation and annihilation of momentum-carrying virtual particles.

\subsection{Scale dependence and renormalization group flow\label{sec:vacuum:renormalization:group}}

From the above 1-loop calculations, we see after subtracting infinities using counter terms, the effects of loop diagrams are shifting parameters in the Lagrangian in an energy-dependent way. For example, when we impose physical normalization condition that $\lambda$ is the coupling constant between particles at rest, then Eq.~(\ref{eq:Lambda}) tells us what the effective coupling $\Lambda(E)$ is at larger particle energy. Alternatively, we could have imposed the physical renormalization condition at some other energy scale $\Lambda(E_0)=\lambda_0$. Then, after renormalizing the loop diagrams, we can similarly determine what the coupling constant becomes at other energy scales. In other words, the coupling constant at one energy scale is related to the coupling constant at another energy scale by renormalization. We can think of $\Lambda(E)$ as a one-parameter flow, where $\Lambda$ evolves as a function of $E$ in this Wilson's picture of renormalization. Unless we specify how parameters are renormalized, two Lagrangians with seemingly different parameters, can in fact be the same Lagrangian if their parameters are along the same flow line $\Lambda(E)$. On the other hand, if parameters of two Lagrangians are not along the same flow line, then these Lagrangians describe two distinct theories, which are not equivalent under renormalization. More generally, if we regard parameters of a theory as coordinates on a manifold, then renormalization induces one-parameter local diffeomorphism $\Phi_E$ of the manifold. The diffeomorphism $\Phi_E$ forms a one-parameter multiplicative group $\Phi_E\Phi_{E'}=\Phi_{E+E'}$. This group is called the renormalization group. The renormalization group flow allows systematic investigation of how parameters, and thereof observables, change when we do experiments at different scales.

The renormalization group flow is described by the Callan-Symanzik equation. Let me use the $\phi^4$ theory as an example to derive the Callan-Symanzik equation for the $n$-point function, which is a fundamnetal quantity in quantum field theory. In terms of the bare values, the $n$-point function $G_B^{(n)}=\langle\Omega|\mathcal{T}\phi_B(x_1)\dots\phi^*_B(x_n)|\Omega\rangle$ is independent of the renormalization scale $M$ in the dimensional regularization scheme Eq.~(\ref{eq:L_phi4_renormalize}). Therefore, the momentum space $n$-point irreducible amplitude $\Gamma_B^{(n)}(p_1,\dots,p_n)$, which is related to $\hat{G}_B^{(n)}$ by the LSZ reduction formula $\hat{G}_B^{(n)}(p_1,\dots,p_n)=\hat{\Gamma}_B^{(n)}(p_1,\dots,p_4)\prod_{i=1}^n\hat{G}_B^{(2)}(p_i)$, is also independent of the renormalization scale $M$. In other words, the bare value of the $n$-point irreducible amplitude satisfies
\begin{equation}
\label{eq:Gamma_B}
\frac{d}{dM}\hat{\Gamma}^{(n)}_B=0.
\end{equation}   
On the other hand, the renormalized $n$-point irreducible amplitude $\hat{\Gamma}^{(n)}$ does depend on the renormalization scale. Using the scaling relations between the bare and renormalized field values [Eq.~(\ref{eq:Zphi})], the LSZ reduction formula gives
\begin{equation}
\hat{\Gamma}^{(n)}=\frac{\hat{G}^{(n)}(p_1,\dots,p_n)}{\prod_{i=1}^n\hat{G}^{(2)}(p_i)}=\frac{Z_\phi^{-n/2}\hat{G}_B^{(n)}(p_1,\dots,p_n)}{\prod_{i=1}^nZ_\phi^{-1}\hat{G}_B^{(2)}(p_i)}=Z_\phi^{n/2}\hat{\Gamma}_B^{(n)}.
\end{equation}
This renormalized $n$-point irreducible amplitude depends on renormalized parameters in the Lagrangian. In the $\phi^4$ theory example, $\hat{\Gamma}^{(n)}=\hat{\Gamma}^{(n)}(M,m,\lambda)$ explicitly depends on the renormalization scale $M$ and two parameters of the Lagrangian $m(M)$ and $\lambda(M)$. Substituting the above expressions into Eq.~(\ref{eq:Gamma_B}), the total derivative
\begin{eqnarray}
\nonumber
0&=&\frac{d}{dM}\Big(Z_\phi^{-n/2}\hat{\Gamma}^{(n)}\Big)\\
\nonumber
&=&-\frac{n}{2}Z_\phi^{-n/2-1}\frac{\partial Z_\phi}{\partial M}\hat{\Gamma}^{(n)} + Z_\phi^{-n/2}\Big( \frac{\partial \hat{\Gamma}^{(n)}}{\partial M} +\frac{\partial \hat{\Gamma}^{(n)}}{\partial \lambda} \frac{\partial \lambda}{\partial M} + \frac{\partial \hat{\Gamma}^{(n)}}{\partial m} \frac{\partial m}{\partial M}\Big),
\end{eqnarray}
where the partial derivatives means the bare parameters $m_B$ and $\lambda_B$ are kept constant. The above equation states that the dependency of the $n$-point irreducible amplitude on the renormalization scale is exactly canceled by the dependencies of the scaling factors on the renormalization scale. In other words, physical observables have scale dependences, because parameters in the Lagrangian change with scales. This scale dependence is summarized by the above Callan-Symanzik equation, which is conventionally written as
\begin{equation}
\label{eq:CS}
\Big(M\frac{d}{dM}-n\gamma\Big)\hat{\Gamma}^{(n)}=\Big( M\frac{\partial}{\partial M} + \beta \frac{\partial}{\partial \lambda} + \gamma_m m\frac{\partial }{\partial m}  -n\gamma\Big)\hat{\Gamma}^{(n)}=0,
\end{equation}
where the advection speed are given by the dimensionless renormalization group parameters
\begin{eqnarray}
\label{eq:renorm_gamma}
\gamma&=&\frac{M}{2} \Big(\frac{\partial \ln Z_\phi}{\partial M}\Big)_{B},\\
\label{eq:renorm_beta}
\beta&=& M\Big(\frac{\partial \lambda}{\partial M}\Big)_{B},\\
\label{eq:renorm_gamma_m}
\gamma_m&=&\frac{M}{m}\Big(\frac{\partial m}{\partial M}\Big)_{B},
\end{eqnarray}
where the subscript $B$ indicates that the partial derivatives hold the bare parameters constant. In the above equations, parameters in the Lagrangian flow with the energy scale $M$. The flow rates of these parameters then determine how fast the physical observable $\hat{\Gamma}^{(n)}$ changes when varying the experimental scale. Suppose we know how the gamma function $\gamma(M)$ depends in the renormalization scale, then the Callan-Symanzik equation Eq.~(\ref{eq:CS}) can be immediately solved:
\begin{eqnarray}
\label{eq:sol_renorm_gamma}
\hat{\Gamma}^{(n)}(M)=\hat{\Gamma}^{(n)}(M_0)\exp\Big[ n\int_{M_0}^{M} d\mu \frac{\gamma(\mu)}{\mu}\Big].
\end{eqnarray}
This solution connects the renormalized $n$-point irreducible amplitude on one energy scale $M$ to the observable on a different scale $M_0$. Similarly, suppose we know the beta function $\beta(\lambda)$, then the renormalization group equation Eq.~(\ref{eq:renorm_beta}) can be symbolically solved by 
\begin{equation}
\label{eq:sol_renorm_beta}
\ln\frac{M}{M_0}=\int_{\lambda_0}^{\lambda}\frac{d \lambda'}{\beta(\lambda')}.
\end{equation} 
Suppose we can invert this implicit function, then the coupling coefficient $\lambda$ at other energy scales $M$ can be determined by its renormalized value $\lambda_0$ at energy scale $M_0$. Finally, suppose we know $\gamma_m(M)$ as a function of the renormalization scale, then Eq.~(\ref{eq:renorm_gamma_m}) can be easily solved to give the renormalization flow of mass
\begin{equation}
\label{eq:sol_renorm_gamma_m}
m(M)=m_0\exp\Big[\int_{M_0}^{M} d\mu \frac{\gamma_m(\mu)}{\mu}\Big].
\end{equation}
This relation tells us how the mass of the particle at energy scale $M$ is related to its mass at a reference scale $M_0$. For example, suppose we know the rest mass of the particle to be $m_0$, then the above renormalization flow gives the effective mass of the particle when it has finite kinetic energy. The Callan-Symanzik equation with its renormalization group equations thus give a systematic description of how observables depend of the energy scale, which is equivalent to spatial and temporal scales in the experiment.

Now let us determine the renormalization group parameters $\gamma$, $\beta$ and $\gamma_m$. In dimensional renormalization Eq.~(\ref{eq:L_phi4_renormalize}), we see the scaling factors $Z_\phi$, $Z_m$, and $Z_\lambda$ have no explicit dependence on the renormalization scale $M$. The dependence is implicit in the coupling coefficient $\lambda$, and we can write $Z=Z(\lambda(M),\epsilon)$. Therefore, the gamma function [Eq.~(\ref{eq:renorm_gamma})], which is related to wave function renormalization, can be written as
\begin{eqnarray}
\label{eq:gamma_Z}
\nonumber
\gamma&=&\frac{M}{2}\frac{d \ln Z_\phi}{d\lambda} \frac{\partial \lambda}{\partial M} \\
&=&\frac{1}{2}\beta \frac{d \ln Z_\phi}{d\lambda}.
\end{eqnarray} 
Similarly, we can express the beta function in terms of the scaling factors. Using the cyclic identity $(\partial x/\partial y)_z(\partial y/\partial z)_x(\partial z/\partial x)_y=-1$, and the scaling relation $\lambda_B=M^\epsilon \lambda Z_\lambda Z_\phi^{-2}$ [Eq.~(\ref{eq:Zlambda})], the beta function [Eq.~(\ref{eq:renorm_beta})], which is related to renormalization of the coupling coefficient, can be expressed as
\begin{eqnarray}
\label{eq:beta_Z}
\nonumber
\beta&=&-M\frac{(\partial \lambda_B/\partial M)_{\lambda,m_B}}{(\partial \lambda_B/\partial\lambda)_{M,m_B}} \\
\nonumber
&=&-\epsilon\Big[\frac{d}{d\lambda}\ln\big(\lambda Z_\lambda Z_\phi^{-2}\big)\Big]^{-1} \\ 
&=&\frac{-\epsilon+4\gamma}{d \ln(\lambda Z_\lambda)/d\lambda}.
\end{eqnarray}
The last line is obtained by expressing the derivative of $Z_\phi$ in terms of the gamma function, and then solving for $\beta$ using the second line. 
Finally, using the scaling relation $m=Z_\phi^{1/2}Z_m^{-1/2} m_B$ [Eq.~(\ref{eq:Zm})], the $\gamma_m$ function [Eq.~(\ref{eq:renorm_gamma_m})], which is related to mass renormalization, becomes
\begin{eqnarray}
\label{eq:gamma_m_Z}
\nonumber
\gamma_m&=&\frac{M}{m}\frac{d}{d\lambda}\Big(Z_\phi^{1/2}Z_m^{-1/2} m_B \Big) \frac{\partial\lambda}{\partial M}\\
\nonumber
&=&\frac{\beta}{2}\Big(\frac{1}{Z_\phi} \frac{d Z_\phi}{d\lambda}-\frac{1}{Z_m} \frac{d Z_m}{d\lambda}\Big)\\
&=&\gamma-\frac{\beta}{2} \frac{d \ln Z_m}{d\lambda}.
\end{eqnarray}
The above results express the renormalization group parameters in terms of the scaling factors. Using perturbation theory, the scaling factors, expanded by asymptotic series Eqs.~(\ref{eq:Z_lambda})-(\ref{eq:Z_phi}), can be computed order by order. Once the expansion coefficients are known, we can then determine the renormalization group flow.

The group flow at 1-loop level can be determined from the renormalization of 1-loop diagrams. In Sec.~\ref{sec:vacuum:renormalization:1-loop}, we have already determined the first-order expansion coefficients, which are $z_1=0$, $a_1=3/8\pi^2\epsilon+O(1)$, and $b_1=1/8\pi^2\epsilon+O(1)$ as Laurent series of $\epsilon$. Regarding $\epsilon$ as a finite number, then as series of $\lambda$, the derivatives $dZ_\phi/d\lambda=O(\lambda)$,  $d(\lambda Z_\lambda)/d\lambda=1+2\lambda a_1+ O(\lambda^2)$, and $dZ_m/d\lambda=b_1+ O(\lambda)$. Now we are ready to compute the renormalization parameters. Using Eq.~(\ref{eq:gamma_Z}), the gamma function $\gamma=\frac{1}{2}\beta O(\lambda)$. Substituting this result into Eq.~(\ref{eq:beta_Z}) and solving for $\beta$, the 1-loop beta function
\begin{eqnarray}
\nonumber
\beta&=&-\epsilon \frac{\lambda+a_1\lambda^2+O(\lambda^2)}{1+2a_1\lambda+O(\lambda^3)}\\
\nonumber
&=&-\epsilon[\lambda-a_1\lambda^2+ O(\lambda^3)]\\
&\rightarrow& -\epsilon\lambda +\frac{3\lambda^2}{8\pi^2}+\dots,
\end{eqnarray}
where the last line in the limit $\epsilon\rightarrow 0$. Notice that only the coefficient of the divergent term in the Laurent series of $a_1$ contributes to the beta function. Having obtained the beta function, the lowest order gamma function
\begin{equation}
\gamma=-\epsilon z_2\lambda^2+\dots,
\end{equation}
where the coefficient $z_2$ needs to be determined by 2-loop calculations. Finally, substituting the 1-loop beta function and gamma function into Eq.~(\ref{eq:gamma_m_Z}), the mass renormalization flow at 1-loop level is given by
\begin{equation}
\gamma_m=\frac{\lambda}{16\pi^2}+\dots,
\end{equation} 
where the contribution only comes from the divergent term in the Laurent series of $b_1$. 
While the group flow for the $n$-point function is trivial at $\lambda$ order [Eq.~(\ref{eq:sol_renorm_gamma})], the renormalized coupling coefficient and the renormalized mass already start to flow at 1-loop level. Integrating the renormalization group equation Eq.~(\ref{eq:sol_renorm_beta}), it is easy to express $\lambda$ in terms of $M$. Then, the coupling coefficient $\hat{\lambda}=3\lambda/8\pi^2$ at scale $\mu=M/M_0$ is related to the coupling coefficient $\hat{\lambda}_0$ at the reference scale $\mu=1$ by the following flow
\begin{equation}
\hat{\lambda}(\mu)=\frac{\hat{\lambda}_0}{\mu^\epsilon+\hat{\lambda}_0\epsilon^{-1}(1-\mu^\epsilon)}\rightarrow \frac{\hat{\lambda}_0}{1-\hat{\lambda}_0\ln\mu},
\end{equation}
where the limit is attained when $\epsilon\rightarrow 0$. We see for positive $\lambda_0$, as the energy scale $\mu\rightarrow 0$, the coupling coefficient goes to zero. On the other hand, then $\mu\rightarrow+\infty$, the coupling coefficient grows until the perturbation theory fails. This is intuitive because as the energy scale increases, more virtual particles can be excited, whose interactions add up and contribute to a larger effective coupling coefficient. Of course, when the coupling coefficient becomes large, contributions from higher-order diagrams, which are ignored so far, will become important. Therefore, the above 1-loop result is self-contained only in the low energy limit. 
Now suppose it is possible to have $\epsilon>0$, for example, in some condensed matter system. Then, the 1-loop beta function has two fixed point $\lambda_1=0$ and $\lambda_2=8\epsilon\pi^2/3$. When $\lambda_0>\lambda_2$, we have $\beta_0=  M \frac{\partial \lambda}{\partial M}|_0>0$, so $\lambda$ will increase as $M$ increases. On the other hand, when  $\lambda_1<\lambda_0<\lambda_2$, we have $\beta_0<0$. Then, $\lambda$ will decrease towards $\lambda_1$ as $M$ increases. Finally, when $\lambda_0<\lambda_1$, we have $\beta_0>0$, so $\lambda$ will increase towards $\lambda_1$ as $M$ increases. Hence, we see the smaller $\lambda_1$ is a stable fixed point of the renormalization group flow, and a point starts nearby will ultimately flow towards the stable fixed point. On the contrary, the larger $\lambda_2$ is an unstable fixed point of the renormalization group flow, and a point starts nearby will ultimately flow away from the unstable fixed point.
Having obtained the normalization flow for the coupling coefficient, we can readily obtain the mass renormalization flow using Eq.~(\ref{eq:sol_renorm_gamma_m}). After carrying out the integral, the mass flows with the change of scale by
\begin{equation}
m(\mu)=m_0\Big[1+\frac{\hat{\lambda}_0}{\epsilon}\big(\mu^{-\epsilon}-1\big)\Big]^{-1/6}\rightarrow m_0(1-\hat{\lambda}_0 \ln\mu)^{-1/6},
\end{equation}
where the limit is again attained when $\epsilon\rightarrow 0$. For positive $\lambda$, we have $\gamma_m=\frac{M}{m}\frac{\partial m}{\partial M}>0$. Therefore, the mass becomes larger at higher energy. This is also intuitive, because when a particle propagates, it drags virtual particles with it, which add to its effective mass. At higher energy, more virtual particles are created, so the effective mass becomes larger.

\chapter{Quantum electrodynamics in plasmas: effective action approach\label{ch:action}}

In this chapter, I will extend QED to model plasma waves using an effective action approach. Plasma waves are fluctuations that involve self-consistent interactions between charged particles and electromagnetic fields. To make contact with vacuum waves, it is helpful to focus on fluctuations of the EM fields, and package charged particle responses into an effective action. 
Due to the charged particle responses, a wave that exist in the vacuum now propagate differently, which is manifested by a different wave dispersion relation between the wavelength and the wave frequency. In addition, due to the presence of the plasma, waves that do not exist in the vacuum now emerge, whose dispersion relation and polarization differ substantially from the vacuum EM waves. 
A detailed understanding of the plasma waves is the first step towards a more complete theory of weakly-interacting QED plasmas, in which collective effects dominate collisional effects so that linear fluctuations are the simplest phenomena. 
As we shall see, the spectrum of linear waves, which are directly observable in experiments, already exhibits interesting modifications as consequences of relativistic quantum effects in high-energy-density plasmas.

The effective action approach used here is related to the Green's function approach commonly adopted in the literature.
As discussed in Sec.~\ref{sec:vacuum:renormalization:regularization}, the effective action is constituted of the action of the free field plus interactions, and the inverse of the effective action is the Green's function of the full theory. 
In other words, to determine the effective action, it is equivalent to compute the Green's function of the full theory. 
The later approach is taken by \cite{Schwinger61} and \cite{Keldysh65}, who developed a nonequilibrium quantum field theory using the closed time path formalism. Combining the nonequilibrium Green's function formalism with QED, \cite{Bezzerides1972quantum} developed a relativistic quantum theory for plasmas. In their theory, the dynamics of the plasma can be described using the Schwinger-Dyson's equations satisfied by the nonequilibrium Green's functions. By taking the classical and the adiabatic limits, the Schwinger-Dyson's equations recover the Boltzmann's equation, which describes the advection and collisions of charged particles, as well as the wave kinetic equation, which describes the propagation, absorption, and emission of waves.
When the plasma is close to equilibrium, fluctuations can also be described by the finite-temperature field theory using the thermal Green's functions \citep{Rojas79,Melrose07,Melrose2012quantum,Kuznetsov13}, where the thermal average is a special case of the average over an arbitrary density matrix.

Despite of the connections, the effective action approach, which treats the problem on the Lagrangian level using path integrals, has a number of important differences from the Green's function approach.
The effective approach separates statistical fluctuations from quantum fluctuations, and thereof does not rely on the interaction picture used in the Green's function approach, which requires that the system is in thermal equilibrium either in the remote past or in the remote future in order for the asymptotic states and operators to be well defined.
In the Green's function approach, the initial and boundary conditions are hided in the density matrix and the Green's function, whereas they are now displayed explicitly in the Lagrangian in the effective action approach. The later treatment is more convenient in practice, which build up successively higher order approximations from below instead of truncating the BBGKY Hierarchy of $N$-point functions from the above.

The machinery for computing the effective action has already been provided by quantum field theory, and the only additional ingredient is that now there exists a background plasma filling up the vacuum. The presence of a background plasma maybe unfamiliar for QED, which has hitherto been developed to incorporate background EM fields only. However, background plasma can be regarded as a background particle field, which is analogous to the background EM fields already included in the strong-field QED. From the perspective of classical field theory, background fields are a set of initial and boundary conditions that differ from the vacuum. Alternatively, from the point of view of second quantization, background fields are initial and final states other than the vacuum states. Finally, in the path integral formulation, background fields are classical field configurations that differ from the vacuum configuration, upon which quantum fluctuations take place. I find the path integral perspective convenient, and I will use it to derive the general theory of wave effective action for plasmas \citep[Sec. II]{Shi16QED} in this chapter.

\section{The general theory\label{sec:action:general}}

The starting point of a relativistic quantum plasma model is the standard action of scalar QED, in which the complex scalar field is coupled to the gauge field through the covariant derivative [Eq.~(\ref{eq:SQED_Lagrangian})]. To focus on interactions between the charged field and the gauge field, I will consider the special theory in which the renormalized $\phi^4$ coupling is zero. In other words, the scalar-QED plasma model is based on the action
\begin{eqnarray}\label{eq:action}
S=\int d^4x\Big[(D_{\mu}\phi)^{*}(D^{\mu}\phi)-m^2\phi^{*}\phi -\frac{1}{4}F_{\mu\nu}F^{\mu\nu}\Big].
\end{eqnarray}
The complex scalar field $\phi$ describes charged spin-0 bosons with mass $m$ and charge $e$. For simplicity, I have only included one scalar field, keeping in mind that additional scalar fields can be added to model a multi-species plasma, in which we can set, for example, mass $m=m_e$ and the fine structure constant $\alpha=e^2/4\pi\approx1/137$ to model electron-like species. 
The real-valued 1-form $A=A_{\mu}dx^{\mu}$ is the gauge field that defines the gauge covariant derivative $D_{\mu}=\partial_{\mu}-ieA_{\mu}$. The covariant derivative has curvature 2-form $F_{\mu\nu}=\partial_{\mu}A_{\nu}-\partial_{\nu}A_{\mu}$, commonly known as the field strength tensor. 
Although I will not deal with the $\phi^4$ term, it is worth mentioning that this coupling is necessary for the theory to be renormalizable. In the strongly coupled regime, the $\phi^4$ nonlinearity can lead to intriguing structures like the Abrikosov vortex \citep{Abrikosov57}. In the weak coupling regime, the $\phi^4$ term can be treated perturbatively and contributes at the 2-loop level. Here, I will focus on the weak coupling regime and study the propagation of the gauge field at the 1-loop level, where we can safely set the renormalized value of $\lambda$ to zero.

\subsection{Plasmas as background fields\label{sec:action:general:background}}

\begin{figure}[!b]
	\centering
	\renewcommand{\figurename}{FIG.}
	\includegraphics[angle=0,width=0.6\textwidth]{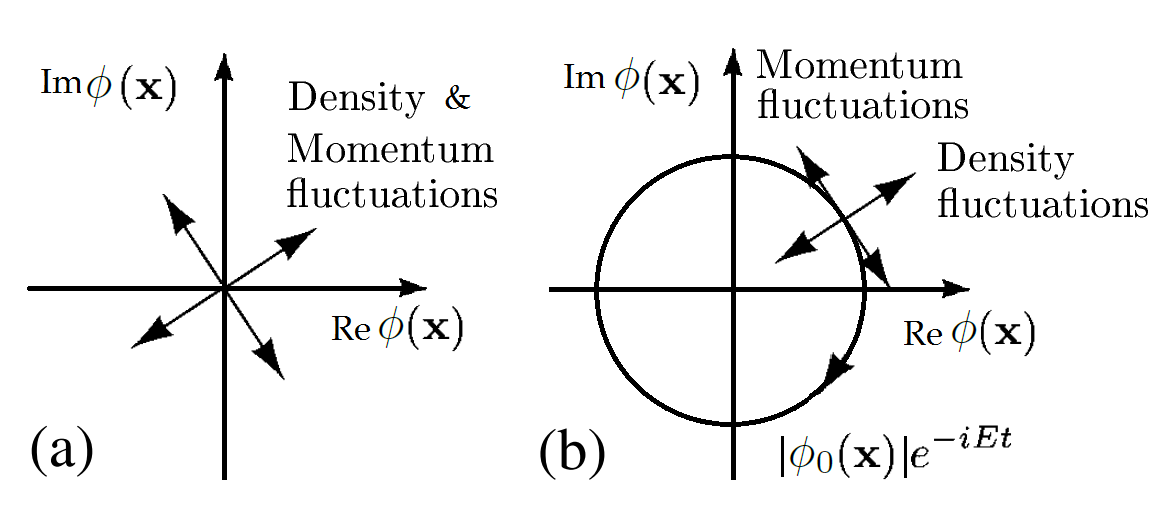}
	\caption[Background fields in quantum field theory]{Comparison between field theory in the vacuum and field theory with dynamical background. 
	When a scalar field $\phi(\mathbf{x})$ fluctuates in the vacuum (a), energy and momentum fluctuations are always accompanied by density fluctuations. On the other hand, when the field fluctuates near some non-vacuum background (b), energy and momentum fluctuations can be orthogonal to density fluctuations.}
	\label{fig:background}
\end{figure}

To describe plasmas, which are constituted of charged particles and their self-consistent EM fields, let us first understand the roles of background fields in quantum field theory. 
In the usual quantum field theory, fields fluctuate near their vacuum expectation values. In finite temperature field theory, fields fluctuate in a thermal bath, which is characterized by two parameters: temperature and chemical potential. More generally, fields fluctuate on some background. The background, which is described by some wave function, can be dynamical and out of thermal equilibrium. The presence of such a non-trivial background adds new ingredients to field theories (Fig.~\ref{fig:background}).  
Mathematically, $\phi$ and $A$ can be decomposed into classical fields and quantum fluctuations
\begin{equation}\label{Expand}
	\phi=\phi_{0}+\varphi,\quad A_{\mu}= \bar{A}_{\mu}+\mathcal{A}_{\mu}.  
\end{equation}
The classical fields $\phi_0$ and $\bar{A}_{\mu}$ account for statistical fluctuations of the system, whereas the quantum fields $\varphi$ and $\mathcal{A}_{\mu}$ account for quantum fluctuations. Notice that the expectation value $\langle \phi\rangle=\phi_0+\langle \varphi\rangle$. In other words, the classical field $\phi_0$ is not the mean field, and the quantum field $\varphi$ is not assumed to have zero expectation value. Similarly, the field $\mathcal{A}_{\mu}$ can have nonzero expectation value and in general depends on $\bar{A}_\mu$ and $\phi_0$.
Vacuum is the trivial case when the background fields $\phi_0$ and $\bar{A}$ are zero. When the background fields are nontrivial, the only condition for the classical fields $\phi_0$ and $\bar{A}$ is that they satisfy the self-consistent classical Euler$-–$Lagrange equations [Eqs.~(\ref{eq:EOM_phi}) and (\ref{eq:EOM_A})]. To emphasize that now it is the background fields that satisfy these classical field equations, we can write 
\begin{eqnarray}
	\label{EOM}
	(\bar{D}_{\mu}\bar{D}^{\mu}+m^2)\phi_{0}&=&0,\\
	\label{EOMA}
	\partial_{\mu}\bar{F}^{\mu\nu}=\bar{J}_{0}^{\nu}.
\end{eqnarray}
In the above equations, $\bar{D}_{\mu}=\partial_{\mu}-ie\bar{A}_{\mu}$ is the background gauge covariant derivative, $\bar{F}_{\mu\nu}=\partial_{\mu}\bar{A}_{\nu} -\partial_{\nu}\bar{A}_{\mu}$ is the background field strength tensor, and $\bar{J}_{0}^{\nu}=\sum_s\bar{J}_{s0}^{\nu}$ is the total background current [Eq.~(\ref{eq:Jmu})], summed over all charged species. It is clear that the above equations are invariant under the background U(1)-gauge transformation of $\phi_{0}$ and $\bar{A}$. The classical equations of motion describe bound states as well as unbound states. When the potential energy is larger than the kinetic energy, as is the case in condensed matter systems, particles are bound by the potential created by other particles. In this case, the wave functions $\phi_0$ and $\bar{A}$ are localized and correlation between particles can be strong. On the other hand, when the kinetic energy is larger than the potential energy, as is the case in plasmas, particles are unbound. In this case, the motion of one particle is weakly correlated with the motion of other particles, except during collisions.

The above background separation scheme is different from the usual BBGKY hierarchy commonly adopted for many-body systems. In the BBGKY scheme, the exact $N$-point function $G^{(N)}(x_1,\dots,x_N)=\langle\phi(x_1)\dots\phi(x_n)\rangle$ satisfies an infinite hierarchy of equations of the form $\hat{\mathcal{O}}[G^{(N)}]=\hat{\mathcal{C}}[G^{(N+1)}]$, where $\hat{\mathcal{O}}$ and $\hat{\mathcal{C}}$ are some operators. In a mathematical language, the evolution of the mean is affected by the standard deviation, and more generally, the evolution of the $N$-th cumulant is affected by the $(N+1)$-th cumulant. Equivalently, in a physical language, the BBGKY hierarchy expresses the mean field in terms of the binary collision operator, and express the two-point correlation function in terms of tertiary collision operator and so on. To solve the infinite set of equations, one has to truncate the hierarchy by imposing some closure conditions. In other words, in order to solve lower-order correlation functions, one has to approximate higher-order correlations, and the approximation propagates from high order to low order. This is in contrast to the separation scheme Eq.~(\ref{Expand}). In this scheme, the lower-order correlation functions satisfy closed equations, such as Eqs.~(\ref{EOM}) and (\ref{EOMA}). Building upon the exact solutions to the closed equations, higher-order correlation functions then pick up information from lower order, and the approximation is thereof bottom-up instead of top-down.

While the background EM field $\bar{F}_{\mu\nu}$ is conceptually simple, the background charged particle field $\phi_0$ needs some clarifications. When the plasma background is constituted of $N$ bosons, the classical background field $\phi_{0}(x)$ is formally related to the properly symmetrized $N$-body wave function $\Phi_{0}(x_1,x_2,\dots,x_N)$ by
\begin{equation}
	\label{eq:phi0}
	\phi_0(x)=\int\sqrt{V}\Phi_0(x,x_2,\dots,x_N).
\end{equation}
Here $V=d^4x_2\wedge\dots\wedge d^4x_N$ is the volume form of the $4(N-1)$-dimensional subspace of the $N$-boson configuration space. The half-form $\sqrt{V}$ is commonly seen in geometric quantization \citep{Bates97}. It is easy to check that the $N$-body wave function has mass dimension $[\Phi_0(x_1,\dots,x_N)]=M^{2N-1}$ and the field $\phi_0(x)$ has mass dimension $[\phi_0(x)]=M$ as expected. When a pair $\phi_{0}(x)\phi^{*}_{0}(x)$ appears in an expression, two half-forms combine into the volume form, and the integration can then be carried out. For example, the 4-current density $\bar{J}_{0}^{\mu}$ of an $N$-body wave function $\Phi_{0}$ can be written explicit as
\begin{equation}\label{MBcurrent}
	\bar{J}_{0}^{\mu}(x)=\frac{e}{i}\!\int\!V[\Phi_{0}^{*}(x,x_2...)\bar{D}^{\mu}(x)\Phi_{0}(x,x_2...)-\text{c.c.}],
\end{equation}
where c.c. denotes the complex conjugation. 
For conciseness, whenever the pair $\phi_{0}(x)\phi^{*}_{0}(x)$ appears in an expression, integration over all other coordinates of $x_2,\dots$ of the many-body wave function $\Phi_0(x,x_2,\dots)$ will be implied.

The background particle field $\phi_{0}$ may be interpreted using notions in the second quantization formulation.
In this formulation, the combination $\phi_{0}(x)\phi^{*}_{0}(x)$ may be understood as a functional representation of the density operator $\hat{\rho}_0=|\hat{\phi}_{0}\rangle\langle\hat{\phi}_{0}|$. From another perspective, the combination $\phi_{0}(x)\phi^{*}_{0}(y)$ may be understood as the 2-point correlation function $F(x,y)=\langle\hat{\phi}_{0}(x)\hat{\phi}_{0}^\dagger(y)\rangle$. The correlation function can be transformed to a phase space distribution using the Wigner-–Weyl transform, which first change variables to $R=(x+y)/2$ and $r=x-y$, and then take Fourier transform $\exp(ipr)$ to obtain $F(R,p)$. The phase space correlation function can be related to the phase space distribution function $f(R,p)$ in the semiclassical limit, where the relativistic quantum scales are well separated from the collective scales.
That being said, interpretation of the background field $\phi_0$ using notions of second quantization is not essential. Here in the path integral formulation, the background field $\phi_0$ is simply a classical field that satisfies the classical field equation, upon which quantum fluctuations take place.

\subsection{Background-reduced action\label{sec:action:general:reduced}}
Having clarified the roles of background fields, we can now study their effects in the field theory. 
When solving the classical field equations, the separation of small fluctuations from the dominant backgrounds allows the equations to be solved perturbatively. This is a viable way to proceed in classical field theory, whereby we can use expansions Eq.~(\ref{Expand}) to obtain linear dispersion relations and higher order wave-wave interactions in a way similar to what I have done in the first part of this thesis. However, at the end of the perturbative solution, we will only know the classical trajectory of the fields. To allow quantum fluctuations to take place, we need to carry out path integrals, which accounts for quantum trajectories that are forbidden classically.

Path integrals of quantum fluctuations are easier to carry out after the classical action is subtracted. Similar to the example in Sec.~\ref{sec:vacuum:pathintegral:formulation}, subtracting the classical contribution allows the fluctuating fields to vanish on the boundaries, whereby functional integrals can be easily computed. 
To subtract classical fields from quantum fluctuations using decomposition Eq.~(\ref{Expand}), we can use the following trick. Suppose at least one of the two functions $h$ and $f$ vanishes at infinity, then the covariant integration by part is given by
\begin{eqnarray}\label{eq:IBT}
	\nonumber
	\int dx h^{*}\bar{D}_{\mu}f&=&\int dx h^{*} (\partial_{\mu}-ie\bar{A}_\mu)f\\
	\nonumber
	&=&\int dx f(-\partial_{\mu}-ie\bar{A}_\mu)h^{*} \\
	&=&	-\int dx f (\bar{D}_{\mu}h)^{*}.
\end{eqnarray} 
Using the classical equations of motion [Eqs.~(\ref{EOM}) and (\ref{EOMA})] to cancel terms linear in the fluctuating fields, the action Eq.~(\ref{eq:action}) can be written as
\begin{eqnarray}\label{ReducedAction}
	\nonumber
	S&=&S_c+\int d^4x\Big[(D_{\mu}\varphi)^{*}(D^{\mu}\varphi)-m^2\varphi^{*}\varphi -\frac{1}{4}\mathcal{F}_{\mu\nu}\mathcal{F}^{\mu\nu} \\
	&&\hspace{50pt}-\bar{\eta}^{\mu} \mathcal{A}_{\mu}+e^2(\phi_{0}\phi_{0}^{*}+\phi_{0}\varphi^{*}+\phi_{0}^{*}\varphi) \mathcal{A}_{\mu}\mathcal{A}^{\mu}\Big], 
\end{eqnarray}
where $S_c$ is the classical action, and  $\mathcal{F}_{\mu\nu}=\partial_{\mu}\mathcal{A}_{\nu}-\partial_{\nu}\mathcal{A}_{\mu}$ is the field strength tensor of the fluctuating field $\mathcal{A}$.
Terms on the first line define the usual strong-field scalar QED. The background gauge field $\bar{A}_\mu$ shows up in the background gauge covariant derivative $\bar{D}_\mu=D_\mu+ie\mathcal{A}_\mu$, which affects the parallel transport of the charged field on the U(1) bundle. Similar to scalar QED in the vacuum, the 4-current density arises from the vacuum excitation is given by
\begin{equation}
\label{eq:jmu}
\bar{\jmath}^{\mu}:=\frac{e}{i}(\varphi^{*}\bar{D}^{\mu}\varphi-\text{c.c.}),
\end{equation}
except now the covariant derivative is $\bar{D}_\mu$. 
What was not included in strong-field QED but now becomes incorporated is the background plasma contribution, which appears on the second line of Eq.~(\ref{ReducedAction}). The background field $\phi_0$ shows up in two places. First, it shows up as the interaction vertex in the background current
\begin{eqnarray}
\label{eq:etamu}
\bar{\eta}^{\mu}&:=&\frac{e}{i}(\phi_{0}^{*}\bar{D}^{\mu}\varphi+\varphi^{*}\bar{D}^{\mu}\phi_{0}-\text{c.c.}),
\end{eqnarray}
through which fluctuations in the plasma medium affect the fluctuating gauge field $\mathcal{A}_{\nu}$. Second, $\phi_0$ shows up in coefficients of the quadratic term $\mathcal{A}_{\mu}\mathcal{A}^{\mu}$, through which the gauge field acquire mass without breaking the local U(1)-gauge symmetry. This mass generation mechanism is similar to the Higgs mechanism \citep{Higgs64}, except now the vacuum expectation values, generated by spontaneous symmetry breaking, is replaced by the physical presence of charged particles \citep{Anderson63}, whereby the symmetry remains unbroken.

The background-reduced action Eq.~(\ref{ReducedAction}) is associated with a Lagrangian density $\mathcal{L}$ that depends on the fluctuating fields $\varphi$ and $\mathcal{A}$. When waves propagate through background plasmas, the background field $\bar{F}_{\mu\nu}$, which is usually generated by some slowly-varying and large-scale external charge current distributions, can be regarded as purely classical. The background charged particle field $\phi_{0}$, which is self-consistent with $\bar{F}_{\mu\nu}$, can also be regarded as purely classical. In this way, all interactions between waves and charged particles are taken into account by the fluctuating fields. Up to some constant terms in the classical action, the Lagrangian density of the fluctuating fields $\mathcal{A}$ and $\varphi$ is
\begin{eqnarray}\label{ReducedLagrangian}
	\nonumber
	\mathcal{L}\enspace=&&\mathcal{L}_{\varphi}+\mathcal{L}_{\mathcal{A}}+\mathcal{L}_{I}\\
	\nonumber
	=&&(\bar{D}_{\mu}\varphi)^{*}(\bar{D}^{\mu}\varphi)-m^2\varphi^{*}\varphi\\
	&-&\frac{1}{4}\mathcal{F}_{\mu\nu}\mathcal{F}^{\mu\nu}+e^2\phi_{0}^{*}\phi_{0}\mathcal{A}_{\mu}\mathcal{A}^{\mu}\\
	\nonumber
	&-&(\bar{\jmath}^{\mu}+\bar{\eta}^{\mu})\mathcal{A}_{\mu}+e^2(\phi_{0}\varphi^{*}+\phi_{0}^{*}\varphi +\varphi^{*}\varphi)\mathcal{A}_{\mu}\mathcal{A}^{\mu}.
\end{eqnarray}
Here $\mathcal{L}_{\varphi}$, $\mathcal{L}_{\mathcal{A}}$, and $\mathcal{L}_{I}$ correspond to terms on the second, the third, and the fourth line, respectively. $\mathcal{L}_{\varphi}$ is the Lagrangian density of the free $\varphi$ field. It should be clarified that $\varphi$ is not free in the sense that its dynamics is influenced by the background field $\bar{A}$, as is manifested by the background gauge covariant derivative $\bar{D}$ acting on $\varphi$. But $\varphi$ is nevertheless free in the sense that it neither interacts with $\mathcal{A}$ nor couples to itself. Similarly, $\mathcal{L}_{\mathcal{A}}$ is the Lagrangian density of the free $\mathcal{A}$ field. Notice that the background field $\phi_0$ endows the gauge field $\mathcal{A}$ with a mass term that can have spatial and temporal dependencies. Finally, the interaction Lagrangian $\mathcal{L}_{I}$ contains interactions between $\varphi$ and $\mathcal{A}$. Some interactions only involve the fluctuating fields $\varphi$ and $\mathcal{A}$ with constant couplings. These interactions happen in plasmas as well as in the vacuum. Other interactions involve the background fields $\phi_0$ and $\bar{A}$ in the coupling. These interactions do not happen unless nontrivial background fields are present.

The Lagrangian density Eq.~(\ref{ReducedLagrangian}) has a number of gauge symmetries. It is obvious that the Lagrangian is invariant under background local U(1)-gauge transformation
\begin{equation}\label{BackgroundGauge}
	\phi_{0}\rightarrow\phi_{0} e^{ie\chi},\quad \varphi\rightarrow\varphi e^{ie\chi} ,\quad \bar{A}_{\mu}\rightarrow \bar{A}_{\mu}+\partial_{\mu}\chi,
\end{equation}
where $\chi$ is an arbitrary real scalar field. These correspond to a local phase rotation of the total particle field $\phi$, where the background gauge field $\bar{A}_{\mu}$ absorbs all the transformations. Alternatively, we can keep the  
background fields $\phi_{0}$ and $\bar{A}$ fixed, and transform the $\varphi$ and $\mathcal{A}_{\mu}$ fields. It is a straightforward calculation to verify that the Lagrangian is invariant under the following transformation of fluctuating fields
\begin{equation}\label{Symmetry}
	\mathcal{A}_{\mu}\rightarrow \mathcal{A}_{\mu}+\partial_{\mu}\chi ,\quad \varphi\rightarrow\varphi e^{ie\chi}+\phi_{0}(e^{ie\chi}-1).
\end{equation}
This can be understood intuitively 
as follows. The local U(1)-gauge transformation (\ref{Gauge}) is a shift in $A$ and a phase rotation in $\phi$. When $\bar{A}$ is fixed the shift is completely absorbed into $\mathcal{A}$. When $\phi_{0}$ is fixed, $\varphi$ has to transform by Eq.~(\ref{Symmetry}) in order to preserve the norm of $\phi$. The conserved symmetry current $\partial_{\mu}\mathcal{J}^{\mu}=0$ is
\begin{equation}
	\mathcal{J}^{\mu}=\bar{\jmath}^{\mu}+\bar{\eta}^{\mu}-2e^2\phi\phi^{*}\mathcal{A}^{\mu}.
\end{equation}
This gauge invariant current $\mathcal{J}$ contains contributions from excitations of the background fields as well as excitations of the vacuum. Through this current, the fluctuations of the charged field can be transmutated to fluctuations of the gauge field. In other words, the $\varphi$ field and the $\mathcal{A}_\mu$ field are mixed by the above symmetry current, and both fields thereof share the same set of quantum numbers.

\subsection{Effective action of gauge bosons\label{sec:action:general:effective}}

So far, no approximation has been made, and the Lagrangian density (\ref{ReducedLagrangian}) is exact. The Lagrangian density describes the free $\varphi$ field, the free $\mathcal{A}$ field, and their interactions. 
When behaviors of the charged particles are of concern, this background-reduced Lagrangian is well suited for describing propagation and collisions of charged particles. The focus here is, however, the behavior of the gauge field.
When the gauge field $\mathcal{A}$ propagates, it interacts with charged particles and becomes dressed by these interactions. After summing up all these dressings, the effective action, which describes the propagation of the dressed $\mathcal{A}$ field, can be obtained. The summation of dressings can be rigorously implemented using the path integral, which can be evaluated perturbatively using the small dimensionless coupling constant $e$, namely, the elementary charge in the natural units, as an expansion parameter. 

Formally, the exponentiated effective action $e^{i\Gamma\bm{[}\mathcal{A}\bm{]}}$ of the $\mathcal{A}$ field is the partially evaluated quantum partition function when the $\varphi$ field is integrated out. The quantum partition function serves a similar role as the statistical partition function. In the statistical case, the average is weighted by the Boltzmann factor $e^{-H/k_\text{B}T}$, while in the quantum case, the average is weighted by the phase factor $e^{iS/\hbar}$. To integrate out the $\varphi$ field, we will need to expand the action exponential $e^{iS}$. It is convenient to group terms in the interaction $S_{I}=\int d^4x\mathcal{L}_{I}$ according to their powers in $e$, $\varphi$, and $\mathcal{A}$. Schematically, we can write
\begin{equation}
	S_{I}=S_{e\varphi\mathcal{A}}+S_{e\varphi^2\mathcal{A}}+S_{e^2\varphi\mathcal{A}^2}+S_{e^2\varphi^2\mathcal{A}^2},
\end{equation}
where each term corresponds to an interaction Lagrangian $\mathcal{L}_{e\varphi\mathcal{A}}=-\bar{\eta}^{\mu}\mathcal{A}_{\mu}$, $\mathcal{L}_{e\varphi^2\mathcal{A}}=-\bar{\jmath}^{\mu}\mathcal{A}_{\mu}$, $\mathcal{L}_{e^2\varphi\mathcal{A}^2}=e^2(\phi_{0}\varphi^{*}+\phi_{0}^{*}\varphi )\mathcal{A}_{\mu}\mathcal{A}^{\mu}$, and $\mathcal{L}_{e^2\varphi^2\mathcal{A}^2}=e^2\varphi^{*}\varphi\mathcal{A}_{\mu}\mathcal{A}^{\mu}$.
Denoting the action of the free $\varphi$ field and the free $\mathcal{A}$ field by $S_\varphi$ and $S_\mathcal{A}$, expanding the action exponential to $e^2$ order, and using properties of Gaussian integrals to eliminate terms that contain odd powers of $\varphi$ in the path integral, the exponentiated effective action
\begin{eqnarray}\label{EffectiveAction}
e^{i\Gamma\bm{[}\mathcal{A}\bm{]}}&:=&\frac{1}{Z_{\varphi}}\int\mathscr{D}\varphi\mathscr{D}\varphi^{*}e^{i(S_{\varphi}+S_{\mathcal{A}}+S_{I})}\\
\nonumber
&=&\frac{e^{iS_{\mathcal{A}}}}{Z_{\varphi}}\int\mathscr{D}\varphi\mathscr{D}\varphi^{*}e^{iS_{\varphi}}\bigg[1+i\Big(S_{e\varphi^2\mathcal{A}}+S_{e^2\varphi^2\mathcal{A}^2}\Big)+\frac{i^2}{2}\Big(S_{e\varphi\mathcal{A}}^2+S_{e\varphi^2\mathcal{A}}^2\Big)+O(e^3)\bigg],
\end{eqnarray}
where $Z_{\varphi}:=\int\mathscr{D}\varphi\mathscr{D}\varphi^{*}e^{iS_{\varphi}}$ is the partition function of the free $\varphi$ field. The term $S_{e\varphi^2\mathcal{A}}$ is linear in $\mathcal{A}$. It serves as the source term that is responsible for the emission, absorption and scattering of gauge bosons. Since the focus here is wave propagation, I will not be concerned with this term. The remaining terms in the expansion (\ref{EffectiveAction}) are quadratic in $\mathcal{A}$ and they are responsible for the propagation of the gauge field. It is worth mentioning that when deriving the full effective action of the gauge field, or equivalently its Green's function in the full theory, the collisional term $S_{e\varphi^2\mathcal{A}}$ should not be disregarded.

To express the effective action in a more illuminating form, we can write the above expansion in terms of quantities that are familiar in quantum field theory. The first quantity is the propagator, or the Green's function, of the free $\varphi$ field
\begin{eqnarray}
\nonumber
G(x,x')&=&\langle\varphi(x)\varphi^*(x')\rangle_{\varphi}\\
&=&\frac{1}{Z_{\varphi}}\int\mathscr{D}\varphi\mathscr{D}\varphi^{*}e^{iS_{\varphi}}\varphi(x)\varphi^{*}(x').
\end{eqnarray}
The Green's function of the free  $\varphi$ field appears when evaluating the Gaussian functional integrals similar to Eqs.~(\ref{eq:path_Gaussian0}) and (\ref{eq:path_Gaussian1}). Using covariant integration by part [Eq.~(\ref{eq:IBT})], the quadratic form of the Gaussian integral is
\begin{eqnarray}
iS_{\varphi}=-\int d^4 x\; \varphi^* i\big(\bar{D}_{\mu}\bar{D}^{\mu}+m^2-i\epsilon\big) \varphi.
\end{eqnarray}
where $-i\epsilon$ is added to make the operator positive definite. The inverse of the quadratic operator is the Feynman Green's function
\begin{equation}\label{eq:SDGreen}
[\bar{D}_{\mu}(x)\bar{D}^{\mu}(x)+m^2]G(x,x')=-i\delta(x-x'),
\end{equation}
where the pole prescription is given by $m^2\rightarrow m^2-i\epsilon$. The equation satisfied by the above Green's function is similar to Eq.~(\ref{eq:dG}), except the partial derivatives are now replaced by background gauge covariant derivatives. When the background field is nontrivial, namely, when $\bar{F}_{\mu\nu}$ is nonzero, $\bar{A}_\mu$ cannot be shifted away by background U(1)-gauge transformation. In this case, the Green's function of the charged field is different from its vacuum value.

The second quantity useful for rewriting the effective action is the gauge invariant polarization tensor $\Pi^{\mu\nu}(x,x')$. The polarization tensor is the current-current correlation function. It is the probability amplitude that a wave excites a current at location $x$, from which the current propagates to another location $x'$, where the current emits another wave and becomes de-excited. 
Using properties of Gaussian integrals to integrate out the $\mathcal{A}$ field, we can evaluate the exact polarization tensor to $O(e^2)$ order
\begin{eqnarray}\label{polarization}
	\nonumber
	\Pi^{\mu\nu}(x,x')&=&\langle\mathcal{J}^{\mu}(x)\mathcal{J}^{\nu}(x')\rangle\\
	\nonumber
	&=&\frac{1}{Z}\int\mathscr{D}\varphi\mathscr{D}\varphi^{*}\mathscr{D}\mathcal{A}e^{iS}\mathcal{J}^{\mu}(x)\mathcal{J}^{\nu}(x')\\
	\nonumber
	&=&\frac{1}{Z_{\varphi}}\int\mathscr{D}\varphi\mathscr{D}\varphi^{*}e^{iS_{\varphi}}(\bar{\eta}^{\mu}\bar{\eta}^{\nu}+\bar{\jmath}^{\mu}\bar{\jmath}^{\nu})+O(e^3)\\
	&=&\Pi^{\mu\nu}_{2,\text{bk}}(x,x')+\Pi^{\mu\nu}_{2,\text{vac}}(x,x')+O(e^3),
\end{eqnarray}
where $Z=\int\mathscr{D}\varphi\mathscr{D}\varphi^{*}\mathscr{D}\mathcal{A}e^{iS}$ is the total partition function of Lagrangian density (\ref{ReducedLagrangian}). 
Notice that integrating the $\mathcal{A}$ field requires gauge fixing, which can be done using the Faddeev--Popov procedure similar to Eq.~(\ref{eq:2pt_A}). However, since the $\mathcal{A}$ field does not contribute to the polarization tensor at $e^2$ order, we do not need to be concerned with gauge fixing at this order.
On the third line of Eq.~(\ref{polarization}), cross terms between $\bar{\eta}$ and $\bar{\jmath}$, which contain odd power of either $\varphi$ or $\varphi*$, vanish upon evaluating the Gaussian path integral. 
The two terms $\Pi^{\mu\nu}_{2,\text{bk}}(x,x')$ and $\Pi^{\mu\nu}_{2,\text{vac}}(x,x')$ are the polarization of the background plasma and the polarization of the vacuum, respectively. They appear from $S_{e\varphi\mathcal{A}}^2$ and $S_{e\varphi^2\mathcal{A}}^2$ after evaluating the path integral (\ref{EffectiveAction}). The subscript ``2" indicates that they are approximate expressions to $e^2$ order in the perturbation series.

In terms of the Green's function and the polarization tensors, the effective action of gauge boson propagation can be written in a concise form. Eliminating the source term $S_{e\varphi^2\mathcal{A}}$ in Eq.~(\ref{EffectiveAction}), the Gaussian path integral can be computed as
\begin{eqnarray}
\nonumber
e^{i\Gamma\bm{[}\mathcal{A}\bm{]}}&=&\frac{e^{iS_{\mathcal{A}}}}{Z_{\varphi}}\int\mathscr{D}\varphi\mathscr{D}\varphi^{*}\Big[1+i\int d^4 x \mathcal{L}_{e^2\varphi^2\mathcal{A}^2} \\
\nonumber
&&+\frac{i^2}{2} \Big(\int d^4 x \mathcal{L}_{e\varphi\mathcal{A}} \int d^4 x' \mathcal{L}^{'}_{e\varphi\mathcal{A}} + \int d^4 x \mathcal{L}_{e\varphi^2\mathcal{A}} \int d^4 x' \mathcal{L}^{'}_{e\varphi^2\mathcal{A}}\Big)+O(e^3)\Big]\\
\nonumber
&=&e^{iS_{\mathcal{A}}}\Big\{1+ i\int d^4 x e^2G(x,x) \mathcal{A}_{\mu}(x)\mathcal{A}^{\mu}(x) \\
\nonumber
&&-\frac{1}{2}\int d^4 x d^4 x' \mathcal{A}_{\mu}(x)\Big[\Pi^{\mu\nu}_{2,\text{bk}}(x,x')+\Pi^{\mu\nu}_{2,\text{vac}}(x,x')\Big]\mathcal{A}_{\nu}(x') +O(e^3)
\Big\}\\
\nonumber
&=&\exp\Big\{ i\int d^4 x\Big[ \mathcal{L}_\mathcal{A}+ e^2G\mathcal{A}_{\mu}\mathcal{A}^{\mu} +\frac{i}{2}\int d^4 x' \mathcal{A}_{\mu}\Big(\Pi^{\mu\nu}_{2,\text{bk}}+\Pi^{\mu\nu}_{2,\text{vac}}\Big)\mathcal{A}'_{\nu} +O(e^3)\Big] \Big\},
\end{eqnarray}
where the expansion is put back into the exponential on the last line. From the above result, we can read out the effective action $\Gamma\bm{[}\mathcal{A}\bm{]}$ from the exponent. Using Eq.~(\ref{ReducedLagrangian}) for the free gauge field $\mathcal{L}_\mathcal{A}$, the $O(e^2)$-order effective action for wave propagation
\begin{eqnarray}\label{1-loop}
\Gamma_2\bm{[}\mathcal{A}\bm{]}&=&\frac{1}{2} \int  d^4x\Bigg[\mathcal{A}_{\mu}(x)(\partial^2g^{\mu\nu}-\partial^{\mu}\partial^{\nu})\mathcal{A}_{\nu}(x) + \int  d^4x'\mathcal{A}_{\mu}(x)\Sigma_2^{\mu\nu}(x,x')\mathcal{A}_{\nu}(x')\Bigg], \hspace{20pt}
\end{eqnarray}
where the first term is the vacuum action, and $\Sigma_2^{\mu\nu}(x,x')$ is the $e^2$-order response tensor, which is also called the self-energy of the gauge boson. The response tensor contains contributions from the background plasma as well as the vacuum
\begin{equation}
	\Sigma_2^{\mu\nu}(x,x')=\Sigma_{2,\text{bk}}^{\mu\nu}(x,x')+\Sigma_{2,\text{vac}}^{\mu\nu}(x,x').
\end{equation}
The response due to the background plasma is constituted of the gauge boson mass term and the plasma polarization term
\begin{fmffile}{bk}
	\begin{eqnarray}\label{bk}
		\Sigma_{2,bk}^{\mu\nu}(x,x')
		&=&\quad
		\begin{gathered}
			\begin{fmfgraph*}(40,15)
				\fmfkeep{mass}
				\fmfleft{i}
				\fmfright{o}
				\fmf{photon}{i,v}
				\fmf{photon}{v,o}
				\fmfdot{v}
				\fmfv{label=$x$,label.angle=90,label.dist=6}{v}
				\fmfv{label=$\mu$,label.dist=0.2}{i}
				\fmfv{label=$\nu$,label.dist=0.2}{o}
			\end{fmfgraph*}
		\end{gathered}
		\quad+\quad
		\begin{gathered}
			\begin{fmfgraph*}(40,15)
				\fmfkeep{line}
				\fmfleft{i}
				\fmfright{o}
				\fmf{plain}{v1,v2}
				\fmfdot{v1,v2}
				\fmfv{label=$x$,label.angle=90,label.dist=8}{v1}
				\fmfv{label=$x'$,label.angle=90,label.dist=8}{v2}
				\fmfv{label=$\mu$,label.dist=0.5}{i}
				\fmfv{label=$\nu$,label.dist=0.5}{o}
				\fmf{photon}{i,v1}
				\fmf{photon}{v2,o}
			\end{fmfgraph*}
		\end{gathered}\\
		\nonumber
		&=&\!2e^2\phi_0\phi^*_0\delta(x-x')g^{\mu\nu}\!+i\Pi^{\mu\nu}_{2,bk}(x,x').
	\end{eqnarray}
\end{fmffile}The first term, corresponding to the first Feynman diagram, is the photon mass term in Lagrangian (\ref{ReducedLagrangian}). The second term, corresponding to the second Feynman diagram, comes from the $\bar{\eta}^{\mu}\bar{\eta}^{\nu}$ term in the path integral (\ref{polarization}). The background plasma responds by particle-hole pair excitation. During this process, a gauge boson is forward scattered, namely, the gauge boson is first absorbed after exciting a plasma particle and then get re-emitted by this particle after its de-excitation. The response due to the vacuum is constituted of the gauge boson mass renormalization and the vacuum polarization 
\begin{fmffile}{vac}
	\begin{eqnarray}\label{vac}
		\Sigma_{2,vac}^{\mu\nu}(x,x')
		&=&\quad
		\begin{gathered}
			\begin{fmfgraph*}(40,25)
				\fmfkeep{hairpin}
				\fmfleft{i}
				\fmfright{o}
				\fmf{photon}{i,v}
				\fmf{photon}{v,o}
				\fmf{plain}{v,v}
				\fmfdot{v}
				\fmfv{label=$x$,label.angle=-90,label.dist=6}{v}
				\fmfv{label=$\mu$,label.dist=0.2}{i}
				\fmfv{label=$\nu$,label.dist=0.2}{o}
			\end{fmfgraph*}
		\end{gathered}
		\quad+\quad
		\begin{gathered}
			\begin{fmfgraph*}(50,25)
				\fmfkeep{bubble}
				\fmfleft{i}
				\fmfright{o}
				\fmf{plain,left=1,tension=0.3}{v1,v2}
				\fmf{plain,right=1,tension=0.3}{v1,v2}
				\fmfdot{v1,v2}
				\fmfv{label=$x$,label.angle=120,label.dist=8}{v1}
				\fmfv{label=$x'$,label.angle=60,label.dist=8}{v2}
				\fmfv{label=$\mu$,label.dist=0.5}{i}
				\fmfv{label=$\nu$,label.dist=0.5}{o}
				\fmf{photon}{i,v1}
				\fmf{photon}{v2,o}
			\end{fmfgraph*}
		\end{gathered}\\
		\nonumber
		&=&2e^2\langle\varphi\varphi*\rangle_{\varphi}\delta(x-x')g^{\mu\nu}+i\Pi^{\mu\nu}_{2,vac}(x,x').
	\end{eqnarray}
\end{fmffile}The first term, corresponding to the first Feynman diagram, is the photon mass renormalization term. It comes from the $S_{e^2\varphi^2\mathcal{A}^2}$ term in the path integral (\ref{EffectiveAction}). The second term, corresponding to the second Feynman diagram, is the vacuum polarization term. It comes from the $\bar{\jmath}^{\mu}\bar{\jmath}^{\nu}$ term in the path integral (\ref{polarization}). The vacuum responds by virtual pair excitation. During this process, a gauge boson first decays into a pair of virtual particle and antiparticle, and then get reproduced when the virtual pair annihilates. The first line of the effective action (\ref{1-loop}) is the same as $\frac{1}{4}\mathcal{F}_{\mu\nu}\mathcal{F}^{\mu\nu}$ after integration by part. This is the action of the $\mathcal{A}$ field in the vacuum. The second line is a nonlocal term that depends on two coordinates $x$ and $x'$. This term describes the dressing of the $\mathcal{A}$ field due to its interactions with the background plasma and the vacuum.

\section{Polarization tensors\label{sec:action:integral}}

Explicit expressions of the polarization tensors $\Pi^{\mu\nu}_{2,\text{bk}}(x,x')$ and $\Pi^{\mu\nu}_{2,\text{vac}}(x,x')$ can be found by evaluating the path integrals in Eq.~(\ref{polarization}). For conciseness, I will abbreviate 1-point functions by $\varphi(x)=\varphi$, $\varphi(x')=\varphi'$, and so on. Similarly, I will abbreviate 2-point functions by $G(x,x')=G$, $G(x',x)=G'$, and so on. It is useful to note since $\mathcal{L}_\varphi$ is quadratic in $\varphi\varphi^*$, the Gaussian integrals $\langle\varphi\varphi'\rangle_{\varphi}=\langle\varphi^*\varphi'^*\rangle_{\varphi}=0$. Moreover, due to the imaginary exponent $e^{iS}$, we have $G^*=-G'$ in the presence of background gauge fields.

\subsection{Polarization of the plasma medium\label{sec:action:integral:plasma}}

Polarization in the plasma medium is similar to polarization in other charged matter. The difference between plasma and normal matter is that particles are bounded and non-relativistic in normal matter, whereas they become unbound and can have relativistic energy in the rest frame of the plasma. Although the states of the matter are different, the physical processes that lead to polarization are the same for condensed matter and plasmas. 
When the medium is perturbed, for example by a photon, charged particles in the medium have some probability to absorb the photon. However, unless special resonance conditions are satisfied, the absorption is virtual. This is because the dispersion relation of the photon usually does not match the dispersion relation of the particle. In this case, the absorption process does not respect energy-momentum conservation. Nevertheless, quantum uncertainty allows such process to happen within 
some time $\Delta t\sim1/\Delta E$. Within this time, the photon can be transiently absorbed and then re-emitted, returning the particle to its original state. The process described above is the lowest order process where the interaction is two-body. It gives the dominant contribution to the polarization tensor when the medium is weakly coupled.

To compute the polarization tensor of the plasma, we need to compute the current-current correlation function, where the current is due to medium excitation. In other words, we need to compute the expectation value $\Pi^{\mu\nu}_{2,\text{bk}}=\langle\bar{\eta}^{\mu}\bar{\eta}^{\nu}\rangle$, where $\bar{\eta}^{\mu}$ is given by Eq.~(\ref{eq:etamu}). In the path integral formulation, the expectation value can be obtained by functional integration in Eq.~(\ref{polarization}). Using properties of Gaussian integrals, the expectation value can be expressed in terms of the background wave function $\phi_0$ and the charged particle Green's function $G$ as follows
\begin{eqnarray}
\nonumber
\langle\bar{\eta}_{\mu}\bar{\eta}_{\nu}\rangle&=&\frac{1}{Z_{\varphi}}\int\mathscr{D}\varphi\mathscr{D}\varphi^{*}e^{iS_{\varphi}}\bar{\eta}_{\mu}\bar{\eta}^{'}_{\nu}\\
\nonumber
&=&\frac{e^2}{Z_{\varphi}}\int\mathscr{D}\varphi\mathscr{D}\varphi^{*}e^{iS_{\varphi}}\Big[ (\varphi\bar{D}^{*}_{\mu}\phi^{*}_0-\phi^{*}_0\bar{D}_{\mu}\varphi) (\varphi^{'*}\bar{D}^{'}_{\nu}\phi_0^{'}-\phi_0^{'}\bar{D}_{\nu}^{'*}\varphi^{'*})+\text{c.c.}\Big]\\
\nonumber
&=&e^2\big[G(\bar{D}^{*}_{\mu}\phi^{*}_0)(\bar{D}^{'}_{\nu}\phi_0^{'}) \!-\! (\bar{D}^{*}_{\mu}\phi^{*}_0)(\phi_0^{'}\bar{D}_{\nu}^{'*}G) \!-\! (\bar{D}^{'}_{\nu}\phi_0^{'})(\phi^{*}_0\bar{D}_{\mu}G) \!+\! \phi^{*}_0\phi_0^{'}\bar{D}_{\mu}\bar{D}_{\nu}^{'*}G \!-\! \text{c.c.}\big]. 
\end{eqnarray}
On the second line, I have used $\langle\varphi\varphi'\rangle=\langle\varphi^*\varphi'^*\rangle=0$, so only cross terms like $\varphi\varphi^{'*}$ contribute to the Gaussian integral. In addition, I have used $(z-\text{c.c.})(w-\text{c.c.})=zw-zw^*+\text{c.c.}$ to compress the expression. On the third line, the four additional terms involve $G'$, which can be written in terms of $G^*=-G'$, giving rise to the negative sign. 
The last line can be factorized using commutations like $[\bar{D}_{\mu},\phi_0']=0$, which holds because $\phi_0'$ is at location $x'$ whereas the derivative $\bar{D}_{\mu}$ acts on a different location $x$. 
From the above calculation, we see the background polarization tensor 
\begin{equation}\label{bkPol}
\Pi^{\mu\nu}_{2,\text{bk}}=e^2\big[\phi_0^*\bar{D}^{\mu}-(\bar{D}^{\mu}\phi_{0})^*\big]\big[\phi_0'\bar{D}^{'*\nu}-(\bar{D}^{'\nu}\phi_{0}')\big]G-\text{c.c.}\hspace{2pt}.
\end{equation}
The background polarization tensor corresponds to the second Feynman diagram in Eq.~(\ref{bk}). 
The incoming photon excites the plasma medium and creates current $\bar{\eta}^{\mu}(x)$. The current is carried by an ``electron-hole" pair, namely, a virtual excitation of the plasma medium, whose propagation from $x$ to $x'$ is given by the charged particle Green's function $G(x,x')$. The transient current de-excites at $x'$, where the current  $\bar{\eta}^{\nu}(x')$ emits an outgoing photon when the ``electron-hole" pair recombines. 
Notice that this process cannot happen unless the occupation number of charged particles is nonzero. In other words, when the background particle field is trivial $\phi_0=0$, background polarization is also zero. This is intuitive because when there is no on-shell particles filling up the vacuum, the plasma medium does not exist and thereof cannot be polarized. 
When the plasma medium does exist, the presence of $\bar{A}_\mu$ affects the excitation and de-excitation of the medium current through covariant derivatives like $\bar{D}_{\mu}$, as well as affects the propagation of the  ``electron-hole" pair through the charged particle Green's function.

\subsection{Vacuum polarization under background fields\label{sec:action:integral:vacuum}}

Regardless of whether the plasma is present or not, the vacuum can be polarized. Vacuum polarization is a genuine relativistic quantum effect: due to the quadratic relativistic $E^2=m^2+p^2$ dispersion relation, antiparticles must exist in addition to particles; due to quantum uncertainty, energy and momentum are conserved only on average. Therefore, when both quantum and relativistic effects are in place, a seemingly empty space is in fact filled with vacuum fluctuations where particle-antiparticle pairs pop out and then disappear spontaneously. When a real photon propagates through the fluctuating vacuum, it may encounter a virtual ``electron-positron" pair and transiently interacts with these charged particles. This process can also be thought of as spontaneously decay of the photon to an ``electron-positron" pair, which then recombines to re-emit the photon. 

Although vacuum polarization knows nothing about the plasma medium at $e^2$ order, it is affected by the presence of background electromagnetic fields, because EM fields affect dynamics of the virtual pairs. 
For example, in strong magnetic field, virtual pairs are magnetized and occupy quantized Landau levels. When a photon interacts with magnetized vacuum, its dispersion relation receives corrections from the Heisenberg--Euler effective action \citep{Heisenberg36}. 
As another example, when intense laser field is present, the linearly polarized laser accelerates charged particles mostly in the direction of the laser electric field. The anisotropic vacuum fluctuations appear to another photon as a birefringent medium, resulting in an effect known as vacuum birefringence \citep{Brezin71}.
These effects of background EM fields can also be though of as multi-photon effects, where the background field can be written as a superposition of coherent virtual photons. Just as plasmas are made of charged particles, which interact through virtual photons, background EM fields are made of photons, which interact through virtual charged particles.

Using path integral formulation, the vacuum polarization tensor can be computed when arbitrary background fields are present. Regardless of whether the background field is static, such as a DC magnetic field, or dynamic, such as a laser field, the general formula given by path integral is the same. 
To compute the polarization tensor of the vacuum, we need to compute the current-current correlation function, where the current is due to vacuum fluctuations. In other words, we need to compute the expectation value $\Pi^{\mu\nu}_{2,\text{vac}}=\langle\bar{\jmath}^{\mu}\bar{\jmath}^{\nu}\rangle$, where $\bar{\jmath}^{\mu}$ is given by Eq.~(\ref{eq:jmu}). In the path integral formulation, the expectation value can be computed by functional integration in Eq.~(\ref{polarization}). Using properties of Gaussian integrals, 
\begin{eqnarray}
\nonumber
\langle\bar{\jmath}_{\mu}\bar{\jmath}_{\nu}\rangle&=&\frac{1}{Z_{\varphi}}\int\mathscr{D}\varphi\mathscr{D}\varphi^{*}e^{iS_{\varphi}}\bar{\jmath}_{\mu}\bar{\jmath}^{'}_{\nu}\\
\nonumber
&=&\frac{e^2}{Z_{\varphi}}\int\mathscr{D}\varphi\mathscr{D}\varphi^{*}e^{iS_{\varphi}}\Big[ (\varphi^*\bar{D}_{\mu}\varphi)(\varphi^{'}\bar{D}^{'*}_{\nu}\varphi^{'*}) - (\varphi^{*}\bar{D}_{\mu}\varphi)(\varphi^{'*}\bar{D}_{\nu}^{'}\varphi^{'})+\text{c.c.}\Big]\\
\nonumber
&=&e^2\big[G'(\bar{D}_{\mu}\bar{D}^{'*}_{\nu}G)- (\bar{D}^{'}_{\nu}G^{'})(\bar{D}_{\mu}G)\big] + \text{c.c.}\hspace{2pt}.
\end{eqnarray}
On the second line, I have again used $(z-\text{c.c.})(w-\text{c.c.})=zw-zw^*+\text{c.c.}$ to compress the expression. On the third line, the Gaussian integral is calculated using properties similar to Eq.~(\ref{eq:path_Gaussian1}). The last line can be factorized using $G^*=-G'$, then the $e^2$-order vacuum polarization tensor can be written as
\begin{equation}\label{vacPol}
\Pi^{\mu\nu}_{2,\text{vac}}=e^2\big[G'\bar{D}^{\mu}-(\bar{D}^{*\mu}G')\big](\bar{D}^{'*\nu}G)+\text{c.c.}\hspace{2pt}.
\end{equation}
The vacuum polarization tensor corresponds to the second Feynman diagram in Eq.~(\ref{vac}). 
The incoming photon decays to a virtual pair, creating a vacuum current $\bar{\jmath}^{\mu}(x)$ at location $x$ through the interaction vertex $e\bar{D}_\mu(x)$. The creation of the virtual pair cannot satisfy energy and momentum conservation, so the pair must annihilate within $\Delta t\sim1/\Delta E$. Within this time allowed by quantum fluctuations, the particle propagates `forward in time" from $x$ to $x'$ by the Green's function $G(x',x)$, while the antiparticle propagates ``backward in time" from $x'$ to $x$ by the Green's function $G(x,x')$. At location location $x'$, the virtual pair annihilates and the current $\bar{\jmath}^{\nu}(x')$ re-emits the photon through the interaction vertex $e\bar{D}_\nu(x')$.
As discussed earlier, this process is affected by the background gauge field through covariant derivatives like $\bar{D}_{\mu}$, whereas it knows little about the presence of the background plasma at $e^2$ order.
Expressions (\ref{1-loop})-(\ref{vacPol}) combined give an explicit formula of the effective action of gauge field propagation to order $e^2$ in the most general setting. To this order, effective action contains all Feynman diagrams of the gauge boson propagator up to 1-loop level, so the $e^2$-order effective action is the same as the 1-loop effective action.

\section{Properties of the effective action\label{sec:action:properties}}

In the previous section, I derived the 1-loop effective action of gauge boson propagation using path integrals. The general formulas for plasma and vacuum responses are applicable when arbitrary background fields $\phi_0$ and $\bar{A}_\mu$ are present. The background fields can be inhomogeneous and dynamical, as long as they satisfy the self-consistent classical field equations. In this section, I will discuss a number of symmetries and conservation properties of the effective action, and point out how the behaviors of the gauge boson are fully encoded in the effective action. 

\subsection{Symmetries and conservation laws\label{sec:action:properties:plasma}}

First, the 1-loop effective action is manifestly Lorentz-invariant. The fluctuating gauge field $\mathcal{A}^\mu$ transforms as a Lorentz vector, and the response tensor $\Sigma_{2}^{\mu\nu}$ transforms as a rank-$(2,0)$ tensor. The effective action, with all indexes properly contracted, is a Lorentz scalar. In other words, the theory I have developed for waves in scalar-QED plasmas is a relativistic theory. Although when the plasma is present, there exist a special reference frame, namely, the plasma rest frame in which particle momentum averages to zero, the plasma wave theory is equally valid in any other inertial frames. Once we compute the response tensor in one reference frame, we can immediately find its expression in boosted frames using Lorentz transformations.

Second, the 1-loop effective action is invariant under the background U(1)-gauge transformation. Therefore, when computing the effective action, we can choose any gauge of convenience, and the final result will be independent of the gauge choice. In practice, it is usually convenient to chose the gauge under which the background 1-point functions and the 2-point Green's function respect other symmetries of the problem. To see the background U(1)-gauge invariance, notice that under transformation Eq.~(\ref{BackgroundGauge}), the photon mass term $\phi_0\phi_0^*$, namely the first term in Eq.~(\ref{bk}), is obviously invariant. Moreover, the Green's function $G(x,x')=\langle\varphi(x)\varphi^*(x')\rangle$ transforms as
\begin{equation}\label{eq:GaugeGreen}
G(x,x')\rightarrow e^{ie\chi(x)}G(x,x')e^{-ie\chi(x')}.
\end{equation} 
While the covariant derivative of the 1-point function is simply transformed by $\bar{D}_\mu(x)\phi_0(x)\rightarrow e^{ie\chi(x)}\bar{D}_\mu(x)\phi_0(x)$, the transformation of the background covariant derivatives of the Green's function is slightly more complicated
\begin{eqnarray}
\bar{D}_\mu(x)G(x,x')&\rightarrow& e^{ie\chi(x)}[\bar{D}_\mu(x)G(x,x')]e^{-ie\chi(x')},\\
\bar{D}^{*}_\mu(x')G(x,x')&\rightarrow& e^{ie\chi(x)}[\bar{D}^{*}_\mu(x')G(x,x')]e^{-ie\chi(x')}.
\end{eqnarray}
Using the above transformation rules, it is a straightforward calculation to verify that the background polarization tensor [Eq.~(\ref{bkPol})], and thereof the plasma response tensor [Eq.~(\ref{bk})] are invariant under background U(1)-gauge transformation. Similarly, the vacuum response tensor [Eq.~(\ref{vac})] is also invariant. Its first term, namely, the mass renormalization term, involves the Green's function $G(x,x)$ at the same point, which is trivially invariant by transformation Eq.~(\ref{eq:GaugeGreen}). Moreover, its second term, namely the vacuum polarization [Eq.~(\ref{vacPol})], involves $G':=G(x',x)=-G^*(x,x')$, which transforms in exactly the opposite way as $G(x,x')$. Therefore, using transformation rules for the covariant derivatives, it is easy to verify that $\Pi^{\mu\nu}_{2,\text{vac}}$, and thereof the 1-loop effective action is invariant under background U(1)-gauge transformation.

Finally, the 1-loop effective action is invariant under local gauge transformation of the fluctuating field $\mathcal{A}_{\mu}\rightarrow\mathcal{A}_{\mu}+\partial_{\mu}\chi$. In other words, the presence of background fields do not break the gauge symmetry. The gauge symmetry of $\mathcal{A}_{\mu}$ is the relic of Eq.~(\ref{Symmetry}), after the fluctuating charged particle field $\varphi$ is integrated out. The relic gauge symmetry ensures that charge is conserved in the effective theory. By direct calculations (Appendix~\ref{ch:append:Ward}), the response tensors satisfy conservation laws
\begin{eqnarray}
\label{bkConservation}
\partial_{\mu}\Sigma_{2,\text{bk}}^{\mu\nu}(x,x')&=&\partial'_{\nu}\Sigma_{2,\text{bk}}^{\mu\nu}(x,x')=0,\\
\label{vacConservation}
\partial_{\mu}\Sigma_{2,\text{vac}}^{\mu\nu}(x,x')&=&\partial'_{\nu}\Sigma_{2,\text{vac}}^{\mu\nu}(x,x')=0.
\end{eqnarray}
These conservation laws are in fact local charge conservation laws, because the linear response tensors are related to currents by the Ohm's law. To be more precise, the perturbation $\mathcal{A}_{\mu}$ creates a plasma current $\bar{\eta}^{\mu}(x)=\int d^4x' \Sigma_{2,\text{bk}}^{\mu\nu}(x,x')\mathcal{A}_{\nu}(x')$ and a vacuum current $\bar{\jmath}^{\mu}(x)=\int d^4x' \Sigma_{2,\text{vac}}^{\mu\nu}(x,x')\mathcal{A}_{\nu}(x')$. 
The plasma current is due to excitation of ``electron-hole" pair. Since charge is conserved during this process, we have $\partial_{\mu} \bar{\eta}^{\mu}=0$. Similarly, the vacuum current is due to creation of ``electron-positron" pair. Although the number of particles is not conserved during this process, the charge is nevertheless conserved $\partial_{\mu} \bar{\jmath}^{\mu}=0$. 
After integration by part, it is clear that the effective action is invariant under the local gauge transformation of the $\mathcal{A}$ field. Identities (\ref{bkConservation}) and (\ref{vacConservation}) indicate that the plasma current and the vacuum current are conserved separately, so the plasma contribution to wave propagation is separable from the vacuum contribution.

\subsection{Experimental observables\label{sec:action:properties:observables}}

The full effective action encodes all properties of the gauge boson. To $e^2$ order, the effective action is quadratic in $\mathcal{A}$, and therefore describes the propagation of the gauge boson. 
From the classical field theory point of view, we can derive the classical equation of motion of the gauge field using variational principle. To $e^2$ order, the resultant equation is a linear hyperbolic PDE, whose solutions are linear waves. Components of the 4-vector equation are simply the Maxwell's equations, in which the 4-current is given self-consistently by the Ohm's law through the response tensors. Solutions to the field equations give the classical behaviors of linear fluctuations on top of the background $\phi_0$ and $\bar{A}$, which can be inhomogeneous and dynamical. 
From the quantum field theory point of view, the quadratic effective action can be inverted to give the Green's function of the free $\mathcal{A}$ field. Notice that the free $\mathcal{A}$ field feels the effects of background fields $\phi_0$ and $\bar{A}$, and is thereof different from the vacuum gauge boson propagator. With the Green's function of the $\mathcal{A}$ field, we can then calculate higher order correlation functions and study interactions mediated by the $\mathcal{A}$ field, such as screened collisions between charged particles inside the plasma medium.

The effective action becomes particularly simple when the background fields are translational invariant. In this case, the response tensor $\Sigma(x,x')$ only depends on the difference between coordinates $r=x-x'$ and is independent of $R=(x+x')/2$. In this case, it is more convenient to work in the momentum space. In general, the momentum space is related to the configuration space by Fourier transforms
\begin{eqnarray}
\mathcal{A}_{\mu}(x)&=&\int\frac{d^4k}{(2\pi)^4}e^{-ikx}\hat{\mathcal{A}}_{\mu}(k),\\
\label{Fourier}
\Sigma_{2}^{\mu\nu}(x,x')&=&\int\frac{d^4k}{(2\pi)^4}\frac{d^4k'}{(2\pi)^4}e^{-ikx}\hat{\Sigma}_{2}^{\mu\nu}(k,k')e^{ik'x'}_{\hspace{15pt}.}
\end{eqnarray}
The configuration space reality condition $\mathcal{A}^{*}(x)=\mathcal{A}(x)$ and the exchange symmetry $\Sigma_{2}^{\mu\nu}(x,x')=\Sigma_{2}^{\nu\mu}(x',x)$ correspond to the momentum space conditions
\begin{eqnarray}
\hat{\mathcal{A}}_{\mu}(k)&=&\hat{\mathcal{A}}^{*}_{\mu}(-k),\\
\label{eq:reality}
\hat{\Sigma}_{2}^{\mu\nu}(k,k')&=&\hat{\Sigma}_{2}^{\nu\mu}(-k',-k).
\end{eqnarray}
Using the above properties, the configuration space 1-loop effective action [Eq.~(\ref{1-loop})] can be transformed to the momentum space
\begin{eqnarray}\label{MS1Loop}
\nonumber
\Gamma_{2}\bm{[}\mathcal{A}\bm{]}&=&\frac{1}{2}\int\frac{d^4k}{(2\pi)^4} \Big[\hat{\mathcal{A}}_{\mu}(-k)(k^{\mu}k^{\nu}-k^2g^{\mu\nu})\hat{\mathcal{A}}_{\nu}(k)\\
&&+\int\frac{d^4k'}{(2\pi)^4}\hat{\mathcal{A}}_{\mu}(-k)\Sigma_2^{\mu\nu}(k,k') \hat{\mathcal{A}}_{\nu}(k')\Big],
\end{eqnarray}
where $k^2=k^{\mu}k_{\mu}$ is the Minkowski inner product. Simplifications can be made when the plasma is translational invariant. In this case, changing variables from $x$ and $x'$ to $r$ and $R$ in Eq.~(\ref{Fourier}), we have $\hat{\Sigma}(k,k')=(2\pi)^4\delta^{(4)}(k-k')\hat{\Sigma}(k)$, where $\hat{\Sigma}^{\mu\nu}(k)=\int d^4re^{ikr}\Sigma^{\mu\nu}(r)=\hat{\Sigma}^{\nu\mu}(-k)$. 
The gauge invariance and current conservation laws in the configuration space [Eqs.~(\ref{bkConservation}) and (\ref{vacConservation})] becomes the Ward--Takahashi identities
\begin{equation}\label{eq:Ward}
k_{\mu}\Sigma_{2,\text{bk}}^{\mu\nu}(k)=k_{\mu}\Sigma_{2,\text{vac}}^{\mu\nu}(k)=0.
\end{equation} 
Moreover, with the extra delta function from $\hat{\Sigma}(k,k')$, the $k'$ integral on the second line of Eq.~(\ref{MS1Loop}) can be easily carried out. The momentum space $e^2$-order effective action can then be simplified as
\begin{equation}
\Gamma_{2}\bm{[}\mathcal{A}\bm{]}=-\frac{1}{2}\!\int\!\frac{d^4k}{(2\pi)^4} \hat{\mathcal{A}}_{\mu}(-k)D^{\mu\nu}(k)\hat{\mathcal{A}}_{\nu}(k).
\end{equation}
The dispersion tensor is constituted of the free field contribution [Eq.~(\ref{eq:Green_A_xi})], as well as plasma and the vacuum responses
\begin{equation}\label{eq:DispersionTensor}
D^{\mu\nu}(k)=k^2g^{\mu\nu}-k^{\mu}k^{\nu}-\Sigma_2^{\mu\nu}(k)=D^{\nu\mu}(-k).
\end{equation}
For given background fields $\phi_0$ and $\bar{A}$, the dispersion tensor $D^{\mu\nu}(k)$ may be inverted after gauge fixing 
$i\tilde{D}_{\mu\nu}(\xi)\hat{\Lambda}_F^{\nu\sigma}(k)=\delta^{\rho}_{\mu}$, from which the momentum space Green's function can be readily obtained. The Green's function $\hat{\Lambda}_F^{\mu\nu}(k)$ can then be used to calculate Feynman diagrams, when substituted into the usual machinery of quantum field theory.

When there is no external source, the classical equation of motion in the momentum space is $D^{\mu\nu}(k)\hat{\mathcal{A}}_{\nu}(k)=0$. The nontrivial solutions are plane waves whose wave 4-momentum $k$ satisfies $\det D^{\mu\nu}(k)=0$. The property $k_{\mu}D^{\mu\nu}(k)=0$ guarantees that one eigenvalue of $D^{\mu\nu}$ is trivial. 
In fact, using the Ward$–-$Takahashi identity and performing elementary row and column operations, it is easy to show that the temporal components of the dispersion tensor (\ref{eq:DispersionTensor}) can be eliminated by matrix similarity. Hence, the dispersion relation of linear waves can be written as
\begin{equation}
\label{dispersion}
\det D_{ij}(k)=0,
\end{equation}
where $D_{ij}$ is the spatial block of the dispersion tensor. In general, the 3-by-3 matrix $D_{ij}$ has three nontrivial eigenvalues, giving relativistic covariant dispersion relations of three waves.
When there exists some external test current $\hat{\mathcal{J}}^{\mu}_{\text{ext}}(k)$, the equation of $\hat{\mathcal{A}}(k)$ is $D^{\mu\nu}(k)\hat{\mathcal{A}}_{\nu}(k)+\hat{\mathcal{J}}_{\text{ext}}^{\mu}(k)=0$. After gauge fixing, the solution to this inhomogeneous equation gives response of the relativistic quantum plasma to external perturbations
\begin{equation}
\hat{\mathcal{A}}=-D^{-1}\hat{\mathcal{J}}_{\text{ext}}=-i\Lambda_F \hat{\mathcal{J}}_{\text{ext}},
\end{equation}
where $\Lambda_F$ is the Feynman Green's function. 
Taking inverse Fourier transform, the linear response of $\mathcal{A}(x)$ to the external test current $\mathcal{J}_{\text{ext}}(x)$ can be easily found. For example, when placing a test charge in the plasma $\mathcal{J}^{\mu}_{\text{ext}}(x)=e\delta^{(3)}(\mathbf{x})(1,0,0,0)$, one can derive Debye screening in the relativistic quantum plasma.

Finally, it is worth pointing out that the configuration space response tensor $\Sigma=\Sigma_{r}+i\Sigma_{i}$ is in general complex, corresponding to the momentum space response tensor $\hat{\Sigma}=\hat{\Sigma}_{H}+i\hat{\Sigma}_{A}$ that contains an antihermitian part. In classical field theory, when one solves the dispersion relation Eq.~(\ref{dispersion}) with $\Sigma_i\ne 0$, the wave 4-momentum $k^{\mu}$ is necessarily complex. So the amplitude of a plane wave either changes in time in an initial value problem, or changes in space in a boundary value problem. In the quantized field theory, the wave 4-momentum $k^{\mu}$ is always real, and it is the number of gauge bosons that change when $\Sigma_i\ne 0$. By the famous optical theorem, the imaginary part $\Sigma_{i}$ is proportional to the total cross section of the gauge boson. In fact, the optical theorem can be heuristically derived as follows. In the configuration space, we can separate the exponentiated action into an oscillatory part and an exponential part
\begin{equation}
e^{i\Gamma}=e^{iA(\nabla+\Sigma)A}=e^{iA(\nabla+\Sigma_r)A}e^{-A\Sigma_iA}.
\end{equation}
When $\Sigma_i=0$, the exponential is purely oscillatory. This corresponds to the simple propagation of the gauge field. When $\Sigma_i>0$, namely, when the matrix is positive definite, the exponential decays. This corresponds to wave damping in the classical theory, and the decay or absorption of gauge bosons in the quantized theory. When $\Sigma_i<0$, namely, when the matrix is negative definite, the exponential grows. This corresponds to instabilities in the classical theory, and the production or emission of gauge bosons in the quantized theory. Finally, when the matrix $\Sigma_i$ is indefinite, some eigenmodes grow while others decay. In this case, the dominant mode of a state of the $\mathcal{A}$ field can convert from one mode to another mode as the state evolves.

\chapter{Waves in unmagnetized scalar-QED plasmas\label{ch:unmag}}

In this chapter, I will demonstrate how to apply the general formalism developed in Ch.~\ref{ch:action} using the example of an unmagnetized plasma \citep[Sec.~III]{Shi16QED}. Formulas for the wave effective action can be evaluated once the self-consistent background fields $\phi_0$ and $\bar{A}$, as well as the Green's function of the free $\varphi$ field are known. 
To study waves in the plasma, there are basically four steps. First, we need to solve the classical field equations [Eqs.~(\ref{EOM}) and (\ref{EOMA})] and find the self-consistent solution that corresponds to conditions of the background plasma. 
Second, we need to solve the Schwinger--Dyson equation [Eq.~(\ref{eq:SDGreen})] to determine the charged particle Green's function. The Green's function knows about the background field $\bar{A}$, whose gauge may be chosen to simplify the expression of the Green's function.  
Third, we need to compute the background plasma response using Eq.~(\ref{bk}), in which the plasma polarization tensor can be evaluated using Eq.~(\ref{bkPol}). Similarly, we can compute the vacuum response using Eq.~(\ref{vac}), where the vacuum polarization tensor is given by Eq.~(\ref{vacPol}).
Finally, having obtained an explicit expression for the wave effective action under specific background fields $\phi_0$ and $\bar{A}$, we can determine properties of the plasma waves. For example, we can treat the fluctuation $\mathcal{A}$ as a classical field, and determine its linear eigenmodes by solving the dispersion relation Eq.~(\ref{dispersion}).
In what follows, I will elaborate on each of these four steps using a homogeneous unmagnetized plasma. Using this example, I will verify the effective action formalism by recovering known results in the literature, which have been obtained by other methods.

\section{Unmagnetized background and Green's functions\label{ch:unmag:basic}}

An unmagnetized and homogeneous plasma background is perhaps the simplest background after to the vacuum. In this case, there is no macroscopic EM fields, and charged particles uniformly fill up the space. A realistic plasma is of course of some finite size. Nevertheless, if the length scale of the plasma is much larger than both the plasma skin depth and the plasma wave length, then near the center of the plasma slab, a homogeneous plasma background is a reasonable approximation.

\subsection{Background fields\label{ch:unmag:basic:background}}

When there is no background EM field, it is convenient to choose the vacuum gauge
\begin{equation}
	\label{eq:A_unmag}
	\bar{A}=0.
\end{equation} 
In this case, the equation of motion of $\phi_0$ reduces to the Klein-Gordon equation in its simplest form. 
As we have seen in Sec.~\ref{sec:vacuum:field:free}, the single-boson solutions to the KG equation are plane waves with the dispersion relation $p^2=p_{\mu}p^{\mu}=m^2$. Since particles are not confined, the background wave functions are not square integrable. To deal with an infinitely large plasma with finite density, it is helpful to first think of a periodic spatial box with size $L$ and a temporal box of length $T$ that contains $N$ particles, and then take the limit $L, T\rightarrow\infty$ while keeping the density $n_0=N/L^3$ fixed. Inside the box, the properly normalized single-boson wave function 
\begin{equation}
	\psi_{\mathbf{p}}^{s}(x)=\frac{e^{is px}}{\sqrt{2mL^3}},
\end{equation}
where $p^{\mu}=(p^0,\mathbf{p})$ is the 4-momentum with $p^0=\sqrt{\mathbf{p}^2+m^2}$. The wave function represents a particle state when $s=+1$ and an anti-particle state when $s=-1$. The wave function is normalized such that the current density $\bar{J}_{0}^{\mu}=s ep^{\mu}/mL^3$ is what one would expect of a single particle. In the periodic box, $\mathbf{p}_{\mathbf{n}}=2\pi\mathbf{n}/L$ is quantized. We can label a single particle state by its wave number $\mathbf{n}$. It is clear that the inner products $\langle\psi^{+}_{\mathbf{n}}|\psi^{+}_{\mathbf{n}'}\rangle =\langle\psi^{-}_{\mathbf{n}}|\psi^{-}_{\mathbf{n}'}\rangle =\delta_{\mathbf{n},\mathbf{n'}}T/2m$ and $\langle\psi^{+}_{\mathbf{n}}|\psi^{-}_{\mathbf{n}'}\rangle=0$, so the single boson wave functions form an orthogonal basis. 

To see how to deal with an infinitely large plasma, let us calculate current density of $N$ bosons contained in a box. Since particles in plasmas are unbound, they interact weakly with each other. To lowest order, using the random phase approximation, the plasma may be treated as a collection of noninteracting particles. Suppose the $N$ bosons occupy $M$ orthogonal states $\psi_1,\dots,\psi_M$, with $N_k$ bosons in the state $k$, then the properly symmetrized and normalized wave function of the boson gas can be approximated by
\begin{equation}
	\Phi_0=\sqrt{\frac{(2m/T)^{N-1}}{(N-1)!N_1!..N_M!}}\sum_{\sigma\in S_N}\prod_{k=1}^{N}\psi_{d_{\sigma(k)}}(x_k).
\end{equation} 
Here $S_N$ is the permutation group of $N$ elements. The index function $d_k$ is defined such that $d_k=1$ for $k=1,\dots,N_1$; $d_k=2$ for $k=N_1+1,\dots,N_1+N_2$; and $d_k=M$ for $k=N-N_M+1,\dots,N$. After carrying out the integrals and summations, the current density [Eq.~(\ref{MBcurrent})] becomes $\bar{J}_{0}^{\mu}=e\sum_{k=1}^{M}N_ks_kp^{\mu}_{k}/mL^3$. More elaborately, the current density can be written as
\begin{equation}
	\bar{J}_{0}^{\mu}(x)=\sum_{s,k}\frac{f^{s}_{k}}{2m}\frac{e}{i}\Big(e^{-is p_kx}\partial^{\mu}e^{i s p_kx}-c.c\Big),
\end{equation}
where $f^{s}_{k}=N_k/L^3$ is the occupation density of the state with quantum numbers $(s,k)$, and the summation runs over all single-boson states. When $L\rightarrow\infty$, the spectrum of $\phi_0$ becomes continuous. In this case, let a single-boson state be labeled by its wave vector $\mathbf{p}$. If we keep the occupation density $f_{s}(\mathbf{p})$ fixed when we take the limit $L,T\rightarrow\infty$, the current density can be written as
\begin{equation}\label{SBcurrent}
	\bar{J}_{0}^{\mu}(x)=\sum_{s=\pm 1}\int\frac{d^3\mathbf{p}}{(2\pi)^3}\frac{e}{i} \Big(\Psi^{s*}_{\mathbf{p}}(x)\partial^{\mu}\Psi^{s}_{\mathbf{p}}(x)-c.c\Big),
\end{equation}
where the properly normalized effective single-boson wave function
\begin{equation}\label{Single}
	\Psi_{\mathbf{p}}^{s}(x)=\sqrt{\frac{f_{s}(\mathbf{p})}{2m}}e^{is px}.
\end{equation}
We see that the current [Eq.~(\ref{SBcurrent})] can be obtained from the many-body current [Eq.~(\ref{MBcurrent})] by replacing $\phi_0(x)$ with the properly normalized effective single-boson wave function [Eq.~(\ref{Single})], followed by summations over discrete labels, and integrations over continuous labels in the Hilbert space of single-boson states. The occupation density $f_{s}(\mathbf{p})$ is the momentum space distribution function of the plasma. In classical plasma physics, one can specify both the location and the momentum of a classical particle, so the distribution function $f_s(\mathbf{x},\mathbf{p})$ is defined on the entire phase space.
However, in a quantum plasma, due to the uncertainty principle, one is not allowed to specify a particle's location once its momentum is known accurately. This is manifested by the wave function Eq.~(\ref{Single}), whose momentum is certain but spatial distribution is completely uncertain.

\subsection{Green's function\label{ch:unmag:basic:green}}

When the background gauge field is trivial, the Green's function for the charged particles is the usual Green's function of scalar field in quantum field theory. In Sec.~\ref{sec:vacuum:field:green}, this Green's function has been discussed in details. 
In its integral representation, the Green's function
\begin{equation}\label{eq:UMGreen}
	G(x,x')=\int\frac{d^4k}{(2\pi)^4}\frac{ie^{-ik(x-x')}}{k^2-m^2},
\end{equation} 
where the pole prescription is given by the replacement $m^2\rightarrow m^2-i\epsilon$, such that the exponentiated action is positive definite and the path integrals can thereof converge. Recall that using the above pole prescription, the integral Eq.~(\ref{eq:UMGreen}) can be evaluated in terms of Bessel functions (Appendix~\ref{ch:append:GreenF}). However, in what follows, the integral representation will be sufficient for determining the response tensor.

\section{Unmagnetized response tensors\label{ch:unmag:polarization}}

Now that we have determined the self-consistent background fields [Eqs.~(\ref{eq:A_unmag}) and (\ref{Single})] and the Green's function [Eq.~(\ref{eq:UMGreen})], we can evaluate formulas of the plasma response tensor [Eq.~(\ref{bk})] and the vacuum response tensor [Eq.~(\ref{vac})]. Notice that the photon mass terms are important in addition to the polarization terms. Without the mass terms, the response tensors in scalar-QED plasmas loss their gauge invariance. After summing the mass terms with the polarization terms, the resultant effective action gives a Lorentz-invariant and gauge-invariant description of wave propagation in unmagnetized scalar-QED plasmas.

\subsection{Plasma dielectric response\label{ch:unmag:polarization:plasma}}

The plasma response $\Sigma_{2,\text{bk}}^{\mu\nu}(x,x')$ can be evaluated by substituting the effective single-boson wave function and the Green's function into the photon mass term and the plasma polarization term, followed by integration and summation over the single-boson Hilbert space. The contribution of each charged species to the mass term of the $\mathcal{A}$ field is
\begin{equation}\label{eq:bkmass}
	2e^2\phi_0\phi^*_0=\sum_{s=\pm 1}\int\frac{d^3\mathbf{p}}{(2\pi)^3}\frac{e^2f_{s}(\mathbf{p})}{m},
\end{equation}
and the plasma polarization tensor [Eq.~(\ref{bkPol})] becomes
\begin{eqnarray}
	\label{eq:bkPol}
	\Pi^{\mu\nu}_{2,\text{bk}}&=&\sum_{s=\pm 1}\int\frac{d^3\mathbf{p}}{(2\pi)^3} e^2[\Psi_{\mathbf{p}}^{s*}\partial^{\mu}-(\partial^{\mu}\Psi_{\mathbf{p}}^{s})^*][\Psi_{\mathbf{p}}^{s'}\partial^{'\nu}-(\partial^{'\nu}\Psi_{\mathbf{p}}^{s'})]G-\text{c.c.}\hspace{2pt}.
\end{eqnarray}
Similar results are shown by \cite{Melrose07}, using the prescription of cutting one charged particle propagator in the vacuum polarization diagram and replacing it by statistical average over the plasma. The path integral formulation developed in Ch.~\ref{ch:action} has thus provided an alternative justification for such a prescription. 

For the purpose of illustrating the general formalism, let us consider the simple example of a cold particle plasma. Denoting the 4-momentum of cold particles by $q^{\mu}$, then the occupation density of the state with quantum numbers $(s,\mathbf{p})$ is
\begin{equation}\label{UmagDOS}
	f_{s}(\mathbf{p})=n_0(2\pi)^3\delta^{(3)}(\mathbf{p}-\mathbf{q})\delta_{s,1}\hspace{2pt},
\end{equation} 
where $n_0$ is the number density of the plasma. The occupation density $f_{s}(\mathbf{p})$ is nothing other than the momentum space distribution function commonly seen in plasma physics. In general, the distribution function can be any integrable function of interest. Here in this simple example, due to the $\delta$-functions, integrals and summations can be evaluated very easily. The current density due to each charged species becomes
\begin{equation}
	\bar{J}_{0}^{\mu}(x)=en_0q^{\mu}/m.
\end{equation} 
This is what one would expect of a cold uniform fluid. To satisfy the background self-consistency $\partial_{\mu}\bar{F}^{\mu\nu}=0$, the plasma needs to be constituted of more than one charged species, such that the total current $\sum_s\bar{J}_{s0}^{\mu}=0$, after summing over all charged species. Using the momentum space distribution function [Eq.~(\ref{UmagDOS})], the mass term of the $\mathcal{A}$ field [Eq.~(\ref{eq:bkmass})] becomes
\begin{equation}\label{UMmass}
	2e^2\phi_0(x)\phi^*_0(x)=\frac{e^2n_{0}}{m}=\omega_{p}^2.
\end{equation}
It is easy to recognize that $\omega_p$ is the plasma frequency in the natural units. In other words, photons become massive particles in plasmas, where the mass is precisely the plasma frequency. In the limit where the plasma density $n_0\rightarrow0$, photons become massless particles as in the usual quantum field theory. Similarly, substituting in the distribution function [Eq.~(\ref{UmagDOS})], the plasma polarization tensor [Eq.~(\ref{eq:bkPol})] becomes
\begin{eqnarray}\label{UMpol}
	\Pi^{\mu\nu}_{2,\text{bk}}(x,x')=\frac{\omega_{p}^2}{2}\int\frac{d^4k}{(2\pi)^4}ie^{-ik(x-x')}\Big[\frac{(2q+k)^{\mu}(2q+k)^{\nu}}{(k+q)^2-m^2} +\frac{(2q-k)^{\mu}(2q-k)^{\nu}}{(k-q)^2-m^2}\Big].
\end{eqnarray} 
The two terms above correspond to the $s$-channel and the $t$-channel Feynman diagrams of the forward scattering of a gauge boson. We see quantum recoil, the change of the 4-momentum of charged particles during forward scattering of the gauge boson, is automatically taken into account. Combining the photon mass term Eq.~(\ref{UMmass}) and the plasma polarization term Eq.~(\ref{UMpol}), and taking Fourier transform, the contribution of each charged species to the momentum space plasma response tensor is
\begin{equation}\label{UMbk}
	\hat{\Sigma}_{2,\text{bk}}^{\mu\nu}=\omega_{p}^2\Big[g^{\mu\nu}-\frac{k^2(4q^{\mu}q^{\nu}\!+k^{\mu}k^{\nu}) -4kq(q^{\mu}k^{\nu}\!+k^{\mu}q^{\nu})}{(k^2)^2-4(kq)^2}\Big].
\end{equation}
Here $k^2=k^{\mu}k_{\mu}$ and $kq=k^{\mu}q_{\mu}$ are Minkowski inner products. The above expression, involving only Lorentz scalars and Lorentz vectors, is manifestly Lorentz covariant. Having obtained the plasma response tensor in the reference frame where the plasma 4-momentum is $q^\mu$, we can boost to any other inertial frames, in which the response tensor takes the same form after the Lorentz transformation. 
Finally, it is straightforward to check that the Ward$–-$Takahashi identity [Eq.~(\ref{eq:Ward})] is satisfied, so the plasma response tensor Eq.~(\ref{UMbk}) is gauge invariant.

\subsection{Vacuum polarization\label{ch:unmag:polarization:vacuum}}

The vacuum response tensor contains two 1-loop diagrams, both of which are divergent. These divergences can be removed by the renormalization procedure discussed in Sec.~\ref{sec:vacuum:renormalization}, using the renormalization condition that photons are massless particles in the vacuum. Multiplicative renormalization introduces counter terms as additional interactions. For the free $\mathcal{A}$ field, whose Lagrangian is given by Eq.~(\ref{ReducedLagrangian}), it is easy to see the Feynman rules for the counter terms are
\begin{fmffile}{Acounter}
	\fmfcmd{
		path quadrant, q[], otimes;
		quadrant = (0, 0) -- (0.5, 0) & quartercircle & (0, 0.5) -- (0, 0);
		for i=1 upto 4: q[i] = quadrant rotated (45 + 90*i); endfor
		otimes = q[1] & q[2] & q[3] & q[4] -- cycle;
	}
	\fmfwizard
	\begin{eqnarray}
	\label{eq:Acounter}
	\begin{gathered}
		\begin{fmfgraph*}(40,25) 
		\fmfkeep{Acounter}
		\fmfleft{i}
		\fmfright{o}
		\fmf{photon,label=k}{i,v}
		\fmf{photon,label=k}{v,o}		
		\fmfv{d.sh=otimes,d.f=empty,d.si=.15w}{v}
		\end{fmfgraph*}
	\end{gathered}
	\hspace{5pt}&=&i\big[(-k^2g^{\mu\nu}+k^\mu k^\nu)\delta_\mathcal{A}+\delta_Mg^{\mu\nu}\big],
	\end{eqnarray}
\end{fmffile}where $\delta_\mathcal{A}$ comes from the wave function renormalization of the $\mathcal{A}$ field, with the scaling factor $Z_\mathcal{A}=1+\delta_\mathcal{A}$. Since photons become massive particles, the mass term of the $\mathcal{A}$ field also needs to be renormalized, where the scaling factor $Z_M=1+\delta_M$. The scaling factors can be expanded using Taylor series using the small charge $e$ as the expansion parameter, and at the same time the scaling factors can be expanded as Laurent series using dimensional regularization to cancel the loop divergences. 

Now let us compute the loop diagrams using dimensional regularization in $d=4-\epsilon$ dimension. The first diagram is the mass renormalization term, which is essentially the same as the scalar-field loop [Eq.~(\ref{eq:phi_1loop_dim})] except for a different interaction vertex [Eq.~(\ref{eq:A2phi2})]. Using previous results, the mass renormalization term is
\begin{fmffile}{Amass}
	\begin{eqnarray}
	\label{Amass}
	\begin{gathered}
		\begin{fmfgraph*}(40,25)
		\fmfkeep{hairpin_k}
		\fmfleft{i}
		\fmfright{o}
		\fmf{photon}{i,v}
		\fmf{photon}{v,o}
		\fmf{plain}{v,v}
		\fmfdot{v}
		\fmfv{label=$\mu$,label.dist=0.2}{i}
		\fmfv{label=$\nu$,label.dist=0.2}{o}
		\end{fmfgraph*}
	\end{gathered}	
	\quad&=& 2ie^2g^{\mu\nu}\int \frac{d^4k}{(2\pi)^4}\frac{i}{k^2-m^2} 
	=2ie^2g^{\mu\nu}\frac{\Gamma(1-d/2)}{(4\pi)^{d/2} (m^2)^{1-d/2}}.
	\end{eqnarray}
\end{fmffile}The second loop diagram is the vacuum polarization term Eq.~(\ref{vacPol}). To compute this diagram, we will need the following symmetry properties of the momentum space integrals:
\begin{eqnarray}
\label{eq:lid1}
&&\int \frac{d^4l}{(2\pi)^4} \frac{l^\mu}{f(l^2)}=0,\\
\label{eq:lid2}
&&\int \frac{d^dl}{(2\pi)^d} \frac{l^\mu l^\nu}{f(l^2)}=\frac{g^{\mu\nu}}{d}\int \frac{d^dl}{(2\pi)^d} \frac{l^2}{f(l^2)}.
\end{eqnarray}
the first identity is similar to $\int x f(x^2)dx =0$, which vanishes because of the antisymmetry $x\rightarrow -x$. The second identity is true for similar reasons, and the coefficient can be easily check by contracting both sides with $g_{\mu\nu}$, where the trace in $d$-dimension $g_{\mu\nu}g^{\mu\nu}=d$. 
Then, in the momentum space, the vacuum polarization tensor can be computed using the usual Feynman rules [Eqs.~(\ref{eq:phi4G}) and (\ref{eq:Aphi2})], which give
\begin{fmffile}{vacpol}
	\begin{eqnarray}\label{vacpol_k}
		\nonumber
		\begin{gathered}
			\begin{fmfgraph*}(55,25)
				\fmfkeep{vacpol_k}
				\fmfleft{i}
				\fmfright{o}
				\fmf{plain,left=1,tension=0.3,label=k+p}{v1,v2}
				\fmf{plain,right=1,tension=0.3,label=p}{v1,v2}
				\fmfdot{v1,v2}
				\fmfv{label=$\mu$,label.dist=0.5}{i}
				\fmfv{label=$\nu$,label.dist=0.5}{o}
				\fmf{photon,label=k}{i,v1}
				\fmf{photon,label=k}{v2,o}
			\end{fmfgraph*}
		\end{gathered}
		\quad&=&(-ie)^2 \int \frac{d^4p}{(2\pi)^4}\frac{i(2p+k)^\mu}{p^2-m^2} \frac{i(2p+k)^\nu}{(p+k)^2-m^2} \\
		\nonumber
		&=&e^2 \int \frac{d^4p}{(2\pi)^4} \int_0^1 dx \frac{(2p+k)^\mu(2p+k)^\nu}{(p^2+2xpk+xk^2-m^2)^2} \\
		\nonumber
		&=&e^2 \int \frac{d^4l}{(2\pi)^4} \int_0^1 dx \frac{l^2g^{\mu\nu}+(1-2x)^2k^\mu k^\nu}{[l^2+x(1-x)k^2-m^2]^2} \\
		\nonumber
		&=&ie^2 \int_0^1 dx \int_0^{+\infty} \frac{S_{d-1} l^{d-1} dl}{(2\pi)^d} \frac{-\frac{4}{d}l^2g^{\mu\nu}+(1-2x)^2k^\mu k^\nu}{(l^2+\Delta^2)^2} \\
		&=&ie^2\frac{\Gamma(2-d/2)}{(4\pi)^{d/2}} \int_0^1 \frac{dx}{(\Delta^2)^{1-d/2}}\Big[ \frac{2g^{\mu\nu}}{d/2-1}  +\frac{(1-2x)^2k^\mu k^\nu }{\Delta^2}\Big].
	\end{eqnarray}
\end{fmffile}On the second line, I have used the Feynman trick Eq.~(\ref{eq:FeynmanTrick}). On the third line, I have changed the integration variable to $l=p+xk$, and used identities Eqs.~(\ref{eq:lid1}) and (\ref{eq:lid2}). On the fourth line, the divergent integral is regulated in $d$-dimension after the Wick rotation $l^0=il_E^0$. The normalization factor $\frac{4}{d}$ is inserted so that the trace of the first term remains the same. In the spherical integral, $S_{d-1}$ is again the area of $(d-1)$-dimensional unit sphere, and I have denoted $\Delta^2=m^2-x(1-x)k^2$. On the last line, the momentum integral is carried out using the beta function Eq.~(\ref{eq:beta}). The above result is a well-known result in quantum field theory.

Having computed the two 1-loop diagrams, the vacuum response tensor can be readily obtained. Substituting the above results into Eq.~(\ref{vac}), it is a straightforward calculation to verify that the Ward$–-$Takahashi identity [Eq.~(\ref{eq:Ward})] is satisfied in the limit $d\rightarrow 4$, after the Feynman parameter $x$ is integrated out. Therefore, the momentum space vacuum polarization tensor can be written in the form
\begin{equation}
	\label{eq:UMvac}
	\hat{\Sigma}_{2,\text{vac}}^{\mu\nu}(k)=\chi_{v}(k^2)(k^{\mu}k^{\nu}-k^2g^{\mu\nu}),
\end{equation}
where $\chi_{v}(k^2)$ is a Lorentz scalar. 
Imposing the renormalization condition that photons remain massless in the vacuum, and subtracting the counter terms [Eq.~(\ref{eq:Acounter})], the renormalized 1-loop vacuum permittivity
\begin{eqnarray}\label{UMvac}
	\nonumber
	\chi_{v}(k^2)&=&e^2\frac{\Gamma(2-d/2)}{(4\pi)^{d/2}}\int_{0}^{1}dx\frac{(1-2x)^2}{(m^2)^{2-d/2}}\Big[\Big(1-x(1-x)\frac{k^2}{m^2}\Big)^{d/2-2}-1\Big]\\
	&=&\frac{2e^2}{3(4\pi)^2}\Big\{\frac{4}{3}-\frac{4m^2}{k^2}+\Big(\frac{4m^2}{k^2}-1\Big)^{3/2}\arctan\Big[\Big(\frac{4m^2}{k^2}-1\Big)^{-1/2}\Big]\Big\}.
\end{eqnarray}
The second line is obtained by taking the limit $d\rightarrow 4$, and then integrating over the Feynman parameter. It is not hard to see that $\hat{\Sigma}_{2,\text{vac}}(k^2)$ is real when $k^2=k_{\mu}k^{\mu}\le 4m^2$, and $\hat{\Sigma}_{2,\text{vac}}(k^2)$ becomes complex with a positive imaginary part when $k^2>4m^2$. The positive imaginary part is proportional to the cross section of the gauge boson, which can decay into a pair of ``electron" and ``positron" when $k^2>4m^2$ is above the mass threshold.

\section{Spectrum of linear waves\label{ch:unmag:dispersion}}

Having calculated the response tensors due to the plasma response [Eq.~(\ref{UMbk})] and the vacuum response [Eqs.~(\ref{eq:UMvac}) and (\ref{UMvac})], we have thus obtained an explicit expression for the 1-loop wave effective action in the momentum space [Eq.~(\ref{MS1Loop})]. Since the plasma is translational invariant, the momentum space classical field equation has a well-defined spectrum of linear eigenmodes. 

\subsection{Dispersion relations in the plasma rest frame\label{ch:unmag:dispersion:rest}}

Since the effective action is Lorentz invariant, we can study linear eigenmodes in any inertial frame. Both the dispersion relation and the eigenmodes are Lorentz covariant. 
The simplest case is when different charged species in the plasma have no relative motion. In this case, there exists an inertial frame in which all background particles are at rest. In this plasma rest frame, the particle 4-momentum $q^{\mu}=(m,0,0,0)$. Let us choose a coordinate system such that the wave 4-momentum $k^{\mu}=(\omega,\mathrm{k},0,0)$. 
Notice that to avoid confusion, I use the italic $k$ for 4-momentum of and the roman $\mathrm{k}=|\bm{k}|$ for the magnitude of the wave vector. While $k^2=k^{\mu}k_{\mu}$ and $kq=k^{\mu}q_{\mu}$ denote the Minkowski inner products, produces such as $\mathrm{k}^2$ and $\omega\mathrm{k}$ are the usual scalar products.
In the special coordinate system, the tensor $k^{\mu}k^{\nu}-k^2g^{\mu\nu}$, which is contained in the vacuum response tensor [Eq.~(\ref{eq:UMvac})], becomes very simple. Moreover, the plasma response tensor [Eq.~(\ref{UMbk})] is also simplified. The nonzero components of the plasma response tensor are
\begin{eqnarray}
	\nonumber
	\hat{\Sigma}_{2,\text{bk}}^{00}&=&\chi_p \mathrm{k}^2,\\
	\nonumber
	\hat{\Sigma}_{2,\text{bk}}^{11}&=&\chi_p \omega^2,\\
	\nonumber
	\hat{\Sigma}_{2,\text{bk}}^{01}&=&\hat{\Sigma}_{2,\text{bk}}^{10}=\chi_p\omega \mathrm{k},\\
	\hat{\Sigma}_{2,\text{bk}}^{22}&=&\hat{\Sigma}_{2,\text{bk}}^{33}=-\omega_{p}^{2},
\end{eqnarray} 
where the total plasma frequency $\omega_{p}^{2}$ and the total plasma permittivity $\chi_p$ are contributed by each charged species
\begin{eqnarray}
	\label{eq:plasma_wp}
	\omega_{p}^{2}&=&\sum_{s}\omega_{ps}^{2},\\
	\label{eq:plasma_chip}
	\chi_p&=&\sum_{s}\frac{\omega_{ps}^{2}(\bm{k}^2-\omega^2+4m_{s}^2)}{(\omega^2-\bm{k}^2)^2-4m_{s}^2\omega^2}.
\end{eqnarray}
The above results have been obtained previously by \cite{Hines78,Kowalenko85,Eliasson11}. Here, using a different approach, namely, the effective action approach, I have thus recovered previously known results.

Using elementary column and row operations, the dispersion matrix $D^{\mu\nu}$ can be diagonalized and the eigenvalue problem can be solved. There are two transverse modes and one longitudinal mode. The two transverse modes are degenerate and electromagnetic with the dispersion relation
\begin{equation}\label{EM}
	(1+\chi_v)(\omega^2-\bm{k}^2)-\omega_{p}^{2}=0.
\end{equation}
From this dispersion relation, it is easy to see that the photon modes are gapped when background plasmas exist. Namely, the wave frequency $\omega\neq 0$ when the wave vector $\mathrm{k}=0$ if $\omega_p\neq 0$. When ignoring the vacuum response, the above result is the familiar dispersion relation $\omega^2=\omega_p^2+ \bm{k}^2$ of EM waves in unmagnetized plasmas. 
The longitudinal mode is purely electrostatic with the dispersion relation
\begin{equation}\label{langmuir}
	1+\chi_v+\chi_p=0.
\end{equation}
In the absence of the vacuum response, the above dispersion relation resembles the result in classical plasmas, except now the plasma permittivity is given by Eq.~(\ref{eq:plasma_chip}). 
Since $\chi_p(\bm{k}=\mathbf{0})=-\omega_{p}^{2}/\omega^2$, there always exists one gapped plasmon mode, known classically as the Langmuir wave. When there are two or more charged species, there also exist nontrivial gapless phonon modes, known classically as the ion acoustic waves. Moreover, due to vacuum fluctuations, now there exist additional modes known as the pair modes. In the pair mode, whose frequency $\omega>2m$, a single gauge boson has enough energy to created ``electron-positron" pairs. As the longitudinal wave oscillates, virtual pairs are constantly being created and annihilated. The pair mode only exists in relativistic quantum plasmas. 

\begin{figure}[!b]
	\renewcommand{\figurename}{FIG.}
	\centering
	\includegraphics[angle=0,width=0.55\textwidth]{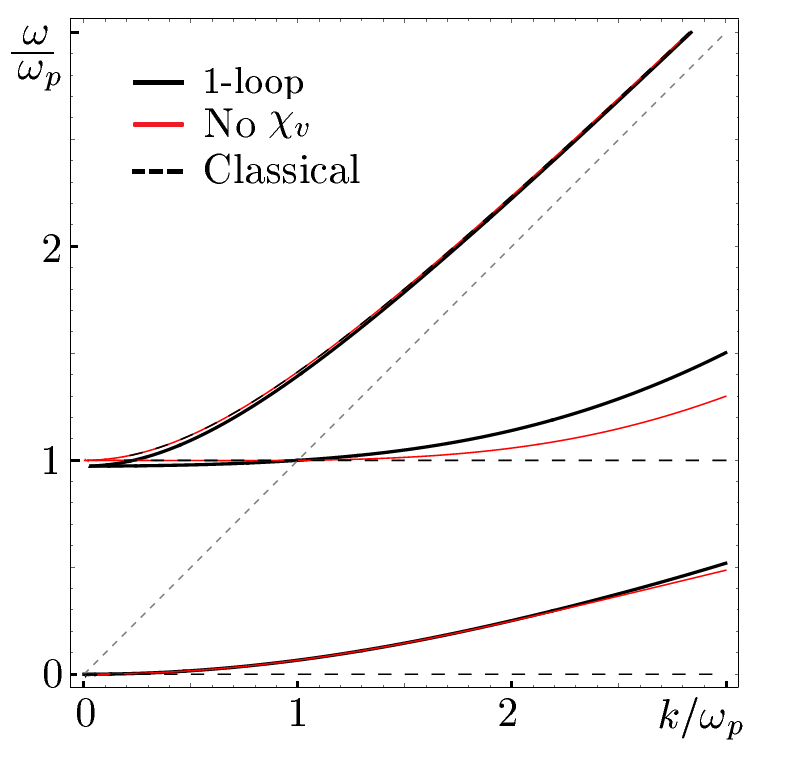}
	\caption[Dispersion relation in unmagnetized cold scalar-QED plasma]{Wave dispersion relations in a cold, unmagnetized, quasineutral, ``electron-ion" plasma. For various effects to be visible on the scale of this figure, parameters used for making this plot are $2e^2/3(4\pi)^2=20$, $m_e/\omega_{pe}=5$, and $m_i/m_e=3$. The solid black curves are the 1-loop dispersion relations. The solid red curves are the dispersion relations that ignore the vacuum polarization. The dashed black curves are wave dispersion relations in a classical plasma. The upper curves are the electromagnetic waves, the middle curves are the Langmuir waves, the bottom curves are the ion acoustic waves, and the dashed gray line across the diagonal represents the light cone. Notice that near the light cone, wave dispersion relations in the relativistic quantum plasma asymptote to wave dispersion relations in the classical plasma. The pair modes, emanating from $\omega=2m_e$ and $\omega=2m_i$, are out of the scale of this figure.}
	\label{fig:UMDispersion}
\end{figure}

An example of wave dispersion relations in a cold, quasineutral, ``electron-ion" plasma is plotted in Fig.~\ref{fig:UMDispersion}. In the figure, the upper curves are the degenerate EM waves, the middle curves are the Langmuir waves, and the lower curves are the ion acoustic waves. There also exist pair modes with $\omega>2m$. These high energy modes readily decay and are not plotted here. 
In Fig.~\ref{fig:UMDispersion}, 
the wave dispersion relations in a classical plasma (dashed black) is modified by tree-level relativistic quantum effects (red), even when the vacuum permittivity $\chi_v$ is ignored. Moreover, when 1-loop effects are included, the wave dispersion relations, solved from Eqs.~(\ref{EM}) and (\ref{langmuir}), receive further corrections (solid black).
To make all effects visible on the scale of this figure, the plasma density is set so high that $\omega_{pe}/m_e=0.2$, such that relativistic quantum effects are comparable to classical collective effects. The ion mass is set artificially low with $m_i/m_e=3$, such that ion effects are comparable to electron effects. The coupling constant is taken to be unphysically strong $2e^2/3(4\pi)^2=20$, such that loop-level effects are comparable to tree-level effects. 
When parameters are more physical, collective plasma effects dominate tree-level relativistic quantum effects, which in turn dominate loop-level effects in the frequency range relevant to contemporary laboratory experiments. 
In fact, as can be seen from the figure, wave dispersion relations in the relativistic quantum plasma are very similar to those in the classical plasma near the light cone, where the effective mass of the gauge boson $m_\mathcal{A}=\omega \sqrt{1-(\partial\omega/\partial k)^2}$ is much smaller than the ``electron" mass. 
However, there are clear distinctions away from the light cone, where the gauge boson becomes very massive. In particular, unlike classical plasma theories,  the relativistic quantum theory predicts that longitudinal waves propagate with nonzero group velocities even when the plasma is cold. This can be understood intuitively, because a longitudinal wave spends a part of its time in the form of an excited charged boson [Eq.~(\ref{bk})]. Due to the recoil effect, the momentum carried by the wave transfers to charged bosons, whose nonzero velocity contributes to the finite group velocity of the wave.

The photon modes and the plasmon mode have the same cutoff frequency $\omega_c$, which is given by the solution to the equation $\omega_c^2(1+\chi_v(\omega_c,\mathrm{k}=0))=\omega_{p}^{2}$. The cutoff frequency $\omega_c$, or the mass gap, is less than the plasma frequency $\omega_p$ due to vacuum polarization. It can be shown that the ratio $\omega_c/\omega_p$ decreases with increasing $\omega_p/2m$. This can be understood intuitively. Vacuum polarization produces virtual pairs near charged particles. These virtual pairs screen the electric field of charged particles, so the effective electric charge of real particles are reduced. For higher plasma densities, the virtual pair density is also higher, resulting in stronger shielding of the electric charge and consequently smaller cutoff frequencies. To get a sense of how small the vacuum polarization effect is, let us approximate $\omega_c$ when $\omega_p/2m=1$. Denoting $g=2e^2/3(4\pi)^2$, since the physical value for electron charge is $g=\alpha/6\pi\ll 1$, where $\alpha\approx 1/137$ is the fine structure constant, the equation for $\omega_c$ can be solved asymptotically. To lowest order, $\omega_c/\omega_p\sim 1-g/6$. We see in comparison, the effect of the vacuum polarization is minuscule. However, when the plasma frequency is large, the absolute value of the relativistic quantum shift can be appreciable.
\subsection{Asymptotics of the dispersion relations\label{ch:unmag:dispersion:asymptotics}}

The above dispersion relations are applicable to all energy range within the scalar-QED model. 
Now let us check that classical dispersion relations can be recovered when taking the classical limit in relativistic quantum results. Since energy of particles are not quantized in unmagnetized plasmas, the non-relativistic low energy limit is the classical limit. In the low energy limit $k^2/m^2\rightarrow 0$, namely, near the light cone, contribution of a relativistic quantum plasma asymptote to that of a classical plasma
\begin{equation}
	\chi_p\sim-\sum_{s}\frac{\omega_{ps}^{2}}{\omega^2}\Big(1+\frac{k^2}{4m_{s}^2}\Big)\Big(1+\frac{k^4}{4m_{s}^2\omega^2}\Big) \rightarrow-\frac{\omega_{p}^{2}}{\omega^2},
\end{equation}
and the contribution of the vacuum response vanishes
\begin{equation}
	\chi_v\sim\sum_{s}\frac{g_s}{5}\frac{k^2}{4m_{s}^2-k^2}\rightarrow 0.
\end{equation}
This can be understood intuitively, because in this limit, the mass of the gauge boson is much smaller than the mass of charged particles. Since the gauge boson do not have sufficient energy to excite ``electron-positron" pairs, it sees little effect of the vacuum polarization. 
In the low energy limit, the next-to-leading order asymptotic dispersion relations of the photon, the plasmon, and the phonon modes are
\begin{eqnarray}\label{eq:UMDispApprix}
	\omega^2&\simeq&\omega_p^2(1-\lambda_D^2\omega_p^2)+\bm{k}^2,\\
	\omega^2&\simeq&\omega_p^2\big[1\!-\!\lambda_D^2(\omega_p^2-\bm{k}^2)\big]\!-\!\Big(\frac{\omega_{pe}^2}{4m_e^2}+\frac{\omega_{pi}^2}{4m_i^2}\Big) \Big(\bm{k}^2-\frac{\bm{k}^4}{\omega_p^2}\Big),\\
	\omega^2&\simeq&\Big(\frac{\omega_{pe}^2}{4m_i^2}+\frac{\omega_{pi}^2}{4m_e^2}\Big) \frac{\bm{k}^4}{\omega_p^2}=\frac{\bm{k}^4}{4m_em_i}.
\end{eqnarray}
Here $\lambda_D^2=\sum_{s}g_s/20m_s^2 $ is the vacuum shielding length due to virtual pair production. Similar results have been obtained using other approaches by \cite{Hines78, Kowalenko85}, in which the phonon mode was not considered. Here, these known results are recovered using the effective action approach.

In the opposite limit $k^2/m^2\rightarrow\infty$, namely, away from the light cone where the gauge boson becomes very massive, the plasma contribution diminishes
\begin{equation}\label{eq:pasym}
	\chi_p\sim-\frac{\omega_{p}^{2}}{k^2}\rightarrow 0.
\end{equation}
This behavior can be understood in the physical picture where gauge bosons are regarded as particles. In the above limit, gauge bosons are infinitely massive, so when they collide with charged particles whose masses are much smaller, the gauge bosons feel little impact. 
Alternatively, the above behavior can be understood when gauge fields are regarded as waves. From the perspective of spatial scales by fixing $\omega$ and letting $\bm{k}$ go to infinity. In this perspective, since the wave length of a high energy gauge boson is much smaller than the typical inter-particle spacing in the plasma, the gauge boson rarely encounters a plasma particle and propagates as if it is in the vacuum. The asymptotic behavior Eq.~(\ref{eq:pasym}) can also be understood from the perspective of time scales by fixing $\bm{k}$ and letting $\omega$ go to infinity. In this perspective, since the wave frequency is much larger than the plasma frequency, the plasma does not have time to respond. Unlike the plasma response, which diminishes when $k^2/m^2\rightarrow\infty$, the real part of the vacuum susceptibility blows up 
\begin{equation}
	\text{Re}(\chi_v)\sim-\sum_{s}\frac{g_{s}}{2}\ln\Big|\frac{k^2}{4m_{s}^2}\Big|\rightarrow\infty,
\end{equation}
This can be understood intuitively. Since the gauge bosons now have sufficient energy, they can easily decay to create virtual ``electron-positron" pairs, and thereof see a large effect of vacuum polarization. Outside the light cone, the imaginary part of $\chi_v$ is always zero. Inside the light cone, when $k^2> 4m^2$, the imaginary part of the vacuum susceptibility 
\begin{equation}
	\text{Im}(\chi_v)\sim\sum_{s}\frac{\pi}{2}\Big(1-\frac{4m_{s}^2}{k^2}\Big)\rightarrow\sum_{s}\frac{\pi}{2}, 
\end{equation} 
This positive imaginary part is proportional to the total decay cross section of a massive gauge boson. The imaginary part is larger when there are more charged species, in which case there are more types of particles that the massive gauge boson can decay into. After its typical life time, a massive gauge boson decays and thereafter stops propagating.

\chapter{Waves in magnetized scalar-QED plasmas\label{ch:mag}}

Classical treatments of plasma waves start to break down when strong magnetic fields beyond gigagauss are present. In fields of such strengths, both relativistic and quantum effects become important.
Quantization effects are relatively well understood. Since charged particles are confined in the perpendicular direction, the perpendicular energy is quantized. Quantization is explicitly included in the usual quantum mechanics, where electrons occupy discrete Landau levels. 
Alternatively, since statistical fluctuations result in similar effects as quantum fluctuations, quantization effects are included in usual plasma physics implicitly when averaging of the distribution function over gyro orbits, which introduces Bessel functions that effectively quantize the angular momentum.
Using either non-relativistic quantum mechanics or classical plasma physics, the response of the plasma medium can be computed, and the wave dispersion relations are well known when magnetic fields are moderate.
However, the usual quantization is altered by relativistic effects in strong magnetic fields. For example, in gigagauss fields, the magnetic energy \mbox{$\epsilon_B=\sqrt{eBc^2\hbar}\sim 10$ keV} starts to be comparable to the electron rest energy $m_ec^2\approx 511$ keV. Consequently, the evenly spaced Landau levels become anharmonic.
In additional to modifying the quantum states, strong magnetic fields also affect transitions between quantum states, which determines the plasma dielectric response. When making transitions, electrons carry recoil momentum on the scale of \mbox{$\Omega_e\hbar\sim10$ eV} in gigagauss field. This recoil momentum can be comparable to the momentum of gauge bosons, unless the wave frequency is orders of magnitude larger. 
In other words, there are two additional effects in strongly magnetized plasmas: magnetic energy is now comparable to the electron rest energy in each quantum state; and momentum change is no longer negligible when electrons make transitions between quantum states. To capture these effects, a relativistic quantum treatment becomes necessary in strongly magnetized plasmas.

In this chapter, I will treat relativistic quantum effects for wave propagation in strongly magnetized plasmas by applying the general theory developed in Ch.~\ref{ch:action}. 
This is yet another example, from which we will see the powerfulness of the general formalism that enables the wave dispersion relations in strongly magnetized scalar-QED plasmas to be determined for the first time. While previous methods were only able to compute a useful dispersion relation parallel to the magnetic field, the effective action formalism can easily treat arbitrary geometry transparently, by simply following the four steps discusses in Ch.~\ref{ch:unmag}.
First, I will solve the classical field equations to determine self-consistent background fields $\phi_0$ and $\bar{A}$. Second, I will solve the Schwinger--Dyson equation to determine the charged particle Green's function. Third, using the general formulas, the vacuum response and the background plasma response can be evaluated. Finally, having obtained an explicit expression for the wave effective action, we can then determine properties of the gauge bosons.
In particular, when wave propagates perpendicular to the magnetic field, Bernstein waves become unevenly spaced \citep[Sec.~IV]{Shi16QED}. The resultant anharmonic cyclotron absorption features have been observed in spectra of X-ray pulsars, whose magnetic fields $\sim 10^{12}$ G.
Although current laboratory techniques can only produce $\sim10^9$ G magnetic fields, it turns out that relativistic quantum effects already become observable through Faraday rotation \citep[Sec.~IV]{Shi2018laser}, where the rotation angle is predicted to have a different frequency dependence than expected classically.


\section{Magnetized background and Green's functions\label{ch:mag:basic}}

To solve for the background wave functions and the Green's function, it is convenient to pick a reference frame and fix a gauge.
The explicit expressions depends on these choices, although the general formula for the wave effective action is Lorentz invariant and gauge invariant.
Notice that EM fields depend on reference frames, and the two Lorentz invariant scalars are 
$F\wedge \star F\propto (\mathbf{B}^2-\mathbf{E}^2)$ 
and 
$F\wedge F\propto\mathbf{B}\cdot\mathbf{E}$. 
When $F\wedge \star F>0$ and $F\wedge F=0$, there exist inertial frames in which the EM fields are purely magnetic. In what follows, I will use such a special reference frame, and the coordinate where the background magnetic field $\bar{\mathbf{B}}=B_0\mathbf{z}$. 
Physically, the magnetic field must be generated by some macroscopic current, such as the current in a long solenoid. We can then fill plasmas inside the solenoid, and wait until dissipative processes damp out the diamagnetic surface current, after which the plasma will become magnetized. Although the solenoid plasma is of finite size, in the regime where the size of the solenoid is much larger than both the plasma skin depth and the wavelength of interest, the center of the solenoid plasma can be well approximated by a uniformly magnetized plasma. 

\subsection{Background fields\label{ch:mag:basic:background}}

Since the solenoid plasma is rotationally symmetric, it is convenient to use the symmetric gauge, in which the 4-potential
\begin{equation}\label{eq:SymmetricGauge}
	\bar{A}^{\mu}=(0,-\frac{1}{2}B_0y,\frac{1}{2}B_0x,0).
\end{equation}
In the symmetric gauge, the background covariant derivatives, with proper signs from the Minkowski metric, can be written as 
\begin{eqnarray}
\bar{D}^{0}&=&\partial_t,\\
\bar{D}^{1}&=&-\partial_x+\frac{ieB_0}{2}y= -\cos\theta\frac{\partial}{\partial r}+\frac{\sin\theta}{r}\frac{\partial}{\partial\theta} \pm\frac{i r}{r_0^2}\sin\theta,\\
\bar{D}^{2}&=&-\partial_y-\frac{ieB_0}{2}x= -\sin\theta\frac{\partial}{\partial r}-\frac{\cos\theta}{r}\frac{\partial}{\partial\theta}\mp\frac{i r}{r_0^2}\cos\theta,\\
\bar{D}^{3}&=&-\partial_z,
\end{eqnarray}
where the Cartesian coordinate $(x,y)$ is related to the cylindrical coordinate $(r,\theta)$ by the usual relations $x=r\cos\theta$ and $y=r\sin\theta$.  
Since $eB_0$ has the units of length squared, let us denote $eB_0/2=\pm 1/r_0^2$, where the upper and lower sign of $\pm$ correspond to $eB_0>0$ and $eB_0<0$ such that $r_0^2>0$ and the length scale $r_0$ is always real. 
Having chosen the background gauge $\bar{A}$, the field $\phi_0$ can be solved from its equation of motion [Eq.~(\ref{EOM})].
In the symmetric gauge, the second-order covariant derivative becomes
\begin{eqnarray}
\bar{D}_\mu\bar{D}^\mu&=&\partial^2+ieB_0(y\partial_x-x\partial_y)+\frac{1}{4}e^2B_0^2(x^2+y^2).
\end{eqnarray}
The equation of motion can be solved as a partial differential equation, where the wave function can be factorized as products of the decoupled $t$ and $z$ wave functions, and the coupled perpendicular wave function.

Perhaps a more illuminating way of solving the equation is using second quantization. In the usual quantum mechanics, the Hamiltonian equation is of the form $H|\psi\rangle=E|\psi\rangle$. In the relativistic case, it is more convenient to write $h^2=H^2-E^2$, then the Hamiltonian equation becomes $h^2|\psi\rangle=0$. In terms of kinetic momentum, $h^2=\Pi_\mu\Pi^\mu+m^2$, 
where $\Pi_\mu=p_\mu-e\bar{A}_\mu=-i\bar{D}_\mu$ is given by the covariant derivative, while the canonical momentum $p_\mu=-i\partial_\mu$ is given by the usual partial derivative. Due to the presence of the background  magnetic field in the $z$ direction, the kinetic momentum in the perpendicular plane do not commute. In Cartesian coordinate, the $x$ and $y$ components of the kinetic momentum 
\begin{equation}
[\Pi_x,\Pi_y]=-im\Omega,
\end{equation}
where $\Omega=eB_0/m$ is the gyro frequency. For ``electrons" with negative charge $\Omega<0$, we can define the rising and lowering operators analogous to what is done for quantum harmonic oscillator,
\begin{eqnarray}
a^\dagger&=&\frac{1}{\sqrt{2m|\Omega|}}(\Pi_x-i\Pi_y)=-i\big(r_0 \partial_\perp-\frac{\bar{w}}{2r_0}\big),\\
a&=&\frac{1}{\sqrt{2m|\Omega|}}(\Pi_x+i\Pi_y)=-i\big(r_0\bar{\partial}_\perp+\frac{w}{2r_0}\big).
\end{eqnarray}
Here, it is convenient to introduce the complex variable $w=x+iy$ due to the rotation symmetry in the perpendicular plane. We see $r_0$ is a fundamental length scale of the wave functions. Restoring full units, the magnetic de Broglie length
\begin{equation}
\label{eq:r0}
r_0=\sqrt{\frac{2\hbar}{|eB_0|}},
\end{equation}
is determined only by the background magnetic field.
Using the complex variable, the perpendicular derivatives $\partial_\perp:=\partial/\partial w$ and $\bar{\partial}_\perp:=\partial/\partial \bar{w}$ satisfy $\partial_\perp w=\bar{\partial}_\perp\bar{w}=1$, while $\partial_\perp\bar{w}=\bar{\partial}_\perp w=0$. 
It is a straightforward calculation to verify that 
\begin{equation}
[a,a^\dagger]=1,
\end{equation}
which satisfies the canonical commutation relation. The rising operator $a^\dagger$ satisfies $[h^2,a^\dagger]=2m|\Omega| a$, and the lowering operator $a$ satisfies $[h^2,a]=-2m|\Omega| a$. In terms of these operators, the squared Hamiltonian can be written as
\begin{eqnarray}
h^2=\partial_t^2-\partial_z^2+m^2+2m|\Omega|\big(a^\dagger a+\frac{1}{2}\big).
\end{eqnarray}
The particle eigenstates are of the form $|\psi\rangle=\exp(iE_{n,p_\parallel}t-ip_\parallel z)|n\rangle$, where $n$ is the principle quantum number and $p_\parallel$ is the parallel momentum. 
The energy of the eigenstate can be easily read out from the Hamiltonian equation
\begin{equation}
\label{eq:Enp}
E_{n,p_\parallel}=\sqrt{p_\parallel^2+m^2+2m|\Omega|(n+\frac{1}{2})}.
\end{equation}
The energy can be written as $E_{n,p_\parallel}=\sqrt{m_n^2+p_\parallel^2}$, where the effective mass of the $n$-th excited state is $m_n=\sqrt{m^2+|eB_0|(2n+1)}$. Notice that the ground state mass $m_0=\sqrt{m^2+|eB_0|}$ is higher than the rest mass of the particle due to the background magnetic field. In the non-relativistic limit, namely when $p_\parallel, |\Omega|\ll m$, the energy level $E_{n,p_\parallel}$ recovers the usual non-relativistic Landau level $E_{n,p_\parallel} \sim m+p_\parallel^2/2m+|\Omega|(n+1/2)$, for which $E_{n,p_\parallel}-E_{n',p_\parallel}=(n-n')|\Omega|$ changes harmonically by integer multiples of $|\Omega|$.
On the other hand, when the magnetic field is strong, energy levels $E_n\sim\sqrt{2m|\Omega|n}$ become anharmonically spaced. 

Since there are two degrees of freedom in the perpendicular plane, the principle quantum number $n$ does not fully characterize quantum states. The principle quantum number determines the perpendicular energy, which is related to how fast particles gyrate in the perpendicular plane, and equivalently the size of gyro the radius. The other degree of freedom is where particles gyrate about, namely, the location of the gyro center. For a charged particle at $(x, y)$ with velocity $(v^x, v^y)$, the coordinates of its gyro center are $X=x+v^y/\Omega=x-\Pi_y/m\Omega$ and $Y=y-v^x/\Omega=y+\Pi_x/m\Omega$, where the minus sign $v^i=-\Pi_i/m$ is due to the Minkowski metric. Since the kinetic momentum is related to the covariant derivative by $\Pi_\mu=-i\bar{D}_\mu$, it is easy to see that the gyro center coordinates do not commute in finite magnetic fields
\begin{equation}
[X,Y]=\frac{i}{m\Omega}.
\end{equation} 
Suppose $\Omega<0$, then analogous to what is done for quantum harmonic oscillators, we can define the ladder operators for gyro centers
\begin{eqnarray}
b^\dagger&=&\sqrt{\frac{m|\Omega|}{2}}\Big(X+iY\Big) 
=\frac{w}{2r_0}-r_0\bar{\partial}_\perp,\\
b&=&\sqrt{\frac{m|\Omega|}{2}}\Big(X-iY\Big)
=\frac{\bar{w}}{2r_0}+r_0\partial_\perp.
\end{eqnarray}
These ladder operators, combined with the $a$ and $a^\dagger$, form a complete set in the perpendicular plane. The gyro center operators satisfy the canonical commutation relation
\begin{eqnarray}
[b,b^\dagger]=1,
\end{eqnarray}
and commute with the Hamiltonian $[h^2,b^\dagger]=[h^2,b]=0$. In other words, the above ladder operators do not change the energy of quantum states, and are associated with another quantum number $l$ that can be measured simultaneously with the energy. 
To see what this quantum number is, notice that the canonical angular momentum $L_z=xp_y-yp_x$. By straightforward calculations, the ladder operators satisfies the commutation relations $[L_z,a^\dagger]=-a^\dagger$, $[L_z,a]=a$, $[L_z,b^\dagger]=b^\dagger$, and $[L_z,b]=-b$. Therefore, the quantum state $|n,l\rangle$ is a simultaneous eigenstate of both the perpendicular energy $H_\perp=(\Pi_x^2+\Pi_y^2)/2m$ and the canonical angular momentum 
\begin{eqnarray}
H_\perp|n,l\rangle&=&|\Omega|(n+\frac{1}{2})\;|n,l\rangle,\\
L_z|n,l\rangle&=&(l-n)\;|n,l\rangle,
\end{eqnarray}
where the quantum numbers $n, l=0,1,2\dots$ can take nonnegative integer values.
It is a straightforward calculation to show that for the eigenstate $|n,l\rangle$, the expectation value of its kinetic angular momentum $\langle x\Pi_y-y\Pi_x\rangle=2n+1$ only depends on the principle quantum number, which determines how fast the charged particle rotates. 
While the expectation value of the perpendicular coordinate $\langle r^2\rangle=r_0^2(n+l+1)$ depends on both the gyro radius and the location of the gyro center.

Using the above two sets of ladder operators, we can create a complete set of wave functions of particle states from the ground state. The ground state satisfies $a|0,0\rangle=b|0,0\rangle=0$. Using the configuration space representation of the ladder operators, the ground state wave function $|0,0\rangle\propto\exp(-w\bar{w}/2r_0^2)$ is a Gaussian wave packet. The wave function for other single-particle states can be created by acting the rising operators on the ground state wave function
\begin{eqnarray}
\nonumber
|n,l\rangle&=&\frac{(a^\dagger)^n}{\sqrt{n!}}\frac{(b^\dagger)^l}{\sqrt{l!}}|0,0\rangle\\
&=&\frac{(-i)^n}{\sqrt{\pi r_0^2 n!l!}}\sum_{k=0}^{n}\binom{n}{k}\frac{l!}{(l-n-k)!}\Big(-\frac{w\bar{w}}{r_0^2}\Big)^k \Big(\frac{w}{r_0}\Big)^{l-n}e^{-w\bar{w}/2r_0^2}.
\end{eqnarray}
In cylindrical coordinate, the complex variable can be written as $w=re^{i\theta}$. Then, the above perpendicular wave function can be expressed as a function of the radius $r$ and the azimuthal angle $\theta$.
Moreover, the above series solution can be written in terms of the generalized Laguerre function $\Lambda_{n}^{(\alpha)}(w) =L_{n}^{(\alpha)}(w)e^{-w/2}$, where the generalized Laguerre polynomial is given by the sum
\begin{equation}
\label{eq:Laguerre}
L_{n}^{(\alpha)}(w)= \sum_{k=0}^{n}\binom{n+\alpha}{n-k}\frac{(-w)^k}{k!}.
\end{equation}
To obtain the full wave function, we can multiply the perpendicular wave function by the temporal and parallel wave functions. Since particles are confined by the magnetic field in the $xy$ plane, the single-boson wave function is already normalizable in the perpendicular plane. Hence, it is only necessary to impose a periodic box in the $z$ direction. Let $L$ be the length of this periodic box. Then, in polar coordinate, the properly normalized single-boson wave function is 
\begin{eqnarray}
\nonumber
\psi_{n,l,p_\parallel}^{s}(x)&=&\sqrt{\frac{n!}{2m_{n}l!\pi r_0^2L}}\;\rho^{l-n} \Lambda_{n}^{(l-n)}(\rho^2)e^{is[E_{n,p_\parallel}t-p_\parallel z\mp(l-n)\theta]}.
\end{eqnarray}
where $\rho=r/r_0$ is the normalized radius. The above derivation using second quantization assumed particle states with $\Omega<0$. Similar derivations can be carried out for $\Omega>0$, as well as for antiparticle states. In the end, these four cases can be accounted for by $s=+1$ and $-1$ for particle and antiparticle states; 
as well as the upper and lower sign of $\mp$ 
for $eB_0>0$ and $eB_0<0$, which account for the fact that positively and negatively charged particles gyrate in opposite directions. 
The above wave functions are relativistic Landau levels, which have been obtained, for example, by \cite{Witte87}. To check the wave functions are properly normalized, one can calculate, for example, the total current in the $z$ direction.

To deal with plasmas that are infinitely large, we can follow procedures in Sec.~\ref{ch:unmag:basic:background}. First, we can consider finite number of particles in a spatial box of size $L$ and temporal box of length $T$. Using the random phase approximation, the many-body wave function can be expressed as the symmetrized products of single-boson wave functions. 
We can carry out calculations using the normalized many-body wave function, and then take the limit $L,T\rightarrow\infty$ while keeping the plasma density fixed.
Results of this formal procedure can be obtained using an alternative method, where we first take the limit, and then carry out calculations using the effective single-boson wave function 
\begin{eqnarray}\label{eq:MagSingleEff}
\Psi_{n,l,p_\parallel}^{s}(x)&=&\sqrt{\frac{n!f^{s}_{n,l}(p_\parallel)}{2m_{n}l!}}\;\rho^{l-n} \Lambda_{n}^{(l-n)}(\rho^2)e^{is[E_{n,p_\parallel}t-p_\parallel z\mp(l-n)\theta]},
\end{eqnarray}
followed by integration over the continuous label $p_\parallel/2\pi$ and summations over the discrete labels $n,l$,and $s$ over the single-boson Hilbert space. In the above expression, $f^{s}_{n,l}(p_\parallel)$ is the occupation density of the state with quantum numbers $(s,n,l,p_\parallel)$. Except for the fact that perpendicular states are now quantized, the occupation density is nothing other than the momentum space distribution function commonly seen in plasma physics. In general, the distribution function can be any integrable function of interest.

\subsection{Green's function\label{ch:mag:basic:green}}

Besides the wave functions, we will also need the Green's function of charged bosons in the uniform magnetic field. The Green's function can either be found by calculating the propagator of the quantized $\varphi$ field, or more directly by solving the Schwinger--Dyson equation [Eq.~(\ref{eq:SDGreen})]. There are many representations of the Green's function, for example, the proper time representation \citep{Schwinger51} and the spectral representation \citep{Melrose2012quantum}. Although these representations are equivalent, for computing the vacuum response tensor, the proper time representations appears to be more convenient. On the other hand, for computing the plasma response tensor, it is more convenient to use the spectral representation, which I shall derive next.

To derive a spectral representation of the Green's function in the uniform magnetic field, it is more convenient to first use the Landau gauge and then transform to the symmetric gauge using Eq.~(\ref{eq:GaugeGreen}). In the Landau gauge,
\begin{equation}\label{eq:LandauGauge}
\bar{A}^{\mu}_L=(0,-B_0y,0,0),
\end{equation}
which is convenient because it only depends on a single coordinate $y$. Then, the Schwinger--Dyson equation for the Green's function $G_L$ becomes
\begin{equation}
\label{eq:SDGreen_Landau}
\big(\partial^2+m^2+e^2B_0^2y^2+2ieB_0y\partial_x\big)G_{L}(x,x')=-i\delta^{(4)}(x-x').
\end{equation}
Since $G_L$ is translational invariant in $t,x$ and $z$ directions, we can take Fourier transform $\exp[iq_0(t-t')-iq_\perp(x-x')-iq_\parallel(z-z')]$. Without loss of generality, suppose $\Omega>0$. Then, denoting the nondimensionalization variable $\xi=\sqrt{m\Omega} y-q_\perp/\sqrt{m\Omega}$, the partially Fourier transformed Green's function satisfies
\begin{equation}
\nonumber
\Big[\frac{d^2}{d\xi^2}+(\mu-\xi^2)\Big]\hat{G}_L=\frac{i}{\sqrt{m\Omega}}\delta\Big(\xi+\frac{q_\perp}{\sqrt{m\Omega}}-\sqrt{m\Omega}y'\Big),
\end{equation} 
where the dimensionless ratio $\mu=(q_0^2-q_\parallel^2-m^2)/m\Omega$, and I have used the property $\delta(ax)=\delta(x)/|a|$ of the delta function. 
The LHS of the above equation is clearly related to the Hermite function $\psi_n$, which satisfies $\ddot{\psi}_n(\xi)+(2n+1-\xi^2)\psi_n=0$. Since the Hermite functions form an orthonormal basis, we can expand the Green's function $\hat{G}_L=\sum_nc_n\psi_n(\xi)$. Taking inner products with $\psi_l$ on both sides of the expansion, the coefficient
\begin{equation}
\nonumber
c_n=\frac{i}{\sqrt{m\Omega}(\mu-2n-1)} \psi_n\Big(\sqrt{m\Omega}y'-\frac{q_\perp}{\sqrt{m\Omega}}\Big),
\end{equation}
where the argument of the Hermite function is enforced by the delta function. 
The above results are derived for $\Omega>0$. Analogous results for $\Omega<0$ can be obtained similarly. Regardless of the sign of $\Omega$, after taking inverse Fourier transform of the spectral expansion, the Green's function in the Landau gauge can always be written as
\begin{eqnarray}\label{eq:LandauGreen}
\nonumber
G_{L}(x,x')&=&\!\frac{\sqrt{2}}{r_0}\sum_{n=0}^{\infty}\int\frac{dq_0dq_\perp dq_\parallel}{(2\pi)^3}\frac{ie^{i[q_0(t-t')-q_\perp(x-x')-q_\parallel(z-z')]}} {q_0^2-q_\parallel^2-m_n^2}\\
&&\times\psi_n\Big[\frac{r_0}{\sqrt{2}}(q_\perp+eB_0y)\Big]\psi_n\Big[\frac{r_0}{\sqrt{2}}(q_\perp+eB_0y')\Big],
\end{eqnarray}
where the characteristic length scale $r_0$ is given by Eq.~(\ref{eq:r0}). 
Using the completeness of the Hermite functions \citep{Olver10} that the infinite sum
\begin{equation}
\sum_{n=0}^{\infty}\psi_n(x)\psi_n(y)=\delta(x-y),
\end{equation}
it is straightforward to check that the Green's function (\ref{eq:LandauGreen}) solves the Schwinger--Dyson equation in the Landau gauge [Eq.~(\ref{eq:SDGreen_Landau})]. This form of the Green's function is expanded by wave functions that are eigenfunctions of the $\mathbf{E}\times\mathbf{B}$ drift. These eigenfunctions are featured by free propagation along the $\mathbf{B}$ field; free propagation in the $\mathbf{E}\times\mathbf{B}$ direction; and harmonic oscillation in the direction of the E-field. These features make the Green's function $G_L$ convenient for studying DC quantum Hall conductivity. But the loss of rotation symmetry in the perpendicular plane makes it inconvenient for studying AC wave phenomena.

To restore the rotation symmetry, we need to make a gauge transformation into the symmetric gauge Eq.~(\ref{eq:SymmetricGauge}). The symmetric gauge is related to the Landau gauge Eq.~(\ref{eq:LandauGauge}) by gauge transformation Eq.~(\ref{BackgroundGauge}), where the scalar field is
\begin{equation}
\chi=-\frac{1}{2}B_0xy.
\end{equation} 
Under this gauge transformation, the Green's function is transformed by Eq.~(\ref{eq:GaugeGreen}) as
\begin{equation}\label{eq:SLGreen}
G(x,x')=e^{-ieB_0(xy-x'y')/2}G_{L}(x,x'),
\end{equation}
where $G$ denotes the Green's function in the symmetric gauge. In the symmetric gauge, the eigenfunctions [Eq.~(\ref{eq:MagSingleEff})] are circular in the perpendicular plane, so the Green's function $G$, which can be expanded by these eigenfunctions, is also invariant under rotations around the $z$ axis.
To put $G$ in a manifestly rotational invariant form, we need to carry out the integral of Hermite functions. Notice that the Hermite function is
\begin{equation}
\psi_n(x)=(2^nn!\sqrt{\pi})^{-1/2}e^{-x^2/2}H_n(x),
\end{equation}
where $H_n(x)$ is the Hermite polynomial and satisfies
\begin{eqnarray}
&&H_n(x+y)=\sum_{k=0}^{n}\binom{n}{k}H_k(x)(2y)^{n-k},\\
&&\int dx H_n(x)H_m(x)e^{-x^2}=2^nn!\sqrt{\pi}\delta_{n,m}\hspace{2pt}.
\end{eqnarray}
Recall that the Laguerre function $\Lambda_n^{(0)}(x):=L_n^{(0)}(x)e^{-x/2}$, where the Laguerre polynomial $L_n^{(0)}(x)$ has the closed series expansion Eq.~(\ref{eq:Laguerre}).
Denoting $w=u+iv$ and $\bar{w}=u-iv$, and changing variable to $p=q+iu$, the integration
\begin{eqnarray}
\label{eq:HermiteLaguerre}
\nonumber
&&\int dq e^{-2iqu}\psi_n(q+v)\psi_n(q-v) \\
\nonumber
&=&\frac{1}{2^nn!\sqrt{\pi}}\int dq e^{-q^2-v^2-2iqu}H_n(q+v)H_n(q-v)\\
\nonumber
&=&\frac{e^{-w\bar{w}}}{2^nn!\sqrt{\pi}}\int dp e^{-p^2}H_n(p-iw)H_n(p-i\bar{w})\\
\nonumber
&=&\frac{e^{-w\bar{w}}}{2^nn!}\sum_{k,l=0}^{n}\binom{n}{k}\binom{n}{l}(-2iw)^{n-k}(-2i\bar{w})^{n-l}2^kk!\delta_{k,l}\\
\nonumber
&=&\Lambda_n^{(0)}[2(u^2+v^2)],
\end{eqnarray}
which can be expressed in terms of the Laguerre function.
Using the above identity, we can carry out the $q_\perp$ integral in the Green's function Eq.~(\ref{eq:LandauGreen}) by identifying $q=r_0[q_\perp+\frac{eB_0}{2}(y+y')]/\sqrt{2}$, $u=(x-x')/\sqrt{2}r_0$, and $v=r_0eB_0(y-y')/2\sqrt{2}$.
After the integration, the Green's function Eq.~(\ref{eq:SLGreen}) can be put into the following rotational invariant form: 
\begin{eqnarray}
\label{eq:MagGreen}
&&G(x,x')=\frac{i}{\pi r_{0}^{2}}e^{ieB_0(xy'-yx')/2}\sum_{n=0}^{\infty}\int\!\frac{dq_0dq_\parallel}{(2\pi)^2}\frac{e^{i[q_{0}(t-t')-q_{\parallel}(z-z')]}}{q_{0}^{2}-q_{\parallel}^{2}-m_n^2}\Lambda_{n}^{(0)}\bm{\Big(}\frac{r^2}{r_0^2}\bm{\Big)},
\end{eqnarray}
where $r^2=(x-x')^2+(y-y')^2$ is the relative separation between the two points in the perpendicular plane, and $m_n$ is again the effective mass of the $n$-th excited state.
Using the completeness of the Laguerre functions \citep{Olver10}, we have
\begin{equation}
\sum_{n=0}^{\infty}\Lambda_{n}^{(0)}(x^2+y^2)=\pi\delta(x)\delta(y).
\end{equation}
It is thereof straightforward to check that the Green's function (\ref{eq:MagGreen}) satisfies the Schwinger--Dyson equation in the symmetric gauge. 
In addition to the rotation symmetry, the above Green's function respects a number of symmetries of the system. 
First, it is invariant under parity $\mathbf{x}\rightarrow-\mathbf{x}$. Second, it is invariant under the joint symmetry action of charge conjugation $e\rightarrow-e$ and time reversal $t\rightarrow-t, B_0\rightarrow-B_0$. 
Notice that the Green's function has poles when $q_0=\pm E_{n,q_\parallel}$. These are nothing other than the dispersion relations of charged particles and antiparticles occupying relativistic Landau levels. When the $\varphi$ field propagates, it can propagate through any of these quantum channels.

\section{Magnetized response tensors\label{ch:mag:polarization}}

The general formulas of the response tensors are Lorentz covariant and gauge invariant. However, a particular explicit expression of the response tensor depends on the choice of the reference frame as well as the gauge. In the reference frame where the background magnetic field is in the $z$ direction and the 4-potential is in the symmetric gauge, the background gauge field $\bar{A}$ is given by Eq.~(\ref{eq:SymmetricGauge}), and the background particle field $\phi_0$ is given by Eq.~(\ref{eq:MagSingleEff}). Combining background fields with the Green's function of the charged field [Eq.~(\ref{eq:MagGreen})], the vacuum response and the plasma response can be evaluated.

\subsection{Vacuum polarization\label{ch:mag:polarization:vacuum}}

The vacuum response is given by the famous Heisenberg–-Euler effective Lagrangian \citep{Heisenberg36,Dunne2012heisenberg}. Here, I will not consider the vacuum response for three reasons. First, as can be seen from Ch.~\ref{ch:unmag}, effects of the vacuum response are minuscule compared to collective plasma effects for low energy waves, especially when the wave effective mass 
and the cyclotron energy 
are much smaller than the electron rest energy. 
Second, due to separate conservation of the plasma current and the vacuum current [Eqs.~(\ref{bkConservation}) and (\ref{vacConservation})], contributions by the plasma background and the vacuum are separable. Ignoring the vacuum response does not break any symmetry of the system, and is thereof allowed. 
Finally, for a practical reason, obtaining a useful expression of the vacuum response is highly nontrivial. Although many formal representations of the vacuum response tensor have been obtained \citep{Witte90,Kuznetsov13}, they can be evaluated analytically to give concrete numbers only in some special limits \citep{Shabad75,Karbstein13}. In more general cases, the vacuum response tensor needs to be evaluated numerically \citep{Kohri2002polarization}.

\subsection{Plasma dielectric response\label{sec:mag:polarization:plasma}}

In order for the plasma background to be consistent with a uniform magnetic field, the plasma needs to fill the entire space uniformly with zero current and charge density everywhere in space. This is achievable by infinite sums of eigenfunctions, which form a complete basis.
After choosing the occupation density appropriately, such that the equation for the background EM field $\partial_{\mu}\bar{F}^{\mu\nu}=0$ is satisfied, we can evaluate the plasma response tensor [Eq.~({\ref{bkPol}})] by plugging in the effective single-boson wave functions [Eq.~(\ref{eq:MagSingleEff})] and the Green's function [Eq.~(\ref{eq:MagGreen})], followed by integration over the continuous parallel momentum $p_\parallel/2\pi$, as well as summations over the discrete principle quantum number $n$, the angular momentum quantum number $l$, and the species index $s$. 
The plasma response tensor is constituted of a mass term and a polarization term. The mass term becomes a constant after summing eigenfunctions to form a uniform plasma background. The polarization term involves excitation and de-excitation of the plasma current, when charged bosons make transitions between relativistic Landau levels. 

\subsubsection{Covariant derivatives}
When evaluating the plasma response tensor, we need covariant derivatives of the background wave function [Eq.~(\ref{eq:MagSingleEff})]. For simplicity, let us abbreviate the effective single-boson wave function as
\begin{equation}
\Psi=Me^{i\Theta}\rho^{l-n}\Lambda_n^{(l-n)},
\end{equation}
where $M$ is a constant amplitude, $\rho=r/r_0$ is the normalized radius, $\Theta=s[Et-pz\mp(l-n)\theta]$ is the phase, and the argument of $\Lambda_n^{(l-n)}$ is omitted. Using the above abbreviated notations, the covariant derivatives of the wave function are
\begin{eqnarray}
\label{eq:D0Psi}
\bar{D}^{0}\Psi&=&is EMe^{i\Theta}\rho^{l-n}\Lambda_n^{(l-n)},\\
\label{eq:D1Psi}
\bar{D}^{1}\Psi&=&\frac{M}{r_0}e^{i\Theta}[2\rho^{l-n+1}\Lambda_{n-1}^{(l-n+1)}\cos\theta+e^{\pm is\theta}\rho^{l-n-1}(\rho^2-l-n)\Lambda_{n}^{(l-n)}],\\
\label{eq:D2Psi}
\bar{D}^{2}\Psi&=&\frac{M}{r_0}e^{i\Theta}[2\rho^{l-n+1}\Lambda_{n-1}^{(l-n+1)}\sin\theta+e^{\pm is(\theta-\pi/2)}\rho^{l-n-1}(\rho^2-l-n)\Lambda_{n}^{(l-n)}],\hspace{10pt}\\
\label{eq:D3Psi}
\bar{D}^{3}\Psi&=&is pMe^{i\Theta}\rho^{l-n}\Lambda_n^{(l-n)}.
\end{eqnarray}
Here I have used the property of the Laguerre function $\Lambda_{n}^{(\alpha)'}(x)=-\Lambda_{n-1}^{(\alpha+1)}(x)-\Lambda_{n}^{(\alpha)}(x)/2$, where $\Lambda_n^{(\alpha)}=0$ whenever $n<0$. From the above expressions, we see $\bar{D}^{2}\Psi$ can be obtained from $\bar{D}^{1}\Psi$ by replacing $\theta\rightarrow\theta-\pi/2$, which is expected from the rotation symmetry.

In addition to the background wave functions, we will also need covariant derivatives of the Green's function [Eq.~(\ref{eq:MagGreen})]. For simplicity, let us abbreviate the Green's function in the symmetric gauge as 
\begin{equation}
G=\Upsilon\mathcal{G}_ne^{iX}\Lambda_n^{(0)},
\end{equation}
where $\Upsilon=1/(\pi r_0^2)\sum_n\int d^2q/(2\pi)^2$ is the summation and integration prefactor, $\mathcal{G}_n(q_0,q_\parallel)=i/(q_0^2-q_\parallel^2-m_n^2)$ is the momentum space propagator in the $tz$ subspace, $X=\pm is(xy'-x'y)/r_0^2+iq_0(t-t')-iq_\parallel(z-z')$ is the phase, and the argument of $\Lambda_n^{(0)}$ is omitted. Write $\eta_1=(x-x')/r_0$ and $\eta_2=(y-y')/r_0$, the covariant derivatives of the Green's function 
\begin{eqnarray}
\label{eq:D0G}
\bar{D}^{0}G&=&\Upsilon\mathcal{G}_niq_0e^{iX}\Lambda_n^{(0)},\\
\label{eq:D1G}
\bar{D}^{1}G&=&\Upsilon\mathcal{G}_n\frac{e^{iX}}{r_0}[2\eta_1\Lambda_{n-1}^{(1)}+(\eta_1\pm is\eta_2)\Lambda_{n}^{(0)}],\\
\label{eq:D2G}
\bar{D}^{2}G&=&\Upsilon\mathcal{G}_n\frac{e^{iX}}{r_0}[2\eta_2\Lambda_{n-1}^{(1)}+(\eta_2\mp is\eta_1)\Lambda_{n}^{(0)}],\\
\label{eq:D3G}
\bar{D}^{3}G&=&\Upsilon\mathcal{G}_niq_\parallel e^{iX}\Lambda_n^{(0)}.
\end{eqnarray} 
The covariant derivatives $\bar{D}^{'\mu}$ with respect to $x'$ can be found by direct calculations. Alternatively, recall $G'=-G^*$. We can also find $\bar{D}^{'\mu}$ derivatives using $\bar{D}^{'\mu}G=-(\bar{D}^{'*\mu}G')^*$. The above are all the background gauge covariant derivatives that are necessary for evaluating the plasma response tensor.

\subsubsection{Distribution function}
For illustrative purpose, let us consider the simple example of a cold particle plasma, in which all charged bosons are condensed in the lowest Landau levels. In the rest frame of the cold plasma, the occupation density 
\begin{equation}\label{eq:MagDOS}
	f^{s}_{n,l}(p_\parallel)=2\pi n_0\delta(p_\parallel)\delta_{n,0}\delta_{s,1}\hspace{2pt},
\end{equation}
where $n_0$ is the number density of the plasma. In this simple example, the three $\delta$-functions make it very easy to carry out the integrations and summations. 
In this simple example, the ground states $\Psi_l=Me^{i(m_0t\mp l\theta)}\rho^l e^{-\rho^2/2}$ are the only relevant states.

To check that the above distribution function is consistent with the uniform background magnetic field, let us compute the current density. 
Using Eqs.~(\ref{eq:D0Psi})-(\ref{eq:D3Psi}), the covariant derivatives of ground states $\bar{D}^{0}\Psi_l=im_0\Psi_l$, $\bar{D}^{1}\Psi_l=\frac{1}{r}e^{\pm i\theta}(\rho^2-l)\Psi_l$, $\bar{D}^{2}\Psi_l=\mp i\bar{D}^{1}\Psi_l$, and $\bar{D}^{3}\Psi_l=0$. After summing over angular momentum quantum number $l$, the background 4-current density 
\begin{eqnarray}
\bar{J}_{0}^{0}&=&en_0\sum_{l=0}^{\infty}\frac{1}{l!}\rho^{2l}e^{-\rho^2}=en_0,\\
\bar{J}_{0}^{1}&=&\pm\frac{en_0\sin\theta}{m_0r}\sum_{l=0}^{\infty}\frac{1}{l!}\rho^{2l} (\rho^2-l)e^{-\rho^2}=0,\\
\bar{J}_{0}^{2}&=&\mp\frac{en_0\cos\theta}{m_0r}\sum_{l=0}^{\infty}\frac{1}{l!}\rho^{2l} (\rho^2-l)e^{-\rho^2}=0,\\
\bar{J}_{0}^{3}&=&0.
\end{eqnarray}
Although each eigenfunction is nonuniform, the sum of all ground state wave functions gives a uniform charge distribution. Similarly, although each gyro orbit carries nonzero current, the sum of all gyro orbits cancels the current density everywhere in space. 
The total 4-current density due to each charged species is $\bar{J}_{s0}^{\mu}=e_sn_{s0}(1,0,0,0)$, which is what one would expect of a uniform cold fluid. The self-consistency condition $\sum_s\bar{J}_{s0}^{\mu}=0$ is thereof satisfied if the plasma is quasi neutral.

\subsubsection{The gauge boson mass term}
Having confirmed that the plasma, with all particles equally occupying the degenerate ground states, is self-consistent with the uniform background magnetic field, let us now compute the plasma response tensor. The contribution of each species to the mass term of the $\mathcal{A}$ field is 
\begin{eqnarray}
2e^2\phi_0\phi_0^*&=&2e^2 \sum_{l=0}^{\infty}\Psi_l\Psi_l^*=2e^2 \sum_{l=0}^{\infty} \frac{n_0}{2m_0 l!}\rho^{2l} e^{-\rho^2}=\frac{m\omega_p^2}{m_0},
\end{eqnarray}
where $\omega_p^2=e^2n_0/m$ is the usual plasma frequency. Due to the background magnetic field, ground states acquire zero-point energy, so that the ground state mass $m_0>m$ is larger than mass of the free particle. The larger particle mass results in a smaller plasma frequency, and consequently a smaller photon mass in strongly magnetized plasmas.

\subsubsection{The plasma polarization tensor}
Now let us compute the $\Pi_{2,\text{bk}}^{00}$ component of the plasma polarization tensor. For simplicity, in addition to abbreviations used previously, I will further abbreviate $\tau=t-t'$, $\zeta=z-z'$, $\bm{k}=(k_1,k_2)$, and $\bm{\eta}=(\eta_1,\eta_2)$. Substituting the occupation density Eq.~(\ref{eq:MagDOS}) into the effective single-boson wave function [Eq.~(\ref{eq:MagSingleEff})] and use it in place of the background field $\phi_0$ in the expression of the background polarization tensor [Eq.~(\ref{bkPol})], the temporal component of the polarization tensor 
\begin{eqnarray}
\nonumber
\Pi_{2,\text{bk}}^{00}(x,x')&=&e^2\sum_{l=0}^{\infty}\Upsilon(q_0+m_0)^2\Psi_l^*\Psi_l'\mathcal{G}_ne^{iX}\Lambda_n^{(0)}-\text{c.c.}\\
\nonumber
&=&\frac{e^2n_0}{2m_0}\Upsilon(q_0+m_0)^2\mathcal{G}_ne^{i[(q_0-m_0)\tau-q_\parallel\zeta]-\bm{\eta}^2/2}\Lambda_n^{(0)}-\text{c.c.}\\
&=&\frac{im\omega_p^2}{2m_0}\Upsilon e^{i(q_0\tau-q_\parallel\zeta)-\bm{\eta}^2/2}\Lambda_n^{(0)}\pi_n^{00}.
\end{eqnarray}
The kernel of the polarization tensor is similar to the unmagnetized case [Eq.~(\ref{UMpol})], except now only the $t$ and $z$ components of the momentum are free
\begin{eqnarray}
\pi_n^{00}(q_0,q_\parallel)&=&\frac{(q_0+2m_0)^2}{(q_0+m_0)^2-q_\parallel^2-m_n^2}+\frac{(q_0-2m_0)^2}{(q_0-m_0)^2-q_\parallel^2-m_n^2}.
\end{eqnarray}
As in the unmagnetized case, the above two terms correspond to the $s$-channel and the $t$-channel forward scattering.
Notice that $\Pi_{2,\text{bk}}^{00}$ only depends on the difference between coordinates \mbox{$r^{\mu}=(x-x')^{\mu}$}. This is expected since the system is translational invariant. The above results are qualitatively similar to those in \cite{Rojas79}, who studied fermion plasmas using the temperature Green's functions.

Since the system is translational invariant, the polarization tensor takes a simpler form in the momentum space. Recall in Sec.~\ref{sec:action:properties:observables}, the Fourier space tensor is of the form $\hat{\Pi}(k,k')=(2\pi)^4\delta^{(4)}(k-k')\hat{\Pi}(k)$. Denoting $\bm{k}\bm{\eta}=k_1\eta_1+k_2\eta_2$, we have
\begin{eqnarray}
\nonumber
\hat{\Pi}_{2,\text{bk}}^{00}(k)&=&\int d^4r e^{ikr}\Pi_{2,\text{bk}}^{00}(r)\\
\nonumber
&=&\frac{im\omega_p^2}{2\pi m_0}\sum_{n=0}^{\infty}\int d\bm{\eta}e^{-ir_0\bm{k}\bm{\eta}-\bm{\eta}^2}L_n^{(0)}(\bm{\eta}^2)\pi_n^{00}(k_0,k_\parallel).
\end{eqnarray}
To calculate the integral, we need a number of identities of special mathematical functions \citep{Olver10}. First, the  
Laguerre polynomial can be split as $L_{n}^{(\alpha+\beta+1)}(x+y)=\sum_{l=0}^{n}L_{l}^{(\alpha)}(x)L_{n-l}^{(\beta)}(y)$. Second, the special value $L_{n}^{(-1/2)}(x^2)=\frac{(-1)^n}{2^{2n}n!}H_{2n}(x)$ is related to the Hermite polynomial,
whose Fourier integral $\int dxe^{-ikx-x^2}H_{2n}(x)=\sqrt{\pi}(-1)^{n}k^{2n}e^{-k^2/4}$.
With these properties, the integral in $\hat{\Pi}_{2,\text{bk}}^{00}(k)$ can be computed 
\begin{eqnarray}
\nonumber
&&\int d\bm{\eta}e^{-i\bm{k}\bm{\eta}-\bm{\eta}^2}L_n^{(0)}(\bm{\eta}^2)\\
\nonumber
&=&\int d\bm{\eta}e^{-i\bm{k}\bm{\eta}-\bm{\eta}^2}\sum_{l=0}^{n}L_{l}^{(-1/2)}(\eta_1^2)L_{n-l}^{(-1/2)}(\eta_2^2)\\
\nonumber
&=&\frac{(-1)^n}{2^{2n}}\sum_{l=0}^{n}\frac{1}{l!(n-l)!}\int d\eta_1e^{-ik_1\eta_1-\eta_1^2}H_{2l}(\eta_1)\int d\eta_2e^{-ik_2\eta_2-\eta_2^2}H_{2(n-l)}(\eta_2)\\
\nonumber
&=&\frac{\pi}{2^{2n}} e^{-\bm{k}^2/4}\sum_{l=0}^{n}\frac{k_1^{2l}k_2^{2(n-l)}}{l!(n-l)!}\\
\nonumber
&=&\frac{\pi}{n!} e^{-\bm{k}^2/4}\Big(\frac{\bm{k}^2}{4}\Big)^{n}.
\end{eqnarray}
Denoting the normalized wave vector $\kappa^{\mu}=r_0k^{\mu}/2$, and the squared perpendicular momentum
$\bm{\kappa}^2=r_0^2\bm{k}^2/4$, the momentum space $\hat{\Pi}_{2,\text{bk}}^{00}(k)$ becomes
\begin{equation}
\nonumber
\hat{\Pi}_{2,\text{bk}}^{00}(k)=\frac{im\omega_p^2}{2m_0}e^{-\bm{\kappa}^2} \sum_{n=0}^{\infty}\frac{(\bm{\kappa}^2)^n}{n!}\pi_n^{00}(k_0,k_\parallel).
\end{equation}
Taking the limit $B_0\rightarrow 0$, the effective mass $m_n\rightarrow m$, so the kernel $\pi_n^{00}$ becomes independent of $n$. Then, the summation can be easily carried out, and the above expression recovers the polarization tensor $\hat{\Pi}_{2,\text{bk}}^{00}$ in the unmagnetized case. Notice that the pole of $\pi_n^{00}$ is weighted by $w_n(\bm{\kappa})=e^{-\bm{\kappa}^2}(\bm{\kappa}^2)^n/n!$, which is proportional to the strength of interaction between the plane wave with 4-momentum $k$ and particles in the $n$-th Landau level. The weighting factor maximizes at $\bm{\kappa}^2=n$. For large $n$, the maximum value scale as $w_n\sim1/\sqrt{2\pi n}$. We see waves couple more strongly to electrons in lower Landau levels.

The summation in the above expression can be carried out using the confluent hypergeometric functions \citep{Olver10}.
For convenience, let us define the $K$-function, which is related to the confluent hypergeometric function ${}_1F_1(a;b;z)$ by 
\begin{eqnarray}\label{eq:Kfunction}
K(x,z)&:=&\frac{1}{x}{}_1F_1(1;1-x;-z)=e^{-z}\sum_{n=0}^{\infty}\frac{z^n}{n!}\frac{1}{x-n}.
\end{eqnarray}
From this expression, it is easy to see that when $x\sim n$, where n is some integer, the $K$-function $K(x,z)\sim z^ne^{-z}/n!(x-n)$ is dominated by the pole at $x=n$. 
Using the $K$-function, 
the temporal component of the momentum space plasma polarization tensor 
\begin{eqnarray}
\label{eq:Pi00}
\hat{\Pi}_{2,\text{bk}}^{00}(k)=\frac{im\omega_p^2}{2m_0}\sum_{\varsigma=\pm 1}(\kappa^0+\varsigma\varrho^0)^2K(\kappa_{\varsigma}^2,\bm{\kappa}^2).
\end{eqnarray}
Here, $\varrho^{\mu}=r_0(m_0,0,0,0)$ is the normalized 4-momentum of plasma particles, the normalized poles of the kernel are $\kappa_\varsigma^2:=\kappa_0^2-\kappa_3^2+\varsigma\varrho_0\kappa_0$, and the summation over $\varsigma=\pm 1$ corresponds to the summation of the $s$-channel and the $t$-channel Feynman diagrams for forward scattering of the gauge boson.

Other components of the plasma response tensor $\hat{\Pi}_{2,\text{bk}}^{\mu\nu}(k)$ can be calculated using similar methods. When calculating other components, one will encounter Fourier integrals, where the identity of the Laguerre polynomial $L_{n}^{(\alpha+1)}(x)=\sum_{k=0}^{n}L_{k}^{(\alpha)}(x)$ is useful. 
Using this property, all Fourier integrals that appear in the calculation of other components of $\hat{\Pi}_{2,\text{bk}}^{\mu\nu}$ can be calculated. For example,
\begin{eqnarray}
\nonumber
\int d\bm{\eta}e^{-i\bm{k}\bm{\eta}-\bm{\eta}^2}\eta_1L_{n-1}^{(1)}(\bm{\eta}^2)
=i\frac{\partial}{\partial k_1}\sum_{l=0}^{n-1}\int d\bm{\eta}e^{-i\bm{k}\bm{\eta}-\bm{\eta}^2}L_l^{(0)}(\bm{\eta}^2)
=-\frac{i\pi }{2}e^{-\bm{\kappa}^2}k_1\frac{(\bm{\kappa}^2)^{n-1}}{(n-1)!}.
\end{eqnarray}
To carry out summations of the kernels of the response tensor, the following recurrence relation of the confluent hypergeometric function $b\enspace{}_1F_1(a;b;z)=b\enspace{}_1F_1(a-1;b;z)+z\enspace{}_1F_1(a;b+1;z)$ is useful.
Using this recurrence relation, summations that appear in the calculation of other components of $\hat{\Pi}_{2,\text{bk}}^{\mu\nu}$ can be simplified. For example,
\begin{eqnarray}
\nonumber
e^{-z}\sum_{n=0}^{\infty}\frac{z^n}{n!}\frac{n}{x+n}&=&\frac{z}{x+1}{}_1F_1(1;x+2;-z)=1-{}_1F_1(1;x+1;-z).
\end{eqnarray}
Finally, in terms of the $K$-function, the recurrence relation becomes $xK(x,y)-yK(x-1,y)=1$, 
which is useful when verifying that the response tensor satisfies the Ward--Takahashi identity. In fact, the Ward--Takahashi identity, which is proven for the most general case in Appendix~\ref{ch:append:Ward}, provides a useful check of calculations.

\subsubsection{The plasma response tensor}
After tedious but otherwise straightforward calculations, all components of the polarization tensors can be computed. Combining the polarization term with the mass term, the contribution of each charged species to the Fourier space plasma response tensor is
\begin{eqnarray}
	\label{MagPol}
	\hat{\Sigma}^{\lambda\sigma}_{2,\text{bk}}(k)&=&\frac{m\omega_p^2}{m_0}\Big\{g^{\lambda\sigma}-\frac{1}{2}\sum_{\varsigma=\pm 1}(\kappa+\varsigma\varrho)^{\lambda}(\kappa+\varsigma\varrho)^{\sigma}K_\varsigma^{(0)}\Big\},\\
	\label{MagPol_latin}
	\nonumber
	\hat{\Sigma}^{ab}_{2,\text{bk}}(k)&=&\frac{m\omega_p^2}{2m_0}\sum_{\varsigma=\pm 1}\Big\{\varepsilon^{ac}\varepsilon^{bd}\kappa^c\kappa^d(2K_\varsigma^{(1)}-K_\varsigma^{(0)})\\
	&&-\kappa_\varsigma^2[\delta^{ab}K_\varsigma^{(1)}\pm i\varsigma\varepsilon^{ab}(K_\varsigma^{(1)}-K_\varsigma^{(0)})]\Big\},\\
	\label{MagPol_cross}
	\nonumber
	\hat{\Sigma}^{\lambda a}_{2,\text{bk}}(k) &=&\hat{\Sigma}^{a\lambda}_{2,\text{bk}}(-k)=\frac{m\omega_p^2}{2m_0}\sum_{\varsigma=\pm 1}(\kappa+\varsigma\varrho)^{\lambda}\\
	&&\times\Big\{-\kappa^aK_\varsigma^{(1)}\pm i\varsigma\varepsilon^{ab}\kappa^b(K_\varsigma^{(1)}-K_\varsigma^{(0)})\Big\}.
\end{eqnarray}
In the above expressions, the Greek indices $\lambda,\sigma=0,3$ correspond to the unconfined directions, and the Latin indices $a,b=1,2$ correspond to the confined directions. On right hand sides, $g^{\lambda\sigma}$ is the metric tensor of the Minkowsi space, $\delta^{ab}$ is the $\delta$-function, and $\varepsilon^{ab}$ is the rank-2 Levi-Civita symbol. The upper and lower sign of $\pm$ in the imaginary parts correspond the the case $eB_0>0$ and $eB_0<0$, respectively. 
For conciseness, I abbreviate $K_\varsigma^{(n)}:= K(\kappa_\varsigma^2-n,\bm{\kappa}^2)$ to suppress arguments of the $K$-function. The wave and plasma 4-momentum are normalized by the magnetic de Broglie length as before.

The plasma response tensor Eqs.~(\ref{MagPol})-(\ref{MagPol_cross}) satisfies a number of symmetry properties. 
First, it satisfies the exchange symmetry $\hat{\Sigma}^{\mu\nu}_{2,\text{bk}}(k)=\hat{\Sigma}^{\nu\mu}_{2,\text{bk}}(-k)$, as required by the reality condition [Eq.~(\ref{eq:reality})]. 
Second, it is invariant under rotations around the $z$-axis, which is a basic symmetry of the coordinate system. 
Third, it transforms properly under time reversal symmetry \mbox{$T^{\mu}_{\nu}=\text{diag}(-1,1,1,1)$} by $\hat{\Sigma}^{\mu\nu}(\omega,\bm{k})|_{B_0}=T^{\mu}_{\alpha}T^{\nu}_{\beta}\hat{\Sigma}^{\alpha\beta}(-\omega,\bm{k})|_{-B_0}$. 
Finally, using the recurrence relation of the confluent hypergeometric function, it is straightforward to check that the Ward--Takahashi identity [Eq.~(\ref{eq:Ward})], which is required by charge conservation and gauge invariance, is satisfied as expected.

In addition to the aforementioned symmetry properties, the plasma response tensor Eqs.~(\ref{MagPol})-(\ref{MagPol_cross}) satisfies a number of asymptotic properties. First, since the confluent hypergeometric function ${}_1F_1(a;b;z)$ has poles whenever $b$ equals to nonpositive integers, the response tensor has poles whenever $\omega=\pm\omega_{n,k_\parallel}^{\pm}$, where the frequency of relativistic quantum cyclotron resonances
\begin{eqnarray}\label{eq:Poles}
	\omega_{n,k_\parallel}^{\pm}&=&E_{n,k_\parallel}\pm m_0
	\sim \left\{ \begin{array}{ll}
		\omega_{n,k_\parallel}^{-}+2(m+|\Omega|/2), & ``+",\\
		k_\parallel^2/2m+n|\Omega|, & ``-".
	\end{array} \right.
\end{eqnarray}
The eigenenergy $E_{n,k_\parallel}$ is given by Eq.~(\ref{eq:Enp}).
The above asymptotic behavior is in the limit $k_\parallel, |\Omega|\ll m$. These resonances have clear physical meanings. The $\omega_{n,k_\parallel}^{-}$ resonance corresponds to the energy it takes to excite a plasma particle from the ground state to the $n$-th Landau level with parallel momentum $k_\parallel$. The $\omega_{n,k_\parallel}^{+}$ resonance corresponds to the aforementioned excitation energy plus the energy it takes to create a pair of new particles in the ground state. 
The second important asymptotic property is when the magnetic field $B_0\rightarrow 0$. In this limit, the ground state mass $m_0$ asymptotes to the vacuum mass $m$. Moreover, using the asymptotic property that ${}_1F_1(1;b;z)\rightarrow b/(b-z)$ when $z,b\rightarrow\infty$ while keeping $b/z$ fixed, we can find the asymptotic behavior of the $K$-function 
\begin{equation}
	\frac{r_0^2}{4}K_{\varsigma}^{(n)}\rightarrow\frac{1}{k^2+2\varsigma m k_0}, \qquad B_0\rightarrow 0.
\end{equation}
Here $k^2=k^{\mu}k_{\mu}$ is the Minkowski inner product. Using the above expression, it is straightforward to check that in the limit $B_0\rightarrow 0$, the response tensor of cold magnetized plasmas asymptotes to the response tensor Eq.~(\ref{UMbk}) of cold unmagnetized plasmas.

\section{Spectrum of magnetized waves\label{ch:mag:dispersion}}

Substituting the plasma response tensor into Eq.~(\ref{1-loop}), we have thus obtained an explicit expression of the tree-level nonlocal wave effective action, using which we can determine properties of linear waves. If we treat the fluctuating gauge field $\mathcal{A}$ as a classical field, then by solving its equation of motion in the momentum space, we can obtain the dispersion relation of waves. The dispersion relation contains many branches. At frequency $\omega<2m$, the spectrum of linear waves in strongly magnetized scalar-QED plasmas is qualitatively similar to that in warm classical plasmas, but quantitatively modified by relativistic quantum effects.

\subsection{Oblique propagation\label{ch:mag:dispersion:oblique}}

For convenience, let us choose a coordinate in which the wave 4-momentum \mbox{$k^{\mu}=(\omega,k_\perp,0,k_\parallel)$}. Since $k_2=0$ in this coordinate system, the plasma response tensor can be simplified. In components, the wave dispersion relation Eq.~(\ref{dispersion}) can be written explicitly
\begin{equation}\label{eq:MagDisp}
\det\!\left(\!\begin{array}{ccc}
\omega^2\!-\!k_{\parallel}^2\!+\!\hat{\Sigma}^{11}\!&\!\hat{\Sigma}^{12}\!&\!k_\perp k_\parallel\!+\!\hat{\Sigma}^{13}\!\\
\hat{\Sigma}^{21}\!&\!\omega^2\!-\!\bm{k}^2\!+\!\hat{\Sigma}^{22}\!&\!\hat{\Sigma}^{23}\!\\
k_\perp k_\parallel\!+\!\hat{\Sigma}^{31}\!&\!\hat{\Sigma}^{32}\!&\!\omega^2\!-\!k_\perp^2\!+\!\hat{\Sigma}^{33}\!
\end{array}\!\right)\!=\!0,
\end{equation}
where the subscripts of the response tensor are omitted.
In this form, it is easy to recognize that the spatial components of the response tensor $\hat{\Sigma}^{ij}=\omega^2\chi^{ij}$ is related to the linear susceptibility. While the dispersion relation is formally identical to that in classical plasmas, relativistic quantum effects are encoded in the response tensor.

In the coordinate system where the background magnetic field is in the $z$ direction, and the wave vector is in the $xz$ plane, the plasma response tensor Eqs.~(\ref{MagPol})-(\ref{MagPol_cross}) is greatly simplified. The diagonal components are
\begin{eqnarray}
\hat{\Sigma}^{11}&=&-\frac{m\omega_p^2}{2m_0}\sum_{\varsigma=\pm 1}\kappa_{\varsigma}^2K_\varsigma^{(1)},\\
\hat{\Sigma}^{22}&=&\hat{\Sigma}^{11}-\frac{m\omega_p^2}{2m_0}\sum_{\varsigma=\pm 1}\kappa_{\perp}^2\Big(K_\varsigma^{(0)}-2K_\varsigma^{(1)}\Big),\\
\hat{\Sigma}^{33}&=&-\frac{m\omega_p^2}{m_0}\Big(1+\frac{1}{2}\sum_{\varsigma=\pm 1}\kappa_{\parallel}^2K_\varsigma^{(0)}\Big),
\end{eqnarray}
where summation over charged species is implied. Similarly, the off-diagonal components of the response tensor are much simplified
\begin{eqnarray}
\hat{\Sigma}^{12}&=&-\hat{\Sigma}^{21}=\mp i\frac{m\omega_p^2}{2m_0}\sum_{\varsigma=\pm 1} \varsigma\kappa_{\varsigma}^2\Big(K_{\varsigma}^{(1)}-K_\varsigma^{(0)}\Big),\\
\hat{\Sigma}^{23}&=&-\hat{\Sigma}^{32}=\pm i\frac{m\omega_p^2}{2m_0}\sum_{\varsigma=\pm 1} \varsigma\kappa_{\perp}\kappa_{\parallel}\Big(K_{\varsigma}^{(1)}-K_\varsigma^{(0)}\Big),\\
\hat{\Sigma}^{31}&=&+\hat{\Sigma}^{13}=-\frac{m\omega_p^2}{2m_0}\sum_{\varsigma=\pm 1}\kappa_{\perp}\kappa_{\parallel}K_{\varsigma}^{(1)} .
\end{eqnarray}
The above six distinct components of the plasma response tensor completely describe how charged bosons, filling up the ground states in a uniform background magnetic field, interact with EM perturbations by making transitions between relativistic Landau levels.

The dispersion relation for oblique propagation at general angles can be solved numerically using the above formulas. 
Since the $K$-function is related to the confluent hypergeometric function by Eq.~(\ref{eq:Kfunction}), it can be readily evaluated by established numerical procedures.
In a single species plasma with a neutralizing background, the spectrum of the dispersion relation contains two non-degenerate EM waves hybridized with the plasma oscillation and the relativistic cyclotron resonances [Eq.~(\ref{eq:Poles})].
When two or more charged species are present, the spectrum contains additional gapped hybrid waves and gapless acoustic waves, whose low-frequency asymptotics give the magnetohydrodynamics waves, modified by relativistic quantum effects. 

\subsection{Parallel and perpendicular propagations\label{ch:mag:dispersion:special}}

The wave dispersion relations become particularly simple when the wave vector is exactly parallel ($k_\perp=0$) or perpendicular ($k_\parallel=0$) to the magnetic field. In these cases, $D_{i3}=D_{3i}=0$ for both $i=1$ and $2$, whereby the dispersion tensor $D_{ij}$ becomes very simple. Therefore, simple analytical expressions of the wave dispersion relation can be obtained at these special angles.

\subsubsection{Parallel propagation}
When waves propagate parallel to the magnetic field, namely, when $k_\perp=0$, the nonvanishing spatial components of the plasma response tensor can be written as
\begin{eqnarray}
	\hat{\Sigma}^{11}&=&\hat{\Sigma}^{22}=\omega^2(S-1),\\
	\hat{\Sigma}^{12}&=&-\hat{\Sigma}^{21}=-i\omega^2D,\\
	\hat{\Sigma}^{33}&=&\omega^2(P-1).
\end{eqnarray}
In the above expressions, $S=(R+L)/2$, $D=(R-L)/2$, and $P$ are the Stix's notations of permittivities typically used in classical plasma physics. Using these notations, the dispersion relations of the right-handed circularly polarized electromagnetic wave (R wave), the left-handed circularly polarized electromagnetic wave (L wave), and the longitudinal electrostatic wave are
\begin{equation}
	\label{eq:para_disp}
	R=n_\parallel^2,\qquad L=n_\parallel^2, \qquad P=0.
\end{equation}
where $n_\parallel=k_\parallel/\omega$ is the refractive index. Although the above dispersion relations are formally identical to those in classical plasmas, the permittivities are modified by relativistic-quantum effects.
For exact parallel propagation, the $K$-functions take special values $K_{\varsigma}^{(n)}=1/(\kappa_{\varsigma}^2-n)$. Then,
writing summations over charged species explicitly, the permittivities are
\begin{eqnarray}
	\label{eq:para_R}
	R&=&1-\sum_{s}\frac{m_s\omega_{ps}^2}{m_{s0}\omega^2}\frac{\omega^2-k_\parallel^2\mp 2m_{s0}\omega}{\omega^2-k_\parallel^2\mp 2(m_{s0}\omega+m_s\Omega_s)},\\
	\label{eq:para_L}
	L&=&1-\sum_{s}\frac{m_s\omega_{ps}^2}{m_{s0}\omega^2}\frac{\omega^2-k_\parallel^2\pm 2m_{s0}\omega}{\omega^2-k_\parallel^2\pm 2(m_{s0}\omega-m_s\Omega_s)},\\
	\label{eq:para_P}
	P&=&1-\sum_{s}\frac{m_s\omega_{ps}^2}{m_{s0}}\frac{\omega^2-k_\parallel^2-4m_{s0}^2}{(\omega^2-k_\parallel^2)^2-4m_{s0}^2\omega^2}.
\end{eqnarray} 
In the expressions of $R$ and $L$, the upper and lower sign of $\mp$ and $\pm$ correspond to $e_sB_0>0$ and $e_sB_0<0$, respectively. Since particle energy is not quantized in the direction parallel to the magnetic field, the low energy limit is the classical limit. In the classical limit $\omega,k_\parallel,|\Omega_e|\ll m_e$, it is clear that the above expressions asymptote to their classical values. Consequently, wave dispersion relations in relativistic-quantum plasmas asymptote to those in classical plasmas when magnetic fields are weak.

\begin{figure}[!b]
	\renewcommand{\figurename}{FIG.}
	\centering
	\includegraphics[angle=0,width=0.6\textwidth]{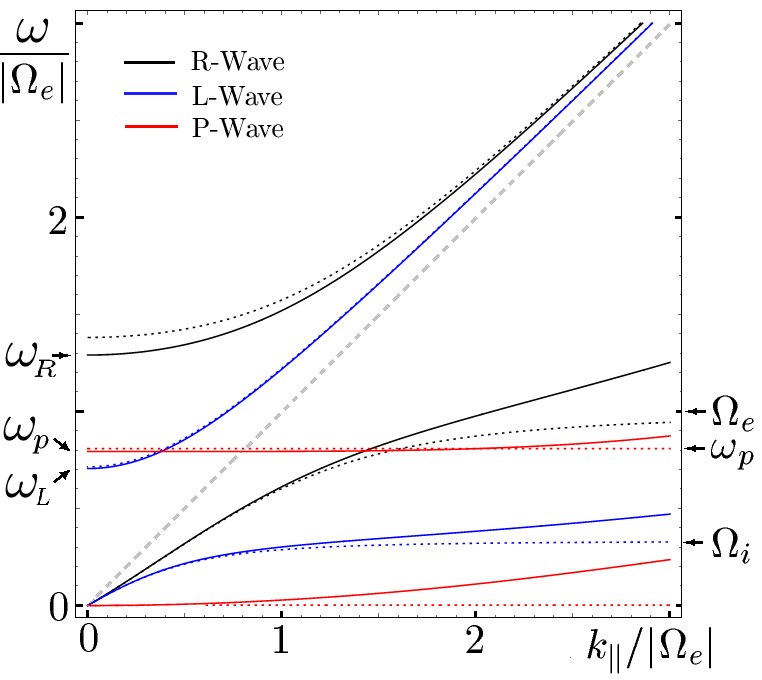}
	\caption[Parallel wave dispersion relations in a cold magnetized scalar-QED plasma]{Parallel wave dispersion relations in a cold, magnetized, quasineutral, ``electron-ion" plasma. The solid curves are waves in a relativistic-quantum plasma and the dashed curves are the corresponding waves in a classical plasma. The black and blue curves are the right- and left-handed circularly polarized electromagnetic waves, respectively. The red curves are the longitudinal electrostatic waves, which include a gapped Langmuir wave (upper) and a gapless acoustic wave (lower). The dashed gray line across the diagonal represents the light cone. 
	For various effects to be visible on the scale of this figure, parameters used for making this plot are $\omega_{pe}/|\Omega_{e}|=0.7$, $|\Omega_{e}|/m_e=0.1$ and $m_i/m_e=3$. 
	Notice that near the light cone, wave dispersion relations in the relativistic-quantum plasma asymptote to wave dispersion relations in the classical plasma. While away from the light cone, relativistic-quantum modifications become appreciable.}
	\label{fig:ParallelDispersion}
\end{figure}

An example of wave dispersion relations for parallel propagation in a quasineutral ``electron-ion" plasma is plotted in Fig.~\ref{fig:ParallelDispersion}. Only low-energy branches with $\omega,k_\parallel\ll m_e$ are plotted, for which effects of the vacuum polarization can be safely ignored. In the figure, the solid curves are wave dispersion relations in a relativistic-quantum plasma and the dashed curves are corresponding wave dispersion relations in a classical plasma. The black and blue curves are the R and L waves, respectively. The red curves are the longitudinal electrostatic waves, which include a gapped Langmuir wave and a gapless acoustic wave. For relativistic effects to be visible, the magnetic field is made strong such that $|\Omega_e|/m_e=0.1$. For ion effects to be visible, the ion mass is chosen to be close to the electron mass with $m_i/m_e=3$. The ratio of the plasma frequency to the gyrofrequency is chosen to be $\omega_{pe}/|\Omega_e|=0.7$. It is easy to see that the relativistic-quantum dispersion relations asymptote to the classical dispersion relations near the light cone. On the other hand, away from the light cone, relativistic-quantum modifications become prominent.

To get a sense of how large relativistic-quantum corrections are, we can calculate the cutoff frequencies, the wave frequencies when the wave vector $\bm{k}=0$. In a single-species plasma, the approximate cutoff frequencies in the limit $\omega_p\sim|\Omega|\ll m$ are
\begin{eqnarray}\label{eq:MCutoff}
	\frac{\omega_{R0}-\omega_R}{\omega_{R0}}&\sim&\frac{|\Omega|}{2m}\Big(1-\frac{\Omega}{\sqrt{\Omega^2+4\omega_p^2}}\Big),\\
	\frac{\omega_{L0}-\omega_L}{\omega_{L0}}&\sim&\frac{|\Omega|}{2m}\Big(1+\frac{\Omega}{\sqrt{\Omega^2+4\omega_p^2}}\Big),\\
	\frac{\omega_{P0}-\omega_P}{\omega_{P0}}&\sim&\frac{|\Omega|}{4m}.
\end{eqnarray} 
Here $\omega_{R0}$ is the cutoff frequency of the R wave in a classical plasma and $\omega_{R}$ is the cutoff frequency of the R wave in a relativistic-quantum plasma. Similar notations are used for the L wave and the longitudinal wave. As expected, relativistic-quantum effects are large when $|\Omega|\sim m$, namely, when the cyclotron energy is comparable to the rest energy of particles.

\subsubsection{Perpendicular propagation}
When waves propagate perpendicular to the magnetic field, namely, when $k_\parallel=0$, the contribution by each charged species to the nonvanishing spatial components of the plasma response tensor are
\begin{eqnarray}
\hat{\Sigma}^{11}&=&-\frac{m\omega_p^2}{2m_0}\sum_{\varsigma=\pm 1}\kappa_{\varsigma}^2K_\varsigma^{(1)},\\
\hat{\Sigma}^{22}&=&\hat{\Sigma}^{11}-\frac{m\omega_p^2}{2m_0}\sum_{\varsigma=\pm 1}\kappa_{\perp}^2(K_\varsigma^{(0)}-2K_\varsigma^{(1)}),\\
\hat{\Sigma}^{12}&=&-\hat{\Sigma}^{21}=\mp i\frac{m\omega_p^2}{2m_0}\sum_{\varsigma=\pm 1} \varsigma\kappa_{\varsigma}^2\Big(K_{\varsigma}^{(1)}-K_\varsigma^{(0)}\Big),\\
\hat{\Sigma}^{33}&=&-\frac{m\omega_p^2}{m_0}.
\end{eqnarray}
Notice that the perpendicular components are the same as in the general case, whereas the parallel component becomes simplified.

The dispersion relations can be easily read out by substituting the above nonvanishing components of the response tensor into Eq.~(\ref{eq:MagDisp}).  When the wave electric field is parallel to the background magnetic field, the wave is purely transverse. The dispersion relation of this linearly polarized ordinary electromagnetic wave (O wave) is
\begin{equation}
\omega^2=\frac{m\omega_p^2}{m_0}+k_\perp^2.
\end{equation}
This is very similar to the dispersion relation of the O wave in classical plasmas, except that the bare mass $m$ is now replaced by the ground state mass $m_0$.
On the other hand, when the wave electric field is perpendicular to the background magnetic field, relativistic-quantum modifications are less trivial. In this case, the longitudinal and transverse components of the wave are mixed by the off-diagonal components of the response tensor. Relativistic quantum cyclotron resonances [Eq.~(\ref{eq:Poles})] then hybridize with the extraordinary electromagnetic wave (X wave) by the dispersion relation
\begin{equation}\label{eq:Xwave}
(\omega^2+\hat{\Sigma}^{11})(\omega^2-k_\perp^2+\hat{\Sigma}^{22})=\hat{\Sigma}^{12}\hat{\Sigma}^{21}.
\end{equation}  
While the X wave is qualitatively captured by classical plasma theories, cyclotron resonances, also known as the Bernstein waves, are absent in classical theories when plasmas are cold \citep{Stix92}. In classical plasmas, charged particles sample wave fields along their gyro orbits. Bernstein resonances arise when the gyro frequencies match the wave frequency. If plasma temperature is zero, cyclotron motion of classical particles stops and Bernstein resonances vanish consequently. However, this is not the case when quantum effects are taken into account. Using the uncertainty principle and the fact that the kinetic momentums $\Pi_{\mu}=-i\bar{D}_{\mu}$ do not commute $[\bar{D}_{\mu},\bar{D}_{\nu}]=-ie\bar{F}_{\mu\nu}$, 
it is easy to see that the gyro motion of a quantum particle never stops. So Bernstein waves persist in a quantum plasma even when it is cold.

\begin{figure}[!t]
	\renewcommand{\figurename}{FIG.}
	\centering
	\includegraphics[angle=0,width=0.6\textwidth]{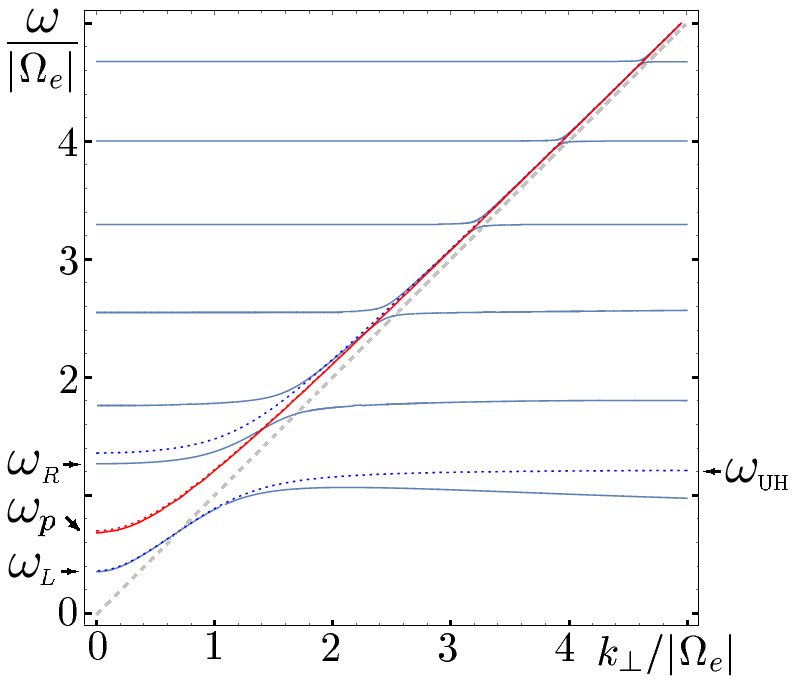}
	\caption[Perpendicular wave dispersion relations in a cold magnetized scalar-QED plasma]{Perpendicular wave dispersion relations in a magnetized cold ``electron" gas with immobile ions as neutralizing background. The solid curves are waves in a relativistic-quantum plasma and the dashed curves are corresponding waves in a classical plasma. The red curves are the ordinary electromagnetic waves. The blue curves are the extraordinary electromagnetic wave hybridized with cyclotron resonances. The dashed gray line across the diagonal represents the light cone. 
	Parameters used for making this plot are $\omega_{pe}/|\Omega_{e}|=0.7$ and $|\Omega_{e}|/m_e=0.1$. 
	Notice that near the light cone, wave dispersion relations in the relativistic-quantum plasma asymptote to wave dispersion relations in the classical plasma. While the classical dispersion relations only capture the upper-hybrid resonance $\omega_{\text{UH}}$, the relativistic-quantum dispersion relations capture all cyclotron resonances even when the plasma is cold. 
	Notice that cyclotron resonances are not harmonically spaced. The fifth resonance occurs near $4\Omega$ instead of $5\Omega$ in this example.}
	\label{fig:PerpendicularDispersion}
\end{figure}

An example of wave dispersion relations for perpendicular propagation is plotted in Fig.~\ref{fig:PerpendicularDispersion}. For the sake of clarity, the ion mass is set to infinity, such that the immobile ions merely serve as a neutralizing background. By doing so, ion cyclotron resonances, gapped hybrid waves, and gapless magnetohydrodynamics waves are removed. What remains in this single species plasma are the O wave, the X wave, and the relativistic-quantum electron-Bernstein waves. 
Only low energy branches with $\omega,k_\parallel\ll m_e$ are plotted here, for which effects of the vacuum polarization can be safely ignored. When making the plot, I choose the ratio $\omega_{pe}/|\Omega_e|=0.7$ on the order of unity, so that collective plasma effects are comparable to the magnetization effect. In addition, for relativistic-quantum effects to be clearly visible, I choose the ratio $|\Omega_e|/m_e=0.1$ not too much smaller than one.
In Fig.~\ref{fig:PerpendicularDispersion}, the solid curves are dispersion relations in a relativistic-quantum plasma, and the dashed curves are dispersion relations in a classical plasma. The solid and the dashed red curves almost overlap, since the dispersion relations of the relativistic-quantum and the classical O wave differ only in their mass gaps by the ratio $m/m_0\lesssim 1$. 
The blue curves are the X waves hybridized with relativistic cyclotron resonances. While the classical dispersion relation only captures the upper-hybrid resonance at $\omega_{\text{UH}}=\sqrt{\Omega^2+\omega_p^2}$, the quantum dispersion relation captures all the cyclotron resonances, which are present even when the plasma is cold. Notice that cyclotron resonances are not harmonically spaced due to relativistic effects. As can be seen from the figure, the fifth resonance occurs near $4\Omega$ instead of $5\Omega$ in this example.

The dispersion relations of relativistic-quantum Bernstein waves may be grossly approximated as follows. For simplicity, let us abbreviate $\omega_n:=\omega_{n,0}^{-}$, where $\omega_{n,k_\parallel}^{-}$ is the lower cyclotron resonance given by Eq.~(\ref{eq:Poles}). Using properties of the $K$-function, the asymptotic behaviors of the plasma response tensor when $\omega\sim\omega_n$ are 
\begin{eqnarray}\label{eq:CyclotronRes}
\hat{\Sigma}^{11}&\sim&-\frac{m\omega_p^2}{m_0}\Big(1+\sigma_n\frac{\kappa_{+}^2}{\kappa_{+}^2-n}\Big),\\
\hat{\Sigma}^{22}&\sim&-\frac{m\omega_p^2}{m_0}\Big(1+\sigma_n\frac{\kappa_{+}^2-\kappa_\perp^2(2-\kappa_\perp^2/n)}{\kappa_{+}^2-n}\Big),\\
\hat{\Sigma}^{12}&=&-\hat{\Sigma}^{21}\sim-i\frac{m\omega_p^2}{m_0}\sigma_n\frac{\kappa_\perp^2-\kappa_{+}^2}{\kappa_{+}^2-n},
\end{eqnarray}
where $\sigma_{n+1}=(\kappa_\perp^2)^n\exp(-\kappa_\perp^2)/n!$. A reasonable approximation of wave dispersion relations may be obtained by substituting the above expressions into Eq.~(\ref{eq:Xwave}), keeping all the even powers of $\omega$ intact, while replacing odd powers $\omega^{2l+1}\rightarrow\omega^{2l}\omega_n$, such that the asymptotic behavior $\omega\sim k_\perp$ near the light cone is respected. To lowest order, the approximate dispersion relation near the resonance $\omega_n$ is
\begin{eqnarray}\label{eq:ApproxBernstein}
\omega^2&\sim&\frac{1}{2}\Big[\Big(\omega_n^2+\xi_n^2+k_\perp^2+\frac{m\omega_p^2}{m_0}\Big)\pm\sqrt{\Big(\omega_n^2+\xi_n^2-k_\perp^2-\frac{m\omega_p^2}{m_0}\Big)^2+4\omega_n^2\xi_n^2}\hspace{4pt}\Big], \hspace{10pt}
\end{eqnarray}
where $\xi_n^2=n|\Omega|m^2\omega_p^2\sigma_n/2\omega_nm_0^2$ is a function of $k_\perp$. 
The ``$+$" branch emanates from $\omega_n$ and asymptotes to the light cone, while the ``$-$"branch emanates from the cutoff of the O wave and asymptotes to the cyclotron resonance $\omega_n$. Of course, the above approximation is only valid near $\omega\sim\omega_n$, where the gaps between branches of relativistic-quantum Bernstein waves are controlled by the factor $\omega_n\xi_n$. Notice that the gaps remain open even when the plasma is cold.

\section{Observable consequences\label{ch:mag:observables}}

From the above discussion of wave dispersion relations, we see relativistic quantum modifications are most prominent away from the light cone. This is expected, because away from the light cone $\omega^2=\mathbf{k}^2$, gauge bosons become massive particles. When massive gauge bosons interact with charged particles, the recoil momentum ignored in classical calculations becomes important. 
There are two directions the dispersion curve can move away from the light cone. First, it can move inside the light cone, whereby wave behaviors near cutoffs are modified. Modifications of this type can be observed, for example, using Faraday rotation of linearly polarized EM waves, as we shall see in Sec.~\ref{ch:mag:observables:para}. 
Alternatively, the dispersion curve can move outside the light cone, whereby the behaviors near resonances are modified. Apart from exciting resonances directly, the modifications can be seen from passive absorptions. In fact, as we shall see in Sec.~\ref{ch:mag:observables:perp}, modified absorptive behaviors have already been observed in spectra of X-ray pulsars.

\subsection{Modifications of Faraday rotation in gigagauss fields\label{ch:mag:observables:para}}

Since the R wave and the L wave of the same frequency have different phase velocities, when they combine to form a linearly polarized wave, the wave polarization vector rotates at a rate
\begin{equation}
\lambda\frac{d\theta}{dz}=\pi\Delta n.
\end{equation}
Here, $\theta$ is the polarization angle, $z$ is the distance of propagation along the magnetic field, $\lambda=2\pi c/\omega$ is the vacuum wavelength, and $\Delta n=n_L-n_R$ is the difference in refractive indexes between the L wave and the R wave of the same frequency. 
In electron-positron plasmas with charge-conjugation symmetry, Faraday rotation remains identically zero as in the classical case.
On the other hand, once charge conjugation symmetry is broken, so is the joint parity and time-reversal symmetry broken, whereby Faraday rotation happens. 
For example, in an electron-ion plasma, since $m_i\gg m_e$, the dominant contribution comes from electrons. Keeping only electron terms in the dispersion relations Eqs.~(\ref{eq:para_disp}), and using the relativistic-quantum permittivities Eqs.~(\ref{eq:para_R}) and (\ref{eq:para_L}), the refractive indexes
\begin{eqnarray}
\label{eq:nRL}
n_{R/L}^2&=&1-\frac{m\Omega}{\omega^2}-\frac{m\omega_p^2}{2m_0\omega^2}\mp\frac{m_0}{\omega} \pm\sqrt{\Big(\frac{m\Omega}{\omega^2}+\frac{m\omega_p^2}{2m_0\omega^2}\pm\frac{m_0}{\omega}\Big)^2\mp\frac{2m\omega_p^2}{\omega^3}}, \hspace{10pt}
\end{eqnarray}
where the upper signs correspond to the R wave and the lower signs correspond to the L wave. It is straightforward to check that in the classical limit $\omega,\omega_p,|\Omega|\ll m$, the above formulas recover the classical results. For waves of given frequency, the phase velocity of the R wave is decreased by a larger amount than the phase velocity of the L wave due to relativistic quantum effects. Consequently, Faraday rotation is reduced in strongly magnetized relativistic quantum plasmas.
 
\begin{figure}[!b]
	\centering
	\includegraphics[angle=0,width=0.55\textwidth]{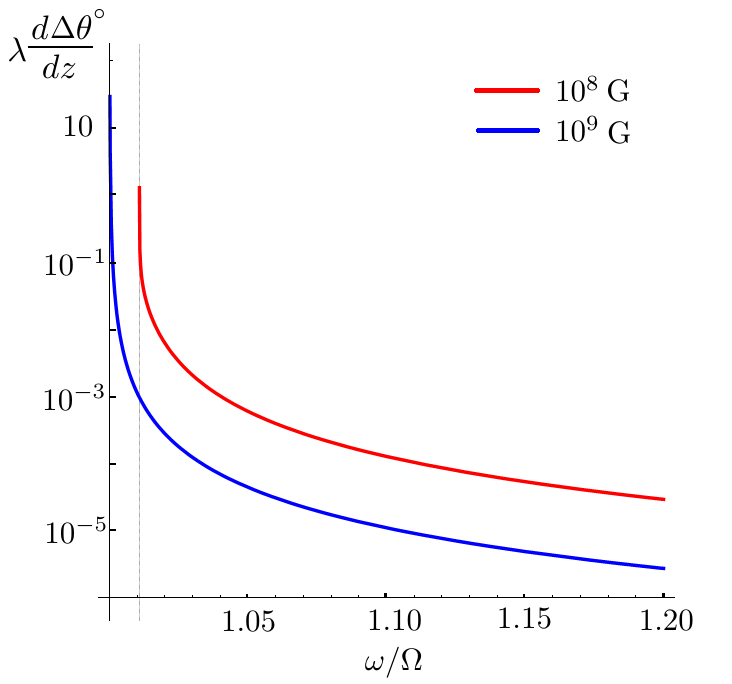}
	\caption[Deviation of Faraday rotation from classical prediction]{Deficiency of Faraday rotation from the classical prediction can be used to measure relativistic-quantum corrections. In a gas jet plasma with density $n_e=10^{19}\, \text{cm}^{-3}$, a 0.1 GG magnetic field (red) leads to a deviation $\Delta\theta$ of $\sim 1^\circ$ after the laser, whose frequency is near the R-wave cutoff, propagates by a vacuum wavelength $\lambda$. In a stronger 1 GG magnetic field (blue), a deviation as large as $\sim 10^\circ/\lambda$ may be observed using a laser whose frequency is slightly above the cutoff. Notice that the deviations fall precipitously when the laser frequency is above the classical cutoff. Therefore, for relativistic-quantum effects to be measurable, the laser frequency must be sufficiently close to the R-wave cutoff.}
	\label{fig:Faraday}
\end{figure}

Although relativistic-quantum modifications remain small in gigagauss magnetic fields, they are boosted near the cutoff frequency of the R wave, where Faraday rotation is maximized. Suppose we measure Faraday rotation by passing multiple linearly-polarized lasers of slightly different frequencies through the same plasma, then the relativistic-quantum formula predicts a different frequency dependence than expected classically. To see the difference, one can subtract measured data from the classical prediction, and plot the discrepancy $\Delta\theta$ as a function of the laser frequency (Fig.~\ref{fig:Faraday}). 
For example, in a gas jet plasma with density $n_e=10^{19}\, \text{cm}^{-3}$, a magnetic field $B_0=10^{8}$ G results in a difference of $\sim 1^\circ/\lambda$ when the laser frequency approaches the R-wave cutoff  $\sim 1.16$ eV (red line). This discrepancy can be resolved if the measurement uncertainty is $\lesssim1.5\%$ at the classical cutoff, and $\lesssim 15$ ppm at $\sim0.1$ eV above the cutoff. 
In a stronger magnetic field $B_0=10^{9}$ G, the difference is as large as $\sim 10^\circ/\lambda$ near the cutoff $\sim 11.5$ eV (blue line). This discrepancy can be resolved if measurement uncertainty is $\lesssim67\%$ at the classical cutoff, and $\lesssim0.13\%$ at $\sim0.1$ eV above the cutoff.
While corrections introduced by a 0.1 GG magnetic field is unlikely to be measurable, much larger corrections introduced by gigagauss magnetic fields might be discernible from noise and inhomogeneities.

\begin{figure}[t]
	\renewcommand{\figurename}{FIG.}
	\centering
	\includegraphics[angle=0,width=0.6\textwidth]{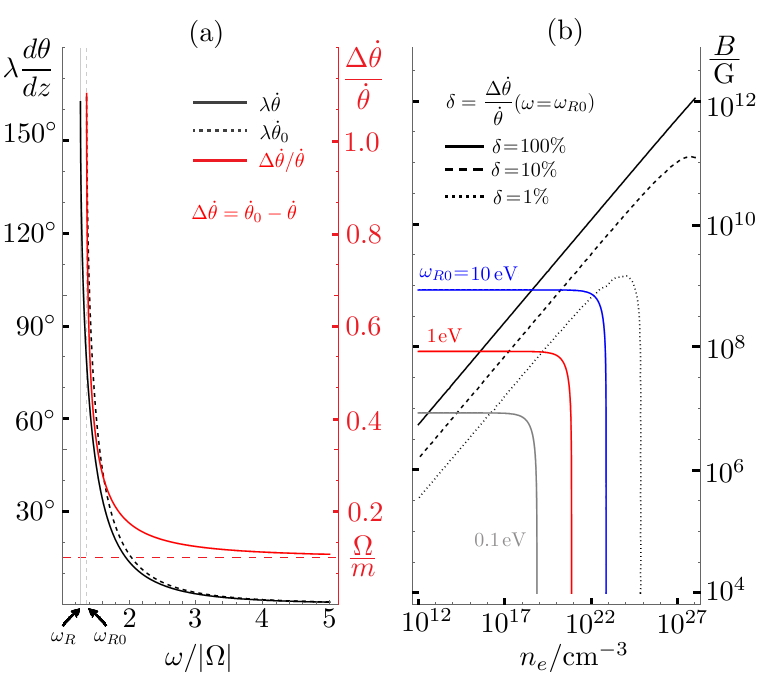}
	\caption[Faraday rotation modified by QED effects]{(a) Faraday rotation per vacuum wavelength in a cold magnetized ``electron" gas. The solid and dashed black curves are Faraday rotations in a relativistic quantum plasma and a classical plasma, respectively. The red curve is their relative difference. Parameters used for making this plot are $\omega_{pe}/|\Omega_{e}|=0.7$ and $|\Omega_{e}|/m_e=0.1$. Notice that near the classical cutoff $\omega_{R0}$, Faraday rotations in the relativistic quantum plasma and the classical plasma differ significantly. (b) Region in the $n_e$--$B_0$ space where relativistic quantum corrections are important. The regions above the solid, large-dashed and small-dashed black contours are regions where $\delta>100\%$, $ 10\%$ and $1\%$, respectively. The blue, red and gray contours are where the classical cutoff $\omega_{R0}=10$ eV, $1$ eV and $0.1$ eV, respectively. These two sets of contours combined can be used to determine how important relativistic quantum corrections are in given conditions. }
	\label{fig:FaradayRotation}
\end{figure}

Generally speaking, relativistic quantum modifications to Faraday rotation are important when magnetic field is strong and density is low. 
A comparison between Faraday rotations in a relativistic quantum plasma and a classical plasma is plotted in Fig.~\ref{fig:FaradayRotation}(a), for parameters $\omega_{pe}/|\Omega_{e}|=0.7$ and $|\Omega_{e}|/m_e=0.1$. In the figure, the left axis is Faraday rotation per vacuum wavelength. The solid black curve is the Faraday rotation $\lambda\dot{\theta}$ in a relativistic quantum plasma and the dashed black curve is the Faraday rotation $\lambda\dot{\theta}_0$ in a classical plasma. The right axis of the figure is the relative difference $\dot{\theta}_0/\dot{\theta}-1$. As can be seen from the figure, while the relative difference asymptotes to a small number $|\Omega_e|/m_e$ when $\omega\gg|\Omega_e|$, it can be of order $1$ near the classical cutoff of the R wave. 
Denoting $\delta$ the relative difference at $\omega=\omega_{R0}$. The region in the $n_e$--$B_0$ space where $\delta$ is of order 1 is plotted in Fig. \ref{fig:FaradayRotation}(b). In the figure, the horizontal axis is the density of the electron gas and the vertical axis is the strength of the magnetic field. The region above the solid black contour is where $\delta>100\%$, the region above the large-dashed black contour is where $\delta>10\%$, and the region above the small-dashed black contour is where $\delta>1\%$. To facilitate reading of the figure, contours of $\omega_{R0}$ are also plotted. The blue contour is where $\omega_{R0}=10$ eV, the red contour is where $\omega_{R0}=1$ eV, and the gray contour is where $\omega_{R0}=0.1$ eV. The contours of $\delta$ and $\omega_{R0}$ combined can be used to determined how important relativistic quantum corrections are in a given situation. For example, the small-dashed black contour and the red contour intersect around $n_e\sim10^{19}\,\text{cm}^{-3}$ and $B\sim10^{8}$ G. This means if laser with photon energy $\hbar\omega\sim 1$ eV is used to diagnose such a plasma, then ignoring relativistic quantum effects will introduce $\sim1\%$ systematic error.

In laser plasma experiments, when lasers with frequencies close to classical cutoffs are used for diagnostics, relativistic quantum corrections of wave dispersion relations need to be taken into account in order to avoid systematic errors. As can be seen from Fig. \ref{fig:FaradayRotation}(a), if one tries to match data points on the relativistic quantum curve by shifting the classical curve, then $|\Omega_e|$ will have to be smaller than its true value, resulting in systematic errors. As the frequency of the diagnostic laser increases, the inferred magnetic field strength approaches its true value from below. This is why the inferred magnetic field appears to increase with the frequency of the diagnostic laser when classical formulas are used. In experiments conducted by \cite{Tatarakis02, Wagner04}, the magnetic field strength is determined from Cotton-Mouton effect, which depends on frequencies of cutoffs just as the Faraday rotation does. It is beyond the scope of this thesis to analyze their experimental details, but the peculiar dependence of the inferred magnetic field strength on the frequencies of diagnostic lasers can already be understood qualitatively as a consequence of relativistic quantum modifications of cutoff frequencies.

\subsection{Anharmonic cyclotron absorptions in X-ray pulsar spectra\label{ch:mag:observables:perp}}

Even stronger magnetic fields can be found near neutron stars, where relativistic quantum effects become more prominent. Since we can only passively observe these stars, relativistic quantum effects shows up in their spectra. In particular, anharmonic cylotron absorption features have been observed using a number of X-ray telescopes, such as \textit{Ginga} \citep{Makishima90}, \textit{BeppoSAX} \citep{Santangelo99}, \textit{RXTE} \citep{Heindl99,Heindl00,Pottschmidt05}, INTEGRAL \citep{Tsygankov06,Tsygankov07,Boldin13}, and \textit{Suzaku} \citep{Pottschmidt2012,Jaisawal15}. 
These anharmonic cyclotron absorption features are observed for accretion powered X-ray pulsars. These neutron stars orbit in close proximity of their companion stars, from which plasmas are accreted. The accreted plasmas flow along the magnetic fields $\sim10^{12}$ G to the polar regions of the neutron stars, where charged particles accelerate and radiate when falling into the deep gravitational potential of the neutron stars.

The observed cyclotron absorption lines are believed to form in some localized regions, otherwise inhomogeneities of neutron stars' dipole magnetic fields would have wiped out the line features.
If the plasma density is low, particles free fall to the surface of the neutron stars and form plasma mounds, which can subsequently collapse to give off localized radiations.
When the plasma density is higher, before reaching the surfaces of neutron stars, the converging plasma flows can create hydrodynamic shocks, where localized radiations are emitted. 
Moreover, in a number of neutron stars, the radiations are so intense that the Eddington limit appears to have been exceeded. In this case, the radiation pressure can balance the plasma pressure, whereby radiation shocks are formed in the plasma falls. 
From these localized layers, photons escape the plasma columns mostly from their sides. Therefore, X rays propagate nearly perpendicular to magnetic fields, passing through colder plasma layers where absorptions occur. 
Although a consensus regarding the exact mechanisms of cyclotron line formation have not been reached, it is widely accepted that the absorption features are due to transitions between relativistic Landau levels.

\begin{figure}[t]
	\centering
	\renewcommand{\figurename}{FIG.}
	\includegraphics[angle=0,width=0.8\textwidth]{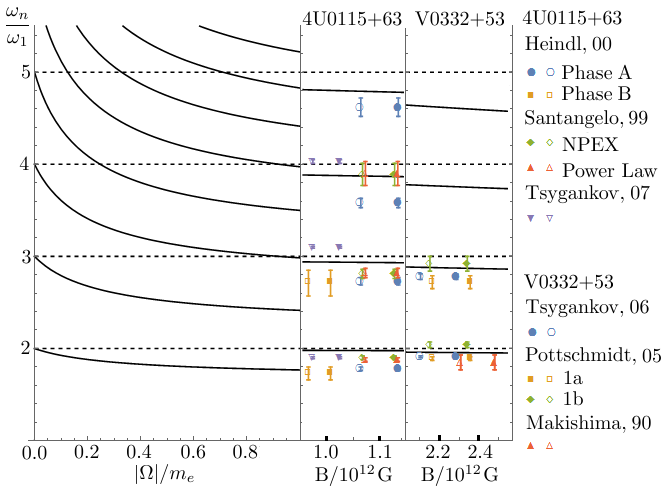}
	\caption[Anharmonic cyclotron absorptions due to QED effects]{Ratios of the frequencies of cyclotron harmonics to the frequency of the fundamental. The solid curves are ratios when relativistic quantum effects are taken into account, and the dashed curves are the classical ratios. Quantum effects sustain cyclotron resonances even when the plasma is cold. Relativistic effects space resonances anharmonically even when the magnetic field is uniform. Notice that the anharmonicity is larger for stronger magnetic field and higher harmonic order.
	In neutron star magnetosphere, the plasma temperature is low $k_BT\sim\Omega\hbar/4$ when compared to the magnetic field, and cyclotron absorptions occur when X-ray photons leave plasma columns perpendicular to the magnetic field in a localized layer. The observed absorption lines (colored symbols) qualitatively agree with the relativistic quantum expectation (black lines) that the $n$-th harmonic occurs at $\omega_n<n\omega_1$ in the strong magnetic field inferred by fitting all cyclotron lines.	
	}
	\label{fig:HarmonicRatio}
\end{figure}

Due to relativistic effects, cyclotron resonances in strongly magnetized plasmas are anharmonically spaced. In uniform magnetic fields, the anharmonicity is due to relativistic effect, which redshifts the cyclotron resonance $\omega_{n}$ from its classical value $n|\Omega|$. The frequencies of line centers are given by $\omega_{n,k_\parallel}^{-}$ in Eq.~(\ref{eq:Poles}). 
The relativistic redshift is significant when either the magnetic field is strong or the cyclotron order $n$ is large. More specifically, the redshift is comparable to the gyrofrequency, namely $n|\Omega|-\omega_n\gtrsim|\Omega|$, when the order
\begin{equation}\label{eq:CycloRedShift}
n\gtrsim\sqrt{\frac{2m}{|\Omega|}}=\sqrt{\frac{2m^2c^2}{eB_0\hbar}}\approx 9.4\times\sqrt{\frac{10^{12}\hspace{3pt}\text{G}}{B_0}}.
\end{equation}
In other words, for X-ray pulsars with $B_0\sim10^{12}$ G, the redshift is of order unity from the ninth resonance.
The line-averaged magnetic field may be determined by fitting the center of absorption lines to Eq.~(\ref{eq:Poles}).
The ratio of the frequency of cyclotron harmonics to the frequency of the fundamental is plotted in Fig.~\ref{fig:HarmonicRatio}. In the figure, the solid curves are ratios when relativistic quantum effects are taken into account, and the dashed lines are classical ratios.
Using expression of $\omega_n^{-}$, it is easy to see when in the weak field limit $|\Omega|\ll m$, cyclotron resonances are harmonically spaced $\omega_n\sim n\omega_1$. While in the strong field limit $|\Omega|\gg m$, cyclotron resonances are anharmonically spaced with $\omega_n\sim \sqrt{n}\omega_1$. The anharmonic line ratios for pulsars 4U011+63 and V0332+53, where more than two cyclotron lines have been observed, are also plotted in Fig.~\ref{fig:HarmonicRatio}. Although data points (colored symbols) are somewhat scattered, the qualitative feature that the $n$-th harmonic occurs at $\omega_n<n\omega_1$ roughly agrees with the relativistic quantum expectation. 

While the line center contains information regarding the magnetic field, the line shape contains information regarding the plasmas. 
When the plasma density is high, transitions between relativistic Landau levels result in collective plasma responses as we have seen in Sec.~\ref{sec:mag:polarization:plasma}. The plasma response modifies the dispersion relation of EM waves, and thereof affects how long it takes for X-ray photons to leave the plasma column. 
Moreover, the plasma response dresses charged particles, so that the absorption cross section is modified by the spectral density function, which affects the line shape of cyclotron absorptions.
Roughly speaking, the width and depth of the absorption lines are correlated with the gaps between branches of Bernstein waves. As can be seen from Fig.~\ref{fig:PerpendicularDispersion}, lower Bernstein branches have larger gaps, resulting in wider absorption lines with larger optical depth. Quantitatively, when plasma density and magnetic field profile are known, the absorption line shapes can be calculated by solving the radiative transfer equations \citep{Meszaros92}, in which photons advection is governed by the dispersion relation, and the absorption cross sections are dressed by collective plasma effects. Conversely, when the absorption line shapes are measured, the plasma and magnetic field profile can be retrieved by solving the inverse problem. Thus, a new era in astrophysics has been opened, in which it is possible to measure the profile of the magnetosphere of an X-ray pulsar while it is accreting materials from its companion star.


\chapter{Plasma simulations using real-time lattice scalar QED\label{ch:simulation}}

In previous chapters, I develop an analytical theory for waves in scalar-QED plasmas. Although the formulation is applicable in the most general cases, to study scenarios where obtaining analytical expression is not practical, we will need simulation schemes that can solve the problem numerically. 
In this chapter, I will develop such a simulation scheme \citep{Shi2018lattice}, by exploiting the fact that tree-level effects dominate loop effects especially when plasmas are present. In the classical-statistical regime, statistical fluctuations dominate quantum fluctuations, and the path integral is dominated by the solution to the classical field equations. The usual lattice QED, which relies on numerical path integrals so that both statistical and quantum fluctuations are captured, can then be simplified to real-time simulations, which retains only the dominant statistical fluctuations. 
Solving the classical field equations in real time is trivial in the usual lattice QED, because the classical fields for the vacuum are simply zero. However, when background fields are present, solutions to the classical field equations already contain rich physics, as we shall see in this chapter.

Real-time lattice QED, which has been used to study strong-field effects in the vacuum, also provides a unique tool for simulating plasmas in the strong-field regime, where collective plasma scales are not well-separated from relativistic-quantum scales. As a toy model, I again focus on scalar QED. 
To solve this model on a computer, I first discretize the action on a spacetime lattice, in a way that respects both the geometric structures of exterior calculus and the U(1)-gauge symmetry. The lattice scalar QED can then be solved, in the classical-statistical regime, by advancing an ensemble of statistically equivalent initial conditions in time. The initial ensemble, which may or may not be a thermal ensemble, is necessary because the exact field configurations cannot be determined uniquely, when only the statistical properties of the initial state are known.
For each realization of the initial condition, the time advance is achieved using classical field equations obtained by extremizing the discrete action.
The numerical scheme I will use is fully explicit and respects local conservation laws, making it efficient and reliable for long-time dynamics. The algorithm is readily parallelized using domain decomposition on modern supercomputers. Moreover, the ensemble may be computed efficiently using quantum parallelism in the future. 
Having advanced the field configurations in time, an statistical observable at a future time can then be computed from its ensemble average. Using this numerical procedure, the accuracy of the observable is expected to decay when it involves higher order correlation functions, because the real-time simulation only captures classical field effects, whose dominance deteriorate in higher order correlation functions.

To demonstrate the capability of the numerical scheme, we apply it to two example problems. The first example is the propagation of linear waves, where analytic wave dispersion relations are recovered using numerical power spectra. The second example is an intense laser interacting with a one-dimensional plasma slab, where natural transition from wakefield acceleration to pair production when the laser amplitude exceeds the Schwinger threshold is demonstrated for the first time.

\section{Simulations beyond classical schemes: lattice QED\label{sec:simulation:latticeQED}}

While lattice simulations may be unfamiliar for plasma physics, they have been used extensively in quantum chromodynamics (QCD) to model the strong interaction, which binds the nucleus \citep{Wilson74} and mediate interactions in quark-gluon plasmas formed during heavy-ion collisions and the Big Bang \citep{Bass99,Satz00}. In conventional lattice-QCD simulations, quantum correlation functions are computed using numerical path integrals, from which observables are extracted as coefficients of scaling laws \citep{Creutz80}. This scheme can be analytically continued to imaginary time to describe statistical systems in thermal equilibrium \citep{Yagi05}. 
For out-of-equilibrium systems, real-time simulations can be carried out using the Schwinger-Keldysh time contours. The above formulations, based on numerical path integrals, are capable of capturing genuine quantum loop effects, but are numerically expensive. 
Fortunately, the computational cost can be dramatically reduced when the occupation numbers of quantum states are high and when the coupling is weak. This is precisely the case for plasma physics, where a large number of particles are present, and the coupling coefficient $e\approx0.3$ is small. In this classical-statistical regime, statistical fluctuations dominate quantum fluctuations \citep{Aarts02,Mueller04,Jeon05,Berges07,Berges14}, and the quantum system can be adequately described by time-advancing the classical field equations with an ensemble of statistically equivalent initial conditions \citep{Aarts99,Polkovnikov03,Borsanyi09,Gelis13}. Based on this approach, lattice spinor-QED simulations have been carried out to demonstrate production of fermion pairs from the vacuum by self-consistent background electric fields \citep{Hebenstreit13,Hebenstreit13a,Kasper14}. However, the role plasmas is usually not considered in lattice simulations, where attention is unnecessarily restricted to fluctuations on the vacuum background.

By incorporating a nonperturbative amount of background particle fields, real-time lattice simulations can be turned into numerical tools useful for plasma physics, 
especially when plasmas are dense or when fields are strong. Under these extreme conditions where collective QED effects are important, the commonly adopted classical plasma kinetic model is no longer sufficient. An example is the production of electron-positron pairs when intense lasers interact with plasma targets \citep{Liang98,Gahn00,Liang15,Sarri15}. To describe such phenomena in the classical framework, source terms must be inserted into kinetic or fluid equations \citep{Berezhiani92,Kluger98,Schmidt98,Roberts02,Hebenstreit10}, which can then be solved by numeric integration \citep{Hebenstreit08,Hebenstreit09} or QED particle-in-cell simulations \citep{Duclous11,Nerush11,Ridgers12}. However, prefabricated source terms take little account of the interplay between coexisting processes \citep{Schutzhold08}, which may interfere quantum mechanically. While classical approximations may be applicable when scales are well separated, large source-term errors are expected when fields, such as those of X-ray lasers, evolve on scales comparable to intrinsic QED scales. Moreover, in classical treatments, there is no obvious way to conserve both energy and momentum, when strong fields produce pairs and when particles radiate high-energy photons. Although errors may be tolerable in some cases, disrespecting energy-momentum conservation will likely have nonphysical consequences. Therefore, lattice QED is in fact an indispensable tool when relativistic-quantum and collective effects are both important.

In the following sections, I will develop an algorithm for solving the Klein-Gordon-Maxwell's equations [Eqs.~(\ref{EOM}) and (\ref{EOMA})], which can be used to model behaviors of scalar-QED plasmas as classical fields. A variational algorithm is derived by first discretizing the scalar-QED action [Eq.~(\ref{eq:action})] in a way that respects the U(1)-gauge symmetry. 
The finite difference equations can then be obtained by taking variations of the discretized action with respect to the discrete fields. The resultant discrete classical equations of motion guarantee that the Bianchi identities, namely, $\nabla\cdot\mathbf{B}=0$ and the Faraday's law, are automatically and exactly satisfied. 
The remaining equations of motions are the discrete Gauss's law, 
which can be used to initialize the simulation; the discrete Klein-Gordon (KG) equation, 
which can be used to advance the charged field; and the discrete Maxwell-Amp\`ere's law, 
which can be used to advance the gauge field. 
After fixing a gauge, explicit schemes for advancing the discrete fields in time can be constructed. 
The variational scheme respects local symmetries and conservation laws, 
and can be easily parallelized using domain decomposition. Moreover, such a numerical scheme can be inherently mimicked by quantum systems with local couplings \citep{Wiese13,Martinez16}, which can be efficiently realized using quantum parallelism \citep{Feynman86,Lloyd96} in the future.

Before going into details, let me first point out 
a number of advantages of the real-time lattice QED scheme, in comparison with conventional methods for simulating plasmas. The two conventional methods that can fully simulate kinetic effects are the particle-in-cell (PIC) scheme and the Vlasov scheme. 
The PIC scheme represents point particles in the continuum and EM fields on a grid. Particles feel EM fields through interpolations, and EM fields feel particles through depositions. Using proper smoothing functions, these two steps can preserve gauge symmetry and symplectic structures, thereby respect local conservation properties when used in geometrical algorithms \citep{Squire12,Xiao13,Xiao15,Qin15}. Nevertheless, interpolation and deposition introduce artificial collisions that are absent in physical systems.
In the alternative Vlasov scheme, EM fields are represented on the three-dimensional space, while particles are represented in the six-dimensional phase space. Particles are directly forced by fields on spatial grids, while fields feel particles though velocity space integrals, which requires resolving three extra dimensions with substantial computational cost. 
In contrast, the lattice QED scheme represents both particles and EM fields on the same grid. Therefore, there is no need for interpolations and depositions as in the case of the PIC scheme, nor is there need for resolving extra velocity space dimensions as in the case of the Vlasov scheme. 
By folding the phase space dynamics of charged particles into the complex plane, lattice QED enables the modeling of relativistic and quantum dynamics in regimes where classical treatments are not applicable.


Of course, the advantages of the real-time lattice QED plasma simulations come at an expense. The expanse comes from the necessity of resolving the relativistic-quantum scales, which can be much smaller than scales that classical plasma physics typically deals with. The coarsest resolution needed in relativistic-quantum plasma simulations is determined by the lowest energy scale of the problem, which is the rest mass of electrons $\sim0.5$ MeV, corresponding to time scale of \mbox{$\sim10^{-21}$ s}, and spatial scale of $\sim 10^{-12}$ m. 
This resolution requirement can be seen from the discrete KG equation, 
in which we must have $m\Delta t\ll 1$ in order for $\delta\phi\ll\phi$. Moreover, since we are solving a system of hyperbolic partial differential equations, 
the Courant--Friedrichs--Lewy (CFL) condition $\Delta t<\Delta x$ must be satisfied, in order for the numerical scheme to be stable. 
Finally, it is worth noting that high resolution is required for large gauge fields. Since the gauge field will appears through the Wilson's lines [Eq.~(\ref{eq:comparator})] in complex exponentials, 
the discrete theory is invariant under the gauge transformation $A\rightarrow A+2\pi/(e\Delta)$. Therefore, 
the discrete gauge field lives on the torus $\mathbb{T}^{1,3}$, which has a very different topology than $\mathbb{R}^{1,3}$. Consequently, the step size must be small enough in order to avoid exciting topological modes that are absent in the continuous theory.

These stringent resolution requirements make lattice plasma simulations excessively expensive for some problems. For example, to simulate $\sim1$-$\mu$m lasers interacting with plasmas, at least $\sim10^6$ grid points are needed in each dimension. In such cases where QED scales are well-separated from classical plasma physics scales, schemes based on semiclassical approximations may be more suitable. However, in other plasma physics problems, the lack of scale separation renders semiclassical approximations invalid. For example, to simulate $\sim50$-keV free-electron lasers interacting with plasmas, the Compton wavelength of electrons is only $\sim1/10$ of the laser wavelength. In such cases where relativistic-quantum scales overlap with plasma physics scales, real-time QED plasma simulations are indispensable.   



\section{Variational algorithm\label{sec:simulation:variation}}

Instead of discretizing classical field equations directly, a better approach is to discretize the action, and then extremize the discrete action to obtain finite difference equations. Algorithms derived in this way are called variational algorithms, which are known to have good conservation properties, by inheriting as many symmetries as possible from the original action.
In fact, a variational algorithm for solving the KGM equations has already been developed in the numerical analysis community \citep{Christiansen11}, which shows superior charge conservation property when gauge symmetry is respected. In what follows, I will rederive the variational algorithm in arbitrary gauge, using local energy conservation to justify the choice of Yee-type action \citep{Yee66} over Wilson-type action \citep{Wilson74}, and emphasize on the application of such an algorithm to plasma physics.

\subsection{Discretization on spacetime manifold\label{sec:simulation:variation:discret}}

To solve the continuous system numerically, let us discretize the spacetime manifold. For convenience, here I will use a rectangular lattice, keeping in mind that other lattices, such as triangular latices, are also viable. The classical scalar field $\phi_0$, namely a 0-form in the language of differential geometry, naturally lives on the vertexes of the discrete manifold
\begin{equation}
	\phi^{n}_{i,j,k}:=\phi_0(t_n,x_i,y_j,z_k),
\end{equation} 
where $(t_n,x_i,y_j,z_k)$ is the coordinate of the vertex. 
Notice that here the classical field $\phi_0$ is treated as a simple function, instead of the half density operator $\sqrt{\rho}$ as in Eq.~(\ref{eq:phi0}), which is related to the many-body wave function. Such a treatment is not valid in general. However, in the classical-statistic regime, where the coupling $e\ll1$ is small and the occupation number $\langle\phi\phi^\dagger\rangle\gtrsim O(1/e^2)$ is large, replacing the many-body wave function by the classical field already captures the dominate behavior of the quantum field. 
One way of seeing this is by comparing the Schwinger-Keldysh's closed time path (CTP) formulation of nonequilibrium quantum fields, with the Martin-–Siggia-–Rose (MSR) formulation \citep{Martin1973statistical} of nonequilibrium classical fields. The Schwinger-Dyson's equations satisfied by the quantum field and the classical field are formally identical, except that the quantum field encounters additional interaction vertexes that are higher order in $\hbar$ \citep{Cooper2001classical,Blagoev2001schwinger}. In other words, a classical vertex is larger than a quantum vertex by a factor of the background occupation number. Therefore, in the regime where the occupation number is large, statistical fluctuations dominate quantum fluctuations, which can be neglected to the lowest order. In this regime, a reasonably good approximation can already be obtained by treating the classical field $\phi_0$ as a simple function.

\begin{figure}[!t]
	\renewcommand{\figurename}{FIG.}
	\centering
	\includegraphics[angle=0,width=0.55\textwidth]{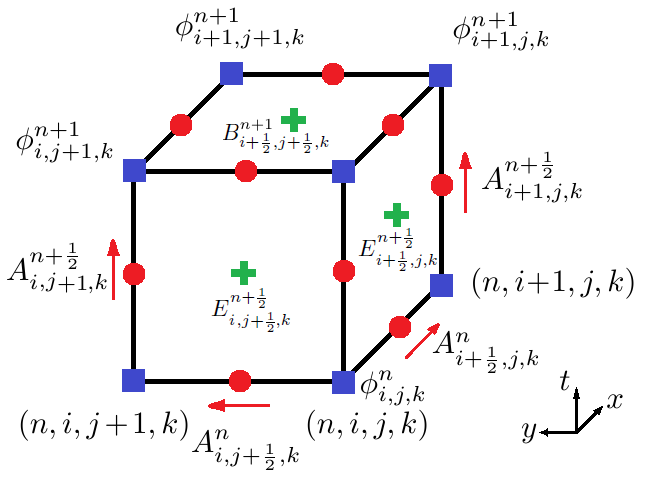}
	\caption[Discretization of scalar-QED on spacetime lattice]{Discretization of the $txy$ submanifold of spacetime using a rectangular lattice. The discrete function $\phi_v$ lives on the vertexes (blue squares). For example, $\phi^{n}_{i,j,k}=\bar{\phi}_0(t_n,x_i,y_j,z_k)$ lives on the vertex $(n,i,j,k)$. 
	The discrete 1-form $A_e$ lives along edges (red circles). For example, the $t$ component $A_{i+1,j,k}^{n+1/2}=\bar{A}^{0}(t_n+\Delta t/2,x_{i+1},y_j,z_k)$ lives along the timelike edge connecting vertexes $(n,i+1,j,k)$ and $(n+1,i+1,j,k)$, and the $x$ component $A_{i+1/2,j,k}^{n}=-\bar{A}^{1}(t_n,x_i+\Delta x/2,y_j,z_k)$ lives along the spacelike edge connecting vertexes $(n,i,j,k)$ and $(n,i+1,j,k)$. 
	The discrete 2-form $F_f$ lives on faces (green crosses). For example, electric field $E_{i+1/2,j,k}^{n+1/2}=E_x(t_n+\Delta t/2,x_i+\Delta x/2,y_j,z_k)$ lives on the timelike face spanned by vertexes $(n,i,j,k), (n+1,i,j,k), (n+1,i+1,j,k)$ and $(n,i+1,j,k)$; magnetic field $B_{i+1/2,j+1/2,k}^{n+1}=B_z(t_{n+1},x_i+\Delta x/2,y_j+\Delta y/2,z_k)$ lives on the spacelike face spanned by vertexes $(n+1,i,j,k), (n+1,i,j+1,k), (n+1,i+1,j+1,k)$ and $(n+1,i+1,j,k)$.
	}
	\label{fig:FieldDisc}
\end{figure}

Similar to the particle field, the gauge field can be well approximated by a simple 1-form in the classical-statistics regime. Upon discretization, the classical gauge 1-form $\bar{A}=\bar{A}_\mu dx^{\mu}$ naturally lives along the edges of the discrete spacetime manifold. For example, the $t$ and $x$ components
\begin{eqnarray}
	A_{i,j,k}^{n+\frac{1}{2}}&:=&+\bar{A}^{0}(t_n+\frac{\Delta t}{2},x_i,y_j,z_k),\\
	A_{i+\frac{1}{2},j,k}^{n}&:=&-\bar{A}^{1}(t_n,x_i+\frac{\Delta x}{2},y_j,z_k),
\end{eqnarray}
where $\Delta t=t_{n+1}-t_n$ and $\Delta x=x_{i+1}-x_i$. The minus sign comes from the Minkowski metric $g_{\mu\nu}$, which lowers the index $\bar{A}_\mu=g_{\mu\nu}\bar{A}^{\nu}$. In the above discretization, a half-integer index indicates which edge does the field resides along. For example, $A_{i,j,k}^{n+1/2}$ resides along the edge connecting vertices $(t_n,x_i,y_j,z_k)$ and $(t_{n+1},x_i,y_j,z_k)$, and is therefore the $\bar{A}_0$ component of $\bar{A}$. 
Notice that since $\bar{A}$ is a 1-form living along edges, only one of its four indexes can take half-integer values, while the other three indexes must take integer values. Moreover, to each edge of the lattice, the discrete 1-form only assigns the component of $\bar{A}$ that is parallel to this edge (Fig.~\ref{fig:FieldDisc}), to which other components of $\bar{A}$ are not assigned.  

Having discretized the fields, the gauge-covariant derivatives can be computed using the Wilson's lines [Eq.~(\ref{eq:comparator})]. Since the covariant derivatives are 1-forms, they also lives along edges when discretized. For example, the $t$ and $x$ components of the first-order pull-back gauge-covariant derivatives are
\begin{eqnarray}
	\label{eq:D0}
	(D_0^<\phi)_{i,j,k}^{n+\frac{1}{2}}\!&=&\!\frac{1}{\Delta t}\Big(\bar{U}_{i,j,k}^{n+1/2}\phi^{n+1}_{i,j,k}-\phi^{n}_{i,j,k}\Big),\\
	\label{eq:D1}
	(D_1^<\phi)_{i+\frac{1}{2},j,k}^{n}\!&=&\!\frac{1}{\Delta x}\Big(\bar{U}_{i+1/2,j,k}^{n}\phi^{n}_{i+1,j,k}-\phi^{n}_{i,j,k}\Big),\hspace{10pt}
\end{eqnarray}
where $U_{i,j,k}^{n+1/2}=\exp(ie\Delta tA_{i,j,k}^{n+\frac{1}{2}})$ and $U_{i+1/2,j,k}^{n}=\exp(ie\Delta xA_{i+\frac{1}{2},j,k}^{n})$ are the infinitesimal Wilson's lines, which are usually called gauge links in lattice field theory.
Here, I denote the complex conjugation of $U$ as $\bar{U}$ in order to avoid pilling up superscripts.
Analogously, one can define push-forward covariant derivatives, which is not needed here. As in the continuous case, the classical gauge field $\bar{A}$ serves as the 1-form defining the connection on the U(1)-bundle along the Wilson's lines [Eq.~(\ref{eq:comparator})], which enables parallel transport of the $\phi_0$ field on the discrete spacetime manifold.
Since the Lagrangian is quadratic in derivatives of the $\phi$ field, the above first-order covariant derivative results in a second-order finite difference approximation to the KG equations. Higher order algorithms may be constructed using higher-order covariant derivatives.

To compute the classical field strength tensor $\bar{F}_{\mu\nu}$, notice that $\bar{F}=d\bar{A}$ is the curvature 2-form and hence lives on faces of the lattice upon discretization. 
To compute the discrete exterior derivative, we can use the Stokes' theorem $\int_{S} d\alpha=\int_{\partial S} \alpha$, where $\alpha$ is a differential $p$-form on some manifold $M$, and $S$ is a $(p+1)$-dimensional submanifold of $M$. Upon discretization, the manifold is made of a chain of complexes, and the differential form $\alpha$ assigns values to each element of the $p$-dimensional complex. Using the Stokes' theorem, the exterior derivative $d\alpha$, which assigns values to each element of the $(p+1)$-dimensional complex, can be computed by $(d\alpha)_i=l_{ij}\alpha_j/S_i$, where $S_i$ is the volume of the $i$-the element of the $(p+1)$-dimensional complex, whose boundaries have volumes $l_{ij}$.
Now that the gauge field is a 1-form, the volumes $l_{ij}$ are simply lengths of edges of the lattice, and the volume $S_i$ is simply the surface area of the $i$-th face of the lattice.
For example, the timelike component $\bar{F}_{01}=E^1$, namely the electric field in the $x$ direction, 
can be computed to first-order accuracy by
\begin{equation}
	E_{i+\frac{1}{2},j,k}^{n+\frac{1}{2}}\!=\!\frac{A_{i+\frac{1}{2},j,k}^{n+1}\!-\!A_{i+\frac{1}{2},j,k}^{n}}{\Delta t}
	\!-\!\frac{A_{i+1,j,k}^{n+\frac{1}{2}}\!-\!A_{i,j,k}^{n+\frac{1}{2}}}{\Delta x}.
\end{equation}
This component lives on the timelike face spanned by four vertices $(n,i,j,k)$, $(n,i+1,j,k)$, $(n+1,i+1,j,k)$, and $(n+1,i,j,k)$. 
Analogously, we can compute the spacelike components of $\bar{F}$. For example, $\bar{F}_{12}=-B^3$ is the magnetic field in the $z$ direction. To first order,
\begin{eqnarray}
	-B_{i+\frac{1}{2},j+\frac{1}{2},k}^{n}&=&\frac{1}{\Delta x}\bigg(A_{i+1,j+\frac{1}{2},k}^{n}\!-\!A_{i,j+\frac{1}{2},k}^{n}\bigg)-\frac{1}{\Delta y}\bigg(A_{i+\frac{1}{2},j+1,k}^{n}\!-\!A_{i+\frac{1}{2},j,k}^{n}\bigg). \hspace{10pt}
\end{eqnarray}
This \textit{z} component of the magnetic field lives on the spacelike face spanned by four vertices $(n,i,j,k)$, $(n,i+1,j,k)$, $(n,i+1,j+1,k)$ and $(n,i,j+1,k)$. 
Notice that the sign of the discrete $\bar{F}$ is determined by the orientation of the face. 
Since the Lagrangian is quadratic in derivatives of the $A$ field, the above exterior derivatives result in second-order finite difference approximations to the Maxwell's equations. Higher order algorithms may be constructed using higher order exterior derivatives, which involve more faces and edges than included in the basic units of the discrete manifold.

Using the discrete gauge-covariant derivatives and the discrete field strength, the action can be discretized and written as the summation 
\begin{eqnarray}
	\label{eq:Sd}
	S_d=\sum_{c}\Delta V \mathcal{L}_d[\phi_v,A_e],
\end{eqnarray}
where $\phi_v$ and $A_e$ are the discrete fields. Here the subscript $v$ denotes vertexes, and $e$ denotes edges. In the discrete action, $\Delta V$ is the volume 4-form, and the summation runs over all cells of the lattice. In each unit cell, the discrete Lagrangian density 
\begin{eqnarray}
	\label{eq:Ld}
	\mathcal{L}_d=(\overline{D_\mu\phi})_e(D^\mu\phi)_e-m^2\bar{\phi}_{v}\phi_v+\frac{1}{2}(E_f^2-B_f^2),
\end{eqnarray}
where summations over unique vertexes $v$, edges $e$, and faces $f$ are implied. 
Here, I change the notation for complex conjugation $\bar{\phi}=\phi^*$ for the classical charged field, in order to avoid double superscripts when writing the indexes of the discrete field explicitly. 
Notice that in favor of local energy conservation, I choose the non-compact $F_{\mu\nu}F^{\mu\nu}$ instead of the standard Wilsonian plaquettes $\text{Re}[1-\exp(ieF_{\mu\nu}\Delta_\mu\Delta_\nu)]$ for the gauge sector. The Wilsonian formulation is numerically convenient, because it uses gauge links $U_\mu=\exp(ieA_\mu\Delta_\mu)$ as the basic variables and thereby avoids computing exponentiations. However, this compact formulation introduces an $O(\Delta^2)$ local energy error, which can be eliminated using the non-compact formulation as we shall see later. 
Since capturing long-time dynamics accurately is what concerns real-time lattice simulations, local energy conservation is more preferable than numerical convenience.

\subsection{Finite difference equations\label{sec:simulation:variation:EOM}}

\begin{figure}[t]
	\renewcommand{\figurename}{FIG.}
	\centering
	\includegraphics[angle=0,width=0.5\textwidth]{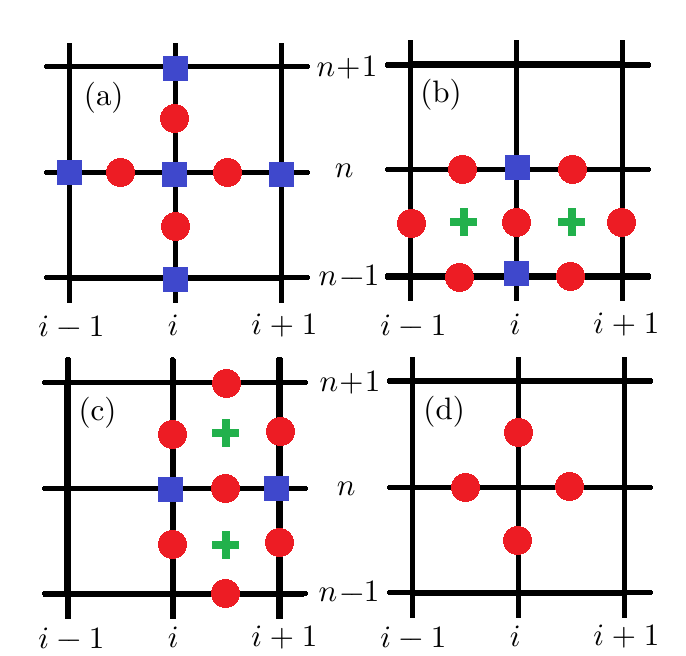}
	\caption[Coupling patterns of finite difference equations]{Coupling pattern of $\phi_v$ (blue squares), $A_e$ (red circles) and $F_f$ (green crosses) in the $tx$ submanifold. (a) The discretized KG equation [Eq.~(\ref{eq:phid})] couples $\phi_{v}$ with its nearest neighbors though $A_e$. (b) The discretized Gauss's law [Eq.~(\ref{eq:Ed})] couples $E_{i-1/2}$ and $E_{i+1/2}$ through $\phi_v$, centered around the common timelike edge. (c) The Maxwell-Amp\`ere's law [Eq.~(\ref{eq:Bd})] couples $E^{n+1/2}$ to $E^{n-1/2}$ through $\phi_v$ and $B^n$ (not depicted here), centered around the common spacelike edge. (d) The Lorenz gauge condition couples $A_e$'s that share the same vertex. }
	\label{fig:Coupling}
\end{figure}

Having discretized the action, the classical equation of motion (EOM) for the discrete field $\phi_v$ can be obtained by extremizing $S_d$. Taking variation with $\bar{\phi}_v$ and set $\delta S_d/\delta \bar{\phi}_{v}=0$, a discrete version of the KG equation [Eq.~(\ref{EOM})] can be written as
\begin{eqnarray}
\label{eq:phid}
\nonumber
&&\frac{1}{\Delta t^2}\bigg(\bar{U}_{s}^{n+\frac{1}{2}}\phi_{s}^{n+1}-2\phi_{s}^{n} +U_{s}^{n-\frac{1}{2}}\phi_{s}^{n-1}\bigg)\\
&=&\frac{1}{\Delta_l^2}\bigg(\bar{U}_{s+\frac{l}{2}}^{n}\phi_{s+l}^{n}-2\phi_{s}^{n} +U_{s-\frac{l}{2}}^{n}\phi_{s-l}^{n}\bigg)\!-\!m^2\phi_{s}^{n},
\end{eqnarray}
where the time index is explicit, the vertex-centered spatial index is abbreviated as $s:=(i,j,k)$, and summations over $l=i,j,k$ directions are implied.
By taking variation with $\phi_v$, we can obtain the EOM for $\bar{\phi}_v$, which is the complex conjugation of the above equation. The finite difference equation (\ref{eq:phid}) is centered around vertexes, and couples $\phi_{v}$ with its eight nearest neighbors though $A_e$, as illustrated by Fig.~\ref{fig:Coupling}(a) in the $tx$ submanifold.

To find the equation for the electric field, which lives on timelike faces, take variation of $S_d$ with respect to the timelike component $A_{s}^{n+1/2}$. By setting $\delta S_d/\delta A_{s}^{n+1/2}=0$, we can obtain a discrete version of the Gauss's law $\nabla\cdot\mathbf{E}=j^0$, centered along timelike edges:
\begin{eqnarray}
	\label{eq:Ed}
	\frac{1}{\Delta_l}\Big(E_{s+\frac{l}{2}}^{n+\frac{1}{2}}-E_{s-\frac{l}{2}}^{n+\frac{1}{2}}\Big)=J^{n+1/2}_{s}.
\end{eqnarray}
The charge density 1-form $J^{n+1/2}_{s}$ is the hodge dual of the charge density 3-form $j_0=\star j^0$, which is given by the following expression:
\begin{equation}
\label{eq:charge}
J^{n+1/2}_{s}=\frac{ie}{\Delta t}\Big(\bar{\phi}_{s}^{n+1}U_{s}^{n+\frac{1}{2}}\phi_{s}^{n}-\text{c.c.}\Big).
\end{equation}
The above discretization of the charge density [Eq.~(\ref{eq:Jmu})] is dictated by the variational algorithm once the discretization of the Lagrangian density is given. 
When there are multiple charged species, the RHS should sum over charge densities of all species. In Fig.~\ref{fig:Coupling}(b), the coupling pattern of the above finite difference equation is illustrated.

To find equations involving components of the magnetic field, we can take variation of $S_d$ with respect to spacelike components $A_{s+l/2}^{n}$. For example, by setting $\delta S_d/\delta A_{i+1/2,j,k}^{n}=0$, we can obtain an equation advancing the electric field $E_i$ in time by
\begin{eqnarray}
	\label{eq:Bd}
	\frac{E_{s+\frac{i}{2}}^{n+\frac{1}{2}}-E_{s+\frac{i}{2}}^{n-\frac{1}{2}}}{\Delta t}=\epsilon_{ijk}\frac{B_{r-\frac{k}{2}}^{n}-B_{r-\frac{k}{2}-j}^{n}}{\Delta_j}+J^{n}_{s+\frac{i}{2}}.
\end{eqnarray}
Here, $r=(i+1/2, j+1/2, k+1/2)$ is the abbreviated index for the body center, $\epsilon_{ijk}$ is the Levi-Civita symbol, and summations over repeated indexes are implied. The current density 1-form $J^{n}_{s+i/2}$ is the hodge dual of the current density 3-form $j_i=\star j^i$. The hodge dual gives rise to a negative sign, so that the \textit{x} component of the current density $-j^x$ is discretized by 
\begin{equation}
\label{eq:current}
J^{n}_{s+\frac{l}{2}}=\frac{ie}{\Delta_l}\Big(\bar{\phi}_{s+l}^{n}U_{s+\frac{l}{2}}^{n}\phi_{s}^{n}-\text{c.c.}\Big).
\end{equation}
Again the above discretization of the current density [Eq.~(\ref{eq:Jmu})] is dictated by the variational algorithm.
The finite difference equation (\ref{eq:Bd}) is the discrete version of the Maxwell-Amp\`ere's law  $\partial_tE_i=\epsilon_{ijk}\partial_jB_k-j^i$ centered around spacelike edges, whose coupling pattern is illustrated in Fig.~\ref{fig:Coupling}(c). When computing the RHS, summation over charged species is implied.

In order to advance the above finite difference equations in time, we need to fix a gauge to eliminate the extra degree of freedom. To see that the discrete action $S_d$ is U(1)-gauge invariant, notice that under the continuous U(1)-gauge transformation [Eq.~(\ref{Gauge})], the discrete fields are transformed by
\begin{eqnarray}
	\label{eq:U1phi}
	\phi_{s}^{n}&\rightarrow&\phi_{s}^{n} e^{ie\alpha_{s}^{n}}, \\
	A_{s}^{n+\frac{1}{2}}&\rightarrow& A_{s}^{n+\frac{1}{2}}+\frac{1}{\Delta t}(\alpha_{s}^{n+1}-\alpha_{s}^{n}),\\
	\label{eq:GaugeAi}
	A_{s+\frac{l}{2}}^{n}&\rightarrow& A_{s+\frac{l}{2}}^{n}+\frac{1}{\Delta_l}(\alpha_{s+l}^{n}-\alpha_{s}^{n}),
\end{eqnarray}
where $\alpha_{s}^n$ is any real-valued function living on vertexes. It is a straightforward calculation to verify that these transformations leave the discrete face-centered field strength tensor $F_f$ invariant, while transforming the pull-back covariant derivative by
\begin{eqnarray}
	\label{eq:Cov}
	(D_\mu^<\phi)_{s}^n\rightarrow e^{ie\alpha_{s}^{n}}(D_\mu^<\phi)_{s}^n.
\end{eqnarray}
Therefore, the discrete Lagrangian density Eq.~(\ref{eq:Ld}) is U(1)-gauge invariant, and we can choose any gauge of convenience. For example, one convenient choice is the Lorenz gauge  $\partial_\mu A^\mu=0$, which becomes
\begin{eqnarray}
\label{eq:Lorenz}
\frac{A_s^{n+1/2}-A_s^{n-1/2}}{\Delta t}=\frac{A^n_{s+l/2}-A^n_{s-l/2}}{\Delta_l},
\end{eqnarray} 
after being discretized. The Lorenz gauge condition allows time advance $A_{s}^{n-1/2}\rightarrow A_{s}^{n+1/2}$ in a very simple way [Fig.~\ref{fig:Coupling}(d)]. Another convenient choice is the temporal gauge $A^0=0$. When discretized, $A_{s}^{n+1/2}$ remains zero on all timelike edges. 

\subsection{Numerical scheme\label{sec:simulation:equations:scheme}}

Having obtained discrete equations and fixed the gauge, an explicit time advance scheme can be constructed. 
The first step is initializing the simulation by giving values of $\phi_s^n$ at both $n=0$ and $n=1$ for every spatial lattice points $s$ in the simulation domain. This is necessary because the KG equation is a second-order partial differential equation and therefore needs two initial conditions. Similarly, we need to give initial values of $A_e$ at $n=0$ and $n=1/2$, because Maxwell's equations are second-order equations when written in terms of the gauge field. 
Although the initial field configurations $\phi^0_s, \phi^1_s, A^0_{s+l/2}$, and $A^{1/2}_s$ can take any values, the initialization step is in fact very crucial. Together with boundary conditions, the initial field configurations determine what physical system will be evolved subsequently during the time advance.

\begin{figure}[!t]
	\renewcommand{\figurename}{FIG.}
	\centering
	\includegraphics[angle=0,width=0.7\textwidth]{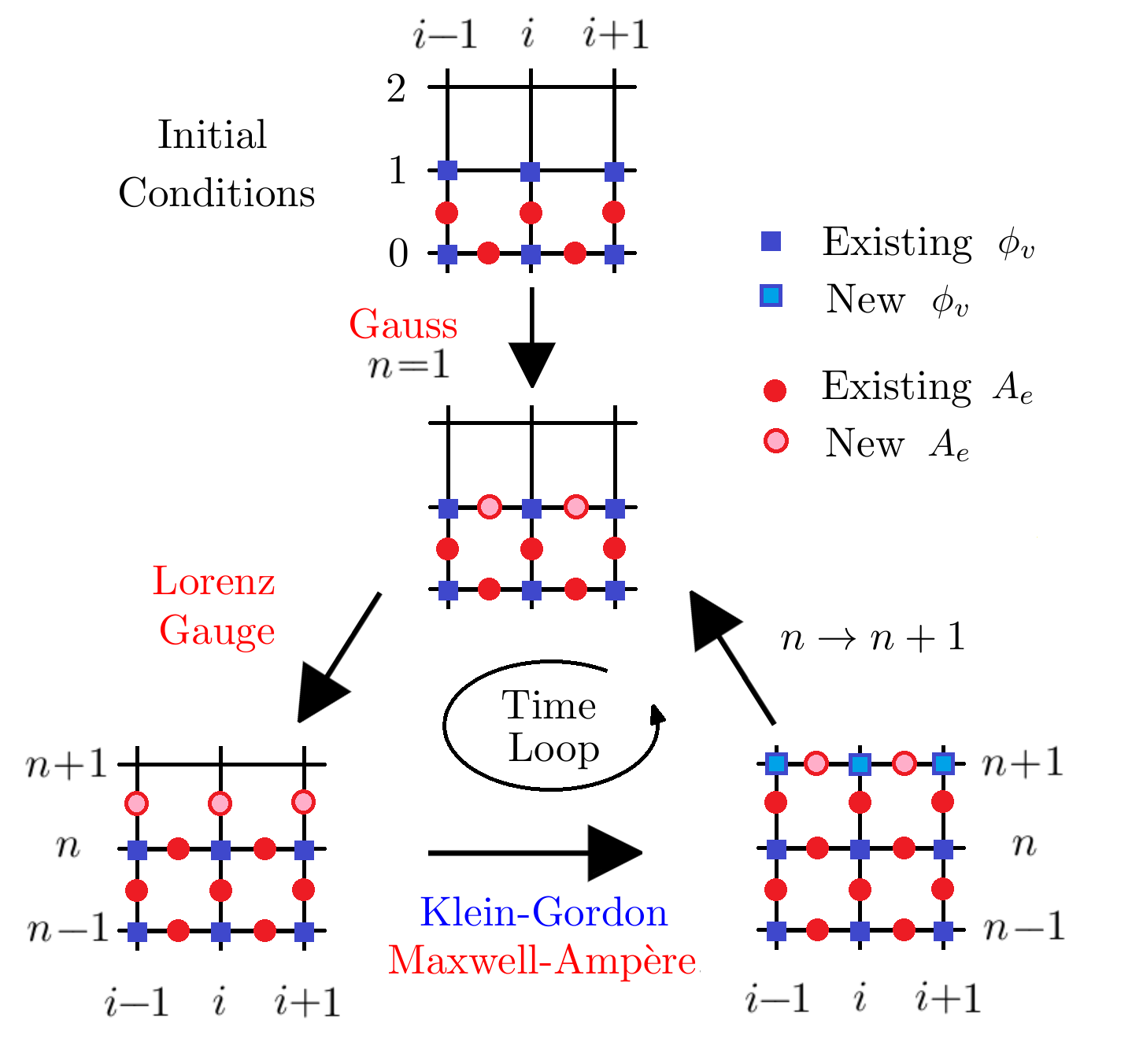}
	\caption[Numerical scheme for real-time lattice QED]{Time evolution scheme for discrete KGM equations using the Lorenz gauge.  As initial conditions, the values of $\phi_v(n=0)$ and $\phi_v(n=1)$ are given (blue squares), so are $A_e(n=0)$ and $A_e(n=1/2)$ (red circles). Then the Gauss's Law [Eq.~(\ref{eq:Ed})] is used to calculate $A_e(n=1)$. On entering the time loop, the first step is to calculate $A^{n+1/2}$ using the Lorenz gauge condition. The second step is to use the KG equation [Eq.~(\ref{eq:phid})] to calculate $\phi^{n+1}$, and concurrently, use the Maxwell-Amp\`ere's law [Eq.~(\ref{eq:Bd})] to calculate $A^{n+1}$. The time loop is advanced by $n\rightarrow n+1$ and then repeat. }
	\label{fig:Schematics}
\end{figure}

The second step is calculating $A^1_{s+l/2}$ using the Gauss's law. This step ensures that the self-consistency of classical fields is satisfied initially. The EOMs then guarantee that the self-consistency between the charged field and the gauge field will always be satisfied at later time. The discrete Gauss's law [Eq.~(\ref{eq:Ed})] is a system of linear equations, which can be rewritten explicitly as
\begin{eqnarray}
\label{eq:Gauss}
\frac{A^1_{s+l/2}-A^1_{s-l/2}}{\Delta_l}&=&\frac{A^0_{s+l/2}-A^0_{s-l/2}}{\Delta_l}+\frac{\Delta t}{\Delta l^2}\Big(A^{1/2}_{s+l}-2A^{1/2}_s+A^{1/2}_{s-l}\Big)+\Delta t J_s^{1/2}.\hspace{15pt}
\end{eqnarray}
Notice that all terms on the RHS are known, and the unknowns are $A^1_{s+l/2}$ at every spatial points in the next time slice. Since the LHS couples only two adjacent $A^1_{s+l/2}$ in each direction [Fig.~\ref{fig:Coupling}(b)], the discrete Gauss's law is easier to solve than the Poisson's equation, which couples three nearest neighbors in each direction.
In fact, the continuous version of the above finite difference equation is $\partial_t\nabla\cdot\mathbf{A}=-\nabla^2A^0-\rho$, where the RHS is known. Because the unknowns on the LHS involve only first-order spatial derivative, the discrete Gauss's law couples less number of points than the discrete Poisson's equation, which involves second-order spatial derivatives.

The third step is advancing the time-component of the gauge field $(A_s^{n-1/2},A_{s+l/2}^{n})\rightarrow A_s^{n+1/2}$. This step depends on the choice of the gauge condition. For example, when the Lorenz gauge is used [Fig.~\ref{fig:Coupling}(d)], the time advance is simply given by
\begin{eqnarray}
\label{eq:Lorenz_time}
A_s^{n+1/2}=A_s^{n-1/2}+C_l\Big(A^n_{s+l/2}-A^n_{s-l/2}\Big),
\end{eqnarray} 
where $C_l=\Delta t/\Delta_l$ is the dimensionless Courant number.
In comparison, when temporal gauge is used instead, $A^{n+1/2}_s=0$ and the time advance is trivial. Using the temporal gauge, one only needs to store values of $A_e$ at integer time steps $t=n$, which is numerically efficient. However, when a background electric field is present, $A_{s+l/2}^{n}$ will grow indefinitely in the temporal gauge. In this case, long-time dynamics may be more accurately computed using the Lorenz gauge instead.

In the fourth step, we can use the discrete KG equation [Eq.~(\ref{eq:phid})] to time advance the charged field $(\phi_s^{n-1}, \phi_s^{n};  A_{s+l/2}^{n}, A_s^{n\pm1/2})\rightarrow\phi_s^{n+1}$. The explicit time advance is given by
\begin{eqnarray}
\label{eq:phin}
\nonumber
\phi_s^{n+1}=U_{s}^{n+\frac{1}{2}}\Big[(2-2C_l^2-\Delta t^2m^2)\phi_{s}^{n} -U_{s}^{n-\frac{1}{2}}\phi_{s}^{n-1}&&\\
+C_l^2\Big(\bar{U}_{s+\frac{l}{2}}^{n}\phi_{s+l}^{n}+U_{s-\frac{l}{2}}^{n}\phi_{s-l}^{n}\Big)\Big],&&
\end{eqnarray}
where all terms on the RHS are known.
For the free $\phi_0$ field, suppose the fluctuation is of the form $\exp(ip_lx^l-iEt)$, then the numerical dispersion relation of the massive particle is
\begin{eqnarray}
\frac{4}{\Delta t^2}\sin^2\frac{E \Delta t}{2}=\frac{4}{\Delta_l^2}\sin^2\frac{p_l \Delta_l}{2}+m^2,
\end{eqnarray}
which is consistent with the continuum energy-momentum relation $E^2=\mathbf{p}^2+m^2$ for relativistic particles when the resolution $\Delta\rightarrow0$. For the numerical solution to be stable, $E$ must be real, which holds if and only if the CFL condition $C_l<1$ is satisfied for all $l=i,j,k$.
Computing $\phi_s^{n+1}$ needs the values of the gauge links, which require exponentiations of $A^n$ and $A^{n+1/2}$ whose values are already known at this step.

Finally, without relying on the values of $\phi_s^{n+1}$, we can use the discrete Maxwell-Amp\`ere's law [Eq.~(\ref{eq:Bd})], concurrently with the KG equation, to advance the spatial component of the gauge field $(A_{s+l/2}^{n-1},A_s^{n\pm1/2},A_{s+l/2}^{n};\phi_s^{n})\rightarrow A_{s+l/2}^{n+1}$. The explicit time advance is given by
\begin{eqnarray}
\label{eq:An}
\nonumber
A_{s+\frac{i}{2}}^{n+1}&=&A_{s+\frac{i}{2}}^{n}+C_i\Big(A_{s+i}^{n+\frac{1}{2}}-A_s^{n+\frac{1}{2}}\Big)\!+\!\Delta t^2J_{s+\frac{i}{2}}^n\\
&+&\!\Delta t\Big[E_{s+\frac{i}{2}}^{n-\frac{1}{2}}+\epsilon_{ijk}C_j\Big(B_{r-\frac{k}{2}}^n-B_{r-\frac{k}{2}-j}^n\Big)\Big],
\end{eqnarray}
where all terms on the RHS is known. For free gauge field, it is straightforward to show that the numerical solution is stable if and only if the CFL condition $C_l<1$ is satisfied.
Notice that the discrete Gauss's Law is preserved during time advance, which is a consequence of the discrete local charge conservation law, which we shall see next. Having computed both $\phi_v$ and $A_e$ at $t=n+1$, we can move forward in the time loop by updating $n\rightarrow n+1$, with proper boundary conditions supplied (Fig.~\ref{fig:Schematics}). In similar fashion, explicit time advance schemes can be constructed when other gauge conditions are used.

\section{Discrete identities and conservation laws\label{sec:simulation:conservation}}

Since the discretization in Sec.~\ref{sec:simulation:variation:discret} respects the structure of exterior calculus, a number of geometric identities are automatically satisfied. Moreover, by the famous Noether's theorem, symmetries of the action results in conservation laws. Although the continuous Poincar\'e group becomes discrete on a spacetime lattice, the continuous gauge symmetry is preserved. Therefore, charge is exactly conserved by the numeric scheme, while energy and momentum have errors that are consistent with the order of the algorithm.   

\subsection{Geometric identities of discrete exterior derivatives\label{sec:simulation:conservation:geometry}}

When discretizing the classical gauge 1-form $\bar{A}$ and calculating the classical field strength 2-form $\bar{F}=d\bar{A}$ in Sec.~\ref{sec:simulation:variation:discret}, geometric structures of discrete exterior calculus are respected. Consequently, the identity $d^2=0$ holds for the discrete exterior derivative. In components, the Bianchi identity can be written as $0=d\bar{F}=(\partial_\sigma \bar{F}_{\mu\nu}+\partial_\mu \bar{F}_{\nu\sigma}+\partial_\nu \bar{F}_{\sigma\mu})dx^{\mu}\wedge dx^{\nu}\wedge dx^{\sigma}/3!$. One nontrivial identity, corresponding to all indexes being spatial, is $\nabla
\cdot\mathbf{B}=0$. When discretized, this identity becomes
\begin{eqnarray}
\label{eq:B0}
\frac{1}{\Delta_l}\Big(B_{r+\frac{l}{2}}^{n}-B_{r-\frac{l}{2}}^{n}\Big)=0.
\end{eqnarray}
In other words, the equation $\nabla\cdot\mathbf{B}=0$ is automatically guaranteed by the variational algorithm.
The other nontrivial identity, corresponding to two spatial indexes and one temporal index, is the Faraday's law $\partial_t\mathbf{B}=-\nabla\times\mathbf{E}$, whose discrete version is
\begin{eqnarray}
\label{eq:E0}
&&\frac{1}{\Delta t}\Big(B_{r-\frac{i}{2}}^{n+1}-B_{r-\frac{i}{2}}^{n}\Big)=\frac{\epsilon_{ijk}}{\Delta_k}\Big(E_{s+\frac{j}{2}+k}^{n+\frac{1}{2}}-E_{s+\frac{j}{2}}^{n+\frac{1}{2}}\Big).
\end{eqnarray}
In other words, the Faraday's law does not need to be solved. Instead, it is automatically satisfied by geometric constructions. This is different from standard electromagnetic algorithms, such as the Yee's algorithm \citep{Yee66}, in which the Faraday's law needs to be solved as a dynamical equation. 
In the standard Yee's algorithm, the gauge invariant electric and magnetic fields, which have six components in total, are solved as dynamical fields using two equations that are first order in time. In comparison, in the variational algorithm, the gauge field, which has three components after gauge fixing, is solved as the only dynamical field using an equation that is second order in time. 
Although the degrees of freedom are the same in both schemes, the Yee's algorithm advances all six degrees of freedom at each time step, while the variational algorithm folds the six degrees of freedom at two time steps, and thereof only advances three field components at each time step.  

\subsection{Charge conservation: continuous U(1)-gauge symmetry\label{sec:simulation:conservation:charge}}

In addition to geometric identities, we also have local conservation laws due to symmetry of the discrete action. In particular, charge is conserved as a direct consequence of local U(1)-gauge symmetry. Using the classical field equation $\delta S_d/\delta\phi_v=0$, we have
\begin{equation}
\frac{\delta S_d}{\delta \phi_{s}^{n}}\delta\phi_{s}^n+\text{c.c.}=0.
\end{equation}
Substituting the infinitesimal transformation $\delta\phi_{s}^n=ie\alpha_{s}^n\phi_{s}^n$ under the local U(1) transformation [Eq.~(\ref{eq:U1phi})] in to the discrete KG equation [Eq.~(\ref{eq:phid})], the above identity becomes
\begin{eqnarray}
\nonumber
0&=&\Delta V ie\alpha_s^n\phi_{s}^n\bigg[\frac{1}{\Delta t^2}\bigg(U_{s}^{n+\frac{1}{2}}\bar{\phi}_{s}^{n+1} +\bar{U}_{s}^{n-\frac{1}{2}}\bar{\phi}_{s}^{n-1}\bigg)-\frac{1}{\Delta_l^2}\bigg(U_{s+\frac{l}{2}}^{n}\bar{\phi}_{s+l}^{n} +\bar{U}_{s-\frac{l}{2}}^{n}\bar{\phi}_{s-l}^{n}\bigg)\bigg]+\text{c.c.}\\
\nonumber
&=&\Delta V ie\alpha_s^n\bigg\{\frac{1}{\Delta t^2} \bigg[\bigg(\bar{\phi}_{s}^{n+1}U_{s}^{n+\frac{1}{2}}\phi_{s}^n-\text{c.c.}\bigg)- \bigg(\bar{\phi}_{s}^{n}U_{s}^{n-\frac{1}{2}}\phi_{s}^{n-1}-\text{c.c.}\bigg)\bigg]\\
\nonumber
&&\hspace{44pt}- \frac{1}{\Delta_l^2} 
\bigg[\bigg(\bar{\phi}_{s+l}^{n}U_{s+l/2}^{n}\phi_{s}^n-\text{c.c.}\bigg)- \bigg(\bar{\phi}_{s}^{n}U_{s-l/2}^{n}\phi_{s-l}^{n}-\text{c.c.}\bigg)\bigg]
\bigg\}.
\end{eqnarray}
It is easy to recognize the above terms are the discrete charge density [Eq.~(\ref{eq:charge})] and the discrete current density [Eq.~(\ref{eq:current})]. Since the above identity holds for all $\alpha_s^n$, we have an exact discrete charge conservation law
\begin{equation}
\label{eq:Jconserv}
\frac{1}{\Delta t} \Big(J_{s}^{n+\frac{1}{2}}-J_{s}^{n-\frac{1}{2}}\Big)=\frac{1}{\Delta_l} \Big(J_{s+\frac{l}{2}}^{n}-J_{s-\frac{l}{2}}^{n}\Big).
\end{equation}
Here, 
the sign is due to the Minkowski metric. It is straightforward to check that the above discrete charge conservation law is compatible with the discrete Gauss's law [Eq.~(\ref{eq:Ed})] and the discrete Maxwell-Amp\`ere's law [Eq.~(\ref{eq:Bd})]. Therefore, once the Gauss's law is satisfied at the initial time, it will be satisfied for all time.

\subsection{Energy error at finite coupling: loss of time-translation symmetry\label{sec:simulation:conservation:energy}}

The discrete action $S_d$ is invariant under translations on the discrete spacetime manifold. 
Although the symmetry group in this case is discrete and hence the Noether's theorem does not immediately apply, we do have local energy conservation laws for the charged field and EM fields separately when their coupling vanishes. Using the classical field equations $\delta S_d/\delta\phi_v=0$ [Eq.~(\ref{eq:phid})] and $\delta S_d/\delta A_{s+l/2}^n=0$ [Eq.~(\ref{eq:Bd})], as well as the geometric identity [Eq.~(\ref{eq:E0})], we have the following identity 
\begin{eqnarray}
\nonumber
0&=&\frac{\delta S_d}{\delta \phi_{s}^{n}}(\mathcal{D}_0\phi)_{s}^{n}+\frac{\delta S_d}{\delta \bar{\phi}_{s}^{n}}(\overline{\mathcal{D}_0\phi})_{s}^{n}\\
&+&\frac{\delta S_d}{\delta A_{s+l/2}^{n}}\frac{1}{2}\Big(E_{s+l/2}^{n+1/2}+E_{s+l/2}^{n-1/2}\Big)\\
\nonumber
&+&B_{r-l/2}^{n}\frac{1}{2}\Big[(d^2A)_{r-l/2}^{n+1/2}+(d^2A)_{r-l/2}^{n-1/2}\Big],
\end{eqnarray}
where the vertex-centered time covariant derivative $(\mathcal{D}_0\phi)_{s}^{n}=\frac{1}{2\Delta t}(\bar{U}_s^{n+\frac{1}{2}}\phi_s^{n+1}-U_s^{n-\frac{1}{2}}\phi_s^{n-1})$.
After rearranging terms, similar to what is done when proving charge conservation, the above identity gives rise to the local energy conservation law
\begin{eqnarray}
\label{eq:PEnergy}
\frac{\mathcal{H}_{s}^{n+1/2}\!-\!\mathcal{H}_{s}^{n-1/2}}{\Delta t}
=\frac{\mathcal{P}_{s+l/2}^{n}\!-\!\mathcal{P}_{s-l/2}^{n}}{\Delta_l}+\mathcal{O}(e\Delta^2), \hspace{15pt}
\end{eqnarray}
where the sign is again due to the Minkowski metric. The energy density can be separated into three terms
\begin{equation}
\mathcal{H}_{s}^{n+1/2}=\mathcal{H}_{s}^{n+1/2}[\phi]+\mathcal{H}_{s}^{n+1/2}[A]+h_{s}^{n+1/2},
\end{equation}
where the energy density of the charged field is
\begin{eqnarray}
\label{eq:Hphi}
\nonumber
\hspace{-2pt}\mathcal{H}_{s}^{n+\frac{1}{2}}[\phi] \!&=&\!\frac{1}{2}\Big[(D_0^<\phi)_s^{n+\frac{1}{2}}\!(\overline{D_0^<\phi})_s^{n+\frac{1}{2}}
\!+\!m^2\phi_s^{n}U_{s}^{n+\frac{1}{2}}\bar{\phi}_s^{n+1}\\
&&\hspace{3pt}+(D_l^<\phi)_{s+\frac{l}{2}}^{n}U_{s}^{n+\frac{1}{2}}(\overline{D_l^<\phi})_{s+\frac{l}{2}}^{n+1}\Big]\!+\!\text{c.c.},
\end{eqnarray}
and the energy density of the EM fields is
\begin{eqnarray}
\label{eq:HA}
\mathcal{H}_{s}^{n+\frac{1}{2}}[A]=\frac{1}{2}\Big[\big(E_{s+\frac{l}{2}}^{n+\frac{1}{2}}\big)^2+B_{r-\frac{l}{2}}^{n+1} B_{r-\frac{l}{2}}^{n}\Big].
\end{eqnarray}
The energy density correction $h=O(e\Delta^2)$ can take many different forms, each has a corresponding error term at finite-resolution. 
As expected, the energy density is U(1)-gauge invariant, so is the momentum density, which can be split into two terms 
\begin{equation}
\mathcal{P}_{s+l/2}^{n}=\mathcal{P}_{s+l/2}^{n}[\phi]+\mathcal{P}_{s+l/2}^{n}[A].
\end{equation}
The momentum density of the charged field is
\begin{equation}
\mathcal{P}_{s+\frac{l}{2}}^{n}[\phi]=(D_l^<\phi)_{s+\frac{l}{2}}^{n} U_{s+\frac{l}{2}}^{n}(\overline{\mathcal{D}_0^<\phi})_{s+l}^{n}+\text{c.c.},
\end{equation}
and the momentum density of the EM fields $\mathcal{P}_i=-\mathcal{P}^i=-(\mathbf{E}\times\mathbf{B})^i$ is
\begin{eqnarray}
\mathcal{P}_{s+\frac{i}{2}}^{n}[A]\!=\epsilon_{ijk}B_{r-\frac{j}{2}}^{n}\frac{1}{2}\Big(E_{s+i+\frac{k}{2}}^{n+\frac{1}{2}}\!+\!E_{s+i+\frac{k}{2}}^{n-\frac{1}{2}}\Big).
\end{eqnarray}
Since the stress-energy tensor $\mathcal{T}^{\mu\nu}$ [Eq.~(\ref{eq:stress-energy})] is not a 2-form, neither the energy density $\mathcal{H}$ nor the momentum density $\mathcal{P}$ is well-defined on the discrete spacetime manifold. Hence, it can be shown, by enumerating combinations of U(1)-gauge invariant basis terms, that the resulting error in the local energy conservation law [Eq.~(\ref{eq:PEnergy})] is always second order. 
A special case is when the coupling $e=0$, where the conservation law becomes exact even at finite spacetime resolutions. This remarkable feature would be lost if we had instead used the Wilsonian plaquettes in the discrete action.

\section{Numerical examples\label{sec:simulation:examples}}

In previous sections, I have developed a second-order algorithm for solving the KGM equations with good conservation properties. Since plasmas are typically in the classical-statistical regime, solving the classical field equations with an ensemble of statistically equivalent initial conditions captures the dominant behaviors of scalar-QED plasmas. 
To extract observables from real-time lattice simulations, one may first compute the distribution function from the classical field using Wigner–-Weyl transform. More elaborately, one may use spectral expansion of the classical field and keep track of the evolution of individual spectral components. However, these additional information is rarely observable in experiments, which may thereof be bypassed.
In this section, I will use two examples to demonstrate the numerical scheme and compute simple observables that can be constructed directly from the classical fields. The first example is the propagation of linear
waves, and the second example is laser-plasma interaction in one spatial dimension. 

\subsection{Linear waves in unmagnetized plasmas\label{sec:simulation:examples:linear}}

To validate the code implementation, we can compare numerical spectra and analytical linear wave dispersion relations \citep{Hines78,Kowalenko85,Eliasson11,Shi16QED}. For small-amplitude waves, the dispersion relation constrains the wave frequency $\omega$ as a function of the wave vector $\mathbf{k}$. In unmagnetized cold scalar-QED plasmas, the dispersion relation of the transverse EM wave is given by Eq.~(\ref{EM}). To tree-level, ignoring the vacuum permittivity, the dispersion relation is simply
\begin{equation}
\label{eq:EMD}
\omega^2=\omega_p^2+\mathbf{k}^2,
\end{equation}
where $\omega_{p}^2=\sum_s\omega_{ps}^2$ is the total plasma frequency, and $\omega_{ps}^2=e_s^2n_{s0}/m_s$ is the plasma frequency of individual charged species $s$. 
The other eigenmode is the longitudinal electrostatic wave, whose dispersion relation is given by Eq.~(\ref{langmuir}). To tree-level, it becomes
\begin{equation}
\label{eq:ESD}
1+\chi_p=0,
\end{equation}
where the susceptibility of a cold scalar-QED plasma is given by Eq.~(\ref{eq:plasma_chip}).
As discussed in Sec.~\ref{ch:unmag:dispersion:rest}, the dispersion relation of the electrostatic wave contains three branches. The gapless branch is the acoustic wave, the gapped low-frequency branch is the Langmuir mode, and the gapped high-frequency branch is the pair mode. While acoustic mode and Langmuir mode exist in classical plasmas, the pair mode only exists in relativistic-quantum plasmas \citep{Fuda82}. The pair mode can be excited when gamma photons $(\omega>2m)$ inelastically scatter in high density plasmas, creating longitudinal oscillations in which virtual pairs are created and annihilated to carry the wave quanta.

\begin{figure}[!b]
	\renewcommand{\figurename}{FIG.}
	\centering
	\includegraphics[angle=0,width=0.8\textwidth]{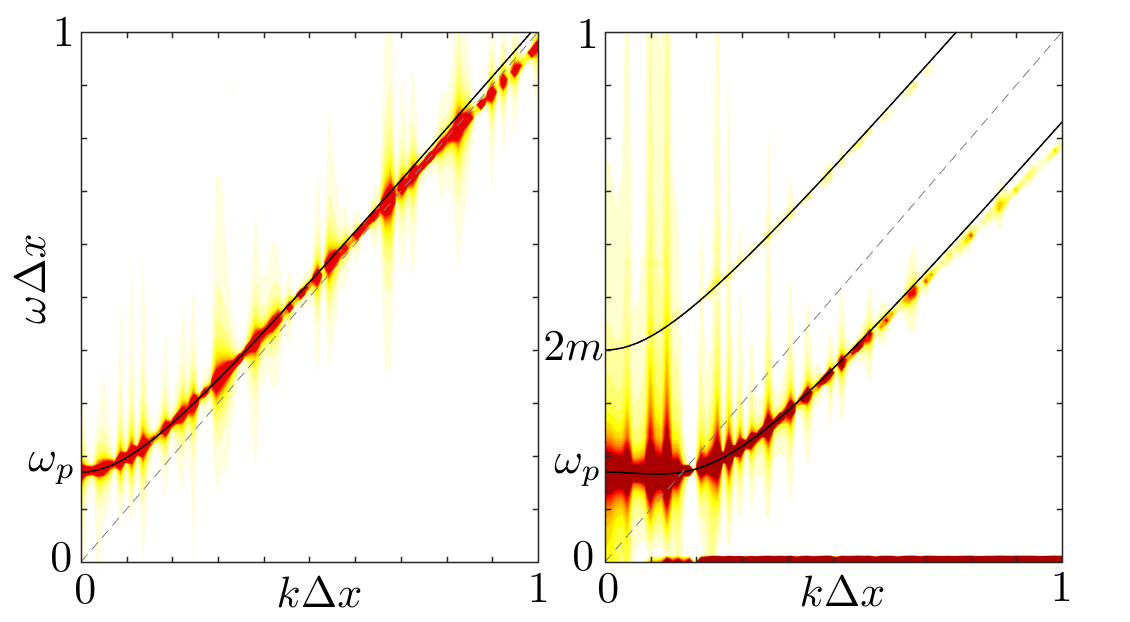}
	\caption[Spectra of linear waves in scalar-QED plasma]{Power spectra (color) of the transverse electric field $E_y$ (a) and the longitudinal electric field $E_x$ (b) are well-traced by tree-level dispersion relations (black lines) up to the grid resolution. The power spectra are averaged over an ensemble of 100 simulations with statistically equivalent initial conditions. In these simulations, immobile ion background is homogeneous. The charge \mbox{$e=0.3$}, such that the fine structure constant $e^2/4\pi\approx1/137$ is physical. The unperturbed background plasma density is extremely high, such that the plasma frequency $\omega_p=0.85m$ can be shown on the same scale as $m$. 
	The resolution $m\Delta x=0.04$ and $m\Delta t=0.02$. The number of spatial grid point is $L=512$, and the total number of time steps, including the initial conditions, is $T=1024$. The dashed gray lines is the light cone.}
	\label{fig:Dispersion}
\end{figure}

Let us compute the numerical spectra in a single species plasma, in which immobile ions serve as homogeneous neutralizing background. To initialize the simulation so that a broad spectrum of linear waves are excited, the initial values of $A_e$ are given using small amplitude white noise with mean $\mu(A_e)=0$ and standard deviation $\sigma(A_e)=10^{-4}m$. Assuming the charged field is initially free, then its initial conditions can be given using the free field expansion Eq.~(\ref{eq:phi0}). 
The expansion coefficients are related to the momentum space distribution functions for particles and antiparticles by $f_a(\mathbf{p})=a_\mathbf{p}^\dagger a_\mathbf{p}$ and the $f_b(\mathbf{p})=b_\mathbf{p}^\dagger b_\mathbf{p}$, respectively. Consider the simple example where the plasma is initially homogeneous and constituted of cold particles, namely, \mbox{$f_a(\mathbf{p})=n_0\delta^{(3)}(\mathbf{p})$} and $f_b(\mathbf{p})=0$, where $n_0$ is the background plasma density. Then, the free charged field $\phi(x)=\sqrt{n_0/2m}\exp(-imt+i\alpha)$, where $\alpha$ is some random phase. When discretized, this free field corresponds to the initial conditions $\phi_s^0=\sqrt{n_0/2m}\exp(i\alpha)$ and $\phi_s^1=\phi_s^0\exp(-im\Delta t)$. An ensemble of statistically equivalent initial conditions can then be constructed by randomly sample the phase $\alpha$ of the charged field, and randomly assign noise to the gauge field.

After advancing the initial conditions in time using periodic boundary conditions, numerical spectra can be read out from simulations by taking discrete Fourier transforms of components of the electric field. Since the unmagnetized plasma is isotropic, it is sufficient to read out the dispersion relation in the \textit{tx} submanifold. In this submanifold, the spectra of either $E_y$ or $E_z$ correspond to the dispersion relation of transverse EM modes, and the spectrum of $E_x$ corresponds to the dispersion relation of longitudinal electrostatic modes. 
The ensemble-averaged power spectrum of $E_y$ [Fig.~\ref{fig:Dispersion}(a)] is indistinguishable from that of $E_z$, and is well-traced by the analytical dispersion relation (black line) of the transverse EM wave [Eq.~(\ref{eq:EMD})], until $k\Delta x\sim 1$ where the spatial resolution is no longer sufficient.  
Similarly, the ensemble-averaged power spectrum of $E_x$ [Fig.~\ref{fig:Dispersion}(b)] is localized near three bands, corresponding to the cold acoustic mode, the Langmuir mode and the pair mode [Eq.~(\ref{eq:ESD})]. That the analytical dispersion relations are recovered by numerical power spectra indicates that our solutions faithfully capture the propagation of linear waves up to the grid resolution.

\subsection{From laser wakefield acceleration to Schwinger pair production\label{sec:simulation:examples:nonlinear}}

Having verified the code implementation, let us study laser-plasma interaction as another example, which can no longer be easily solve analytically. Laser-plasma interactions cannot be described self-consistently under the classical framework once the laser wavelength becomes too short or the field strength becomes too large. For illustrative purposes only, as opposed to suggesting a futuristic device, let us use the example of gamma lasers to show that lattice QED now enables simulations in a regime that was not accessible through previous methods.

Before discussing simulations in the relativistic-quantum regime, it is helpful to recall what happens in the classical regime \citep{Kruer88}. Classically, when the plasma slab is under-dense, namely when the laser frequency $\omega>\omega_p$, much of the laser will travel through the plasma slab, with some reflection and inverse Bremsstrahlung absorption. In an initially quiescent slab, the laser will propagate uneventfully, if its frequency stays away from the two-plasmon-decay resonance, and its intensity is not strong enough to grow instabilities within the pulse duration. 
Beyond nonlinear wave instabilities, when the laser field becomes relativistically strong, namely when the normalized field [Eq.~(\ref{eq:ak})] $a\approx eE/m\omega\gtrsim1$, the ponderomotive force of a short laser pulse can expels a significant fraction of plasma electrons and form wakefield \citep{Pukhov02}. The wakefield can then accelerate particles, generating energetic beams of particles and radiations trailing the laser pulse. When the beams are energetic enough, they may produce gamma photons through synchrotron radiation or Bremsstrahlung. The virtual gamma photons may then decay into electron-positron pairs through the trident process \citep{Bjorken67}. Alternatively, the on-shell gamma photons may produce pairs when interacting with ion potentials through the Bethe-Heitler process \citep{Bethe34}, or interacting with other photons through the Breit-Wheeler process \citep{Breit34}. Finally, when the laser field becomes even stronger, namely when $eE/m^2\gtrsim 1$, pairs may also be produced directly through the Schwinger process \citep{Schwinger51}. 

Many aspects of laser-plasma interaction can be studied using real-time lattice QED. Here, to validate that the numerical scheme in Sec.~\ref{sec:simulation:equations:scheme} can capture genuine relativistic-quantum effects, parameters can be selected in 1D simulations to demonstrate transition from wakefield acceleration to Schwinger pair production as the laser intensity increases. Notice that in 1D, the phase space is highly constrained. 
Using periodic boundary conditions in directions transverse to laser propagation, Schwinger pair production by laser fields is suppressed. This is because when transverse fields try to pull $e^-/e^+$ pairs apart, their wave functions are enforced to be the same by the periodic boundary condition, which prevents pairs from emerging out of vacuum fluctuations. Therefore, in 1D simulations, Schwinger pair production requires longitudinal field $E_x$. To generate $E_x$ beyond the Schwinger field $E_c=m^2/e$ through plasma wakefield, the plasma density must be extremely high. Heuristically, to produce on-shell pairs, the critical electric field needs to separate the pair by Compton wavelength $1/m$ within the Compton time $T\sim\pi/m$, namely, $eE_xT^2/m\gtrsim1/m$. In the wavebreaking regime, $E_x\simeq a m\omega_p/e$, so the inequality requires that the plasma density be high enough such that the plasma frequency $\omega_p/m\gtrsim1/a\pi^2$. In reality, at those densities, it is necessary to treat the electron Fermi degeneracy to capture the full physical effects. However, simulating instead a high-density bosonic plasma is just a toy model that tests real-time lattice simulations, with the density picked so high that we can already see laser Schwinger pair production in 1D simulations.

\begin{figure}[t]
	\renewcommand{\figurename}{FIG.}
	\centering
	\includegraphics[angle=0,width=0.8\textwidth]{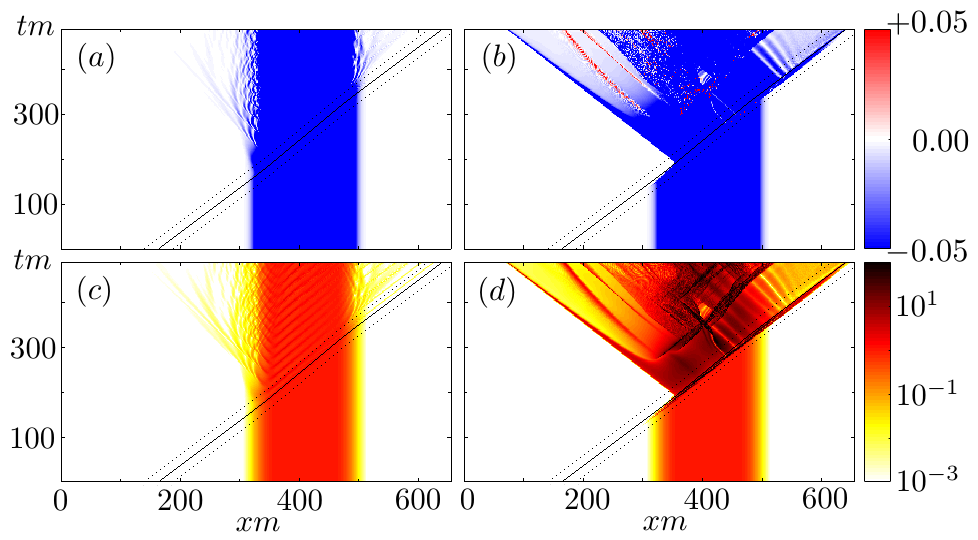}
	\caption[Charge and energy density of $\phi$ field]{Charge density (a, b) and energy density (c, d) of the $\phi_0$ field. 
	When the gamma-ray laser ($\omega_0=0.7m$) is relativistic $(a\approx1)$, but not strong enough to produce Schwinger pairs $(E_x\approx0.3E_c)$, ``electrons" are expelled by the laser ponderomotive force, accelerated by the wakefield, and splashed from the plasma boundaries (a, c). 
	On the other hand, when the laser field exceeds the Schwinger threshold $(a\approx16, E_x\approx5E_c)$, copious pairs are produced when laser interacts with plasma waves (b, d). The spin-0 ``electrons" are initially confined by a smooth immobile neutralizing background, with a density plateau $n_0=m^3$ and a Gaussian off-ramp $\sigma =20/m$. The trajectories of the pulse center (black lines) and the pulse half widths (dashed lines) are well traced by geometric optics. Both the charge density (normalized by $em^3$) and the energy density (normalized by $m^4$) are averaged over an ensemble of size 200. The resolutions are such that $m\Delta x=0.04$ and $m\Delta t=0.005$.}
	\label{fig:part}
\end{figure}

With this basic understanding of how laser pair production happens in 1D, we can choose setups to suppress the trident and Bethe-Heitler processes, by treating ions as immobile homogeneous neutralizing background, so that there is no spiky ion potentials from which energetic ``electrons" and gamma photons can scatter. The smooth ion background provides an electrostatic potential that initially confines the ``electrons". 
The charged boson wave function can be initialized according to $\phi(x)=\sqrt{n_0(x)/2m}\exp(-imt)$, where $n_0(x)$ is the background ion density with a plateau of width $L\approx100/m$ and Gaussian off-ramps with $\sigma=20/m$. For density of the bosonic plasma to be high enough to enable pair production, let us pick $n_0=m^3$ so that the plasma frequency $\omega_p=0.3m$ is enormous. The above wave function is a linear superposition of many eigenstates of the system. 
In the simulations, the wave function is allowed to evolve to statistically stationary states through phase mixing, before samples are drawn at random time intervals. The sampled wave functions are then used as initial conditions for $\phi_v$, which are combined with the initial values $A_e$ of a Gaussian pulse to construct an ensemble. 
The linearly-polarized Gaussian pulse is initialized in the vacuum region with zero carrier phase $A_y\propto\exp(-\xi^2/2\tau^2)\cos\omega\xi$, where $\xi=x-t$ and $\tau=20/m$. For the laser to be able to transmit the high-density plasma slab, we can pick the frequency of the gamma-ray laser above the plasma frequency $\omega_0=0.7m$, for which classical treatments are far from valid. The laser envelope is slowly varying ($\omega_0\tau=14$), and has full width at half maximum about twice the plasma skin depth. When the intense laser pulse propagates, it can excite plasma waves, from which the laser can be Raman scattered.

\begin{figure}[!b]
	\renewcommand{\figurename}{FIG.}
	\centering
	\includegraphics[angle=0,width=0.8\textwidth]{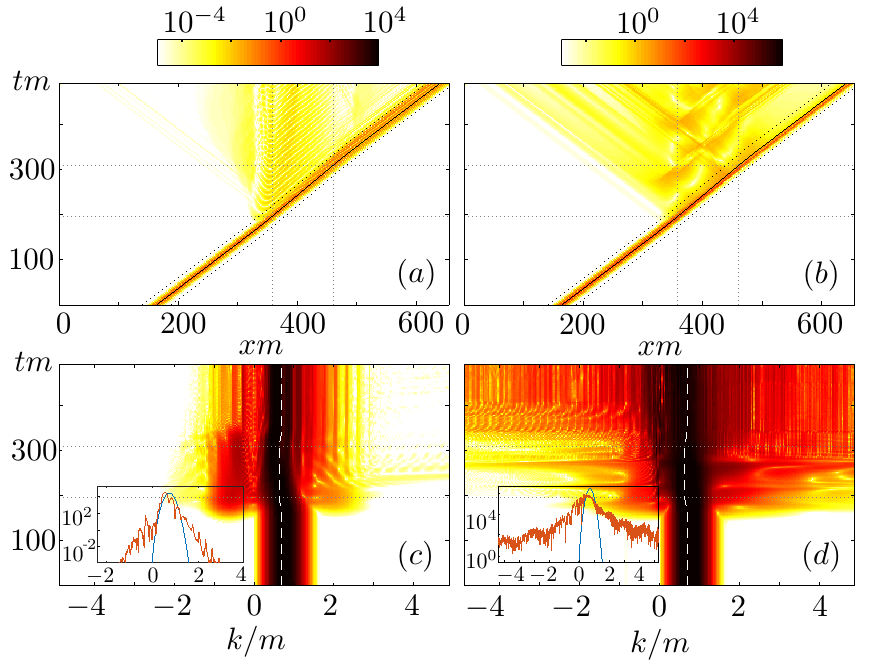}
	\caption[Energy density of EM fields and their transverse spectra]{Total energy density of EM fields (a, b), and the power spectral density of its transverse components (c, d). The inserts show the initial (blue) and final (red) spectra of EM waves. 
	When $a\approx1$ ($E_x\approx0.3E_c$) is below the Schwinger field, the laser excites plasma waves and is Raman scattered (a, c). The time evolution of the main pulse is well-traced by geometric optics (dashed lines). 
	On the other hand, when the laser field $a\approx16$ ($E_x\approx5E_c$) is above the Schwinger field, a noticeable amount of energy is lost due to pair production (b), and the $k$ spectrum is substantially broadened (d). 
	The field energy density is normalized by the Schwinger field $E_c^2$, and are averaged over an ensemble of size 200. The resolutions are such that $m\Delta x=0.04$ and $m\Delta t=0.005$. The dotted gray lines mark where the geometric-optics trajectory of the pulse center crosses the plasma plateau boundaries.}
	\label{fig:field}
\end{figure}

\begin{figure}[!b]
	\renewcommand{\figurename}{FIG.}
	\centering
	\includegraphics[angle=0,width=0.7\textwidth]{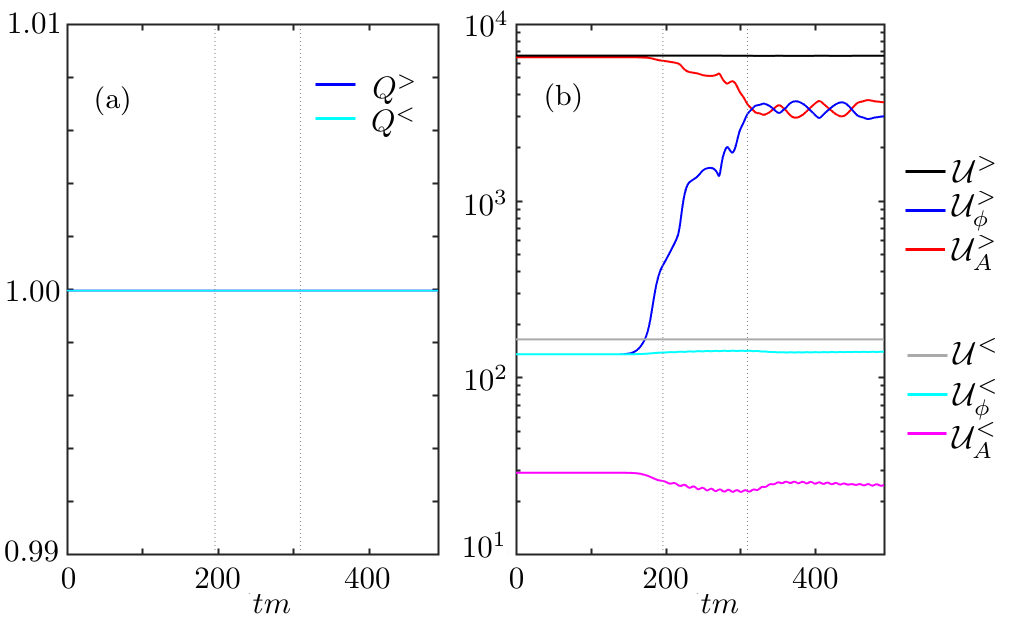}
	\caption[Evolution of total charge and total energy]{Evolution of total charge (a) and total energy (b), when periodic boundary conditions are used. The total charge remains constant up to the machine precision, both when $E<E_c$ (cyan), where little pairs are produced, and when $E>E_c$ (blue), where copious pairs are produced. When $E<E_c$ is below the Schwinger field, a small amount of energy is transfered from the electromagnetic field (magenta) to the charged field (cyan) due to wakefield acceleration and plasma wave excitation, while the total energy (gray) remains constant. In contrast, when $E>E_c$, a large amount of laser energy (red) is consumed by pair production. The energy of the charged field (blue) significantly increases until pair production and annihilation roughly balance. During this process, the total energy (black) remains constant. The total charge $Q^{n+1/2}=\sum_s J_{s}^{n+1/2}$ is normalized by the total ion charge, and the total energy $\mathcal{U}^{n+1/2}=\sum_s\mathcal{H}_{s}^{n+1/2}$ is normalized by $m^3/\Delta x$. The vertical dashed gray lines mark the time when the laser pulse center enters and leaves the plasma plateau boundaries.}
	\label{fig:total}
\end{figure}

With the above setup, the laser pulse simply travels through the plasma with some refraction and reflections when the laser field is weak ($a\ll 1$). More interesting phenomena happen when the laser field becomes strong. For example, when $a\approx1$ is relativistically strong but the resulting $E_x\approx0.3E_c$ is below the Schwinger field, the simulation recovers what happens in classical plasmas \citep{McKinstrie96,Naumova04,Geyko09}. First, let us look at what happens to charged particles. After the laser enters the plasma, beams of ``electrons" are formed in the forward direction by both ponderomotive snow-plow and laser wakefield acceleration. At the same time, some ``electrons" are splashed in the backward direction from strongly-driven plasma boundaries (Fig.~\ref{fig:part}a,~c). 
Next, for the laser pulse, its center (solid black lines) and half widths (dotted black lines) are well-traced by geometric optics in the $xt$ space (Fig.~\ref{fig:field}a), as well as in the $kt$ space (Fig.~\ref{fig:field}c, dashed white line), because the background plasma is smooth on the laser wavelength scale. Beyond geometric optics, as the laser travels through the plasma slab, ponderomotive expulsion of ``electrons" cause the laser pulse to adiabatically loose a small amount of energy in the form of frequency redshift $\omega<\omega_0$ (Figs.~\ref{fig:field}a,~c and \ref{fig:total}b). In addition, the laser excites plasma waves, from which the laser is Raman-scattered in both forward and backward directions. In the insert of Fig.~\ref{fig:field}c, the final spectrum (red) shows distinctive Raman scattering peaks at $\omega+n\omega_p$ up to $n=8$, and second harmonics peaks $2\omega$ and $2\omega+\omega_p$ in the forward direction. In the backward direction, peaks at $\omega-\omega_p, \omega, \omega+\omega_p$ and $2\omega$ can also be identified unambiguously.

When laser field is increased beyond the Schwinger threshold ($a_c=m/\omega$). For example, when $a\approx16$ ($E_x\approx5E_c$), a large amount of $e^-/e^+$ pairs are produced (Figs.~\ref{fig:part}b,~d). A very small fraction of pairs are produced and trapped in the laser wakefield, forming low-luminosity ``electron" (negative charge density, blue) and ``positron" (positive charge density, red) beams that leave the plasma slab from its right boundary. On the other hand, a much larger fraction of pairs are produced when the backscattered EM wave, whose intensity is near the Schwinger threshold (Fig.~\ref{fig:field}b), interacts with forward-propagating plasma waves. ``Positrons" produced in this way form high-luminosity collimated beams, leaving the plasma slab from its left boundary. Apart from these beams, many ``positrons" never manage to leave the plasma slab. These trapped ``positrons" have large probabilities to annihilate with ``electrons" in the highly constrained 1D phase space. 
Due to pair creation and particle acceleration, the laser initially looses a significant amount of energy, until pair creation and annihilation roughly balance (Figs.~\ref{fig:field}b,~c and \ref{fig:total}b). At that point, the $k$ spectrum of the laser is substantially broadened (Fig.~\ref{fig:field}d). Such a spectral broadening is expected from general wave action considerations \citep{Wilks88,Dodin10}, which predict frequency upshift due to pair creation, and frequency downshift due to pair annihilation and plasma expulsion. In the insert of Fig.~\ref{fig:field}d, the final EM wave spectrum (red) shows distinctive annihilation bumps near integer multiples of ``electron" rest mass. These annihilation peaks are very broad since ``electrons" and ``positrons" annihilate with large kinetic energy. Finally, notice that no pair is produced when the laser travels through the vacuum region, which is expected in 1D. It is remarkable that very rich physics can already be captured by simply solving the classical field equations with proper initial and boundary conditions.

To extract observables from simulations, the charge density (Figs.~\ref{fig:part}a,~b) is computed using Eq.~(\ref{eq:charge}), which includes no contribution from background ions. Therefore, negative charge (blue) indicates ``electron" density in excess of ``positron" density, whereas positive charge (red) indicates the contrary. The energy density of the charged field (Figs.~\ref{fig:part}c,~d) and the EM fields (Figs.~\ref{fig:field}a,~b) are computed using Eqs.~(\ref{eq:Hphi}) and (\ref{eq:HA}), respectively. To compute the $k$ spectra of EM waves (Figs.~\ref{fig:field}c,~d), notice that a monochromatic EM wave satisfies $k_xE_y=\omega B_z$. 
Upon discretization, this relation remains exactly satisfied if we take $k_x=\sin(k\Delta x)/\Delta x$ and $\omega=2\tan(\omega_k\Delta t/2)/\Delta t$, where $\omega_k>0$ is the positive solution of the local numerical dispersion relation $4\sin^2(\omega_k \Delta t/2)/\Delta t^2=4\sin^2(k \Delta x/2)/\Delta x^2$. In the discrete version of $k_xE_y=\omega B_z$, it is necessary that we take $E_y=E_{s+j/2}^{n+1/2}$, and center $B_z$ on time-like faces $B_{r-k/2-i/2}^{n+1/2}=(B_{r-k/2}^{n}\!+\!B_{r-k/2-i}^{n}\!+\!B_{r-k/2}^{n+1}\!+\!B_{r-k/2-i}^{n+1})/4$. A similar relation holds for the $E_z$ and $B_y$ components, which are subdominant now that the laser is linearly polarized. Using these momentum-space Faraday's law, the $k$ spectrum of right-propagating EM waves ($k>0$) and left-propagating EM waves ($k<0$) can be separated from the spatial Fourier transforms of electric and magnetic fields.

Results presented in Figs.~\ref{fig:part}-\ref{fig:total} are averaged over an ensemble of 200 simulations with statistically equivalent initial conditions. The ensemble average starts to show convergence for tens of realizations. In these simulations, temporal gauge $A^0=0$ is used, and periodic boundary conditions are employed for both $\phi_v$ and $A_e$. The resolutions $mdx=0.04$ and $mdt=0.005$ are chosen high enough so that the fastest dynamics is resolved and the simulation results converge. The 1D box is large enough such that the laser does not transit the spatial domain before the simulations are terminated.

In the above numeric examples, the total charge $Q^{n+1/2}=\sum_i J^{n+1/2}_i$ is constant up to the machine precision (Fig.~\ref{fig:total}a), both when the laser field is below ($Q^<$) and above ($Q^>$) the Schwinger field. 
Although the total energy $\mathcal{U}^{n+1/2}\!=\!\sum_i\! \mathcal{H}^{n+1/2}_i$, whose error 
is of order $O(en\Delta t^2)$, is not exactly conserved, the resolution is chosen high enough such that 
the total energy fluctuates up to 6 ppm and $0.2 \%$ when the laser field is below ($\mathcal{U}^<$) and above ($\mathcal{U}^>$) the Schwinger field, respectively. The roughly constant amount of energy is redistributed among the classical fields $\phi_0$ and $\bar{A}$ (Fig.~\ref{fig:total}b) when the laser interact with the plasma.
By solving the classical field equations, the transition from laser wakefield acceleration to Schwinger pair production has thus been demonstrated for the first time.

\bookmarksetup{startatroot}
\chapter{Conclusion and discussion\label{ch:conclusion}}

\section{Thesis highlights}

In the first part of this thesis, I study the effects of strong background magnetic fields on three-wave interactions, which are important when electron gyro frequency is not negligible when compared to the plasma frequency. 
A comprehensive understanding is obtained by solving the fluid-Maxwell's equations to second order using a multiscale expansion. The resultant second-order electric-field equation [Eq.~(\ref{eq:E2s})] takes a very intuitive form, in which quasimodes develop and linear eigenmodes evolve due to three-wave interactions.   
Since resonant three-wave interactions conserve wave actions, 
the second-order electric-field equation can be reduced to the three-wave amplitude equations. 
Although the three-wave equations are well-known, it contains an essential coupling coefficient, whose general formula was not known when background magnetic fields are present. In this thesis, a convenient formula for the coupling coefficient is obtained for the first time [Eq.~(\ref{eq:coupling})], which can be readily evaluated for any three resonant waves propagating at arbitrary angles in the magnetic field (e.g. Fig.~\ref{fig:Stokes}).
In addition to its practical significance, the general formula is also aesthetically satisfying. Using the Lagrangian formulation [Eq.~(\ref{eq:Lagrangian})], I demonstrate, for the first time, that the scattering strength 
can be represented as $3!=6$ ways of contacting a single Feynman diagram [Eq.~(\ref{eq:w3})], which is nonvanishing only when background plasmas are present. 

As an application of three-wave interactions in magnetized plasmas, I consider laser pulse amplification mediated by magnetized plasma waves. For example, when mediated by the upper-hybrid wave, it is possible to use more controllable background magnetic fields to replace the less  controllable internal plasma density to achieve better performance of plasma-based laser amplification. 
Although the amplification rate is reduced in less dense plasmas, what is of critical importance is that the competing effects are reduced even more (Sec.~\ref{sec:compression:limits}). 
In particular, the modulational instability is relatively suppressed, resulting in longer allowable amplification time and thereof higher output pulse intensity. Moreover, both collisionless and collisional dampings are relatively suppressed, which enable efficient pulse compression also for shorter-wavelength lasers.
In other words, using magnetized plasma mediation, we can significantly expand the operation window and achieve efficient pulse compression for higher-frequency and lower-intensity pumps (Fig.~\ref{fig:regime}) to produce laser pulses of higher final intensity (Table~\ref{table:parameters}).
Even for lasers that can already be compressed using unmagnetized plasmas, applying a magnetic field improves pulse intensity (Fig.~\ref{fig:PIC}) and relaxes the engineering requirements of producing high and uniform plasma densities.

In the second part of this thesis, I develop a relativistic quantum theory for plasmas, when fields are so strong that classical plasma models become invalid. 
In this new regime, I extend quantum field theory to incorporate plasma effects by adding the extra ingredient of dynamical background fields [Eq.~(\ref{ReducedLagrangian})]. 
In the extended theory, the lowest-order phenomena are linear waves, which can be described using an effective action approach. 
The wave effective action is computed to 1-loop level using path integrals [Eq.~(\ref{1-loop})], and a general formula is obtained for the first time whereby both the plasma response [Eq.~(\ref{bkPol})] and the vacuum response [Eq.~(\ref{vacPol})] can be described. 
Using this new formalism, the known dispersion relation in unmagnetized QED plasmas are recovered. 
Moreover, the effective action approach enables a useful general dispersion relation in strongly magnetized plasmas to be determined for the first time (Sec.~\ref{ch:mag:dispersion:oblique}). 
Relativistic-quantum modifications contained therein have already been observed near X-ray pulsars, where anharmonic cyclotron absorption features can now be associated with relativistic Bernstein waves (Fig.\ref{fig:PerpendicularDispersion}). 
More excitingly, laboratory tests of strong-field effects may already become possible in gigagauss magnetic fields, 
where Faraday rotation is predicted to have a different frequency dependence than expected classically (Fig.~\ref{fig:FaradayRotation}).
My thesis thereof provides a theoretical basis, when QED plasma effects become relevant in the presence of strong fields.

Beyond analytical theory, to 
simulate effects such as laser pair production, I extend real-time lattice QED to become a unique tool for plasma physics for the first time. 
In the classical-statistic regime, the behaviors of relativistic quantum plasmas are adequately described by solving the classical field equations. By discretizing the action in a way that respects both the local U(1)-gauge symmetry and the structures of discrete exterior calculus (Sec.~\ref{sec:simulation:variation:discret}), I develop a variational algorithm for solving the classical field equations with good conservation properties (Sec.~\ref{sec:simulation:conservation}). 
By affording a much higher resolution than needed classically, the numerical scheme is much simpler than standard methods for simulating classical plasmas, and may be parallelizable using quantum computing in the future. 
The numerical scheme easily recovers the spectrum of linear ways including the pair mode (Fig.~\ref{fig:Dispersion}). Moreover, the scheme can be used to simulate laser-plasma interactions (Figs.~\ref{fig:part}-\ref{fig:total}). 
When the laser intensity is relativistically strong, the scheme recovers well-known phenomena, such as parametric instability, harmonic generation, and wakefield acceleration. 
Beyond the applicability of classical models, when the laser reaches quantum strength, my scheme naturally captures new phenomena, such as Schwinger pair production by strong electric fields and gamma-ray lasing during recollisions of electron-positron pairs.

\section{Future Work}


In the classical regime, both the fundamental wave-wave interactions and their implications are open grounds for further investigations. 
As a fundamental physical phenomenon, magnetized wave-wave interactions remain to be thoroughly charted using theories, simulations, and experiments. 
In terms of theory, an obvious next step for three-wave interactions is to incorporate thermal effects, using the warm fluid model and then the kinetic model. It remains to be verified that the general formula for three-wave coupling coefficient, which is expressed in terms of the linear susceptibility, remains valid. 
To the next-order, four-wave interactions in magnetized plasmas can be studied either by solving equations or expanding the Lagrangian to the next order.  
In another direction, wave-wave interaction can be analyzed analytically in inhomogeneous medium, either when there is a weak gradient or when there are statistical fluctuations. 
The aforementioned analytical theories remain to be confirmed by detailed numerical simulations in the multidimensional parameter space, which provide valuable verifications that the effects being considered in theories are the dominant effects in the problems. 
Ultimately, wave-wave interactions in magnetized plasmas should be studied experimentally, which is particularly relevant to laser-driven inertial fusion where magnetic fields are imposed to enhance the confinement.
In applications such as magnetized inertial confinement, wave-wave interactions are usually considered deleterious effects that need to be mitigated. However, by exploiting the effects, one may be able to utilize them to design experiments such that laser plasma coupling are optimized.
Moreover, wave-wave interactions can be utilized to produce powerful lasers beyond the attainment of current technologies. For pulse compression, mediations by other hybrid waves, the MHD waves, and the Bernstein waves remains to be analyzed and compared. 
Beyond pulse compression, magnetized plasmas can be used to mediate four-wave mixing and harmonic generation. These interactions have particularly large cross section using cyclotron resonances. The existence of multiple tunable resonances and the capability of sustaining high power make multi-species magnetized plasmas promising media for next-generation lasers.

In the relativistic-quantum regime, strong-field plasma physics is again open ground for novel theories, numerical schemes, and experimental tests. 
An obvious extension to what has been done in this thesis is to study spinor-QED plasmas, which are constituted of fermions instead of bosons. The Fermi statistics changes the nature of the background fields, whereby the fields become anti-commuting. 
In addition to changing particle statistics, the QED plasma theory can be solved to higher orders to describe effects beyond linear waves. For example, due to the presence of plasmas, virtual photons that mediate collisions between particles are modified. Consequently, interactions between charged particles are altered by the plasma-dressing effects, and the modified cross sections of many phenomena, such as pair annihilation, remains to be calculated.
Apart from the dressing effects during particle interactions, wave-wave interactions in relativistic quantum plasmas also remain to be studied. 
In parallel to analytical theory, simulation capabilities remain to be developed to capture nonperturbative effects in QED plasmas, which can then be applied to study many interesting phenomena. 
One direction is to develop higher-order algorithms, which relax the resolution requirement for given error tolerance so that long-time dynamics in three-dimensional space are affordable. 
Another direction is to develop schemes that can capture next-to-leading-order effects in path integrals, which are not incorporated by simply solving the classical field equations. 
Moreover, the numerical schemes, which runs on nowadays supercomputers, remains to be adapted to quantum computers, by exploiting the fact that lattice QED can be intrinsically mimicked by a lattice of quantum particles in the quantum computer. 
Relativistic quantum effects predicted by theory and simulations remains to be tested by observations and experiments.
In particular, for neutron stars where spectral data is being collected, quantitative connections between QED plasma theory and observed spectral features remains to be built by developing radiative transfer models, which may enable remote sensing of neutron star atmospheres in the future.
In laboratory conditions where gigagauss magnetic fields become feasible, tests of basic predictions of QED plasma theory remain to be conducted. 
Last but not least, relativistic-quantum plasmas need not be limited to regular plasmas, in which interactions are electromagnetic. Beyond quark-gluon plasmas \citep{Berges2015nonequilibrium}, where interactions are mediated by the strong force, and neutrino plasmas \citep{Kuznetsov13}, where interactions are mediated by the weak force, it is plausible to study dark matter plasmas, in which interactions are mediated by yet unknown forces. 
If dark matter really exists, as evidenced by astrophysical observations, then we may be living inside a dark-matter plasma, which may have observable consequences that are yet to be discovered.  
\section{Suggested experiments}

Although this thesis focuses on theories and simulations of plasma physics in the strong-field regime, I would also like to speculate a number of experiments that might be feasible in the near future. A series of experiments will be necessary to confirm or refute basic phenomena predicted in this thesis. Once these fundamental phenomena are understood, they can then be exploited in various applications.

\subsubsection{Laser scattering from magnetized targets}
\begin{figure}[!t]
	\renewcommand{\figurename}{FIG.}
	\centering
	\includegraphics[angle=0,width=0.55\textwidth]{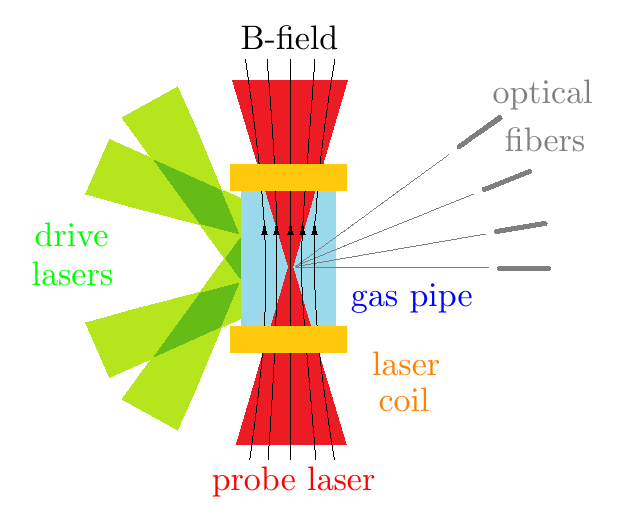}
	\caption[Experimental setup measuring laser scattering in magnetized plasma]{An experimental setup for measuring laser scattering in a magnetized plasma. The plasma target is an imploding gas pipe, which is driven by azimuthally symmetric drive lasers and magnetized by a pair of laser coils. The probe laser, whose pulse duration is much shorter than the implosion time scale, is focused at the center of the plasma to measure spatially and temporally localized scattering. The angle-dependent scattering signals are collected by an array of optical fibers and transmit to a spectrometer, which is not depicted.}
	\label{fig:Exp_scattering}
\end{figure}

To design implosion experiments where both magnetic fields and lasers are present, it is imperative to understand how lasers might scatter from the magnetized plasma target. A general formula for laser scattering at arbitrary angles is provided in this thesis, whose validity should now be tested by taking measurements in well-controlled environments. 
A basic experimental setup (Fig.~\ref{fig:Exp_scattering}) involves a magnetized plasma target, a pump laser, and a spectrometer. 
The magnetized plasma target may be produced by pulsed power devices or laser-driven coils, whereby the magnetic field is tunable, and the plasma parameters can be characterized. 
Suppose the plasma and the magnetic field are uniform and stationary on the scale of the pump laser, then Eq.~(\ref{eq:coupling}) is applicable. 
Theoretical predictions of the scattering spectra can then be compared with the experimental spectra, taken at various angles under a set of plasma conditions. 

For example, consider a plasma target produced by imploding a magnetized gas pipe, which is driven radially by 351-nm lasers with $\sim 1$ ns duration, where a seed magnetic field of $\sim 0.1$ MG is provided by a pair of laser-driven coils in a quasi-Helmholtz geometry.
During the implosion, the gas is ionized and the plasma is compressed, which amplifies the frozen-in magnetic field by roughly the convergence ratio squared. Suppose the convergence ratio is $\sim 10$, then the gas pipe with initial diameter $\sim 1$ mm is compressed to $\sim 0.1$ mm in size, and the magnetic field is amplified to $\sim 10$ MG. With an initial fill pressure $\sim 10$ Torr at room temperature, the final plasma density is $\sim 10^{19}\, \text{cm}^{-3}$ and the temperature is \mbox{$\sim 10$ eV}.  
In this final state, the plasma frequency \mbox{$\omega_p\sim 0.1$ eV}, the Debye length \mbox{$\lambda_D\sim10$ nm}, and the electron gyro frequency $\Omega_e\sim 0.1$ eV is on the same order of the plasma frequency.
 
In such a magnetized plasma target, coherent scattering can be measured using a \mbox{1053-nm} probe laser with picosecond durations, for which the plasma is transparent, uniform, and stationary. 
The scattering is coherent because the laser wavelength is much larger than the Debye length. Moreover, the plasma is quasi-uniform in the absence of fine-scale structures, because the laser wavelength is much smaller than the size of the plasma. Finally, the implosion is quasi-stationary, because the pulse duration is much shorter than the implosion time scale. 
To localize the scattering signal, we can propagate the probe laser along the axis of the gas pipe and focus the laser on the plasma center. To collect the scattered light, an array of optical fibers can be placed in the far field and pointed at various angles towards the focal region. The scattered light is then transmitted to a spectrometer, whose necessary spectral range is from $\sim$ 600 nm to $\sim$ 1500 nm in order to capture electron-scale features, and resolution is $\sim 0.1$ nm in order to resolve ion-scale features.  
Since features of low-mass ions are easier to resolve, the filling gas is preferably helium or hydrocarbon. 
The measured spectra can then be compared with predictions from the analytical formula. Evaluating of the formula requires diagnosing the plasma parameters. The magnetic field may be measured using Zeeman effect of ionic lines, and the plasma density and temperature may be measured using Stark broadening together with an X-ray framing camera.

Measurement errors of local plasma parameters translate to uncertainties of the analytical spectra. Roughly speaking, the scattering intensity is proportional to the plasma density, so the intensities of spectral lines are sensitive to density by $\delta n_0/n_0$. 
The frequency shifts of the spectral lines are algebraic functions of $\omega_p$ and $\Omega_s$. When $\omega_p$ and $\Omega_e$ are of the same order, then, roughly speaking, electron-scale lines are sensitive to both the density by $\sqrt{\delta n_0/n_0}$ and to the magnetic field by $\delta B_0/B_0$, while ion-scale lines are only sensitive to the magnetic field by $\delta B_0/B_0$. 
Finally, since the plasma temperature is low, thermal effects on the order of $v_Tk\sim10^{-2}$ eV are small. Therefore, the spectra are unlikely to be sensitive to uncertainties of the temperature measurements.

The largest source of noise perhaps comes from fine-scale structures, which may evolve on faster time scales than the implosion process. The analytical formula is not applicable, whenever the homogeneous and stationary assumptions are not satisfied. Nevertheless,  using the X-ray framing camera, one may be able to select shots where the formula is indeed applicable. To suppress fine scale structures, the implosion may be controlled by changing the gas fill pressure, the seed magnetic field, and the drive laser pulse shape. 
Alternatively, without changing the implosion process, we can launch a train of probe pulses during the implosion, and select the time slices where the plasma is uniform and stationary on the probe laser scales. 
Given an implosion trajectory, the noise may be reduced using a shorter probe pulse focused more tightly inside the plasma target, and aligning all diagnostics well within the focal region.

\subsubsection{Interaction of two lasers in magnetized plasmas}
In the second experiment, we can stimulate three-wave interactions with a seed laser, unlike in the first experiment where the scattering is spontaneous. 
When the seed has comparable duration as the pump, this scenario gives the simplest setup for studying cross-beam energy transfer, which commonly occurs in laser-driven inertial confinement experiments. 
On the other hand, when the seed has much shorter duration than the pump, this setup naturally leads to the application of laser pulse compression, during which the weak seed pulse gains energy from the intense pump laser.
While interactions between two lasers in unmagnetized plasmas have been investigated intensively, what happens in a magnetized plasma remains largely unknown until work presented in this thesis, whose theoretical predictions should now be tested experimentally.

\begin{figure}[!t]
	\renewcommand{\figurename}{FIG.}
	\centering
	\includegraphics[angle=0,width=0.95\textwidth]{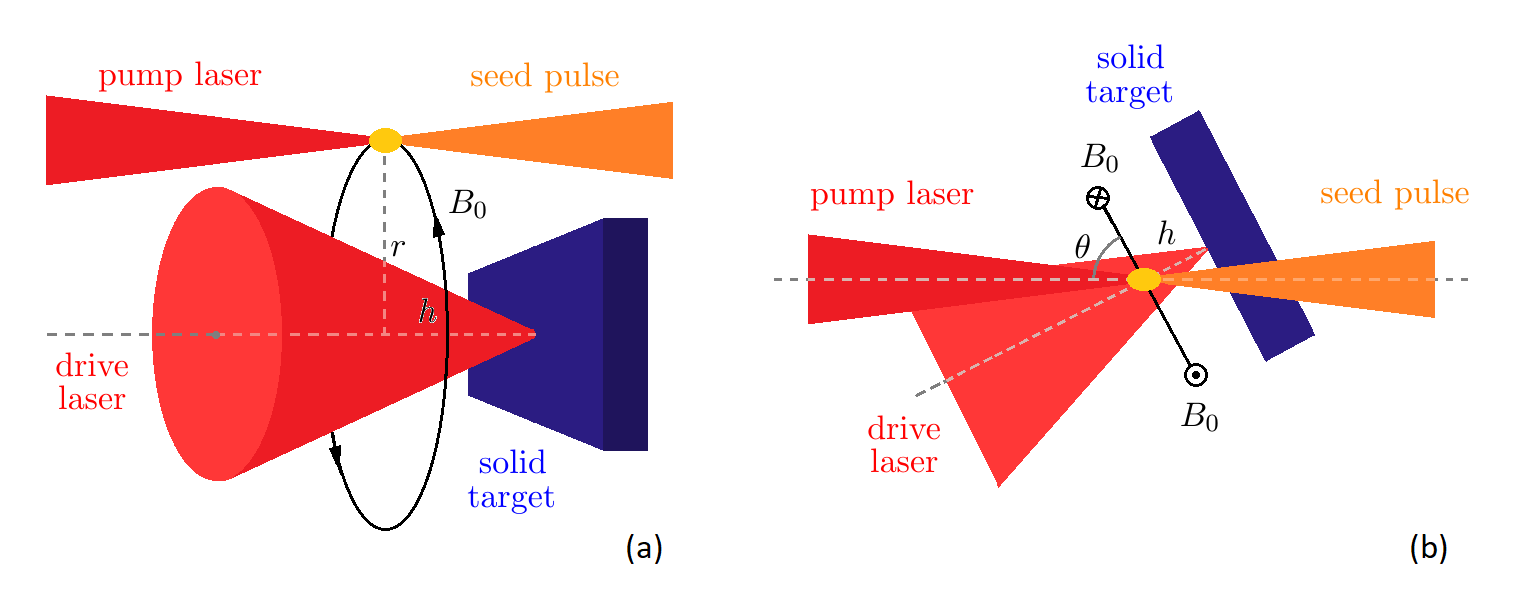}
	\caption[Experimental setup for laser amplification in magnetized plasma]{Side view (a) and top view (b) of an experimental setup for laser amplification in a magnetized plasma. 
		The plasma is produced by ablating a solid target with a drive laser at time $t_0$. In the coronal plasma (not depicted here), a region can then be chosen to mediate energy transfer from a pump laser to a seed pulse, which overlap in the interaction region at time $t_1$. 
		For given laser and target conditions, plasma parameters may be tuned to match the resonance condition by either changing the time delay $t_1-t_0$ or moving the solid target, which is of radial distance $r$ and axial distance $h$ away from the interaction region. 
		Plasma parameters including density and magnetic field may be diagnosed using interferometry and polarimetry. 
		For fixed plasma parameters, the optimal angle $\theta$ for achieving the maximum coupling may be computed analytically. In experiments, this angle may be scanned by rotating the solid target.}
	\label{fig:Exp_amplification}
\end{figure}

Consider laser pulse amplification, whose basic experimental setup involves a magnetized plasma, a long pump laser, and a short seed pulse (Fig.~\ref{fig:Exp_amplification}). 
A well-conditioned plasma target may be provided by imploding a magnetized gas pipe as discussed before. Alternatively, a much cheaper and lower-quality target may be provided by ablating a solid surface with a single drive laser. 
When the drive laser impinges on a planar solid surface, a dense plasma jet can form in the backward direction, which is spontaneously magnetized in the azimuthal direction. For example, by focusing a 100-J and 1-ps laser to a $\sim 10\,\mu$m spot, the drive laser reaches an intensity of $\sim10^{20}\,\text{W/cm}^2$. When such an intense laser hits a solid surface, a coronal plasma of $\sim 100\,\mu$m in size can form \citep{Borghesi98,Chatterjee2017micron}. After the initial rapid expansion, the plasma becomes uniform on 10-$\mu$m scale and stationary on 10-ps scale. 
The plasma density decays from $\sim10^{22}\,\text{cm}^{-3}$ near the solid surface to $\sim10^{19}\,\text{cm}^{-3}$ at $\sim 10\,\mu$m away from both the axis of the plasma jet and the solid surface. Around the same toroidal region, the spontaneously generated magnetic field peaks at $\sim 100$ MG, and the plasma temperature is $\sim 10$ eV.
The exact plasma density and magnetic field may be diagnosed using a combined interferometry and polarimetry technique for a given target and drive laser. Thereafter, we may select a volume within the coronal plasma as the interaction region.
Although such an interaction volume is not sufficiently uniform and stationary for efficient laser pulse compression, it may already be sufficient to demonstrate energy transfer from a long pump laser to a short seed pulse.

The aforementioned plasma target can be used to amplify lasers in the 1-$\mu$m range. 
Suppose we focus a 1054-nm pump laser with $\sim$1 ns duration and $\sim1$ J energy in a $\sim 10\,\mu$m interaction region, then the pump intensity is $\sim 10^{15}\,\text{W/cm}^2$. 
The picosecond seed pulse can have much lower energy $\sim 0.1$ mJ. After focusing it in the same interaction region, the seed intensity $\sim 10^{14}\,\text{W/cm}^2$ is much smaller than to the pump intensity.
There are four choices of the seed wavelength such that the two lasers couple resonantly through the four waves in a two-species plasma. The required wavelength shift of the seed pulse depends on plasma parameters.
For example, in an interaction region where the coronal plasma has density $\sim10^{19}\, \text{cm}^{-3}$, temperature $\sim 10$ eV, and magnetic field $\sim 100$ MG, coupling through the slow MHD wave and the kinetic Alfv\'en wave require little wavelength shift. On the other hand, coupling though the lower-hybrid wave requires a wavelength shift of \mbox{$\sim 100$ nm}, which can be achieved using a Raman cell. Coupling through the upper-hybrid wave requires a large shift of $\sim 1000$ nm, which may be achieved using a down-conversion crystal combined with a Raman cell. 
For a given frequency shift, the resonance condition may be satisfied by scanning plasma parameters. The parameter scan may be achieved by either choosing a different region or a different time delay in the expanding coronal plasma where the interaction takes place. 

Since the pump and seed meet only in a very small region, their interaction is largely linear. The short seed grows exponentially in the long pump by an amount determined by the linear growth rate, which can be calculated analytically and compared with experimental results. 
For example, in the plasma condition considered above, theory predicts that the largest growth rate for slow MHD wave mediation occurs at $\theta\approx80^\circ$, where $\theta$ is the angle between the counter-propagating lasers and the background magnetic field. The maximum growth rate is about half the Raman growth rate in a unmagnetized plasma with the same density, which is $\sim 10\, \text{ps}^{-1}$. 
Second, for Alfv\'en wave mediation, the largest growth occurs over a wide range of angles around $\theta\approx20^\circ$, with the growth rate about five times that of Raman. 
Third, for lower-hybrid wave mediation, the growth rate peaks near $\theta\approx60^\circ$ at about one third of the Raman growth rate.
Finally, for upper-hybrid wave mediation, the maximum growth rate is comparable to that of Raman, and peaks near $\theta\approx50^\circ$.
The exact growth rates depend rather sensitively on plasma parameters, which need to be diagnosed in the experiment and are likely to have large uncertainties. 
Although a quantitative comparison between the theory and experiments may be overly ambitious in this setup, qualitative results of laser pulse amplification in magnetized plasmas may already be interesting as a proof-of-principle demonstration of a new technology. 

\subsubsection{Measure relativistic-quantum modifications through Faraday rotation}
When hundreds-megagaus or even gigagauss magnetic fields become available, we can use them to test strong-field relativistic quantum effects. For example, this thesis predicts that Faraday rotation can be modified substantially in strong fields (Fig.~\ref{fig:FaradayRotation}). 
Consider an experiment where we prepare a quasi-static and quasi-uniform plasma target, and pass multiple linearly-polarized lasers with different frequencies along the magnetic field to measure Faraday rotation. 
Since the lasers are collimated and pass the same plasma at the same time, results of the magnetic-field-strength measurements should be independent of the frequency of the diagnostic lasers. 
However, this will not be the case when the classical formula is used to interpret the results in the relativistic-quantum regime, where the correct interpretations should be given using Eq.~(\ref{eq:nRL}) instead. 

\begin{table}[b]
	\centering
	\begin{tabular}{c|c|c|ccc}
		\toprule
		\multirow{2}{*}{$B_0$ (GG)}& \multirow{2}{*}{$n_e (\text{cm}^{-3})$} & \multirow{2}{*}{$\omega_R$ (eV)} &\multicolumn{3}{c}{$(\lambda\partial_z\theta,\delta)\;@\; \Delta\omega$}\\
		& & &0 eV &$10^{-4}$ eV &$10^{-3}$ eV\\
		\midrule
		\multirow{3}{*}{0.1} 
		& $10^{19}$ & 1.16 & $(176.90^\circ,1.48\%)$ & $(163.12^\circ,0.36\%)$ & $(130.00^\circ,0.04\%)$ \\
		& $10^{17}$ & 1.15 & $(154.50^\circ,16.50\%)$ & $(59.38^\circ,4.15\%)$ & $(10.28^\circ,0.13\%)$ \\
		& $10^{15}$ & 1.15 & $(41.44^\circ,3.34)$ & $(1.10^\circ,13.02\%)$ & $(0.11^\circ,0.13\%)$ \\
		\hline
		\multirow{3}{*}{\textcolor{ForestGreen}{0.2}} 
		& $10^{19}$ & 2.31 & $(172.55^\circ,4.24\%)$ & $(156.14^\circ,1.81\%)$ & $(112.63^\circ,0.25\%)$ \\
		& \textcolor{ForestGreen}{$10^{17}$} & \textcolor{ForestGreen}{2.30} & \textcolor{ForestGreen}{$(114.19^\circ,57.63\%)$} & \textcolor{ForestGreen}{$(37.01^\circ,21.85\%)$} & \textcolor{ForestGreen}{$(5.35^\circ,0.52\%)$} \\
		& $10^{15}$ & 2.30 & $(9.16^\circ,18.67)$ & $(0.53^\circ,51.39\%)$ & $(0.06^\circ,0.52\%)$ \\
		\hline
		\multirow{3}{*}{0.4} 
		& $10^{19}$ & 4.60 & $(159.46^\circ,12.87\%)$ & $(142.64^\circ,8.05\%)$ & $(90.15^\circ,1.33\%)$ \\
		& $10^{17}$ & 4.60 & $(52.94^\circ,2.40)$ & $(19.05^\circ,1.02)$ & $(2.70^\circ,2.08\%)$ \\
		& $10^{15}$ & 4.60 & $(1.33^\circ,134.71)$ & $(0.23^\circ,2.04)$ & $(0.03^\circ,2.07\%)$ \\
		\bottomrule		
	\end{tabular}
	\caption[Example parameters using Faraday rotation to measure QED effects]{Relativistic-quantum corrections to Faraday rotation are larger for higher field $B_0$ and lower density $n_e$, and decreases rapidly when the frequency of the probe laser increases $\Delta\omega$ above the \textit{R} wave cutoff $\omega_R$. For parameters discussed in the text (green), Faraday rotation per vacuum wavelength $\lambda\partial_z\theta$ deviates from classical expectation by $\delta\gtrsim 10\%$, when the laser frequency is \mbox{$\lesssim 10^{-4}$ eV} above the cutoff.}
	\label{table:Faraday}
\end{table}

To see what experimental conditions are necessary to observe relativistic-quantum corrections, 
notice that the feasible magnetic field is $B_0\lesssim 1$ GG with current experimental techniques. Such a magnetic field strength makes the relativistic parameter $\Omega_e\hbar/m_ec^2\lesssim 10^{-5}$ a rather small number. Since the corrections diminish rapidly towards $\Omega_e\hbar/m_ec^2$ when the EM wave frequency increases above the cutoff frequency $\omega_R$, 
the best way to observe the corrections is perhaps by employing a probe laser whose frequency is right above $\omega_R$.
Near $\omega_R$, relative-quantum corrections are boosted and can be of order unity, especially when the magnetic field $B_0$ is large and the plasma density $n_e$ is small. 
Consider the example parameters highlighted in Table~\ref{table:Faraday}, where $B_0\sim0.2$ GG and $n_e\sim10^{17}\,\text{cm}^{-3}$. Then, right at the cutoff $\Delta\omega=0$ eV, Faraday rotation per vacuum wavelength $\lambda\partial_z\theta\approx 114.19^\circ$, and the relativistic-quantum correction is as large as $\delta\sim 57.63\%$. Such a large correction is measurable if the experimental uncertainty $\lesssim 10\%$, provided that we know the values of $B_0$ and $n_e$ exactly.

\begin{figure}[!t]
	\renewcommand{\figurename}{FIG.}
	\centering
	\includegraphics[angle=0,width=0.8\textwidth]{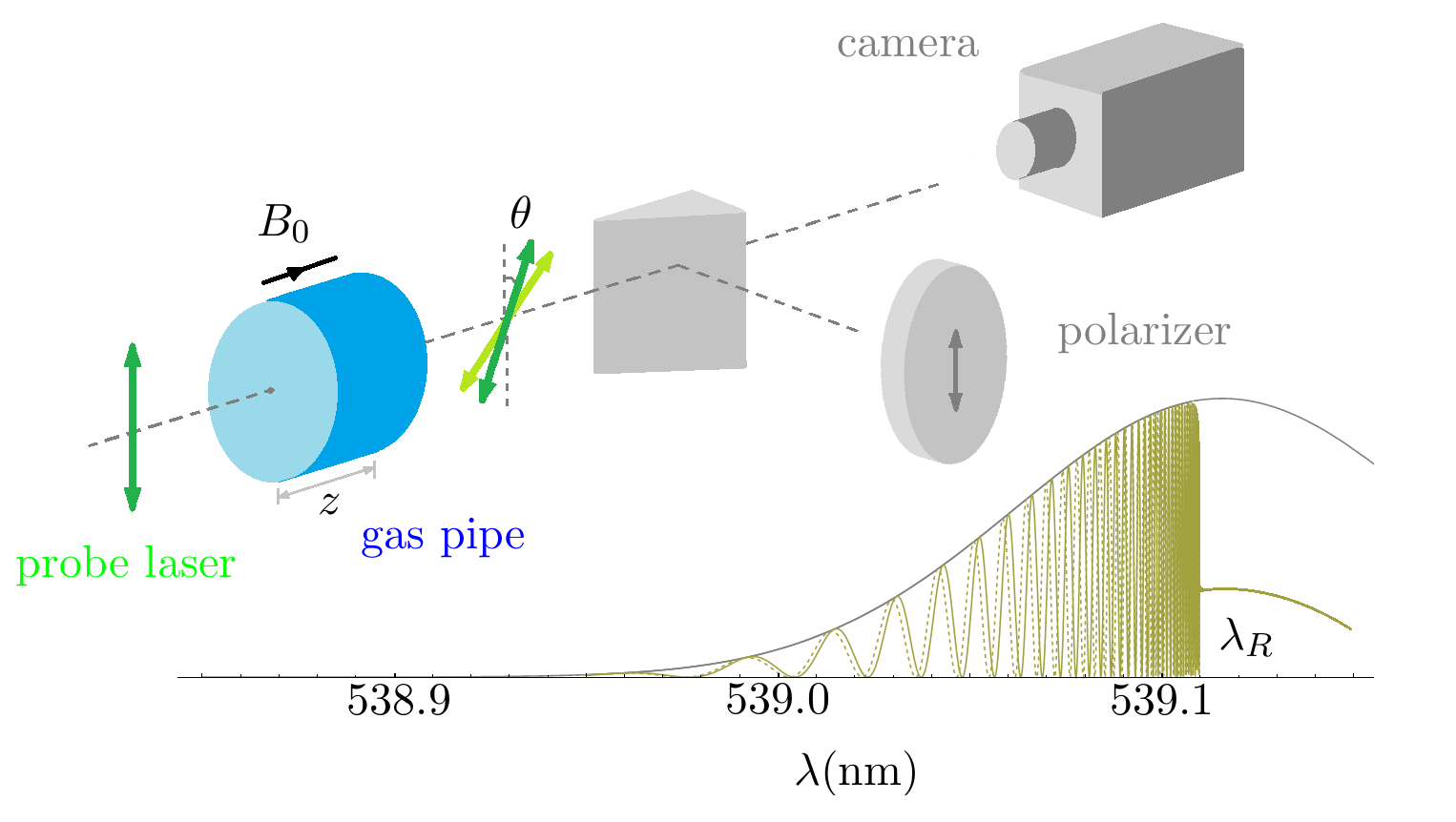}
	\caption[Experimental setup measuring QED corrections to Faraday rotation]{Faraday rotation near the R-wave cutoff $\lambda_R$ may be used to measure relativistic-quantum corrections to the wave dispersion relation in strongly magnetized plasmas. During the flux compression, the plasma becomes opaque to the probe laser when the magnetic field and plasma density exceed their critical values. Right before the plasma becomes opaque, the spectrum of the transmitted laser, which is sent through a polarizer, shows a fringe patten. In the regime $\Omega_e\gg\omega_p$, the cutoff $\lambda_R$ is mostly sensitive to the background magnetic field $B_0$, while the spacing between the fringes depends on both $B_0$ and the electron density $n_e$. For a given distance of propagation inside the plasma $z$, one may fit the fringe pattern to determine the line-averaged $B_0$ and $n_e$. For given plasma parameters, the measured spectrum (solid green) deviates systematically from the classical spectrum (dashed green). }
	\label{fig:Exp_Faraday}
\end{figure}


In practice, both $B_0$ and $n_e$ need to be diagnosed. This can be achieved, for example, by utilizing the entire bandwidth of the probe laser. Consider a solid state probe laser with central wavelength $539.1$ nm and bandwidth $\sim 0.1$ nm. Then, within the bandwidth $\Delta\omega\sim10^{-4}$ eV, Faraday rotation diminishes above the cutoff to $\lambda\partial_z\theta\sim37.01^\circ$, which is significantly different from the rotation exactly at the cutoff. Moreover, the correction reduces to $\delta\sim 21.85\%$, which is still measurable within $\lesssim 10\%$ experimental uncertainty.
After sending the linearly polarized probe laser through the magnetized plasma of $z\sim50\,\mu$m in length, the transmitted laser, after passing through a polarizer, will have a spectrum similar to what is shown in Fig.~\ref{fig:Exp_Faraday}. The spectral intensity suddenly drops beyond $\lambda_R$, where the plasma is opaque to the \textit{R} wave, so that only the \textit{L} component of the linearly polarized probe laser can transmit the plasma. 
On the other hand, below $\lambda_R$, both the \textit{R} and the \textit{L} waves can transmit, whose phase velocity difference leads to Faraday rotation of the linearly polarized EM wave. Slightly below $\lambda_R$, Faraday rotation $\lambda\partial_z\theta$ changes rapidly with $\lambda$. Therefore, after propagating by a distance $z$, the EM wave with $\lambda_0$ may have rotated by $N$ cycles, while the wave with $\lambda_0-\Delta\lambda$ may have only rotated by $N-1$ cycles.  
Such a sensitive $\lambda$ dependence leads to a fringe patter in the spectrum, which becomes less densely spaced when $\lambda$ is further below $\lambda_R$. In the regime where $\Omega_e\gg\omega_p$, the fringe pattern is sensitive to both $B_0$ and $n_e$, while the cutoff $\lambda_R$ is only sensitive to $B_0$. Therefore, by fitting the fringe pattern, which can be resolve by a spectrometer with $\lesssim 1$-pm resolution in the range $\lambda\lesssim \lambda_R-50$ pm, both $B_0$ and $n_e$ can be determined experimentally. 
If the classical formula is used for the fitting, one will find that the transmitted EM wave has a narrower underlying line shape than the incident probe laser, unveiling systematic errors in the classical formula.

The required experimental precision is relaxed using a less dense plasma with a stronger magnetic field (bottom left corner of Table~\ref{table:Faraday}). On the other hand, the requisite precision is higher for a more dense plasma with a weaker field (top right corner of Table~\ref{table:Faraday}). 
The highlighted plasma parameters in Table~\ref{table:Faraday} might be feasible with the flux compression technique discussed earlier. For example, consider a gas disk with height $z\sim50\,\mu$m and radius $r\sim1$ mm. Suppose the initial gas fill is $\sim 10$ mTorr, then the plasma density reaches $\sim 10^{17}\,\text{cm}^{-3}$ after a radial implosion with convergence ratio $\sim 30$. 
Moreover, suppose we impose an initial magnetic field of $\sim 0.2$ MG with a pair of laser coils, then the final magnetic field is $\sim 200$ MG after the flux compression.

The experimental procedure is something like the following. We first drive the laser coils, and then implode the gas pipe. We can continuously monitor the implosion process, which happens on $\sim 1$ ns time scale, by illuminating the target with the probe laser along the $z$ direction and monitor the $\sim 1$ mm spot of the probe laser using a high speed camera (Fig.~\ref{fig:Exp_Faraday}).  
At the initial stage of the compression, the entire probe laser spot is bright. However, as $B_0$ and $n_e$ increase above their critical values, the center of the laser spot, where the compressed gas pipe is located, will become dark after the plasma is no longer transparent to the probe laser. 
Right before the plasma becomes opaque, we can record the spectrum with a spectrometer, after passing the transmitted light through a polarizer. The high-resolution spectrum then contains information necessary for measuring relativistic-quantum corrections.

To reduce noise and increase the experimental sensitivity, only light from a $\sim 10\,\mu$m region near the center should be sent to the spectrometer, while the background light outside the plasma target should be blocked. 
In addition, to reduce temporal blurring, the spectrum should be taken within a time period much shorter than the compression time scale. 
Moreover, to reduce Faraday rotation in the ambient plasma, the background pressure in the vacuum chamber should be kept low such that the plasma density outside the target is negligible. 
The influence of the wall material of the gas disk may be subtracted by comparing experiments with and without a gas fill. 
Finally, since the fringe pattern close to the cutoff is unlikely to be resolvable, it is favorable to use a probe laser with a fat but otherwise stable intensity profile. 


 
\section{Summary}

In this thesis, I study strong-field effects in plasma physics, in both the classical and the relativistic quantum regimes. 
In Ch.~\ref{ch:multiscale}, I review the multiscale-expansion technique, using which secular terms can be systematically removed from perturbative solutions of differential equations.
Multiscale expansion is then used to solve the cold fluid-Maxwell's equations to study the effects of strong magnetic fields on wave-wave interactions. To the first order (Ch.~\ref{ch:fluid-1st}), linear eigenmodes in magnetized plasmas are recovered from the modern perspective of linear operators. 
This perspective then leads to a tractable second-order solution (Ch.~\ref{ch:fluid-2nd}), based on which a general and yet convenient description of three-wave interactions is obtained for the first time. 
Having obtained the magnetized three-wave coupling coefficient in arbitrary geometry, the behaviors of resonant waves can then be found by solving the three-wave equations (Ch.~\ref{ch:3wave}).
As an application, I consider laser pulse compression mediated by the upper-hybrid wave (Ch.~\ref{ch:compression}), which provides a promising way of producing high-intensity short-wavelength pulses beyond the attainment of current methods.

In even stronger fields, plasma physics enters the relativistic-quantum regime. 
In Ch.~\ref{ch:vacuum}, I give a self-contained review of quantum field theory, a powerful tool not commonly used in plasma physics. 
By allowing for dynamical background fields, I extend scalar QED to a model for plasma physics using a new formalism based on the effective action, where the dispersion relation of linear waves can be computed using path integrals (Ch.~\ref{ch:action}).
When applied to unmagnetized plasmas (Ch.~\ref{ch:unmag}), the effective action formulation recovers the known dispersion relation in unmagnetized scalar-QED plasmas.
Moreover, the same formula, given by the effective action approach, also enables a convenient wave dispersion relation in magnetized scalar-QED plasma to be determined for the first time (Ch.~\ref{ch:mag}). Strongly magnetized plasma waves, modified by relativistic quantum effects, account for cyclotron absorptions in spectra of X-ray pulsars, and lead to observable corrections of Faraday rotation.
Finally, beyond the perturbative regime, I extends lattice QED to simulate relativistic quantum plasmas (Ch.~\ref{ch:simulation}).  Using real-time lattice simulations for laser-plasma interactions, the transition from wakefield acceleration to pair production is demonstrated for the first time when laser intensity exceeds the Schwinger limit. 

\bookmarksetupnext{level=part}
\begin{appendices}{}
	\renewcommand{\thesubsection}{\Alph{subsection}}
	
	\chapter{Properties of the cold forcing operator\label{ch:append:F}}

In this appendix, I will prove three nontrivial properties of the cold forcing operator $\mathbb{F}$, which is defined in Sec.~\ref{sec:fluid-1st:model:operatorF}. First, let us derive the formula for $\mathbb{F}$. Recall that the forcing operator is defined to solve the cold momentum equation of the form
\begin{equation}
\label{eq:vF}
\hat{\mathbf{v}}=\mathcal{E}+i\beta\hat{\mathbf{v}}\times\mathbf{b},
\end{equation}
such that $\hat{\mathbf{v}}$ is the image of $\mathcal{E}$ under the linear map $\hat{\mathbf{v}}=\mathbb{F}\mathcal{E}$. To find the formula for $\mathbb{F}$, what we need to do is to solve this vector equation. Taking inner product with $\mathbf{b}$ on both sides,
\begin{equation}
\nonumber
\hat{\mathbf{v}}\cdot\mathbf{b}=\mathcal{E}\cdot\mathbf{b}.
\end{equation}
Next, taking cross product with $\mathbf{b}$ on both sides, and using the above result,
\begin{eqnarray}
\nonumber
\hat{\mathbf{v}}\times\mathbf{b}&=&\mathcal{E}\times\mathbf{b}+i\beta[(\mathcal{E}\cdot\mathbf{b})\mathbf{b}- \hat{\mathbf{v}}].
\end{eqnarray}
Substituting this into Eq.~(\ref{eq:vF}), we immediately find
\begin{equation}
(1-\beta^2)\hat{\mathbf{v}}=\mathcal{E}+i\beta(\mathcal{E}\times\mathbf{b})-\beta^2(\mathcal{E}\cdot\mathbf{b})\mathbf{b},
\end{equation}
which expresses $\hat{\mathbf{v}}$ in terms of $\mathcal{E}$. After identifying $\gamma^2=1/(1-\beta^2)$, the formula Eq.~(\ref{eq:F}) can be read out from the above solution.

Second, let us prove the identity $\mathbb{F}^2=\mathbb{F}-\omega\partial\mathbb{F}/\partial\omega$. Apart from a straightforward calculation, this identity can also be proven using the following trick. Notice that the inverse operator satisfies $\hat{\mathbf{v}}=\mathbb{F}^{-1}\hat{\mathbf{v}}+i\beta\hat{\mathbf{v}}\times\mathbf{b}$. Since neither $\hat{\mathbf{v}}$ nor $\mathbf{b}$ depends on $\omega$, after taking $\partial/\partial\omega$ derivative on both sides, we immediately find
\begin{equation}
\omega\frac{\partial \mathbb{F}^{-1}}{\partial\omega}\mathbf{z}=i\beta\mathbf{z}\times\mathbf{b},
\end{equation}
where $\mathbf{z}\in\mathbb{C}^{3}$ is any complex vector. Next, taking derivative on both side of the identity $\mathbb{I}=\mathbb{F}^{-1}\mathbb{F}$, we have $0=\partial\mathbb{F}^{-1}/\partial\omega\mathbb{F} +\mathbb{F}^{-1}\partial\mathbb{F}/\partial\omega$. Acting this identity on vector $\mathbf{z}$, we have
\begin{eqnarray}
\nonumber
\omega\frac{\partial \mathbb{F}}{\partial\omega}\mathbf{z}&=&-\mathbb{F}\Big(\omega\frac{\partial \mathbb{F}^{-1}}{\partial\omega}\Big)\mathbb{F}\mathbf{z}\\
\nonumber
&=&-\mathbb{F}\big[i\beta\big(\mathbb{F}\mathbf{z}\big)\times\mathbf{b}\big]\\
&=&(\mathbb{F}-\mathbb{F}^2)\mathbf{z}.
\end{eqnarray}
To obtain the last equality, I have used the vector identity Eq.~(\ref{eq:Fvector}). Since the above relation holds for all $\mathbf{z}\in\mathbb{C}^{3}$, we have thus proven the identity Eq.~(\ref{eq:F2}).


Finally, we can use the same trick to prove the quadratic identity $(\beta_1-\beta_2)\mathbb{F}_1\mathbb{F}_2=\beta_1\mathbb{F}_1-\beta_2\mathbb{F}_2$. To avoid going through tedious algebra, instead of computing $\mathbb{F}_1\mathbb{F}_2$ directly, let us compute the following:
\begin{eqnarray}
\nonumber
\beta_2\mathbb{F}_1^{-1}\big(\mathbb{F}_2\mathbf{z}\big)&=&\beta_2\big[\mathbb{F}_2\mathbf{z} -i\beta_1\big(\mathbb{F}_2\mathbf{z}\big)\times\mathbf{b}\big]\\
\nonumber
&=&\beta_2\mathbb{F}_2\mathbf{z} -\beta_1\big[i\beta_2\big(\mathbb{F}_2\mathbf{z}\big)\times\mathbf{b}\big]\\
&=&(\beta_2-\beta_1)\mathbb{F}_2\mathbf{z}+\beta_1\mathbf{z},
\end{eqnarray}
where I have again used the vector identity Eq.~(\ref{eq:Fvector}). Now, acting the linear operator $\mathbb{F}_1$ on both sides, we immediately obtain $(\beta_1-\beta_2)\mathbb{F}_1\mathbb{F}_2\mathbf{z}=\beta_1\mathbb{F}_1\mathbf{z}-\beta_2\mathbb{F}_2\mathbf{z}$. Since this relation holds for all $\mathbf{z}\in\mathbb{C}^{3}$, we have thus proven the identity Eq.~(\ref{eq:F12}).
	\chapter{Resonances in magnetized cold electron-ion plasma\label{ch:append:resonance}}

\begin{figure}[t]
	\centering
	\includegraphics[angle=0,width=0.6\textwidth]{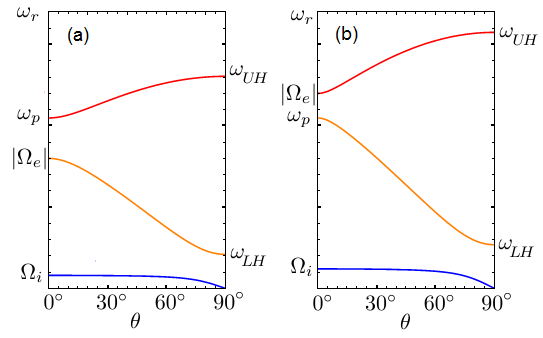}
	\caption[Resonance frequencies in magnetizd electron-ion plasma]{Resonance frequencies in electron-ion plasma with $m_i/m_e=10$. In over-dense plasma, e.g.~$|\Omega_e|/\omega_p=0.8$ (a), as $\theta$ increases from $0^\circ$ to $90^\circ$, the upper resonance (red) increases from $\omega_p$ to $\omega_{UH}$; the lower resonance (orange) decreases from $|\Omega_e|$ to $\omega_{LH}$; and the bottom resonance (blue) decreases from $\Omega_i$ to zero. In under-dense plasma, e.g.~$|\Omega_e|/\omega_p=1.2$ (b), as $\theta$ increases from $0^\circ$ to $90^\circ$, the upper resonance (red) increases from $|\Omega_e|$ to $\omega_{UH}$; the lower resonance (orange) decreases from $\omega_p$ to $\omega_{LH}$; and the bottom resonance (blue) decreases from $\Omega_i$ to zero. 
	}
	\label{fig:Resonance}
\end{figure}

The resonance frequencies are the finite asymptotic values of $\omega$ when $ck\rightarrow\infty$ in a cold plasma. Along a dispersion branch, as the frequency approaches the resonance frequencies from below, the refractive index $n^2\rightarrow+\infty$. Using Eq.~(\ref{eq:disp}), we can find $\omega_r$ by solving $A(\omega_r^2)=0$. In electron-ion plasma, this equation can be written explicitly as
\begin{eqnarray}
	\label{eq:resonance}
	0&=&\omega_r^6-\omega_r^4(\omega_p^2+\Omega_e^2+\Omega_i^2)-\omega_p^2\Omega_e^2\Omega_i^2\cos^2\theta\\
	\nonumber
	&+&\omega_r^2[\omega_p^2(\Omega_e^2+\Omega_i^2)\cos^2\theta-\omega_p^2\Omega_e\Omega_i\sin^2\theta+\Omega_e^2\Omega_i^2],
\end{eqnarray}
where I have removed the poles to convert the equation to a polynomial form. This cubic equation for $\omega_r^2$ has three positive roots (Fig.~\ref{fig:Resonance}), which can be ordered from large to small as the 
upper ($\omega_u$, red), lower ($\omega_l$, orange), and bottom ($\omega_b$, blue) resonances. In a given plasma with fixed plasma parameters, the resonance frequency $\omega_r$ is a function of the propagation angle $\theta$. Although expressions of the three roots can be found using the cubic formula, they are not more illuminating than obtaining numerically solutions of the polynomial equation. In what follows, I will only list the asymptotic expressions of the three resonance frequencies in the parallel and the perpendicular limits, in a form that is more accurate than what is given by \cite{Aleksandrov1978principles}. 

When $\sin\theta\rightarrow 0$, the resonance frequencies approaches $\omega_p, |\Omega_e|$, and $\Omega_i$. Keeping the next-order angular dependence, the three resonance frequencies can be approximated by
\begin{eqnarray}
	\label{eq:wu_para}
	\frac{\omega_r^2}{\omega_p^2}&\simeq&1-\frac{\Omega_e^2\sin^2\theta}{\Omega_e^2(2-\cos^2\theta)-\omega_p^2},\\
	\label{eq:wl_para}
	\frac{\omega_r^2}{\Omega_e^2}&\simeq&1-\frac{\omega_p^2\sin^2\theta}{\omega_p^2(2-\cos^2\theta)-\Omega_e^2},\\
	\label{eq:wb_para}
	\frac{\omega_r^2}{\Omega_i^2}&\simeq&1-\frac{\Omega_i}{|\Omega_e|}\tan^2\theta.
\end{eqnarray}
In the other limit, $\cos\theta\rightarrow0$, the resonance frequencies approach the upper-hybrid frequency $\omega_{UH}$, the lower hybrid frequency $\omega_{LH}$, and $0$. The upper $(+)$ and lower $(-)$ hybrid frequencies are
\begin{eqnarray}
\omega_\pm^2=\frac{1}{2}\big[(\Omega_e^2+\Omega_i^2+\omega_p^2)\pm\sqrt{(\Omega_e^2-\Omega_i^2)^2+2\omega_p^2(\Omega_e+\Omega_i)^2+\omega_p^4}\big].
\end{eqnarray} 
Since the ion mass is much larger than the electron mass, the gyro frequencies $\Omega_i\ll|\Omega_e|$. Therefore, to a good approximation, we have
\begin{eqnarray}
\label{eq:w_UH}
\omega_{UH}^2&\simeq&\omega_p^2+\Omega_e^2,\\
\label{eq:w_LH}
\omega_{LH}^2&\simeq&\frac{\omega_p^2}{\omega_{UH}^2}|\Omega_e|\Omega_i.
\end{eqnarray}
Away from the perpendicular angle, keeping the next-order angular dependence, the resonance frequencies can be approximated by
\begin{eqnarray}
	\label{eq:wu_perp}
	\frac{\omega_u^2}{\omega_{UH}^2}&\simeq&1- \frac{\omega_p^2\Omega_e^2\cos^2\theta}{(\omega_p^2+\Omega_e^2)^2+\omega_p^2\Omega_e^2\cos^2\theta},\\
	\label{eq:wl_perp}
	\frac{\omega_l^2}{\omega_{LH}^2}&\simeq&1+\frac{\Omega_e^2\cos^2\theta}{\Omega_e^2\cos^2\theta+|\Omega_e|\Omega_i(1+\cos^2\theta)},\\
	\label{eq:wb_perp}
	\frac{\omega_b^2}{\Omega_i^2}
	&\simeq&\frac{|\Omega_e|\cos^2\theta}{\Omega_i+|\Omega_e|\cos^2\theta}.
\end{eqnarray}
The above asymptotic expressions for resonance frequency $\omega_r$ are useful when we approximate the scattering strength and wave energy coefficients.

	\chapter{Secular-free identity of three-wave scattering tensor\label{ch:append:scattering}}

The secular-free identity $\mathbf{S}^{s}_{\mathbf{q},-\mathbf{q}}=\mathbf{0}$ of the scattering tensor is an important identity that justifies the multiscale solution \textit{a posteriori}. This identity is obvious using formula Eq.~(\ref{eq:S23}), together with the self-adjoint property of the forcing operator Eq.~(\ref{eq:Fadj}), because now $\omega_3=-\omega_2$ and $\omega_1=\omega_2+\omega_3=0$. In this appendix, I will give an alternative proof using a straightforward calculation. 

First, substituting Eqs.~(\ref{eq:L}), (\ref{eq:T}), and (\ref{eq:C}) into the formula for the quadratic response [Eq.~(\ref{eq:R})], we have
\begin{equation}
\label{eq:R_explicit}
\omega_{\mathbf{q}}\omega_{\mathbf{q}'}\mathbf{R}_{\mathbf{q},\mathbf{q}'}=\mathbb{F}_{\mathbf{q}+\mathbf{q}'} \big[F_{\mathbf{q}}\times\big(\mathbf{q}'\times\mathcal{E}_{\mathbf{q}'}\big)+F_{\mathbf{q}}\big(\mathbf{q}\cdot F_{\mathbf{q}'}\big)\big]+\Big(1+\frac{\omega_\mathbf{q}}{\omega_{\mathbf{q}'}}\Big)F_{\mathbf{q}}\big(\mathbf{q}'\cdot F_{\mathbf{q}'}\big),
\end{equation}
where $F_{\mathbf{q}}:=\mathbb{F}_{\mathbf{q}}\mathcal{E}_\mathbf{q}$, and I have suppressed the species index with the implied understanding that all terms are associated with the same species. Then, using notations (\ref{eq:notationz}) and (\ref{eq:notationa}), the secular quadratic responses
\begin{eqnarray}
\label{eq:Rpm}
-\omega_{\mathbf{q}}^2\mathbf{R}_{\mathbf{q},-\mathbf{q}}&=&\mathbb{F}_{\mathbf{0}} \big[F_{\mathbf{q}}\times\big(-\mathbf{q}\times\mathcal{E}_{\mathbf{q}}^*\big)+F_{\mathbf{q}}\big(\mathbf{q}\cdot F_{\mathbf{q}}^*\big)\big],\\
\label{eq:Rmp}
-\omega_{\mathbf{q}}^2\mathbf{R}_{-\mathbf{q},\mathbf{q}}&=&\mathbb{F}_{\mathbf{0}} \big[F_{\mathbf{q}}^*\times\big(\mathbf{q}\times\mathcal{E}_{\mathbf{q}}\big)+F_{\mathbf{q}}^*\big(-\mathbf{q}\cdot F_{\mathbf{q}}\big)\big],
\end{eqnarray}
where $\mathbb{F}_{\mathbf{0}}=\mathbb{F}(\omega=0)$ is the zero-frequency forcing operator. Since $\omega_{-\mathbf{q}}=-\omega_{\mathbf{q}}$, the last term on the RHS of Eq.~(\ref{eq:R_explicit}) does not contribute.

Next, let us find an expression for the zero-frequency forcing operator $\mathbb{F}_{\mathbf{0}}$. In magnetized plasmas, when $\omega\rightarrow 0$, the magnetization ratio $\beta\rightarrow\infty$. Therefore, the magnetization factor \mbox{$\gamma^2=1/(1-\beta^2)\rightarrow0$}, while the product $\gamma^2\beta^2\rightarrow-1$. Using formula for the forcing operator [Eq.~(\ref{eq:F})], we see the zero-frequency operator is simply the projection operator along the background magnetic field
\begin{equation}
\mathbb{F}_\mathbf{0}=\mathbf{b}\mathbf{b}.
\end{equation}
This is intuitive, because at zero frequency, charged particles stays along the same magnetic field line.  
In unmagnetized plasmas, there is no well-defined direction $\mathbf{b}$. Nevertheless, since the forcing operator is the identity operator, $\mathbf{R}_{\mathbf{q},-\mathbf{q}}+\mathbf{R}_{-\mathbf{q},\mathbf{q}}=\mathbf{0}$ is trivially satisfied in the unmagnetized case.

Now let us compute the secular scattering strength $\mathbf{S}_{\mathbf{q},-\mathbf{q}}$. Using the vector identity Eq.~(\ref{eq:Fvector}), we have $\mathcal{E}_{\mathbf{q}}=F_{\mathbf{q}}-i\beta F_{\mathbf{q}}\times\mathbf{b}$. Substituting this into Eqs.~(\ref{eq:Rpm}) and (\ref{eq:Rmp}), then the scattering strength [Eq.~(\ref{eq:S})] is proportional to
\begin{eqnarray}
\nonumber
-\omega_{\mathbf{q}}^2\big(\mathbf{R}_{\mathbf{q},-\mathbf{q}}+\mathbf{R}_{-\mathbf{q},\mathbf{q}}\big)&=& \mathbf{b}\mathbf{b}\big[-\mathbf{q}\big(F_{\mathbf{q}}\cdot\mathcal{E}_{\mathbf{q}}^*\big)+\mathcal{E}_{\mathbf{q}}^*\big(\mathbf{q}\cdot F_{\mathbf{q}}\big)+F_{\mathbf{q}}\big(\mathbf{q}\cdot F_{\mathbf{q}}^*\big)\\
\nonumber
&&\hspace{32pt} \mathbf{q}\big(F_{\mathbf{q}}^*\cdot\mathcal{E}_{\mathbf{q}}\big)-\mathcal{E}_{\mathbf{q}}\big(\mathbf{q}\cdot F_{\mathbf{q}}^*\big)-F_{\mathbf{q}}^*\big(\mathbf{q}\cdot F_{\mathbf{q}}\big)\big]\\
\nonumber
&=&i\beta\mathbf{b}\mathbf{b}\big\{\mathbf{q}\big[F_{\mathbf{q}}\cdot\big(F_{\mathbf{q}}^*\times\mathbf{b}\big)\big]-\big(\mathbf{q}\cdot F_{\mathbf{q}}\big)\big(F_{\mathbf{q}}^*\times\mathbf{b}\big) +\text{c.c.}\big\}\\
&=&0.
\end{eqnarray}
To obtain the last equality, notice that $F_{\mathbf{q}}\cdot\big(F_{\mathbf{q}}^*\times\mathbf{b}\big)$ is of the form $\mathbf{z}\cdot(\mathbf{z}^*\times\mathbf{b})=\mathbf{z}_i\epsilon_{ijk}\mathbf{z}_j^*b_k=-\mathbf{z}_j^*\epsilon_{jik}\mathbf{z}_ib_k=-\mathbf{z}^*\cdot(\mathbf{z}\times\mathbf{b})$. Since this term is purely imaginary, $F_{\mathbf{q}}\cdot\big(F_{\mathbf{q}}^*\times\mathbf{b}\big)+$c.c.$=0$ vanishes. As for the second term, since $F_{\mathbf{q}}^*\times\mathbf{b}$ is perpendicular to $\mathbf{b}$, its parallel projection along $\mathbf{b}\mathbf{b}$ is trivially zero. I have thus proven the secular-free identity.
	\chapter{Feynman Green's function\label{ch:append:GreenF}}

In this appendix, I calculate the Feynman Green's function by evaluating its integral representation [Eq.~(\ref{eq:GF})] in terms of Bessel functions \citep[Ch.~10]{NIST:DLMF}. First, integrating along the $p_0$ direction
\begin{eqnarray}
\label{eq:GFD}
\nonumber
G_{F}(x,x')&=&\int \frac{d^3\mathbf{p}}{(2\pi)^3}ie^{i\mathbf{p}\cdot(\mathbf{x}-\mathbf{x}')} \int \frac{d p_0}{2\pi}
\frac{e^{-i p_0(t-t')}}{p_0^2-\mathbf{p}^2-m^2+i\epsilon}\\
\nonumber
&=&\int \frac{d^3\mathbf{p}}{(2\pi)^3}ie^{i\mathbf{p}\cdot(\mathbf{x}-\mathbf{x}')} i\Big[-\frac{e^{-iE_{\mathbf{p}}(t-t')}}{2E_{\mathbf{p}}}\theta(t-t') + \frac{e^{-iE_{\mathbf{p}}(t'-t)}}{-2E_{\mathbf{p}}}\theta(t'-t)\Big]\\
&=&\theta(t-t')D(x-x')+\theta(t'-t)D(x'-x),
\end{eqnarray} 
where $\theta$ is the Heaviside step function. To obtain the second line, when $t>t'$, we can take closure of the integration contour in the lower half of the complex plane, whereby the pole at $p_0=E_{\mathbf{p}}$ contributes. Analogously, when $t'>t$, we can take closure in the upper half of the complex plane, whereby the pole at $p_0=-E_{\mathbf{p}}$ contributes. Here, $E_\mathbf{p}=\sqrt{\mathbf{p}^2+m^2}$ is again the positive energy associated with momentum $\mathbf{p}$. Since the Green's function is invariant under translational symmetry, it is natural to introduce the correlation function
\begin{equation}
D(x)=\int \frac{d^3\mathbf{p}}{(2\pi)^3}\frac{e^{-iE_{\mathbf{p}}t+i\mathbf{p}\cdot\mathbf{x}}}{2E_{\mathbf{p}}},
\end{equation}
so that both $D(x-x')$ and $D(x'-x)$, which appear on the last line of Eq.~(\ref{eq:GFD}), only depend on the separation between the two points.

Under Lorentz transform $x^\mu\rightarrow x^{'\mu}=\Lambda^{\mu}_{\nu}x^\nu$, the correlation function is transformed by the pullback $D(x)\rightarrow D'(x)=D(\Lambda^{-1}x)$. In particular, under boost in $\mathbf{b}$ direction, the coordinates are transformed by
\begin{eqnarray}
\label{eq:boost_t}
t'&=&\gamma(t-\beta\mathbf{b}\cdot\mathbf{x}), \\
\label{eq:boost_x}
\mathbf{x}'&=&\gamma(\mathbf{x}-\beta\mathbf{b}t)+(\gamma-1)(\mathbf{x}\times\mathbf{b})\times\mathbf{b},
\end{eqnarray}
where $\mathbf{b}$ is a unit spatial vector, $\beta=v/c<1$ is the boost speed, and $\gamma=1/\sqrt{1-\beta^2}$ is the relativistic factor. In order to calculate $D(x)$ for a general separation $x$, we can first calculate expressions for timelike and spacelike separations, and then boost back to the general reference frame.

When $x$ is timelike ($t^2>\mathbf{x}^2$), there exists an inertial frame in which the separation $x'^\mu=(t',\mathbf{0})$ is purely time. To transform to this reference frame, we can boost with $\beta\mathbf{b}=\mathbf{x}/t$, with the relativistic factor $\gamma=|t|/\sqrt{t^2-\mathbf{x}^2}$. Then, in the new reference frame, the spatial separation $\mathbf{x}'=\mathbf{0}$ vanishes, and the time separation $t'=\text{sgn}(t)\sqrt{t^2-\mathbf{x}^2}$ preserves the sign of the original time separation. 
In the boosted reference frame, integrating using the spherical coordinate in the momentum space, we have
\begin{eqnarray}
\label{eq:GF_Dt}
\nonumber
D'(t')&=&\frac{1}{(2\pi)^2}\int_0^{+\infty}dp\frac{p^2}{\sqrt{p^2+m^2}}e^{-it'\sqrt{p^2+m^2}}\\
\nonumber
&=&\frac{1}{(2\pi)^2}\int_m^{+\infty}dE\sqrt{E^2-m^2}e^{-iEt'}\\
&=&\frac{im}{8\pi t'}H_{1}^{(2)}(mt'),
\end{eqnarray}
where 
$H_\nu^{(2)}(z)$ is the Hankel function of the second kind. 
When the temporal separation $t'\rightarrow\infty$, using the asymptotic expressions for the Hankel function, the correlation function $D'(t')\simeq -\frac{m^2}{2} (2\pi mt')^{-3/2} e^{-i(mt'-\pi/4)}$. We see the correlation between two points decay with their time separation as $\propto (mt')^{-3/2}e^{-imt'}$, with a phase related to wave propagation from one point to the other.

\begin{figure}[h]
	\centering
	\includegraphics[angle=0,width=0.7\textwidth]{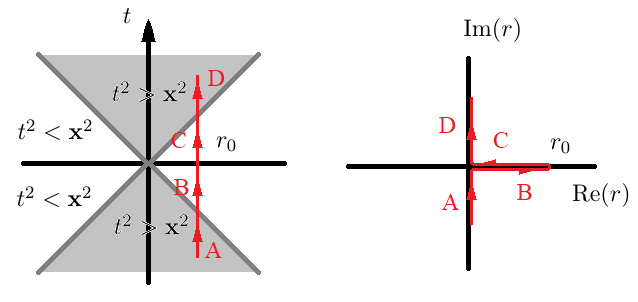}
	\caption[Analytic continuation of Feynman Green's function]{The formulas for Feynman Green's function inside [Eq.~(\ref{eq:GF_Dt})] and outside [Eq.~(\ref{eq:GF_Dx})] the light cone are connected by analytic continuation. When moving along a trajectory (red lines), which first exits and then reenter the light cone, the argument of the modified Bessel function $r=\sqrt{\mathbf{x}^2-t^2}$ first moves along the negative imaginary axis towards the origin, then moves along the positive real axis towards the maximum $r_0$, then returns to the origin along the Re$(r)$ axis, and finally moves along the Im$(r)$ axis away from the origin.}
	\label{fig:Analytic_Continuation}
\end{figure}

When $x$ is spacelike ($\mathbf{x}^2>t^2$), there exists an inertial frame in which the separation $x'^\mu=(0,\mathbf{x}')$ is purely space. To transform to this reference frame, we can boost with $\beta\mathbf{b}=t\hat{\mathbf{x}}/|\mathbf{x}|$, where $\hat{\mathbf{x}}$ is the unit vector in the $\mathbf{x}$ direction and $|\mathbf{x}|$ is the norm of the spatial separation. With such a boost, the relativistic factor $\gamma=|\mathbf{x}|/\sqrt{\mathbf{x}^2-t^2}$, the time separation $t'=0$, and the spatial separation $\mathbf{x}'=\hat{\mathbf{x}}\sqrt{\mathbf{x}^2-t^2}$. 
In the boosted reference frame, integrating using the spherical coordinate and denoting $r'=|\mathbf{x}'|$, we have
\begin{eqnarray}
\label{eq:GF_Dx}
\nonumber
D'(\mathbf{x}')&=&\frac{1}{2(2\pi)^2}\int_0^{+\infty}dp\frac{p^2}{\sqrt{p^2+m^2}}\int_0^{\pi}d\theta \sin\theta e^{ipr'\cos\theta}\\
\nonumber
&=&\frac{1}{(2\pi)^2}\int_0^{+\infty}dp\frac{p\sin(pr')}{r'\sqrt{p^2+m^2}}\\
&=&\frac{m}{4\pi^2 r'}K_{1}(mr'),
\end{eqnarray}
where $K_\nu(z)$ is the modified Bessel functions of the second kind. When the spatial separation $r'\rightarrow+\infty$, using the asymptotic expressions for the modified Bessel functions, the correlation function $D'(r')\simeq \frac{m^2}{2} (2\pi mr')^{-3/2} e^{-mr'}$. We see the correlation between two spatially separated points $\propto (mr')^{-3/2}e^{-mr'}$ is exponentially suppressed. Using analytical continuation (Fig.~\ref{fig:Analytic_Continuation}), the modified Bessel function is related to the Hankel function by $K_1(iz)=-\frac{\pi}{2}H_1^{(2)}(z)$. Hence, it is easy to verify the Eqs.~(\ref{eq:GF_Dx}) and (\ref{eq:GF_Dt}) are in fact the same formula

	\chapter{Energy, momentum, and charge operators of free $\phi$ field\label{ch:append:HPQ}}

In this appendix, I compute three quantum operators in terms of the creation and annililation operators. First, the energy operator, namely the Hamiltonian [Eq.~(\ref{eq:H0})], can be expressed using the spectral expansion of $\phi$ [Eq.~(\ref{eq:phi0_opt})], which gives
\begin{eqnarray}
\label{eq:energy_opt}
\nonumber
H_0&=& \int d^3\mathbf{x} \Big[ \pi \pi^\dagger + \nabla\phi^\dagger \cdot\nabla\phi  +m^2\phi^\dagger \phi  \Big]\\
\nonumber
&=&\int \frac{d^3\mathbf{x}}{2\sqrt{E_\mathbf{q} E_\mathbf{p}}} \frac{d^3\mathbf{q}}{(2\pi)^3} \frac{d^3\mathbf{p}}{(2\pi)^3} \Big[ \Big(E_\mathbf{q} E_\mathbf{p}+\mathbf{p}\cdot\mathbf{q}+m^2\Big)\Big(a_\mathbf{q}^\dagger a_\mathbf{p}e^{i(q-p)x} +b_\mathbf{q}b_\mathbf{p}^\dagger e^{-i(q-p)x} \Big) \\
\nonumber
&&\phantom{\int \frac{d^3\mathbf{x}}{\sqrt{E_\mathbf{q} E_\mathbf{p}}} \frac{d^3\mathbf{q}}{(2\pi)^3} \frac{d^3\mathbf{p}}{(2\pi)^3}}-\Big(E_\mathbf{q} E_\mathbf{p}+\mathbf{p}\cdot\mathbf{q}-m^2\Big) \Big(a_\mathbf{q}^\dagger b_\mathbf{p}^\dagger e^{i(q+p)x} +b_\mathbf{q}a_\mathbf{p} e^{-i(p+q)x} \Big)\Big]\\
&=&\int\frac{d^3\mathbf{p}}{(2\pi)^3} E_\mathbf{p} (a^\dagger_{\mathbf{p}}a_{\mathbf{p}}+b_{\mathbf{p}}b^\dagger_{\mathbf{p}}).
\end{eqnarray}
Here I have used the fact that $E_{-\mathbf{p}}=E_{\mathbf{p}}=\sqrt{\mathbf{p}^2+m^2}$, which leads to the simplifications on the last line, after carrying out the $\mathbf{x}$ and $\mathbf{q}$ integrals. Using the commutation relation Eq.~(\ref{eq:ab}), we can write $b_{\mathbf{p}}b^\dagger_{\mathbf{p}}=b^\dagger_{\mathbf{p}}b_{\mathbf{p}} +(2\pi)^2 \delta^{(3)}(\mathbf{0})$. Then, the Hamiltonian can be written in terms of the number operators $a^\dagger_{\mathbf{p}}a_{\mathbf{p}}$ and $b^\dagger_{\mathbf{p}}b_{\mathbf{p}}$ after a shift by an infinity. The infinity comes from the ground state energy of particles of type $b$, namely the antiparticles. In the Dirac-sea picture, the vacuum is filled up by antiparticles, and thereof contains infinite energy, which can be removed by redefining the origin on the energy axis.

Next, let us compute the momentum operator. The total momentum $\mathbf{P}$ is the spatial integral of the momentum density $\mathbf{P}^i=\int d^3\mathbf{x} \mathcal{P}^i$, where the momentum density $\mathcal{P}^i$ is the $\mathcal{T}^{0i}$ component of the stress energy tensor [Eq.~(\ref{eq:stress-energy})]. Using the metric tensor $g_{\mu\nu}$ to lower the index of the spatial derivative, substituting in the spectral expansion of $\phi$, we have
\begin{eqnarray}
\label{eq:momentum_opt}
\nonumber
\mathbf{P}&=&\int d^3\mathbf{x} \big(-\dot{\phi}\nabla\phi^\dagger-\dot{\phi}^\dagger\nabla\phi\big) \\
\nonumber
&=&\int \frac{d^3\mathbf{x}}{2\sqrt{E_\mathbf{q} E_\mathbf{p}}} \frac{d^3\mathbf{q}}{(2\pi)^3} \frac{d^3\mathbf{p}}{(2\pi)^3} \Big[ 
-\Big(\mathbf{q}E_\mathbf{p}+\mathbf{p}E_\mathbf{q}\Big) \Big(a_\mathbf{p}^\dagger b_\mathbf{q}^\dagger e^{i(p+q)x} + b_\mathbf{p}a_\mathbf{q}e^{-i(p+q)x} \Big) \\
\nonumber
&&\hspace{70pt}+\mathbf{p}E_\mathbf{q}\Big(a_\mathbf{p}^\dagger a_\mathbf{q} +b_\mathbf{q}b_\mathbf{p}^\dagger  \Big)e^{i(p-q)x}+ \mathbf{p}E_\mathbf{q}\Big(b_\mathbf{p} b_\mathbf{q}^\dagger +a_\mathbf{q}^\dagger a_\mathbf{p}  \Big)e^{i(q-p)x} \Big]\\
&=&\int\frac{d^3\mathbf{p}}{(2\pi)^3} \mathbf{p} (a^\dagger_{\mathbf{p}}a_{\mathbf{p}}+b_{\mathbf{p}}b^\dagger_{\mathbf{p}}),
\end{eqnarray}
where I have used $E_{-\mathbf{p}}=E_{\mathbf{p}}$ to cancel the cross terms, where $\mathbf{q}=-\mathbf{p}$ is enforced by the delta function after the $\mathbf{x}$ integral.
Using the commutation relation, we can replace  $b_{\mathbf{p}}b^\dagger_{\mathbf{p}}$ by the number operator $b^\dagger_{\mathbf{p}}b_{\mathbf{p}}$, up to a shift of the total momentum by infinity.

Finally, let us compute the charge operator. The total charge is the spatial integral of the charge density $Q=\int d^3\mathbf{x} \rho$, where the charge density $\rho=J^{0}$ is the time component of the 4-current density $J^\mu$ [Eq.~(\ref{eq:Jmu})]. In the absence of the gauge field, the covariant derivatives $D^\mu$ is simply the partial derivative $\partial^\mu$. Using the spectral expansion of $\phi$,
\begin{eqnarray}
\label{eq:charge_opt}
\nonumber
Q&=&\int d^3\mathbf{x} \;\frac{e}{i}\;\big[\phi^\dagger\dot{\phi}-\dot{\phi}^\dagger\phi\big] \\
\nonumber
&=&e\int \frac{d^3\mathbf{x}}{2\sqrt{E_\mathbf{q} E_\mathbf{p}}} \frac{d^3\mathbf{q}}{(2\pi)^3} \frac{d^3\mathbf{p}}{(2\pi)^3} \Big[ 
\Big(E_\mathbf{p}-E_\mathbf{q}\Big) \Big(a_\mathbf{q}^\dagger b_\mathbf{p}^\dagger e^{i(p+q)x} + b_\mathbf{p}a_\mathbf{q}e^{-i(p+q)x} \Big) \\
\nonumber
&&\hspace{50pt}+E_\mathbf{p}\Big(-a_\mathbf{q}^\dagger a_\mathbf{p} +b_\mathbf{p}b_\mathbf{q}^\dagger  \Big)e^{i(q-p)x}+ E_\mathbf{p}\Big(b_\mathbf{q} b_\mathbf{p}^\dagger -a_\mathbf{p}^\dagger a_\mathbf{q}  \Big)e^{i(p-q)x} \Big]\\
&=&e\int\frac{d^3\mathbf{p}}{(2\pi)^3} (-a^\dagger_{\mathbf{p}}a_{\mathbf{p}}+b_{\mathbf{p}}b^\dagger_{\mathbf{p}}),
\end{eqnarray}
where I have again used the fact that $E_{-\mathbf{p}}=E_{\mathbf{p}}$ to cancel the cross terms. 
Using the commutation relation, we can replace  $b_{\mathbf{p}}b^\dagger_{\mathbf{p}}$ by the number operator $b^\dagger_{\mathbf{p}}b_{\mathbf{p}}$, up to a shift of the total charge by infinity.
	\chapter{Separate conservation of plasma and vacuum currents\label{ch:append:Ward}}

In this appendix, I prove by direct calculation that the plasma and vacuum currents in the 1-loop effective action are separately conserved. First, to prove the conservation of plasma current, we need to show the configuration space Ward--Takahashi identity Eq.~(\ref{bkConservation}). Since $\Sigma_{2,\text{bk}}^{\mu\nu}(x,x')=\Sigma_{2,\text{bk}}^{\nu\mu}(x',x)$, it is sufficient to show $\partial_{\mu}\Sigma_{2,\text{bk}}^{\mu\nu}(x,x')=0$. To directly compute the LHS, we can use Eq.~(\ref{EOM}) for the background field $\phi_0$ and Eq.~(\ref{eq:SDGreen}) for the Green's function $G$. The derivative of the plasma polarization tensor contains
\begin{eqnarray}
\nonumber
\partial_{\mu}\big[\phi_0^*\bar{D}^{\mu}-(\bar{D}^{\mu}\phi_{0})^*\big] G
&=&(\partial_{\mu}\phi_0^*)(\bar{D}^{\mu}G) -G(\partial_{\mu}\bar{D}^{\mu}\phi_{0})^* \\
\nonumber
&&-(\bar{D}^{\mu}\phi_{0})^*(\partial_{\mu} G) + \phi_0^*(\partial_{\mu}\bar{D}^{\mu}G) \\
\nonumber
&=&\phi_0^*(\bar{D}_{\mu}\bar{D}^{\mu}G)-G(\bar{D}_{\mu}\bar{D}^{\mu}\phi_0)^*\\
\nonumber
&=&-i\phi_0^*\delta,
\end{eqnarray}
where the arguments of the functions are omitted for brevity. The second equality is obtained by completing partial derivatives to covariant derivatives. 
To compute partial derivatives of polarization tensors, we need the following properties of the $\delta$ function, which can be shown from its integral definition
\begin{eqnarray}
\label{eq:d1delta}
f(x)\delta'(x)&=&-\delta(x)f'(x),
\end{eqnarray}
where $f'(x)$ denotes the derivative of $f(x)$. 
To see how to take derivatives of the delta function when both $x$ and $x'$ are present, we can change variables to $r=x-x'$ and $R=\frac{1}{2}(x+x')$. Then, the partial derivatives $\partial=\partial_r+\frac{1}{2}\partial_R$ and  $\partial'=-\partial_r+\frac{1}{2}\partial_R$, and the inverse relations are $\partial_R=\partial+\partial'$ and $\partial_r=\frac{1}{2}(\partial-\partial')$. The derivative of delta function
\begin{eqnarray}
\nonumber
f(x,x')\partial\delta(x-x')&=&f(r,R)\Big(\partial_r+\frac{1}{2}\partial_R\Big)\delta(r)\\
\nonumber
&=&-\delta(r)\partial_r f(r,R)\\
\nonumber
&=&\frac{1}{2}\delta(x-x')(\partial'-\partial)f(x,x').
\end{eqnarray}
When $x$ is a vector, different components of $x$ are independent, and the above identity can be analogously written for partial derivatives
\begin{eqnarray}
\label{eq:d2delta}
f(x,x')\partial_{\mu}\delta(x-x')&=&\frac{1}{2}\delta(x-x')(\partial^{'}_{\mu}-\partial_{\mu})f(x,x'),
\end{eqnarray} 
where $f$ is an arbitrary differentiable function. With the above property of the delta function, the partial derivative of the background polarization tensor [Eq.~(\ref{bkPol})] becomes
\begin{eqnarray}
\label{eq:d_bkPol}
\nonumber
\partial_{\mu}\Pi^{\mu\nu}_{2,\text{bk}}&=&-ie^2\phi_0^* \big[\phi_0'\bar{D}^{'*\nu}-(\bar{D}^{'\nu}\phi_{0}')\big]\delta-\text{c.c.}\\
\nonumber
&=&-ie^2\phi_0^* \big[\phi_0'\partial^{'\nu}-(\partial^{'\nu}\phi_{0}')\big]\delta-\text{c.c.}\\
\nonumber
&=&-ie^2\delta \Big[\frac{1}{2}\big(\partial^\nu-\partial^{'\nu}\big)\phi_0^*\phi_0' -\phi_0^*(\partial^{'\nu}\phi_{0}')\Big]-\text{c.c.}\\
\nonumber
&=&ie^2\delta\big(\phi_0\partial^{'\nu}\phi_{0}^{'*} + \phi_{0}^{'*}\partial^{\nu}\phi_{0}\big)\\
&=&2ie^2\partial^\nu\big(\phi_0 \phi_0^{'*}\delta \big).
\end{eqnarray}
On the second line, covariant derivatives become partial derivatives because the gauge part is purely real and is thereof canceled when subtracting the complex conjugate. On the third and the last line, I have used Eq.~(\ref{eq:d2delta}) to compute derivatives of the delta function. Due to the delta function, we have $f(x)g(x')\delta(x-x')= f(x')g(x)\delta(x-x')$, which is used on the fourth line when taking complex conjugation. From the above result Eq.~(\ref{eq:d_bkPol}) and the formula for the plasma response tensor [Eq.~(\ref{bk})], it is obvious that the Ward--Takahashi identity [Eq.~(\ref{bkConservation})] is satisfied, so is the conservation of the plasma current.

Next, to prove the conservation of vacuum current, we need to show the configuration space Ward--Takahashi identity Eq.~(\ref{vacConservation}). Again, due to symmetry of the response tensor, it is sufficient to show $\partial_{\mu}\Sigma_{2,\text{vac}}^{\mu\nu}(x,x')=0$. The LHS contains the derivative of the vacuum polarization tensor 
\begin{eqnarray}
\label{eq:d_vacPol}
\nonumber
\partial_{\mu}\Pi^{\mu\nu}_{2,\text{vac}}
&=&e^2\big[(\partial_{\mu}G')(\bar{D}^{\mu}\bar{D}^{'\nu*}G)+G'(\partial_{\mu}\bar{D}^{\mu}\bar{D}^{'\nu*}G)\\
\nonumber
&&-(\partial_{\mu}\bar{D}^{\mu*}G')(\bar{D}^{'\nu*}G)- (\bar{D}^{\mu*}G')(\partial_{\mu}\bar{D}^{'\nu*}G)\big] +\text{c.c.}\\
\nonumber
&=&e^2\big[(\bar{D}^*_{\mu}G')(\bar{D}^{\mu}\bar{D}^{'\nu*}G)+G'(\bar{D}_{\mu}\bar{D}^{\mu}\bar{D}^{'\nu*}G)\\
\nonumber
&&-(\bar{D}^*_{\mu}\bar{D}^{\mu*}G')(\bar{D}^{'\nu*}G)- (\bar{D}^{\mu*}G')(\bar{D}_{\mu}\bar{D}^{'\nu*}G)\big] +\text{c.c.}\\
\nonumber
&=&ie^2\big[\delta (\bar{D}^{'\nu*}G)-G' (\bar{D}^{'\nu*}\delta)\big]+\text{c.c.}\\
\nonumber
&=&ie^2\big[\delta (\partial^{'\nu}G)-G' (\partial^{'\nu}\delta)\big]+\text{c.c.}\\
\nonumber
&=&ie^2\delta\partial^{\nu}(G+G')\\
&=&2ie^2\partial^{\nu}(G\delta).
\end{eqnarray}
The first equality directly follows from Eq.~(\ref{vacPol}), where the partial derivatives can be completed to covariant derivatives to obtain the second equality. Using Eq.~(\ref{eq:SDGreen}) for the Green's function $G$, we can then obtain the third equality, in which the gauge terms cancel in the presence of the delta function. The last two lines are obtained using Eq.~(\ref{eq:d2delta}) to take derivatives of the delta function. Using the above result Eq.~(\ref{eq:d_vacPol}) and the formula for the vacuum response tensor [Eq.~(\ref{vac})], it is obvious that the Ward--Takahashi identity [Eq.~(\ref{vacConservation})] is satisfied. Consequently, the conservation of vacuum current is also satisfied.

\end{appendices}

\bookmarksetup{startatroot}
\singlespacing
\bibliographystyle{authoryear}

\cleardoublepage
\ifdefined\phantomsection
  \phantomsection  
\else
\fi
\addcontentsline{toc}{chapter}{Bibliography}

\bibliography{thesis}

\end{document}